\documentclass[
headinclude=true, % treat headings as part of the text rather than the margin
headsepline=true,
twoside=true,
BCOR=10mm, % binding correction cf KOMA documentation p30
    fontsize=10 pt,
paper=17cm:24cm, % paper=B5, % cf KOMA documentation p45
pagesize, % cf KOMA documentation p46
DIV=14, %  DIV=calc gives DIV=13 for 17x24cm, BCOR=5mm and 11pt (see top of log file) % cf KOMA documentation p31
headings=normal, % cf KOMA documentation p93
appendixprefix=true,
draft=false,
numbers=noendperiod, % no dot after section (etc) number
toc=bibliography, % Bibliography appears in Table of Contents (without a number)
parskip=false, %no inline spacing between paragraphs
captions=bottombeside,
version=last % Use latest version of the KOMA-Script
]{scrbook}
 
\usepackage{ifdraft}
\ifdraft{
	\usepackage[notref,notcite]{showkeys}
	\usepackage[obeyspaces,hyphens,spaces]{url}
	\renewcommand*{\showkeyslabelformat}[1]{%
		\fbox{\parbox{1.15cm}{\normalfont\small\ttfamily\url{#1}}}}
}{}

\addtokomafont{caption}{\small}

\renewenvironment{thebibliography}[1]{
  \begin{oldthebibliography}{#1}
    \setlength{\itemsep}{.3em}
    \setlength{\parskip}{0em}
}
{
  \end{oldthebibliography}
}

\makeatletter
\DeclareRobustCommand*{\bfseries}{%
   \not@math@alphabet\bfseries\mathbf
   \fontseries\bfdefault\selectfont
   \boldmath
}
\makeatother

\usepackage{tabularx}

\makeatletter
\renewcommand{\@chapapp}{}% Not necessary...

\makeatother

\usepackage{subfiles}
\usepackage[dutch,american]{babel}
\usepackage[T1]{fontenc}
\usepackage{lmodern, microtype}
\usepackage{wrapfig}
\setcapindent{0pt}
\usepackage[labelfont=bf]{caption}

\usepackage{mathrsfs}
\usepackage[hang, flushmargin]{footmisc}
\usepackage[pdftex]{hyperref}
\usepackage{graphicx}
\usepackage{amssymb}
\usepackage{amsmath}
\usepackage{verbatim}
\usepackage{bm}
\usepackage[separate-uncertainty=true,multi-part-units=single]{siunitx}
\usepackage{cancel}
\usepackage{cite}
\usepackage{makecell}
\usepackage{booktabs}
\usepackage{arydshln}
\usepackage{braket}
\usepackage{multirow}

\mathchardef\mhyphen="2D
\newcommand{\onlyinsubfile}[1]{#1}
\newcommand{\notinsubfile}[1]{}

\usepackage{footmisc} 
\usepackage{appendix}
\usepackage{chngcntr}
\usepackage{etoolbox}
\AtBeginEnvironment{subappendices}{
\section*{Appendices}
\counterwithin{figure}{chapter}
\counterwithin{table}{chapter}
}

\def\blfootnote{\xdef\@thefnmark{}\@footnotetext}
\long\def\symbolfootnote[#1]#2{\begingroup%
\def\thefootnote{\fnsymbol{footnote}}\footnote[#1]{#2}\endgroup}

\usepackage[dvipsnames,xcdraw]{xcolor}

\DeclareOldFontCommand{\rm}{\normalfont\rmfamily}{\mathrm}
\DeclareOldFontCommand{\sf}{\normalfont\sffamily}{\mathsf}
\DeclareOldFontCommand{\tt}{\normalfont\ttfamily}{\mathtt}
\DeclareOldFontCommand{\bf}{\normalfont\bfseries}{\mathbf}
\DeclareOldFontCommand{\it}{\normalfont\itshape}{\mathit}
\DeclareOldFontCommand{\sl}{\normalfont\slshape}{\@nomath\sl}
\DeclareOldFontCommand{\sc}{\normalfont\scshape}{\@nomath\sc}

\newcommand{\beq}{\begin{equation}}  \newcommand{\eeq}{\end{equation}}
\newcommand{\bal}{\begin{aligned}}   \newcommand{\eal}{\end{aligned}}
\newcommand{\bea}{\begin{eqnarray}}  \newcommand{\eea}{\end{eqnarray}}

\newcommand{\bmat}{\left(\begin{array}}
\newcommand{\emat}{\end{array}\right)}

\newcommand{\bbC}{\mathbb{C}}
\newcommand{\bbR}{\mathbb{R}}

\newcommand{\nn}{\nonumber}

\newcommand{\cO}{\mathcal{O}}

\newcommand{\cN}{\mathcal{N}}

\usepackage{cancel}

\newcommand{\be}{\begin{equation}}
\newcommand{\ee}{\end{equation}}

\newcommand{\half}{\frac{1}{2}}

\newcommand{\slt}{\mathfrak{sl}(2)}

\newcommand{\bbZ}{\mathbb{Z}}

\usepackage{graphbox}
\usepackage{bbm}

\newcommand{\dd}{\mathrm{d}}

\DeclareMathOperator{\ad}{ad}
\newcommand*{\Scale}[2][4]{\scalebox{#1}{$#2$}}
  
\usepackage{ytableau}

\usepackage{dsfont}

\usepackage{tensor}
\usepackage{amsthm}
\usepackage{quiver}
\usepackage{tikz-cd}
\usetikzlibrary{babel}
\usepackage{mathtools}
\usepackage{adjustbox}
\usepackage{epigraph}

\usepackage[margin=0pt,font=small,labelfont=normalfont,skip=22pt]{subcaption}

\DeclareMathOperator\arctanh{arctanh}
\DeclareMathOperator{\arccot}{arccot}

\newtheorem{lem}{Lemma}[section]
\newtheorem{cor}{Corollary}[section]
\newtheorem{thm}{Theorem}[section]

\newtheorem{conjecture}{Conjecture}[section]

\usepackage{enumerate}
\usepackage{tkz-euclide}
\usepackage{pifont}
\usepackage[many]{tcolorbox}

\tcbset {
	base/.style={
		arc=0mm, 
		bottomtitle=0.5mm,
		boxrule=0mm,
		colbacktitle=black!10!white, 
		coltitle=black, 
		fonttitle=\bfseries, 
		left=2.5mm,
		leftrule=1mm,
		right=3.5mm,
		title={#1},
		toptitle=0.75mm,
		breakable
	}
}

\definecolor{brandblue}{rgb}{0.34, 0.7, 1}
\definecolor{carmine}{rgb}{0.59, 0.0, 0.09}

\newtcolorbox{redbox}[1]{
	colframe=carmine, 
	base={#1}
}

\newtcolorbox{mainbox}[1]{
	colframe=brandred, 
	base={#1}
}

\newtcolorbox{subbox}[1]{
	colframe=black!30!white,
	base={#1}
}

\begin{document}

\renewcommand{\onlyinsubfile}[1]{}
\renewcommand{\notinsubfile}[1]{#1}

\thispagestyle{empty}
\vspace*{4cm}

{\centering
{\huge \textbf{Asymptotic Hodge Theory in String Compactifications and Integrable Systems}}\vspace{.2cm} \\
%{\large } \vspace{1.45cm} \\
\vspace{1.45cm}
{\LARGE Jeroen Monnee}\\
}
\newpage~\thispagestyle{empty}
\vfill
\noindent 
PhD thesis, Utrecht University, July 2024
\vspace{1cm}

\noindent ISBN: 978-94-6510-049-4
\vspace{1cm}

\noindent DOI: 10.33540/2451
\vspace{1cm}

\noindent \textbf{Cover design by:} Vladimir Barvinko
\vspace{0.5cm}

\noindent\textbf{About the cover:} The illustration depicts an artistic impression of the landscape of string theory. The background features a rugged, mountainous terrain that symbolizes the wild and complicated profile of the scalar potential that arises in string compactifications. The latter is intricately related to the mathematical properties of the internal compactified dimensions, here depicted by a complicated geometric shape floating in the sky -- a so-called Calabi--Yau manifold. Far in the distance, the landscape flattens out and becomes much simpler to describe, in accordance with deep mathematical theorems in asymptotic Hodge theory. 

\newpage

\thispagestyle{empty}
\vspace*{0.2cm}

\begin{center}
{\huge \textbf{Asymptotic Hodge Theory in String Compactifications and Integrable Systems}}\vspace{.2cm} \\
\vspace{1.45cm}

{\huge Asymptotische Hodge-Theorie in Snaarcompactificaties en Integreerbare Systemen}\vspace{.2cm}  \\

(met een samenvatting in het Nederlands)\vspace{2.2cm}\\
 
{\LARGE Proefschrift}\vspace{0.8cm}\\
\end{center}

{\large{{ \noindent ter verkrijging van de graad van doctor aan de Universiteit Utrecht op gezag van de rector magnificus, prof.~dr.~H.R.B.M.~Kummeling, ingevolge het besluit van het College voor Promoties in het openbaar te verdedigen op}}}\vspace{0.4cm}\\
{\centering
{\large dinsdag 27 augustus 2024 des middags te 2.15 uur}\vspace{0.6cm}\\
{\large door}\vspace{0.6cm}\\
{\LARGE Jeroen Monnee}\vspace{.6cm}\\
{\large geboren op 18 oktober 1997\\ te Eindhoven}\par}

{
 \newpage \thispagestyle{empty}
 {\raggedright
{\Large \textbf{Promotoren:}}\\
{\large Prof.~dr.~T.W.~Grimm}\\
{\large Prof.~dr.~S.J.G.~Vandoren}}

\vspace{2cm}
\noindent
{\Large \textbf{Beoordelingscommissie:}}\\
{\large Dr.~L.~Anderson}\\
{\large Prof.~dr.~R.A.~Duine}\\ 
{\large Dr.~U.~Gursoy}\\
{\large Dr.~D.~Schuricht}\\
{\large Prof.~dr.~D.~Thompson}\\
\vfill
%\newpage

%\thispagestyle{empty}
%\vspace*{5cm}

%{\centering
%{\Large \textit{To xxx,}}\vspace{.25cm} \\ 
%{\Large \textit{I wish xxx}}\vspace{1.15cm}\\
%}
%\newpage~\thispagestyle{empty}
%\

\frontmatter
\chapter*{Publications}
\textbf{Part I} of this thesis provides a general introduction as well as a review of string compactifications in order to set the stage for the remainder of the thesis. 
\\

\noindent\textbf{Part II} of this thesis gives a review of asymptotic Hodge theory. It contains both mathematical details regarding the proofs of the main theorems, as well as explicit examples and concrete computational algorithms to compute physical observables in the context of low-energy effective theories coming from string compactifications. It is based, in part, on the following publication:
\begin{itemize}
	\item[\cite{Grimm:2021ikg}] Thomas W. Grimm, Jeroen Monnee, Damian van de Heisteeg: \emph{Bulk Reconstruction in Moduli Space Holography}, \textbf{JHEP 05 (2022) 010}, \\ \href{https://arxiv.org/abs/2103.12746}{\textbf{[arXiv: 2103.12746]}}
\end{itemize}
and additionally draws upon parts of the works \cite{Grimm:2023lrf,Grimm:2021idu,Grimm:2022ajb} listed below. This part of the thesis also contains additional results which have not appeared elsewhere.
\\

\noindent\textbf{Part III} of this thesis is devoted to the application of the mathematical machinery described in Part II to the F-theory flux landscape. It is based on the work:
\begin{itemize}
	\item[\cite{Grimm:2023lrf}] Thomas W. Grimm, Jeroen Monnee: \emph{Finiteness Theorems and Counting Conjectures for the Flux Landscape}, \href{https://arxiv.org/abs/2311.09295}{\textbf{[arXiv: 2311.09295]}}
\end{itemize}
\noindent\textbf{Part IV} of this thesis is concerned with a reformulation of some of the results of Part II in the language of non-linear sigma-models and applications in certain two-dimensional integrable field theories. It is based on the publications: 
\begin{itemize}
	\item[\cite{Grimm:2021idu}] Thomas W. Grimm, Jeroen Monnee: \emph{Deformed WZW models and Hodge theory. Part I}, \textbf{JHEP 05 (2022) 103},  \href{https://arxiv.org/abs/2112.00031}{\textbf{[arXiv: 2112.00031]}}
	\item[\cite{Grimm:2022ajb}] Thomas W. Grimm, Jeroen Monnee: \emph{Bi-Yang--Baxter models and Sl(2)-orbits}, \textbf{JHEP 11 (2023) 123},  \href{https://arxiv.org/abs/2212.03893}{\textbf{[arXiv: 2212.03893]}} 
\end{itemize}
as well as the work \cite{Grimm:2021ikg} listed above.

\tableofcontents

\mainmatter

\addchap{Introduction}
A beautiful feature of physics is that it tends to organize itself into various regimes, each describing physical processes that occur at a particular energy/length scale in a mutually consistent way. Perhaps even more striking is the fact that the possible length scales range from around $10^{26}\,\mathrm{m}$ (corresponding to the size of the observable universe) to around $10^{-35}\,\mathrm{m}$ (corresponding to the Planck length), thus traversing a sheer 61 orders of magnitude! Certainly most spectacular is the fact that observations have managed to probe physics from the largest length scales all the way down to around $10^{-20}\,\mathrm{m}$. Ultimately, the goal of physics is to be able to describe Nature along the \textit{entire} spectrum of length scales, all the way down to the Planck scale. As theoretical physicists, we are thus faced with the challenge of formulating consistent laws of physics that describe Nature at energy scales which are not yet accessible to experiments. \\

\noindent At first sight, it may appear that the areas of physics that focus on different length scales are effectively independent. For example, in order to describe the macroscopic movement of the water in the oceans it is not necessary (and completely unpractical) to know about the detailed microscopic interactions between the zillions of water molecules. Fortunately for us, this apparent independence is not quite true. Indeed, it can happen that a microscopic theory leaves particular imprints at bigger length scales which are experimentally testable, and which would be difficult to explain from a purely macroscopic perspective. A beautiful example of this is Brownian motion, which describes the motion of a ``large'' particle suspended in a medium in terms of its collisions with a large collection of ``small'' particles that make up the molecules in the surrounding gas/liquid. The combination of its initial observation by Brown in 1827, its mathematical formulation by Bachelier in 1900, and its physical explanation by Einstein in 1905 together served as convincing evidence for the existence of atoms/molecules. The latter was then verified experimentally by Perrin, confirming the atomic nature of matter and resulting in a Nobel Prize in Physics in 1926. \newpage

\noindent In this thesis, the analogous microscopic theory we will be concerned with is \textbf{string theory}, which is a candidate theory to describe physics at the tiniest length scales. The fundamental constituents of the theory consist of one-dimensional relativistic strings whose length is expected to be of the order of the Planck length.\footnote{It should be noted, however, that in the presence of (relatively) large extra dimensions, the string length may be significantly larger than the Planck length.} At these scales effects of both quantum mechanics and gravity must be taken into account, which is an extremely non-trivial task. Strikingly, string theory manages to capture these effects in a unified and mathematically consistent manner, thus forming a theory of \textbf{quantum gravity}. 

\subsubsection*{The landscape of string theory}
It is, at present, not at all clear whether string theory actually describes the Universe we find ourselves in. In light of the fact that the string scale is likely to be inaccessible to direct experimental observation in the foreseeable future, it is thus important to search for the imprints of string theory in low-energy effective theories coupled to gravity. Here we are faced with the reality that, although string theory itself is remarkably constrained (and, in the supersymmetric case, believed to be essentially unique), it can give rise to a myriad of low-energy effective theories: the \textbf{string landscape}. Different theories within the string landscape may differ wildly in their particle spectrum and couplings, and at present there is no clear selection principle that singles out any particular theory.  On the one hand, this may lead to the expectation that ``anything goes'' in string theory, and that our Universe just happens to be described by one of these theories, leading to a potential lack of predictive capability. On the other hand, it must be stressed that it has proven remarkably difficult to find well-controlled string theory constructions that reproduce both the correct particle spectrum of the Standard Model as well as the correct cosmology corresponding to an expanding universe. Thus, before attempting to make solid low-energy predictions from string theory, it is necessary to have a deep understanding of what is and is not possible in string theory. \\

\noindent In this thesis, we will be concerned with addressing this question from a ``top-down'' perspective. This means that we will consider a specific corner of string theory, flux compactifications of type IIB string theory and F-theory to be precise, and study the possible low-energy effective theories that arise in this setting. From this point of view, the vastness of the string landscape originates from the fact that string theory is inherently a ten-dimensional theory, rather than a four-dimensional one. This discrepancy is not by design. Rather, it is forced upon us by the mathematical consistency of the theory. As a result, while the ten-dimensional theory is effectively unique, its four-dimensional low-energy effective description -- obtained through a process called \textbf{compactification} -- depends heavily on what is going on in the additional six internal dimensions (which are assumed to be compact and small), for which there are typically many possibilities. For example, the number of light particles in the low-energy effective theory is dictated by the \textit{topology} of the internal space, while features pertaining to the dynamics, such as the values of couplings (which determine the strength of the interactions between the light particles) and the scalar potential, depend on the \textit{geometry} of the internal space. In this manner, the problem of characterizing the imprints of string theory on low-energy effective theories is turned into a problem of both physics and mathematics.\\

\noindent In full generality, this is an incredibly difficult problem to tackle, simply due to the abundance of six-dimensional compact manifolds to use in the compactification process. A helpful guiding principle in physics is to use symmetries to simplify a problem. In this work, we will follow this principle and investigate those compactifications which preserve a minimal amount of \textit{supersymmetry} in four dimensions, allowing for the possibility of spontaneous breaking of supersymmetry at low energies. The preservation of supersymmetry at the level of the effective theory imposes a number of restrictions on the geometric properties of the internal space, such that it has to be a so-called \textbf{Calabi--Yau manifold}. Thus, coming back to our initial goal, we are tasked with describing the topology and geometry of Calabi--Yau manifolds, and understanding how exactly this relates to the physical properties of the four-dimensional low-energy effective theory. This brings us to the central mathematical tool of this work: \textbf{(asymptotic) Hodge theory}.\footnote{It should be stressed, however, that the tools we discuss in this work apply to far more general settings than just Calabi--Yau manifolds.} 

\subsubsection*{(Asymptotic) Hodge theory}
One of the central insights of Hodge theory is that one can describe families of complex algebraic varieties, of which Calabi--Yau manifolds form a particular subset, in terms of analytic methods and differential equations. The idea is that the geometric properties of these spaces are encoded in the way certain functions called ``periods'' change as one varies in the family. This dependence is, in turn, described in terms of a set of special differential equations that, mathematically, define a so-called \textbf{variation of Hodge structure}. In physics, the parameters that parametrize this variation correspond to scalar fields in the four-dimensional low-energy effective theory and are referred to as \textbf{moduli}. Furthermore, the functional dependence of the aforementioned periods directly determines the kinetic couplings and scalar potential of the effective theory. Thus, it is key to understand, in as much generality as possible, the properties of variations of Hodge structure. This is the central topic of part \ref{part2} of the thesis, and forms the foundation for parts \ref{part3} and \ref{part4}. \\

\noindent Following the seminal works of Cattani, Kaplan, and Schmid, who built on earlier works of Deligne and Griffiths, there is by now an extraordinarily deep understanding of variations of Hodge structure. Roughly speaking, this understanding comes from studying the possible singularities of the differential equations that are satisfied by periods, or, more precisely, the possible degenerations of Hodge structures and the type of limiting structures they give rise to. Physically, this corresponds to probing the features of the low-energy effective theory in asymptotic regions of the moduli space, where the underlying Calabi--Yau manifold becomes singular, for example due to the pinching of some cycle. Consequently, the framework in which these matters are studied is also called \textbf{asymptotic Hodge theory}. It will be the main purpose of this thesis to provide an in-depth description of this framework, accompanied with illustrative examples, and show how it can be applied to study the features of low-energy effective theories coming from string compactifications in great generality. In particular, in part \ref{part3} of the thesis we will apply it to investigate the finiteness and geometric structure of the landscape of type IIB/F-theory flux compactifications. 

\subsubsection*{Hodge theory and the landscape of two-dimensional integrable field theories}
So far, this has been a tale of only a single type of landscape, namely that of string theory, and how Hodge theory plays a central role in determining its features. Now we come across another beautiful feature of physics, namely that the same equations tend to appear in completely different contexts! In part \ref{part4} of the thesis we will encounter an instance of this phenomenon by showing that the defining equations of variations of Hodge structure and their (asymptotic) solutions make an appearance in certain classes of \textbf{two-dimensional integrable field theories}. On the one hand, this allows us to use the results of part \ref{part2} of the thesis to construct very general classes of solutions to these models. On the other hand, and perhaps more intriguingly, this opens up a new perspective on Hodge theory, in which its essential features are encoded in terms of an auxiliary field theory formulated \textit{on the moduli space}. Our results provide a first step in this relatively unexplored direction.

%One of the appealing features of string theory is the fact that it has no free parameters. Indeed, the value of any coupling in a low-energy effective description of string theory is determined dynamically, for example through the minimization of an effective potential. This is to be contrasted with, say, the theory of elementary particle physics known as the Standard Model, which contains 19 free parameters (such as the electron mass) whose value is determined by experimental measurement. 

\RedeclareSectionCommand[ 
beforeskip=12ex,% afterskip=1sp%
            ]{part}
%%%%%%%%%%%%%%%%%%%%%%%%%%%%%%%%%%%%%%%%%%%%%%%%%%%%%%
\setpartpreamble[u][\textwidth]{
\vspace*{1cm}
\hrulefill 
\vspace*{0.5cm}

In this part of the thesis we describe the essentials of string compactifications in order to set the stage for the remainder of the thesis. We provide a brief overview of the basic aspects of (perturbative) string theory, after which we focus on the particular setting of type IIB string theory and its formulation in terms of F-theory. Our goal is to describe how various features of the four-dimensional low-energy effective theories that arise from (flux) compactifications of these theories are described in terms of geometrical properties of the underlying internal geometry. The latter will then be studied extensively in the rest of the thesis.

\vspace*{0.5cm}
\hrulefill }

\part{Preliminaries}\label{part1}
\chapter{Essentials of string compactifications}\label{chap:string}

In this chapter we will review some fundamental aspects of string theory and string compactifications. The goal of this chapter is two-fold. First, we aim to introduce the relevant concepts that are required to motivate and understand the rest of this work. Second, we hope to provide the necessary context in order to understand in which particular corner of the vast realm of string theory our work takes place. 
\\

\noindent In section \ref{sec:string_theory_basics} we begin with a rather general point of view on perturbative string theory from a worldsheet perspective, emphasizing the features of string theory which make it different from and more restrictive than quantum field theory. We then move to the target space formulation of a very rich class of string backgrounds known as non-linear $\sigma$-models and give an overview of the resulting five 10-dimensional superstring theories that arise in this setting, together with the putative 11-dimensional M-theory formulation. In section \ref{sec:IIB_effective} we focus on one of these five superstring theories which will play a central role throughout this thesis: type IIB string theory. In particular, we discuss its low-energy effective description in terms of IIB supergravity, including localized sources such as D-branes and O-planes. In section \ref{sec:IIB_compactification} we review some aspects of Calabi--Yau threefold compactifications of type IIB to four dimensions. In particular, we emphasize the dependence of various physical quantities in the four-dimensional low-energy effective theory on the geometric properties of the Calabi--Yau manifold. A proper understanding of the latter naturally leads to the study of (asymptotic) Hodge theory, which forms the central pillar of this work and will be discussed at length in part \ref{part2}. In section \ref{sec:F-theory} we discuss some basic aspects of F-theory, which provides a beautiful framework to describe more complicated (flux) compactifications of type IIB that include non-perturbative effects in a geometric fashion. The latter is discussed in section \ref{sec:F-theory_flux}, which will ultimately lead us to the landscape of four-dimensional $\mathcal{N}=1$ type IIB/F-theory compactifications, which will play a central role in part \ref{part3} of the thesis. We close this chapter with an outlook for the rest of the thesis. \newpage

\noindent It must be said that this chapter is far from a comprehensive review on the incredibly rich field of string theory. For a more in-depth and pedagogical treatment of the matters discussed in this chapter, we refer the reader to the textbooks \cite{Polchinski_I:1998, Polchinski_II:1998,Johnson:2002,Becker:2006,Blumenhagen:2012,Green_I:2012,Green_II:2012} as well as the lecture notes \cite{Kiritsis:1997hj,Johnson:2000ch,Tong:2009np,Agmon:2022thq}.

\section{Basics of string theory}\label{sec:string_theory_basics}
One of the core principles of string theory is the idea that (spatially) extended objects could play a crucial role in describing the fundamental constituents of Nature. In this sense, the name ``string theory'' is a slight misnomer, as a typical endeavour in the field usually involves (or even requires!) both one-dimensional objects (strings) as well as higher-dimensional objects (membranes). Nevertheless, as far as the current formulation of the theory is concerned, strings -- as opposed to membranes -- do play a particularly special role. Indeed, it is an amazing fact that first quantization of a string yields a discrete spectrum of states which admit an interpretation as different particles. In contrast, first quantization of a membrane is expected to give rise to a continuous spectrum of states. As a result, while a classical theory of $p$-dimensional membranes exists for all non-negative integers $p$, a complete quantum-mechanical formulation remains elusive for $p>1$. At the same time, however, even if one starts with a theory of just strings, one is quickly confronted with the reality that one must include these higher-dimensional membranes as well -- for example to describe non-perturbative effects. For similar reasons, a full non-perturbative description of string theory remains elusive as well. 

\subsection{Perturbative string theory (1): Worldsheet perspective}
Thus, let us (for the moment) focus on the perturbative description of strings in string theory.  At its core, \textbf{perturbative string theory} is defined as an S-matrix theory, which means that it consists of a set of rules to compute scattering amplitudes between quantum states. Such a scattering amplitude is interpreted as the probability amplitude for a given initial state to evolve into a (possibly different) final state. In ordinary quantum field theory, one typically imagines this as some initial collection of particles flying about in space -- tracing out a world-line in spacetime -- and subsequently colliding with each other to produce some final collection of particles. In a perturbative description, the possible ways in which this process can take place are then described in terms of certain topologically distinct one-dimensional graphs: Feynman diagrams. In string theory, the situation is very similar, but there are some important differences. 
\\

\noindent The first difference is that, because strings trace out a 2-dimensional \textit{worldsheet} in spacetime, scattering amplitudes are instead expressed in terms of two-dimensional surfaces. There are two different kinds of string scattering processes to consider, owing to the fact that a compact one-dimensional string can have two different topologies: either it is a circle (closed string) or it is a compact interval (open string). An important point, however, is that both types of scattering processes, the interactions between strings, such as the splitting of one string into two strings as depicted in figure \ref{fig:string_interaction}, have a smooth geometrical description. This is to be contrasted with the interactions between particles, which are ``localized'' at a single vertex of a Feynman diagram. As a result, string scattering amplitudes typically enjoy a much better behaviour in the ultraviolet regime, corresponding to small length scales.  
\\

\begin{figure}[t]
	\centering
	\includegraphics[scale=0.4]{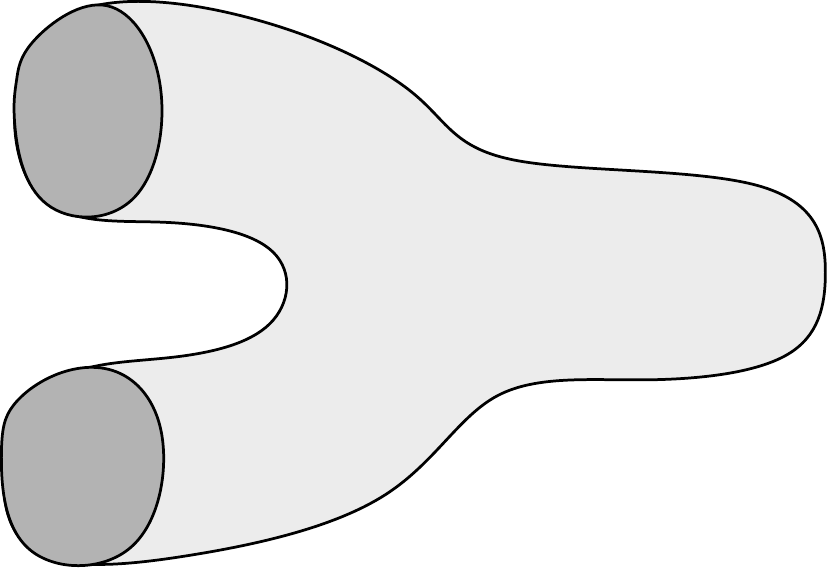}
	\caption{A depiction of a tree-level closed string interaction, which can be interpreted as the splitting of a single closed string into two different closed strings.}
	\label{fig:string_interaction}
\end{figure}

\noindent A second difference is that, while the possible allowed particle interactions depend greatly on the kind of quantum field theory one is considering, the possible interactions between strings are, in a sense, rather simple. Indeed, when working in Euclidean signature, the possible geometries of string scattering diagrams are effectively described by (punctured) Riemann surfaces, with boundary if one considers open string scattering. Topologically distinct Riemann surfaces are enumerated simply in terms of the number of holes and boundaries of the worldsheet. To emphasize this point, note that in particle physics there can already be many different kinds of one-loop diagrams, while in string theory there is really only one diagram at each loop order. One point of concern may be that, as quantum theory dictates, one should also sum over all possible geometries, i.e.~metrics, of the Riemann surfaces. Fortunately, the ``moduli spaces'' of Riemann surfaces, which one can think of as parametrizing the space of possible metrics, are very well-understood. Note that the same cannot be said about arbitrary higher-dimensional surfaces. 
\\

\noindent A third difference concerns the choice of background. As any perturbative theory, perturbative string theory is defined as a perturbative expansion around a given background. In contrast to ordinary quantum field theory, there are many more restrictions on the possible backgrounds in which a string scattering process can be described in a quantum-mechanically consistent manner. To be precise:
\begin{subbox}{String background/vacuum}
	A \textbf{(string) background} or \textbf{(string) vacuum} of perturbative string theory is a modular invariant two-dimensional (super)conformal field theory of central charge $c=26$ in the bosonic case, or $c=15$ in the supersymmetric case.
\end{subbox}
\noindent Classically, one can think of the choice of background CFT as determining the dynamics of the string, which in turn affect the details of the scattering process. Let us briefly discuss the various conditions listed above:
\begin{itemize}
	\item \textbf{Conformal invariance}: \\
	At the classical level, string theory is invariant under local diffeomorphisms of the worldsheet coordinates, as well as local Weyl transformations of the worldsheet metric. After gauge-fixing the metric, one is still left with an infinite dimensional residual symmetry group: the two-dimensional conformal transformations. Precisely because in string theory the worldsheet metric is considered to be a dynamical field (we say that the theory is ``coupled to gravity'') these conformal transformations originate from local gauge transformations, i.e.~redundancies in the description of the theory. Therefore, in string theory it is vital that conformal symmetry remains unbroken at the quantum level in order to avoid gauge anomalies, which is a non-trivial condition. This is to be contrasted with theories in which the metric is non-dynamical. For such theories, conformal invariance should be thought of as a genuine physical symmetry which may or may not be preserved at the quantum level. 
	\item \textbf{Critical central charge}:  \\
	Any two-dimensional CFT defined on a curved background suffers from the Weyl anomaly, which is the statement that the expectation value of the trace of the energy-momentum tensor satisfies
	\begin{equation}
		\langle \tensor{T}{^\alpha_\alpha}\rangle = - \frac{c^{\text{tot}}}{12}R\,,
	\end{equation}
	where $R$ denotes the two-dimensional Ricci scalar and $c^{\text{tot}}$ denotes the total central charge of the CFT in question. In order for the theory to remain conformal at the quantum level, it is necessary that $\langle \tensor{T}{^\alpha_\alpha}\rangle=0$. For CFT's defined in a fixed flat background, this is of no concern. However, in string theory we must consider all possible backgrounds, and hence the consistency of the theory imposes that $c^{\text{tot}}=0$. At the same time, for any string background there is a universal part of the total CFT which is described by the ghost fields which incorporate the gauge-fixing of the local diffeomorphism and Weyl symmetries via the Faddeev--Popov method. Taking into account their contribution to the total central charge, one finds that
	\begin{equation}
		\text{bosonic}:\quad c^{\text{tot}}=c-26\,,\qquad \text{supersymmetric}:\quad c^{\text{tot}}=c-15\,,
	\end{equation}
	thus the vanishing of the Weyl anomaly fixes the central charge of the background CFT to be 26 in the bosonic case, and 15 in the supersymmetric case. 
	\item \textbf{Modular invariance}: \\
	The string perturbation series (which will be discussed in a moment) involves the integration of CFT correlators over the moduli space of various Riemann surfaces, in particular the torus. It is thus a necessary consistency condition that these correlators behave properly under the group $\mathrm{SL}(2,\mathbb{Z})$ of large diffeomorphisms of the torus. More precisely, the statement is that the torus partition function must be invariant under modular transformations. If this is not the case, then equivalent descriptions of the torus would give rise to different values for the partition function. Modular invariance places very strong restrictions on the spectrum of the CFT. Furthermore, it can be shown that to establish modular invariance of the full all-loop string partition function it suffices to check this up to two loops. This is to be contrasted with ordinary quantum field theory, in which it typically suffices to check for the presence of anomalies at one loop.  
\end{itemize}
It is important to stress that, despite the stringent consistency conditions mentioned above, there is currently no complete classification of the set of all possible string backgrounds. Moreover, there is also no clear selection principle which favours one choice of string background over another, and it is not at all obvious whether string theory in its current form is secretly background-independent, or what this would even mean. For completeness, let us mention that there is an attempt at such a manifestly background-independent formulation known as \textbf{string field theory}, but we will not make use of this perspective in this thesis. 

\subsubsection*{The string perturbation series}
For each choice of string background, the equation that defines perturbative string theory in that background is given by a formal power series, referred to as the \textbf{string perturbation series}, in which one sums certain correlators of the two-dimensional background CFT over all possible Riemann surfaces, as mentioned earlier. Schematically, the string perturbation series thus takes the form
\begin{equation}\label{eq:string_amplitude}
	\text{probability amplitude} = \sum_{g,b} g_s^{-\chi}\int_{\text{moduli space of $\Sigma_{g,b}$}} \text{correlators on $\Sigma_{g,b}$}\,,
\end{equation} 
where $\Sigma_{g,b}$ denotes a punctured Riemann surface of genus with $g$ handles and $b$ boundaries. The contributions from topologically distinct worldsheets are weighted by a dimensionless expansion parameter $g_s$, referred to as the \textbf{string coupling},  and the exponent is given by the Euler character
\begin{equation}
	\chi = 2-2g-b\,,
\end{equation}
of $\Sigma_{g,b}$. Finally, the number of punctures is simply given by the total number of ingoing and outgoing strings. To close this part of the discussion, let us again emphasize one more crucial difference between quantum field theory and perturbative string theory. Namely, for the former the perturbative computation in terms of Feynman diagrams is mostly a practical tool to approximate some observables which, in principle, have a non-perturbative description. In contrast, at present string theory does not have a fully non-perturbative formulation.

\subsection{Perturbative string theory (2): Target space perspective}
\label{subsec:pert_string_target}
As we have stressed above, in principle any CFT with the right properties serves as a valid string background. In this work, we will focus our attention on a broad class of string backgrounds: \textbf{(non-linear) $\sigma$-models}. 
\\

\noindent Non-linear $\sigma$-models focus on a geometrical description of strings moving inside spacetime and are, in a sense, the most natural generalization of how particles are described in ordinary quantum field theory. The embedding of the string worldsheet is specified by a map
\begin{equation}
	X:\quad \Sigma\rightarrow M_D\,,
\end{equation}
from a given two-dimensional Riemann surface $\Sigma$ to a $D$-dimensional target space $M_D$, giving rise to a set of $D$ worldsheet scalars $X^\mu$, for $\mu=1,\ldots, D$. In superstring theory, one additionally considers a collection of worldsheet fermions $\psi^\mu$ which correspond to the superpartners of the scalars $X^\mu$. For simplicity, we will restrict our discussion to the bosonic case, deferring the description of superstrings to the next section. 

\subsubsection*{The Polyakov action}
Classically, the dynamics of the string are described by the Polyakov action:
\begin{subbox}{Polyakov action}
	\begin{equation}\label{eq:non-linear_sigma_model}
		S = \frac{1}{2\pi \alpha'}\int_\Sigma G_{\mu\nu}\, \mathrm{d}X^\mu\wedge\star\,\mathrm{d}X^\nu+ B_{\mu\nu}\, \mathrm{d}X^\mu\wedge\,\mathrm{d}X^\nu+\alpha'\Phi\,R\star 1\,,
	\end{equation}
	where $\star$ and $R$ denote the Hodge star and two-dimensional Ricci scalar on $\Sigma$, respectively. Additionally, $\alpha'$ is a parameter of mass dimension $[\alpha']=-2$ and thus defines a length scale 
	\begin{equation}
		\ell_s = 2\pi\sqrt{\alpha'}\,,
	\end{equation} 
	known as the \textbf{string length}. The factor $T=\frac{1}{2\pi\alpha'}$ can be interpreted as the tension of the string. 
\tcblower 
\textbf{Note:}\\
If the string worldsheet $\Sigma$ has a boundary, i.e.~for open strings, there is an additional piece 
\begin{equation*}
	\frac{1}{2\pi\alpha'}\int_{\partial\Sigma} A_\mu \mathrm{d}X^\mu\,.
\end{equation*}
\end{subbox}
\noindent The fields $G_{\mu\nu}$, $B_{\mu\nu}$, and $\Phi$ that appear in the Polyakov action have the interpretation of a metric, anti-symmetric 2-form and scalar on the target space $M$, respectively. In the case of open strings, the field $A^\mu$ carries the interpretation of a gauge boson. From the worldsheet point of view, these fields are instead interpreted as coupling constants which, generically, will run with the energy scale. In order for the theory defined by \eqref{eq:non-linear_sigma_model} to define a CFT, the $\beta$-functions of these couplings must vanish, so that the theory is scale-invariant. To first order in $\alpha'$, the $\beta$-functions are given by the following expressions\footnote{For simplicity, we neglect the contributions coming from the open string gauge field $A_\mu$.}
\begin{align*}
	\beta_{\mu\nu}(G) &=\alpha' \left(R_{\mu\nu} + 2\nabla_\mu\nabla_\nu\Phi - \frac{1}{4}H_{\mu\lambda\kappa}\tensor{H}{_\nu^{\lambda\kappa}}\right)+\mathcal{O}((\alpha')^2)\,,\\
	\beta_{\mu\nu}(B) &= \alpha'\left(-\frac{1}{2}\nabla^\lambda H_{\lambda\mu\nu}+(\nabla^\lambda\Phi)H_{\lambda\mu\nu}\right)+\mathcal{O}((\alpha')^2)\,,  \\
	\beta(\Phi) &= \frac{D-26}{6} - \alpha'\left(\frac{1}{2}\nabla^2\Phi+\nabla_\mu\Phi \nabla^\mu \Phi - \frac{1}{24}H_{\mu\nu\lambda}H^{\mu\nu\lambda}\right)+\mathcal{O}((\alpha')^2)\,,
\end{align*}
where $H_3=\mathrm{d}B_2$ is the field-strength of the 2-form $B_2$. 

\subsubsection*{An exact background}
Clearly, there are many possible configurations which can lead to $\beta_{\mu\nu}(G)=\beta_{\mu\nu}(B)=\beta(\Phi)=0$. The simplest solution is to set
\begin{equation}
	G_{\mu\nu} = \eta_{\mu\nu}\,,\qquad B_2 = 0\,,\qquad \Phi = \Phi_0\,,\qquad D=26\,,
\end{equation}
where $\eta_{\mu\nu}$ denotes the flat Minkowski metric and $\Phi_0$ is a constant. Classically, this particular solution describes the propagation of a bosonic string in a flat 26-dimensional target space (with vanishing $B_2$). Furthermore, the on-shell value of the dilaton determines the value of the string coupling to be
\begin{equation}
	g_s = e^{\Phi_0}\,,
\end{equation}
due to the fact that the two-dimensional Einstein--Hilbert term in \eqref{eq:non-linear_sigma_model} is topological and precisely integrates to the Euler characteristic of the string worldsheet.\\

\noindent In fact, one can verify that this is an \textit{exact} background of string theory, which means, in particular, that it has vanishing $\beta$-functions to all orders in $\alpha'$. Furthermore, one can show that the resulting CFT indeed has central charge equal to 26, and that it has a modular invariant partition function. 

\subsubsection*{More general backgrounds}
Clearly, the simple background discussed above leaves much to be desired when it comes to describing a phenomenologically reasonable model. A more interesting background arises if one allows the metric $G_{\mu\nu}$ to be more general, while keeping $H_3=0$ and $\Phi$ constant, in which case the $\beta$-equations reduce to\footnote{Another possibility is that $\Phi$ depends linearly on one of the $X^\mu$ as $\Phi = Q X$. In this case, there exists a CFT description which is valid to all orders of $\alpha'$, known as the \textit{linear dilaton CFT}, which has central charge $c=D+6\alpha'Q^2$. }
\begin{equation}
	D = 26\,,\qquad R_{\mu\nu}=0\,.
\end{equation}
Again, one finds that the target space must be 26-dimensional, with a metric that is Ricci-flat, which means that it satisfies \textit{Einstein's equations} in the absence of additional sources of stress-energy. It is a remarkable property of string theory that one of the massless \textit{quantum} excitations of the closed string is, to leading order in $\alpha'$ described by the laws of general relativity! Still, we will see in section \ref{sec:IIB_effective} that we will have to consider even more general backgrounds and include configurations for which $H_3$ is non-zero. 

\begin{redbox}{String backgrounds}
	From this point onwards, we will take the phrase ``string theory'' to mean critical string theory, i.e.~string theory in the background of a non-linear $\sigma$-model whose target space has dimension 26 in the bosonic case, or 10 in the supersymmetric case. In particular, we will restrict to this class of backgrounds for the remainder of this work. 
\end{redbox}

\subsubsection*{Low-energy effective description}
There is another way to interpret the vanishing of the $\beta$-functions. Namely, one can regard them as equations of motion for the \textit{target space} fields $G_{\mu\nu}, B_2$ and $\Phi$. It is then a natural question to find an action principle whose extremization reproduces these equations of motion. That action is given by 
\begin{equation}\label{eq:action_NSNS}
	S = \frac{1}{2\kappa^2}\int_{M_{26}} e^{-2\Phi}\left( R\star 1+4\,\mathrm{d}\Phi\wedge\star\,\mathrm{d}\Phi-\frac{1}{2} H_3\wedge\star\,H_3\right)\,,
\end{equation}
where we stress that the integration is now performed over the target space $M_{26}$ and we have specialized to the critical dimension $D=26$.
\\

\noindent The action \eqref{eq:action_NSNS} is referred to as the \textbf{low-energy effective action} of critical bosonic string theory. The reason for this is that it should be viewed as an effective field theory that describes the massless excitations of the string, which is valid at length scales much greater than the string length $\ell_s$. In other words, the action \eqref{eq:action_NSNS} is precisely such that it reproduces the low-energy limit of string scattering amplitudes at leading order in the string-coupling $g_s$.

\subsection{The 5 critical superstring theories}
There are a number of issues with the bosonic string. Most notably, the ground state of the bosonic string is tachyonic, which signals that the theory has been formulated around an unstable vacuum. Another problem is that the spectrum of the bosonic string does not contain any fermionic excitations, which are clearly necessary in order to describe matter. Both of these issues are addressed by adding additional fermionic fields $\psi^\mu$ to the worldsheet action in a supersymmetric fashion. It should be noted, however, that it is not strictly necessary to work with a spacetime supersymmetric theory, as happens for example for the $\mathrm{SO}(16)\times\mathrm{SO}(16)$ heterotic string (though this theory does again suffer from tachyonic instabilities). Nevertheless, the presence of supersymmetry typically leads to additional simplifications which allow for more precise computations, for example through non-renormalization theorems, as well as better control regarding the stability of certain solutions. For these reasons, we will restrict our attention to supersymmetric string theories.
\\

\noindent In contrast to the bosonic case, there are multiple possible superstring theories that can be constructed in this way. To be precise, there are five critical string theories which have spacetime supersymmetry. Throughout most of this work we will focus our attention on one of these theories: the type IIB superstring. However, for completeness we summarize below the main features of the other theories as well. 

\begin{itemize}
	\item \textbf{Type IIA / IIB:}\\
	The type II theories are characterized by the fact that they have extended supersymmetry in spacetime, having 32 supercharges. At the technical level, the difference between the two theories arises from the different choices of GSO projection\footnote{Actually, there are two more possibilities, which lead to the non-supersymmetric type 0A/0B string.}. As a result, the two theories differ in their spectrum and spacetime supercharges. Indeed, type IIA string theory has target space $\mathcal{N}=(1,1)$ supersymmetry and is non-chiral, while type IIB string theory has target space $\mathcal{N}=(2,0)$ supersymmetry and is chiral. The low-energy description of the type II superstring theories is given by the two corresponding 10-dimensional $\mathcal{N}=2$ supergravity theories.
	\item \textbf{Type I:}\\
	The type I string theory is a chiral theory of unoriented strings, and has $\mathcal{N}=1$ supersymmetry. It can be thought of as an orientifold of type IIB string theory, in which certain left-moving and right-moving excitations of the string are identified under reflection with respect to an $(9+1)$-dimensional orientifold plane or O9-plane. Cancellation of the R-R tadpole (see also section \ref{subsec:tadpole_cancellation}) then requires the presence of an additional 32 coincident D$9$-branes, leading to an $\mathrm{SO}(32)$ gauge group on the worldvolume of the brane stack. At low energies, the theory is described by type I supergravity coupled to 10-dimensional $\mathcal{N}=1$ supersymmetric Yang--Mills theory with $\mathrm{SO}(32)$ gauge group. 
	\item \textbf{Heterotic $E_8\times E_8\,/\,\mathrm{SO}(32)$:}\\
	The two heterotic theories are both theories of only closed strings, having $\mathcal{N}=1$ supersymmetry. They are constructed by combining the right-moving sector of closed superstrings together with the left-moving sector of the bosonic string. Roughly speaking, one interprets the 16 additional left-moving bosons as parameters of a compactified 16-dimensional torus with radius $\sqrt{\alpha'}$. Modular invariance and anomaly cancellation then greatly restrict the part of the spectrum coming from these additional states. In particular, the massless states must correspond to gauge bosons of either an $E_8\times E_8$ or an $\mathrm{SO}(32)$ gauge group. Similarly to the type I string, the low-energy description of the heterotic string theories is given by type I supergravity coupled to 10-dimensional supersymmetric Yang--Mills theory with the corresponding gauge group.   
\end{itemize}

\subsection{M-theory}
In contrast to the five supersymmetric (perturbative) \textit{string} theories which were discussed in the previous section, \textbf{M-theory} is instead supposed to be a theory of \textit{membranes}. It is an attempt at a more fundamental, non-perturbative framework in which the five superstring theories are unified into a single 11-dimensional description \cite{Witten:1995ex}. While, at present, a complete formulation of the theory is still lacking, many important features of M-theory, and its relation to the other superstring theories, can be understood through its low-energy effective description. This corresponds to the unique 11-dimensional supergravity theory, for which the bosonic part of the action is given by
\begin{equation}\label{eq:action_M-theory}
	S_{\mathrm{M}} = \frac{1}{2\kappa_{11}^2}\int_{M_{11}} R\star 1- \frac{1}{2}G_4\wedge\star\,G_4-\frac{1}{6}C_3\wedge G_4\wedge G_4\,,
\end{equation}
where $G_4=\mathrm{d}C_3$ is the field-strength of a 3-form $C_3$, which is sourced through M2-branes and M5-branes. For example, there is by now a large collection of evidence that the strong-coupling limit $g_s\rightarrow\infty$ of type IIA should be described as M-theory compactified on a circle of radius $g_s\ell_s/2\pi$, in which the D0-branes of type IIA arise as Kaluza--Klein modes. Note that this relation fixes the 11-dimensional Planck scale in terms of the string length $\ell_s$ and the string coupling $g_s$. The action \eqref{eq:action_M-theory} will also play an important role in the F-theory formulation of type IIB superstring theory, to which we now turn our attention. 

\section{Type IIB at low energies}\label{sec:IIB_effective}
In this section we focus our attention on the low-energy effective description of type IIB string theory from a 10-dimensional spacetime perspective. In section \ref{subsec:IIB_SUGRA} we first summarize the main features of the massless bosonic excitations of the closed type IIB string, as well as their effective description in terms of the 10-dimensional $\mathcal{N}=2$ type IIB supergravity theory. Subsequently, in section \ref{subsec:sol_no_sources} we consider a broad class of classical solutions in the absence of localized sources and recover the Maldacena--Nu\~nez no-go theorem. This particular class of solutions will play an important role in section \ref{sec:IIB_compactification}. Then, in section \ref{subsec:D-branes_O-planes}, we discuss some essential non-perturbative objects -- D-branes and O-planes -- which play an important role in the effective description of the massless excitations of the open string. Additionally, the inclusion of these localized sources is crucial in order to evade the Maldacena--Nu\~nez no-go theorem and leads to a phenomenologically much more appealing class of solutions. These will play a vital role in section \ref{sec:F-theory_flux} as well as part \ref{part3} of the thesis, where we discuss flux compactification of type IIB/F-theory. Importantly, the inclusion of such objects leads to additional consistency conditions through the requirement of the cancellation of the R-R tadpoles, which are discussed in section \ref{subsec:tadpole_cancellation}. \newpage

\subsection{Type IIB supergravity}\label{subsec:IIB_SUGRA}
As discussed in section \ref{subsec:pert_string_target}, in the low-energy description of a string we focus on an effective description of its massless degrees of freedom. Of course, this is a drastic simplification of the full theory, as we are restricting ourselves to only a small part of the full spectrum of string excitations. Nevertheless, such a description is expected to provide a good approximation at length scales which are much longer than the string length $\ell_s$ or, equivalently, at energy scales which are much below the mass of the higher excited states. 

\subsubsection*{The massless spectrum of type IIB}
Let us first consider the massless states of the closed string. In the NS-NS sector, one finds the fields
\begin{equation}
	\text{NS-NS}:\qquad G_{MN}\,,\qquad B_2\,,\qquad \Phi\,,
\end{equation}
which already appeared in our discussion of the bosonic string. For the superstring, there are additional massless bosonic states in the R-R sector, which are given by $p$-form gauge fields
\begin{equation}
	\text{R-R}:\qquad C_p\,,\qquad p=0,2,4\,.
\end{equation}
For completeness, we should mention that there are also massless fermionic states coming from the NS-R and R-NS sectors, but we will not explicitly consider those. For each of the gauge fields we introduce the usual field strengths
\begin{equation}
	H_3 = \mathrm{d}B_2\,,\qquad F_{p+1} = \mathrm{d}C_p\,,
\end{equation}
which are referred to as \textbf{fluxes}.

\subsubsection*{The low-energy effective action of type IIB (1): democratic formulation}
In the following, we will introduce two equivalent low-energy effective action principles that describe the dynamics of the massless excitations of the closed type IIB string. 
\\

\noindent A first elegant way to write down the low-energy effective action is to use the so-called \textit{democratic formulation} in which one doubles the number of degrees of freedom in the R-R sector by also including higher $p$-forms $C_6$ and $C_8$, together with their corresponding field strengths, which are Hodge dual to $C_2$ and $C_0$, respectively. In order to recover the original number of degrees of freedom, one subsequently imposes a number of duality conditions which will be written down shortly. It will be convenient to define the following generalized field strengths
\begin{equation}
	\hat{F}_{p+1} = \mathrm{d}C_p - H_3\wedge C_{p-2}\,.
\end{equation}
In terms of these generalized field strengths, the 10-dimensional low-energy effective action of type IIB is given as follows.

\begin{subbox}{10D low-energy effective action of type IIB (democratic formulation)}
	\begin{equation}\label{eq:action_IIB_democratic}
		S_{\mathrm{IIB}} = S_{\text{NS-NS}}+S_{\text{R-R}}\,,
	\end{equation}
	where
	\begin{align}
		\label{eq:action_IIB_NS-NS}
		S_{\text{NS-NS}} &= \frac{2\pi}{\ell_s^8}\int_{M_{10}} e^{-2\Phi}\left(R\star 1+4\,\mathrm{d}\Phi\wedge\star\,\mathrm{d}\Phi-\frac{1}{2}H_3\wedge\star\, H_3 \right)\\
		\label{eq:action_IIB_R-R}
		S_{\text{R-R}} &=\frac{2\pi}{\ell_s^8}\int_{M_{10}} - \frac{1}{4}\sum_{p=1,3,5,7,9} \hat{F}_p\wedge\star\,\hat{F}_p\,,
	\end{align} 
\tcblower 
\textbf{Note:} \\
In order to reproduce the correct equations of motion as well as to recover the original number of degrees of freedom, one must additionally impose the duality conditions
\begin{equation}\label{eq:R-R_duality}
	\hat{F}_p = \star\,\hat{F}_{10-p}\,.
\end{equation}
\end{subbox}
\noindent The actions \eqref{eq:action_IIB_NS-NS} and \eqref{eq:action_IIB_R-R}  respectively describe the NS-NS sector and the R-R sector. Note that the action $S_{\text{NS-NS}}$ is exactly the same as we found for the bosonic string in equation \eqref{eq:action_NSNS}. Finally, let us also remark that for $p=5$ the duality conditions \eqref{eq:R-R_duality} imply that the 5-form flux is self-dual.

\subsubsection*{The low-energy effective action of type IIB (2): $\mathrm{SL}(2,\mathbb{Z})$-invariant formulation}
There is a second formulation for the low-energy effective action of type IIB in which a particular $\mathrm{SL}(2,\mathbb{Z})$-symmetry becomes manifest. In this formulation we do \textit{not} include the dual $p$-form fields $C_6$ and $C_8$. In order to write down the action, we first introduce the following combinations
\begin{equation}\label{eq:axio-dilaton_G3}
	\tau = C_0+i e^{-\Phi}\,,\quad G_3 = F_3-\tau H_3\,,\quad \tilde{F}_5 = F_5-\frac{1}{2}C_2\wedge H_3+\frac{1}{2}B_2\wedge F_3\,,
\end{equation}
where the complex field $\tau$ is commonly referred to as the \textbf{axio-dilaton}. After performing a Weyl rescaling of the metric to pass to the Einstein frame the low-energy effective action for type IIB string theory can be organized as follows.\footnote{To be precise, the relation between the string-frame metric and Einstein-frame metric is given by
\begin{equation}
	G_{MN}^{\text{(string-frame)}} = e^{\Phi/2}G_{MN}^{\text{(Einstein-frame)}}\,.
	\end{equation}}

\begin{subbox}{10D low-energy effective action of type IIB ($\mathrm{SL}(2,\mathbb{Z})$-invariant formulation)}
	\begin{equation}\label{eq:action_IIB_SL2}
		S_{\mathrm{IIB}} = S_{\mathrm{kinetic}}+S_{\mathrm{CS}}\,,
	\end{equation}
	where 
	\begin{align}
		S_{\mathrm{kinetic}} &= \frac{2\pi}{\ell_s^8}\int_{M_{10}} R\star 1 - \frac{1}{2}\frac{\mathrm{d}\tau\wedge\star\,\mathrm{d}\bar{\tau}}{\left(\mathrm{Im}\,\tau\right)^2}-\frac{1}{2}\frac{G_3\wedge\star\,\bar{G}_3}{\mathrm{Im}\,\tau}-\frac{1}{4}\tilde{F}_5\wedge\star\,\tilde{F}_5\,,\\
		S_{\text{CS}}&=\frac{2\pi}{\ell_s^8}\int_{M_{10}} \frac{1}{4i}\frac{C_4\wedge G_3\wedge\bar{G}_3}{\mathrm{Im}\,\tau}\,,
	\end{align}
\tcblower 
\textbf{Note:} \\
In order to reproduce the correct equations of motion as well as to recover the original number of degrees of freedom, one must additionally impose the duality condition
\begin{equation}\label{eq:Fhat_duality}
	\tilde{F}_5 =\star\,\tilde{F}_{5}\,.
\end{equation}
\end{subbox}
\noindent The action \eqref{eq:action_IIB_SL2} is manifestly invariant under the following $\mathrm{SL}(2,\mathbb{R})$ transformation
\begin{equation}
	\begin{pmatrix}
		a & b \\
		c & d
	\end{pmatrix}\in\mathrm{SL}(2,\mathbb{R}):\qquad \tau\mapsto \frac{a \tau+ b}{c\tau +d}\,,\qquad G_3\mapsto \frac{G_3}{c \tau+d}\,,
\end{equation}
which is broken to the discrete group $\mathrm{SL}(2,\mathbb{Z})$ at the quantum level due to charge quantization of the fluxes. This underlying symmetry signals the presence of an underlying elliptic curve of which $\tau$ is the modular parameter, which will be discussed in more detail in section \ref{sec:F-theory}, where we discuss the F-theory description of type IIB.

\subsection{Classical solutions without localized sources}\label{subsec:sol_no_sources}
In order to obtain a candidate string background (at least to first order in $\alpha'$) it is necessary to obtain a solution to the classical equations of motion induced by the low-energy effective action \eqref{eq:action_IIB_SL2}. For the moment we will consider a class of solutions that arises in the absence of localized sources. The more general case will be discussed in section \ref{sec:F-theory_flux}. 

\subsubsection*{Assumptions}
As our starting point, we make three simplifying assumptions. 
\begin{enumerate}
	\item \textbf{Warped compactification ansatz:}\\
	We assume the 10-dimensional target space to admit a (warped) product structure
	\begin{equation}\label{eq:spacetime_decomposition}
		M_{10} = M_4\times_A M_6\,,
	\end{equation}
	where $M_4$ is the $(3+1)$-dimensional spacetime and is referred to as the \textit{external space}, while $M_6$ is assumed to be a compact 6-dimensional manifold and is referred to as the \textit{internal space}. We also assume that $M_4$ is maximally symmetric, so that it corresponds to four-dimensional de Sitter, Minkowski or anti de Sitter space. 
	\item \textbf{4D Poincar\'e invariance:}\\
	We assume that the solution preserves four-dimensional Poincar\'e invariance. This has a number of implications:
	\begin{itemize}
		\item Denoting by $x^\mu,\,\mu=0,\ldots, 3$ and $y^m,\,m=1,\ldots, 6$ two sets of local coordinates on $M_4$ and $M_6$, respectively, the most general ansatz for the metric that is compatible with the (warped) product structure \eqref{eq:spacetime_decomposition} is given by
		\begin{equation}\label{eq:metric_warped}
			ds^2_{10} = e^{2A(y)}\tilde{g}_{\mu\nu}^{(4)} \mathrm{d}x^\mu\otimes \mathrm{d}x^\nu+e^{-2A(y)}\tilde{g}_{mn}^{(6)}\mathrm{d}y^m\otimes\mathrm{d}y^n\,,
		\end{equation}
		where $A(y)$ is referred to as the \textbf{warp factor}. We stress that the warp factor can only depend on the internal coordinates in order to respect four-dimensional Poincar\'e invariance in the external space. 
		\item The self-dual 5-form flux $\tilde{F}_5$ must be of the form
		\begin{equation}
			\tilde{F}_5 = (1+\star)\,\mathrm{d\alpha}\wedge \mathrm{d}x^0\wedge\mathrm{d}x^1\wedge\mathrm{d}x^2\wedge\mathrm{d}x^3\,,
		\end{equation}
		where $\alpha=\alpha(y)$ is an arbitrary function of the internal coordinates and we have the self-duality property \eqref{eq:Fhat_duality} manifest.
		\item The 3-form flux $G_3$ can only have internal components. 
		\item The axio-dilaton $\tau=\tau(y)$ can only depend on the internal coordinates.  
	\end{itemize}	
	\item \textbf{No localized sources:}\\
	We assume that there are no localized sources for the fluxes. 
\end{enumerate}
To summarize, we consider warped compactifications of type IIB supergravity down to four dimensions that preserve Poincar\'e invariance, in the absence of localized sources. Although this appears to be a relatively large class of potential solutions (note that we have not imposed any supersymmetry conditions), it turns out that the actual solutions satisfying these assumptions are rather limited. To see this explicitly, one considers the  $(\mu\nu)$ components of the (trace-reversed) Einstein equations for the metric ansatz \eqref{eq:metric_warped}. Subsequently taking the trace yields
\begin{equation}\label{eq:IIB_no-go}
	\tilde{\Delta}_{(6)} e^{4A} = e^{-2A}\tilde{R}^{(4)}+ \frac{e^{2A}}{2\,\mathrm{Im}\,\tau}|G_3|^2 + e^{-6A}\left[(\partial\alpha)^2+(\partial e^{4A})^2 \right]\,,
\end{equation}
where $\tilde{R}^{(4)}$ denotes the Ricci scalar of the metric $\tilde{g}_{\mu\nu}^{(4)}$. Notably, for Minkowski and de Sitter space this quantity is non-negative. Therefore, since the left-hand side of \eqref{eq:IIB_no-go} integrates to zero on the \textit{compact} internal space $M_6$, one arrives at the following conclusion.
\begin{subbox}{Maldacena--Nu\~nez no-go theorem \cite{Maldacena:2000mw}}
	Under the stated assumptions, the only solutions to the type IIB supergravity equations of motion are anti-de Sitter and Minkowski space. Furthermore, in the latter case the $F_3$, $H_3$ and $F_5$ fluxes must have vanishing expectation value, and the warp factor must be constant.
\end{subbox}
\noindent From this point onwards, we only consider the Minkowski background. In this case the remaining equations of motion drastically simplify to
\begin{equation}\label{eq:eom_IIB_no-fluxes}
	\mathrm{d}\star\mathrm{d}\tau = \frac{1}{2}\frac{\mathrm{d}\tau\wedge\star\,\mathrm{d}\bar{\tau}}{\tau-\bar{\tau}}\,,\qquad R_{MN} = \frac{1}{2}\frac{\partial_M\tau\,\partial_N\bar{\tau}}{(\mathrm{Im\,\tau})^2}\,.
\end{equation}
A particularly simple solution is thus obtained by setting the axio-dilaton $\tau$ to be constant, in which case we are left with the condition that the internal space is Ricci-flat. The four-dimensional low-energy effective description of such solutions will be discussed in section \ref{sec:IIB_compactification}. More general solutions with a non-trivial profile for the axio-dilaton will instead be discussed in section \ref{sec:F-theory} and form the basis for introducing the framework of F-theory.

\subsection{Localized sources: D-branes and O-planes}\label{subsec:D-branes_O-planes}
As a result of the Maldacena--Nu\~nez no-go theorem, we conclude that, in order to consider more general compactifications that involve non-zero fluxes and/or non-trivial warpings, it is necessary to include localized sources which contribute to the energy-momentum tensor in a suitable fashion. Luckily, string theory naturally hosts a large collection of such localized objects with both positive and negative tension: \textbf{D-branes} and \textbf{O-planes}. Additionally, one of the great insights of the second superstring revolution is that these objects carry R-R charge, which means that they are sources for the various $p$-form fields $C_p$. As a result of these features, D-branes and O-planes play a central role in the study of the string landscape.

\subsubsection*{D-branes}
At the most basic level, $p$-branes correspond to extended $(p+1)$-dimensional surfaces inside the target spacetime. There is a particular subset of those branes called \textbf{D$p$-branes}, to which the endpoints of fundamental open strings can attach themselves. Such D$p$-branes are special, because their properties are dictated by the excitations and interactions of the open strings that are attached to them. For example, by computing the one-loop annulus diagram of an open string stretched between two D-branes, one finds that the physical tension $\tau_{\mathrm{D}p}$ of a D$p$-brane is given by
\begin{equation}\label{eq:tension_Dp}
	\tau_{\mathrm{D}p} = \frac{T_{\mathrm{D}p}}{g_s} = \frac{1}{g_s}\frac{2\pi}{\ell_s^{p+1}}\,.
\end{equation}
Note, in particular, that in the limit of small string-coupling $g_s\ll 1$, the physical tension of a D$p$-brane becomes very large and thus the brane is effectively a rigid object. Therefore, in perturbative string theory, it is the strings which comprise the fundamental degrees of freedom of the theory, while D-branes should be regarded as \textit{non-perturbative} solitonic objects. At the same time, one also sees that in the strong-coupling regime $g_s\gg 1$ it becomes essential to include D-branes as the fundamental degrees of freedom. 
\\

\noindent One of the great insights in the study of D-branes is that the open string excitations can be described in terms of the dynamics of the brane itself. Said differently, while the actions \eqref{eq:action_IIB_democratic} and \eqref{eq:action_IIB_SL2} provide a low-energy effective description of the massless closed string excitations, one can describe the dynamics of the massless open string excitations in terms of an action on the \textit{worldvolume} of the D-brane. To leading order in the string coupling, this action consists of the following two terms
\begin{equation}\label{eq:action_Dp}
	S_{\mathrm{D}p} = S_{\mathrm{D}p, \mathrm{DBI}}+S_{\mathrm{D}p, \mathrm{CS}}\,.
\end{equation}
The first term corresponds to the Dirac--Born--Infeld (DBI) action, which describes the coupling of the massless open string fields to the massless closed string fields in the NS-NS sector. For a single D$p$-brane, the bosonic part of the DBI action is given (in string frame) as follows.

\begin{subbox}{Dirac--Born--Infeld (DBI) action D$p$-brane}
\begin{equation}\label{eq:action_Dp_DBI}
		S_{\mathrm{D}p, \mathrm{DBI}} = -T_{\mathrm{D}p}\int_{\mathrm{D}p} d^{p+1}\xi\,e^{-\Phi}\sqrt{-\mathrm{det}\left(G_{\alpha\beta}+2\pi\alpha' \mathscr{F}_{\alpha\beta} \right)}\,.
\end{equation}
\tcblower 
The string-frame tension $T_{\mathrm{D}p}$ was introduced in equation \eqref{eq:tension_Dp}, the integral is taken over the $(p+1)$-dimensional worldvolume of the D$p$-brane, with local coordinates $\xi^\alpha$ for $\alpha=0,\ldots, p$, and we have introduced the notation
\begin{equation}
	2\pi \alpha' \mathscr{F}_2 = B_2 + 2\pi\alpha' F_2\,,
\end{equation}
for the gauge-invariant combination of the Kalb--Ramond two-form field $B_2$ and the field strength $F_2=\mathrm{d}A_1$ of the open string gauge field $A_1$. 
\end{subbox}
\noindent Assuming a constant dilaton and vanishing worldvolume flux $\mathscr{F}_2$, the DBI-action is simply proportional to the volume of the D$p$-brane, measured in terms of the metric induced by $G_{\alpha\beta}$. If the worldvolume flux $\mathscr{F}_2$ is non-trivial, then to first order in $\alpha'$ this is described by a $\mathrm{U}(1)$ gauge-theory on the brane. 
\\

\noindent The second term in \eqref{eq:action_Dp} corresponds to the Chern--Simons action for a D$p$-brane, which describes the coupling of the massless open string fields to the massless closed string fields in the R-R sector, namely the $p$-form gauge fields $C_p$.

\begin{subbox}{Chern--Simons (CS) action D$p$-brane}
	\begin{equation}\label{eq:action_Dp_CS}
			S_{\mathrm{D}p, \mathrm{CS}} = T_{\mathrm{D}p}\int_{\mathrm{D}p} \mathrm{ch}\left(-i\ell_s^2\mathscr{F}_2\right)\wedge \sqrt{\frac{\hat{A}(\ell_s^2R_T)}{\hat{A}(\ell_s^2R_N)}}\wedge \bigoplus_{q} C_q\,.
	\end{equation}
\tcblower 
Here $\mathrm{ch}(F)$ denotes the Chern character
\begin{equation}
	\mathrm{ch}(F) = \sum_{n=0}^\infty \mathrm{ch}_n(F)\,,\qquad \mathrm{ch}_n(F) = \frac{1}{n!}\mathrm{Tr}\left[\left(\frac{i F}{2\pi} \right)^n \right]\,.
\end{equation}
Furthermore, $R_T$ and $R_N$ respectively denote the curvature 2-forms on the tangent bundle and normal bundle of the brane worldvolume, and $\hat{A}(R)$ denotes the A-roof genus
\begin{equation}\label{eq:A_roof}
	\hat{A}(R) = 1 -\frac{p_1(R)}{24}+\cdots = 1+\frac{1}{24}\frac{\mathrm{Tr}\left(R\wedge R\right)}{8\pi^2}+\cdots \,.
\end{equation}
\end{subbox}
\noindent It is important to stress that the CS action \eqref{eq:action_Dp_CS} involves a (formal) sum over \textit{all} gauge fields $C_q$. Effectively, this means that in a non-trivial background a D$p$-brane not only couples to $C_{p+1}$, but possibly also to other $C_q$ with $q<p$. Indeed, by expanding the action \eqref{eq:action_Dp_CS}, one finds that the following terms arise:
\begin{align}
	S_{\mathrm{D}p,\mathrm{CS}}=\qquad\, T_{\mathrm{D}p}&\int_{\mathrm{D}p} C_{p+1} \label{eq:Dp_coupling}\\
	+T_{\mathrm{D}(p-2)}&\int_{\mathrm{D}p} C_{p-1}\wedge \mathrm{ch}_1\left(-i\mathscr{F}_2\right) \label{eq:D(p-2)_coupling}\\
	-T_{\mathrm{D}(p-4)}&\int_{\mathrm{D}p} C_{p-3}\wedge \left[\mathrm{ch}_2(\mathscr{F}_2) +\frac{p_1(R_T)-p_1(R_N)}{48} \right]+\cdots \label{eq:D(p-4)_coupling}\,,
\end{align}
where we leave out higher-order terms as they will not play an important role. Let us briefly comment on the interpretation of these first three terms. 
\begin{itemize}
	\item \textbf{Interpretation of \eqref{eq:Dp_coupling}:} \\
	This term simply states that a D$p$-brane is a source for $C_{p+1}$.
	\item \textbf{Interpretation of \eqref{eq:D(p-2)_coupling}:} \\
	This term implies that if a D$p$ brane wraps a 2-cycle in the internal space which carries a $\mathrm{U}(1)$-bundle with a non-trivial first Chern class, then (from the four-dimensional perspective) this is interpreted as a contribution to the effective D$(p-2)$ brane charge.
	\item \textbf{Interpretation of \eqref{eq:D(p-4)_coupling}:} \\
	This term consists of two qualitatively different contributions. The first contribution implies that if a non-trivial instanton configuration is excited on a 4-cycle that is wrapped by the D$p$-brane, then this provides exactly one unit of negative D$(p-4$)-brane charge. Additionally, from the second contribution one sees that if the wrapped 4-cycle has a non-trivial first Pontryagin class $p_1$, then there is also a curvature-induced D$(p-4$)-brane charge. For example, wrapping a D$p$-brane on a K3 surface also yields exactly one unit of negative D$(p-4)$-brane charge.
\end{itemize}

\subsubsection*{O-planes}
Just like a D$p$-brane, an O$p$-plane corresponds to an extended $(p+1)$-dimensional surface inside the target spacetime. However, in constrast to D$p$-branes, O$p$-planes correspond to the fixed loci of certain involutive actions on the target space. As a result, they are not dynamical objects (in particular, they do not host a gauge theory on their worldvolume). Nevertheless, they turn out to have a (negative) tension, which means that they couple to gravity and thus backreact on the geometry, and they additionally turn out to be a source for the R-R fields $C_p$. The action for O$p$-planes is very similar to that of D$p$-branes, again consisting of two terms
\begin{equation}\label{eq:action_Op}
	S_{\mathrm{O}p} = S_{\mathrm{O}p, \mathrm{DBI}}+S_{\mathrm{O}p, \mathrm{CS}}\,.
\end{equation}
Let us discuss these two terms in turn. The first term again corresponds to a DBI-like action
\begin{subbox}{Dirac--Born--Infeld (DBI) action O$p$-plane}
\begin{equation}\label{eq:action_Op_DBI}
	S_{\mathrm{O}p, \mathrm{DBI}}=-T_{\mathrm{O}p}\int_{\mathrm{O}p} d^{p+1}\xi\, e^{-\Phi}\sqrt{-\det\left(G_{\alpha\beta}\right)}\,.
\end{equation}
\tcblower
Here the integral is taken over the $(p+1)$-dimensional worldvolume of the O$p$-plane, and the tension $T_{\mathrm{O}p}$ of an O$p$ plane is related to that of a D$p$-brane by
\begin{equation}\label{eq:tension_Op}
	T_{\mathrm{O}p} = - 2^{p-5} T_{\mathrm{D}p}\,.
\end{equation}
In particular, O-planes have a \textit{negative} tension.
\end{subbox}
\noindent Note, however, that the action \eqref{eq:action_Op_DBI} does not include the worldvolume flux $\mathscr{F}_2$ hence an O-plane indeed does not host a gauge-theory on its worldvolume.  
\\

\noindent The CS action for an O$p$-plane also takes a form similar to that of a D$p$-brane.
\begin{subbox}{Chern--Simons (CS) action O$p$-plane}
\begin{equation}\label{eq:action_Op_CS}
		S_{\mathrm{O}p, \mathrm{CS}} = T_{\mathrm{O}p}\int_{\mathrm{O}p} \sqrt{\frac{L(\ell_s^2R_T/4)}{L(\ell_s^2R_N/4)}}\wedge\bigoplus_q C_q\,.
\end{equation}
Here the integral is taken over the $(p+1)$-dimensional worldvolume of the O$p$-plane, $R_T$ and $R_N$ respectively denote the curvature 2-forms on the tangent bundle and normal bundle of the O$p$-plane worldvolume, and  $L(R)$ denotes the Hirzebruch $L$-polynomial
\begin{equation}
	L(R) = 1 - \frac{1}{48\pi^2}\mathrm{Tr}\left(R\wedge R\right)+\cdots\,,
\end{equation}
where the dots denote higher orders in the curvature.
\end{subbox}
\noindent Just like for a D$p$-brane, one finds that an O$p$-plane is not just a source for $C_{p+1}$, but can also give rise to effective lower-dimensional R-R charges by including the $\alpha'$ corrections to the Chern--Simons action. Explicitly expanding the action \eqref{eq:action_Op_CS} and using the relation \eqref{eq:tension_Op} one finds
\begin{align}
	S_{\mathrm{O}p,\mathrm{CS}} =\qquad  -2^{p-5}T_{\mathrm{D}p}&\int_{\mathrm{O}p} C_{p+1}\\
	 - 2^{p-6} T_{\mathrm{D}(p-4)}&\int_{\mathrm{O}p}C_{p-3}\wedge \frac{p_1(R_T)-p_1(R_N)}{48}+\cdots\,,
\end{align}
where we have again left out the higher-order terms. The interpretation of the first correction is similar to the curvature correction for D$p$-branes, though it comes with a slightly different prefactor. Importantly, note that there is no term involving $C_{p-1}$ (as there was for a D$p$-brane) due to the fact that O-planes do not couple to the worldvolume flux $\mathscr{F}_2$. As an example, while a D5-brane can induce an effective D3-brane charge, an O5-plane cannot.

\subsection{Tadpole cancellation conditions}\label{subsec:tadpole_cancellation}
Combining the results of the previous two subsections, we conclude that the full low-energy effective action of type IIB string theory is really given by a combination
\begin{equation}
	S_{\mathrm{IIB+sources}} = S_{\mathrm{IIB}}+S_{\mathrm{sources}}\,,
\end{equation}
where $S_{\mathrm{IIB}}$ is given by \eqref{eq:action_IIB_democratic} and describes the massless excitations of the closed string, while $S_{\mathrm{sources}}$ describes the massless excitations of the open string through the combined backreaction of various D-branes and O-planes given by the actions \eqref{eq:action_Dp} and \eqref{eq:action_Op}. One point which deserves further attention is how the inclusion of the Chern--Simons terms for the D-branes and O-planes affects the equations of motion of the R-R $p$-form fields $C_p$. To be a bit more general, we consider compactifications down to $d$ dimensions and write
\begin{equation}
	M_{10} = M_{10-d}\times M_{d}\,,
\end{equation}
with $M_d$ the $d$-dimensional internal compact space. Furthermore, in order to preserve Poincar\'e invariance we assume all D$p$-branes and O$p$-planes to be spacetime-filling with respect to the $(10-d)$-dimensional external space (when $d+p\geq 9$). Then the equations of motion of $C_{p+1}$ take the form
\begin{equation}\label{eq:tadpole_local}
	\mathrm{d}\star\mathrm{d}C_{p+1}+ \text{non-localized}+\ell_s^{7-p}\sum_{i}Q^{(i)}_p\, \mathrm{PD}[\Gamma_{d+p-9}^{(i)}]=0\,,
\end{equation}
where the non-localized contribution refers to any additional terms which arise from the variation of the closed-string effective action, while the remaining terms arise from D-branes and/or O-planes which wrap an internal $(d+p-9)$-cycle $\Gamma_{d+p-9}^{(i)}$ and thereby induce an effective D$p$-brane charge $Q_p^{(i)}$, as explained in the previous section. For example, a D3-brane induces $Q_{3}=+1$ while an O3-plane induces $Q_{3}=-\frac{1}{4}$. Furthermore, $\mathrm{PD}[\Gamma_{d+p-9}^{(i)}]$ denotes the Poincar\'e dual of $\Gamma_{d+p-9}^{(i)}$ with respect to the internal $d$-dimensional manifold, which yields a $(9-p)$-form. 
\\

\noindent The importance of \eqref{eq:tadpole_local} is it can lead to global restrictions upon integrating the equation over the space transverse to the cycles $\Gamma_{d+p-9}^{(i)}$. This is similar in spirit to how we arrived at the Maldacena--Nu\~nez no-go theorem in section \ref{subsec:sol_no_sources}. Indeed, if the space one is integrating over is \textit{compact}, then by Stokes' theorem the contribution coming from the exact form $\mathrm{d}\star\mathrm{d}C_{p+1}$ will vanish. One is then typically left with a global constraint on the charges $Q_p^{(i)}$ and possibly the fluxes coming from the non-localized term. Such relations are referred to as \textbf{tadpole cancellation conditions}. In this way, one finds (in the context of type IIB compactifications) tadpole cancellation conditions for D3/D5/D7/D9-branes. In the most general setting one should take into account the presence of all possible branes and orientifold planes, see for example \cite{Plauschinn:2008yd} for a detailed overview. However, in this thesis we choose to make the following simplification.

\begin{redbox}{D-branes and O-planes}
	From this point onwards, we will only consider type IIB compactifications which include D3/D7-branes and O3/O7-planes, unless stated otherwise.
\end{redbox}

\subsubsection*{D3 tadpole cancellation}
Let us consider the case $d=6$ and $p=3$, for which the transverse space is the whole 6-dimensional internal space $M_6$. For simplicity, we only include the contribution coming from spacetime filling D3-branes and O3-planes. Then one finds
\begin{equation}
	\frac{2\pi}{\ell_s^8}\int_{M_6}\frac{1}{2}\left(\mathrm{d}\star\tilde{F}_5 +F_3\wedge H_3\right)+\frac{2\pi}{\ell_s^4}\left(N_{\mathrm{D}3}-\frac{1}{4}N_{\mathrm{O}3} \right)=0\,,
\end{equation}
where we have used \eqref{eq:tension_Op} for $p=3$ and we denote by $N_{\mathrm{D}3}$ and $N_{\mathrm{O}3}$ the total number of D3-branes and O3-planes, respectively. Using the fact that $M_6$ is compact, Stokes' theorem implies that the contribution from $\tilde{F}_5$ will vanish, and hence one is left with the following. 
\begin{subbox}{D3 tadpole cancellation condition}
\begin{equation}\label{eq:D3_tadpole}
	N_{\mathrm{D}3}+\frac{1}{2\ell_s^4}\int_{M_6} F_3\wedge H_3 = \frac{1}{4}N_{\mathrm{O}3}\,.
\end{equation}
\tcblower
\textbf{Note:} \\
There can be additional contributions to \eqref{eq:D3_tadpole} coming from D5/D7/D9-branes and non-trivial configurations of the worldvolume flux $\mathscr{F}_2$ on the branes cf.~\eqref{eq:action_Dp_CS}, as well as from O7-planes cf.~\eqref{eq:action_Op_CS}. For simplicity of exposition these have not been included explicitly.
\end{subbox}
\noindent Intuitively, this relation states that the positive contributions to the D3-charge (corresponding to the left-hand side) must match the negative contributions to the D3-charge (corresponding to the right-hand side), such that the net charge is equal to zero, as it must be on a compact space. Let us also recall that in string theory (as opposed to just supergravity) fluxes have to satisfy the following quantization condition.
\begin{subbox}{Dirac quantization}
	\begin{equation}\label{eq:Dirac_quantization}
		\frac{1}{\ell_s^p}\int_{\Gamma_{p+1}}F_{p+1}\in\mathbb{Z}\,,
	\end{equation}
	for any $(p+1)$-cycle $\Gamma_{p+1}$. 
	\tcblower
	\textbf{Note:}\\
	In particular, this implies that a $p$-form flux $F_p$ has to be an element of the integer cohomology group $H^p(M_6,\mathbb{Z})$ of the compact internal space.
\end{subbox}

\subsubsection*{D7 tadpole cancellation}

For completeness, let us also record the \textbf{D7 tadpole cancellation condition}, which is given by (here we again take $d=6$)
\begin{subbox}{D7 tadpole cancellation condition}
\begin{equation}\label{eq:D7_tadpole}
	\sum_{\mathrm{D}7_a }N_{\mathrm{D7}_a}\, \mathrm{PD}\left[\Gamma_{\mathrm{D}7_a}\right] = 4\sum_{\mathrm{O}7_b} \mathrm{PD}\left[\Gamma_{\mathrm{O}7_b}\right]\,.
\end{equation}
Here have included the possibility that there are multiple stacks (labeled by the index $a$) of $N_{\mathrm{D}7_a}$ D7-branes wrapping internal 4-cycles $\Gamma_{\mathrm{D}7_a}$, as well as multiple O7-planes (labeled by the index $b$) wrapping various internal 4-cycles $\Gamma_{\mathrm{O}7_b}$.
\tcblower 
\textbf{Note:}\\
There can be additional contributions to \eqref{eq:D7_tadpole} coming from D9-branes with a non-trivial worldvolume flux $\mathscr{F}_2$ cf.~\eqref{eq:action_Dp_CS}.
\end{subbox}
\noindent As an example, if one considers a single O7-plane, then in order for the D7 tadpole cancellation condition to be satisfied, one has to include exactly 4 D7-branes in the same homology class.

\section{Type IIB Calabi--Yau compactifications}\label{sec:IIB_compactification}
In section \ref{subsec:sol_no_sources} we have found that compactifications of type IIB supergravity down to four dimensions in the absence of localized sources are relatively simple to describe, in particular they involve a Ricci-flat internal space (assuming a constant axio-dilaton). In this section, we present an alternative point of view on such solutions by first performing the dimensional reduction of the ten-dimensional type IIB supergravity action and subsequently studying the resulting four-dimensional effective theory. For simplicity we focus on compactifications that preserve some amount of supersymmetry. In section \ref{subsec:KK_reduction}, we present a brief motivation and explanation for how the process of Kaluza--Klein compactification yields an effective four-dimensional description of the theory. Additionally, we recall the supersymmetry conditions which ultimately require the internal space to be a Calabi--Yau manifold. In section \ref{subsec:CY_moduli} we discuss how this typically gives rise to families of solutions due to the fact that the Calabi--Yau manifold can vary in moduli. Finally, in section \ref{subsec:D=4_N=2_SUGRA} we present the important features of the resulting four-dimensional $\mathcal{N}=2$ supergravity theory that arises from performing the compactification. We pay special attention to how the geometry of the internal Calabi--Yau manifold dictates the physical couplings of the effective theory, as this will play a central role in the remainder of the thesis.

\subsection{Motivation}
\label{subsec:KK_reduction}
It is time to address the elephant in the room: we have been considering string backgrounds which involve a \textit{ten-dimensional} target space, while our world certainly appears to be only four-dimensional! From an abstract point of view, this is not as big of a deal as it may seem. Indeed, one could take the stance that the actual CFT defining the string background in consideration really consists of two pieces, one describing the propagation of a string in some four-dimensional target space which we identify with the physical spacetime in which we live, while the other is some compact CFT that happens to have a geometric description.
\\

\noindent It would, however, be more natural if the additional internal dimensions do in fact correspond to real, physical dimensions. In that case, in order not to be in tension with observations, one would have to conclude that the typical size of the extra dimensions is too small to be accessible to current-age measurements. This is admittedly a large assumption to make, but it also comes with some practical benefits. Indeed, if the size of the internal manifold is small compared to the length scales we would like to describe, it is not necessary to take into account the full ten-dimensional description of the theory. Instead, it suffices to use a four-dimensional effective description, which we briefly describe in the following.

\subsubsection*{Dimensional reduction}
Whenever a given field $\phi(x,y)$ has a momentum component along the internal directions, this is perceived from the four-dimensional point of view as an effective mass term. In the case of a free scalar field, this can be seen explicitly by writing its equation of motion as
\begin{equation}
	\Delta_{(10)}\phi= \Delta_{(4)} \phi + \Delta_{(6)}\phi=0\,.
\end{equation}
Indeed, decomposing $\phi(x,y)$ in terms of eigenmodes of the six-dimensional Laplacian $\Delta_{(6)}$, which amounts to a Fourier decomposition into different momentum modes
\begin{equation}\label{eq:KK_decomp}
	\phi(x,y) = \sum_{k} \phi_k(x) u_k(y)\,,\qquad \Delta_{(6)}u_k = k^2 u_k\,,
\end{equation}
one finds that each of the momentum modes or \textbf{Kaluza--Klein  (KK) modes} $\phi_k(x)$ satisfies the equation
\begin{equation}
	\left(\Delta_{(4)}+k^2\right)\phi_k(x) = 0\,,
\end{equation}
such that the eigenvalue $k$ acts as an effective mass term for the four-dimensional field $\phi_k(x)$. Explicit computation of the eigenvalues of the six-dimensional Laplacian can be very involved, as this requires detailed knowledge of the internal metric. A rough estimate of the typical mass scale of the Kaluza--Klein modes is that it scales inversely with the size of the internal manifold, and is hence large in the regime where the internal dimensions are small. For example, in the case of a one-dimensional reduction on a circle of radius $R$, the mass scale is given by
\begin{equation}
	m_{\mathrm{KK},n} = \frac{n}{R}\,,\qquad n\in\mathbb{N}\,.
\end{equation}
Therefore, if we are considering energy scales much below this typical mass scale, it suffices to consider those field configurations which do not have any internal momentum. In other words, this corresponds to considering those configurations which are the zero-modes of the six-dimensional Laplacian. This can equivalently be captured in the dimensional reduction of the higher-dimensional action
\begin{equation}\label{eq:KK_EFT}
	S = \int_{M_{10}}\mathrm{d}\phi\wedge\star\,\mathrm{d}\phi = \int_{M_{4}}\mathrm{d}\phi_0\wedge\star\,\mathrm{d}\phi_0 +\cdots\,,
\end{equation}
which follows from inserting the decomposition \eqref{eq:KK_decomp} into the ten-dimensional action and performing the integration over the internal manifold $M_6$ using the orthogonality of the eigenfunctions $u_k(y)$ and neglecting the massive higher-momentum modes which are contained in the dots. The leading term in \eqref{eq:KK_EFT} is what is meant by the \textit{four-dimensional} low energy effective action. To summarize:
\begin{subbox}{Dimensional reduction}
 	The dimensional reduction of a $D$-dimensional theory of a field $\phi(x,y)$ over a $d$-dimensional compact space is a $(D-d)$-dimensional effective theory that describes those configurations which are associated with (internal) \textit{harmonic} fields:
 	\begin{equation}
 		\Delta_{(d)}\phi = 0\,.
 	\end{equation}
 	Physically, these configurations are the ones which have no internal momentum and thus remain massless in the $(D-d)$-dimensional effective theory.
\end{subbox}
\noindent Above we have exemplified the idea of dimensional reduction in the case of a scalar field, but the same logic also applies to higher tensor fields such as the various $p$-form fields of type IIB supergravity, as we will see shortly.  

\subsubsection*{Supersymmetry conditions}
At this point, one may be tempted to simply apply the above recipe to the ten-dimensional type IIB supergravity action in order to obtain an effective four-dimensional description. However, without making some additional assumptions it is unfeasible to do this in practice. For completely general backgrounds it is very difficult to say anything about the solutions to the Laplace equation (for higher-forms at least) and thus it is not even clear how many degrees of freedom the lower-dimensional effective theory will contain. Additionally, while the starting point of type IIB supergravity has a great deal of supersymmetry (32 supercharges), a generic configuration will typically break supersymmetry and thus the resulting theory can be rather wild.
\\

\noindent One guiding principle which allows one to keep a handle on things is to impose that the resulting theory does preserve some amount of supersymmetry. To be precise, we will consider compactifications down to four dimensions in which $\frac{1}{4}$ of the supercharges is preserved (leaving 8 residual supercharges in total), which thus give rise to four-dimensional $\mathcal{N}=2$ supergravity theories. While such theories have too much supersymmetry to be phenomenologically viable, they will serve as an important first step. Indeed, the motivation for considering supersymmetric solutions is analogous to the reason we study superstrings. Namely, such solutions typically give rise to stable configurations, which are furthermore amenable to concrete computations due to their restricted nature. At the same time, there are well-controlled scenarios in which supersymmetry is further broken upon inclusion of non-zero fluxes, as will be discussed in part \ref{part3} of the thesis. The resulting four-dimensional $\mathcal{N}=1$ theories provide an excellent arena to study potentially interesting phenomenological implications of string theory.
\\

\noindent 
The following is a well-known result which relies on a detailed analysis of the supersymmetry transformations of type IIB supergravity:
\begin{subbox}{Supersymmetric compactifications of type IIB}
	Consider warped compactifications of type IIB supergravity on 
	\begin{equation}
		M_{10} = M_{4}\times M_{6}\,,
	\end{equation}
	with metric
	\begin{equation}
		ds^2_{10} = e^{2A(y)}g_{\mu\nu}\,\mathrm{d}x^\mu\otimes\mathrm{d}x^\nu + g_{mn}\,\mathrm{d}y^m\otimes\mathrm{d}y^n\,,
	\end{equation}
	where $g_{\mu\nu}$ is maximally symmetric, i.e.~either de Sitter, Minkowski or anti-de Sitter. Then \textit{in the absence of fluxes} this compactification preserves supersymmetry if and only if the following conditions are satisfied
	\begin{itemize}
		\item $A(y)$ is constant,
		\item $M_4$ is Minkowski,
		\item $M_6$ has holonomy contained in $\mathrm{SU}(3)$ .
	\end{itemize}
\end{subbox}
\noindent If the holonomy group is exactly $\mathrm{SU}(3)$, then indeed $\frac{1}{4}$ of the supercharges is preserved. If the holonomy group reduces even further, then more supersymmetry is preserved. Manifolds with $\mathrm{SU}(3)$ holonomy are known as Calabi--Yau manifolds. Thus, the only possible supersymmetry-preserving solutions (in the absence of fluxes) consist of an unwarped external Minkowski spacetime together with an internal compact Calabi--Yau threefold. It should be noted that the celebrated theorem of Yau ensures that Calabi--Yau manifolds admit a Ricci-flat metric, so that the conditions coming from supersymmetry are compatible with the supergravity equations of motion if one furthermore takes the axio-dilaton to be constant cf.~\eqref{eq:eom_IIB_no-fluxes}. Indeed, note that the resulting solution is a special case of the one discussed in section \ref{subsec:sol_no_sources}. 

\subsection{Moduli spaces of Calabi--Yau manifolds}
\label{subsec:CY_moduli}
Before moving toward the four-dimensional effective theory that results from compactifying type IIB on a Calabi--Yau threefold, let us first discuss a very important point. Namely, from the 10-dimensional point of view, in principle any choice of Calabi--Yau threefold will provide a valid solution to the equations of motion. Typically, Calabi--Yau manifolds come in \textit{families}, meaning that they can depend on additional parameters which will be referred to as \textit{moduli}. In the four-dimensional effective theory, such parameters will correspond to massless scalar fields. The fact that they are massless is roughly due to the fact that there is no energetic obstruction to deforming shape and size of the underlying Calabi--Yau manifold. The following discussion closely follows \cite{Candelas:1991} as well as the lecture notes \cite{Greene:1996cy,Bouchard:2007ik}.
\\

\noindent One way to see how such additional moduli can arise is by noting that there may exist small deformations of the metric $g\mapsto g+\delta g$ which preserve the Ricci-flatness condition. There are two possible types of such deformations, corresponding to the two different index structures:
\begin{equation}
	\delta g = \delta g_{ij}\,\mathrm{d}y^i\otimes \mathrm{d}y^j+ \delta g_{i\bar{\jmath}}\,\mathrm{d}y^i\otimes \mathrm{d}y^{\bar{\jmath}}+\text{complex conjugate}\,.
\end{equation}
The deformations corresponding to $\delta g_{ij}$ (and its complex conjugate) are referred to as \textit{complex structure deformations}, while those corresponding to $\delta g_{i\bar{\jmath}}$ are referred to as \textit{K\"ahler deformations}. Let us discuss these in turn. For simplicity, we restrict to the case of Calabi--Yau threefolds, which will be denoted by $Y_3$, but the statements naturally generalize to arbitrary Calabi--Yau manifolds.  

\subsubsection*{Complex structure deformations}
The condition that the deformation induced by $\delta g_{\alpha\beta}$ (and its complex conjugate) preserves the Ricci-flatness is very restrictive. To be precise, it is equivalent to the statement that the following $(2,1)$-form 
\begin{equation}
	\tensor{\Omega}{_{\gamma\delta}^{\bar{\beta}}} \delta g_{\bar{\alpha}\bar{\beta}} \,\mathrm{d}y^\gamma\wedge\mathrm{d}y^\delta\wedge \mathrm{d}\bar{y}^{\bar{\alpha}}\,,
\end{equation}
is harmonic. Here $\Omega = \Omega_{\alpha\beta\gamma}\mathrm{d}y^\alpha\wedge\mathrm{d}y^\beta\wedge\mathrm{d}y^\gamma$ denotes the (up to scaling) unique $(3,0)$-form on the Calabi--Yau threefold. Roughly speaking, one can think of the harmonicity condition as the generalization of the arguments given for the scalar field $\phi(x,y)$ in previous section to the metric. One of the crucial properties of compact Calabi--Yau manifolds (or, more generally, compact K\"ahler manifolds) is that harmonic $(2,1)$-forms are in one-to-one correspondence with cohomology classes in the Dolbeault cohomology group $H^{2,1}(Y_3)$. As a result, the number of independent complex structure deformations is determined by the Hodge number $h^{2,1}=\mathrm{dim}\,H^{2,1}(Y_3)$, which is a topological quantity. Correspondingly, the $h^{2,1}$ complex parameters which parameterize these deformations are referred to as \textit{complex structure moduli}, and will be denoted by $z^i$, for $i=1,\ldots, h^{2,1}$. To be precise, the relation between $\delta g_{\alpha\beta}$ and $z^i$ is given by
\begin{equation}\label{eq:cs_deformation}
	\delta g_{\alpha\beta} = i\sum_{i=1}^{h^{2,1}} \bar{z}^i(x) \left(\frac{\left(\bar{b}_i\right)_{\alpha\bar{\gamma}\bar{\delta}}\tensor{\Omega}{^{\bar{\gamma}\bar{\delta}}_{\beta}}}{||\Omega||^2} \right)\,,\qquad ||\Omega||^2 = \frac{1}{3!}\Omega_{\alpha\beta\gamma}\bar{\Omega}^{\alpha\beta\gamma}\,,
\end{equation}
where $b_i$ is a basis of harmonic $(2,1)$-forms. Note that because the metric deformations $\delta g_{\alpha\beta}$ are non-Hermitian (with respect to the original complex structure), one can alternatively interpret these deformations as deformations of the complex structure of the Calabi--Yau threefold, hence the name. 
\\

\noindent  A theorem of Bogomolov--Tian--Todorov \cite{Bogomolov:1978,Tian:1987,Todorov:1989} ensures that the infinitesimal deformations considered above can be integrated to finite deformations -- it is said that the deformation space is unobstructed. This gives rise to a \textit{complex structure moduli space} $\mathcal{M}_{\mathrm{cs}}$ of complex dimension $h^{2,1}$ in which the complex structure moduli take value. In fact, the complex structure moduli space carries the structure of a K\"ahler manifold, which implies that it admits a metric $G^{\mathrm{cs}}_{i\bar{\jmath}}$ which is locally expressed in terms of a K\"ahler potential
\begin{equation}\label{eq:metric_cs}
	G^{\mathrm{cs}}_{i\bar{\jmath}} = \partial_i \partial_{\bar{\jmath}} K^{\mathrm{cs}}\,,\qquad K^{\mathrm{cs}} = -\log\left[i\int_{Y_3}\Omega\wedge\bar{\Omega} \right]\,,
\end{equation}
where the derivatives are taken with respect to the complex structure moduli. 

\subsubsection*{K\"ahler deformations}
The deformations parametrized by $\delta g_{\alpha\bar{\beta}}$ correspond to deformations of the K\"ahler class 
\begin{equation}
	J = i g_{\alpha\bar{\beta}}\,\mathrm{d}y^\alpha\wedge\mathrm{d}\bar{y}^{\bar{\beta}}\,,
\end{equation}
of the Calabi--Yau manifold, which is a harmonic $(1,1)$-form. Similarly to the complex structure deformations, the K\"ahler deformations can thus be parametrized by $h^{1,1}=\mathrm{dim}\,H^{1,1}(Y_3)$ \textit{real} parameters $v^A$, for $A=1,\ldots, h^{1,1}$, whose relation to $\delta g_{\alpha\bar{\beta}}$ is given by
\begin{equation}\label{eq:Kahler_deformation}
	\delta g_{\alpha\bar{\beta}} = i\sum_{A=1}^{h^{1,1}} v^A (\omega_A)_{\alpha\bar{\beta}}\,,
\end{equation}
where $\omega_A$ is a basis of harmonic $(1,1)$-forms.
\\

\noindent As for the complex structure deformations, the K\"ahler deformations can be integrated to finite deformations, giving rise to the \textit{K\"ahler moduli space} $\mathcal{M}_{\mathrm{ks}}(Y_3)$ of real dimension $h^{1,1}$. Typically, one considers the complexification of the K\"ahler moduli space by defining the complexified K\"ahler moduli
\begin{equation}
	T^A = b^A+ i v^A\,,
\end{equation}
where $b^A$ label the components of $B_2$ in the basis $\omega_A$. The advantage of this complexification is that the resulting complexified K\"ahler moduli space becomes itself a K\"ahler manifold, just like the complex structure moduli space, with a K\"ahler metric given by
\begin{equation}
	G_{A\bar{B}} = \partial_A \bar{\partial}_{\bar{B}} K^{\mathrm{ks}}\,,\qquad K^{\mathrm{ks}} = -\log\left[\mathcal{K}_{ABC}v^A v^B v^C \right] = -\log\left[6\mathcal{V}_{Y_3}\right]\,,
\end{equation}
where
\begin{equation}
	\mathcal{K}_{ABC} = \int_{Y_3}\omega_A\wedge\omega_B\wedge\omega_C\,,
\end{equation}
denote the triple-intersection numbers of the Calabi--Yau threefold and $\mathcal{V}_{Y_3}$ denotes its volume. 

\subsection{$D=4$, $\mathcal{N}=2$ SUGRA from type IIB}
\label{subsec:D=4_N=2_SUGRA}
We are now ready to discuss the dimensional reduction of type IIB supergravity on Calabi--Yau threefolds. We summarize the essential features that will be relevant for us, and refer the reader to \cite{Strominger:1990,Candelas:1991,Andrianopoli:1996cm,Craps:1997gp} for more extensive reviews. The main strategy will be to decompose all the supergravity fields in terms of the appropriate harmonic forms on the Calabi--Yau threefold. Below we first introduce some convenient bases for the various cohomology groups. For the convenience of the reader, we have summarized the various restrictions on the Hodge numbers of a Calabi--Yau threefold, summarized in the Hodge diamond in figure \ref{fig:Hodge}. 
\begin{figure}[h!]
	\centering
	\begin{tikzpicture}[baseline={([yshift=-.5ex]current bounding box.center)},scale=1,cm={cos(45),sin(45),-sin(45),cos(45),(15,0)}]
		\draw[step = 1, gray, ultra thin] (0, 0) grid (3, 3);
		\draw[fill] (3, 0) circle[radius=0.04] node[above]{\small $1$};
		\draw[fill] (0, 0) circle[radius=0.04] node[above]{\small $1$};
		\draw[fill] (2, 1) circle[radius=0.04] node[above]{\small $h^{2,1}$};
		\draw[fill] (1, 1) circle[radius=0.04] node[above]{\small $h^{1,1}$};
		\draw[fill] (2, 2) circle[radius=0.04] node[above]{\small $h^{1,1}$};
		\draw[fill] (1, 2) circle[radius=0.04] node[above]{\small $h^{2,1}$};
		\draw[fill] (3, 3) circle[radius=0.04] node[above]{\small $1$};
		\draw[fill] (0, 3) circle[radius=0.04] node[above]{\small $1$};
	\end{tikzpicture}
	\caption{Hodge diamond of a Calabi--Yau threefold.
	\label{fig:Hodge}}
\end{figure}
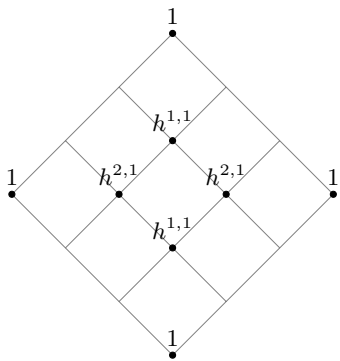

\subsubsection*{Cohomology bases}
We have already introduced the basis $\omega_A$ of harmonic $(1,1)$-forms. In a similar fashion, we denote by  $\tilde{\omega}^A$ the dual basis of harmonic $(2,2)$-forms, which can be chosen such that
\begin{equation}
	\int_{Y_3}\omega_A\wedge\tilde{\omega}^B = \delta_A^B\,.
\end{equation}
Furthermore, while we have already introduced the basis $b_i$ of harmonic $(2,1)$-forms, we will also need a basis of the full middle cohomology group $H^{3}(Y_3)$, which additionally contains the two harmonic $(3,0)$ and $(0,3)$ forms. In the following, it will be useful to employ the standard symplectic basis $\alpha_I,\beta^I$, for $I=0,\ldots, h^{2,1}$, defined by the relations
\begin{equation}
	\int_{Y_3} \alpha_I\wedge\beta^J = \delta_I^J\,,
\end{equation}
with all other pairings vanishing. 

\subsubsection*{Decomposition of 10-dimensional supergravity fields}
In order to perform the compactification, it is necessary to expand the various 10-dimensional supergravity fields according to the factorization $M_{10}=M_{4}\times Y_3$. In terms of the various bases described above, the most general decomposition in terms of harmonic forms is given by\footnote{Here we have already used the self-duality of $C_4$ to simplify its decomposition.}
\begin{align}
	\Phi(x,y)&= \Phi(x)\,, \nonumber \\
	B_2(x,y) &=B_2(x)+ b^A(x)\omega_A(y)\, \nonumber\\
	\label{eq:KK_decomp_IIB}
	C_0(x,y) &=C_0(x)\,,\\
	C_2(x,y) &=C_2(x)+c^A(x)\omega_A(y)\,,\nonumber\\
	C_4(x,y) &=V^I(x)\wedge\alpha_I(y) +\rho_A(x)\tilde{\omega}^A(y)\,,\nonumber
\end{align}
where $\Phi(x), C_0(x),b^A(x), c^A(x),\rho_A(x)$ are spacetime scalars, $V^I(x)$ are spacetime 1-forms and $B_2(x)$ is a spacetime 2-form. For future reference, let us note that these resulting 4-dimensional fields can be collected into supergravity multiplets as summarized in table \ref{table:multiplets}.
\begin{table}[h!]
\centering
\begin{tabular}{|c|c|c|}
	\hline
	\textbf{Gravity multiplet:} & $g_{\mu\nu}^{(4)}, V^0$  & \\
	\textbf{Vector multiplets:} & $V^i, z^i$ & $i=1,\ldots, h^{2,1}$\\
	\textbf{Hypermultiplets:} & $T^A, b^A, c^A,\rho_A$ & $A=1,\ldots, h^{1,1}$\\
	\textbf{Tensor multiplet:} &$ B_2, C_2, \Phi, C_0$ & \\ \hline
\end{tabular}
\caption{Overview of the various four-dimensional $\mathcal{N}=2$supergravity multiplets that arise from compactification of type IIB supergravity on a Calabi--Yau threefold. }
\label{table:multiplets}
\end{table}

\subsubsection*{Four-dimensional supergravity action}
For the remainder of this section, we will restrict our attention to the gravity multiplet and the vector multiplets. By inserting the expansions \eqref{eq:KK_decomp_IIB} into the low-energy effective action of type IIB, doing the integration over the internal Calabi--Yau threefold, and performing an appropriate Weyl rescaling of the metric to pass to the Einstein frame, one finds the action of four-dimensional $\mathcal{N}=2$ (ungauged) supergravity:

\begin{subbox}{$D=4,\mathcal{N}=2$ SUGRA from type IIB/$\mathrm{CY}_3$ (gravity+vector multiplets)}
	\begin{equation}\label{eq:action_IIB_4d}
		S_{\mathrm{IIB}}^{(4)} = \int\frac{1}{2}R\star 1-G^{\mathrm{cs}}_{i\bar{\jmath}}\,\mathrm{d}z^i\wedge\star\,\mathrm{d}\bar{z}^{\bar{\jmath}}+\frac{1}{4}\mathcal{I}_{IJ} F^I\wedge\star F^J - \frac{1}{4}\mathcal{R}_{IJ} F^I\wedge F^J\,.
	\end{equation}
\tcblower 
Here $R$ denotes the four-dimensional Ricci-scalar and $G^{\mathrm{cs}}_{i\bar{\jmath}}$ is the metric \eqref{eq:metric_cs} on the complex structure moduli space parametrized by the complex scalar fields $z^i$. Furthermore, we have denoted by $F^I=\mathrm{d}V^I$ the field strengths of the $\mathrm{U}(1)$ gauge fields.
\end{subbox}
\noindent The gauge-kinetic coupling matrices appearing in \eqref{eq:action_IIB_4d} can be recovered from the internal geometry by the following formula
\begin{align}
	\int_{Y_3} \begin{pmatrix}
		\alpha \\
		\beta
	\end{pmatrix} \wedge\star\begin{pmatrix}
	\alpha & \beta
	\end{pmatrix} = \begin{pmatrix}
	-\mathcal{I}-\mathcal{R}\mathcal{I}^{-1}\mathcal{R} & -\mathcal{R}\mathcal{I}^{-1}\\
	-\mathcal{I}^{-1}\mathcal{R} & -\mathcal{I}^{-1}
	\end{pmatrix}\,.
\end{align}
Notably, this means that the gauge-kinetic couplings are effectively determined by the Hodge inner product on the Calabi--Yau threefold. As a result, all of the couplings $G_{i\bar{\jmath}},\mathcal{I}_{IJ},\mathcal{R}_{IJ}$ appearing in the four-dimensional effective action are a \textit{function} of the complex structure moduli. Therefore, in order to understand the generic features of four-dimensional effective actions that originate from a UV complete theory of quantum gravity such as type IIB string theory, it is vital get a handle on this moduli dependence. 

\subsubsection*{Periods}
In order to make the moduli dependence of the various supergravity couplings more manifest, it is useful to introduce the notion of \textbf{periods}. Given an integral basis $\gamma_I\in H^{3}(Y_3,\mathbb{Z})$, the periods of the holomorphic 3-form $\Omega$ are a collection of functions $\Pi^I$ which are defined by
\begin{equation}
	\Omega = \Pi^I \gamma_I\,,\qquad \Pi^I = \int_{\mathrm{PD}[\gamma_I]}\Omega = \int_{Y_3}\Omega\wedge \gamma_I\,.
\end{equation}
In particular, the periods $\Pi^I(z)$ carry all the dependence on the complex structure moduli. In terms of the periods, the K\"ahler potential for the complex structure moduli can be written as
\begin{equation}
	K^{\mathrm{cs}} = -\log\left[i \Pi^I(z) S_{IJ}\bar{\Pi}^J(\bar{z}) \right]\,,
\end{equation}
where we have introduced the intersection form 
\begin{equation}
	S_{IJ} = \int_{Y_3} \gamma_I\wedge \gamma_J\,.
\end{equation}
Thus, through the relation \eqref{eq:metric_cs}, once the periods $\Pi^I(z)$ are known it is straightforward to compute the metric $G_{i\bar{\jmath}}^{\mathrm{cs}}$ that describes the kinetic coupling between the various complex structure moduli. Furthermore, assuming that the periods are expressed in a symplectic basis, such that
\begin{equation}
	\mathbf{\Pi} = \begin{pmatrix}
		X^I\\ -\mathcal{F}_I
	\end{pmatrix}\,,\qquad S  = \begin{pmatrix}
	 0 & +\mathbb{I}\\
	 -\mathbb{I} & 0
\end{pmatrix}\,,
\end{equation}
the gauge-kinetic coupling matrices $\mathcal{R}_{IJ}$ and $\mathcal{I}_{IJ}$ can be expressed as
\begin{equation}
	\mathcal{N}_{IJ} = \mathcal{R}_{IJ}+i\,\mathcal{I}_{IJ}\,,\qquad \mathcal{N}_{IJ} = \Big(
		\mathcal{F}_I, \overline{D_i\mathcal{F}_I}\Big)\,\left(X^J, \overline{D_i X^J} \right)^{-1}\,,
\end{equation}
where $D_i$ denotes the standard K\"ahler--Weyl covariant derivative.\newpage

\section{Basics of F-theory}\label{sec:F-theory}
In the previous section we have focused on compactifications of type IIB supergravity on relatively simple backgrounds. In particular, we have neglected the contributions from fluxes and localized sources. In this section we introduce the basics of F-theory \cite{Vafa:1996xn}, which is a framework which naturally incorporates the effects of D7-branes and a varying axio-dilaton in a geometric way. Additionally, it will serve as the basis for our discussion in section \ref{sec:F-theory_flux}, where we also include the effects from D3-branes and non-trivial 3-form fluxes. For further details and references we refer the reader to \cite{Denef:2008wq,Weigand:2010wm,Weigand:2018rez}.

\subsection{Motivation}

\subsubsection*{Type IIB with varying axio-dilaton}
Let us recall that, in the absence of fluxes and localized sources, the equations of motion of type IIB supergravity reduce to
\begin{align}
	\mathrm{d}\star\mathrm{d}\tau = \frac{1}{2}\frac{\mathrm{d}\tau\wedge\star\,\mathrm{d}\bar{\tau}}{\tau-\bar{\tau}}\,,\qquad R_{MN} = \frac{1}{2}\frac{\partial_M\tau\,\partial_N\bar{\tau}}{(\mathrm{Im\,\tau})^2}\,.
\end{align}
In section \ref{sec:IIB_compactification} we have focused on the case where the axio-dilaton is constant, and the internal manifold is Ricci-flat. However, when the axio-dilaton is not constant, one finds that the internal manifold is no longer Ricci-flat. Rather, it will have positive Ricci curvature. 
\\

\noindent In the presence of a D7-brane, one expects that the axio-dilaton will develop a non-trivial profile in the directions perpendicular to the D7-brane. Locally, the transverse directions can be parametrized by a complex coordinate $z$ and its conjugate $\bar{z}$, chosen such that the location of the D7-brane is at $z=0$. Taking $\tau=\tau(z,\bar{z})$ and considering a region away from the D7-brane, one finds that the equations of motion for the axio-dilaton reduce to
\begin{equation}\label{eq:eom_tau_complex}
	\partial \bar{\partial}\tau - 2 \frac{\partial\tau \,\bar{\partial}\tau}{\tau-\bar{\tau}}=0\,,
\end{equation}
where the derivatives are taken with respect to $z$ and $\bar{z}$. Clearly, any holomorphic function $\bar{\partial}\tau = 0$ will satisfy \eqref{eq:eom_tau_complex}. Furthermore, it turns out that such solutions are supersymmetric (to be precise, they preserve 16 supercharges). Additionally, inserting a general holomorphic solution into Einstein's equations, one finds that the Ricci-curvature is given by
\begin{equation}\label{eq:EE_holomorphic}
	R_{z\bar{z}} = -\frac{1}{2}\partial\bar{\partial}\log\mathrm{Im}\,\tau= \frac{1}{2}\partial\bar{\partial}\Phi\,.
\end{equation}
Note that the internal metric is K\"ahler.

\subsubsection*{D7-brane backreaction}
In order to give a more precise expression for $\tau(z)$ near the D7-brane, we recall that the Bianchi identity for the field-strength $F_9$ implies the following relation
\begin{equation}
	\int_{S^1}\star\, F_9 = \int_{S^1}\mathrm{d}C_0 = 1\,,
\end{equation}
where $S^1$ is a circle that winds around the location $z=0$ of the D7-brane once. As a result, one sees that the presence of a D7-brane induces a monodromy transformation $\tau\mapsto \tau+1$ upon encircling the point $z=0$. Combined with the fact that $\tau$ should depend holomorphically on $z$, this implies that locally $\tau$ takes the following form
\begin{equation}
	\tau(z) = \tau_0 +\frac{1}{2\pi i}\log z+\cdots\,,
\end{equation}
where the dots denote regular terms in $z$. Alternatively, one can also arrive at this result by noting that the inclusion of the D7-brane leads to the Poisson equation with a non-trivial source term: $\partial\bar{\partial}\tau = -\delta^2(z)$. 
\\

\noindent Note, in particular, that near the D7-brane the axio-dilaton diverges to $i\infty$, such that the effective string coupling $g_s$ goes to zero. In other words, near the D7-brane one expects that the perturbative type IIB description in terms of fundamental strings is valid. However, due to the logarithmic dependence, far away from the brane there will be regions where the effective string-coupling is no longer small, and it is necessary to go to a different duality frame of type IIB in order to recover the appropriate light degrees of freedom.\footnote{This can be made more precise by explicity computing the backreaction of the D7-brane on the metric. One finds that, at large distances from the D7-brane, the metric develops a conical singularity due to the appearance of a deficit angle.} One purpose of F-theory is to provide a unifying framework in which these effects can be formulated in purely geometric terms, thus providing a possible non-perturbative description of type IIB string theory.

\subsection{Geometric perspective (1): Elliptic fibrations}\label{ssec:Weierstrass}
There are two complementary perspectives one can take on F-theory, both of which will be useful to discuss to some extent. Both perspectives are built on the idea that the axio-dilaton $\tau$ of type IIB should be viewed as the modular parameter of an auxiliary torus/elliptic curve. One motivation for this is that the action of the $\mathrm{SL}(2,\mathbb{Z})$ symmetry on $\tau$, which is a symmetry of the type IIB supergravity action, exactly corresponds to the action of the modular group on the torus. In the following, we will describe how this idea can be made more precise.

\subsubsection*{Weierstrass model}
An elliptic curve $\mathbb{E}_\tau$ with complex structure parameter $\tau$ can be realized as a hypersurface in weighted projective space $\mathbb{P}^{2,3,1}$ with homogeneous coordinates $[x:y:z]$ by the Weierstrass equation
\begin{equation}
	\mathbb{E}_\tau:\qquad P \equiv y^2 - x^3 - f xz^4 - gz^6=0\,,
\end{equation}
for some constants $f,g\in\mathbb{C}$ which are given in terms of $\tau$ by the expressions
\begin{equation}
	f(\tau) = -4^{1/3} g_2(\tau)\,,\qquad g(\tau) = -4 g_3(\tau)\,,
\end{equation}
where $g_i(\tau)$ is the $i$-th Eisenstein series. Such a realization of an elliptic curve is referred to as a \textbf{Weierstrass model}. Conversely, given a Weierstrass model one can recover its complex structure parameter $\tau$ via the relations
\begin{equation}\label{eq:j_Delta}
	j(\tau) = 4 \frac{1728 f^3}{\Delta}\,,\qquad \Delta = 4f^3+27 g^2\,,
\end{equation}
where $j(\tau)$ is the Jacobi $j$-function which enjoys the expansion
\begin{equation}\label{eq:j_tau}
	j(\tau) = e^{-2\pi i \tau}+744+196884 e^{2\pi i \tau}+\cdots\,.
\end{equation}
The significance of the discriminant $\Delta$ is that it indicates when the elliptic curve develops a singularity. To be precise, the hypersurface defined by the Weierstrass equation is singular when $P = \mathrm{d}P=0$, which corresponds precisely to the vanishing of the discriminant $\Delta$. At the same time, combining the relations \eqref{eq:j_Delta} and \eqref{eq:j_tau} one finds that this corresponds to $\tau\rightarrow i\infty$, for which indeed one finds that the torus pinches and thus develops a singularity.

\subsubsection*{Elliptic fibrations and singular fibres}
The Weierstrass model provides a very useful way to describe the situation where the axio-dilaton varies over the internal space. Indeed, this can be accomplished by promoting the coordinates $[x:y:z]$ as well as the constants $f,g$ in the Weierstrass model to sections of a fibration over some base space. In other words, we view the complex three-dimensional internal space on which type IIB is compactified as the base $B_3$ of a torus fibration
\begin{equation}\label{eq:elliptic_fibration}
	\mathbb{E}_\tau\rightarrow Y_{4}\rightarrow B_3\,,
\end{equation}
for some complex four-dimensional space $Y_{4}$. It turns out that the combination of supersymmetry preservation and Einstein's equations leads to some remarkable restrictions on the possible fibrations:
\begin{itemize}
	\item $B_3$ is K\"ahler,
	\item $\tau$ varies holomorphically,
	\item $Y_4$ is a Calabi--Yau fourfold. 
\end{itemize}
Let us briefly comment on the last point, which can be roughly understood as follows. Since $\tau$ varies holomorphically one can describe the elliptic fibration \eqref{eq:elliptic_fibration} in terms of a holomorphic line bundle $\mathcal{L}$ over the base $B_3$, whose first Chern class $c_1(\mathcal{L})$ is given by the curvature of the connection 1-form $\frac{i}{2}\left(\bar{\partial}\Phi-\partial\Phi\right)$, see \cite{Bianchi:2011qh} for a more in-depth discussion. Recalling Einstein's equation \eqref{eq:EE_holomorphic}, this can be shown to be equivalent to the relation
\begin{equation}
	c_1(B_3) = c_1(\mathcal{L})\,,
\end{equation}
where $c_1(B_3)$ is the first Chern class of the tangent bundle of $B_3$, which is nothing but the Ricci 2-form. Finally, using standard results in algebraic geometry, one finds that
\begin{equation}
	c_1(Y_4) = c_1(B_3) - c_1(\mathcal{L}) = 0\,,
\end{equation}
so that the first Chern class of $Y_4$ vanishes, hence $Y_4$ is Calabi--Yau! Roughly speaking, the positive curvature of the base manifold has been cancelled by the negative curvature of the additional torus. The fact that the geometrization of the axio-dilaton of type IIB allows us to remain within the realm of Calabi--Yau compactifications is one of the great advantages of the F-theory framework. 
\\

\noindent Furthermore, recalling that the vanishing of the discriminant $\Delta$ in \eqref{eq:j_Delta} corresponds locally to sending $\tau\rightarrow i\infty$ one sees that the co-dimension one loci in the base $B_3$ over which the elliptic fibre is singular correspond exactly to the locations of D7-branes in the type IIB language. To be precise, one actually finds more general types of 7-branes, depending on how exactly the elliptic fibre degenerates. Roughly speaking, if the degeneration is due to the shrinking of the A-cycle of the torus, the resulting brane will be the standard D7-brane. However, more generally the degeneration may be due to a linear combination $p A+q B$ of the A-cycle and B-cycle that shrinks. The resulting 7-brane is called a \textbf{$(p,q)$ 7-brane}. Correspondly, the open strings whose endpoints lie on such $(p,q)$ 7-branes are called \textbf{$(p,q)$-strings}, and can roughly be thought of as bound states of $p$ F1-strings and $q$ D1-branes. Of course, these different types of branes and strings in principle exist within the type IIB picture, as they are related through the $\mathrm{SL}(2,\mathbb{Z})$-duality. However, F-theory provides a beautiful geometrization of these various objects through the different types of degenerations of the elliptic fibration. 
\\

\noindent As a final comment, let us remark that a more general study of the possible degenerations, in particular including effects from shrinking 2-cycles, gives rise to an incredibly rich framework to describe the emergence of non-abelian gauge symmetries. Although this will not play a role in this thesis, it is important to stress this point in view of the phenomenological viability of F-theory. Indeed, following the Kodaira classification of singular fibres in elliptic surfaces, one for example finds that exceptional gauge groups $E_6,E_7,E_8$ can be realized as enhanced gauge symmetries in F-theory, leading to the possibility of GUT models. We refer the reader to \cite{Weigand:2010wm,Weigand:2018rez} for further details and references.

\subsection{Geometric perspective (2): F-/M-theory duality}

Although the perspective taken in the previous subsection provides a natural motivation for studying F-theory, a more useful approach for the purpose of this thesis is a dual description in terms of M-theory. Let us recall the low-energy effective description of M-theory in terms of 11-dimensional supergravity
\begin{subbox}{11D M-theory effective action}
	\begin{equation}\label{eq:action_M-theory_corrected}
		S_M = \frac{2\pi}{\ell_M^9}\int_{M_{11}} R\star 1- \frac{1}{2}G_4\wedge\star\,G_4-\frac{1}{6}C_3\wedge G_4\wedge G_4+\ell_M^6\,C_3\wedge I_8\,.
	\end{equation} 
Here we have included a topological higher-curvature correction
\begin{equation}\label{eq:R4-correction_topological}
	I_8 = \frac{1}{(2\pi)^4}\left(\frac{1}{192}\mathrm{tr}R^4-\frac{1}{768}\left(\mathrm{tr}R^2\right)^2 \right) = \frac{1}{192}\left(p_1(R)^2-4 p_2(R) \right)\,.
\end{equation}
Furthermore, the 11-dimensional Planck length $\ell_M$ is related to the string length $\ell_s$ via
\begin{equation}
	2\pi\ell_M = g_s^{1/3}\ell_s\,. 
\end{equation}
\end{subbox}
\noindent Regarding the higher-curvature correction, let us remark that from the type IIA perspective this correction can be obtained by computing the one-loop scattering amplitude of four gravitons and the Kalb--Ramond field $B_2$, while from the heterotic $E_8\times E_8$ perspective it is required for anomaly cancellation \cite{Becker:2006}.

\subsubsection*{M-theory / type IIB duality}
As a first step, let us briefly sketch the duality between M-theory and type IIB string theory. Our starting point is to compactify M-theory on a 2-torus, which is parametrized by two circles $S^1_{R_A}\times S^1_{R_B}$ with radii $R_A$ and $R_B$, respectively. The duality then consists of two steps:
\begin{itemize}
	\item \textbf{Step 1: Reduce M-theory on $S^1_{R_A}$}\\
	M-theory compactified on a \textit{small} circle gives type IIA string theory in the \textit{weak}-coupling regime. To be precise, the relation between the circle radius and the string-coupling is given by
	\begin{equation}\label{eq:Mcircle_IIA}
		R_{A} =\sqrt{\alpha'}g_{s,\mathrm{IIA}}\,,
	\end{equation}
	and we are considering the regime where $R_A$, and hence $g_{s,\mathrm{IIA}}$ is small.
	\item \textbf{Step 2: T-duality along $S^1_{R_B}$}\\
	It is well-known that type IIA string theory and type IIB string theory are T-dual to each other. To be precise, this means that type IIA string theory compactified on a circle $S^1_{R_B}$ is dual to type IIB string theory compactified on a circle $S^1_{\alpha'/R_B}$. For our purpose, it will be most important to keep track of the relation between the type IIA string coupling $g_{s,\mathrm{IIA}}$ and the type IIB string coupling $g_{s,\mathrm{IIB}}$, which is given by
	\begin{equation}\label{eq:gsIIB_gsIIA}
		g_{s,\mathrm{IIB}} = \frac{\sqrt{\alpha'}}{R_B} g_{s,\mathrm{IIA}}\,.
	\end{equation}
\end{itemize}
Combining the two steps described above, we arrive at the following relation between M-theory and type IIB string theory:

\begin{subbox}{M-theory / type IIB duality}
	\begin{equation}
		\text{M-theory}\,/\,S_{R_A}^1\times S_{R_B}^1 \qquad \longleftrightarrow\qquad \text{Type IIB}\,/\,S^1_{\alpha'/R_B}\,.
	\end{equation}
\tcblower
Furthermore, combining equations \eqref{eq:Mcircle_IIA} and \eqref{eq:gsIIB_gsIIA}, one finds that the type IIB string coupling is given by the ratio
\begin{equation}
	g_{s,\mathrm{IIB}} = \frac{R_A}{R_B}\,.
\end{equation}
\end{subbox}

\subsubsection*{The F-theory limit}
In order to recover the full 10-dimensional formulation of type IIB, one would like to take the limit in which $R_B\rightarrow 0$, such that the circle direction on the type IIB side is decompactified. To be precise, one should take the limit in which the volume of the 2-torus
\begin{equation}
	V = R_A R_B\,,
\end{equation}
goes to zero while the ratio $R_A/R_B$ stays finite, such that there is a well-defined value for the IIB string coupling \eqref{eq:gsIIB_gsIIA}. In fact, note that the inverse ratio of the circle radii exactly corresponds to the imaginary part of the complex structure parameter $\tau$ of the two-torus, such that one can make the identification
\begin{equation}\label{eq:gsIIB_tau}
	\mathrm{Im}\,\tau = \frac{R_B}{R_A} = g_{s,\mathrm{IIB}}^{-1}\,.
\end{equation}
In other words, we arrive at the following relation
\begin{subbox}{F-theory / type IIB duality (trivial fibration)}
	\begin{equation}
		\text{M-theory}\,/\,\mathbb{E}_{\tau}\Big|_{\mathrm{Vol}(\mathbb{E}_\tau)\rightarrow 0} \qquad \longleftrightarrow\qquad \text{Type IIB}\,,
	\end{equation}
where the type IIB string coupling is determined by the relation \eqref{eq:gsIIB_tau}.
\end{subbox}
\noindent More generally, one can proceed along exactly the same lines and consider M-theory compactified on an elliptic fibration 
\begin{equation}
	\mathbb{E}_\tau\rightarrow Y_{n+1}\rightarrow B_n\,,
\end{equation}
over some base space $B_n$. Recall that supersymmetric configurations correspond to the case where $Y_{n+1}$ is Calabi--Yau and $B_n$ is K\"ahler. Then the more general relation becomes
\begin{subbox}{F-theory / type IIB duality (non-trivial fibration)}
	\begin{equation}
	\text{M-theory}\,/\,Y_{n+1}\Big|_{\mathrm{Vol}(\mathbb{E}_\tau)\rightarrow 0} \qquad \longleftrightarrow\qquad \text{Type IIB}\,/\, B_n\,,
\end{equation}
where the type IIB string coupling is again determined by the relation \eqref{eq:gsIIB_tau}, which now varies over the base space $B_n$.
\tcblower 
\textbf{Note:}\\
The resulting theory on the left-hand side of this relation could be viewed as the definition of F-theory.
\end{subbox}
\noindent The M-theory perspective on F-theory is especially beautiful, as it suggests that the elliptic curve should really be thought of as being part of the (external!) geometry, as opposed to being some abstract book-keeping device to describe the axio-dilaton. Furthermore, M-theory provides a useful framework in which to describe the various other fluxes and branes that are present in type IIB, as we will discuss now.

\subsection{Fluxes, branes and the tadpole cancellation condition}
In order to make the duality between M-/F-theory and type IIB more precise, one should also describe how the various fluxes and branes on the two sides are related. Here we will briefly recall the standard dictionary, and refer the reader to \cite{Denef:2008wq} for a more in-depth discussion. 

\subsubsection*{Branes}
In 11-dimensional supergravity, there is only a single $p$-form field $C_3$, which is electrically sourced by M2-branes. The action of an M2-brane takes a form which is very similar to that of a D$p$-brane
\begin{equation}
	S_{\mathrm{M}2} = \frac{2\pi}{\ell_M^3}\int_{\mathrm{M}2}d^3\xi\sqrt{-\mathrm{det}\left(G_{\alpha\beta}\right)} + \frac{2\pi}{\ell_M^3}\int_{\mathrm{M}2}C_3\,,
\end{equation}
where the first term is again a DBI-like term which is simply the worldvolume of the M2-brane, and the second term describes the coupling to $C_3$. Naturally, there is also a corresponding action for the dual M5-branes, which provide a magnetic source for $C_3$. 
\\

\noindent M2-branes and M5-branes can have different interpretations from the type IIB perspective, depending on which (if any) of the cycles of the elliptic fibre are wrapped. 
\begin{itemize}
	\item \textbf{M2-brane:}
	\begin{itemize}
		\item If the M2-brane is spacetime-filling, it gives rise to a spacetime filling D3-brane.\footnote{Here spacetime-filling is meant in the M-theory sense, namely that the M2-brane is completely in the $\mathbb{R}^{1,2}$ direction.}
		\item If the M2-brane wraps $p$ times around $S^1_A$ and $q$ times around $S^1_B$, this gives rise to a $(p,q)$-string, which was discussed earlier in section \ref{ssec:Weierstrass}. Two notable cases are 
		\begin{itemize}
			\item $(p,q)=(1,0)$: F1-string,
			\item $(p,q)=(0,1)$: D1-brane,
		\end{itemize}
		where by the F1-string we mean the fundamental type IIB superstring.
	\end{itemize}
	\item \textbf{M5-brane:}\\
	For an M5-brane the possible configurations are simply the magnetic duals of the ones described above. Let us go through them for completeness:
	\begin{itemize}
		\item If the M5-brane wraps $p$ times around $S^1_A$ and $q$ times around $S^1_B$, this gives rise to a so-called $(p,q)$ $5$-brane. Again, two notable cases 
		\begin{itemize}
			\item $(p,q)=(1,0)$: D5-brane,
			\item $(p,q)=(0,1)$: NS5-brane,
		\end{itemize}
		where we recall that a D5-brane is the magnetic dual of a D1-brane and an NS5-brane is the magnetic dual of an F1-string.
		\item If the M5-brane wraps the two-cycle $S^1_A\times S^1_B$ this corresponds to a D3-brane. 
	\end{itemize}
\end{itemize}
In table \ref{tab:Mbranes} we have summarized the various possibilities discussed above.

\begin{table}[h!]
	\centering
	\begin{tabular}{|c|c|c|}
		\hline
		 & Wrapped cycle &  IIB interpretation \\ \hline
		& $S^1_A$ & F1-string\\
		M2-brane &  $S^1_B$ & D1-brane\\
		&$p$ times around $S^1_A$ and $q$ times around $S^1_B$ & $(p,q)$-string \\ \hline 
		& $S^1_A$ & D5-brane\\ 
		M5-brane & $S^1_B$ & NS5-brane\\
		& $S^1_A\times S^1_B$ & D3-brane \\ \hline
	\end{tabular}
	\caption{Summary of how the various branes in type IIB arise from different wrappings of the M2/M5-branes in M-theory.}
	\label{tab:Mbranes}
\end{table}

\subsubsection*{The tadpole cancellation condition revisited}
Just like we found in section \ref{subsec:tadpole_cancellation}, the inclusion of a non-zero $G_4$ flux and/or M2-branes will lead to a non-trivial global constraint when the internal space is compact. To be precise, the corresponding \textbf{M2 tadpole cancellation condition} reads
\begin{subbox}{M2 tadpole cancellation condition}
\begin{equation}\label{eq:M2_tadpole}
		N_{\mathrm{M2}}+\frac{1}{2\ell_M^6}\int_{Y_4} G_4\wedge G_4 = \frac{\chi(Y_4)}{24}\,,
\end{equation}
where $N_{\mathrm{M}2}$ is the number of spacetime-filling M2-branes, and $\chi(Y_4)$ denotes the Euler character of the Calabi--Yau fourfold $Y_4$. 
\end{subbox}
\noindent The term on the right-hand side of \eqref{eq:M2_tadpole} exactly arises from integrating the contribution of the topological higher-curvature correction $I_8$ to the Bianchi identity of $C_3$. To see how this relates to the D3 tadpole cancellation condition in the type IIB setting, one decomposes the $G_4$ flux in terms of 3-form fluxes along the two legs of the elliptic fibre as
\begin{equation}
	G_4 = \frac{\ell_M^3}{\ell_s^2}\left(F_3\wedge \mathrm{d}x+H_3\wedge\mathrm{d}y\right)\,,
\end{equation} 
which results in the familiar condition (after performing the integral over the elliptic curve)
\begin{equation}
	N_{\mathrm{D}3}+\frac{1}{2\ell_s^4}\int_{B_3}F_3\wedge H_3 = \frac{\chi(Y_4)}{24}\,.
\end{equation}
Comparing this with the D3 tadpole cancellation condition found in equation \eqref{eq:D3_tadpole}, one sees that the contributions coming from the D7-branes and O7-planes have been geometrized into the Euler characteristic of the Calabi--Yau fourfold. It should be noted, however, that it is not necessarily the case that $\chi(Y_4)>0$.\footnote{As a simple counter-example, one may consider a $T^8/\mathbb{Z}_2\times\mathbb{Z}_2$ orbifold, which has $\frac{\chi}{24}=-8$, see also \cite{Sethi:1996es}.} In this way, we find another manifestation of the fact that the F-theory language provides a beautiful description of many different aspects of type IIB string theory. 
\\

\noindent As a side remark, let us also mention that in M-theory the M2 tadpole cancellation condition also places non-trivial restrictions on the allowed Calabi--Yau fourfolds one may consider, at least in the presence of M2-branes and/or non-trivial $G_4$ flux. This is because the left-hand side of \eqref{eq:M2_tadpole} is clearly an integer, while the right-hand side might not be. In contrast, in F-theory this turns out to be of no concern. This is because elliptically fibered Calabi-Yau fourfolds (with a section) automatically have an Euler characteristic $\chi(Y_4)$ that is divisible by 72, see e.g.~\cite{Sethi:1996es}. In fact, if $Y_4$ is non-singular, then one can write the Euler characteristic purely in terms of objects pertaining to the base $B_3$ as
\begin{equation}\label{eq:Euler_character_72}
	\chi(Y_4) = 72\int_{B_3} \frac{1}{2}c_1c_2+15 c_1^3\,,
\end{equation}
where $c_i=c_i(B_3)$ are the $i$-th Chern classes of $B_3$. The first term in \eqref{eq:Euler_character_72} can be shown to correspond to the D3-brane charge induced by spacetime-filling D7-branes through anomaly in-flow arguments \cite{Bershadsky:1993cx,Vafa:1995fj}, while the second term roughly arises from 7-branes wrapped on the intersections of the irreducible components of the discriminant locus of the Weierstrass model. 

\section{Type IIB / F-theory flux compactifications}\label{sec:F-theory_flux}
In the previous section we have argued that the F-theory perspective is useful (and arguably necessary) to describe configurations with 7-branes and a non-trivial profile for the axio-dilaton. In this section, we additionally include the effects of non-trivial three-form fluxes in type IIB, or alternatively four-form fluxes in M-/F-theory, to study the landscape of four-dimensional low-energy effective theories coming from type IIB/F-theory flux compactification. In section \ref{ssec:GKP} we first reconsider the 10-dimensional type IIB point of view and explain how the inclusion of appropriate local sources allows us to evade the Maldacena--Nu\~nez no-go theorem disucssed in section \ref{sec:IIB_effective}. The resulting class of solutions was found by Giddings--Kachru--Polchinski \cite{Giddings:2001yu} and will be referred to as the GKP solution. In section \ref{ssec:M-theory_CY4} we briefly review the dual M-theory solution found by Becker--Becker \cite{Becker:1996gj}. Finally, in subsection \ref{ssec:N=1-sugra} we describe the four-dimensional low-energy effective $\mathcal{N}=1$ supergravity theory that results from compactifying the latter on a Calabi--Yau fourfold and subsequently taking the F-theory limit. 

\subsection{The GKP solution}\label{ssec:GKP}
In the following we review the analysis of \cite{Giddings:2001yu}, to which we refer the reader for additional details. In the presence of local sources, the right-hand side of equation \eqref{eq:IIB_no-go} receives additional corrections coming from the associated energy-momentum tensor. To be precise, the correction is given by
\begin{equation}\label{eq:IIB_no-go_correction}
\frac{\ell_s^8}{8\pi} e^{2A}\times \left(\tensor{T}{^m_m}-\tensor{T}{^\mu_\mu} \right)^{\mathrm{loc}}\,,\qquad T_{MN}^{\mathrm{loc}} = \frac{-2}{\sqrt{-g}}\frac{\delta S_{\mathrm{loc}}}{\delta g_{MN}}\,,
\end{equation}
where $S_{\mathrm{loc}}$ is the action of the localized sources. Importantly, if the term in brackets is \textit{negative}, it is possibly to evade the argument that lead to the Maldacena--Nu\~nez no-go theorem. For spacetime-filling D$p$-branes/O$p$-planes wrapping a $(p-3)$-cycle $\Gamma_{p-3}$, the contribution to the energy-momentum tensor comes from their DBI action.
For the sake of exposition, let us assume for the moment that the dilaton is constant, and that there are no worldvolume fluxes on the brane. Then to leading order in $\alpha'$ one readily finds
\begin{equation}\label{eq:energy-momentum_sources}
	\left(\tensor{T}{^m_m}-\tensor{T}{^\mu_\mu} \right)^{\mathrm{D}p/\mathrm{O}p} = (7-p)T_{\mathrm{D}p/\mathrm{O}p} \delta^{(9-p)}(\Gamma_{p-3})\,,
\end{equation}
where $\delta^{(9-p)}(\Gamma_{p-3})$ denotes the $(9-p)$-dimensional $\delta$-function localized on the cycle $\Gamma_{p-3}$.\footnote{Since the equations of motion \eqref{eq:IIB_no-go} were written in Einstein-frame, the tension $T_p$ we are referring to here corresponds to the Einstein-frame tension, as opposed to the string-frame tension that appears in the DBI actions \eqref{eq:action_Dp_DBI} and \eqref{eq:action_Op_DBI}. The relation between the two is given by
	\begin{equation}
		T_p^{\text{(string-frame)}} = e^{\Phi(p-3)/4} T_p^{\text{(Einstein-frame)}}\,.
\end{equation}} In particular, note that e.g.~for O3-planes their contribution is indeed negative! 
\\

\noindent In turns out that, even with the additional complication of having included localized sources, one can still combine the various equations motion in such a way that an elegant class of solutions presents itself. Indeed, consider the modified Bianchi identity for the five-form flux cf.~\eqref{eq:tadpole_local}
\begin{equation}
	\mathrm{d}\star\tilde{F}_5 + F_3\wedge H_3 + \ell_s^4 \sum_{i} Q_3^{(i)}\,\mathrm{PD}[\Gamma_0^{(i)}]=0\,,
\end{equation}
which now includes the local D3 charge density coming from all localized sources such as D3-branes and O3-planes, as well as possibly wrapped D5/D7-branes and O7-planes. Then combining this together with the relation \eqref{eq:IIB_no-go} including the correction \eqref{eq:IIB_no-go_correction}, one can arrive at the following condition
\begin{align}\label{eq:warp_factor_GKP}
	\tilde{\Delta}_{(6)}\left(e^{4A}-\alpha\right) &= \frac{e^{2A}}{\mathrm{Im}\,\tau}\left|i G_3-\star \,G_3\right|^2+e^{-6A}\left|\partial(e^{4A}-\alpha)\right|^2 \nonumber \\
	&\qquad +2\kappa^2 e^{2A}\left[\frac{1}{4}\left(\tensor{T}{^m_m}-\tensor{T}{^\mu_\mu} \right)^{\mathrm{loc}}-T_{\mathrm{D}3} \rho_3^{\mathrm{loc}} \right]\,,
\end{align}
where $\rho_3^{\mathrm{loc}}$ is given by 
\begin{equation}
	\rho_3^{\mathrm{loc}} = \sum_{i} Q_3^{(i)}\,\delta^{(6)}(\Gamma_0^{(i)})\,.
\end{equation}
The central insight of GKP is that many of the local sources in string theory satisfy the condition
\begin{equation}\label{eq:BPS-condition_GKP}
	\frac{1}{4}\left(\tensor{T}{^m_m}-\tensor{T}{^\mu_\mu} \right)^{\mathrm{loc}}-T_{\mathrm{D}3} \rho_3^{\mathrm{loc}}\geq 0\,.
\end{equation}
As a result, one can again conclude that the left-hand side in \eqref{eq:warp_factor_GKP} is positive, so that upon integrating the equation over the compact internal space one concludes that all individual terms must vanish. 
\\

\noindent Indeed, as an important example, the condition \eqref{eq:BPS-condition_GKP} is satisfied (in fact, saturated) by D3-branes and O3-planes, while it is violated by O5-planes. It is also saturated by wrapped D7-branes, though this requires taking into account $\alpha'$ corrections to the DBI action, since at tree level a D7-brane does not contribute to the energy-momentum tensor cf.~\eqref{eq:energy-momentum_sources}. 
\\

\noindent To summarize the above considerations, we are led to the following class of solutions to the IIB supergravity equations of motion in the presence of localized sources.
\begin{subbox}{The GKP solution \cite{Giddings:2001yu}}
	In the presence of localized sources that satisfy the condition
	\begin{equation}\label{eq:assumption_sources}
		\frac{1}{4}\left(\tensor{T}{^m_m}-\tensor{T}{^\mu_\mu} \right)^{\mathrm{loc}}-T_{\mathrm{D}3} \rho_3^{\mathrm{loc}}\geq 0\,.
	\end{equation}
	all warped compactifications of type IIB supergravity down to $\mathbb{R}^{1,3}$ that preserve 4-dimensional Poincar\'e invariance must satisfy the following conditions:
	\begin{itemize}
		\item $G_3$ is imaginary self-dual (ISD): $\star\,G_3 = iG_3$,
		\item $\alpha = e^{4A}$,
		\item The bound \eqref{eq:assumption_sources} is saturated.
	\end{itemize}
	\tcblower
	\textbf{Note:} \\
	Examples of localized sources which saturate the bound \eqref{eq:assumption_sources} are D3-branes, O3-planes as well as wrapped D7-branes. 
\end{subbox}
\noindent It remains to discuss the dynamics of the axio-dilaton and the internal metric. Allowing for the presence of 7-branes, we again take the F-theory perspective cf.~section \ref{sec:F-theory} corresponding to the case in which the axio-dilaton varies holomorphically over a complex three-dimensional internal space $B_3$, described by an elliptic fibration $Y_4\rightarrow B_3$. Then it turns out that the remaining equations of motion for the internal metric are simply
\begin{equation}
	\tilde{R}_{m\bar{n}} = \partial_m \bar{\partial}_{\bar{n}}\Phi\,,
\end{equation} 
such that the metric $\tilde{g}_{mn}^{(6)}$ is K\"ahler in complete analogy to the discussion in section \ref{sec:F-theory}. In particular, we conclude that the total space $Y_4$ is still a Calabi--Yau fourfold up to the overall warp factor (it is said to be \textit{conformally} Calabi--Yau). 

\subsection{M-theory on Calabi--Yau fourfolds}\label{ssec:M-theory_CY4}
Let us now discuss the dual M-theory formulation of the type IIB GKP solution. Thus, we consider warped compactifications of M-theory down to $\mathbb{R}^{1,2}$ preserving Poincar\'e invariance. We again make the most general ansatz for the metric compatible with these conditions
\begin{equation}
	ds^2_{11} = \Delta(y)^{-1}\eta_{\mu\nu}^{(4)}\,\mathrm{d}x^\mu\otimes\mathrm{d}x^\nu + \Delta(y)^{1/2}\tilde{g}_{mn}^{(8)}\,\mathrm{d}y^m\otimes \mathrm{d}y^n\,,
\end{equation}
where we have now employed a new notation for the warp factor $\Delta(y)$ to distinguish it from the one used in the type IIB compactifications, and we stress that now the internal manifold is eight-dimensional. Furthermore, Poincar\'e invariance imposes the following decomposition of the four-form flux
\begin{equation}\label{eq:decomp_G4}
	G_4 = \mathrm{d}f\wedge\mathrm{d}x^0\wedge\mathrm{d}x^1\wedge\mathrm{d}x^2+\hat{G}_4\,,
\end{equation} 
where $f=f(y)$ is an arbitrary function of the internal coordinates and $\hat{G}_4$ is a purely internal four-form. The equations of motion of the four-form flux yield
\begin{equation}
	\mathrm{d}\star G_4 +\frac{1}{2}G_4\wedge G_4 + \ell_M^6 I_8=0\,,
\end{equation}
whose integration over the compact eight-dimensional internal space recovers the M2 tadpole cancellation condition \eqref{eq:M2_tadpole} (in the absence of M2-branes).
\\

\noindent In \cite{Becker:1996gj,Becker:2001pm} the resulting equations of motion were analyzed under the assumption that the eight-dimensional metric $g_{mn}^{(8)}$ is K\"ahler by performing a large-volume expansion.\footnote{It is important to note that this analysis requires the inclusion of two additional $R^4$ corrections to the 11-dimensional supergravity action, besides the topological correction \eqref{eq:R4-correction_topological}. See \cite{Becker:2001pm} for further details.} Roughly speaking, this means that one performs an expansion in $\ell_{M}/v^{1/8}$, where $v$ denotes the volume of the eight-dimensional internal space. In other words, one assumes that the typical size of the compactification manifold is large compared to the eleven-dimensional Planck scale. Here we will not repeat the derivation, but simply state the following result.

\begin{subbox}{M-theory on a K\"ahler four-fold}
	At leading order in the large-volume expansion and in the absence of M2-branes, the solutions to the equations of motion and Bianchi identity for M-theory compactified on a K\"ahler four-fold $Y_4$ have to satisfy the following conditions:
	\begin{enumerate}
		\item The function $f(y)$ appearing in \eqref{eq:decomp_G4} is constant, so that $G_4$ only has an internal component. In particular, we may write
		\begin{equation}
			G_4 = \hat{G}_4\,,
		\end{equation} 
		and will thus suppress the hat in the following. 
		\item The flux $G_4$ (which, by the previous point, is purely internal) has to be \textit{self-dual} and satisfy the M2 tadpole cancellation condition: 
		\begin{equation}\label{eq:M-theory_solution}
			G_4 = \tilde{\star}\,G_4\,,\qquad 
				\frac{1}{2\ell_M^6}\int_{Y_4} G_4\wedge G_4 = \frac{\chi(Y_4)}{24}\,.
		\end{equation}
		\item The metric $\tilde{g}_{mn}^{(8)}$ is Ricci-flat and thus describes a Calabi--Yau fourfold, which will be denoted by $Y_4$.
	\end{enumerate}
\end{subbox}
\noindent At the next to leading order the external component of $G_4$ need not vanish, and will be related to the warp factor in a similar fashion as in the GKP solution. Furthermore, at the next to leading order the metric will no longer be Ricci flat. Rather strikingly, It turns out that Einstein's equations are of such a nature that the solution can still be described by a Calabi--Yau manifold (though equipped with a metric that is not Ricci-flat). This solution is especially nice since supersymmetry will thus preserved by the background metric, though it can be broken by the fluxes, as we will discuss in the next section.

\subsection{$D=4$, $\mathcal{N}=1$ SUGRA from F-theory}\label{ssec:N=1-sugra}
To close this section, let us discuss the four-dimensional effective description of F-theory compactified on an elliptically fibered Calabi--Yau fourfold. This proceeds in two steps. First, one performs the dimensional reduction of M-theory on the Calabi--Yau fourfold $Y_4$, see \cite{Haack:2001jz}, which is computationally similar to the dimensional reduction of type IIB supergravity on a Calabi--Yau threefold, which was discussed in section \ref{sec:IIB_compactification}. The resulting theory will again be some supergravity theory which, in particular, contains a number of scalar fields corresponding to the complex structure and K\"ahler deformations of the fourfold. In the orientifold or weak-coupling limit, the complex structure deformations of $Y_4$ collectively describe the complex structure deformations of the Calabi-Yau threefold $Y_3$ that is a double cover of $B_3$, as well as the deformations of the D7-branes and the type IIB axio-dilaton $\tau$. Subsequently, one has to lift the resulting three-dimensional theory to four dimensions by performing the F-theory limit in which the volume of the elliptic fiber goes to zero. This is a rather delicate procedure, as it involves a careful analysis of the full KK-tower associated to the decompactified circle. Here we will not discuss this analysis in detail, but rather refer the reader to \cite{Grimm:2010ks}. \\

\noindent The resulting action describes a theory of four-dimensional $\mathcal{N}=1$ supergravity which will contain various multiplets coming from the reduction of the M-theory flux as well as the moduli of the Calabi--Yau fourfold, just like we found in the reduction of type IIB supergravity. In the following, we will focus our attention on the $h^{3,1}(Y_4)$ chiral multiplets coming from the complex structure deformations of the Calabi--Yau fourfold.

\begin{subbox}{Four-dimensional effective action of F-theory on a Calabi--Yau fourfold}
	\begin{equation}
		S_{\mathrm{F}}^{(4)} = \int \frac{1}{2}R\star 1-G_{i\bar{\jmath}}^{\mathrm{cs}}\,\mathrm{d}z^i\wedge\star\,\mathrm{d}\bar{z}^{\bar{\jmath}}-V(z,G_4)\star 1\,,
	\end{equation}
\tcblower
Here $z^i$, for $i=1,\ldots, h^{3,1}$, denote the complex structure moduli of $Y_4$, and the scalar potential $V(z,G_4)$ is given by
\begin{equation}\label{eq:scalar_potential}
	V(z,G_4) = \frac{1}{\mathcal{V}_b^2}\int_{Y_4} G_4\wedge\star\, G_4 - G_4\wedge G_4\,,
\end{equation}
and $\mathcal{V}_b$ denotes the volume of the base $B_3$. We stress that the appearance of the internal Hodge star induces the dependence of the scalar potential on the complex structure moduli.\footnote{In the preceeding section the metric on $Y_4$ was denoted with a tilde, which we now suppress for notational clarity.}
\end{subbox}
\noindent Importantly, comparing the above result with the four-dimensional $\mathcal{N}=2$ effective action that we found in section \ref{sec:IIB_compactification}, we see that the inclusion of a non-trivial flux $G_4$ induces a scalar potential for the complex structure moduli. Crucially, the potential acts as an effective mass term for these moduli and thus (generically) prescribes a fixed vacuum expectation value for them. In other words, the inclusion of fluxes allows us to \textit{stabilize} the moduli. Below we discuss the precise vacuum conditions in more detail.

\subsubsection*{Moduli stabilization: Hodge theory formulation}

The scalar potential \eqref{eq:scalar_potential} is positive semi-definite and attains a global minimum whenever the four-form flux is \textit{self-dual}, i.e.
\begin{equation}
	\label{eq:self_dual}
	G_4=\star\,G_4\,.
\end{equation}
Note that this agrees with the 11-dimensional analysis \eqref{eq:M-theory_solution}, which was reviewed in the previous section. Note also that a self-dual vacuum corresponds to a Minkowski vacuum, since $V=0$, in agreement with the fact that, to leading order in the large-volume expansion, the warp factor was found to be constant. \\

\noindent Since the fluxes should be viewed as elements of the integral cohomology $H^4(Y_4,\mathbb{Z})$, one should regard the condition $G_4=\star\,G_4$ as a condition in cohomology.\footnote{To be precise, the quantization condition of the $G_4$ flux is actually shifted and reads
\begin{equation*}
	G_4 - \frac{p_1(Y_4)}{4}\in H^4(Y_4,\mathbb{Z})\,,
\end{equation*}
where $p_1(Y_4)$ denotes the first Pontryagin class of $Y_4$ \cite{Witten:1996md}. This shift will not be relevant in this thesis, so we will assume for notational simplicity that $G_4$ itself is integral.} To elucidate the self-duality condition \eqref{eq:self_dual}, we recall that the middle cohomology of $Y_4$ admits a Hodge decomposition
\begin{equation}\label{eq:Hodge_decomp}
	H^4\left(Y_4,\mathbb{C}\right) = H^{4,0}\oplus H^{3,1}\oplus H^{2,2}\oplus H^{1,3}\oplus H^{0,4}\,,
\end{equation}
into harmonic $(p,q)$-forms. One can show that the self-duality condition \eqref{eq:self_dual} implies that $G_4$ has no $(3,1)$ component. In other words, $G_4$ has a decomposition (recall that $G_4$ is real)
\begin{equation}
	G_4 = \left(G_4\right)^{4,0}+\left(G_4\right)^{2,2}+\left(G_4\right)^{0,4}\,.
\end{equation}
The self-duality condition therefore comprises $h^{3,1}$ complex equations for the $h^{3,1}$ complex structure moduli and hence one expects that a generic choice of $G_4$ stabilizes all moduli. It is, however, not at all obvious whether this holds true if $G_4$ is constrained by the tadpole cancellation condition \eqref{eq:M-theory_solution}. In fact, it was recently suggested that indeed this naive expectation may fail when $h^{3,1}$ becomes sufficiently large \cite{Bena:2020xrh}, leading to the so-called tadpole conjecture. See also \cite{Braun:2020jrx,Bena:2021wyr,Marchesano:2021gyv,Lust:2021xds,Plauschinn:2021hkp,Grana:2022dfw,Lust:2022mhk,Coudarchet:2023mmm,Braun:2023pzd} for related works. We will return to this point in chapter \ref{chap:finiteness}.

\subsubsection*{Hodge vacua}
A self-dual vacuum will be referred to as a \textit{Hodge vacuum} if, in addition, $G_4$ only has a (2,2)-component and is primitive. The latter means that 
\begin{equation}\label{eq:primitive}
	J\wedge G_4=0\,,
\end{equation}
where $J$ denotes the K\"ahler $(1,1)$-form on $Y_4$. In mathematics, cohomology classes of this type are referred to as Hodge classes. As will be elaborated upon in chapter \ref{chap:finiteness}, such classes play a very special role in Hodge theory. 

\subsubsection*{Moduli stabilization: superpotential formulation}
Let us briefly describe how the above results fit inside the general formulation of four-dimensional $\mathcal{N}=1$ supergravity theories. For any four-dimensional $\mathcal{N}=1$ supergravity theory, the F-term contribution to the scalar potential can be written as
\begin{equation}
	\label{eq:scalar_potential_W}
	V = e^K\left(G^{I\bar{J}}D_I W \overline{D_J W}-3|W|^2\right)\,,\qquad D_IW = \left(\partial_I+\partial_I K\right) W\,,
\end{equation}
where $K$ is a K\"ahler potential that determines a K\"ahler metric $G_{I\bar{J}}$ and $W$ is the holomorphic superpotential. In the context of F-theory compactifications, the indices $I,\bar{J}$ in \eqref{eq:scalar_potential_W} run over both the complex structure moduli and the K\"ahler moduli. To clarify the relation between the scalar potentials \eqref{eq:scalar_potential} and \eqref{eq:scalar_potential_W} we need to specify the K\"ahler potential and superpotential.
\begin{itemize}
	\item \textbf{K\"ahler potential:}\\
	The K\"ahler potential $K$ is given by
	\begin{equation}
		\label{eq:Kahler_potential}
		K = -2\log\mathcal{V}_b-\log \int_{Y_4}\Omega\wedge\overline{\Omega}\,.
	\end{equation}
	The first term is the tree-level K\"ahler potential for the complex coordinates $T^A$ that depend on the K\"ahler moduli.\footnote{To avoid potential confusion, these are not quite the same as the complexified K\"ahler moduli of type IIB, which were considered in subsection \ref{subsec:CY_moduli}.} The second term is the K\"ahler potential for the complex structure moduli, depending on the holomorphic $(4,0)$-form $\Omega(z)$. The tree-level K\"ahler potential enjoys the no-scale property
	\begin{equation}\label{eq:no_scale}
		G^{\alpha\bar{\beta}}\partial_\alpha K \overline{\partial_\beta K}=3\,.
	\end{equation}
	It is important to stress, however, that $K$ receives both perturbative corrections, coming e.g.~from $\alpha'$ corrections to the ten-dimensional IIB supergravity action, as well as non-perturbative corrections coming from worldsheet instantons. These corrections will generically break the no-scale structure \eqref{eq:no_scale} of the K\"ahler potential. 
	\item \textbf{Superpotential:}\\
	The superpotential $W$ is given by the Gukov--Vafa--Witten superpotential $W_{\mathrm{flux}}$ \cite{Gukov:1999ya,Haack:2001jz}, where 
	\begin{equation}
		\label{eq:superpotential}
		W_{\mathrm{flux}}(z) = \int_{Y_4}G_4\wedge \Omega(z)\,.
	\end{equation}
	In contrast to the K\"ahler potential $K$, the superpotential $W$ is perturbatively exact and only receives non-perturbative corrections coming e.g.~from Euclidean D3-brane instantons and gaugino condensation. Note that, since our discussion is restricted to the perturbative level, $W$ does not depend on the K\"ahler moduli. In particular, we have
	\begin{equation}\label{eq:DW_Kahler}
		D_\alpha W_{\mathrm{flux}} = \left(\partial_\alpha K\right)W_{\mathrm{flux}}\,,
	\end{equation}
	where again $\alpha$ runs over the complex coordinates involving the K\"ahler moduli. 
\end{itemize}
Combining the no-scale condition \eqref{eq:no_scale} together with the simplification \eqref{eq:DW_Kahler}, the scalar potential reduces to
\begin{equation}
	\label{eq:scalar_potential_noscale}
	V = e^K G^{i\bar{\jmath}}D_i W_{\mathrm{flux}}\overline{D_jW_{\mathrm{flux}}}\,,
\end{equation}
where $i,\bar{\jmath}$ run over the complex structure moduli only. In particular, note that the $-3|W|^2$ term has dropped out. As a result, the scalar potential \eqref{eq:scalar_potential_noscale} is positive semi-definite and can be seen to be equivalent to \eqref{eq:scalar_potential}. 
\\

\noindent In the superpotential formulation, the global minima are given by those configurations for which $D_iW_{\mathrm{flux}}=0$. Again, one sees that this comprises $h^{3,1}$ complex equations for the $h^{3,1}$ complex structure moduli. Using the properties of the K\"ahler covariant derivatives, this is seen to be equivalent to the condition that $G_4$ has no $(3,1)$ nor $(1,3)$ component in the Hodge decomposition \eqref{eq:Hodge_decomp}. If $G_4$ is primitive, as we will assume throughout this work, this is in turn equivalent to the self-duality condition \eqref{eq:self_dual}. If additionally $W_{\mathrm{flux}}=0$ then $G_4$ also has no $(4,0)$ and $(0,4)$ components, so $G_4$ is purely of type $(2,2)$. In particular, in this case the vacuum corresponds to a Hodge vacuum. This is summarized in table \ref{tab:vacuum_conditions}.

\begin{table}[t!]
	\centering
	\begin{tabular}{|c|c|c|}
		\hline
		\rule[-.15cm]{0cm}{.55cm}                   & \quad Hodge decomposition of $G_4$ \quad  & superpotential \\ \hline
		\rule[-.15cm]{0cm}{.55cm} Hodge vacuum     & $(2,2)$ & $D_i W_{\mathrm{flux}}=W_{\mathrm{flux}}=0$               \\ \hline
		\rule[-.15cm]{0cm}{.55cm} self-dual vacuum &  $(4,0)+(2,2)+(0,4)$ & $D_i W_{\mathrm{flux}}=0\,,W_{\mathrm{flux}}\neq 0$\\
		\hline
	\end{tabular}
	\caption{Overview of the conditions for a Hodge/self-dual flux vacuum in terms of the Hodge decomposition of $G_4$, see \eqref{eq:Hodge_decomp}, and the flux-induced superpotential $W_{\mathrm{flux}}$.}
	\label{tab:vacuum_conditions}
\end{table}

\section{Summary and outlook}
In this final section we briefly summarize the important observations made in this chapter in light of the remainder of the thesis. This will, in particular, motivate much of the discussion in part \ref{part2}, where we discuss the framework of asymptotic Hodge theory. A central lesson that follows from sections \ref{sec:IIB_compactification} and \ref{sec:F-theory_flux} is that the features of four-dimensional effective theories that arise from dimensional reduction of type IIB/M/F-theory are far from generic, but are rather inherited from the geometric properties of the internal space. In well-controlled settings, where one considers compactifications on (conformal) Calabi--Yau manifolds, these geometric properties are captured in terms of \textit{Hodge-theoretic} objects. To reiterate, we have encountered the following objects:
\begin{itemize}
	\item \textbf{Field space metric $G_{i\bar{\jmath}}^{\mathrm{cs}}$:}\\
	The Weil--Petersson metric on the complex structure moduli space of a Calabi--Yau $D$-fold is given by
	\begin{equation}
		G^{\mathrm{cs}}_{i\bar{\jmath}} = \partial_i \partial_{\bar{\jmath}} K^{\mathrm{cs}}\,,\qquad K^{\mathrm{cs}} = -\log\left[i^{-D}\int_{Y_D}\Omega\wedge\bar{\Omega} \right]\,,
	\end{equation}
	which is expressed in terms of the periods of the holomorphic $(D,0)$-form $\Omega$. Recalling that the action of the Hodge star operator on $\Omega$ is $\star\,\Omega = i^D\Omega$, one can equivalently express the K\"ahler potential as a Hodge norm
	\begin{equation}
		K^{\mathrm{cs}} = -\log\left[\int_{Y_D}\Omega\wedge\star\,\bar{\Omega} \right]\,.
	\end{equation}
	\item \textbf{Gauge-kinetic coupling matrices $\mathcal{I}_{IJ}$ and $\mathcal{R}_{IJ}$:}\\
	The gauge-kinetic coupling matrices $\mathcal{I}_{IJ}$ and $\mathcal{R}_{IJ}$ that describe the quadratic terms of the $\mathrm{U}(1)$-gauge fields that appear in the four-dimensional low-energy effective action of type IIB are given in terms of a Hodge norm as
	\begin{align}
		\int_{Y_3} \begin{pmatrix}
			\alpha \\
			\beta
		\end{pmatrix} \wedge\star\begin{pmatrix}
			\alpha & \beta
		\end{pmatrix} = \begin{pmatrix}
			-\mathcal{I}-\mathcal{R}\mathcal{I}^{-1}\mathcal{R} & -\mathcal{R}\mathcal{I}^{-1}\\
			-\mathcal{I}^{-1}\mathcal{R} & -\mathcal{I}^{-1}
		\end{pmatrix}\,.
	\end{align}
	\item \textbf{Flux scalar potential $V$:}\\
	In the four-dimensional low-energy effective theory obtained from F-theory flux compactification, the scalar potential induced by the $G_4$-flux is given by
	\begin{equation}\label{eq:scalar-potential_2}
		V(z,G_4) = \frac{1}{\mathcal{V}_b^2}\int_{Y_4} G_4\wedge\star\, G_4 - G_4\wedge G_4\,,
	\end{equation}
	in which again all the dependence on the complex structure moduli is captured by the Hodge star operator on the internal manifold. 
\end{itemize}
Therefore, in order to properly describe those four-dimensional effective theories that can arise from string compactifications, it is necessary to have a deep understanding, both conceptually and computationally, of these Hodge-theoretic objects. Furthermore, it would be desirable to do so in full generality, i.e.~without restricting to a particular choice of the underlying Calabi--Yau geometry. This is exactly what will be done in part \ref{part2} of the thesis, by employing the powerful tools of \textbf{asymptotic Hodge theory}. Furthermore, in part \ref{part3} of the thesis we will put these tools to work by studying the flux scalar potential \eqref{eq:scalar-potential_2} and its critical points in detail. Finally, in part \ref{part4} we will discuss a different application in the context of integrable field theories, for which a more detailed motivation will be provided at the start of chapter \ref{chap:WZW}.

\setpartpreamble[u][\textwidth]{
	\vspace*{1cm}
	\hrulefill 
	\vspace*{0.5cm}
	
This part of the thesis is devoted to describing the main mathematical framework which is employed in our work, namely \textbf{asymptotic Hodge theory}. The purpose of this part is two-fold. First, it serves as an introduction to this fascinating field of mathematics, providing a balance between a formal and a practical point of view. Second, it works towards two of the most foundational results in this field: the \textbf{nilpotent orbit theorem} and the \textbf{Sl(2)-orbit theorem}.
\\

\noindent Chapter \ref{chap:Hodge} begins with a general introduction to Hodge theory, starting from the perspective of period integrals of Calabi--Yau manifolds. It then works towards to the more abstract formulation in terms of variations of Hodge structure and ends with a description in terms of the period map. 
\\

\noindent Chapter \ref{chap:asymp_Hodge_I} is the first of two chapters devoted to the study of degenerations of Hodge structures. The main purpose of this chapter is to introduce a progressive approximation scheme with which the asymptotic behaviour of a variation of Hodge structure can be analysed to various degrees of precision. On the formal side, it covers key concepts such as mixed Hodge structures, the Hodge--Deligne splitting and the emergence of commuting $\mathfrak{sl}(2,\mathbb{R})$ symmetries near the boundary of the moduli space.   	
\\

\noindent Chapter \ref{chap:asymp_Hodge_II} is the second of two chapters devoted to the study of degenerations of Hodge structures. The central result of this chapter is a concrete algorithmic description of the ``bulk reconstruction procedure'' that was first described in the seminal work of Cattani, Kaplan, and Schmid, which constitutes the main content of the author's first publication \cite{Grimm:2021ikg}. 
	
	\vspace*{0.5cm}
	\hrulefill }
 
\part{Asymptotic Hodge Theory}\label{part2}
\chapter{Hodge Theory}\label{chap:Hodge}
In this chapter we introduce the basics of Hodge theory from three complementary perspectives -- period integrals, abstract variations of Hodge structure, and the period mapping. In section \ref{sec:periods} we start with the most hands-on point of view, namely period integrals, and illustrate the typical strategy to compute them in simple but explicit examples of Calabi--Yau manifolds. The same examples will be used in the subsequent chapters and will serve as a useful guide to pass between the abstract formalism and concrete realizations in string compactifications. In section \ref{sec:VHS}, we introduce the more general framework of variations of Hodge structure. This will be done in two steps: we first introduce the main definitions for a fixed Hodge structure and then describe the generalization to Hodge structures which vary over some parameter space. The latter case is of particular importance in the context of string compactifications. Finally, in section \ref{sec:period_map} we present the most formal formulation in terms of the so-called period map. For introductory texts, we refer the reader to the books \cite{Voisin:2002,Carlson:2017}.

\section{Perspective (1): Periods}
\label{sec:periods}
\subsection{Preliminaries}
In the general context of algebraic geometry, \textbf{periods} correspond to numbers which arise as integrals of rational functions over certain domains \cite{Kontsevich2001}. Every algebraic number is a period, while not all transcendental numbers are periods. For example, while the number $\pi$ can be obtained as a period via
\begin{equation}
\pi = \int_0^1 \frac{4}{x^2+1}\,\mathrm{d}x\,,
\end{equation}
it is currently not known whether $e$ or $\frac{1}{\pi}$ are periods. In fact, in contrast to the set of all transcendental numbers, the set of all periods is countable. One of the deep problems in number theory/algebraic geometry is to find an algorithm which determines whether a given (transcendental) number is equal to a period. 
\\

\noindent For our purposes, an especially interesting situation arises when the function or differential form one is integrating depends on an additional set of parameters. In this case, the resulting period is not just a fixed number, but rather a function of said parameters.  Typically, such periods satisfy a set of linear differential equations with algebraic coefficients, called the \textbf{Picard--Fuchs equations}. This leads to an intricate interplay between the study of periods and the theory of linear differential equations.

\subsubsection*{In Calabi--Yau compactifications}
\noindent In the particular context of Calabi--Yau compactifications, the term ``periods'' typically refers to the integrals of the holomorphic top-form over a basis of integral homology cycles. To be precise, let $Y_D$ be a Calabi--Yau $D$-fold with holomorphic $(D,0)$-form $\Omega$, and let $\gamma_I\in H_{D}\left(Y_D,\mathbb{Z}\right)$ be an integral homology basis. Then the periods of $\Omega$ are given by
\begin{equation}
\label{eq:def_periods_Omega}
\Pi^I = \int_{\gamma_I}\Omega\,.
\end{equation}
As we have seen in subsection \ref{subsec:CY_moduli}, in string compactifications one typically considers not just a fixed Calabi--Yau background, but rather a family of Calabi--Yau manifolds varying in moduli. As a result, the periods $\Pi^I$ become holomorphic functions of the complex structure moduli of the Calabi--Yau manifold. When working with a concrete family of Calabi--Yau manifolds, it is usually not so difficult to write down an expression for the holomorphic top-form. For example, whenever the Calabi--Yau is defined as some hypersurface within an ambient projective/toric variety, $\Omega$ can be straightforwardly related to the defining equations of the hypersurface. However, explicit computation of the integrals \eqref{eq:def_periods_Omega} can be very complicated. Instead, the typical strategy to compute them is to construct the aforementioned set of Picard--Fuchs differential equations that the periods should satisfy. These equations originate from the fact that the set
\begin{equation}
\{\Omega, \partial\Omega,\ldots, \partial^{D+1}\Omega\}\,,
\end{equation} 
where $\partial$ denotes the partial derivative with respect to a complex structure modulus, must be linearly dependent in cohomology. Hence, there must exist some linear combination of $\Omega$ and its derivatives which is exact. After integration this yields a linear differential equation of order $D+1$ for the periods. While we will not employ it in this work, let us mention that there exists a systematic way to perform this procedure in practice known as Dwork--Griffiths reduction \cite{Dwork:1966,Griffiths:1969,Griffiths:1969_II}, see also \cite{Movasati:2007,Lairez_2015} for reviews as well as some concrete computer implementations. We also refer the reader to \cite{Hosono:1993qy,Hosono:1994ax} for applications in the context of toric geometry and complete intersection Calabi--Yau manifolds, as well as mirror symmetry \cite{Candelas:1990rm}.
\\

\noindent In the remainder of this section, we provide some examples to illustrate the abstract statements made above and hint toward further features which will be discussed in due time. The examples are chosen such that they are simple enough to admit explicit computations, while at the same time rich enough to illustrate some generic features of Calabi--Yau periods. As such, the same examples will be used in later chapters of the thesis as well. For simplicity, we restrict here to periods which only depend on one parameter. Later in the thesis we will also provide examples with multiple parameters. 

\subsection{Example: The elliptic curve}

\subsubsection{The flat torus}
As our first example, let us discuss the (unique) Calabi--Yau 1-fold: the torus $T^2$. The simplest representation of the torus is a quotient of $\mathbb{C}$ by a lattice:
\begin{equation}
T^2 = \mathbb{C}/(\mathbb{Z}+\tau\mathbb{Z})\,,\qquad \mathrm{Im}\,\tau>0\,,
\end{equation}
where the quotient is taken with respect to the equivalence relation
\begin{equation}
z\sim z+n+m\tau\,,\qquad z\in\mathbb{C}\,,\quad m,n\in\mathbb{Z}\,.
\end{equation}
The parameter $\tau$ denotes the complex structure modulus of the torus and takes values in the complex upper half-plane $\mathbb{H}$. However, some values of $\tau$ may give rise to equivalent tori. To be precise, the complex structure moduli space $\mathcal{M}$ parametrizing inequivalent complex structures on $T^2$ is given by the quotient space
\begin{equation}
\mathcal{M} = \mathrm{PSL}(2,\mathbb{Z})\backslash \mathbb{H}\,,
\end{equation}
where the modular group $\mathrm{PSL}(2,\mathbb{Z})=\mathrm{SL}(2,\mathbb{Z})/\mathbb{Z}_2$ acts on the upper half-plane $\mathbb{H}$ as
\begin{equation}
\begin{pmatrix}
a & b\\
c & d
\end{pmatrix}\cdot \tau = \frac{a \tau+b}{c \tau+d}\,,\qquad \begin{pmatrix}
a & b\\
c & d
\end{pmatrix}\in\mathrm{SL}(2,\mathbb{Z})\,.
\end{equation}

\subsubsection*{Periods}
The holomorphic $(1,0)$-form is simply given by
\begin{equation}
\label{eq:Omega_torus}
\Omega = \mathrm{d}z\,.
\end{equation}
In the standard homology basis of $T^2$, corresponding to an ``A-cycle'' and a ``B-cycle'', each parametrizing one of the circles of the torus, the periods of $\Omega$ can be collected in terms of the following period vector
\begin{equation}\label{eq:periods_flat-torus}
\mathbf{\Pi} = \begin{pmatrix}
1\\ \tau
\end{pmatrix}\,.
\end{equation}
Let us remark that, in the limit $\tau\rightarrow i\infty$, the torus pinches and becomes a singular manifold. This is an example of a \textbf{large complex structure} (LCS) point. The singular behaviour is also reflected in the period vector. 

\subsubsection{An elliptic curve as a cubic in $\mathbb{P}^2$}
An alternative representation of the torus, which will serve as a useful warm-up for the more complicated Calabi--Yau threefold example, is as a codimension one hypersurface in the complex projective space $\mathbb{P}^2$. Introducing homogeneous coordinates $[x_0:x_1:x_2]\in\mathbb{P}^2$, we consider the family of complex one-dimensional surfaces defined by the equation
\begin{equation}
\mathcal{E}_\psi:\qquad P=x_0^3+x_1^3+x_2^3 - 3\psi\, x_0 x_1 x_2=0\,,
\end{equation}
parametrized by $\psi\in\mathbb{P}^1$. The surface $\mathcal{E}_\psi$ is singular precisely when $\psi^3=1$, for which $P=\mathrm{d}P=0$ at the point $[1:1:1]$, or in the limit $|\psi|\rightarrow\infty$.

\subsubsection*{Periods}
Typically, the holomorphic form on the Calabi--Yau can be straightforwardly obtained from the defining polynomial(s). Indeed, in a patch where $x_0=1$ the holomorphic $(1,0)$-form can be written as
\begin{equation}
\Omega = -\frac{3\psi}{2\pi i} \frac{\mathrm{d}x_1}{\partial P/\partial x_2}=-\frac{3\psi}{(2\pi i)^2}\int \frac{\mathrm{d}x_1\wedge\mathrm{d}x_2}{P}\,,
\end{equation}
where in the second step we have rewritten the expression for $\Omega$ in terms of a residue integral, and we have chosen a particular normalization for future convenience. 
\\

\noindent The fundamental period $\varpi_0$ is obtained by integrating $\Omega$ over a product of sufficiently small loops encircling the points $x_i=0$. A short computation yields
\begin{align*}
\varpi_0 = \int\Omega &= -\frac{3\psi}{(2\pi i)^2}\int \frac{\mathrm{d}x_1\wedge\mathrm{d}x_2}{P}\\
&=\frac{1}{(2\pi i)^2}\sum_{n=0}^\infty\int \frac{dx_1}{x_1}\wedge \frac{dx_2}{x_2}\left[\frac{1+x_1^3+x_2^3}{3\psi x_1 x_2} \right]^n\\
&=\sum_{n=0}^\infty \frac{(3n)!}{(n!)^3}(3\psi)^{-3n}\\
&={}_2F_1\left[\left\{\frac{1}{3},\frac{2}{3},1\right\},\frac{1}{\psi^3}\right]\,,
\end{align*}
which satisfies the Picard--Fuchs equation
\begin{equation}
\left[\theta^2-\mu z\left(\theta+\frac{1}{3}\right)\left(\theta+\frac{2}{3}\right)\right]\varpi_0=0\,,\qquad \mu z=\psi^{-3}\,,\qquad \mu=3^3\,,
\end{equation}
where $\theta=z\frac{\mathrm{d}}{\mathrm{d}z}$. With the Picard--Fuchs operator at hand, it is straightforward to find the other period $\varpi_1$, which is again given by a hypergeometric function
\begin{equation}
\varpi_1 = {}_2F_1\left[\left\{\frac{1}{3},\frac{2}{3},1\right\},1-\mu z\right]\,.
\end{equation}
In the standard homology basis, the period vector (after appropriate normalization) is then given by the same expression \eqref{eq:periods_flat-torus} we found earlier
\begin{equation}
\mathbf{\Pi} = \begin{pmatrix}
1 \\ \tau(\psi)
\end{pmatrix}\,,
\end{equation}
where now the complex structure parameter $\tau(\psi)$ is given in terms of $\psi$ as
\begin{equation}
\tau(\psi) = \frac{i}{\sqrt{3}}\frac{{}_2F_1\left[\left\{\frac{1}{3},\frac{2}{3},1\right\},1-\mu z\right]}{{}_2F_1\left[\left\{\frac{1}{3},\frac{2}{3},1\right\},\mu z\right]} = \frac{\log z}{2\pi i}+\mathcal{O}(z)\,.
\end{equation}
Note that the limit $|\psi|\rightarrow\infty$ (or $z\rightarrow 0$) corresponds to the limit where $\tau\rightarrow i\infty$, which is the large complex structure point. On the other hand, in the limit $\psi^3\rightarrow 1$ (or $z\rightarrow \mu^{-1}$) one finds that $\tau(\psi)$ approaches zero along the imaginary axis. Using the S-duality transformation 
\begin{equation}
S:\quad \tau\mapsto -\frac{1}{\tau}\,,
\end{equation}
to map $\tau(\psi)$ back into the fundamental domain, one sees that the limit is S-dual to the LCS point.

\subsection{Example: The mirror bicubic $\mathbb{P}^5[3,3]$}
Let us now turn to an example of a Calabi--Yau threefold or, rather, a one-parameter family $X_\psi$ of such manifolds: the mirror bicubic \cite{Libgober:1993}. It is defined as the zero-locus of the following two polynomials 
\begin{equation}
P_1 = x_0^3+x_1^3+x_2^3-3\psi\,x_3 x_4 x_5\,,\qquad P_2 = x_3^3+x_4^3+x_5^3-3\psi\, x_0 x_1 x_2\,,
\end{equation}
inside the five-dimensional complex projective space $\mathbb{P}^5$. One can see that $X_\psi$ develops a singularity at $\psi^6=1$, at $\psi=0$ and in the limit $|\psi|\rightarrow\infty$. The moduli space can thus be thought of as a thrice-punctured Riemann sphere. 

\subsubsection*{Periods}
In complete analogy to the previous example, the holomorphic $(3,0)$-form can be expressed in terms of a residue integral involving the two defining polynomials $P_1$ and $P_2$. Its periods were originally computed in \cite{Berglund:1994}. The expression for the fundamental period follows from a calculation that is analogous to the the we performed in the previous example:
\begin{align*}
\varpi_0 &= \frac{1}{(2\pi i)^6}\int (3\psi)^2 \frac{\mathrm{d}x_0\wedge\cdots\wedge\mathrm{d}x_5}{P_1P_2}\\
&=\frac{1}{(2\pi i)^6}\int (3\psi)^2 \frac{\mathrm{d}x_0\wedge\cdots \wedge\mathrm{d}x_5}{x_0\cdots x_5}\sum_{k=0}\frac{\left[x_0^3+x_1^3+x_2^3\right]^k}{(3\psi x_3 x_4 x_5)^k}\sum_{l=0}\frac{\left[x_3^3+x_4^3+x_5^3\right]^l}{(3\psi x_0 x_1 x_2)^k}\\
&=\sum_{n=0}^\infty \frac{((3n)!)^2}{(n!)^6}z^n\,,\qquad z=(3\psi)^{-6}\,,\\
&={}_4F_3\left(\frac{1}{3}, \frac{1}{3}, \frac{2}{3},\frac{2}{3}; 1,1,1; \mu z\right)\,,\qquad \mu=3^6\,.
\end{align*}
The fundamental period satisfies the Picard--Fuchs equation
\begin{equation}\label{eq:PF_mirror-bicubic}
\left[\theta^4-\mu z\left(\theta+a_1\right)\left(\theta+a_2\right)\left(\theta+a_3\right)\left(\theta+a_4\right)\right]\varpi_0=0\,,
\end{equation}
where $\theta= z\frac{\mathrm{d}}{\mathrm{d}z}$ and 
\begin{equation}\label{eq:indices_mirror_bicubic}
\left(a_1,a_2,a_3,a_4\right) = \left(\frac{1}{3},\frac{1}{3},\frac{2}{3},\frac{2}{3}\right)\,.
\end{equation}
The remaining periods can be computed by finding the general solution to the Picard--Fuchs equation \eqref{eq:PF_mirror-bicubic}, and can be expressed in terms of Meijer-G functions. Note that the singularities of the Calabi--Yau are again reflected in the singularities of the Picard--Fuchs equation, which has 3 regular singular points at $z=0,1/\mu,\infty$. Let us also mention that this example falls into a class of similar Calabi--Yau threefold examples whose periods are all described by Picard--Fuchs equations of the form \eqref{eq:PF_mirror-bicubic}, and thus have a similar singularity structure. This will be discussed in more detail in chapter \ref{chap:asymp_Hodge_I}.

\section{Perspective (2): Hodge structures}
\label{sec:VHS}
For a given family of Calabi--Yau manifolds, if one manages to determine the Picard--Fuchs operator, analytical and numerical methods provide powerful tools to compute periods and thus analyze the behaviour of physical observables such as the K\"ahler potential and gauge couplings across the moduli space. However, as the set of all possible, say, Calabi--Yau manifolds is vast (possibly infinite), such techniques may not be suitable to make statements about the general structure of such observables, or may run into computational difficulties for complicated Calabi--Yau manifolds with many moduli. In addition, it is important to stress that the periods alone may not contain all the necessary information. Indeed, while it is true that 
\begin{equation}
\Omega\in H^{D,0}\left(Y_D,\mathbb{C}\right)\,,\quad \partial_i\Omega\in H^{D,0}\left(Y_D,\mathbb{C}\right)\oplus H^{D-1,1}\left(Y_D,\mathbb{C}\right)\,,\quad \text{etcetera}\,,
\end{equation}
it is not true that the derivatives of $\Omega$ necessarily span the full cohomology $H^D\left(Y_D,\mathbb{C}\right)$. For Calabi--Yau $D$-folds with $D\leq 3$ one can in fact recover the full cohomology, while for $D\geq 4$ one only recovers the so-called ``horizontal'' part of the cohomology. Therefore, this issue is of particular relevance in the setting of F-theory on a Calabi--Yau fourfold.
\\

\noindent In this section we will introduce a framework which instead focuses on the abstract properties of the Hodge structure on the middle cohomology $H^D(Y_D,\mathbb{C})$, in particular how it varies under changes of the complex structure moduli: \textbf{variations of Hodge structure}.

\subsection{Hodge structures (1): Basics}
We start by introducing the machinery to describe a fixed Hodge structure, i.e. without any additional moduli dependence. There are three useful and equivalent descriptions of a Hodge structure, which we now discuss in turn. The starting ingredient is a free abelian group of finite rank $H_{\mathbb{Z}}$, whose associated real vector space will be denoted by $H_{\mathbb{R}}:=H_{\mathbb{Z}}\otimes\mathbb{R}$. We will similarly write  $H_{\mathbb{C}}:=H_{\mathbb{Z}}\otimes\mathbb{C}$ for its complexification. Lastly, $D$ denotes a fixed integer.
\subsubsection*{Description 1: Hodge decomposition}
\begin{subbox}{Definition: Hodge structure}
A \textbf{Hodge structure} of \textbf{weight} $D$ on $H_{\mathbb{Z}}$ is a decomposition of its complexification $H_{\mathbb{C}}$ into $D+1$ complex subspaces
\begin{equation}\label{eq:Hodge_decomp_general}
H_{\mathbb{C}} = H^{D,0}\oplus \cdots \oplus H^{0,D}=\bigoplus_{p+q=D}H^{p,q}\,,
\end{equation}
satisfying $H^{p,q}=\overline{H^{q,p}}$ with respect to complex conjugation.
\tcblower 
The real dimensions of the $H^{p,q}$ spaces will be denoted by $h^{p,q}$ and will be referred to as the \textbf{Hodge numbers}. 
\end{subbox}
\noindent A very important class of examples is provided by the Dolbeault cohomology of a compact K\"ahler manifold $X$. This applies in particular to arbitrary smooth complex projective varieties. Indeed, each cohomology group $H^k(X,\mathbb{C})$ admits a Hodge structure of weight $k$ given by the decomposition of harmonic $k$-forms into $(p,q)$-forms, having $p$ holomorphic components and $q$ anti-holomorphic components. We will sometimes refer to such Hodge structures as ``geometrical'' or ``coming from geometry''. Of course, the case where $X$ is a compact Calabi--Yau $D$-fold is of course a special case within this class of examples.

\subsubsection*{Description 2: Hodge filtration}
\begin{subbox}{Definition: Hodge filtration}
A \textbf{Hodge filtration} is a decreasing filtration of vector spaces
\begin{equation}
0\subseteq F^D\subseteq F^{D-1}\subseteq \cdots \subseteq F^0=H_{\mathbb{C}}\,,
\end{equation}
such that $H_{\mathbb{C}}=F^p\oplus \overline{F^{D+1-p}}$. 
\tcblower
The real dimensions of the $F^p$ spaces will be denoted by $f^p$.
\end{subbox}
\noindent One can pass between the two formulations by using the relations
\begin{equation}
\label{eq:decomp_filtration}
H^{p,q} = F^p\cap\overline{F}^q\,,\qquad F^p=\bigoplus_{k=p}^D H^{k, D-k}\,.
\end{equation}

\subsubsection*{Description 3: Charge operator}
It will often be convenient to describe the decomposition \eqref{eq:Hodge_decomp_general} in terms of an operator $Q$ that acts on $H_{\mathbb{C}}$, which we will refer to as the \textbf{charge operator}. In this formulation, the $H^{p,q}$ are defined as the eigenspaces of $Q$ as follows
\begin{equation}
Qv= \frac{1}{2}(p-q)v\,,\qquad v\in H^{p,q}\,,\qquad p+q=D\,,
\end{equation}
such that $H^{p,q}$ is spanned by vectors with eigenvalue $\frac{1}{2}(p-q)$. In particular, this implies that the adjoint action of $Q$ has an integer spectrum. Such an operator is also called a \textit{grading operator}. We will refer to the eigenvectors of $Q$ as charge eigenstates, and to its eigenvalues as charges. The possible charges range from $D/2$ to $-D/2$, with eigenspaces corresponding to $H^{D,0}$ and $H^{0,D}$, respectively. Furthermore, the property $\overline{H^{p,q}} = H^{q,p}$ implies that $\overline{Q} = -Q$, so that $Q$ is a purely imaginary operator. Regarding the Hodge filtration, one sees that each $F^p$ is spanned by states whose charge is greater or equal to $p-D/2$. Throughout the text, we will often switch between the description of the Hodge structure in terms of the decomposition, the filtration and the charge operator. 

\subsubsection*{Polarization}
Throughout this work we will only consider so-called \textbf{polarized} Hodge structures. This means that $H_{\mathbb{Z}}$ is endowed with a $(-1)^D$-symmetric bilinear form
\begin{equation}
(\cdot,\cdot):\,H_{\mathbb{Z}}\times H_{\mathbb{Z}}\rightarrow \mathbb{Z}\,,
\end{equation}
satisfying the following Hodge--Riemann bilinear relations with respect to the Hodge decomposition\footnote{For simplicity of notation, we will employ the same notation for the complexification of the pairing $\left(\cdot, \cdot\right)$.}
\begin{align}
&\text{(i)}:\qquad \left(H^{p,q},H^{r,s}\right)=0\,,\qquad \text{unless $(p,q)=(s,r)$\,,}\\
&\text{(ii)}:\quad \hspace{0.6cm}i^{p-q}\left(v,\bar{v}\right)>0\,,\qquad \text{for $v\in H^{p,q}$ and $v\neq 0$\,.}
\end{align}
We will often refer to $(\cdot,\cdot)$ as the intersection form. The first condition can equivalently be expressed in terms of the Hodge filtration as
\begin{equation}
\text{(i)}:\qquad \left(F^p, F^{D+1-p}\right)=0\,.
\end{equation}
Finally, in terms of the charge operator, the first condition simply states that the pairing between two charge eigenstates vanishes unless their charges add to zero. 
\\

\noindent In the geometrical setting, a polarized Hodge structure is provided by restricting to the \textit{primitive} part of the Dolbeault cohomology. To be precise, let $X$ be a compact K\"ahler manifold of dimension $2n$. Then for each $0\leq k\leq 2n$, the pairing on the cohomology group $H^k(X,\mathbb{C})$ is given by
\begin{equation}
\left(v, w\right) = (-1)^{k(k-1)/2}\int_X J^{n-k}\wedge v\wedge w\,,
\end{equation}
where $J$ denotes the K\"ahler $(1,1)$-form on $X$. For the particular types of string compactifications that will be considered in this work, let us remark that the middle cohomology of a Calabi--Yau threefold is automatically primitive, while in the fourfold case the primitivity condition amounts to the constraint that $J\wedge v=0$.

\subsubsection*{Weil operator}
Next, we introduce one of the central objects of this thesis: the \textbf{Weil operator}. It will be denoted by $C$, and is defined to act on the various components of the Hodge decomposition as 
\begin{equation}\label{eq:def-Weil}
Cv = i^{p-q}v\,,\qquad v\in H^{p,q}\,.
\end{equation}
In the geometric setting, the Weil operator exactly corresponds to the action of the Hodge star operator $\star$ on the middle cohomology. In general, the Weil operator satisfies $C^2=(-1)^D$, hence its eigenvalues are $\pm 1$ when $D$ is even, and $\pm i$ when $D$ is odd. Correspondingly, we employ the following terminology for its eigenvectors:
\begin{itemize}
\item \textbf{(anti) self-dual:} $Cv=\pm v$\,,
\item \textbf{imaginary (anti) self-dual:} $Cv=\pm i v$.
\end{itemize}
Note also that the Weil operator and the charge operator are straightforwardly related by 
\begin{equation}
C = (-1)^Q\,.
\end{equation}
It is important to stress, however, that the Weil operator itself does not suffice to describe the full Hodge decomposition. Instead, the main relevance of the Weil operator is that it induces a natural inner product on $H_\mathbb{C}$ that is compatible with the Hodge decomposition. Indeed, as a result of the second polarization condition, one finds that
\begin{equation}\label{eq:Hodge_inner_product}
\langle v,w\rangle:=\left(v, C\bar{w}\right)\,,\qquad ||v||^2:=\langle v,v\rangle\,,
\end{equation}
respectively define an inner product and a norm on $H_{\mathbb{C}}$. Furthermore, as a consequence of the first polarization condition, the Hodge decomposition \eqref{eq:Hodge_decomp_general} is orthogonal with respect to this Hodge inner product. 

\subsubsection*{Symmetry groups/algebras}
Let us write $G_{\mathbb{R}}$ for the real automorphism group of the pairing $\left(\cdot,\cdot\right)$, and denote its algebra by $\mathfrak{g}_{\mathbb{R}}$. This means that
\begin{align}
&g\in G_{\mathbb{R}}:\qquad \left(gv, gw\right) = (v,w)\,,\\
&X\in\mathfrak{g}_{\mathbb{R}}:\qquad \left(Xv, w\right)+\left(v,Xw\right)=0\,,
\end{align}
for all $v,w\in H_{\mathbb{R}}$. The complexification of $G_{\mathbb{R}}$ and $\mathfrak{g}_{\mathbb{R}}$ will be denoted by $G_{\mathbb{C}}$ and $\mathfrak{g}_{\mathbb{C}}$, respectively. 
Since $(\cdot,\cdot)$ is assumed to be either symmetric or skew-symmetric, one finds
\begin{equation}
G_{\mathbb{R}} = \begin{cases}
\mathrm{Sp}(2m,\mathbb{R})\,, & \text{$D$ odd,}\qquad \\
\mathrm{SO}(r,s)\,, & \text{$D$ even,}
\end{cases}\,,\qquad G_{\mathbb{C}} = \begin{cases}
\mathrm{Sp}(2m,\mathbb{C})\,, & \text{$D$ odd,}\qquad \\
\mathrm{SO}(2m)\,, & \text{$D$ even,}
\end{cases}\,,
\end{equation} 
where
\begin{equation}
2m = \sum_{p} h^{p,D-p} = \mathrm{dim}\,H_{\mathbb{R}}\,,
\end{equation}
and
\begin{equation}
r = \sum_{\text{$p$ even}} h^{p,D-p}\,,\quad s= \sum_{\text{$p$ odd}} h^{p,D-p}\,,\qquad r+s=2m\,.
\end{equation}
Similarly, we introduce $K$ as the group of real transformations that preserve the inner product $\langle\cdot,\cdot\rangle$ and denote its algebra by $\mathfrak{k}$. In other words, $K$ consists of unitary operators with respect to the given inner product. In particular, this implies that $K$ is a maximal compact subgroup of $G_{\mathbb{R}}$ and is given by
\begin{equation}
K = \begin{cases}
\mathrm{U}(m)\,, & \text{$D$ odd,}\\
\mathrm{SO}(2m)\cap\left(\mathrm{O}(r)\times \mathrm{O}(s)\right)\,, & \text{$D$ even.}
\end{cases}
\end{equation}
Furthermore, from their definition it is clear that $C\in K$ and $Q\in i \mathfrak{k}$ (recall that $Q$ is purely imaginary). Lastly, let us note that the quotient $G/K$ of a semisimple Lie group $G$ by its maximal compact subgroup $K$ always yields a symmetric space. 

\subsubsection*{Adjoint operation}
Finally, we introduce the notation
\begin{equation}\label{eq:def_adjoint}
	g^\dagger := C\bar{g}^{-1}C^{-1}\,,\qquad X^\dagger := - C\bar{X} C^{-1}\,,
\end{equation}
to denote the adjoint of $g\in G_{\mathbb{C}}$, respectively $X\in\mathfrak{g}_{\mathbb{C}}$, with respect to the Hodge inner product $\langle\cdot,\cdot\rangle$, such that the relation
\begin{equation}
	\langle v, gw\rangle = \langle g^\dagger v, w\rangle\,,
\end{equation}
holds. Note, in particular, that the subgroup $K$ consisting of unitary operators is thus generated by elements $g\in G_{\mathbb{R}}$ satisfying
\begin{equation}
	g g^\dagger = 1\,.
\end{equation}
Let us also note that this is equivalent to the condition that $g$ commutes with the Weil operator $C$. 

\subsection{Hodge structures (2): Variations of Hodge structure}\label{subsec:VHS}
We speak of a variation of Hodge structure when the decomposition \eqref{eq:Hodge_decomp_general} varies over some parameter space $\mathcal{M}$ in a particular way which will be specified in a moment. For example, in the F-theory setting the parameter space $\mathcal{M}$ corresponds to the complex structure moduli space of the underlying Calabi--Yau fourfold. Since a variation of the complex structure changes the notion of what we call holomorphic and anti-holomorphic, this induces a variation of the decomposition \eqref{eq:Hodge_decomp_general}. The properties of a variation of Hodge structure are neatly encoded in terms of the Hodge filtration. Indeed, given a set of local coordinates $z^i$ on $\mathcal{M}$, the filtration must satisfy the following conditions
\begin{align}
\label{eq:transversality}	\text{Griffiths transversality}:\qquad \frac{\partial}{\partial z^i} F^{p}&\subseteq F^{p-1}\,,\\
\label{eq:holomorphicity}	\text{holomorphicity}:\qquad \frac{\partial}{\partial \bar{z}^i} F^{p}&\subseteq F^{p}\,.
\end{align}
The former condition implies that when taking a holomorphic derivative of a vector in $F^p$, the resulting vector ends up at most one step down in the filtration. The latter condition means that the Hodge filtration varies \textit{holomorphically} as a function of the moduli. This is in contrast with the Hodge decomposition $H^{p,q}$, for which only $H^{D,0}$ varies holomorphically while the rest of the components do not. These two conditions are motivated by the Hodge decomposition of the primitive middle de Rham cohomology of a K\"ahler manifold, or more generally of algebraic families of algebraic varieties, for which they were shown to hold by Griffiths \cite{Griffiths:1968_I,Griffiths:1968_II}. Any family of polarized Hodge structures satisfying \eqref{eq:transversality} and \eqref{eq:holomorphicity} is referred to as a \textbf{variation of polarized Hodge structure} \cite{Griffiths:1970}. It should be stressed that such variations of polarized Hodge structure need not have a geometric origin in the form of K\"ahler manifolds (or algebraic varieties) and can be studied at an abstract level. 
\\

\noindent Abstracting slighty, one should think of a variation of Hodge structure as being defined in terms of the Hodge bundle
\begin{equation}\label{eq:Hodge_bundle}
E\rightarrow \mathcal{M}\,,
\end{equation}
The fibres of the bundle \eqref{eq:Hodge_bundle} are the vector space $H_{\mathbb{C}}$, and the fibration encodes the variation of the $(p,q)$-decomposition of $H_{\mathbb{C}}$ as one moves in the base space $\mathcal{M}$. Locally, one may think of points in $E$ as a pair $(z^i,v)$, with $z^i\in\mathcal{M}$ and $v\in H_{\mathbb{C}}$. 

\section{Perspective (3): The period map}\label{sec:period_map}
In the preceding section we have first introduced the notion of a fixed polarized Hodge structure on a given vector space, and then generalized this to a family of polarized Hodge structures varying over some parameter space. It is then natural to organize all the different polarized Hodge structures one can put on a given vector space into a so-called \textbf{classifying space}. Subsequently, one may then view a given variation of Hodge structure as one particular embedding inside this classifying space. This results in a slightly more geometric way to think about Hodge structures, which will play a crucial role in \ref{part4} of this thesis. 

\subsection{The classifying space of Hodge structures}\label{subsec:classifying_space}
For a fixed $H_{\mathbb{Z}}$, weight $D$, and bilinear pairing $(\cdot,\cdot)$ one may consider the set of all polarized Hodge structures of weight $D$ on $H_{\mathbb{Z}}$ of fixed type, meaning that we fix the Hodge numbers $h^{p,q}$. The resulting set is denoted by $\mathcal{D}$ and is commonly referred to as the \textbf{period domain}. Concretely, it consists of all possible Hodge filtrations $F^p$, whose dimensions are fixed in terms of the Hodge numbers, which additionally satisfy the polarization conditions. Without going into too much detail, it can be realized as an open, complex subvariety of a product of Grassmannians, see for example \cite{schmid,CKS} for further details.
\\

\noindent For our purposes, a more concrete description of $\mathcal{D}$ can be given as follows. The idea is to first consider some reference Hodge structure $H^{p,q}_{\mathrm{ref}}$ and to realize that, for any other decomposition $H^{p,q}$, there always exist a real operator $h\in G_{\mathbb{R}}$ such that\footnote{Note that $h$ is a real operator by virtue of the property $\overline{H^{p,q}}=H^{q,p}$.}
\begin{equation}
\label{eq:def_h}	H^{p,q} = h\cdot H^{p,q}_{\mathrm{ref}}.
\end{equation}
In other words, the group $G_{\mathbb{R}}$ acts transitively on $\mathcal{D}$. Note, however, that \eqref{eq:def_h} only defines $h$ up to local right multiplication by group elements that leave $H^{p,q}_{\mathrm{ref}}$ invariant. Concretely, we introduce the subgroup
\begin{equation}
V = \{g\in G_{\mathbb{R}}: g H^{p,q}_{\mathrm{ref}} = H^{p,q}_{\mathrm{ref}},\,\forall p,q\}\,,
\end{equation}
which is the stabilizer of all the $H^{p,q}_{\mathrm{ref}}$ spaces. Equivalently, $V$ is generated by operators which commute with the charge operator $Q_{\mathrm{ref}}$, which implies that $V$ is contained in $K$. Explicitly, one finds
\begin{equation}
V = \begin{cases}
\prod_{p\leq l} \mathrm{U}(h^{p,D-p})\,, & D=2l+1\,,\\
\mathrm{SO}(h^{l,l})\times \prod_{p<l} \mathrm{U}(h^{p,D-p})\,,  & D=2l\,.
\end{cases}
\end{equation}
In short, we have argued that the period domain can also be represented as a homogeneous space 
\begin{equation}
\mathcal{D}\cong G_{\mathbb{R}}/V\,.
\end{equation}

\subsubsection*{Monodromy}
A central concept, which will play a crucial role in chapters \ref{chap:asymp_Hodge_I} and \ref{chap:asymp_Hodge_II}, is that of \textbf{monodromy}. This is based on the fact that, in practice, the parameter space $\mathcal{M}$ is not simply-connected. Indeed, a typical local model for $\mathcal{M}$ is a product of punctured disks, in which the puncture corresponds to a value of the parameters where the Hodge structure degenerates. As a result, it may happen that, as one considers the parallel transport of some reference Hodge structure along a path which encirlces a singularity, the Hodge structure transforms non-trivially. Due to the properties of variations of Hodge structure, the transformation only depends on the homotopy class of the path in question.\footnote{To be precise, this uses fact that the so-called Gauss--Manin connection is flat.} 
%In practice, choosing local coordinates $z^i$ such that the punctures are located at the position $z^i=0$, it thus suffices to consider how the Hodge structure changes under the monodromy transformation
%\begin{equation}
%	z^i\mapsto z^i e^{2\pi i}\,.
%\end{equation}
The corresponding transformation is captured by a representation of the fundamental group 
\begin{equation}\label{eq:fundamental_group}
\rho: \pi_1\left(\mathcal{M}\right)\rightarrow G_{\mathbb{Z}}\,,
\end{equation}
whose image will be denoted by $\Gamma$, and is referred to as the \textbf{monodromy group}. 

\subsubsection*{The period map and its lift}
The upshot of the above discussion is that every point $z\in\mathcal{M}$ determines a polarized Hodge structure of weight $D$, i.e.~a point in the period domain $\mathcal{D}$, \textit{up to monodromy}. In other words, all the information about a given variation of Hodge structure can alternatively be described in terms of a map
\begin{equation}
\Phi:\quad \mathcal{M}\rightarrow \Gamma\backslash\mathcal{D} \cong \Gamma\backslash G_{\mathbb{R}}/V\,,
\end{equation}
which is known as the \textbf{period map}. It assigns to each point in the parameter space $\mathcal{M}$ an element in the double coset space $\Gamma\backslash G_{\mathbb{R}}/V$, which in turn provides an equivalence class of Hodge structures via \eqref{eq:def_h}, all of which are related by a monodromy transformation. In practice, it is useful to lift the period map to the universal covering space of $\mathcal{M}$, which will be denoted by $\widetilde{\mathcal{M}}$. This results in a lift of the period map
\begin{equation}
\tilde{\Phi}:\quad \widetilde{\mathcal{M}}\rightarrow\mathcal{D}\,,
\end{equation}
which transforms as
\begin{equation}
\tilde{\Phi}(\gamma\cdot t) = \rho(\gamma)\cdot \tilde{\Phi}(t)\,,
\end{equation}
for $\gamma\in\pi_1(\mathcal{M})$, where we recall that $\rho$ is the representation of the fundamental group, see \eqref{eq:fundamental_group}, and we have introduced local coordinates $t^i$ on $\widetilde{M}$. 

\subsection{Horizontality of the period mapping}
The conditions \eqref{eq:transversality}--\eqref{eq:holomorphicity} that define how a Hodge structure is allowed to vary over the parameter space $\mathcal{M}$ impose a set of differential equations on the period map. Concretely, they imply that $\Phi$ is \textit{holomorphic}, and that its differential takes values in only a restricted part of the tangent bundle, namely the so-called \textit{horizontal tangent bundle}. The purpose of this section is to rephrase these somewhat abstract conditions into concrete differential equations satisfied by a $G_{\mathbb{R}}$-valued field. To this end, let us again consider a fixed reference Hodge structure $H^{p,q}_{\mathrm{ref}}$, and parametrize a given variation of Hodge structure as
\begin{equation}
F^p = h\cdot F^p_{\mathrm{ref}}\,,
\end{equation}
where we use the formulation in terms of Hodge filtrations for future convenience. We stress that the filtration $F^p$ can vary over the parameter space $\mathcal{M}$, and this dependence is fully captured in the group-valued field $h$. At each point $t\in\widetilde{\mathcal{M}}$ one may think of $h(t,\bar{t})$ as a choice of representative for the equivalence class $\tilde{\Phi}(t)$ inside $G_{\mathbb{R}}$. The fact that this is well-defined, i.e.~independent of the choice of representative, is then reflected in the fact that the equations that $h$ satisfies enjoy a gauge symmetry. 
\\

\noindent The conditions \eqref{eq:transversality} and \eqref{eq:holomorphicity} can now be written as differential equations for $h$ as follows. To simplify the notation, we restrict to the case where $\mathcal{M}$ is one-dimensional. The generalization to higher-dimensional moduli spaces is straightforward. Then the conditions \eqref{eq:transversality} and \eqref{eq:holomorphicity} read
\begin{equation}\label{eq:horizontality_h}
(h^{-1}\partial h)F_{\mathrm{ref}}^p \subseteq F_{\mathrm{ref}}^{p-1}\,,\qquad  (h^{-1}\bar{\partial} h)F_{\mathrm{ref}}^p \subseteq F_{\mathrm{ref}}^p\,,
\end{equation}
where now the derivatives are taken with respect to the covering space coordinate $t$. These equations are still somewhat difficult to work with, as they are vector space equations. Therefore, we use the characterization of the Hodge structure in terms of eigenspaces of the reference charge operator $Q_{\mathrm{ref}}$. Recall that $F^p_{\mathrm{ref}}$ is spanned by states with charge greater or equal to $p-D/2$. Therefore, the first condition in \eqref{eq:horizontality_h} states that the operator $h^{-1}\partial h$ can lower the charge of a state by at most one, while $h^{-1}\bar{\partial} h$ may not lower the charge at all. To describe this more quantitatively, one introduces the charge-decomposition of an operator $\mathcal{O}\in \mathfrak{g}_{\mathbb{C}}$ as\footnote{Since $Q$ is not a real operator, this decomposition necessarily requires one to move to the complexification $\mathfrak{g}_\mathbb{C}$ of $\mathfrak{g}$.}
\begin{equation}
\mathcal{O} = \sum_{q} \mathcal{O}_q\,,
\end{equation}
where the charge modes $\mathcal{O}_q$ are defined via
\begin{equation}\label{eq:q_mode}
[Q_{\mathrm{ref}}, \mathcal{O}_q] = q\, \mathcal{O}_q\,,\qquad q=-D,\ldots, D\,.
\end{equation}
Clearly then, the action of $\mathcal{O}_q$ on a charge eigenstate raises its charge by $q$. For convenience, we also introduce the notation
\begin{equation}\label{eq:def_B}
\mathbf{B}=h^{-1}\dd h\,,\qquad B = h^{-1}\partial h\,,\qquad \bar{B} = h^{-1}\bar{\partial}h\,,
\end{equation}
Then the statement that $B$ lowers the charge by at most one and $\bar{B}$ may not lower the charge at all translates to
\begin{equation}\label{eq:Qmodes_t}
B_q = 0\,,\quad q< -1\,,\qquad \bar{B}_q = 0\,,\quad q< 0\,.
\end{equation} 
By taking the complex conjugate of \eqref{eq:Qmodes_t} and recalling that $\bar{Q}_{\mathrm{ref}} = -Q_{\mathrm{ref}}$, we arrive at a similar condition
\begin{equation}\label{eq:Qmodes_tbar}
\bar{B}_q = 0\,,\quad q> 1\,,\qquad B_q = 0\,,\quad q> 0\, .
\end{equation} 
Finally, combining \eqref{eq:Qmodes_t} and \eqref{eq:Qmodes_tbar} one obtains the following condition.
\begin{subbox}{The horizontality condition}
	\begin{equation}\label{eq:horizontality_B}
			B_{-q} = 0 = 
			\bar{B}_q \,,\quad \text{unless $q=0$ or $q=1$}\,.\quad 
	\end{equation}
\end{subbox}
\noindent In summary, we have rephrased the condition \eqref{eq:horizontality_h} into the statement that only components of $B$ and $\bar{B}$ with a particular charge can be non-zero. We will refer to \eqref{eq:horizontality_B} as the \textbf{horizontality condition}. 
\\

\noindent At first sight, it appears that the non-zero components of $B$ and $\bar{B}$ remain unrestricted by the horizontality condition. This is, however, not the case. Indeed, note that $B$ and $\bar{B}$ satisfy the following no-curvature condition
\begin{equation}\label{eq:no_curvature_B}
\partial\bar{B} - \bar{\partial}B+[B,\bar{B}]=0\,,
\end{equation}
as is apparent from their definition. Then by projecting \eqref{eq:no_curvature_B} onto its various charge components and using \eqref{eq:horizontality_B} one obtains a set of differential equations that constrain the various components of $B$ and $\bar{B}$ as follows
\begin{align}
\label{eq:charge_plus1}	&\text{charge $+1$}:\qquad \partial\bar{B}_{+1}+[B_0,\bar{B}_{+1}] = 0\,,\\
\label{eq:charge_0}	&\text{charge $0$}:\qquad \partial\bar{B}_0-\bar{\partial}B_0 +[B_{-1},\bar{B}_{+1}] = 0\,,\\
\label{eq:charge_min1}	&\text{charge $-1$}:\qquad \bar{\partial}B_{-1}+[\bar{B}_{0},B_{-1}] = 0\,.
\end{align}
In appendix \ref{app:horizontality} we will show how one can relate the above equations to the well-known set of Nahm's equations, which will be of central importance in chapters \ref{chap:asymp_Hodge_I} and \ref{chap:asymp_Hodge_II}. 

\subsubsection*{The charge operator revisited}
Before closing this section, it is worthwhile to rephrase the description of a variation of Hodge structure yet again, which will be useful in part \ref{part4}. Recall that the Hodge structure $H^{p,q}_{\mathrm{ref}}$ can be encoded as the eigenspaces of the reference charge operator $Q_{\mathrm{ref}}$. Naturally, one can also encode the full variation of Hodge structure $H^{p,q}(z,\bar{z})$ via a coordinate-dependent charge operator $Q$, which is naturally related to $Q_{\mathrm{ref}}$ via 
\begin{equation}\label{eq:charge-operator_bulk}
Q(t,\bar{t})= h Q_{\mathrm{ref}} h^{-1}\,,
\end{equation}
as can be inferred from \eqref{eq:def_h}. We will sometimes refer to $Q(t,\bar{t})$ as the `bulk charge operator'. Of course, the coordinate-dependence of $Q(t,\bar{t})$ is restricted by the horizontality condition, which can be translated into the following condition\footnote{To derive this, one simply evaluates $\partial Q = \partial(h Q_{\mathrm{ref}}h^{-1}) = h[h^{-1}\partial h, Q_{\mathrm{ref}}]h^{-1}=h B_{-1} h^{-1}$ and then computes the commutator with $Q$, where in the last equality we have used the horizontality condition \eqref{eq:horizontality_B}.}
\begin{equation}\label{eq:horizontality_Q}
\text{horizontality condition}:\qquad [Q,\partial Q] = -\partial Q\,,\qquad [Q,\bar{\partial}Q] = \bar{\partial}Q\,.
\end{equation}
This concludes the discussion on variations of Hodge structures and the period mapping. In short, we have shown that the data of a variation of Hodge structure can be equivalently formulated in terms of the period mapping $h$, satisfying the condition \eqref{eq:horizontality_B}, which furthermore restricts the non-zero components charge components of $B$ and $\bar{B}$ with respect to $Q_{\mathrm{ref}}$ to satisfy \eqref{eq:charge_plus1}-\eqref{eq:charge_min1}. Alternatively, it can be captured by a purely imaginary grading operator $Q(t,\bar{t})$ that respects the condition \eqref{eq:horizontality_Q}.

\begin{subappendices}
\section{Horizontality and Nahm's equations}\label{app:horizontality}
In this section we provide some additional details regarding the horizontality conditions of the period map, recall equations \eqref{eq:transversality}--\eqref{eq:holomorphicity}, or equivalently \eqref{eq:horizontality_h} or \eqref{eq:horizontality_B}. In subsection \ref{subsec:horizontality} we write down an equivalent formulation of these conditions in terms of a set of commutation relations, which will be of use in part \ref{part2} and part \ref{part4} of the thesis. Subsequently, in subsection \eqref{subsec:Nahm} we uncover a well-known set of equations called Nahm's equations from the horizontality conditions, after imposing a certain periodicity condition. We describe various useful formulations of Nahm's equations, as well as a geometric action principle from which they can be recovered as the equations of motion. The latter will play an especially important role in part \ref{part4} of the thesis.

\subsection{Rewriting the horizontality condition}\label{subsec:horizontality}
Recall from \eqref{eq:horizontality_h} and \eqref{eq:horizontality_B} that the horizontality condition of the period map imposes that
\begin{equation}\label{eq:horizontality_app}
B = B_0 + B_{-1}\,,\qquad \bar{B} = \bar{B}_0+\bar{B}_{+1}\,,
\end{equation}
where the subscript denotes the charge with respect to a chosen reference charge operator $Q_{\mathrm{ref}}$, and $B = h^{-1}\partial h$. 

\subsubsection*{A gauge-fixing condition}
It is important to recall that the period map enjoys a gauge symmetry
\begin{equation}
h\mapsto h g\,,\qquad g\in V\,,
\end{equation}
where we recall that $V$ is the stabilizer of the reference Hodge structure, which is generated by operators which commute with $Q_{\mathrm{ref}}$. In order to gauge-fix the above gauge symmetry, we proceed as follows. First, it will be convenient to work in real coordinates $t=x+iy$, such that
\begin{equation}\label{eq:B_complex-to-real}
	B = \frac{1}{2}(B_x - B_y)\,,\qquad \bar{B}=\frac{1}{2}(B_x+i B_y)\,.
\end{equation}
See also subsection \ref{subsec:nilpotent-orbit} in the next chapter for a more detailed discussion on the choice of local coordinates. Next, we consider an infinitesmial gauge transformation generated by
\begin{equation}
g = e^\xi\,,\qquad \xi = \xi_0\,,
\end{equation}
for which we find
\begin{equation}
\left(B_y\right)_0\mapsto \left(B_y\right)_0 + [\left(B_y\right)_0, \xi_0]+\partial_y\xi_0+\mathcal{O}\left(\xi_0^2\right)\,.
\end{equation}
In particular, by an appropriate choice of $\xi$ we may choose to work in a gauge in which $(B_y)_0=0$, or equivalently $B_0 = \bar{B}_0$. Note that this does not fully fix the gauge freedom, since it only restricts the $y$-dependence of $\xi$. The $x$-dependence will be fixed in the next subsection.

\subsubsection*{Rewriting the horizontality condition}
Having chosen a convenient gauge, we will now show how the horizontality conditions \eqref{eq:horizontality_app} can be rephrased as a simple set of commutation relations involving the reference charge operator $Q_{\mathrm{ref}}$. To this end, we perform the following computation.

\begin{align*}
[Q_{\mathrm{ref}}, B_x]&\stackrel{\text{(a)}}{=} [Q_{\mathrm{ref}},B+\bar{B}]\\
&\stackrel{\text{(b)}}{=} [Q_{\mathrm{ref}}, 2B_0+\bar{B}_{+1}+B_{-1}]\\
&= \bar{B}_{+1}-B_{-1}\\
&=\left(B_0+\bar{B}_{+1}\right)-\left(B_0-B_{-1}\right)\\
&\stackrel{\text{(c)}}{=}\bar{B}-B\\
&\stackrel{\text{(d)}}{=}i B_y\,,
\end{align*}
where in steps (a) and (d) we have employed the relation \eqref{eq:B_complex-to-real} between the complex and real variables, and in steps (b) and (c) we have used the horizontality conditions \eqref{eq:horizontality_app}. In a similar fashion, one may compute the following commutator
\begin{align*}
[Q_{\mathrm{ref}}, B_y]&=-i[Q_{\mathrm{ref}},\bar{B}-B]\\
&=-i[Q_{\mathrm{ref}}, B_0+\bar{B}_{+1}-B_0-B_{-1}]\\
&=-i\left(\bar{B}_{+1}+B_{-1}\right)\\
&=-\frac{i}{2}\left[\left(-2B_0+\bar{B}_{+1}+B_{-1} \right)+\left(2B_0+\bar{B}_{+1}+B_{-1} \right)\right]\\
&\stackrel{\text{(a)}}{=}-\frac{1}{2}\left[\left(B+\bar{B}\right)^\dagger + \left(B+\bar{B}\right)\right]\\
&=-\frac{i}{2}\left[B_x^\dagger + B_x\right]\,,
\end{align*}
where in step (a) we have used the definition of the adjoint \eqref{eq:def_adjoint}, which acts on a charge eigenstate as
\begin{equation}
\left(B_{-q}\right)^\dagger = -(-1)^{\mathrm{ad}\,Q_{\mathrm{ref}}}\bar{B}_q=(-1)^{1+q}\bar{B}_q\,.
\end{equation}
In summary, we have shown that the horizontality conditions can be rephrased in terms of the following commutation relations with respect to a chosen reference charge operator. 
\begin{subbox}{Horizontality conditions (commutator version)}
\begin{align}	
	\label{eq:Q_constraint_x}
	[Q_{\mathrm{ref}}, h^{-1}\partial_x h] &= i h^{-1}\partial_y h\,,\\ 
	\label{eq:Q_constraint_y}
	[Q_{\mathrm{ref}}, h^{-1}\partial_y h] &= -\frac{i}{2}\left[\left(h^{-1}\partial_x h\right)^\dagger + h^{-1}\partial_x h \right]\,.
\end{align}
\tcblower 
\textbf{Note:}\\
The relations \eqref{eq:Q_constraint_x}--\eqref{eq:Q_constraint_y} will sometimes be referred to as the \textbf{$Q$-constraint}. 
\end{subbox}
\noindent The reader is invited to check that, conversely, the commutation relations \eqref{eq:Q_constraint_x}--\eqref{eq:Q_constraint_y} imply the horizontality conditions \eqref{eq:horizontality_app}, so that the two sets of equations are equivalent. \newpage

\subsection{Nahm's equations and an action principle}\label{subsec:Nahm}
The equations \eqref{eq:Q_constraint_x} and \eqref{eq:Q_constraint_y} thus comprise all the constraints coming from the horizontality of the period map. In the following, we will show how, under a certain simplifying assumption, these equations give rise to the so-called Nahm's equations. The latter will be of great relevance in chapter \ref{chap:asymp_Hodge_II}. 

\subsubsection*{Derivation of Nahm's equations}

The following discussion can be viewed as an alternative derivation of Lemma 9.8 of \cite{schmid}. The central idea will be to combine the horizontality conditions together with the flatness condition of the Maurer--Cartan form $B$, which reads
\begin{equation}
\partial_x B_y - \partial_y B_x +[B_x, B_y]=0\,.
\end{equation}
Additionally, we will make one simplifying assumption regarding the $x$-dependence of the period map.
\begin{redbox}{Assumption}
From this point onwards, we will assume that
\begin{equation}\label{eq:assumption_period_map}
	\partial_x B_x  = \partial_x B_y = 0\,.
\end{equation}
\tcblower
\textbf{Note}:\\
This assumption will be satisfied in our cases of interest, namely when the period map $h$ satisfies
\begin{equation}\label{eq:assumption_period_map_2}
	h(x,y) = e^{xN}\tilde{h}(y)\,,
\end{equation}
with $N$ and $\tilde{h}(y)$ independent of $x$. As will be explained in chapter \ref{chap:asymp_Hodge_I}, the period map will take the form \eqref{eq:assumption_period_map_2} whenever one is working in the so-called nilpotent orbit approximation. 
\end{redbox}
\noindent Note that, in order for our initial gauge-fixing to be compatible with this assumption, we must have $\partial_x \xi=0$. In particular, after imposing this condition there is no residual gauge symmetry. \\

\noindent As a result of the assumption \eqref{eq:assumption_period_map}, the flatness condition simplifies to
\begin{equation}
\partial_y B_x = [B_x, B_y]\,.
\end{equation}
In particular, together with the fact that $\partial_x B_x=0$, this completely fixes the coordinate dependence of $B_x$ in terms of $B_y$. Since we also assume $\partial_x B_y=0$, it remains to find an equation which fixes $\partial_y B_y$. This is done as follows.
\begin{align*}
-i\partial_y B_y &\stackrel{\text{(a)}}{=}[Q_{\mathrm{ref}}, \partial_y B_x]\\
&\stackrel{\text{(b)}}{=}[Q_{\mathrm{ref}}, [B_x, B_y]]\\
&\stackrel{\text{(c)}}{=}-[B_x, [B_y, Q_{\mathrm{ref}}]] - [B_y, [Q_{\mathrm{ref}}, B_x]]\\
&\stackrel{\text{(d)}}{=}\frac{i}{2}[B_x, B_x+B_x^\dagger] - i[B_y, B_y]\\
&=\frac{i}{2}[B_x, B_x^\dagger]\,,
\end{align*}
where in steps (a) and (d) we have employed the Q-constraint \eqref{eq:Q_constraint_x}--\eqref{eq:Q_constraint_y}, in step (b) we have used the simplified flatness condition, and in step (c) we have used the Jacobi identity. For completeness, let us also remark that
\begin{equation*}
\partial_y B_x^\dagger = \left[B_x, B_y\right]^\dagger = -[B_x^\dagger, B_y]\,,
\end{equation*}
where we have used the fact that $B_y^\dagger = B_y$. In summary, we find the following set of commutation relations.
\begin{equation}\label{eq:Nahm_B}
	\partial_y B_x^\dagger = -[B_x^\dagger, B_y]\,,\qquad \partial_y B_y = \frac{1}{2}[B_x^\dagger, B_x]\,,\qquad		
	\partial_y B_x = [B_x, B_y]\,.
\end{equation}
It is important to remark that the equations \eqref{eq:Nahm_B} still involve the reference charge operator through the dagger $\dagger$. Alternatively, one can view $B_x^\dagger$ as an independent field, so that effectively \eqref{eq:Nahm_B} can be viewed as a set of three coupled differential equations for three different fields. For future reference, it will be useful to introduce the following notation
\begin{align}
\mathcal{N}^+(y)&:=\left(h^{-1}\partial_x h\right)^\dagger = B_x^\dagger\,,\\
\mathcal{N}^0(y)&:= -2h^{-1}\partial_y h = -2 B_y\,,\\
\mathcal{N}^-(y)&:=h^{-1}\partial_x h = B_x\,.
\end{align}
In terms of these fields \eqref{eq:Nahm_B} can be written as follows.  
\begin{subbox}{Nahm's equations}
	\begin{equation}\label{eq:Nahm_N}
		\partial_y\mathcal{N}^\pm = \pm \frac{1}{2}[\mathcal{N}^\pm,\mathcal{N}^0]\,,\qquad \partial_y\mathcal{N}^0 = -[\mathcal{N}^+,\mathcal{N}^-]\,.
	\end{equation}
\end{subbox}
\noindent The equations \eqref{eq:Nahm_N} are known as \textbf{Nahm's equations} \cite{Donaldson:1984}. These equations will play an important role in the context of asymptotic Hodge theory, as will be explained in chapters \ref{chap:asymp_Hodge_I} and \ref{chap:asymp_Hodge_II}. Let us already mention an important result due to Hitchin \cite{Hitchin:1983}, namely that the solutions to Nahm's equations naturally give rise to an $\mathfrak{sl}(2,\mathbb{R})$-triple at the poles of $\mathcal{N}^\bullet(y)$. \\

\noindent For later convenience, let us also record the form of the $Q$-constraint in terms of the fields $\mathcal{N}^\bullet(y)$.
\begin{subbox}{Q-constraint}
\begin{equation}\label{eq:Q_constraint_N}
	[Q_{\mathrm{ref}},\mathcal{N}^\pm] = -\frac{i}{2}\mathcal{N}^0\,,\qquad [Q_{\mathrm{ref}},\mathcal{N}^0] = i\left(\mathcal{N}^++\mathcal{N}^- \right)
\end{equation}
\end{subbox}

\subsubsection*{Complex version}
The above discussion has been formulated purely in terms of real operators. For future reference, it will also be convenient to describe the conditions \eqref{eq:Q_constraint_N} and \eqref{eq:Nahm_N} in terms of a set of three complex fields $\{\mathcal{L}_{+1},\mathcal{L}_0,\mathcal{L}_{-1}\}$ which are defined by
\begin{align}
\mathcal{L}_{\pm 1} &:= \frac{1}{2}\left(\mathcal{N}^+ + \mathcal{N}^- \mp i \mathcal{N}^0\right)\,,\\
\mathcal{L}_0 &:= i\left(\mathcal{N}^- - \mathcal{N}^+\right)\,.
\end{align}
In terms of these fields, Nahm's equations \eqref{eq:Nahm_N} take the very similar form
\begin{equation}
\partial_y\mathcal{L}_{\pm 1} = \pm \frac{1}{2} [\mathcal{L}_{\pm 1},\mathcal{L}_0]\,,\qquad \partial_y\mathcal{L}_0 = -[\mathcal{L}_{+1},\mathcal{L}_{-1}]\,,
\end{equation}
while the $Q$-constraint \eqref{eq:Q_constraint_N} can elegantly be written as
\begin{equation}
[Q_{\mathrm{ref}},\mathcal{L}_q] = q\mathcal{L}_q\,,\qquad q=+1,0,-1\,.
\end{equation}
In other words, the fields $\mathcal{L}_q$ are precisely eigenvectors of charge $q$ under the adjoint action of the reference charge operator $Q_{\mathrm{ref}}$. In particular, note that $\mathcal{L}_0$ commutes with $Q_{\mathrm{ref}}$.

\subsubsection*{An action principle for Nahm's equations}
A real one-dimensional action principle associated to Nahm's equations was already described some time ago in \cite{Donaldson:1984}. In \cite{Grimm:2020cda}, see also \cite{Cecotti:2020rjq,Cecotti:2020uek}, this action principle was generalized to a real two-dimensional action principle, which can be interpreted as a non-linear $\sigma$-model from a worldsheet to the Lie group $G_{\mathbb{R}}$. The action reads as follows. 

\begin{subbox}{Action principle for Nahm's equations}
	\begin{equation}\label{eq:action_Nahm}
		S_{\mathrm{Nahm}}[h] = \frac{1}{4}\int_{\mathcal{M}}\mathrm{Tr}\left|h^{-1}\mathrm{d}h+\left(h^{-1}\mathrm{d}h\right)^\dagger \right|^2\,,
	\end{equation}
	where we have employed the notation $|A|^2=A\wedge\star A$. Furthermore, we recall that
	\begin{equation}
		h:\mathcal{M}\rightarrow G_{\mathbb{R}}\,,
	\end{equation}
	describes (a lift of) the period map describing a variation of Hodge structure over a complex one-dimensional base space $\mathcal{M}$, and $\dagger$ denotes the adjoint with respect to an arbitrary reference charge operator, recall equation \eqref{eq:def_adjoint}.
\end{subbox}
\noindent  Let us point out that this action has two sets of symmetries, which we will discuss in turn. 
\begin{itemize}
	\item \textbf{Global $g_{L}$ symmetry:}\\
	Firstly, we see that \eqref{eq:action_Nahm}
	has a global invariance under left-multiplication 
	$h(\sigma) \rightarrow g_L h(\sigma) $ with $g_L \in G_{\bbR}$.
	This global symmetry yields  a conserved current of the form
	\begin{equation}
		\label{eq:current}
		J_L =  \star\, h\big[h^{-1}\text{d}h+\left(h^{-1}\text{d}h\right)^\dagger \big]h^{-1}\ ,\qquad \mathrm{d}J_L = 0\,.
	\end{equation}
	\item \textbf{Local $g_{R}$ symmetry:}\\
	Secondly, one checks that \eqref{eq:action_Nahm} has a local invariance under right-multiplication 
	$h(\sigma) \rightarrow h(\sigma) g_R(\sigma)$, with $g_R^\dagger(\sigma) = g^{-1}_{R}(\sigma)$. 
	The presence of this gauge symmetry shows that the action \eqref{eq:action_Nahm} actually 
	describes fields in a coset $G/K$, where we recall that $K$ is the subgroup of elements $g_R$ that satisfy the unitarity condition
	$g_R^\dagger = g^{-1}_{R}$ with respect to the inner product induced by the choice of reference Hodge structure.
\end{itemize}
Let us now turn to the derivation of the equations of motion induced by the action \eqref{eq:action_Nahm}. Considering the variation under $h\mapsto h+\delta h$, we find
\begin{align*}
	\delta S_{\mathrm{Nahm}}&\stackrel{\text{(a)}}{=}\int_{\mathcal{M}}\mathrm{Tr}\left\{\delta(h^{-1}\mathrm{d}h)\wedge\star\left(h^{-1}\mathrm{d}h+\left(h^{-1}\mathrm{d}h\right)^\dagger \right)\right\}\\
	&\stackrel{\text{(b)}}{=}\int_{\mathcal{M}}\mathrm{Tr}\left\{\left(-[h^{-1}\delta h, h^{-1}\mathrm{d}h]+\mathrm{d}\left(h^{-1}\delta h\right)\right)\wedge\star\left(h^{-1}\mathrm{d}h+\left(h^{-1}\mathrm{d}h\right)^\dagger \right)\right\}\\
	&\stackrel{\text{(c)}}{=}-\int_{\mathcal{M}}\mathrm{Tr}\left\{h^{-1}\delta h\left([h^{-1}\mathrm{d}h\stackrel{\wedge}{,} \star\left(h^{-1}\mathrm{d}h\right)^\dagger]+\mathrm{d}\star\left(h^{-1}\mathrm{d}h+\left(h^{-1}\mathrm{d}h\right)^\dagger\right) \right)\right\}\,,
\end{align*}
where in step (a) we used the fact that $\mathrm{Tr}(AB^\dagger)=\mathrm{Tr}(A^\dagger B)$ which follows from the cyclicity of the trace and the definition of the adjoint \eqref{eq:def_adjoint}, in step (b) we used the relation
\begin{equation}
	\delta\left(h^{-1}\mathrm{d}h\right) = -[h^{-1}\delta h, h^{-1}\mathrm{d}h]+\mathrm{d}\left(h^{-1}\delta h\right)\,,
\end{equation}
and in step (c) we performed an integration by parts and again used the cyclicity of the trace to write $\mathrm{Tr}([A,B]C)=\mathrm{Tr}(A[B,C])$. Note also that $[A\stackrel{\wedge}{,}\star A]=0$ by symmetry. Setting $\delta S_{\mathrm{Nahm}}=0$ we thus obtain the equation of motion
\begin{equation}\label{eq:eom_h}
	\mathrm{d}\star\left(h^{-1}\mathrm{d}h+\left(h^{-1}\mathrm{d}h\right)^\dagger\right)+\left[h^{-1}\mathrm{d}h\stackrel{\wedge}{,} \star\left(h^{-1}\mathrm{d}h\right)^\dagger\right]=0\,,
\end{equation}
which is in fact equivalent to the conservation of the current $J_L$. To recover Nahm's equations, we recall that
\begin{align}
	h^{-1}\mathrm{d}h &= \mathcal{N}^-(y)\,\mathrm{d}x-\frac{1}{2}\mathcal{N}^0(y)\, \mathrm{d}y\,,\\
	\left(h^{-1}\mathrm{d}h\right)^\dagger &= \mathcal{N}^+(y)\,\mathrm{d}x-\frac{1}{2}\mathcal{N}^0(y)\, \mathrm{d}y\,,
\end{align}
where we have imposed the additional condition that the fields $\mathcal{N}^\bullet$ are independent of $x$. Inserting these relations into \eqref{eq:eom_h} we find\footnote{We are using conventions in which $\star\mathrm{d}x = \mathrm{d}y$ and $\star\mathrm{d}y=-\mathrm{d}x$.}
\begin{align*}
	0&=\mathrm{d}\star\left(h^{-1}\mathrm{d}h+\left(h^{-1}\mathrm{d}h\right)^\dagger\right)+\left[h^{-1}\mathrm{d}h\stackrel{\wedge}{,} \star\left(h^{-1}\mathrm{d}h\right)^\dagger\right]\\
	&=\mathrm{d}\left((\mathcal{N}^++\mathcal{N}^-)\,\mathrm{d}y+\mathcal{N}^0\,\mathrm{d}x \right)+\left[\mathcal{N}^-\,\mathrm{d}x-\frac{1}{2}\mathcal{N}^0\, \mathrm{d}y\stackrel{\wedge}{,}\mathcal{N}^+\,\mathrm{d}x-\frac{1}{2}\mathcal{N}^0\, \mathrm{d}y \right]\,,\\
	&=-\left(\partial_y\mathcal{N}^0+[\mathcal{N}^+, \mathcal{N}^-]\right)\mathrm{d}x\wedge\mathrm{d}y\,.
\end{align*}
As desired, we obtain the equation on the right-hand side of \eqref{eq:Nahm_N}. As already mentioned above, the remainder of Nahm's equations immediately follow from the flatness of the Maurer--Cartan form $h^{-1}\mathrm{d}h$. Thus, we conclude that the dynamics encoded in the action principle \eqref{eq:action_Nahm} indeed correspond precisely to Nahm's equations, under the additional assumption that the $x$-dependence is trivial. \\

\noindent There is another characterization of Nahm's equations in terms of an action principle which will be relevant in part \ref{part4} of the thesis. In order to write it down, let us recall the definition of the bulk Weil operator
\begin{equation}\label{eq:def-g_Nahm}
	g := C=h C_{\mathrm{ref}}h^{-1}\,.
\end{equation}
In terms of $g$, one can equivalently write \eqref{eq:action_Nahm} as\footnote{To see this, one simply computes 
\begin{equation*}
	g^{-1}\mathrm{d}g = - h\left(h^{-1}\mathrm{d}h+\left(h^{-1}\mathrm{d}h\right)^\dagger \right)h^{-1}\,.
	\end{equation*}}
\begin{equation}
	S_{\mathrm{Nahm}}[g] = \frac{1}{4}\int_{\mathcal{M}}\mathrm{Tr}\left|g^{-1}\mathrm{d}g\right|^2\,,
\end{equation}
which is nothing but the standard action for the so-called \textit{principal chiral model}. One may verify that the corresponding equations of motion read
\begin{equation}
	\mathrm{d}\star \left(g^{-1}\mathrm{d}g\right)=0\,,
\end{equation}
and are indeed equivalent to \eqref{eq:eom_h}. Rather interestingly, the description of Nahm's equations takes an especially simple form in terms of the bulk Weil operator $g$, in contrast to the period map $h$. Note also that from the point of view of the Weil operator the symmetry properties are especially apparent, since \eqref{eq:def-g_Nahm} is manifestly invariant under local right-multiplication $h\mapsto h g_R$ with $g_R\in K$. These matters will be revisited in part \ref{part4} of the thesis.

%\subsubsection*{The Hodge metric and a geometric point of view}
%Let us make one final observation regarding the on-shell value of the action \eqref{eq:action_Nahm}, which gives some elegant geometric intuition for Nahm's equation. Following the computations in \cite{Grimm:2020cda}, one can show that the on-shell action can be written as
%\begin{equation}
%	S_{\mathrm{Nahm}}[h]\Big|_{\text{on-shell}} = \int_{\mathcal{M}} \mathrm{Tr}\left(h^{-1}\mathrm{d}h\wedge\star \left(h^{-1}\mathrm{d}h\right)^\dagger\right)\,.
%\end{equation}

\end{subappendices}

%%%%%%%%%%%%%%%%%%%%%%%%%%%%%%%%%%%%%%%%%%%%%%%%%%%%%%%%%%%%%%%%%%%%

\chapter{Asymptotic Hodge Theory I: Bulk to Boundary}
\label{chap:asymp_Hodge_I}

This chapter is the first of two chapters on asymptotic Hodge theory. From a practical point of view, asymptotic Hodge theory provides a set of tools to find approximate expressions for a general variation of Hodge structure which are valid close to certain asymptotic regions in the moduli space. In section \ref{sec:asymp_Hodge_motivation} we give some general motivation for why this is an interesting regime to study. Then, in section \ref{sec:nilp_orbit_theorem}, we introduce the first big theorem of asymptotic Hodge theory: the\textbf{ nilpotent orbit theorem}. This theorem gives a first characterization of the universal behaviour of a variation of Hodge structure near the boundary of the moduli space. Subsequently, this leads to the first approximation of a variation of Hodge structure: the nilpotent orbit approximation. In order to obtain finer information about how the underlying Hodge structure degenerates, we introduce the notion of mixed Hodge structures in section \ref{sec:MHS}. The relation between nilpotent orbits and mixed Hodge structures lies at the heart of the second big theorem of asymptotic Hodge theory: the \textbf{$\mathrm{Sl}(2)$-orbit theorem}. In section \ref{sec:SL2_orbit_one_variable} we explain this relation first in the case of a one-parameter variation of Hodge structure, and defer the general discussion to section \ref{sec:SL2_orbit_multi_variable}. Subsequently, this will give rise to the second approximation of a variation of Hodge structure: the $\mathrm{Sl}(2)$-orbit approximation. We will discuss how this approximation can be used to obtain general asymptotic expressions for various physical couplings such as the K\"ahler potential and the Hodge norm. Finally, appendix \ref{app:nilp_orbit_proof} contains some additional discussion on the proof of the nilpotent orbit theorem, and appendix \ref{sec:asymp_Hodge_examples} contains a number examples which will be referred to in the next chapter. \\

\noindent Regarding the literature, most of the results discussed in this chapter can be found in the seminal works of Schmid \cite{schmid} and Cattani, Kaplan, and Schmid \cite{CKS}. Furthermore, we refer the reader to \cite{Li:2022mhy,vandeHeisteeg:2022gsp} for complementary perspectives as well as additional applications and worked out examples.  

\section{Motivation}\label{sec:asymp_Hodge_motivation}
Recall that our interest in Hodge theory is motivated by the fact that physical couplings appearing in four-dimensional effective theories coming from string theory typically depend on Hodge-theoretic objects. For example, the K\"ahler potential, gauge-kinetic couplings matrices, and the scalar potential depend on the complex structure moduli through the Hodge star operator on the relevant Calabi--Yau manifold. \\

\noindent From a physical point of view, the purpose of \textit{asymptotic} Hodge theory is to study the moduli-dependence of these various couplings in certain asymptotic regimes of the complex structure moduli space, in which we are close to loci where the underlying geometry degenerates. From a mathematical point of view, this corresponds to the study of degenerations of Hodge structure. Indeed, typically the limits we are interested in are of such a nature that the underlying geometry of the Calabi--Yau manifold becomes singular, such that its middle cohomology no longer admits a Hodge structure. Instead, it admits a so-called \textit{mixed Hodge structure}, whose details in turn encode precisely the asymptotic behaviour of the physical couplings. \\

\noindent Below we give three motivations for why these asymptotic limits are of interest to us from a physical point of view.

\subsubsection*{Motivation (1): Computability vs.~generality}
When given an explicit Calabi--Yau geometry, there are plenty of analytic techniques available in order to compute the periods to a high degree of accuracy, and thus obtain detailed expressions for the physical couplings one is interested in. However, given that already the number of different Calabi--Yau threefolds is very large, it is unfeasible to expect to be able to extract general results from an example-by-example analysis of this kind. Instead, one could ask whether there are certain features of periods/variations of Hodge structure that are \textit{universal} and can thus be used to make general statements about the physical couplings. It turns out that this is exactly what happens in the asymptotic regime of the moduli space. This is captured by the two foundational theorems of asymptotic Hodge theory: the \textbf{nilpotent orbit theorem} and the \textbf{$\mathrm{Sl}(2)$-orbit theorem}. Roughly speaking, the former states that any variation of Hodge structure takes a universal form as one approaches the boundary of the moduli space, while the latter gives a more detailed characterization of this asymptotic behaviour in terms of so-called limiting mixed Hodge structures. The latter effectively carry all the information about how the Hodge structure degenerates in a given limit. It will be the purpose of this chapter to explain these statements in detail and illustrate how these powerful theorems can be used to make very general statements about the asymptotic form of physical couplings such as the K\"ahler potential and the Hodge norm. 

\subsubsection*{Motivation (2): The Swampland program}
One area of research in which the above mentality has featured prominently over the last years is the so-called \textbf{Swampland program} \cite{Vafa2005,Ooguri2007}, see \cite{Brennan:2017rbf,Palti:2019pca,vanBeest:2021lhn,Grana:2021zvf,Agmon:2022thq} for reviews. In its most ambitious form, the goal of this program is to identify properties of low-energy effective theories coupled to gravity that must be satisfied in order for the theory to admit a UV completion. A typical strategy to identify candidate properties is to investigate a particular corner of the string landscape, for example the set of four-dimensional $\mathcal{N}=2$ supergravity theories coming from type IIB compactifications on Calabi--Yau threefolds, and search for patterns within this class of theories. If sufficient evidence is found that a particular pattern is present in multiple different corners of the string landscape, this is formulated into a \textbf{Swampland conjecture}.\footnote{Another strategy is take a more bottom-up approach using ingredients that are certainly part of any theory of quantum gravity, such as black holes, in order to provide evidence which is independent of string theory.} Over the years, this has lead to a large collection of interrelated conjectures which are believed to capture deep properties of low-energy descriptions of quantum gravity.\\

\noindent For our purposes, an important feature of many of the Swampland conjectures is that they deal with moduli spaces of effective theories and, notably, their asymptotics. An important example of this is the \textbf{Swampland distance conjecture} \cite{Ooguri2007}, which dictates the behaviour of the low-energy effective theory upon traversing a large distance inside the moduli space. More precisely, it states that (1) any moduli space of a low-energy effective theory coupled to gravity admits an infinite distance limit, and (2) whenever one approaches such an infinite distance limit an infinite tower of states in the original UV theory becomes light. The latter effectively signals a breakdown of the effective theory. Importantly, in the context of type IIB compactifications, these infinite distance points are examples of loci in the complex structure moduli space where the Hodge structure degenerates. Thus, in order to investigate the Swampland distance conjecture in this corner of the string landscape, it is necessary to have detailed knowledge about the asymptotic behaviour of the metric on the moduli space, as well as the degeneration of the underlying geometry (as this dictates the nature of light tower of states). Asymptotic Hodge theory provides a valuable tool to achieve this, as was pioneered in the works \cite{Grimm:2018ohb,Grimm:2018cpv}. Since then, asymptotic Hodge theory has been applied to study numerous other Swampland conjectures as well, see for example \cite{Corvilain:2018lgw,Grimm:2019ixq,Grimm:2019bey,Grimm:2019wtx,Calderon-Infante:2020dhm,Bastian:2020egp,Gendler:2020dfp,Lanza:2020qmt,Palti:2021ubp,Castellano:2021yye,Grimm:2021vpn,Calderon-Infante:2022nxb,Grimm:2022xmj}. For more complete lists of references to works on the various Swampland conjectures we refer the reader to the aforementioned reviews. 

\subsubsection*{Motivation (3): Going beyond the large complex structure lamppost}
There is one particular limit in the complex structure moduli space which has received an enormous amount of attention: the large complex structure point. For simplicity, let us for the moment specify to the setting of Calabi--Yau threefolds. Then this region can be identified with the large volume regime in the K\"ahler moduli space of the mirror Calabi--Yau threefold. In this regime the period vector take the well-known form
\begin{equation}\label{eq:periods_LCS}
\mathbf{\Pi} = \left(1, t^i, \frac{1}{6}\mathcal{K}_{ijk}t^i t^j t^k+ \frac{i\chi \zeta(3)}{8\pi^3},-\frac{1}{2}\mathcal{K}_{ijk}t^j t^k \right)\,,
\end{equation}
plus exponentially suppressed terms. Here $\mathcal{K}_{ijk}$ and $\chi$ respectively denote the triple-intersection numbers and the Euler characteristic of the mirror Calabi--Yau threefold. Importantly, in order to write down an approximate expression for the period vector one only needs to know the topological data of the mirror manifold, without having to go through the Picard--Fuchs equations. At the same time, computing the exponential corrections is also of great interest. This is because these corrections correspond to instanton corrections of type IIA string theory coming from genus zero worldsheets that wrap certain two-cycles in the mirror Calabi--Yau threefold. In other words, through mirror symmetry one can compute quantum corrections on the type IIA side by computing classical periods on the type IIB side. As a result of these considerations, much of the physics literature has been focused on this particular regime, leading to the ``large complex structure lamppost''. \\

\noindent It must be stressed, however, that there are many other kinds of asymptotic limits in the complex structure moduli space, in which the behaviour of the physical couplings is different than in the large complex structure regime. It is thus vital to build lampposts around other asymptotic boundary as well. Fortunately, asymptotic Hodge theory exactly provides us with the necessary ingredients to do so.

\section{The nilpotent orbit theorem}\label{sec:nilp_orbit_theorem}
Before delving into a detailed description of the abstract features of asymptotic Hodge theory, let us first illustrate what is meant by the ``asymptotic'' behaviour of periods, by returning to the example of the mirror bicubic. This will serve as a motivation for one of the big theorems of asymptotic Hodge theory, the nilpotent orbit theorem, and at the same time will be used to exemplify some of the constructions used in asymptotic Hodge theory in appendix \ref{sec:asymp_Hodge_examples}. 
\subsection{Example: The mirror bicubic revisited}\label{subsec:asymp_periods_mirror_bicubic}
Recall that the periods of the mirror bicubic are governed by a Picard--Fuchs equation of hypergeometric form:
\begin{equation}\label{eq:PF_hypergeometric}
\left[\theta^4-\mu z\left(\theta+a_1\right)\left(\theta+a_2\right)\left(\theta+a_3\right)\left(\theta+a_4\right)\right]\varpi=0\,,
\end{equation}
for a certain choice of $a_1,a_2,a_3,a_4$, which we leave arbitrary for now. The differential equation \eqref{eq:PF_hypergeometric} has three regular singularities at $z=0,1/\mu,\infty$. Our goal is to analyse the local solutions around each of these singularities. For differential equations of Fuchsian type, such as \eqref{eq:PF_hypergeometric}, there is a standard procedure to obtain a power series solution around a given singularity, known as the \textit{Frobenius method}. It proceeds as follows. First, around a given singularity, say $z=0$, one makes the Frobenius series ansatz
\begin{equation}
\varpi= z^\alpha\sum_{n=0}^\infty c_n z^n\,,
\end{equation}
for some $\alpha\in\mathbb{C}$ which is to be determined. Plugging the ansatz into the differential equation and considering the leading order term in $z$ yields an equation for $\alpha$ known as the \textit{indicial equation}. Its roots are referred to as \textit{indices} or \textit{local exponents}, and will be denoted by $\alpha_i$. For roots with multiplicity one it simply remains to determine the power series coefficients $c_n$ by recursion. If, however, a root has multiplicity greater than one, the additional linearly independent solutions are obtained by appending a logarithmic term:
\begin{equation}
\omega_i= \omega_j\frac{\log z}{2\pi i}+z^{\alpha_i}\sum_{n=0}^\infty c_{i,n} z^n\,,
\end{equation}
for $\alpha_i=\alpha_j$. For the fourth-order equation \eqref{eq:PF_hypergeometric}, this procedure will yield in total four linearly independent solutions. Together, they can be collected into a vector
\begin{equation}\label{eq:periods_Frobenius}
\varpi = \begin{pmatrix}
\varpi_0 \\ \varpi_1 \\ \varpi_2 \\ \varpi_3
\end{pmatrix}\,,
\end{equation}
which gives the periods of the holomorphic 3-form $\Omega$ in some possibly complex basis of $H^3(X_{3,3},\mathbb{C})$. Typically, the resulting period vector is not single-valued and should be analytically continued along closed paths that encircle the singularities. This is described via the representation of the fundamental group, i.e.~the monodromy group. In the following, we will exemplify these matters for each of the three singularities $z=0,1/\mu,\infty$ of \eqref{eq:PF_hypergeometric}.

\subsubsection*{Solutions around $z=0$}
Around $z=0$ the indicial equation reads
\begin{equation}
\alpha^4=0\,,
\end{equation}
which has a single root $\alpha=0$ of order four. Hence the  Frobenius method tells us that the fundamental system is given by
\begin{align}
\varpi_0 &= f_0\,,\\
\varpi_1 &=f_0\frac{\log z}{2\pi i}+f_1\,,\\
\varpi_2 &=\frac{1}{2}f_0\left(\frac{\log z}{2\pi i}\right)^2+f_1\frac{\log z}{2\pi i}+f_2\,,\\
\varpi_3 &=-\frac{1}{6}f_0\left(\frac{\log z}{2\pi i}\right)^3-\frac{1}{2}f_1\left(\frac{\log z}{2\pi i}\right)^2-f_2\frac{\log z}{2\pi i}+f_3\,,
\end{align}
for some some holomorphic power series $f_0,f_1,f_2,f_3$, where we have chosen a particularly convenient normalization to facilitate comparison with later results. Assembling these solutions into a vector as in \eqref{eq:periods_Frobenius}, the resulting period vector undergoes a monodromy
\begin{equation}
\varpi(ze^{2\pi i}) = T\varpi(z)\,,\qquad T=e^{N}\,,\qquad N = \begin{pmatrix}
0 & 0 & 0& 0\\
1 & 0 & 0& 0\\
0 & 1 & 0 & 0\\
0 & 0 & -1 & 0
\end{pmatrix}\,.
\end{equation}
In particular, one finds that $T$ is maximally unipotent, hence the point $z=0$ corresponds to a point of \textit{maximal unipotent monodromy}, and is therefore referred to as a MUM point. Furthermore, the period vector can be nicely written as
\begin{equation}
\varpi = \mathrm{exp}\left[\frac{\log z}{2\pi i}N\right]\cdot \begin{pmatrix}
f_0 \\ f_1 \\ f_2 \\ f_3
\end{pmatrix}\,,
\end{equation}
such that all of the singular behaviour of the period vector around $z=0$ is captured by an exponential factor acting on a holomorphic vector. 

\subsubsection*{Solutions around $z=1/\mu$}
Changing variables such that the point $z=1/\mu$ is centered around zero and accordingly transforming the Picard--Fuchs operator, one finds that the indicial equation around this singularity becomes
\begin{equation}
\alpha(\alpha-1)(\alpha-2)(\alpha-3+a_1+a_2+a_3+a_4)=0\,.
\end{equation}
Therefore, the roots are $\alpha=0,1,1,2$, so the Frobenius method tells us that the fundamental system is given by
\begin{align}
\varpi_0 &= z f_0\,,\\
\varpi_1 &=f_1\,,\\
\varpi_2 &=z^2f_2\,,\\
\varpi_3 &=-z f_0 \frac{\log z}{2\pi i}+z f_3\,,
\end{align}
for some holomorphic power series $f_0,f_1,f_2,f_3$. Here we have chosen a particular ordering of the fundamental system, as well as some conventional minus signs, in order to facilitate comparison with later results. Assembling these solutions into a vector as in \eqref{eq:periods_Frobenius}, the resulting period vector undergoes a monodromy
\begin{equation}
\varpi(z e^{2\pi i}) = T\varpi(z)\,,\qquad T=e^{N}\,,\qquad N = \begin{pmatrix}
0 & 0 & 0& 0\\
0 & 0 & 0& 0\\
0 & 0 & 0 & 0\\
-1 & 0 & 0 & 0
\end{pmatrix}\,.
\end{equation}
The monodromy is again unipotent. As in the previous case, the period vector can be nicely written as
\begin{equation}
\varpi = \mathrm{exp}\left[\frac{\log z}{2\pi i}N\right]\cdot \begin{pmatrix}
z f_0 \\ f_1 \\ z^2 f_2 \\ z f_3
\end{pmatrix}\,,
\end{equation}
such that again all of the singular behaviour of the period vector is captured by an exponential factor acting on a holomorphic vector. 

\subsubsection*{Solutions around $z=\infty$}
Let us now turn to the last singular point, which has an additional feature which did not appear in the previous two cases. Changing variables such that the point $z\rightarrow\infty$ is centered around zero and accordingly transforming the Picard--Fuchs operator, one finds that the indicial equation around this singularity becomes
\begin{equation}
(\alpha-a_1)(\alpha-a_2)(\alpha-a_3)(\alpha-a_4)=0\,.
\end{equation}
Therefore, the roots are $\alpha=a_1,a_2,a_3,a_4$. Recall that for the case of the mirror bicubic the indices are fractional and given by \eqref{eq:indices_mirror_bicubic}. In particular, there are two pairs of equal roots, so the Frobenius method tells us that the fundamental system is given by
\begin{align}
\varpi_0 &=z^{1/3}f_0\,,\\
\varpi_1 &=z^{2/3}f_1\,,\\
\varpi_2 &=z^{1/3}f_0\frac{\log z}{2\pi i}+z^{1/3}f_2\,,\\
\varpi_3 &=z^{2/3}f_1\frac{\log z}{2\pi i}+z^{2/3}f_3\,,
\end{align}
where we have again chosen a particularly convenient ordering. Assembling these solutions into a vector as in \eqref{eq:periods_Frobenius}, the resulting period vector undergoes a monodromy
\begin{equation}
\varpi(z e^{2\pi i}) = T\varpi(z)\,,
\end{equation}
which is \textit{not} unipotent, but rather decomposes as $T= T_{ss}\cdot T_{u}$ into a semi-simple part of finite order
\begin{equation}
T_{ss} = \begin{pmatrix}
e^{\frac{2\pi i}{3}} & 0 & 0& 0\\
0 & e^{\frac{4\pi i}{3}} & 0& 0\\
0 & 0 & e^{\frac{2\pi i}{3}} & 0\\
0 & 0 & 0 & e^{\frac{4\pi i}{3}}
\end{pmatrix}\,,\qquad T_{ss}^3 = \mathbb{I}\,,
\end{equation}
and a unipotent part
\begin{equation}
T_{u}=e^{N}\,,\qquad N = \begin{pmatrix}
0 & 0 & 0& 0\\
0 & 0 & 0& 0\\
1 & 0 & 0 & 0\\
0 & 1 & 0 & 0
\end{pmatrix}\,.
\end{equation}
It is said that the monodromy is \textit{quasi-unipotent}. One can choose to simplify the expressions by performing the coordinate redefinition $z\mapsto z^3$, which removes the semi-simple part of the monodromy and transforms the unipotent part by $N\mapsto 3N$. After doing so, the period vector can again be written as
\begin{equation}
\varpi= \mathrm{exp}\left[\frac{3\log z}{2\pi i}N\right]\cdot \begin{pmatrix}
zf_0 \\ z^2 f_1 \\ z f_2 \\ z^2 f_3
\end{pmatrix}\,,
\end{equation}
Once again, the structure of the periods around the singularity $z\rightarrow\infty$ is very similar to the other singularities, which hints towards the fact that this may be a general feature. The fact that this is indeed the case is the result of the celebrated nilpotent orbit theorem, which discuss in detail in a moment.

\subsubsection*{Some closing remarks}
The strategy that is outlined above is applicable to study the periods of a wide range of Calabi--Yau manifolds. In this regard, a convenient notation that is often used to summarize the singularity structure of a given Fuchsian equation is the so-called \textit{Riemann symbol}, which takes the form
\begin{equation}
\mathcal{P}\left\{\begin{tabular}{ccc}
0 & $1/\mu$ & $\infty$ \\ \hline
0 & 0 & $a_1$\\
0 & 1 & $a_2$\\
0 & 1 & $a_3$\\
0 & 2 & $a_4$
\end{tabular}\right\}\,,
\end{equation}
for the example of the mirror bicubic, as well as for the other 13 hypergeometric models. Around each singularity, it collects the roots of the corresponding indicial equation and hence the local exponents for the periods as well as the logarithmic structure. Naturally, for more complicated Picard--Fuchs systems there may be more than three singularities, leading to further columns in the Riemann symbol. Additionally, considering Calabi--Yau threefolds with more than one complex structure modulus, or moving to higher-dimensional Calabi--Yau manifolds generically increases the order of the Picard--Fuchs differential operator, and hence introduces additional rows in the Riemann symbol. 
\\

\noindent For any one-dimensional limit in the complex structure moduli space of a Calabi--Yau threefold, the type of singularity can be classified in terms of the local exponents as follows. There are four types of singularities:
\begin{itemize}
\item MUM point: $a_1=a_2=a_3=a_4$.
\item K-point: $a_1=a_2$ and $a_3=a_4$.
\item Conifold point: $a_2=a_3$, but $a_1\neq a_4$.
\item F-point: all exponents are rational and $a_1\neq a_2\neq a_3\neq a_4$.
\end{itemize} 
It is for this reason that we chose to focus on the case of the mirror bicubic $X_{3,3}$, as this particular example exhibits the (for our purposes) three most interesting singularities. The F-point, which does not arise for the mirror bicubic, is a point which only has a finite order monodromy, which can always be removed by a coordinate redefinition.   
\\

\noindent A priori, it is not at all obvious that these four singularity types exhaust all possibilities. For example, it turns out that the case $a_1=a_2=a_3\neq a_4$ cannot occur. This becomes even more non-trivial when the periods depend on multiple complex structure moduli, such that the types of singularites will additionally depend on the order of limits that is taken. The fact that these are indeed the only possibilities will become apparent once we have introduced the notion of mixed Hodge structures. However, before we can do so we must first discuss one of the central results in asymptotic Hodge theory. 

\subsection{The nilpotent orbit theorem}\label{subsec:nilpotent-orbit}
One of the key lessons from the previous section is that the local behaviour of the period vector around a singularity is characterized in terms of its monodromy, through a factor $e^{\frac{\log z}{2\pi i}N}$ acting on some holomorphic vector. The purpose of this subsection is to state a result, known as the \textbf{nilpotent orbit theorem}, which effectively implies that this behaviour holds generally for any variation of polarized Hodge structure.
\\

\noindent Let us consider an abstract variation of polarized Hodge structure over some parameter space $\mathcal{M}$, which we will assume to be quasi-projective, having singular divisors which are at worst normal crossings. This is indeed the case for the complex structure moduli space of Calabi--Yau manifolds, after possibly performing a resolution of the singularities \cite{Viehweg,Hironaka}. The fact that $\mathcal{M}$ is quasi-projective means that its closure $\overline{\mathcal{M}}$ can be embedded in a complex projective space. The central goal of this section is to understand the behaviour of a variation of Hodge structure near the singular region $\overline{\mathcal{M}}/\mathcal{M}$, which will also be referred to as the ``asymptotic regime''. Since we are interested in a local description of $\mathcal{M}$ in this asymptotic regime, we may assume that $\mathcal{M}$ is given by the direct product of $r$ punctured disks $\Delta^*$ and $m-r$ disks $\Delta$, where $m$ denotes the complex dimension of $\mathcal{M}$ and $r$ denotes the number of coordinates that approach the boundary. We choose local coordinates $z^i$ on the punctured disks such that the punctures are located at $z^i=0$ for $i=1\ldots, r$, corresponding to the locations of singular divisors in the moduli space. Furthermore, we introduce local coordinates $\zeta^i$ for $i=r+1,\ldots, m$ on the remaining filled disks. In other words, we consider the period map
\begin{equation}\label{eq:period_map_local}
\Phi:\quad (\Delta^*)^{r}\times \Delta^{m-r}\rightarrow \Gamma\backslash \mathcal{D}\,,
\end{equation}
where we recall that $\mathcal{D}\cong G_{\mathbb{R}}/V$ denotes the period domain, which parametrizes the space of all polarized Hodge structures. We denote by
\begin{equation}
t^i = x^i+i y^i=\frac{1}{2\pi i}\log z^i\,,
\end{equation}
the coordinates on the universal covering space of $(\Delta^*)^r$. The $t^i$ coordinates each take value in the complex upper half-plane $\mathbb{H}$, and the singularities are located at $\mathrm{Im}\,t^i\rightarrow\infty$. The reals and imaginary parts of $t^i$ are referred to as axions and saxions, respectively. Note that, due to the periodic nature of the axionic coordinates, a choice of fundamental domain for the $x^ i$ is the bounded interval $[0,1]$. The two descriptions of the near-boundary regime of $\mathcal{M}$ and its universal cover $\widetilde{\mathcal{M}}$ are illustrated in figure \ref{fig:disc} in the case of a single complex structure modulus. Correspondingly, the period map \eqref{eq:period_map_local} lifts to a map
\begin{equation}
\widetilde{\Phi}:\quad \mathbb{H}^r\times\Delta^{m-r}\rightarrow \mathcal{D}\,,
\end{equation}
whose transformation under the monodromy group will be described in a moment. In the remainder of our discussion, we will assume without loss of generality that $m=r$, meaning that all moduli are taken towards the boundary. 
\begin{figure}[t!]
\begin{center}
\includegraphics[width=0.8\textwidth]{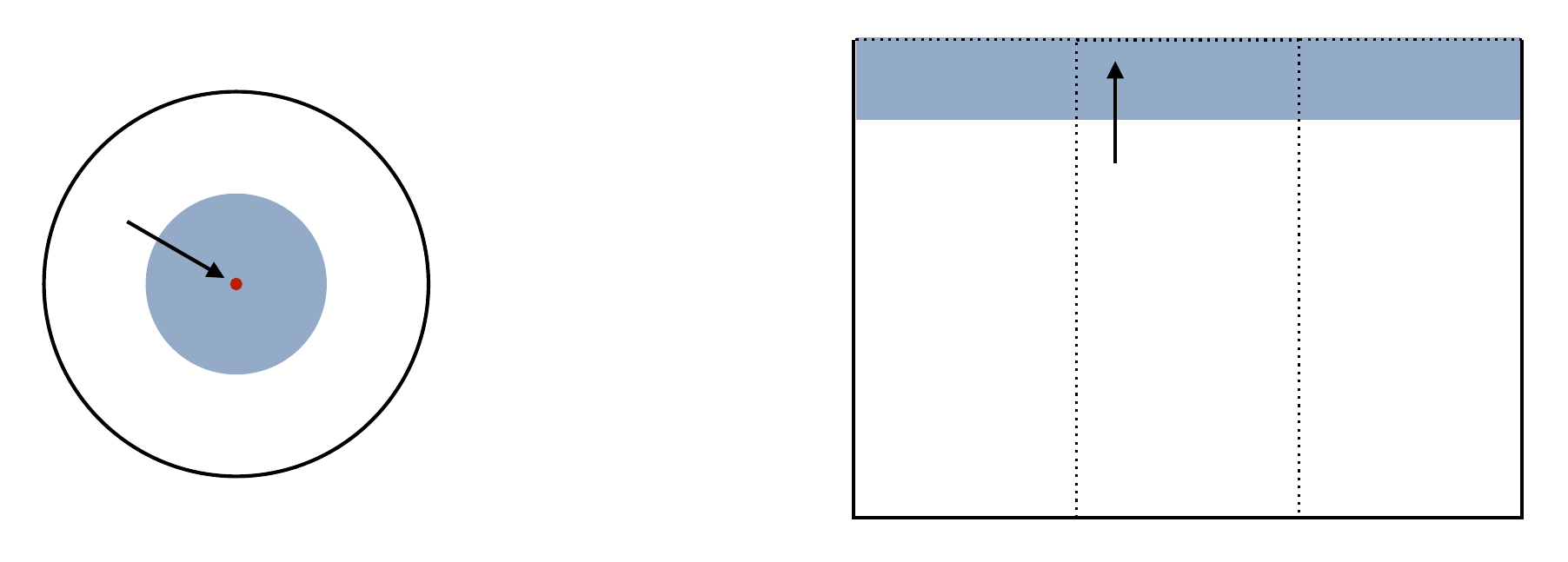}
\vspace*{-1cm}
\end{center}
\begin{picture}(0,0)
\put(70,73){\rotatebox{-31}{\small $z \rightarrow 0$}}
\put(280,80){\small $y \rightarrow \infty$}
\put(10,90){(a)}
\put(170,90){(b)}
\end{picture}

\caption{Two local descriptions of the near-boundary regime of the moduli space $\mathcal{M}$. Figure (a): Poincar\'e disc with the singularity located at $z=0$. Figure (b): upper half-plane, with the singularity located at $y\rightarrow\infty$.\label{fig:disc}}

\end{figure}
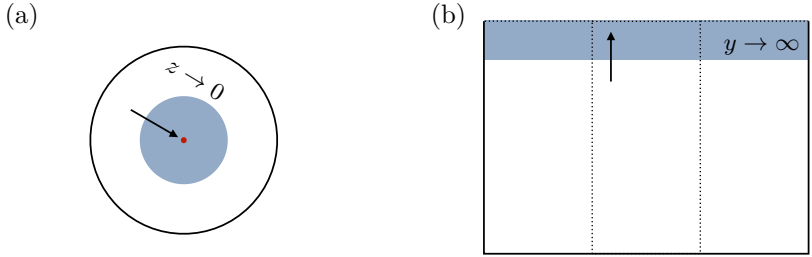

\subsubsection*{Monodromy}
Of vital importance is the local monodromy behaviour of the variation of Hodge structure when encircling the singularity. As discussed in section \ref{subsec:classifying_space}, the parallel transport of a Hodge structure along a path which encircles a given singularity only depends on the homotopy class of the path. Therefore, it suffices to consider the transformations $z^i\mapsto z^i e^{2\pi i }$ or equivalently $t^i\mapsto t^i+1$ and to ask how the Hodge filtration transforms under this map. This yields $m$ monodromy operators $T_i\in \Gamma$, which determine the transformation of the period map as
\begin{equation}
\label{eq:Hodge_monodromy}
\widetilde{\Phi}(t_i+1) = T_i\circ \widetilde{\Phi}(t_i)\,.
\end{equation}
To be more concrete, by the action of $T_i$ on a given filtration we simply mean the action of $T_i$, as a matrix, on the vectors that span that filtration. One of the central results, originally due to Borel, is the following (see e.g.~Lemma 4.5 of \cite{schmid}).
\begin{subbox}{Quasi-unipotent monodromy}
The monodromy operators $T_i$ are \textbf{quasi-unipotent}.
\tcblower This means that for each $T_i$ there exist non-negative integers $k_i,l_i$ such that
\begin{equation}
\left(T_i^{k_i}-\mathbb{I}\right)^{l_i+1}=0\,.
\end{equation}
Equivalently, this means that the eigenvalues of the monodromy operators are roots of unity. 
\end{subbox}\newpage
\noindent This is exactly what we observed when we studied the monodromy transformations of the periods of the mirror bicubic in section \ref{subsec:asymp_periods_mirror_bicubic}. Practically, it means that the local exponents of the periods can be at worst rational. Furthermore, as was exemplified by the K-point of the mirror bicubic, one can always perform the coordinate redefinition $z_i\mapsto z_i^{k_i}$ to remove the finite order semi-simple piece of each monodromy operator $T_i$. For the remainder of this work, we will always assume that this has been done.\footnote{It should be noted, however, that the semi-simple part of the monodromy actually contains a great deal of information regarding the integral structure of the variation of Hodge structure in question. In particular, by removing this piece via a coordinate redefinition, some of this information may be lost. We refer the reader to \cite{Bastian:2023shf} for a more in-depth discussion on these matters.} Therefore, each monodromy operator will be taken to be unipotent and hence of the form 
\begin{equation}
\label{eq:log_monodromy}
T_i = e^{N_i}\,,\qquad [N_i,N_j]=0\,,
\end{equation}
where $N_i\in\mathfrak{g}_{\mathbb{R}}$ are commuting nilpotent operators, whose nilpotency degree lies between $0$ and the weight $D$ of the Hodge structure. They will be referred to as the \textbf{log-monodromy matrices}. The fact that they commute originates from the fact the singular divisors are at worst normal crossings. 

\subsubsection*{The nilpotent orbit theorem}
From the preceding discussions of (asymptotic) Hodge theory, we would like to highlight two important features of the Hodge filtration $F^p$. Namely, (1) it is holomorphic, recall equation \eqref{eq:holomorphicity}, and (2) it undergoes a monodromy transformation when encircling a singularity in the moduli space, recall equation \eqref{eq:Hodge_monodromy}. Intuitively, the simplest types of Hodge filtrations that exhibit these features are of the form
\begin{equation}
\label{eq:nilpotent_orbit}
F_{\mathrm{nil}}^p = \mathrm{exp}\left[\sum_{i=1}^m t^i N_i \right]F_{(m)}^p\,,
\end{equation}
where $F_{(m)}^p$ is some moduli-independent filtration. The transversality condition \eqref{eq:transversality} then simply translates into the statement that all the log-monodromy matrices satisfy the condition
\begin{equation}\label{eq:log-monodromy_transversality}
N_i F^p_{(m)}\subset F_{(m)}^{p-1}\,,
\end{equation}
Hodge filtrations of the form \eqref{eq:nilpotent_orbit}, with the log-monodromy matrices satisfying \eqref{eq:log-monodromy_transversality}, are referred to as \textbf{nilpotent orbits}, since they correspond to the orbit of some fixed filtration under the action of the nilpotent operators $N_i$. 
\\

\noindent This brings us to one of the two central theorems in asymptotic Hodge theory: the nilpotent orbit theorem. It is originally due to Schmid \cite{schmid} and was later refined by Cattani, Kaplan, and Schmid in \cite{CKS}. Below we state a somewhat informal version of the theorem.
\begin{subbox}{Nilpotent orbit theorem}
For $\text{Im}\,t^i$ sufficiently large, for $i=1\ldots, r$, any polarized variation of Hodge structure $F^p(t)$ can be approximated as  
\begin{equation}\label{eq:nilpotent_orbit_theorem}
F^p(t,\zeta) = \mathrm{exp}\left[\sum_{i=1}^r t^i N_i \right]\left( F^p_{(r)}(\zeta)+\mathcal{O}\left(e^{2\pi i t^i}\right)\right)\,,
\end{equation}
for some limiting filtration $F^p_{(r)}$ which is independent of $t^i$. The leading term in \eqref{eq:nilpotent_orbit_theorem} is referred to as the \textbf{nilpotent orbit approximation} of the full Hodge structure. Furthermore, the log-monodromy matrices satisfy
\begin{equation}
N_i F^p_{(r)}\subset F^{p-1}_{(r)}\,.
\end{equation}
\tcblower
Here we recall that $\zeta^i$ denote the ``spectator moduli'', i.e.~the moduli which are not sent to the boundary of the moduli space. Importantly, the filtration $F^p_{(m)}$ can still depend on these moduli. For our purposes, however, we are typically interested in the limit where all the moduli are sent to the boundary, hence we will not explicitly take into account the dependence on the spectator moduli and effectively set $r=m$. 
\end{subbox}
\noindent In other words, the nilpotent orbit theorem states that any polarized Hodge filtration asymptotes to a nilpotent orbit as one approaches a singularity in the moduli space. Furthermore, the sub-leading corrections are exponentially suppressed in the covering space coordinate $t^i$. In appendix \ref{app:nilp_orbit_proof} we have provided some additional background regarding the proof of the nilpotent orbit theorem.

\subsubsection*{The limiting filtration $F_{(m)}^p$}
The filtration $F_{(m)}^p$ is commonly referred to as the \textit{limiting filtration}. In general, it can be obtained from the full filtration $F^p$ by the following limit procedure
\begin{equation}
F^p_{(m)} = \lim_{t^i\rightarrow i\infty} e^{-t^i N_i}F^p\,.
\end{equation}
It is crucial to stress that, generically, the limiting filtration $F^p_{(m)}$ does \textit{not} define a polarized Hodge filtration. In other words, the degeneration is too severe for the limit to still admit a Hodge structure. More geometrically, this means that the value of the period map $\Phi(z)$ at the singular locus lies outside of the period domain $\mathcal{D}$. Instead, the limiting filtration is said to define a \textbf{mixed Hodge structure}. The purpose of the next sections is to explain what this means, and how this indicates a deeper structure through the emergence of multiple $\mathfrak{sl}(2,\mathbb{R})$ symmetries as one approaches the boundary of the moduli space.

\section{Mixed Hodge structures}\label{sec:MHS}
The nilpotent orbit theorem suggests that the degeneration of the variation of Hodge structure is captured by two ingredients: the log-monodromy matrices $N_i$ and the limiting filtration $F_{(m)}^p$. A crucial result is that these two ingredients together define a so-called mixed Hodge structure. In this section, we will give a basic introduction to mixed Hodge structures without making any reference to an underlying variation of Hodge structure. In the next section we will then explain how exactly the log-monodromy matrices and the limiting filtration make up a mixed Hodge structure.\\

\noindent In order to explain the notion of a mixed Hodge structure, we must first discuss the concept of a weight filtration.

\begin{subbox}{Definition: weight filtration}
For any nilpotent endomorphism $N$ of $H_{\mathbb{R}}$, the \textbf{weight filtration} $W_{\ell}$ of weight $D$ associated to $N$ is the unique increasing filtration of vector spaces
\begin{equation}
W_{-1}:=0\subset W_0\subset \cdots \subset W_{2D-1}\subset W_{2D}=H_{\mathbb{C}}\,,
\end{equation}
such that the following two conditions are satisfied:
\begin{enumerate}
\item $N W_{\ell}\subset W_{\ell-2}$ for all $\ell$. 
\item $N^j:\mathrm{Gr}_{D+j}\rightarrow \mathrm{Gr}_{D-j}$ is a linear isomorphism, where
\begin{equation}
	\mathrm{Gr}_\ell:= W_{\ell}/W_{\ell-1}\,,
\end{equation}
are the so-called \textit{graded spaces}.
\end{enumerate} 
\end{subbox}
\noindent Note that this definition makes sense for any nilpotent endomorphism $N$, it does not necessarily have to be associated to some monodromy transformation. The first condition simply states that the action of $N$ moves one downwards in the weight filtration. The second condition can be interpreted as providing an interpolation between the ``upper half'' and the ``lower half'' of the weight filtration, as is illustrated below.\\

\noindent Practically, when given an explicit nilpotent endomorphism $N$, one may use the following formula to compute the various constituents of the weight filtration
\begin{equation}\label{eq:def-weight_filtration}
W_{\ell+D} = \sum_{j\geq \text{max}(-1,\ell)}\text{ker}\,N^{j+1}\cap \text{im}\,N^{j-\ell}\,. 
\end{equation}
To emphasize the dependence on $N$, we will sometimes also write $W_{\ell}(N)$ for the weight filtration associated to $N$. 

\subsubsection*{Mixed Hodge structures}
Recall that a (pure) Hodge structure is described in terms of a single decreasing filtration $F^p$ satisfying a number of conditions. Instead, a mixed Hodge structure will be described in terms of a suitable combination of the weight filtration and another filtration which together give rise to a pure Hodge structure. The precise definition is as follows.

\begin{subbox}{Definition: mixed Hodge structure (MHS)}
A \textbf{mixed Hodge structure} on $H_{\mathbb{R}}$ is given by a tuple $(W, F)$ consisting of a weight filtration $W_{\ell}$ defined over $\mathbb{R}$, as described above, together with a decreasing filtration
\begin{equation}
0\subset F^D\subset \cdots \subset F^0 = H_{\mathbb{C}}\,,
\end{equation}
such that the filtration
\begin{equation}\label{eq:Hodge_decomp_mixed}
\left(F^p\cap W_{\ell}\right)/\left(F^p\cap W_{\ell-1}\right)\,,
\end{equation}
defines a Hodge structure of weight $\ell$ on each graded space $\mathrm{Gr}_{\ell}$.
\tcblower
When the weight filtration is induced by a nilpotent operator $N$, the filtration $F^p$ must satisfy the following compatibility condition
\begin{equation}
N F^p\subset F^{p-1}\,.
\end{equation}
\end{subbox}
\noindent Just like a Hodge structure, a general mixed Hodge structure need not come with the notion of a polarization. However, when a limiting mixed Hodge structure comes from a variation of \textit{polarized} Hodge structure (as will be explained in the next section), this polarization naturally carries over.

\subsubsection*{Polarized mixed Hodge structures}
Recall that in the geometric case one needs to restrict to the primitive cohomology in order for the variation of Hodge structure to be polarized. Similarly, there is a notion of primitivity in the context of mixed Hodge structures. Indeed, the corresponding \textit{primitive subspaces} $P_{\ell}$ of $\mathrm{Gr}_{\ell}$ are defined by
\begin{equation}
P_{D+\ell} = \mathrm{ker}\left(N^{\ell+1}:\, \mathrm{Gr}_{D+\ell}\rightarrow \mathrm{Gr}_{D-\ell-2} \right)\,.
\end{equation}
Naturally, the Hodge structure on the graded pieces descends to the primitive subspaces, and so we write
\begin{equation}\label{eq:Hodge_decomp_mixed_primitive}
P_{\ell} = \bigoplus_{p+q=\ell} P_\ell^{p,q}\,,\qquad \ell\geq D\,.
\end{equation}
for the corresponding Hodge decomposition. We then say that the mixed Hodge structure $(W,F)$ is polarized if the Hodge structure \eqref{eq:Hodge_decomp_mixed_primitive} on each $P_{D+\ell}$ is polarized with respect to the bilinear form $\left(\,\cdot\,,\,N^{\ell}\,\cdot\,\right)$. To be explicit, this means in particular that the following positivity condition is satisfied
\begin{equation}
v\in P^{p,q}_{D+\ell}:\qquad i^{p-q}\left(v,N^\ell\bar{v}  \right) >0\,.
\end{equation}
From this point onwards, we will always assume a mixed Hodge structure to be polarized in this sense.

\subsubsection*{Deligne splitting}
Unfortunately, the above characterization of a mixed Hodge structure is not the most illuminating, nor is it particularly easy to work with. Luckily, there is a clever way of repackaging the data of a mixed Hodge structure which is both easier to handle, and additionally reveals a hidden symmetry. First, we introduce the notion of a splitting.

\begin{subbox}{Definition: splitting of a MHS}
A \textbf{splitting} of a mixed Hodge structure $(W,F)$ is a bigrading
\begin{equation}
H_{\mathbb{C}} = \bigoplus_{p,q} I^{p,q}\,,
\end{equation}
in terms of complex vector spaces $I^{p,q}$, such that 
\begin{equation}\label{eq:Deligne_weight+limiting}
W_{\ell} = \bigoplus_{p+q=\ell} I^{p,q}\,,\qquad F^p = \bigoplus_{s}\bigoplus_{r\geq p} I^{r,s}\,.
\end{equation}
\tcblower Furthermore, the dimensions of the various components of the bigrading are denoted as
\begin{equation}
i^{p,q} = \mathrm{dim}_\mathbb{C} I^{p,q}\,.
\end{equation}
\end{subbox} 
\noindent One can conveniently package the information in such a splitting using the so-called \textbf{Hodge--Deligne diamond}. This is a diagram in which the various $I^{p,q}$ spaces are depicted in a square as in figure \ref{fig:Deligne}. In such a Hodge--Deligne diamond, a dot signifies that the corresponding vector space is non-trivial. Using the relations \eqref{eq:Deligne_weight+limiting}, one readily sees that the spaces $W_{\ell}$ correspond to the first $\ell$ horizontal rows of the Hodge--Deligne diamond when counting from below, while the spaces $F^p$ correspond to the first $D+1-p$ diagonal columns when counting from the top-left. \\

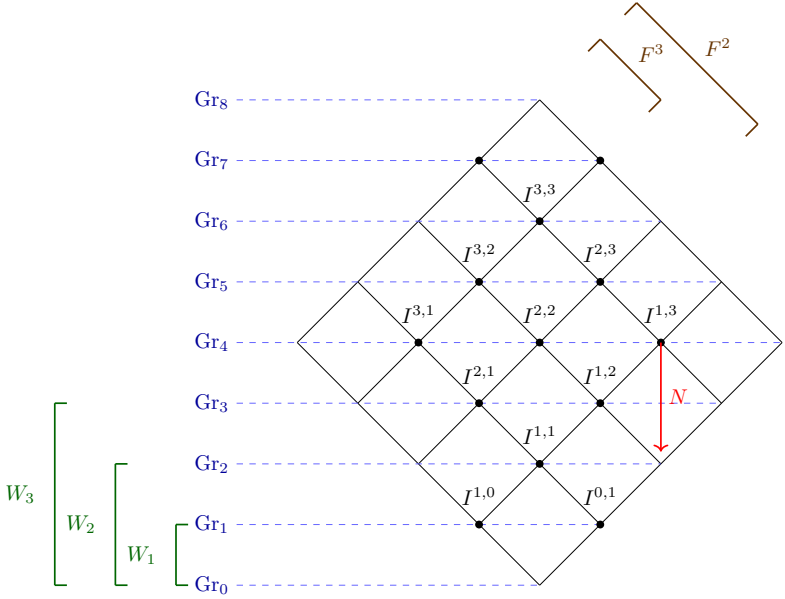
\begin{figure}[t!]

\scalebox{0.8}{\begin{tikzpicture}

\foreach \x in {0,1,2,3,4}
\draw (\x, -\x) -- (4+\x, 4-\x);
\foreach \y in {0,1,2,3,4}
\draw (\y,\y) -- (4+\y,-4+\y);

%   Graded Spaces
\foreach \x in {0,1,2,3,4,5,6,7,8}
\tkzDefPoint(-1,-4+\x){Gr\x};

%   Weight Filtrations
\foreach \x in {0,1,2}
\tkzDefPoint(-2.2-\x,-3.5+0.5*\x){W\x};

% Hodge Filtrations
\tkzDefPoint(5.5,4.5){F3};
\tkzDefPoint(6.6,4.6){F2}

% Label Graded Spaces
\foreach \x in {0,1,2,3}
\draw[dashed,blue!60!white,thin] (-1, 4-\x) -- (4+\x,4-\x);
\foreach \x in {4,5,6,7,8}
\draw[dashed,blue!60!white,thin] (-1, 4-\x) -- (12-\x,4-\x);

\foreach \x in {0,1,2,3,4,5,6,7,8}
\tkzLabelPoint[left](Gr\x){\textcolor{blue!60!black}{$\mathrm{Gr}_{\x}$}};

% Label Weight Filtrations

\foreach \x in {0,1,2}    
\draw[thick,green!40!black] (-2-\x,-4) -- (-2-\x,-3+\x);
\foreach \x in {0,1,2}    
\draw[thick,green!40!black] (-2-\x,-4) -- (-1.8-\x,-4);
\foreach \x in {0,1,2}    
\draw[thick,green!40!black] (-2-\x,-3+\x) -- (-1.8-\x,-3+\x);

\foreach\x in {0,1,2}
\pgfmathtruncatemacro{\y}{int(\x+1)}
\tkzLabelPoint[left](W\x){\textcolor{green!40!black}{$W_{\y}$}};

% Label Hodge Filtrations

\draw[thick,orange!40!black] (5,5) -- (6,4);
\draw[thick,orange!40!black] (5,5) -- (4.8,4.8);
\draw[thick,orange!40!black] (6,4) -- (5.8,3.8);

\draw[thick,orange!40!black] (5.6,5.6) -- (7.6,3.6);
\draw[thick,orange!40!black] (5.6,5.6) -- (5.4,5.4);
\draw[thick,orange!40!black] (7.6,3.6) -- (7.4,3.4);

\tkzLabelPoint[above right](F3){\textcolor{orange!40!black}{$F^3$}};

\tkzLabelPoint[above right](F2){\textcolor{orange!40!black}{$F^2$}};
% Nodes 

\node at (2,0)[circle,fill,inner sep=1.3pt]{};
\node at (4,0)[circle,fill,inner sep=1.3pt]{};
\node at (6,0)[circle,fill,inner sep=1.3pt]{};
\node at (3,1)[circle,fill,inner sep=1.3pt]{};
\node at (4,2)[circle,fill,inner sep=1.3pt]{};
\node at (5,1)[circle,fill,inner sep=1.3pt]{};
\node at (3,-1)[circle,fill,inner sep=1.3pt]{};
\node at (4,-2)[circle,fill,inner sep=1.3pt]{};
\node at (5,-1)[circle,fill,inner sep=1.3pt]{};

\node at (3,3)[circle,fill,inner sep=1.3pt]{};
\node at (5,3)[circle,fill,inner sep=1.3pt]{};

\node at (3,-3)[circle,fill,inner sep=1.3pt]{};
\node at (5,-3)[circle,fill,inner sep=1.3pt]{};

\tkzLabelPoint[above](5,-2.8){$I^{0,1}$};
\tkzLabelPoint[above](3,-2.8){$I^{1,0}$};
\tkzLabelPoint[above](4,-1.8){$I^{1,1}$};
\tkzLabelPoint[above](3,-0.8){$I^{2,1}$};
\tkzLabelPoint[above](5,-0.8){$I^{1,2}$};
\tkzLabelPoint[above](2,0.2){$I^{3,1}$};
\tkzLabelPoint[above](4,0.2){$I^{2,2}$};
\tkzLabelPoint[above](6,0.2){$I^{1,3}$};
\tkzLabelPoint[above](3,1.2){$I^{3,2}$};
\tkzLabelPoint[above](5,1.2){$I^{2,3}$};
\tkzLabelPoint[above](4,2.2){$I^{3,3}$};

\draw[->,thick,red] (6,0) -- (6,-1.8);
\tkzLabelPoint[right](6,-0.9){\textcolor{red}{$N$}};

\end{tikzpicture}}
\caption{Example of a Hodge--Deligne diamond for a splitting associated to a weight $D=4$ mixed Hodge structure. Here we have also indicated the various components of the filtrations $W_{\ell}$ and $F^p$ via the relation \eqref{eq:Deligne_weight+limiting}, as well as the graded spaces via the relation \eqref{eq:graded_Deligne}. Furthermore, the action of $N$ is indicated by the red arrow and follows from the relation \eqref{eq:log-monodromy_Deligne}.}
\label{fig:Deligne}
\end{figure}
\noindent The reason for introducing this additional notion of a splitting is that there is a one-to-one correspondence between mixed Hodge structures and those splittings which satisfy the following relation with respect to complex conjugation
\begin{equation}\label{eq:Deligne_conjugation}
\overline{I^{p,q}} = I^{q,p}\quad \text{mod}\,\bigoplus_{r<q,s<p}I^{r,s}\,.
\end{equation}
In words, this means that the spaces $I^{p,q}$ and $I^{q,p}$ are related to each other by complex conjugation, modulo elements which are positioned lower in the splitting. The property \eqref{eq:Deligne_conjugation} is illustrated in figure \ref{fig:R_split}. Given a mixed Hodge structure $(W,F)$, there is a unique splitting associated to it that satisfies \eqref{eq:Deligne_conjugation} and, importantly, it can be written down explicitly.

\begin{subbox}{Deligne splitting}
For a given mixed Hodge structure $(W,F)$, the unique splitting $I^{p,q}$ of $(W,F)$ that satisfies \eqref{eq:Deligne_conjugation} is given by the formula
\begin{equation}\label{eq:Deligne_splitting}
I^{p,q} = F^p \cap W_{p+q}\cap \Big(\bar{F}^q\cap W_{p+q}+\sum_{j\geq 1}\bar{F}^{q-j}\cap W_{p+q-j-1}\Big)\,,
\end{equation}
and is referred to as the \textbf{Deligne splitting}.
\end{subbox}

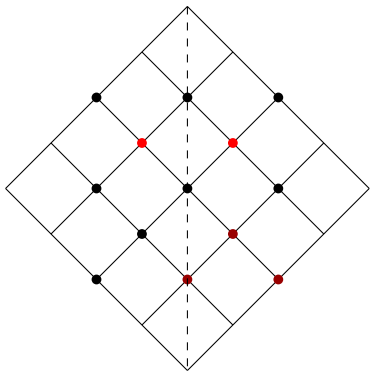
\begin{figure}[t]
\centering
\begin{tikzpicture}[scale=0.6]

\foreach \x in {0,1,2,3,4}
\draw(\x, -\x) -- (4+\x, 4-\x);
\foreach \y in {0,1,2,3,4}
\draw (\y,\y) -- (4+\y,-4+\y);

\node at (2,0)[circle,fill,inner sep=1.3pt]{};
\node at (4,0)[circle,fill,inner sep=1.3pt]{};
\node at (6,0)[circle,fill,inner sep=1.3pt]{};
\node at (3,1)[circle,fill,inner sep=1.3pt,red]{};
\node at (4,2)[circle,fill,inner sep=1.3pt]{};
\node at (5,1)[circle,fill,inner sep=1.3pt,red]{};
\node at (3,-1)[circle,fill,inner sep=1.3pt]{};
\node at (4,-2)[circle,fill,inner sep=1.3pt,black!40!red]{};
\node at (5,-1)[circle,fill,inner sep=1.3pt,black!40!red]{};

\node at (2,2)[circle,fill,inner sep=1.3pt]{};
\node at (2,-2)[circle,fill,inner sep=1.3pt]{};
\node at (6,2)[circle,fill,inner sep=1.3pt]{};
\node at (6,-2)[circle,fill,inner sep=1.3pt,black!40!red]{};

\draw[dashed] (4,4) -- (4,-4);

\end{tikzpicture}
\caption{The property \eqref{eq:Deligne_conjugation} is exemplified for a weight $D=4$ mixed Hodge structure. We have highlighted the spaces $\textcolor{red}{I^{3,2}}$ and $\textcolor{red}{\overline{I^{3,2}}=I^{2,3}}\;\mathrm{mod}\left(\textcolor{black!40!red}{I^{0,2}\oplus I^{1,1}\oplus I^{1,2}}\right)$, where the latter is obtained by reflecting $\textcolor{red}{I^{3,2}}$ in the vertical axis and modding out the lower spaces.}
\label{fig:R_split}
\end{figure}

\noindent Let us now discuss some additional properties of the Deligne splitting which greatly constrain the different number of possible Hodge--Deligne diamonds one can write down. First, using the fact that $NW_{\ell}\subset W_{\ell-2}$ and $NF^p\subset F^{p-1}$, together with the explicit relation \eqref{eq:Deligne_splitting}, it follows that
\begin{equation}\label{eq:log-monodromy_Deligne}
N I^{p,q}\subset I^{p-1,q-1}\,.
\end{equation}
In other words, in the Hodge--Deligne diamond the nilpotent operator $N$ sends an element in a given horizontal row exactly two rows downwards, while staying in the same vertical column. This is illustrated in figure \ref{fig:Deligne}. Second, one can show that the following relations must hold:
\begin{equation}\label{eq:ipq_restriction}
i^{p,q}=i^{q,p}=i^{D-q,D-p}\,,\qquad i^{p-1,q-1}\leq i^{p,q}\,,\quad \text{for $p+q<D$}.
\end{equation}
The first relation simply states that the Hodge--Deligne diamond must be symmetric under reflections about the vertical and horizontal axes, while the second relation gives a constraint on the dimensions of the  $I^{p,q}$ spaces which lie strictly below the central horizontal row. 
\\

\noindent Finally, let us note that the graded spaces and their primitive parts can be recovered from the Deligne splitting using the following relation
\begin{equation}\label{eq:graded_Deligne}
\mathrm{Gr}_\ell = \bigoplus_{p+q=\ell} I^{p,q}\,,\qquad P^{p,q}_{D+\ell} = I^{p,q}\cap \mathrm{ker}\,N^{\ell+1}\,.
\end{equation}
In terms of the Hodge--Deligne diamond, the graded spaces $\mathrm{Gr}_{\ell}$ therefore correspond to the $\ell$-th horizontal row when counting from below. It is important to stress that due to the conjugation property \eqref{eq:Deligne_conjugation} of the Deligne splitting, the decomposition \eqref{eq:graded_Deligne} generically does not define a Hodge decomposition, as opposed to the decomposition induced by \eqref{eq:Hodge_decomp_mixed}. There is, however, a special case in which the decomposition \eqref{eq:graded_Deligne} does define a Hodge structure, namely when the conjugation property \eqref{eq:Deligne_conjugation} happens to simplify to 
\begin{equation}
\overline{I^{p,q}} = I^{q,p}\,,
\end{equation}
without the need for the additional lower-lying elements. If a splitting satisfies this condition it is said to be $\mathbb{R}$-split. In this case, the formula \eqref{eq:Deligne_splitting} simplifies drastically to
\begin{equation}
I^{p,q}_{\text{$\mathbb{R}$-split}} = F^p\cap \bar{F}^q\cap W_{p+q}\,.
\end{equation}

\subsubsection*{The phase operator}
A crucial insight is that, even if a given Deligne splitting is not $\mathbb{R}$-split, one can always perform a $G_{\mathbb{C}}$ basis transformation which effectively makes it $\mathbb{R}$-split. Furthermore, this transformation turns out to be rather restricted. To be precise, one can show that there always exists a real operator $\delta\in\mathfrak{g}_{\mathbb{R}}$ called the \textbf{phase operator} (in the physics literature), such that the filtration
\begin{equation}
\tilde{F}^p := e^{-i\delta}F^p\,,
\end{equation}
together with the original weight filtration $W_\ell$ defines another (polarized) mixed Hodge structure $(W,\tilde{F})$ whose associated Deligne splitting $\tilde{I}^{p,q}$ is $\mathbb{R}$-split. Furthermore, $\delta$ commutes with all the morphisms of the mixed Hodge structure and thus, in particular, commutes with the nilpotent operator:
\begin{equation}\label{eq:delta_commutes_N}
[\delta, N]=0\,.
\end{equation}
Note that, because of the relation \eqref{eq:delta_commutes_N}, the new Deligne splitting is again compatible with the nilpotent operator, in the sense that $N \tilde{I}^{p,q}\subset \tilde{I}^{p-1,q-1}$. Additionally, the phase operator enjoys a decomposition
\begin{equation}\label{eq:delta_Deligne}
\delta = \sum_{p,q\geq 1} \delta_{-p,-q}\,,\qquad \delta_{-p,-q} I^{r,s}\subset I^{r-p,s-q}\,,
\end{equation}
where we stress the restriction to $p,q\geq 1$ in the summation, which originates from the fact that $\delta$ is constructed in such a way that it `fixes' the bad behaviour of the old $I^{p,q}$ under complex conjugation, which only involves lower-positioned elements. 
\\

\noindent Importantly, let us mention that $\delta$ can always be determined algorithmically, as is explained and exemplified in detail in \cite{vandeHeisteeg:2022gsp}. For our purposes, we will typically work in the opposite direction. Namely, we start from a given $\mathbb{R}$-split mixed Hodge structure, and then simply construct the most general phase operator $\delta$ that satisfies the above conditions. This will be exemplified in appendix \ref{sec:asymp_Hodge_examples}.

\subsubsection*{An emergent $\mathfrak{sl}(2,\mathbb{R})$-symmetry}
Earlier it was mentioned that the formulation in terms of the Deligne splitting reveals a hidden symmetry, let us discuss this now. The crucial observation is that to each $\mathbb{R}$-split mixed Hodge structure $\tilde{I}^{p,q}$ one can naturally associate a real $\mathfrak{sl}(2,\mathbb{R})$-triple in the following way. First, one can define a \textbf{weight operator} $N^0\in\mathfrak{g}_{\mathbb{R}}$ by demanding that it satisfies
\begin{equation}
N^0 v = (p+q-D)v\,,\qquad v\in \tilde{I}^{p,q}.
\end{equation}
Note that precisely because $\tilde{I}^{p,q}$ is $\mathbb{R}$-split, this definition is well-behaved under complex conjugation and thus defines a real operator $N^0$. Furthermore, using the fact that $N \tilde{I}^{p,q}\subset \tilde{I}^{p-1,q-1}$, one immediately sees that 
\begin{equation}
[N^0, N] = -2 N\,,
\end{equation}
from which one may recognize $N$ as a lowering operator. From here on, one can find a unique corresponding raising operator $N^+$ by imposing the remaining $\mathfrak{sl}(2,\mathbb{R})$ commutation relations 
\begin{equation}
[N^0, N^+] = 2N^+\,,\qquad [N^+, N] = N^0\,.
\end{equation}
As a result, many of the properties of $\mathbb{R}$-split mixed Hodge structures can be understood in terms of the representation theory of certain $\mathfrak{sl}(2,\mathbb{R})$-triples. 

\subsubsection*{The $\zeta$-operator}
There is one final operator we need to introduce which is of a more technical nature. As will be explained in the next section, in the general case of an $m$-parameter variation of Hodge structure, one generically finds multiple $\mathfrak{sl}(2,\mathbb{R})$-triples which can be made to commute by performing another basis transformation to transform from the $\mathbb{R}$-split mixed Hodge structure $\tilde{I}^{p,q}$ to the so-called \textbf{$\mathrm{sl}(2)$-split} mixed Hodge structure $\hat{I}^{p,q}$. Concretely, we perform another rotation
\begin{equation}\label{eq:def_zeta}
\hat{F}^p :=e^{\zeta}\tilde{F}^p = e^\zeta e^{-i\delta}F^p\,,
\end{equation}
where $\zeta\in\mathfrak{g}_{\mathbb{R}}$ is fixed uniquely in terms of $\delta$. At this point we choose to not yet give the precise relation in the most general case, which will be deferred to section \ref{subsec:h_reconstruction}. Nevertheless, for practical purposes it is useful to record the precise relation in the case of a mixed Hodge structure of weight 3 (or less), which is relevant in the setting of Calabi--Yau threefolds\footnote{Note that while the individual components of $\zeta$ need not be real, the sum is real.}
\begin{align}
\zeta_{-1,-2} &= -\frac{i}{2}\delta_{-1,-2}\,,\nonumber\\
\zeta_{-1,-3} &= -\frac{3i}{4}\delta_{-1,-3}\,,\nonumber\\
\zeta_{-2,-3} &=-\frac{3i}{8}\delta_{-2,-3} - \frac{1}{8}\left[\delta_{-1,-1},\delta_{-1,-2} \right]\,,\nonumber\\
\label{eq:delta_zeta_threefold}
\zeta_{-3,-3} &= -\frac{1}{8}\left[\delta_{-1,-1},\delta_{-2,-2} \right]\,.
\end{align}
Here we stress that $\zeta_{p,q}$ denote the $(p,q)$ components of $\zeta$ with respect to the $\mathbb{R}$-split mixed Hodge structure $\tilde{I}^{p,q}$. All other components that are not listed are either vanishing or related via complex conjugation as $\zeta_{p,q} = \overline{\zeta_{q,p}}$.

\section{The $\mathrm{Sl}(2)$-orbit theorem (1): single-variable}\label{sec:SL2_orbit_one_variable}

Having discussed some general aspects of mixed Hodge structures, let us now return to the problem at hand and explain why this mathematical machinery is useful in the study of asymptotic Hodge theory. Recall that, according to the nilpotent orbit theorem, the asymptotic behaviour of any variation of polarized Hodge structure is effectively characterized in terms of a set of log-monodromy matrices $N_i$, together with the limiting filtration $F_{(m)}^p$. The central point is that this data in turn defines a mixed Hodge structure. Consequently, the type of mixed Hodge structure that arises in a given asymptotic limit contains a lot of information about how the Hodge structure is degenerating in that limit. More practically, the type of mixed Hodge structure will dictate the asymptotic form of e.g.~the K\"ahler potential and the Hodge star operator. In this section, we will discuss this in detail in the one-parameter case. The general case will be discussed in the next section.

\subsection{Perspective (1): Limiting mixed Hodge structures}
Consider a  one-parameter variation of Hodge structure which, by the nilpotent orbit theorem, is well-approximated by a nilpotent orbit of the form
\begin{equation}
F^p(t) \approx F_{\mathrm{nil}}^p(t) = e^{t N}F_{0}^p\,,
\end{equation}
in the regime where $\mathrm{Im}\,t\gg 1$. Note that we have adjusted our notation slightly by denoting the limiting filtration by $F^p_0$. The central result is the following (see Theorem 6.16 in \cite{schmid})
\begin{subbox}{Limiting mixed Hodge structure (LMHS)}
The pair $(W(N), F_{0})$ forms a mixed Hodge structure, which will be referred to as the \textbf{limiting mixed Hodge structure} associated to the limit $y\rightarrow\infty$.
\tcblower
The weight filtration $W(N)$ is typically referred to as the \textbf{monodromy weight filtration} associated to the limit $y\rightarrow\infty$.
\end{subbox}
\noindent Thus, the data of a one-parameter nilpotent orbit naturally induces a limiting mixed Hodge structure $(W, F_{0})$, which can in turn be equivalently described in terms of the associated Deligne splitting $I^{p,q}$ via \eqref{eq:Deligne_splitting}. Importantly, since the Hodge numbers of $F_{0}^p$ are the same as those of $F^p$, one finds the additional condition
\begin{equation}\label{eq:ipq_relation_hpq}
h^{p,D-p} = \sum_{q=0}^D i^{p,q}\,,
\end{equation}
which relates the $i^{p,q}$ to the Hodge numbers of the original pure Hodge structure. In special cases, for example when the underlying geometry is a Calabi--Yau manifold, this can lead to a considerable reduction on the possible Hodge--Deligne diamonds.

\subsubsection*{Intermezzo: Hodge--Deligne diamonds for Calabi--Yau manifolds}
Recall that a central property of a Calabi--Yau $D$-fold $Y_D$ is that $h^{D,0}=h^{0,D}=1$. As a result of the condition \eqref{eq:ipq_relation_hpq}, together with the symmetries of the Hodge--Deligne diamond \eqref{eq:ipq_restriction}, one finds that
\begin{equation}
i^{D,d}=i^{d,D}=i^{0,D-d}=i^{D-d,0}=1\,,
\end{equation}
for some integer $0\leq d\leq D$, while all other spaces lying on the outer edge of the Hodge--Deligne diamond are empty. This suggests a natural classification of Hodge--Deligne diamonds for Calabi--Yau manifolds, where different singularity types are distinguished by the value of the integer $d$. For example, in the case of Calabi--Yau threefolds, this results in four types labeled as
\begin{alignat*}{2}
&\text{Type I}:\quad &d=0\,,\\
&\text{Type II}:\quad &d=1\,,\\
&\text{Type III}:\quad &d=2\,,\\
&\text{Type IV}:\quad &d=3\,.
\end{alignat*}
Similarly, for Calabi--Yau fourfolds there is the additional Type V singularity. It is important to stress, however, that the integer $d$ does not fully fix the allowed Hodge--Deligne diamond, as we still have a lot of freedom in the inner part of the diamond. For example, in the case of Calabi--Yau threefold, we still need to specify the dimensions $i^{1,1},i^{1,2},i^{2,1},i^{2,2}$. Due to the symmetry property \eqref{eq:ipq_restriction} only two of these are independent. Furthermore, the relation \eqref{eq:ipq_relation_hpq} imposes an additional restriction on the sum of all the $i^{p,q}$, such that we are left with only one free parameter. Conventionally, we choose the free parameter to be $i^{2,2}$ and attach it as a subscript to the type labeling, writing 
\begin{equation}\label{Deligne_type_labeling}
\mathrm{I}_{i^{2,2}}\,,\quad \mathrm{II}_{i^{2,2}}\,,\quad \mathrm{III}_{i^{2,2}}\,,\quad \mathrm{IV}_{i^{2,2}}\,.
\end{equation}
By taking into account the bounds on the possible values of $i^{2,2}$ arising from the constraint \eqref{eq:ipq_relation_hpq} one finds in total $4h^{2,1}$ possible Hodge--Deligne diamonds associated to limiting mixed Hodge structures of Calabi--Yau threefolds (one of which, namely $I_0$, is a trivial mixed Hodge structure), see also \cite{Grimm:2018cpv} as well as \cite{Grimm:2019ixq} for an extension of this discussion to the Calabi--Yau fourfold case. In appendix \ref{sec:asymp_Hodge_examples} we will return to the above classification in detail for the case $h^{2,1}=1$.  \\

\noindent To close this intermezzo, let us also remark that, in addition to the allowed Hodge--Deligne diamonds, one can also classify the (conjugacy classes of) associated nilpotent endomorphism $N$ in terms of signed Young diagrams \cite{Djokovic:1982,McGovern:1993}. This is useful for practical purposes as it allows one to construct the most general form of $N$, as well as the intersection form, from a set of simple building blocks or ``normal forms''. We refer the reader to \cite{Grimm:2018ohb,vandeHeisteeg:2022gsp} for more details and illustrative examples. 

\subsubsection*{The $\mathrm{Sl}(2)$-orbit}
Let us now leave the above intermezzo behind and return to the general setting. Generically, the Deligne splitting $I^{p,q}$ that arises from a limiting mixed Hodge structure associated to a nilpotent orbit need not be $\mathbb{R}$-split. Following the discussion in section \ref{sec:MHS}, we can always perform a rotation 
\begin{equation}
\hat{F}^p_{0}:= e^\zeta e^{-i\delta}F_{0}^p\,,
\end{equation}
using the phase operator $\delta$ and the associated operator $\zeta$, such that $(W, \hat{F}_{0})$ defines an $\mathbb{R}$-split limiting mixed Hodge structure. We denote the corresponding Deligne splitting by $\hat{I}^{p,q}$. Importantly, due to the fact that the latter is $\mathbb{R}$-split, we may introduce the canonical grading operator $N^0$ by\footnote{Note that this grading operator is defined with respect to the $\mathrm{sl}(2)$-split Deligne splitting $\hat{I}^{p,q}$, as opposed to the $\mathbb{R}$-split Deligne splitting $\tilde{I}^{p,q}$ that was used in section \eqref{sec:MHS}. The two are straightforwardly related by
\begin{equation}
N^0_{\hat{I}^{p,q}} = e^\zeta N^0_{\tilde{I}^{p,q}} e^{-\zeta}\,.
\end{equation}}
\begin{equation}
N^0v = (p+q-D)v\,,\qquad v\in \hat{I}^{p,q}\,.
\end{equation}
A natural question is whether this new mixed Hodge structure $(W, \hat{F}_{0})$ can itself be interpreted as the limiting filtration of some other nilpotent orbit. Indeed, one might write down the following candidate filtration
\begin{equation}
\hat{F}^p_{\mathrm{nil}}(t):= e^{t N}\hat{F}^p_{0}\,.
\end{equation}
As it turns out, this indeed defines a nilpotent orbit. Furthermore, it is very closely related to the original nilpotent orbit we started with. To be precise, one has the following result (see Lemma 3.12 of \cite{CKS}).
\begin{subbox}{$\mathrm{Sl}(2)$-orbit approximation}
For $\mathrm{Im}\,t>0$, the filtration $\hat{F}^p_{\mathrm{nil}}(t)$ defines a nilpotent orbit (in particular, it defines a variation of polarized Hodge structure). Furthermore, it agrees with the original nilpotent orbit $F^p_{\mathrm{nil}}(t)$ to first order in the regime $y\gg 1$, so that
\begin{equation}
F^p_{\mathrm{nil}}(t)\approx \hat{F}^p_{\mathrm{nil}}(t)\,.
\end{equation}
\tcblower
\textbf{Note:}\\
The filtration $\hat{F}^p_{\mathrm{nil}}(t)$ is typically referred to as the \textbf{$\mathrm{Sl}(2)$-orbit approximation} of $F^p_{\mathrm{nil}}(t)$. 
\end{subbox}
\noindent Let us explain where the terminology ``$\mathrm{Sl}(2)$-orbit approximation'' comes from. This relies on the following crucial identity
\begin{equation}\label{eq:identity_Sl2-orbit}
\boxed{
\rule[-.15cm]{0cm}{.55cm} \quad e^{iyN}\hat{F}^p_{0} = y^{-\frac{1}{2}N^0} F_\infty^p\,, }
\end{equation}
where we have introduced a fixed filtration
\begin{equation}
F_{\infty}^p:= e^{iN}\hat{F}_{0}^p\,.
\end{equation}
The relation \eqref{eq:identity_Sl2-orbit} simply follows from the $\mathfrak{sl}(2,\mathbb{R})$ commutation relations and the fact that $N^0$ preserves the filtration $\hat{F}^p_{0}$. This identity will feature prominently in the multi-parameter case as well, as will be discussed in the next section. As an immediate consequence of the identity, we find that
\begin{equation}\label{eq:Sl2_approx_single}
\hat{F}^p_{\mathrm{nil}}(t) = \left[e^{x N}y^{-\frac{1}{2}N^0}\right]F_{\infty}^p\,,
\end{equation}
A point which is worth emphasizing is that $F_{\infty}^p$ is in fact a (polarized) Hodge filtration! This is to be contrasted with the limiting filtrations $F_{0}^p$ and $\hat{F}_{0}^p$, which are generically not Hodge filtrations. The rough intuition for why this is the case is that $F_{\infty}^p$ corresponds to evaluating the nilpotent orbit $\hat{F}_{\mathrm{nil}}^p(t)$ at the point $t=i$. The Hodge structure defined by $F_{\infty}^p$ will be referred to as the \textbf{boundary Hodge structure}. This terminology is somewhat misleading, as it is not the case that $F_{\infty}$ naturally lies on the boundary of the period domain. Rather, one should think of it as a Hodge structure which can naturally be associated to the limit $y\rightarrow\infty$ and from which the original nilpotent orbit can be approximated.\\

\noindent Indeed, the main point is that the relation \eqref{eq:Sl2_approx_single} tells us that one can view $\hat{F}^p_{\mathrm{nil}}(t)$ as the orbit of the boundary Hodge structure $F_{\infty}^p$ under the action of an element in $\mathrm{SL}(2,\mathbb{R})$. Furthermore, since the filtration $\hat{F}^p_{\mathrm{nil}}(t)$ agrees with $F^p_{\mathrm{nil}}(t)$ to first order in the regime $y\gg 1$, this justifies the name ``$\mathrm{Sl}(2)$-orbit approximation''. For this reason, we will usually write
\begin{equation}
F^p_{\mathrm{Sl}(2)}(t):= \hat{F}^p_{\mathrm{nil}}(t)\,.
\end{equation}

\subsection{Perspective (2): Approximating the period map}
Let us place the above observation in the context of section \ref{sec:period_map}, in which we discussed the notion of the period map. Recall that, since the group $G_{\mathbb{R}}$ acts transitively on the period domain, one can always find a $G_{\mathbb{R}}$-valued map that interpolates between a given ``reference Hodge structure'' and any other Hodge structure. The above discussion indicates that, near a given boundary in the moduli space, there is a natural choice for such a reference Hodge structure, namely the boundary Hodge structure $F_\infty^p$. Indeed, let us consider the period map $h(x,y)$ which satisfies the relation\footnote{To be precise, we are choosing a representative in the equivalence class $[h]\in G/V$, c.f.~the discussion in section \ref{sec:period_map}.}
\begin{equation}
F^p = h(x,y) F^p_\infty\,.
\end{equation}
Then our preceding analysis shows that, to first order in $1/y$, the full period map can be approximated by 
\begin{equation}
h(x,y)\approx h_{\mathrm{Sl}(2)}(x,y):= e^{xN}y^{-\frac{1}{2}N^0}\,,
\end{equation}
which satisfies
\begin{equation}
F^p_{\mathrm{Sl}(2)} = h_{\mathrm{Sl}(2)}(x,y)F^p_\infty\,.
\end{equation}
As a consistency check, one can in fact show that $h_{\mathrm{Sl}(2)}$ itself satisfies Nahm's equations \eqref{eq:Nahm_N}. Recall that the latter comprise part of the horizontality conditions of the full period map. To see this, let us first recall that the grading operator $N^0$ and the nilpotent endomorphism $N$ can be completed into a real $\mathrm{sl}(2)$-triple
\begin{equation}
\{N^+, N^0, N^-\}\,,\qquad N^-:=N\,.
\end{equation}
Then one readily computes
\begin{align}
\mathcal{N}^0 &= -2 h^{-1}\partial_y h = \frac{N^0}{y}+\mathcal{O}(y^{-3/2})\,,\\
\mathcal{N}^-&=h^{-1}\partial_x h= \frac{N^-}{y}+\mathcal{O}(y^{-3/2})\,,
\end{align}
where the higher-order terms correspond to corrections that go beyond the $\mathrm{Sl}(2)$-orbit approximation. Comparing to \eqref{eq:Nahm_N} and recalling the $\mathrm{sl}(2)$ commutation relations, we see that Nahm's equations are solved (to leading order in $1/y$) when additionally setting
\begin{equation}
\mathcal{N}^+=\left(h^{-1}\partial_x h\right)^\dagger = \frac{N^+}{y}+\mathcal{O}(y^{-3/2})\,,
\end{equation}
where the dagger is taken with respect to the boundary charge operator $Q_\infty$ c.f.~\eqref{eq:def_adjoint}. Recall that the latter gives an alternative characterization of the boundary Hodge structure via
\begin{equation}
Q_\infty v = \frac{1}{2}(p-q)v\,,\qquad v\in H^{p,q}_\infty\,,
\end{equation}
with $H^{p,q}_\infty$ the Hodge decomposition induced by $F^p_\infty$. 

\subsubsection*{Horizontal $\mathrm{sl}(2)$-triple}
It is important to stress, however, that Nahm's equations \eqref{eq:Nahm_N} comprise a part of the horizontality conditions of the period map, but are not equivalent to it. Indeed, the complete horizontality conditions are instead captured in terms of the Q-constraint \eqref{eq:Q_constraint_N}. Evaluating this constraint in the $\mathrm{Sl}(2)$-orbit approximation, one finds that the $\mathrm{sl}(2)$ triple must satisfy an additional set of commutation relations with respect to the boundary charge operator. 
\begin{subbox}{Horizontal $\mathfrak{sl}(2,\mathbb{R})$-triple}
The $\mathrm{sl}(2)$-triple must satisfy the following commutation relations
\begin{equation}\label{eq:horizontal_sl2_real}
[Q_\infty, N^0] = i\left(N^++N^- \right)\,,\qquad [Q_\infty, N^{\pm}] = -\frac{i}{2}N^0\,.
\end{equation}
Such an $\mathrm{sl}(2)$-triple is said to be \textbf{horizontal} with respect to $Q_\infty$. Furthermore, one readily checks that the commutation relations \eqref{eq:horizontal_sl2_real} imply that
\begin{equation}\label{eq:dagger_sl2_real}
\left(N^0\right)^\dagger = N^0\,,\qquad \left(N^-\right)^\dagger = N^+\,.
\end{equation}
\tcblower
Alternatively, a horizontal $\mathrm{sl}(2)$-triple is also said to be \textbf{Hodge at} $F^p_\infty$.
\end{subbox}

\subsubsection*{Beyond the $\mathrm{Sl}(2)$-orbit approximation}
Let us look ahead slightly and briefly discuss a central result of asymptotic Hodge theory, which will be covered at length in chapter \ref{chap:asymp_Hodge_II}. Following the above logic, there should also exist a map $h_{\mathrm{nil}}(x,y)$ which includes the polynomial corrections that have been neglected in the $\mathrm{Sl}(2)$-orbit approximation, and interpolates between the boundary Hodge structure and the nilpotent orbit approximation. In other words, it should satisfy
\begin{equation}
F^p_{\mathrm{nil}} = h_{\mathrm{nil}}(x,y)F^p_{\infty}\,.
\end{equation}
The resulting map $h_{\mathrm{nil}}$ is then referred to as the \textbf{nilpotent orbit approximation of the period map}. The main question, then, is whether one can actually \textit{compute} $h_{\mathrm{nil}}(x,y)$. Strikingly, this can indeed be done in a completely algorithmic way following the seminal work of Cattani, Kaplan, and Schmid \cite{CKS}. Describing the mechanism of this algorithm is the main purpose of chapter \ref{chap:asymp_Hodge_II}. However, let us already state the general form of the result, to give an idea. One finds that $h_{\mathrm{nil}}$ takes the form
\begin{equation}
h_{\mathrm{nil}}(x,y) = e^{xN}g(y) y^{-\frac{1}{2}N^0}\,,
\end{equation}
where $g(y)$ admits an infinite series expansion
\begin{equation}
g(y) = 1 + \frac{g_1}{y}+\frac{g_2}{y^2}+\cdots\,,
\end{equation}
which can be calculated explicitly from the data
\begin{equation}\label{eq:boundary_data_one_variable}
\{Q_\infty, N^+, N^-, N^0, \delta\}\,,
\end{equation}
which will be termed the \textbf{boundary data}. Furthermore, the coefficients $g_i$ appearing in the expansion will satisfy some rather stringent conditions, which play a crucial role in the finiteness results that will be discussed in chapter \ref{chap:finiteness}. Note that indeed in the limit where $y\gg 1$ the leading behaviour of $h_{\mathrm{nil}}$ is precisely captured by $h_{\mathrm{Sl}(2)}$. 

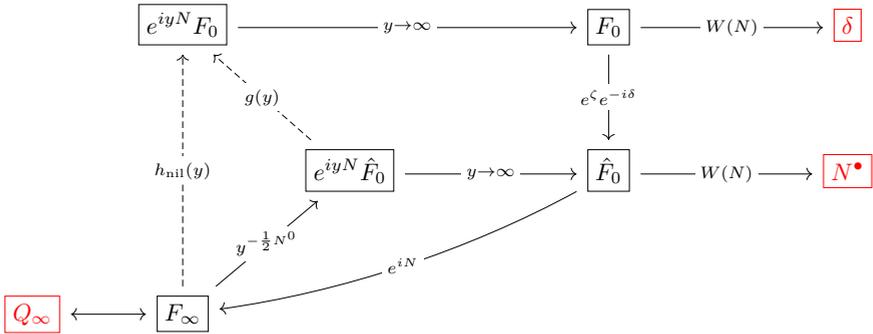
\begin{figure}[t!]
% https://q.uiver.app/?q=WzAsOCxbMSwwLCJcXGJveGVke2Vee2l5Tl4tfUZ9Il0sWzUsMCwiXFxib3hlZHtGfSJdLFs1LDIsIlxcYm94ZWR7XFx0aWxkZXtGfX0iXSxbMiwyLCJcXGJveGVke2Vee2l5Tl4tfVxcdGlsZGV7Rn19Il0sWzEsNCwiXFxib3hlZHtGX1xcaW5mdHl9Il0sWzAsNCwiXFx0ZXh0Y29sb3J7cmVkfXtcXGJveGVke1FfXFxpbmZ0eX19Il0sWzgsMCwiXFx0ZXh0Y29sb3J7cmVkfXtcXGJveGVke1xcZGVsdGF9fSJdLFs4LDIsIlxcdGV4dGNvbG9ye3JlZH17XFxib3hlZHtOXntcXGJ1bGxldH19fSJdLFswLDEsInlcXHJpZ2h0YXJyb3dcXGluZnR5IiwxXSxbMSwyLCJlXntcXHpldGF9ZV57aVxcZGVsdGF9IiwxXSxbMywyLCJ5XFxyaWdodGFycm93XFxpbmZ0eSIsMV0sWzMsMCwiZyh5KSIsMSx7InN0eWxlIjp7ImJvZHkiOnsibmFtZSI6ImRhc2hlZCJ9fX1dLFs0LDMsInleey1cXGZyYWN7MX17Mn1OXjB9IiwxXSxbNCwwLCJoKHkpIiwxLHsic3R5bGUiOnsiYm9keSI6eyJuYW1lIjoiZGFzaGVkIn19fV0sWzIsNCwiZV57aU5eLX0iLDEseyJjdXJ2ZSI6LTJ9XSxbNSw0LCIiLDEseyJzdHlsZSI6eyJ0YWlsIjp7Im5hbWUiOiJhcnJvd2hlYWQifX19XSxbMSw2LCJXKE5eLSkiLDFdLFsyLDcsIlcoTl4tKSIsMV1d
\[\adjustbox{scale=0.9}{\begin{tikzcd}
	& {\boxed{e^{iyN}F_0}} &&&& {\boxed{F_0}} &&& {\textcolor{red}{\boxed{\delta}}} \\
	\\
	&& {\boxed{e^{iyN}\hat{F}_0}} &&& {\boxed{\hat{F}_0}} &&& {\textcolor{red}{\boxed{N^{\bullet}}}} \\
	\\
	{\textcolor{red}{\boxed{Q_\infty}}} & {\boxed{F_\infty}}
	\arrow["y\rightarrow\infty"{description}, from=1-2, to=1-6]
	\arrow["{e^{\zeta}e^{-i\delta}}"{description}, from=1-6, to=3-6]
	\arrow["y\rightarrow\infty"{description}, from=3-3, to=3-6]
	\arrow["{g(y)}"{description}, dashed, from=3-3, to=1-2]
	\arrow["{y^{-\frac{1}{2}N^0}}"{description}, from=5-2, to=3-3]
	\arrow["{h_{\mathrm{nil}}(y)}"{description}, dashed, from=5-2, to=1-2]
	\arrow["{e^{iN}}"{description}, curve={height=-12pt}, from=3-6, to=5-2]
	\arrow[tail reversed, from=5-1, to=5-2]
	\arrow["{W(N)}"{description}, from=1-6, to=1-9]
	\arrow["{W(N)}"{description}, from=3-6, to=3-9]
\end{tikzcd}}\]
\caption{An overview of the various maps and filtrations that have been discussed in this section. In red the boundary data \eqref{eq:boundary_data_one_variable} is highlighted.}
\label{fig:bulk_reconstruction_one_variable}
\end{figure}

\subsubsection*{Summary: successive approximations}
Let us briefly summarize the current state of affairs, see also figure \ref{fig:bulk_reconstruction_one_variable} for an overview of the relations between the various spaces we have been considering. We started off with a completely general one-parameter variation of Hodge structure, which is parametrized by a Hodge filtration $F^p(t)$. From there, we have found that, for large $\mathrm{Im}\,t$ one can make two successive approximations:

\begin{subbox}{Successive approximations of variations of Hodge structure}
\begin{itemize}
\item \textbf{Approximation (1): Nilpotent orbit}\\
In the first approximation, we write
\begin{equation}
	F^p(t)\approx F^p_{\mathrm{nil}}(t) = e^{tN}F_{0}^p = h_{\mathrm{nil}}(x,y)F^p_\infty\,,
\end{equation}
in which exponential corrections of the form $e^{-2\pi y}$ are ignored. \newpage
\item \textbf{Approximation (2): $\mathrm{Sl}(2)$-orbit}\\
In the second approximation, we write
\begin{equation}
	F^p_{\mathrm{nil}}(t)\approx F^p_{\mathrm{Sl}(2)}(t) =\left[e^{x N}y^{-\frac{1}{2}N^0}\right]F_{\infty}^p=h_{\mathrm{Sl}(2)}(x,y)F^p_\infty\,, 
\end{equation}
in which additionally polynomial corrections of the form $1/y$ are ignored. In particular, this is a courser approximation than the nilpotent orbit approximation. 
\end{itemize}

\tcblower
\textbf{Note:}\\
The regime in which the nilpotent orbit approximation is a good approximation, i.e.~$e^{-2\pi y}\ll 1$, is referred to as the \textbf{asymptotic regime}. \\

\noindent Similarly, the regime in which the $\mathrm{Sl}(2)$-orbit approximation is a good approximation, i.e.~$1/y\ll 1$, is referred to as the \textbf{strict asymptotic regime}.
\end{subbox}

\subsection{$\mathrm{Sl}(2)$-orbit approximation of physical couplings}\label{ssec:couplings_one-parameter}
After this rather long discussion on purely mathematical matters, let us return to more physical questions. As we will illustrate, the $\mathrm{Sl}(2)$-orbit approximation is an incredibly useful tool to describe the leading behaviour of physical couplings that appear in the four-dimensional effective $\mathcal{N}=2$ compactifications of type IIB, as well as the scalar potential of four-dimesional effective $\mathcal{N}=1$ compactifications of F-theory. Let us explain this in some detail.

\subsubsection*{Approximating Hodge norms}
First, we introduce some additional notation. To each of the Hodge structures $F^p$, $F^p_{\mathrm{nil}}$, $F^p_{\mathrm{Sl}(2)}$ and $F_{\infty}$ we may associate respective Weil operators $C$, $C_{\mathrm{nil}}$, $C_{\mathrm{Sl}(2)}$ and $C_{\infty}$, which in turn induce various Hodge norms $||\cdot||$, $||\cdot||_{\mathrm{nil}}$, $||\cdot||_{\mathrm{Sl}(2)}$ and $||\cdot||_{\infty}$. Then one of the central results of asymptotic Hodge theory is that for any vector $v\in H_{\mathbb{C}}$ one has \cite{schmid,CKS}
\begin{equation}
||v||^2\sim ||v||^2_{\mathrm{Sl}(2)}\,.
\end{equation}
Here the symbol $\sim$ has a specific meaning, namely that the two norms above are \textit{mutually bounded}. Concretely, this means that, in a region where $y>\gamma$, there exist positive constants $\alpha,\beta$ such that
\begin{equation}
\alpha ||v||_{\mathrm{Sl}(2)}^2 \leq ||v||^2 \leq \beta ||v||_{\mathrm{Sl}(2)}^2\,.
\end{equation}
The constants $\alpha,\beta$ may depend on $\gamma$, but are independent of the choice of $v$. In particular, as long as one is concerned with the approximate scaling behaviour of the full Hodge norm $||\cdot||$, one actually infer this from the $\mathrm{Sl}(2)$-orbit approximation. Importantly, the latter is very straightforward to evaluate. Indeed, one has
\begin{align*}
||v||_{\mathrm{Sl}(2)}^2 &= \left(v, C_{\mathrm{Sl}(2)}\bar{v}\right)\\
&=\left(v, \left[e^{x N}y^{-\frac{1}{2}N^0}\right]C_{\infty}\left[e^{x N}y^{-\frac{1}{2}}\right]^{-1}\bar{v}\right)\\
&=\left(\left[e^{x N}y^{-\frac{1}{2}N^0}\right]^{-1}v,C_{\infty}\left[e^{x N}y^{-\frac{1}{2}N^0}\right]^{-1}\bar{v}\right)\\
&=\left(y^{\frac{1}{2}N^0}\hat{v}, C_{\infty} y^{\frac{1}{2}N^0}\bar{\hat{v}}\right)\\
&=||y^{\frac{1}{2}N^0} \hat{v}||^2_{\infty}
\end{align*}
where in the second step we have used the relation \eqref{eq:Sl2_approx_single}, in the third step we have used the invariance of the pairing $(gv,gw)=(v,w)$ for $g\in G_{\mathbb{R}}$, and in the final step we have simply employed the definition of $||\cdot||_{\infty}$. Furthermore, we have introduced the axion-dependent vector
\begin{equation}
\hat{v} := e^{-x N}v\,.
\end{equation}
Since the axions are always bounded, they do not play an important role in the leading order behaviour of the Hodge norm as $y\rightarrow\infty$, which is what we are considering at the moment. However, it should be noted that as soon as one includes higher-order corrections, one should take the axions into account.\\

\noindent As a final step, we employ the decomposition of $\hat{v}$ into eigenvectors of $N^0$ by writing
\begin{equation}
\hat{v} = \sum_{\ell} \hat{v}_{\ell}\,,\qquad N^0\hat{v}_{\ell} = \ell\hat{v}_{\ell}\,.
\end{equation}
Using the fact that this eigenspace decomposition is orthogonal with respect to the boundary Hodge norm $||\cdot||_{\infty}$\footnote{To see this, note that
\begin{equation}
\ell' \langle v_\ell, v'_{\ell'}\rangle = \langle v_\ell, N^0v'_{\ell'}\rangle =\left\langle \left(N^0\right)^\dagger v_\ell, v'_{\ell'}\right\rangle =\langle N^0 v_\ell, v'_{\ell'}\rangle =\ell \langle v_\ell, v'_{\ell'}\rangle\,,
\end{equation}
so that $\ell=\ell'$. Here we have used the fact that $\left(N^0\right)^\dagger=N^0$, which follows from the horizontality conditions \eqref{eq:horizontal_sl2_real}.}, one arrives at the following
\begin{subbox}{$\mathrm{Sl}(2)$-orbit approximation of the Hodge norm (one-parameter)}
\begin{equation}\label{eq:growth_theorem_single}	
	||v||^2\sim ||v||_{\mathrm{Sl}(2)}^2 = \sum_{\ell} y^{\ell} ||\hat{v}_{\ell}||^2_{\infty}\,.
\end{equation}
\end{subbox}
\noindent The growth theorem is a beautiful example of the importance of the underlying $\mathrm{sl}(2)$-symmetry of limiting mixed Hodge structures. Simply put, it implies that the Hodge norm of a vector $v$ grows/shrinks as $y\rightarrow\infty$ according to the value of its weights with respect to the $\mathrm{sl}(2)$ grading operator.

\subsubsection*{Approximating the K\"ahler potential}
As an application of the above results, let us compute the $\mathrm{Sl}(2)$-orbit approximation of the K\"ahler potential
\begin{equation}
K^{\mathrm{cs}} = -\log\left[i^{-D}\int_{Y_D}\Omega\wedge\bar{\Omega} \right]\,,
\end{equation}
where we have generalized the expression \eqref{eq:metric_cs} to Calabi--Yau $D$-folds, and we recall that $\Omega$ denotes the holomorphic $(D,0)$-form. There are two important observations to make regarding the latter. First, since $\Omega\in H^{D,0}$ and thus $\bar{\Omega}\in H^{0,D}$, the Weil operator acts on the latter as $C\bar{\Omega} = i^{-D}\bar{\Omega}$. Therefore, we can write the K\"ahler potential as a Hodge norm
\begin{equation}
K^{\mathrm{cs}} = -\log||\Omega||^2\,.
\end{equation}
In order to infer the scaling of $||\Omega||^2$, we need to know more about its maximal $\mathrm{sl}(2)$-weight. This can be related to the type of limiting mixed Hodge structure as follows. First, recall that to leading order in $1/y$, we may approximate
\begin{equation}
F^p(t) = e^{tN}\hat{F}_0^p+\cdots\,,
\end{equation}
with the dots denoting higher-order terms. Since $\Omega\in H^{D,0}$, we have that $\Omega\in F^D$. In particular, we may write
\begin{equation}
\Omega = e^{tN}\hat{a}_0+\cdots\,,
\end{equation}
for some $\hat{a}_0\in \hat{F}^D_{0}$. Importantly, since $\hat{F}^D_0$ is one-dimensional in the Calabi--Yau setting, we have that $\tilde{a}_0\in \hat{I}^{D,d}$, for some $d=0,\ldots, D$. Recall that the integer $d$ determines the principal type of the limiting mixed Hodge structure following the discussion around \eqref{Deligne_type_labeling}. In particular, $\hat{a}_0$ has $\mathrm{sl}(2)$-weight equal to $d$. One can then make the following computation
\begin{align*}
y^{\frac{1}{2}N^0}\hat{\Omega}
&=y^{\frac{1}{2}N^0} e^{iyN}\hat{a}_0+\cdots\\
&=\sum_{k=0}^d \frac{(i y)^k}{k!} y^{\frac{1}{2}N^0} N^k \hat{a}_0\\
&=\sum_{k=0}^d \frac{(i y)^k}{k!} y^{\frac{1}{2}(d-2k)}N^k \hat{a}_0\\
&=y^{\frac{d}{2}} e^{iN}\hat{a}_0\,,
\end{align*}
where in the third step we have used the fact that, since $\hat{a}_0\in I^{D,d}$, the combination $N^k \hat{a}_0$ has weight $d-2k$. As a result, we find
\begin{equation}
||\Omega||^2\sim ||y^{\frac{1}{2}N^0}\hat{\Omega} ||^2_{\infty} \sim y^{d}\,||\Omega_{\infty}||^2_{\infty}\,,\qquad \Omega_{\infty}:= e^{iN}\hat{a}_0\in F_{\infty}^D\,.
\end{equation}
To conclude, we find the following result:
\begin{subbox}{$\mathrm{Sl}(2)$-orbit approximation of the K\"ahler potential (one-parameter)}
\begin{equation}\label{eq:Kahler_potential_sl2_one-parameter}
K^{\mathrm{cs}} \sim -\log\left[y^{d}||\Omega_{\infty}||^2_{\infty}\right]\sim -d\log y\,.
\end{equation}
\end{subbox} 
\noindent It is important to stress that the above result is only meaningful when $d\neq 0$. In other words, for a type $\mathrm{I}_1$ singularity (recall that we are in the one-parameter setting, so $i^{2,2}=1$) the $\mathrm{Sl}(2)$-orbit approximation does not suffice to describe the leading behaviour of the K\"ahler potential. Instead, it is then necessary to include corrections coming from the nilpotent orbit expansion of the period map. \\

\noindent If $d\neq 0$, we can also infer the approximate form of the Weil--Petersson metric on the complex structure moduli space from the K\"ahler potential and find
\begin{equation}
ds^2_{\mathrm{WP}} = \left[d+\mathcal{O}(y^{-1})\right]\frac{dx^2+dy^2}{y^2}\,,
\end{equation}
which, to leading order in $y^{-1}$, corresponds to the usual Poincar\'e metric on the complex upper half-plane. \newpage

\section{The $\mathrm{Sl}(2)$-orbit theorem (2): multi-variable}\label{sec:SL2_orbit_multi_variable}
In this section we generalize the results of the previous section to an $m$-parameter variation of Hodge structure. On the one hand, the general case is considerably more involved. This can be traced back to the fact that in the presence of multiple moduli one finds different limiting mixed Hodge structures depending on the hierarchy between the saxions as one approaches a given boundary in the moduli space. On the other hand, the actual results follow quite naturally from the one-parameter case due to a beautiful inductive structure that was discovered by Cattani, Deligne, and Kaplan in \cite{CKS}. We hope to illustrate the central ideas behind the construction, but defer the reader to the original paper for some of the more delicate details. 

\subsection{A web of limiting mixed Hodge structures}
Let us consider an $m$-parameter variation of polarized Hodge structure $F^p$, which can be approximately described, in the regime where $y^1,\ldots, y^m\gg 1$, by a nilpotent orbit
\begin{equation}\label{eq:nilp_orbit_m-parameter}
F^p \approx F_{\mathrm{nil}}^p = \mathrm{exp}\left[i\sum_{i=1}^m y^i N_i\right]F^p_{(m)}\,,
\end{equation}
for some limiting filtration $F^p_{(m)}$. For simplicity we have set all the axions $x^i=0$, as they can always be recovered using the monodromy factor. For a first reading, the reader may find it helpful to set $m=2$ in the discussion below and follow along with the various steps of the construction using figure \ref{fig:bulk_to_boundary_part1}.\\

\noindent As a first thought, one might try to associate some limiting mixed Hodge structure to the limit where one of the saxions, say $y^1$, approaches the boundary. However, it is not at all obvious this might be done. Instead, it is more natural to choose a clever parametrization such that we can view the $m$-parameter nilpotent orbit \eqref{eq:nilp_orbit_m-parameter} as an effective one-parameter nilpotent orbit, such that we can apply the results of the previous section. The way to do this is to perform the following clever rewriting
\begin{equation}
\mathrm{exp}\left[i\sum_{i=1}^m y_i N_i\right]F_{(m)}^p = \mathrm{exp}\left[iy_m\left(\sum_{i=1}^{m-1}\frac{y_i}{y_m}N_i+N_{m}\right)\right]F_{(m)}^p\,.
\end{equation}
Indeed, for fixed values of the ratios $\frac{y_1}{y_m},\ldots, \frac{y_{m-1}}{y_m}$ the right-hand side can be viewed as a one-parameter nilpotent orbit, in which
\begin{equation}
y = y_m\,,\qquad N = \sum_{i=1}^{m-1}\frac{y_i}{y_m}N_i+N_{m}\,,\qquad F^p_0 = F^p_{(m)}\,,
\end{equation}
following the notation in the previous section. Importantly, we thus find a limiting mixed Hodge structure associated to the limit $y_m\rightarrow\infty$. Following the logic described in the one-parameter case, the corresponding weight filtration would be given by
\begin{equation}
W\left(\sum_{i=1}^{m-1}\frac{y_i}{y_m}N_i+N_{m}\right)\,.
\end{equation}
The fact that the ratios $\frac{y_1}{y_m},\ldots, \frac{y_{m-1}}{y_m}$ appear in the weight filtration may seem problematic. Fortunately, weight filtrations associated to sums of nilpotent endomorphisms have a very useful property which takes care of this. Namely, it turns out that
\begin{equation}
W\left(\lambda_1 N_1+\cdots +\lambda_i N_i\right) = W(N_1+\cdots +N_i)\,,
\end{equation}
for any $1\leq i\leq m$, as long as the coefficients $\lambda_i$ are all strictly positive, so $\lambda_i>0$ for all $i$, see Theorem 3.3 of \cite{Cattani1982}. Introducing the short-hand
\begin{equation}
N_{(i)}:=N_1+\cdots+N_i\,,
\end{equation}
and noting that all the ratios $\frac{y_1}{y_m},\ldots, \frac{y_{m-1}}{y_m}$ are indeed positive, we thus conclude the following.
\begin{subbox}{Limiting mixed Hodge structure associated to $y_m\rightarrow\infty$}
The pair $(W(N_{(m)}), F_{(m)})$ forms a limiting mixed Hodge structure associated to the limit $y_m\rightarrow\infty$.
\tcblower
\textbf{Note:}\\
We will typically denote the monodromy weight filtration $W(N_{(m)})$ as $W^{(m)}$.
\end{subbox}
\noindent More informally, one sees that actually the limit we are considering corresponds to sending $y_1,\ldots, y_m\rightarrow\infty$. In other words, the limiting mixed Hodge structure $(W^{(m)},F_{(m)})$ captures the behaviour of the variation of Hodge structure at the intersection of the singular divisors $z_1=\ldots=z_m=0$.
\\

\noindent Continuing as in the one-parameter case, we may equivalently describe the limiting mixed Hodge structure $(W^{(m)}, F_{(m)})$ using the associated Deligne splitting $I^{p,q}_{(m)}$, which we may as usual rotate to the $\mathrm{sl}(2)$-split version via the relation
\begin{equation}
\hat{F}^p_{(m)}:=e^{\zeta_m}e^{-i\delta_m}F^p_{(m)}\,,
\end{equation}
using the corresponding phase operator $\delta_m$ and the induced operator $\zeta_m$. Again, we may construct the $\mathrm{sl}(2)$-grading operator associated to the $\mathrm{sl}(2)$-split Deligne splitting $\hat{I}^{p,q}_{(m)}$ via
\begin{equation}
N^0_{(m)}v = (p+q-D)v\,,\qquad v\in\hat{I}^{p,q}_{(m)}\,.
\end{equation}
Then by exactly the same logic as in the one-parameter case, one can construct a one-parameter nilpotent orbit whose limiting mixed Hodge structure is exactly the rotated filtration $\hat{F}^p_{(m)}$.
\begin{subbox}{A first approximation at $y_m\gg 1$}
For $y_m>0$ and fixed values of $\frac{y_1}{y_m},\ldots, \frac{y_1}{y_m}>0$, the filtration 
\begin{equation*}
\mathrm{exp}\left[i\left(\sum_{i=1}^{m}y_iN_i\right)\right]\hat{F}^p_{(m)}=\mathrm{exp}\left[iy_m\left(\sum_{i=1}^{m-1}\frac{y_i}{y_m}N_i+N_{m}\right)\right]\hat{F}_{(m)}^p\,,
\end{equation*}
defines a one-parameter nilpotent orbit (in particular, it defines a variation of polarized Hodge structure). Furthermore, it agrees with the original nilpotent orbit $F^p_{\mathrm{nil}}$ to first order in the regime $y_m\gg 1$, so that
\begin{equation}\label{eq:Fnil_approx_ym}
F^p_{\mathrm{nil}}=g_m(y_m)\,\mathrm{exp}\left[i\left(\sum_{i=1}^{m}y_iN_i\right)\right]\hat{F}^p_{(m)}\,,
\end{equation}
where $g_{m}(y_m)$ enjoys an expansion
\begin{equation}
g_m(y_m) = 1+\frac{g_{m,1}}{y_m}+\cdots\,.
\end{equation}
\end{subbox}
\noindent As in the one-parameter case, the rotated filtration $\hat{F}^p_{(m)}$ satisfies the following crucial identity
\begin{equation}\label{eq:identity_multi-parameter}
\boxed{
\mathrm{exp}\left[i\left(\sum_{i=1}^{m}y_iN_i\right)\right]\hat{F}^p_{(m)} = y_m^{-\frac{1}{2}N^0_{(m)}}\mathrm{exp}\left[i\left(\sum_{i=1}^{m-1}\frac{y_i}{y_m}N_i\right)\right]F^p_{(m-1)}\,,}
\end{equation}
where on the right-hand side we have introduced the filtration 
\begin{equation}
F^p_{(m-1)} = e^{iN_m}\hat{F}^p_{(m)}\,.
\end{equation}
To make the comparison absolutely clear, note that this is completely identical to the discussion in the one-parameter case, where now the ``boundary Hodge structure'' is given by
\begin{equation}\label{eq:boundary_HS_ym}
\text{``$F^p_\infty$''} = \mathrm{exp}\left[i\left(\sum_{i=1}^{m-1}\frac{y_i}{y_m}N_i\right)\right]F^p_{(m-1)}\,.
\end{equation}
The important difference from the one-parameter case is that in the multi-parameter case this ``boundary Hodge structure'', instead of being a fixed Hodge structure, now corresponds to an $(m-1)$-parameter nilpotent orbit, since it still depends on the ratios $\frac{y_1}{y_m},\ldots,\frac{y_{m-1}}{y_m}$! As the reader may already anticipate, this is rather indicative of an inductive procedure. We return to this point in a moment. \\

\noindent Inserting the identity \eqref{eq:identity_multi-parameter} into \eqref{eq:Fnil_approx_ym}, we can thus rewrite the original nilpotent orbit as
\begin{equation}\label{eq:hnil_recursive}
\boxed{
\mathrm{exp}\left[i\sum_{i=1}^m y_i N_i\right]F_{(m)}^p =h_{\mathrm{nil},m}(y_m)\,\mathrm{exp}\left[i\left(\sum_{i=1}^{m-1}\frac{y_i}{y_m}N_i\right)\right] F^p_{(m-1)}\,,}
\end{equation}
where $h_{\mathrm{nil},m}$ is exactly the nilpotent orbit approximation of the period map which interpolates between ``boundary Hodge structure'' \eqref{eq:boundary_HS_ym} and the original nilpotent orbit. In particular, following the discussion in the one-parameter case, it enjoys an expansion of the form
\begin{equation}
h_m(y_m) =g_m(y_m)y_m^{-\frac{1}{2}N^0_{(m)}}= \left(1+\frac{g_{m,1}}{y_m}+\cdots \right)y_m^{-\frac{1}{2}N^0_{(m)}}\,,
\end{equation}
which, to leading order in $1/y_m$, can be approximated by the $\mathrm{Sl}(2)$-orbit $y_m^{-\frac{1}{2}N^0_{(m)}}$. \\

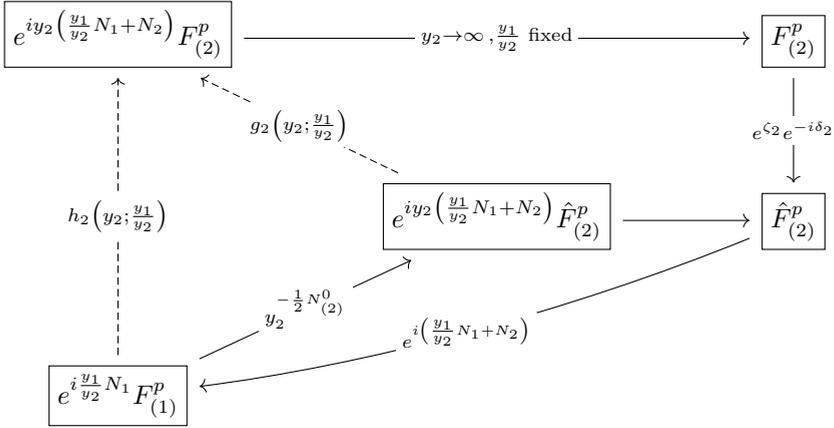
\begin{figure}[t!]
\centering
% https://q.uiver.app/#q=WzAsNSxbNCwwLCJcXGJveGVke0ZecF97KDIpfX0iXSxbNCwyLCJcXGJveGVke1xcaGF0e0Z9XnBfeygyKX19Il0sWzAsMCwiXFxib3hlZHtlXntpeV8yXFxsZWZ0KFxcZnJhY3t5XzF9e3lfMn1OXzErTl8yXFxyaWdodCl9Rl5wX3soMil9fSJdLFswLDQsIlxcYm94ZWR7ZV57aVxcZnJhY3t5XzF9e3lfMn1OXzF9Rl5wX3soMSl9fSJdLFsyLDIsIlxcYm94ZWR7ZV57aXlfMlxcbGVmdChcXGZyYWN7eV8xfXt5XzJ9Tl8xK05fMlxccmlnaHQpfVxcaGF0e0Z9XnBfeygyKX19Il0sWzAsMSwiZV57XFx6ZXRhXzJ9ZV57LWlcXGRlbHRhXzJ9IiwxXSxbMiwwLCJ5XzJcXHJpZ2h0YXJyb3dcXGluZnR5XFwsLFxcZnJhY3t5XzF9e3lfMn1cXHRleHR7IGZpeGVkfSIsMV0sWzEsMywiZV57aVxcbGVmdChcXGZyYWN7eV8xfXt5XzJ9Tl8xK05fMlxccmlnaHQpfSIsMSx7ImN1cnZlIjotMn1dLFs0LDFdLFs0LDIsImdfMlxcbGVmdCh5XzI7XFxmcmFje3lfMX17eV8yfVxccmlnaHQpIiwxLHsic3R5bGUiOnsiYm9keSI6eyJuYW1lIjoiZGFzaGVkIn19fV0sWzMsMiwiaF8yXFxsZWZ0KHlfMjtcXGZyYWN7eV8xfXt5XzJ9XFxyaWdodCkiLDEseyJzdHlsZSI6eyJib2R5Ijp7Im5hbWUiOiJkYXNoZWQifX19XSxbMyw0LCJ5XzJeey1cXGZyYWN7MX17Mn1OXjBfeygyKX19IiwxXV0=
\[\begin{tikzcd}
{\boxed{e^{iy_2\left(\frac{y_1}{y_2}N_1+N_2\right)}F^p_{(2)}}} &&&& {\boxed{F^p_{(2)}}} \\
\\
&& {\boxed{e^{iy_2\left(\frac{y_1}{y_2}N_1+N_2\right)}\hat{F}^p_{(2)}}} && {\boxed{\hat{F}^p_{(2)}}} \\
\\
{\boxed{e^{i\frac{y_1}{y_2}N_1}F^p_{(1)}}}
\arrow["{e^{\zeta_2}e^{-i\delta_2}}"{description}, from=1-5, to=3-5]
\arrow["{y_2\rightarrow\infty\,,\frac{y_1}{y_2}\text{ fixed}}"{description}, from=1-1, to=1-5]
\arrow["{e^{i\left(\frac{y_1}{y_2}N_1+N_2\right)}}"{description}, curve={height=-12pt}, from=3-5, to=5-1]
\arrow[from=3-3, to=3-5]
\arrow["{g_2\left(y_2;\frac{y_1}{y_2}\right)}"{description}, dashed, from=3-3, to=1-1]
\arrow["{h_2\left(y_2;\frac{y_1}{y_2}\right)}"{description}, dashed, from=5-1, to=1-1]
\arrow["{y_2^{-\frac{1}{2}N^0_{(2)}}}"{description}, from=5-1, to=3-3]
\end{tikzcd}\]
\caption{A flowchart depicting the first step of the inductive procedure in the case where $m=2$.}
\label{fig:bulk_to_boundary_part1}
\end{figure}

\noindent Let us pause to reflect on what we have just accomplished, as it is the central point of the multi-parameter discussion. First, we started off with an $m$-parameter nilpotent orbit as given in \eqref{eq:nilp_orbit_m-parameter} and found a clever way to extract a limiting mixed Hodge structure from the limit $y_m\rightarrow\infty$. Then, using the general properties of this limiting mixed Hodge structure, which we discussed at length in the one-parameter version, we arrived at the expression \eqref{eq:hnil_recursive}. Rather strikingly, the right-hand side of this expression takes the form of an $(m-1)$-parameter nilpotent orbit, namely
\begin{equation}
\mathrm{exp}\left[i\left(\sum_{i=1}^{m-1}\frac{y_i}{y_m}N_i\right)\right] F^p_{(m-1)}\,,
\end{equation}
up to the additional factor of $h_m(y_m)$. This suggests an iterative process in which we inductively apply the above procedure to obtain additional limiting mixed Hodge structures. In figure \ref{fig:bulk_to_boundary_part1} we give an overview of the relations between the various spaces in this first part of the inductive procedure.

\subsubsection*{More limiting mixed Hodge structures}
Let us briefly describe the next step in the inductive analysis, so that the general pattern will become clear. We perform a similar trick as before, now applied to the $(m-1)$-parameter nilpotent orbit, by writing
\begin{equation}
\mathrm{exp}\left[i\left(\sum_{i=1}^{m-1}\frac{y_i}{y_m}N_i\right)\right] F^p_{(m-1)} = \mathrm{exp}\left[i \frac{y_{m-1}}{y_m}\left( \sum_{i=1}^{m-2}\frac{y_i}{y_{m-1}}N_i+N_{m-1}\right)\right]F^p_{(m-1)}\,.
\end{equation}
For a fixed value of the ratios $\frac{y_1}{y_{m-1}},\ldots, \frac{y_{m-2}}{y_{m-1}}$, the right-hand side can again be interpreted as a one-parameter nilpotent orbit, in which
\begin{equation}
y = \frac{y_{m-1}}{y_m}\,,\qquad N = \sum_{i=1}^{m-2}\frac{y_i}{y_{m-1}}+N_{m-1}\,,\qquad F^p_{0} = F^p_{(m-1)}\,.
\end{equation}
Repeating the above analysis, this gives rise to a limiting mixed Hodge structure $(W^{(m-1)},F_{(m-1)})$ associated to the limit where  $\frac{y_{m-1}}{y_m}\rightarrow\infty$. Note, in particular, that this means we are considering a limit within the regime where $y_{m-1}\gg y_m$. Intuitively, one can think of this limiting mixed Hodge structure as being associated to the intersection of the divisors $z_1=\ldots=z_{m-1}=0$.  \\

\noindent At this point, the general pattern should be quite apparent, and thus we summarize below the general result. 

\begin{subbox}{Multiple limiting mixed Hodge structures}
Consider an $m$-parameter nilpotent orbit
\begin{equation}
F^p_{\mathrm{nil}}(t) = \mathrm{exp}\left[\sum_{i=1}^m t_i N_i\right]F^p_{(m)}\,.
\end{equation}
Then one can inductively associate to this nilpotent orbit a collection of limiting mixed Hodge structures $(W^{(k)}, F_{(k)})$, for $k=m,\ldots, 1$, each corresponding to the limit where
\begin{equation*}
\frac{y_{k}}{y_{k+1}}\rightarrow\infty\,,\qquad y_{m+1}:=1\,.
\end{equation*} 
At each step, the weight filtration is given by
\begin{equation}
W^{(k)} = W(N_{(k)}) = W(N_1+\cdots +N_k)\,,
\end{equation}
and the limiting filtration is inductively defined by
\begin{equation}
F^p_{(k-1)} = e^{iN_k}\hat{F}^p_{(k)}\,,\qquad \hat{F}^p_{(k)} = e^{\zeta_k}e^{-i\delta_k}F^p_{(k)}\,.
\end{equation}
\tcblower
\textbf{Note:}\\
In the final step $k=1$, one arrives at the filtration
\begin{equation}\label{eq:boundary-HS_multi}
F^p_{\infty} := F^p_{(0)} \,,
\end{equation}
which is in fact a Hodge filtration, as in the one-parameter case. The resulting Hodge structure is again referred to as the \textbf{boundary Hodge structure}.
\end{subbox}

\subsubsection*{Growth sectors, enhancement chains and even more limiting mixed Hodge structures}
It is important to stress that, in the preceding inductive procedure, we have made a particular choice in the ordering of the coordinates. For example, in the very first step of the argument, we first chose to view the $m$-parameter nilpotent orbit as a one-parameter nilpotent orbit in the variable $y_m$, while keeping the ratios $\frac{y_1}{y_m},\ldots,\frac{y_{m-1}}{y_m}$ fixed. In the next step, we chose to view the $(m-1)$-parameter nilpotent orbit as a one-parameter nilpotent orbit in the variable $y_{m-1}$, and so on. Because of this, the resulting expansions for the period maps $h_{\mathrm{nil},i}$ become power series in inverse powers of $\frac{y_i}{y_{i+1}}$, which will only be sensible in the region where $\frac{y_i}{y_{i+1}}>1$. To formalize this, we introduce the notion of a \textbf{growth sector}
\begin{equation}\label{eq:growth_sector}
R_{12\cdots m} = \left\{t_i = x_i+i y_i\,|\, y_m, \frac{y_{m-1}}{y_m},\ldots, \frac{y_1}{y_2}>1\,,|x_i|<1\right\}\,.
\end{equation}
Note that this is the same as working in a regime where $y_1>y_2>\ldots>y_m$. If one additionally chooses to work within the courser $\mathrm{Sl}(2)$-orbit approximation by dropping all sub-leading polynomial corrections, one can only expect this to be a good approximation for $\frac{y_i}{y_{i+1}}\gg 1$, and hence one may need to work with a refinement of the growth sector. \\

\noindent As a result of the above discussion, the recursive construction gives rise to a so-called \textbf{enhancement chain} of limiting mixed Hodge structures within each growth sector, which is typically denoted by a sequence
\begin{equation}
\text{Type $\mathrm{A}_{1}$}\xrightarrow{y^1\rightarrow\infty}\cdots \xrightarrow{y^m\rightarrow\infty} \text{Type $\mathrm{A}_{(m)}$}\,,
\end{equation}
where the type $\mathrm{A}_{(i)}$ singularity denotes the limiting mixed Hodge structure associated to the limit where $y^1,\ldots, y^i\rightarrow\infty$ within the hierarchy $y^1>\ldots>y^i$. It turns out that there are a number of restrictions on the allowed enhancement chains, whose origin can be roughly understood as follows. On the one hand, one can associate a limiting mixed Hodge structure $I^{p,q}_{(2)}$ to the limit $y^1,y^2\rightarrow\infty$ as we have described above. On the other hand, one can also view the enhancement of a type $\mathrm{A}_{(1)}$ singularity to a type $\mathrm{A}_{(2)}$ singularity in terms of the primitive subspaces of $I^{p,q}_{(1)}$. Each of these defines a Hodge structure which itself degenerates in the limit $y^2\rightarrow\infty$, thus giving rise to a bunch of smaller limiting mixed Hodge structures. It is then a non-trivial consistency condition that these smaller limiting mixed Hodge structures (together with their descendants) can be combined into $I^{p,q}_{(2)}$. This may not always be possible, essentially due to non-trivial constraints coming from the polarization conditions. An example of the kind of restriction one finds on the enhancement chain is that the singularity type can never decrease across the chain, so that a type $\mathrm{II}$ singularity can never enhance to a type $\mathrm{I}$ singularity. Effectively, this is saying that the singularity can only become ``worse'' across the chain. Interestingly, it turns out that the enhancement rules are not transitive. For example, while the enhancement $\mathrm{II}_0\rightarrow\mathrm{II}_1\rightarrow\mathrm{IV}_2$ is allowed, the enhancement $\mathrm{II}_0\rightarrow\mathrm{IV}_2$ is not \cite{Kerr2017}. For a more detailed discussion on the various rules, we refer the reader to \cite{Grimm:2018cpv}, see also \cite{vandeHeisteeg:2022gsp}. \\

\noindent An important takeaway of the above discussion is that one can also choose to work in different growth sectors, which amounts to performing the same inductive argument as we gave before, but with a different ordering of the coordinates. Crucially, this is necessary if one wants to understand the complete singularity structure. As a simple example, if one considers a 2-parameter variation of Hodge structure and takes the two possible orderings of the coordinates into account, this gives rise to \textit{three} limiting mixed Hodge structures, corresponding to the two divisors $z_1=0$ and $z_2=0$, and their intersection $z_1=z_2=0$, see also figure \ref{fig:divisors} for a pictorial description. The resulting set of limiting mixed Hodge structures is typically written as a so-called 2-cube $\langle A_1|A_{(2)}|A_2\rangle$. Alternatively, one could also describe this in terms of two different enhancement chains, namely 
\begin{equation}
\text{Type $\mathrm{I}_0$}\xrightarrow{y^1\rightarrow\infty}	\text{Type $\mathrm{A}_{1}$}\xrightarrow{y^2\rightarrow\infty}\text{Type $\mathrm{A}_{(2)}$}\,,
\end{equation}
or 
\begin{equation}
\text{Type $\mathrm{I}_0$}\xrightarrow{y^2\rightarrow\infty}	\text{Type $\mathrm{A}_{2}$}\xrightarrow{y^1\rightarrow\infty}\text{Type $\mathrm{A}_{(2)}$}\,.
\end{equation}

\begin{figure}
\centering
\begin{tikzpicture}[scale=0.7]

\draw[very thick](2,5) to[out=20,in=160] (10,5);
\draw[very thick](4,3) to[out=80,in=230] (7,8);

\node at (5.15,5.75)[circle,fill,inner sep=1.5,blue]{};

\draw[very thick, ->, red] (9.3,5.4) arc (0:300:0.5cm and 0.7cm);

\tkzLabelPoint[left](2,5){\large $z_1=0$};
\tkzLabelPoint[below](4,3){\large $z_2=0$};

\draw[thick,dashed] (5.4,5) ellipse[x radius = 7cm, y radius = 3.5cm, start angle = 30,
end angle = 150];

\fill[gray,opacity=0.1] (5.4,5) ellipse[x radius = 7cm, y radius = 3.5cm, start angle = 30,
end angle = 150];

\tkzLabelPoint[right](10,5){\large $\Delta_1$};
\tkzLabelPoint[right](7,8){\large $\Delta_2$};
\tkzLabelPoint[above left](5.15,5.75){\large \textcolor{blue}{$\Delta_{12}$}};
\tkzLabelPoint[above](8.8,6.1){\large \textcolor{red}{$T_1$}};
\end{tikzpicture}
\caption{A local patch in $\mathcal{M}_{\mathrm{cs}}$ containing two divisors $\Delta_1$ and $\Delta_2$ which intersect at a point $\Delta_{12}$. In red, a monodromy transformation $T_1$ around $\Delta_1$ is depicted. }
\label{fig:divisors}
\end{figure}
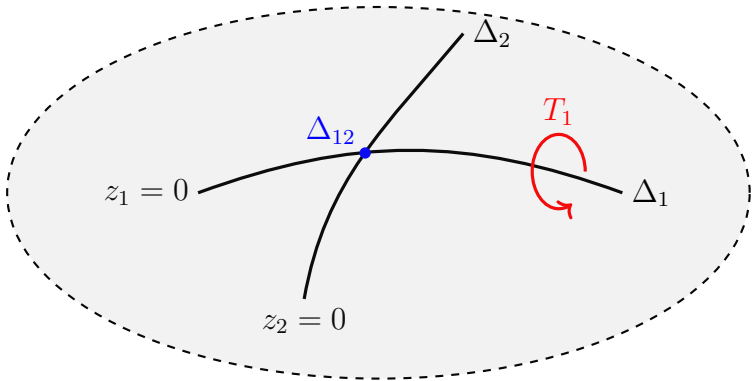
\newpage

\subsection{The $\mathrm{Sl}(2)$-orbit approximation}
By inductive application of the relation \eqref{eq:hnil_recursive}, we find that the original nilpotent orbit can be written succinctly as
\begin{equation}
F^p_{\mathrm{nil}} = h_{\mathrm{nil}}(y_1,\ldots, y_m) F^p_{\infty}\,,
\end{equation}
where
\begin{equation}
h_{\mathrm{nil}} = \prod_{i=m}^1 h_{\mathrm{nil},i}=\prod_{i=m}^1 g_{\mathrm{nil},i}\cdot \left(\frac{y_i}{y_{i+1}}\right)^{-\frac{1}{2}N^0_{(i)}}\,,
\end{equation}
and each individual $g_{\mathrm{nil},i}$ enjoys an expansion
\begin{equation}
g_i = 1+ g_{i,1}\left(\frac{y_i}{y_{i+1}}\right)^{-1}+\cdots = \sum_{k_i=0}^\infty g_{i,k_i}\left(\frac{y_i}{y_{i+1}}\right)^{-k_i} \,.
\end{equation}
In other words, we have managed to show that the full nilpotent orbit can be recovered from the boundary Hodge structure $F^p_\infty$ through a recursive application of maps $h_{\mathrm{nil},i}$, which can be computed explicitly (as will be explained in the next chapter). Furthermore, in a regime where all the ratios satisfy $\frac{y_i}{y_{i+1}}\gg 1$, one may neglect the subleading corrections in the $h_{\mathrm{nil},i}$, which gives rise to the $\mathrm{Sl}(2)$-orbit approximation in the $m$-parameter case.

\begin{subbox}{$\mathrm{Sl}(2)$-orbit approximation (multi-variable case)}
To leading order in the inverse ratios $\left(\frac{y_i}{y_{i+1}}\right)^{-1}$, the \textbf{$\mathrm{Sl}(2)$-orbit approximation}
\begin{equation}
F^p_{\mathrm{Sl}(2)}:=\left[\prod_{i=1}^m \left(\frac{y_i}{y_{i+1}}\right)^{-\frac{1}{2}N^0_{(i)}}\right] F^p_{\infty}\,,
\end{equation}
agrees with the original nilpotent orbit $F^p_{\mathrm{nil}}$. Note that we have set the axions $x_i=0$ for simplicity.  
\tcblower
Correspondingly, we will denote by 
\begin{equation}\label{eq:hsl2_sum}
h_{\mathrm{Sl}(2)}:=\prod_{i=1}^m \left(\frac{y_i}{y_{i+1}}\right)^{-\frac{1}{2}N^0_{(i)}}\,,
\end{equation}
the $\mathrm{Sl}(2)$-orbit approximation of the period map, such that
\begin{equation}
F^p_{\mathrm{Sl}(2)} = h_{\mathrm{Sl}(2)}\,F^p_\infty\,.
\end{equation}
\end{subbox}

\subsection{Commuting $\mathrm{sl}(2)$-triples and the boundary data}

\subsubsection*{Commuting $\mathrm{sl}(2)$-triples}
So far we have mostly been concerned with the grading operators $N^0_{(i)}$ that can be associated to the limiting mixed Hodge structures $(W^{(i)},\hat{F}_{(i)})$. From the general discussion in section \ref{sec:MHS}, we know, however, that these grading operators, together with the relevant nilpotent endomorphism, can be completed into real $\mathrm{sl}(2)$-triples. Let us now discuss in more detail how this happens in the $m$-parameter case. \\

\noindent To simplify the notation, let us first consider the case where $m=2$, corresponding to the 2-parameter case. For concreteness, let us recall that the complete nilpotent orbit can be written as
\begin{equation}
F^p_{\mathrm{nil}} = h_{\mathrm{nil},2}\cdot h_{\mathrm{nil},1}F^p_\infty\,.
\end{equation}
Let us now first consider the limiting mixed Hodge structure $(W^{(1)}, \hat{F}_{(1)})$, for which the nilpotent endomorphism is simply the first log-monodromy matrix
\begin{equation}
N = N_1\,.
\end{equation}
Naturally, together with the grading operator $N^0_{(1)}$, this can be completed to a real $\mathrm{sl}(2)$-triple
\begin{equation}\label{eq:sl2-triple_hnil1}
\{N^+_{(1)}, N^0_{(1)}, N^-_{(1)}\}\,,\qquad N^-_{(1)}:=N_1\,.
\end{equation}
Furthermore, the resulting $\mathrm{sl}(2)$-triple is \textit{horizontal} with respect to the boundary Hodge structure $F_\infty^p$. This is exactly as in the one-parameter case. Things are slightly different, however, when considering the limiting mixed Hodge structure $(W^{(2)},\hat{F}_{(2)})$, for which the nilpotent endomorphism is given by
\begin{equation}\label{eq:lowering_hnil1}
N = \frac{y_1}{y_2}N_1+N_2\,.
\end{equation}
It is certainly true that this operator acts as a lowering operator with respect to $N^0_{(2)}$, and in principle this can be completed uniquely with a raising operator. However, the resulting $\mathrm{sl}(2)$-triple would be horizontal with respect to the Hodge structure
\begin{equation}
h_{\mathrm{nil},1}F^p_\infty\,,
\end{equation}
as opposed to $F^p_\infty$. This is precisely because the above Hodge structure is what plays the role of the ``boundary Hodge structure'' in this step of the recursive argument, as explained earlier. In order to construct a $\mathrm{sl}(2)$-triple which is horizontal with respect to $F^p_\infty$, one just needs to ``transport'' the triple induced by the lowering operator \eqref{eq:lowering_hnil1} using the (inverse) adjoint action of $h_{\mathrm{nil},1}$. In other words, one should instead consider
\begin{equation}
h_{\mathrm{nil},1}^{-1}\left(\frac{y_1}{y_2}N_1+N_2\right) h_{\mathrm{nil},1}\,,\qquad \text{and}\qquad h_{\mathrm{nil},1}^{-1}N^0_{(2)} h_{\mathrm{nil},1}\,.
\end{equation}
Fortunately, one can actually compute what these objects should be, using the horizontality properties of the period map $h_{\mathrm{nil},1}$. The computation is rather involved and is presented in Lemma 4.37 of \cite{CKS}. Here we simply state the result:
\begin{align}
h_{\mathrm{nil},1}^{-1}\left(\frac{y_1}{y_2}N_1+N_2\right) h_{\mathrm{nil},1} &= N_1 + \left(N_{2}\right)_{0}\,,\\
h_{\mathrm{nil},1}^{-1}N^0_{(2)} h_{\mathrm{nil},1}&= N^0_{(2)}
\end{align}
where $\left(N_{2}\right)_{0}$ is the component of $N_2$ that commutes with the grading operator $N^0_{(1)}$. In particular, note that $N^0_{(2)}$ commutes with $h_{\mathrm{nil},1}$, so that it is unchanged under this ``transportation''. In summary, we arrive at the following result.
\begin{subbox}{Horizontal sl(2)-triples $(m=2)$ }
The operators
\begin{equation*}
N^0_{(2)}\,,\qquad\text{and}\qquad N^-_{(2)}:=N_1 + \left(N_{2}\right)_{0}\,,
\end{equation*}
can be completed into a real $\mathrm{sl}(2)$-triple, which is furthermore horizontal with respect to the boundary Hodge structure $F^p_\infty$. In particular, the raising operator can either be computed by solving the sl(2) commutation relations, or it can alternatively be computed via the relation
\begin{equation}
N^+_{(2)} = \left(N^-_{(2)}\right)^{\dagger}\,,
\end{equation}
where the dagger is taken with respect to the boundary Hodge inner product $\langle\cdot,\cdot\rangle_\infty$, c.f.~\eqref{eq:def_adjoint}.
\end{subbox}
\noindent Together with the triple \eqref{eq:sl2-triple_hnil1} associated to the first limiting mixed Hodge structure, we thus find two separate real sl(2)-triples which are both horizontal with respect to the boundary Hodge structure. Having illustrated the logic in the two-parameter case, let us simply state how this generalizes to the $m$-parameter case. 

\begin{subbox}{Commuting horizontal $\mathrm{sl}(2)$-triples}
For each $i=1,\ldots, m$, define 
\begin{itemize}
\item \textbf{Lowering operators}:
\begin{equation}
	N^-_{(i)}:=N_1+ \left(N_2\right)_{0}+\cdots +\left(N_i\right)_{0}\,,
\end{equation}
where $\left(N_i\right)_{0}$ is the component of $N_i$ that commutes with the first $(i-1)$ grading operators $N^0_{(1)},\ldots, N^0_{(i-1)}$.
\item \textbf{Raising operators:}
\begin{equation}
	N^+_{(i)}:=\left(N^-_{(i)}\right)^\dagger\,,
\end{equation} 
where again the dagger is taken with respect to the boundary Hodge inner product. 
\end{itemize}
Then $N^-_{(i)}$ agrees with the ``transportation'' of the relevant nilpotent endormorphism to $F^p_\infty$, i.e.
\begin{equation}
\mathrm{Ad}\left[h_{\mathrm{nil},i-1}\cdots h_{\mathrm{nil},1}\right]^{-1}\left(\sum_{k=1}^{i}\frac{y_{k}}{y_{i}}N_k\right) = N^-_{(i)}\,,
\end{equation}
and furthermore	the triples
\begin{equation}
\{N^+_{(i)}, N^0_{(i)}, N^-_{(i)}\}\,,
\end{equation}
all define real $\mathrm{sl}(2)$-triples, which are horizontal with respect to the boundary Hodge structure $F^p_\infty$.
\tcblower 
Furthermore, putting
\begin{equation}
N^\bullet_{i}:=N^\bullet_{(i)} - N^\bullet_{(i-1)}\,,\qquad \bullet=+,0,-\,,
\end{equation}
the triples 
\begin{equation}
\{N^+_{i}, N^0_{i}, N^-_{i}\}\,,
\end{equation}
form a set of \textit{commuting} $\mathrm{sl}(2)$-triples, which are again horizontal with respect to the boundary Hodge structure $F^p_\infty$.
\end{subbox}

\subsubsection*{Boundary data}
Over the course of this section, we have encountered a number of objects which are naturally associated to a given boundary of the moduli space (within a chosen growth sector). As in the one-parameter case, we will refer to these objects as the \textbf{boundary data}. As will be explained in chapter \ref{chap:asymp_Hodge_II}, this data in fact suffices to recover the full nilpotent orbit approximation near the boundary in question. This equivalence between nilpotent orbits and their associated boundary structure is the most striking consequence of the Sl(2)-orbit theorem of Cattani, Kaplan, and Schmid. Thus, in preparation for the next chapter, let us briefly summarize the essential properties of the boundary data. 

\begin{subbox}{Boundary data}
	\begin{itemize}
		\item \textbf{Boundary charge operator: $Q_\infty$}\\
		The eigenspace decomposition of the boundary charge operator encodes the boundary Hodge structure $F^p_\infty$ in terms of which the Sl(2)-orbit approximation $F^p_{\mathrm{Sl}(2)}$ and nilpotent orbit approximation $F^p_{\mathrm{nil}}$ can be reconstructed.
		\item \textbf{Commuting horizontal real sl(2)-triples: $\{N^+_{i}, N^0_{i}, N^-_{i}\}$}\\
		For $i,j=1,\ldots, m$ these real sl(2)-triples satisfy the commutation relations
		\begin{equation}
			[N^0_i, N^{\pm}_j] = \pm 2 N^\pm_i \delta_{ij}\,,\qquad [N^+_i, N^-_j] = N^0 \delta_{ij}\,,
		\end{equation}
		as well as the horizontality conditions
		\begin{equation}
			[Q_\infty, N_j^\pm] = -\frac{i}{2}N^0_j\,,\qquad [Q_\infty, N^0_j] = i\left(N^+_j+N^-_j\right)\,.
		\end{equation}
		\item \textbf{Phase operators: $\delta_i$}\\
		The phase operators encode how the various limiting mixed Hodge structures $(W^{(i)}, F_{(i)})$ are rotated into their $\mathbb{R}$-split and sl(2)-split counterparts. As will be explained in chapter \ref{chap:asymp_Hodge_II}, they effectively encode the corrections to the Sl(2)-orbit approximation which are required in order to recover the full nilpotent orbit approximation.
	\end{itemize}
\end{subbox}
\newpage

\subsection{$\mathrm{Sl}(2)$-orbit approximation of physical couplings}

In this section we perform a similar analysis to the one in section \ref{ssec:couplings_one-parameter}, and compute the $\mathrm{Sl}(2)$-orbit approximation of the Hodge norm and the K\"ahler potential, but now in the general $m$-parameter case.

\subsubsection*{Approximating the Hodge norm}

As in the one-parameter case, the full Hodge norm and the $\mathrm{Sl}(2)$ Hodge norm are mutually bounded \cite{schmid,CKS}
\begin{equation}
||v||^2\sim ||v||^2_{\mathrm{Sl}(2)}\,,
\end{equation}
within the chosen growth sector. To compute the $\mathrm{Sl}(2)$-approximation of the Hodge norm, we perform a similar computation as in the one-variable case by writing
\begin{align*}
	||v||_{\mathrm{Sl}(2)}^2 &= \left(v, C_{\mathrm{Sl}(2)}\bar{v}\right)\\
	&\stackrel{\text{(a)}}{=}\left(v, h_{\mathrm{Sl}(2)}C_{\infty}h_{\mathrm{Sl}(2)}^{-1}\bar{v}\right)\\
	&\stackrel{\text{(b)}}{=}\left(h_{\mathrm{Sl}(2)}^{-1}v,C_{\infty}h_{\mathrm{Sl}(2)}^{-1}\bar{v}\right)\\
	&\stackrel{\text{(c)}}{=}||h_{\mathrm{Sl}(2)}^{-1}v||^2_{\infty}
\end{align*}
where step (a) we have used the definition of $C_{\mathrm{Sl}(2)}$, in the step (b) we have used the invariance of the pairing $(gv,gw)=(v,w)$ for $g\in G_{\mathbb{R}}$, and in the step (c) we have simply employed the definition of $||\cdot||_{\infty}$. Next, we perform the decomposition of a given vector $v$ in terms of eigenvectors with respect to the commuting grading operators $N^0_{(i)}$, i.e.
\begin{equation}\label{eq:sl2_decomp_vector}
v = \sum_{\ell} v_{\ell}\,,\qquad \ell = (\ell_1,\ldots, \ell_m)\,,\qquad [N^0_{(i)},v_\ell] = \ell_i\,v_\ell\,.
\end{equation}
Recalling the general form of the $\mathrm{Sl}(2)$-orbit approximation of the period map \eqref{eq:hsl2_sum} and reintroducing the axion-dependence, one finds
\begin{equation}
h^{-1}_{\mathrm{Sl}(2)}v_\ell = \left[\prod_{i=1}^m \left(\frac{y_i}{y_{i+1}}\right)^{\frac{1}{2}\ell_i}\right] \hat{v}_\ell\,,\qquad \hat{v} := \mathrm{exp}\left[-\sum_{i=1}^m x^i N_i\right]\cdot v\,,
\end{equation}
Finally, using the fact that the eigenspace decomposition is orthogonal with respect to the boundary Hodge norm $||\cdot||_{\infty}$, one arrives at the following result.
\begin{subbox}{$\mathrm{Sl}(2)$-orbit approximation of the Hodge norm ($m$-parameter)}
\begin{equation}\label{eq:growth_theorem_multi}	
||v||^2\sim ||v||_{\mathrm{Sl}(2)}^2 = \sum_{\ell} \left[\prod_{i=1}^m \left(\frac{y_i}{y_{i+1}}\right)^{\ell_i}\right] ||\hat{v}_\ell||^2_\infty\,.
\end{equation}
\end{subbox}
\noindent Again, we see that the underlying $\mathrm{sl}(2)$-symmetry of limiting mixed Hodge structures is crucial in determining the asymptotic behaviour of the Hodge norm. There is, however, an important aspect of the general $m$-parameter case which did not play a role in the one-parameter case. Namely, it is not at all obvious from \eqref{eq:growth_theorem_multi} which component $v_{\ell}$ actually determines the leading growth of the Hodge norm. This is because, even though one might be working within a growth sector where $y^1>y^2$ (putting $m=2$ for the moment), it is not necessarily the case that also $\frac{y_1}{y_2}>y_2$, for example. Hence it is not clear whether the component $v_{(1,0)}$ or the component $v_{(0,1)}$ determines the leading growth. This will depend on the particular subregion of the growth sector one is considering. This subtlety will play an important role in chapter \ref{chap:finiteness}. 

\subsubsection*{Approximating the K\"ahler potential}
Recall that for Calabi--Yau $D$-folds the K\"ahler potential can be written as
\begin{equation}
e^{-K^{\mathrm{cs}}} =  ||\Omega||^2 \sim  ||\Omega||_{\mathrm{Sl}(2)}^2=||h^{-1}_{\mathrm{Sl}(2)}\Omega||_\infty^2\,.
\end{equation}
To compute the right-hand side, we recall that within the $\mathrm{Sl}(2)$-orbit approximation one can write
\begin{equation}
\Omega= \mathrm{exp}\left[\sum_{i=1}^m x^i N_i \right]\cdot\mathrm{exp}\left[i\sum_{i=1}^m y^i N^-_i \right]\hat{a}_0+\cdots\,,
\end{equation}
for some $\hat{a}_0\in \hat{F}_{(m)}$, with the dots denoting subleading corrections. Recall that the position of $\hat{a}_0$ along the various Deligne splittings $\hat{I}^{p,q}_{(i)}$ dictates the principal types the limiting mixed Hodge structures appearing in the enhancement chain. To be explicit, we denote by $d_i$ the highest weight of $\hat{a}_0$ with respect to $N^0_{(i)}$, or equivalently $d_i-d_{i-1}$ denotes the highest weight with respect to $N^0_i$, where $d_0\equiv 0$. Then one can do the following computation.
\begin{align*}
h^{-1}_{\mathrm{Sl}(2)}\Omega &= h^{-1}_{\mathrm{Sl}(2)}\mathrm{exp}\left[\sum_{i=1}^m x^i N_i \right]\cdot\mathrm{exp}\left[i\sum_{i=1}^m y^i N^-_i \right]\hat{F}_{(m)}+\cdots\,,\\
&=\prod_{i=1}^m y_i^{\frac{1}{2}N^0_i}\sum_{k=0}^\infty \frac{(i y_i)^k}{k!} \left(N^-_i\right)^k \hat{a}_0\\
&\stackrel{\text{(a)}}{=}\prod_{i=1}^m y_i^{\frac{1}{2}(d_i-d_{i-1})}\sum_{k=0}^\infty \frac{(i)^k}{k!} \left(N^-_i\right)^k \hat{a}_0\\
&=\left[\prod_{i=1}^m y_i^{\frac{1}{2}(d_i-d_{i-1})}\right] e^{i(N_1^-+\cdots +N_m^-)}\hat{a}_0\\
&\stackrel{\text{(b)}}{=}\left[\prod_{i=1}^m y_i^{\frac{1}{2}(d_i-d_{i-1})}\right] \Omega_\infty\,,
\end{align*}
where in the step (a) we have used the fact that $\left(N_i^-\right)^k\hat{a}_0$ has highest weight $d_i-d_{i-1}-2k$ with respect to $N^0_i$ and in step (b) we have used the relation
\begin{equation}
e^{iN_{(m)}^-}\hat{F}_{(m)} = F_\infty\,.
\end{equation}
As a result, we find the following.
\begin{subbox}{$\mathrm{Sl}(2)$-orbit approximation of the K\"ahler potential ($m$-parameter)}
\begin{align}\label{eq:Kahler_SL2_multi}
K_{\mathrm{Sl}(2)}^{\mathrm{cs}} &= -\log\left[\left(\frac{y_1}{y_2}\right)^{d_1}\cdots \left(\frac{y_{m-1}}{y_m}\right)^{d_{m-1}}y_m^{d_m}\,||\Omega_\infty||^2_\infty \right]\,,\\
&=-\sum_{i=1}^m (d_i - d_{i-1}) \log y_i+\mathrm{constant}\,.
\end{align}
\end{subbox}
\noindent As in the one-parameter case, it is important to stress that the above result only gives rise to a well-defined positive-definite metric if $(d_i-d_{i-1})$ is never equal to zero. This means that the singularity type must strictly increase along the enhancement chain. Clearly, as soon as one is considering a limit which involves more than $D+1$ moduli, this requirement cannot be satisfied, since there are at most $D+1$ singularity types for a Calabi--Yau $D$-fold. In other words, in many cases the approximation \eqref{eq:Kahler_SL2_multi} does not suffice to properly describe the leading behaviour of the K\"ahler potential, and hence one is forced to include corrections coming from the more general nilpotent orbit expansion of the period map. 

\begin{subappendices}

\section{On the proof of the nilpotent orbit theorem}\label{app:nilp_orbit_proof}

In this section we discuss some aspects of the proof of the nilpotent orbit theorem, following the original work of Schmid \cite{schmid}. Most notably, our goal is to give an idea of the important steps involved in showing the quasi-unipotency of the monodromy operators, and how properties such as the horizontality of the period map play a central role in this regard. To this end, let us briefly recall the characterization of a one-parameter variation of Hodge structure on a punctured disk through the period map
\begin{equation}
	\Phi:\qquad \Delta^*\rightarrow \Gamma\backslash \mathcal{D}\,,
\end{equation}
with $\mathcal{D}=G_{\mathbb{R}}/V$ the period domain, see also the discussion in section \ref{sec:period_map}. 

\subsubsection*{Uniform boundedness}
The first step of the proof involves translating the horizontality property of the period map $\Phi$ into a certain distance-decreasing property. This relies on the following result.
\begin{subbox}{Lemma 3.16 of \cite{schmid}}
	There exists a $G_{\mathbb{R}}$-invariant Hermitian metric on $\mathcal{D}$ whose holomorphic sectional curvatures in the directions of the horizontal tangent bundle on $\mathcal{D}$ are negative and bounded away from zero. 
\end{subbox}
\noindent Informally, this lemma states that the period domain $\mathcal{D}$ is a hyperbolic manifold (i.e. having strictly negative curvature) but only in the horizontal directions. However, the period map is itself a horizontal map, meaning that its differential maps exactly into the horizontal tangent bundle. Thus, for all matters pertaining to the period mapping, one can think of $D$ as being a hyperbolic manifold. \\

\noindent This observation allows one to use some results in hyperbolic complex analysis, which is the study of holomorphic mappings into negatively curved complex manifolds. Let $f:\Delta\rightarrow M$ be such a mapping, then $f$ satisfies two very useful properties \cite{Kobayashi:2005}.
\begin{itemize}
	\item \textbf{Property (1):}\\
	$f^*(ds^2_M)$ defines a (pseudo)-distance 
	\begin{equation}
		f^*(ds^2_M)=h(z,\bar{z})\, dz\,d\bar{z}\,.
	\end{equation}
	on $\Delta$ with (after suitable normalization) curvature 
	\begin{equation}
		K(h)=-\frac{1}{2h}\partial\bar{\partial}\log h\leq -1\,.
	\end{equation}
	\item \textbf{Property (2): (Ahlfors' lemma)}\\
	The function $h(z,\bar{z})$ satisfies
	\begin{equation}
		h(z,\bar{z}) \leq \frac{1}{(1-|z|^2)^2}\,,
	\end{equation}
	i.e. it is bounded by the Poincar\'e metric on the unit disk.
\end{itemize}
The first property simply states that the pull-back of a negatively curved metric by a holomorphic map gives again a negatively curved metric. The second statement furthermore bounds the latter in terms of the Poincar\'e metric. \\

\noindent Applied to the (lifting of the) period mapping, combined with Lemma 3.16 of \cite{schmid} stated above, this yields the following corollary.
\begin{subbox}{Corollary 3.17 of \cite{schmid}}
	The (lifting of the) period map 
	\begin{equation*}
		\tilde{\Phi}:\mathbb{H}\rightarrow \mathcal{D}\,,
	\end{equation*}
	which maps the upper half-plane into $\mathcal{D}$, is \textbf{uniformly bounded} with respect to the Poincar\'e metric on $\mathbb{H}$ and any $G_{\mathbb{R}}$-invariant Hermitian metric on $\mathcal{D}$.
\end{subbox}
\noindent After a suitable normalization, uniform boundedness simply means that $\tilde{\Phi}$ is distance-decreasing, i.e.
\begin{equation}\label{eq:uniform_bounded}
	d(\tilde{\Phi}(t_1),\tilde{\Phi}(t_2))\leq d_{\mathrm{Poincare}}(t_1,t_2)\,,
\end{equation}
for any $G_{\mathbb{R}}$-invariant metric $d$ on $D$. This property of the period map is absolutely central to the proof of the nilpotent orbit theorem. Note that it relies on both the holomorphicity and horizontality of the period map.

\subsubsection*{Quasi-unipotent monodromy}

As a first application of the uniform boundedness of the period map, we show how it is used to prove the fact that the local monodromy is quasi-unipotent. Let $\gamma\in G_{\mathbb{Z}}$ be such that
\begin{equation}
	\label{eq:monodromy}
	\tilde{\Phi}(t+1) = \gamma\circ\tilde{\Phi}(t)\,,
\end{equation}
then we aim to show that the eigenvalues of $\gamma$ are roots of unity. We follow the proof of Lemma 4.5 in \cite{schmid}, which is originally due to Borel. \\

\noindent First, we consider the following sequence of points
\begin{equation}
	t_n = n\cdot i\,,\qquad n\in\mathbb{N}\,,
\end{equation}
which progressively moves upwards in the complex upper half-plane. With respect to the Poincar\'e metric, the distance between the points $t_n$ and $t_n+1$ is given by\footnote{Recall that on the upper half-plane, we have
	\begin{equation}
		ds^2_{\mathrm{Poincare}} = \frac{dx^2+dy^2}{y^2}\,,\qquad t=x+iy\,.
\end{equation}}
\begin{equation}
	d_{\mathrm{Poincare}}(t_n, t_n+1) = \frac{1}{n}\,.
\end{equation}
On the other hand, letting $g_n\in G_{\mathbb{R}}$ such that
$\tilde{\Phi}(t_n)\in D\cong G_{\mathbb{R}}/V$ corresponds to the coset $g_n V$, we have
\begin{align*}
	d(\tilde{\Phi}(t_n), \tilde{\Phi}(t_n+1)) &\stackrel{\eqref{eq:monodromy}}{=} d(\tilde{\Phi}(t_n), \gamma\circ\tilde{\Phi}(t_n))\stackrel{\text{$G_{\mathbb{R}}$-invariance}}{=}d\left(eV, g_n^{-1}\gamma g_n V\right)\,,
\end{align*}
Then, by virtue of the uniform boundedness property \eqref{eq:uniform_bounded}, we find
\begin{equation}
	d\left(eV, g_n^{-1}\gamma g_n V\right)\leq \frac{1}{n}\,.
\end{equation}
This implies that the sequence $g_n^{-1}\gamma g_n$ accumulates in $V$, as $n\rightarrow\infty$. Since $V$ is compact, this implies that the eigenvalues of $\lambda$ have modulus one. \\

\noindent To conclude that the eigenvalues are in fact roots of unity, we apply the following argument due to Kronecker. Since $\gamma\in G_{\mathbb{Z}}$, clearly also $\gamma^m\in G_{\mathbb{Z}}$ for any positive integer $m\in\mathbb{N}$. Denote the eigenvalues of $\gamma$ by $\lambda_1,\ldots, \lambda_r$, all of which have modulus one by the preceding argument, such that the characteristic polynomial of $\gamma^m$ reads
\begin{equation}
	p_m(x) = (x-\lambda_1^m)\cdots (x-\lambda_r^m)\,.
\end{equation}
Since $\gamma^m\in G_{\mathbb{Z}}$, each $p_m$ is in fact a monic polynomial with integer coefficients. Furthermore, the coefficient of $x^k$ in $p_m$ is bounded by $\binom{r}{k}$, since $|\lambda_i^m|=1$. Note that this bound is independent of $m$. In particular, in the infinite sequence of polynomials $\{p_m\}$ only contains finitely many different polynomials. This means that there exist integers $k,l$ such that $p_k=p_l$, for which the lists of roots $\{\lambda_1^k, \ldots, \lambda_r^k\}$ and $\{\lambda_1^l, \ldots, \lambda_r^l\}$ must be related by a permutation. By iterating this process, this permutation must eventually be the trivial permutation\footnote{One way to see this is to write
	\begin{equation}
		\lambda_{i_1}^{k_1} = \lambda_{i_2}^{k_2} = \lambda_{i_3}^{k_3}=\cdots\,.
	\end{equation}
	Since there are only finitely many possible lower indices $i_s$, it must be that at some point $\lambda_{i_1}^{k_1} = \lambda_{i_1}^{k_s}$.}, so that
\begin{equation}
	\lambda_i^s = 1\,,
\end{equation}
for some integer $s$. This proves that all $\lambda_i$ are roots of unity, hence $\gamma$ is quasi-unipotent. 

\subsubsection*{Rest of the proof}
While we will not explain the remaining steps in the proof of the nilpotent orbit theorem in detail, let us briefly describe how the quasi-unipotency of the monodromy together with the uniform boundedness property of the period map play a central role. To this end, let us write
\begin{equation}
	\tilde{\Phi}(t) = e^{tN}\tilde{\Psi}(t)\,,
\end{equation}
where we have factored out the monodromy operator, which can be taken to be unipotent. The goal, then, is to show that 
\begin{equation}\label{eq:nop_proof}
 h_M\left(\tilde{\Psi}_*\left(\frac{\partial}{\partial t}\right),\tilde{\Psi}_*\left(\frac{\partial}{\partial t}\right)\right)^{1/2} \leq C e^{-\epsilon\,\mathrm{Im}\,t}\,,
\end{equation}
for some suitably defined metric $h_M$ and constants $C,\epsilon$, see \cite[(8.22)]{schmid}. If this is achieved, then it follows that $\tilde{\Psi}$ is regular as $\mathrm{Im}\,t\rightarrow\infty$ and thus drops to a map $\Psi$ that extends over the puncture of the disk at $z=0$. Furthermore, thanks to the exponential suppression one then finds that
the distance between the original period map and the nilpotent orbit $e^{\frac{\log{z}}{2\pi i}N}\Psi(0)$ is exponentially small. \\

\noindent 
In order to achieve the bound \eqref{eq:nop_proof}, one first writes
\begin{equation}\label{eq:nop_proof_2}
	\tilde{\Psi}_*\left(\frac{\partial}{\partial t}\right) = l\left(e^{-tN}\right)_* \left(\tilde{\Phi}\left(\frac{\partial}{\partial t}\right)- N\left(\tilde{\Phi}(t) \right)\right)\,,
\end{equation}
where $l$ denotes left-multiplication. The intuition, then, is that the first term on the right-hand side of \eqref{eq:nop_proof_2} is bounded by a polynomial in $t$ due to the nilpotency of $N$, while it turns out that the second term can be argued to be exponentially small thanks to the uniform boundedness of $\tilde{\Phi}$ and the general properties of certain periodic holomorphic functions that are bounded on a strip \cite[Lemma (8.17)]{schmid}.

\section{Examples: one-parameter families of Calabi--Yau threefolds}\label{sec:asymp_Hodge_examples}
In this section we exemplify some of the abstract constructions appearing in the study of mixed Hodge structures. A natural class of examples to consider are those limiting mixed Hodge structures that arise in one-parameter families of Calabi--Yau threefolds, having Hodge numbers $h^{3,0}=h^{2,1}=1$. In this case, we would have
\begin{equation}
\mathrm{dim}_{\mathbb{R}} H_{\mathbb{R}} = 4\,,\qquad D=3\,,\qquad G_{\mathbb{R}} = \mathrm{Sp}(4,\mathbb{R})\,,
\end{equation}
In other words, we will be considering polarized mixed Hodge structures of weight three on a four-dimensional vector space. Using the fact that $h^{3,0}=h^{2,1}=1$, together with the various relations \eqref{eq:ipq_restriction} and \eqref{eq:ipq_relation_hpq}, one quickly finds that there are only three possible non-trivial Hodge--Deligne diamonds, which are depicted in figure \ref{fig:Deligne_one-modulus}. In principle, there is a fourth possibility, corresponding to the type $\mathrm{I}_0$ singularity, which will have trivial unipotent monodromy, so that $N=0$. For our purposes, this is not an interesting case to consider. Below we discuss the three non-trivial cases in some detail. For each of the examples, we describe the most general $\mathbb{R}$-split mixed Hodge structure, and its associated $\mathfrak{sl}(2,\mathbb{R})$-triple. We also write down the most general phase operator. Importantly, we will see that these three cases exactly cover the three types of asymptotic behaviours for the periods of the mirror bicubic, which was discussed in section \ref{subsec:asymp_periods_mirror_bicubic}. Our discussion largely follows \cite{Bastian:2023shf} and also uses the same conventions. Further computational details can be found there. 

\begin{figure}[h!]
	\centering
	\begin{subfigure}{0.3\textwidth}
		\centering
		\begin{tikzpicture}[scale=0.8,cm={cos(45),sin(45),-sin(45),cos(45),(15,0)}]
			\draw[step = 1, gray, ultra thin] (0, 0) grid (3, 3);
			\draw[fill] (3, 0) circle[radius=0.04];
			\draw[fill] (2, 2) circle[radius=0.04];
			\draw[fill] (0, 3) circle[radius=0.04];
			\draw[fill] (1, 1) circle[radius=0.04];
		\end{tikzpicture}	
		\caption{Type $\mathrm{I}_1$.}
		\label{fig:I1}
	\end{subfigure}
	\begin{subfigure}{0.3\textwidth}
		\centering
		\begin{tikzpicture}[scale=0.8,cm={cos(45),sin(45),-sin(45),cos(45),(15,0)}]
			\draw[step = 1, gray, ultra thin] (0, 0) grid (3, 3);
			\draw[fill] (3, 1) circle[radius=0.04];
			\draw[fill] (2, 0) circle[radius=0.04];
			\draw[fill] (1, 3) circle[radius=0.04];
			\draw[fill] (0, 2) circle[radius=0.04];
		\end{tikzpicture}
		\caption{Type $\mathrm{II}_0$.}
		\label{fig:II0}
	\end{subfigure}
	\centering
	\begin{subfigure}{0.3\textwidth}
		\centering
		\begin{tikzpicture}[scale=0.8,cm={cos(45),sin(45),-sin(45),cos(45),(15,0)}]
			\draw[step = 1, gray, ultra thin] (0, 0) grid (3, 3);
			\draw[fill] (3, 3) circle[radius=0.04];
			\draw[fill] (2, 2) circle[radius=0.04];
			\draw[fill] (1, 1) circle[radius=0.04];
			\draw[fill] (0, 0) circle[radius=0.04];
		\end{tikzpicture}
		\caption{Type $\mathrm{IV}_1$.}
		\label{fig:IV1}
	\end{subfigure}
	\caption{The three non-trivial weight $D=3$ limiting mixed Hodge structures with $h^{3,0}=h^{2,1}=1$. Here all dots represent one-dimensional vector spaces. }
	\label{fig:Deligne_one-modulus}
\end{figure}
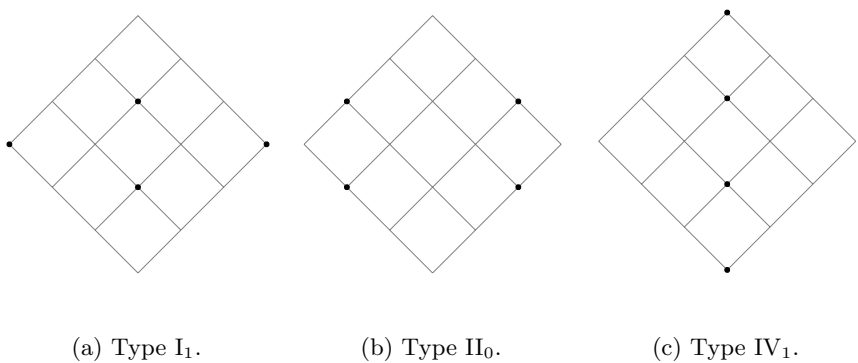

\subsection{Type $\mathrm{I}_1$}
Geometrically, the type $\mathrm{I}_1$ singularity will correspond to a type of conifold singularity, in which a three-cycle shrinks to zero size. One can represent the three-cycle as a three-sphere $S^3$ that is quotiented by a finite discrete group of order $k$.

\subsubsection*{Monodromy weight basis, pairing and log-monodromy}
Without loss of generality, we choose to work in a basis in which 
\begin{equation}
e_4\in W_2\,, \quad e_3,e_2\in W_3\,,\quad e_1\in W_4\,,
\end{equation}
where $e_i$ denotes the standard $i$-th unit basis vector. With respect to this basis, one finds the following:
\begin{itemize}
\item The polarization form can be chosen to be represented by the matrix 
\begin{equation}
S = \begin{pmatrix}
	0 & 0 & 0 & 1\\
	0 & 0 & 1 & 0\\
	0 & -1 & 0 & 0\\
	-1 & 0 & 0 & 0
\end{pmatrix}\,,
\end{equation}
such that $\langle e_1, e_4\rangle = \langle e_2, e_3\rangle = 1$, with all other pairings vanishing.
\item The most general log-monodromy matrix that preserves the above pairing, satisfies the polarization conditions and additionally acts on the weight filtration as $N W_\ell\subset W_{\ell-2}$ is given by
\begin{equation}
N = \begin{pmatrix}
	0 & 0 & 0 & 0\\
	0 & 0 & 0 & 0\\
	0 & 0 & 0 & 0\\
	-k & 0 & 0 & 0
\end{pmatrix}\,,
\end{equation}
with $k\in\mathbb{N}$ strictly positive. It turns out that $k$ corresponds precisely to the order of the discrete group that quotients the $S^3$, as mentioned earlier. 
\end{itemize}

\subsubsection*{$\mathbb{R}$-split MHS}
After imposing the transversality constraints \eqref{eq:log-monodromy_Deligne} and possibly performing a coordinate shift, the most general $\mathbb{R}$-split type $\mathrm{I}_1$ mixed Hodge structure is given by
\begin{align}
\tilde{I}^{2,2} &= \mathrm{span}\left[\left(1,\gamma,\delta,0 \right)\right]\,,\nonumber\\
\tilde{I}^{3,0} &= \mathrm{span}\left[\left(0,1,\tau, \delta-\gamma\tau \right)\right]\,,\qquad \tilde{I}^{0,3} = \overline{\tilde{I}^{3,0}}\,,\\
\tilde{I}^{1,1} &=\mathrm{span}\left[\left(0,0,0,1\right)\right]\,,\nonumber
\end{align}
for some parameters $\tau=\tau_1+i\tau_2\in\mathbb{C}$ with $\tau_2>0$, and $\gamma,\delta\in\mathbb{R}$. Let us briefly comment on these parameters.
\begin{itemize}
\item \textbf{Rigid period: $\tau=\tau_1+i\tau_2$}\\
The parameter $\tau$, taking values in the complex upper half-plane $\tau_2>0$, carries the interpretation of a \textit{rigid period} that is associated with the degenerating geometry. In fact, one can show that a certain subgroup $\mathrm{SL}(2,\mathbb{Z})\subset \mathrm{Sp}(4,\mathbb{Z})$ of the integral transformations that preserve the weight filtration acts on $\tau$ as the modular group. 
\item \textbf{Extension data: $\gamma,\delta$}\\
The parameters $\gamma,\delta$ are referred to as \textit{extension data}, and play an important role in the integral structure of the limiting mixed Hodge structure. One can show that there exists another subgroup of $\mathrm{Sp}(4,\mathbb{Z})$ which acts as a shift on these parameters:
\begin{equation}
\gamma\mapsto \gamma+b_1\,,\qquad \delta\mapsto \delta+b_2\,,\qquad b_1,b_2\in\mathbb{Z}\,.
\end{equation}
This may be used to restrict $\gamma,\delta$ to the internal $[0,1)$.
\end{itemize}

\subsubsection*{$\mathfrak{sl}(2,\mathbb{R})$-triple}
In order to present $\mathfrak{sl}(2,\mathbb{R})$-triple associated to the above $\mathbb{R}$-split mixed Hodge structure in a transparent form, let us introduce the following $\mathrm{Sp}(4,\mathbb{R})$-valued matrix
\begin{equation}\label{eq:transform_I1}
\Lambda = \begin{pmatrix}
1 & 0 & 0 & 0\\
\gamma & 1 & 0 & 0\\
\delta & 0 & 1 & 0\\
0 & \delta & -\gamma & 1
\end{pmatrix}\cdot\begin{pmatrix}
-1 & 0 & 0 & 0\\
0 & 0 & -\frac{1}{\sqrt{\tau_2}} & 0\\
0 & \sqrt{\tau_2} & -\frac{\tau_1}{\sqrt{\tau_2}} & 0\\
0 & 0 & 0 & -1
\end{pmatrix}\cdot\begin{pmatrix}
\frac{1}{\sqrt{k}} & 0 & 0 & 0\\
0 & 1 & 0 & 0\\
0 & 0 & 1 & 0\\
0  & 0 & 0 & \sqrt{k}
\end{pmatrix}\,,
\end{equation}
which will carry all the dependence on the rigid period and the extension data and acts as a rotation from the integer basis to the Hodge basis. In terms of this matrix, the $\mathfrak{sl}(2,\mathbb{R})$-triple is given by
\begin{equation}\label{eq:I1-sl2}
\left\{N^+, N^0, N^-\right\}= \Lambda\left\{\scalebox{0.8}{$\begin{pmatrix}
	0 & 0 & 0 & -1\\
	0 & 0 & 0 & 0\\
	0 & 0 & 0 & 0\\
	0 & 0 & 0 & 0
\end{pmatrix}, \begin{pmatrix}
	1 & 0 & 0 & 0\\
	0 & 0 & 0 & 0\\
	0 & 0 & 0 & 0\\
	0 & 0 & 0 & -1
\end{pmatrix},\begin{pmatrix}
	0 & 0 & 0 & 0\\
	0 & 0 & 0 & 0\\
	0 & 0 & 0 & 0\\
	-1 & 0 & 0 & 0
\end{pmatrix}$} \right\}\Lambda^{-1}\,,
\end{equation}
where $N^- = N$. This induces a decomposition of the four-dimensional underlying vector space into irreducible representations of $\mathfrak{sl}(2,\mathbb{R})$ as
\begin{equation}
\mathbf{4} = \mathbf{1}\oplus\mathbf{2}\oplus\mathbf{1}\,,
\end{equation}
which could already be anticipated based on the associated Hodge--Deligne diamond in figure \ref{fig:I1}. 

\subsubsection*{Phase operator}
It turns out that for the type $\mathrm{I}_1$ mixed Hodge structure described above, the possible phase operators which could make it non $\mathbb{R}$-split are very limited. In fact, imposing the conditions \eqref{eq:delta_commutes_N} and \eqref{eq:delta_Deligne} one is left with only one possibility, namely that it is proportional to the log-monodromy matrix, so that
\begin{equation}\label{eq:I1-delta}
\delta =\Lambda\begin{pmatrix}
0 & 0 & 0 & 0\\
0 & 0 & 0 & 0\\
0 & 0 & 0 & 0\\
-c & 0 & 0 & 0
\end{pmatrix}\Lambda^{-1}\,,
\end{equation}
for some constant $c\in\mathbb{R}$, where we have chosen the sign for future convenience. Note that the rotation by $\delta$ only affects $\hat{I}^{2,2}$ and transforms it to
\begin{equation}
I^{2,2} = e^{i\delta}\tilde{I}^{2,2} = \mathrm{span}\left[ (1,\gamma,\delta, -ic)\right]\,,
\end{equation}
which is indeed manifestly no longer $\mathbb{R}$-split. Furthermore, the component that breaks the $\mathbb{R}$-split property is indeed given by the lower-lying piece $I^{1,1}$ in accordance with the generalized conjugation property \eqref{eq:Deligne_conjugation}. Notably, if the mixed Hodge structure arises as the limiting mixed Hodge structure associated to some nilpotent orbit, one can effectively set $c=0$ by performing a coordinate shift $t\mapsto t+ic$. In other words, one may assume without loss of generality that the type $\mathrm{I}_1$ limiting mixed Hodge structure is $\mathbb{R}$-split.
\\

\noindent Finally, since the phase operator is proportional to $N$, one immediately concludes that
\begin{equation}
\delta = \delta_{-1,-1}\,,
\end{equation}
with respect to the decomposition \eqref{eq:delta_Deligne}. In particular, using the relations \eqref{eq:delta_zeta_threefold}, one finds that
\begin{equation}
\zeta = 0\,,
\end{equation}
for the type $\mathrm{I}_1$ singularity. In other words, the $\mathbb{R}$-split mixed Hodge structure $\tilde{I}^{p,q}$ is automatically $\mathrm{sl}(2)$-split.

\subsubsection*{Boundary charge operator}
The final ingredient of the boundary data is the boundary charge operator, which encodes the boundary Hodge structure given by
\begin{equation}
F_\infty^p = e^{iN}\hat{F}^p_{0}\,,
\end{equation}
and can be straightforwardly computed using the relation \eqref{eq:Deligne_weight+limiting} between the Deligne splitting and the corresponding limiting filtration. The resulting charge operator is given by
\begin{equation}\label{eq:I1-Q}
Q_\infty = \Lambda\begin{pmatrix}
0 & 0 & 0 & \frac{i}{2}\\
0 & 0 & -\frac{3i}{2} & 0\\
0 & \frac{3i}{2} & 0 & 0\\
-\frac{i}{2} & 0 & 0 & 0
\end{pmatrix}\Lambda^{-1}\,.
\end{equation}	

\subsection{Type $\mathrm{II}_0$} 
Geometrically, the type $\mathrm{II}_0$ singularity corresponds to a K-point. When such a point is realized by a Tyurin degeneration \cite{Tyurin:2003} it can be thought of as arising from the intersection of two threefolds in a K3 surface. In the one-modulus case this K3 surface is rigid, and its geometrical properties such as its intersection form and its (rigid) period will appear in the limiting mixed Hodge structure via the log-monodromy matrix.

\subsubsection*{Monodromy weight basis, pairing and log-monodromy}
Without loss of generality, we choose to work in a basis in which
\begin{equation}
e_3,e_4\in W_2\,,\quad e_1,e_2\in W_4\,,
\end{equation}
where again $e_i$ denotes the standard $i$-th unit basis vector. With respect to this basis, one finds the following:
\begin{itemize}
\item There are effectively two choices for the polarization form, which amount to a different ordering of the basis of $W_2$. In the following we follow the conventions of \cite{Bastian:2023shf} and choose it to be represented by the matrix
\begin{equation}
S = \begin{pmatrix}
	0 & 0 & 1 & 0\\
	0 & 0 & 0 & 1\\
	-1 & 0 & 0 & 0\\
	0 & -1 & 0 & 0
\end{pmatrix}\,,
\end{equation}
such that $\langle e_1,e_3\rangle = \langle e_2, e_4\rangle = 1$, with all other pairings vanishing.
\item The most general log-monodromy matrix that preserves the above pairing, satisfies the polarization conditions and additionally acts on the weight filtration as $NW_{\ell}\subset W_{\ell-2}$ is given by
\begin{equation}
N = \begin{pmatrix}
	0 & 0 & 0 & 0\\
	0 & 0 & 0 & 0\\
	a & b & 0 & 0\\
	b & c & 0 & 0
\end{pmatrix}\,,\qquad a,b,c\in\mathbb{Z}\,,ac-b^2>0\,.
\end{equation}
\end{itemize}

\subsubsection*{$\mathbb{R}$-split MHS}
Imposing the transversality constraints and possibly performing a coordinate redefinition, the most general $\mathbb{R}$-split type $\mathrm{II}_0$ mixed Hodge structure can be taken to be
\begin{align*}
\tilde{I}^{3,1} &= \mathrm{span}\left[\left(1,\tau, \delta+\gamma\tau, \gamma\right)\right]\,,\qquad 
\tilde{I}^{2,0}=\mathrm{span}\left[\left(0,0,-\tau,1\right)\right]\,,
\end{align*}
with the remaining spaces determined by complex conjugation. Here $\gamma,\delta\in\mathbb{R}$ are arbitrary parameters and
\begin{equation}
\tau = \frac{-b\pm i\sqrt{d}}{c}\,,\qquad d = ac-b^2>0\,,
\end{equation}
where the sign can be chosen such that $\mathrm{Im}\,\tau>0$. For definiteness, we will assume that $c>0$ such that we take the plus sign. Correspondingly, in order to have $d>0$ we must then also take $a>0$. The parameters $\gamma,\delta$ again play the role of extension data, as for the $\mathrm{I}_1$ singularity (see \cite{Bastian:2023shf} for a more in-depth discussion), while the parameter $\tau$ corresponds to the rigid period of the K3. 

\subsubsection*{$\mathfrak{sl}(2,\mathbb{R})$-triple}
In order to present $\mathfrak{sl}(2,\mathbb{R})$-triple associated to the above $\mathbb{R}$-split mixed Hodge structure in a transparent form, let us introduce the following $\mathrm{Sp}(4,\mathbb{R})$-valued matrix
\begin{equation}\label{eq:transform_II0}
\Lambda = \begin{pmatrix}
1 & 0 & 0 & 0\\
0 & 1 & 0 & 0\\
\delta & \gamma & 1 & 0\\
\gamma & 0 & 0 & 1
\end{pmatrix}\cdot\begin{pmatrix}
-\frac{b}{\sqrt{ad}} & \frac{1}{\sqrt{a}} & 0 & 0\\
\sqrt{\frac{a}{d}} & 0 &0 & 0\\
0 & 0 & 0 & \sqrt{a}\\
0 & 0 & \sqrt{\frac{d}{a}} & \frac{b}{\sqrt{a}}
\end{pmatrix}\,,
\end{equation}
which will carry all the dependence on the rigid period and the extension data and acts as a rotation from the integer basis to the Hodge basis. In terms of this matrix, the $\mathfrak{sl}(2,\mathbb{R})$-triple associated to the type $\mathrm{II}_0$ singularity is given as follows:
\begin{equation}
\{N^+,N^0,N^-\} = \Lambda\scalebox{0.9}{$\left\{\begin{pmatrix}
	0 & 0 & 1 & 0\\
	0 & 0 & 0 & 1\\
	0 & 0 & 0 & 0\\
	0 & 0 & 0 & 0
\end{pmatrix}\,,\begin{pmatrix}
	1 & 0 & 0 & 0\\
	0 & 1 & 0 & 0\\
	0 & 0 & -1 & 0\\
	0 & 0 & 0 & -1
\end{pmatrix}\,,\begin{pmatrix}
	0 & 0 & 0 & 0\\
	0 & 0 & 0 & 0\\
	1 & 0 & 0 & 0\\
	0 & 1 & 0 & 0
\end{pmatrix}\right\}$}\Lambda^{-1}\,.
\end{equation}
This induces a decomposition of the four-dimensional underlying vector space into irreducible representations of $\mathfrak{sl}(2,\mathbb{R})$ as
\begin{equation}
\mathbf{4} = \mathbf{2}\oplus\mathbf{2}\,,
\end{equation}
which could already be anticipated based on the associated Hodge--Deligne diamond in figure \ref{fig:II0}. 

\subsubsection*{Phase operator}
It turns out that also for the type $\mathrm{II}_0$ mixed Hodge structure described above, the possible phase operators which could make it non $\mathbb{R}$-split are very limited. In fact, imposing the conditions \eqref{eq:delta_commutes_N} and \eqref{eq:delta_Deligne} one is again left with only one possibility, namely that it is proportional to the log-monodromy matrix, so that
\begin{equation}\label{eq:delta_II0}
\delta = \Lambda\scalebox{0.9}{$\begin{pmatrix}
	0 & 0 & 0 & 0\\
	0 & 0 & 0 & 0\\
	-\alpha & 0 & 0 & 0\\
	0 & -\alpha & 0 & 0
\end{pmatrix}$}\Lambda^{-1}\,,\qquad \alpha\in\mathbb{R}\,.
\end{equation}
for some constant $\alpha\in\mathbb{R}$, where we have chosen the sign for future convenience.\\

\noindent Finally, since the phase operator is proportional to $N$, one immediately concludes that
\begin{equation}
\delta = \delta_{-1,-1}\,,
\end{equation}
with respect to the decomposition \eqref{eq:delta_Deligne}. In particular, using the relations \eqref{eq:delta_zeta_threefold}, one finds that
\begin{equation}
\zeta = 0\,,
\end{equation}
for the type $\mathrm{II}_0$ singularity. In other words, the $\mathbb{R}$-split mixed Hodge structure $\tilde{I}^{p,q}$ is automatically $\mathrm{sl}(2)$-split. 

\subsubsection*{Boundary charge operator}
Finally, the boundary charge operator associated to the type $\mathrm{II}_0$ singularity is given by
\begin{equation}
Q_\infty = \Lambda\scalebox{0.9}{$\begin{pmatrix}
	0 & i & -\frac{i}{2} & 0\\
	-i & 0 & 0 & -\frac{i}{2}\\
	\frac{i}{2} & 0 & 0 & i\\
	0 & \frac{i}{2} & -i & 0
\end{pmatrix}$}\Lambda^{-1}\,.	
\end{equation}
\subsection{Type $\mathrm{IV}_1$} 
Geometrically, the type $\mathrm{IV}_1$ singularity corresponds to the large complex structure point or MUM point of a one-parameter family of Calabi--Yau manifolds. On the mirror side, this corresponds to sending the volume of the mirror manifold to infinity. 

\subsubsection*{Monodromy weight basis, pairing and log-monodromy}
Without loss of generality, we choose to work in a basis in which 
\begin{equation}
e_4\in W_0\,,\quad e_3\in W_2\,,\quad e_2\in W_4\,,\quad e_1\in W_6\,,
\end{equation}
where again $e_i$ denotes the standard $i$-th unit basis vector. With respect to this basis, one finds the following:
\begin{itemize}
\item The polarization form can be chosen to be represented by the matrix
\begin{equation}
S = \begin{pmatrix}
	0 & 0 & 0 & 1\\
	0 & 0 & 1 & 0\\
	0 & -1 & 0 & 0\\
	-1 & 0 & 0 & 0
\end{pmatrix}\,,
\end{equation}
such that $\langle e_1,e_4\rangle = \langle e_2, e_3\rangle = 1$, with all other pairings vanishing.
\item The most general log-monodromy matrix that preserves the above pairing, satisfies the polarization conditions and additionally acts on the weight filtration as $NW_{\ell}\subset W_{\ell-2}$ is given by
\begin{equation}
N = \begin{pmatrix}
	0 & 0 & 0 & 0\\
	a & 0 & 0 & 0\\
	e & b & 0 & 0\\
	f & e & -a & 0
\end{pmatrix}\,,
\end{equation}
with 
\begin{equation}
a,b\in\mathbb{Z}\,,\qquad e- ab/2\in\mathbb{Z}\,,\qquad f-a^2b/6\in\mathbb{Z}\,,
\end{equation}
and additionally $a\neq 0$ and $b>0$.
\end{itemize}

\subsubsection*{$\mathbb{R}$-split MHS}
Imposing the transversality constraints and possibly performing a coordinate redefinition, the most general $\mathbb{R}$-split type $\mathrm{IV}_1$ mixed Hodge structure can be taken to be
\begin{align*}
\tilde{I}^{3,3} &= \mathrm{span}\left[ \left(1,0,\frac{f}{2a},\gamma\right)\right]\,,\\
\tilde{I}^{2,2} &=\mathrm{span}\left[\left(0,1,\frac{e}{a},\frac{f}{2a}\right)\right]\,,\\
\tilde{I}^{1,1} &=\mathrm{span}\left[(0,0,1,0)\right]\,,\\
\tilde{I}^{0,0} &=\mathrm{span}\left[(0,0,0,1)\right]\,,
\end{align*}
where $\gamma$ is an arbitrary parameter. In section \ref{reconstructing_examples} we will recall how the parameters $a,b,e,f$ and $\gamma$ relate to topological properties of the mirror Calabi--Yau threefold. 

\subsubsection*{$\mathfrak{sl}(2,\mathbb{R})$-triple}
In order to present $\mathfrak{sl}(2,\mathbb{R})$-triple associated to the above $\mathbb{R}$-split mixed Hodge structure in a transparent form, let us introduce the following $\mathrm{Sp}(4,\mathbb{R})$-valued matrix
\begin{equation}\label{eq:basis-transform_IV1}
\Lambda =  \begin{pmatrix}
1 & 0 & 0 & 0\\
0 & 1 & 0 & 0\\
\frac{f}{2a} & \frac{e}{a} & 1 & 0\\
\gamma & \frac{f}{2a} & 0 & 1
\end{pmatrix}\cdot\begin{pmatrix}
\frac{1}{a\sqrt{b}} & 0 & 0 & 0\\
0 & \frac{1}{\sqrt{b}} & 0 & 0\\
0 & 0 & \sqrt{b} & 0 \\
0 & 0 & 0 & a \sqrt{b}
\end{pmatrix}\,.
\end{equation}
In terms of this matrix, the $\mathfrak{sl}(2,\mathbb{R})$-triple associated to the type $\mathrm{IV}_1$ singularity is given by
\begin{equation}\label{eq:sl2-triple_IV1}
\{N^+, N^0, N^-\} = \Lambda\scalebox{0.9}{$\left\{\begin{pmatrix}
	0 & 3 & 0 & 0\\
	0 & 0 & 4 & 0\\
	0 & 0 & 0 & -3\\
	0 & 0 & 0 & 0
\end{pmatrix}\,,\begin{pmatrix}
	3 & 0 & 0 & 0\\
	0 & 1 & 0 & 0\\
	0 & 0 & -1 & 0\\
	0 & 0 & 0 & -3
\end{pmatrix}\,,\begin{pmatrix}
	0 & 0 & 0 & 0\\
	1 & 0 & 0 & 0\\
	0 & 1 & 0 & 0\\
	0 & 0 & -1 & 0
\end{pmatrix}\right\}$}\Lambda^{-1}\,.
\end{equation}
This induces a decomposition of the four-dimensional underlying vector space into irreducible representations of $\mathfrak{sl}(2,\mathbb{R})$ as
\begin{equation}
\mathbf{4} = \mathbf{4}\,,
\end{equation}
which could already be anticipated based on the associated Hodge--Deligne diamond in figure \ref{fig:IV1}. 

\subsubsection*{Phase operator}
It turns out that also the type $\mathrm{IV}_1$ mixed Hodge structure described above, the set of possible phase operators is richer than for the type $\mathrm{I}_1$ and $\mathrm{II}_0$ singularities. Indeed, one finds that the most general phase operator is given by
\begin{equation}\label{eq:delta_IV1_1}
\delta = \delta_{-1,-1}+\delta_{-3,-3}\,,
\end{equation}
where $\delta_{-1,-1}$ is again proportional to the log-monodromy matrix $N$, while $\delta_{-3,-3}$ is given by
\begin{equation}\label{eq:delta_IV1_2}
\delta_{-3,-3} =  \Lambda\begin{pmatrix}
0 & 0 & 0 & 0\\
0 & 0 & 0 & 0\\
0 & 0 & 0 & 0\\
\chi & 0 & 0 & 0
\end{pmatrix}\Lambda^{-1}\,,
\end{equation}
for some arbitrary parameter $\chi\in\mathbb{R}$. Note that actually $\delta_{-3,-3}$ commutes with $\Lambda$. Notably, one cannot remove the $\delta_{-3,-3}$ piece by a coordinate shift, hence the most general type $\mathrm{IV}_1$ limiting mixed Hodge structure is not necessarily $\mathbb{R}$-split. Using the relations \eqref{eq:delta_zeta_threefold}, one does find that again
\begin{equation}
\zeta = 0\,.
\end{equation}

\subsubsection*{Boundary charge operator}
Finally, the boundary charge operator associated to the type $\mathrm{IV}_1$ singularity is given by
\begin{equation}\label{eq:charge-operator_IV1}
Q_\infty = \Lambda\begin{pmatrix}
0 & -\frac{3i}{2} & 0 & 0\\
\frac{i}{2} & 0 & -2i& 0\\
0 & \frac{i}{2} & 0 & \frac{3i}{2}\\
0 & 0 & -\frac{i}{2} & 0
\end{pmatrix}\Lambda^{-1}\,.
\end{equation}

\end{subappendices}

%%%%%%%%%%%%%%%%%%%%%%%%%%%%%%%%%%%%%%%%%%%%%%%%%%%%%%%%%%%%%%%%%%%%
\chapter{Asymptotic Hodge Theory II: Boundary to Bulk}
\label{chap:asymp_Hodge_II}
\epigraph{This chapter is based on: Thomas W. Grimm, Jeroen Monnee, Damian van de Heisteeg: \emph{Bulk Reconstruction in Moduli Space Holography},\\ \textbf{JHEP 05 (2022) 010}, \href{https://arxiv.org/abs/2103.12746}{\textbf{[arXiv: 2103.12746]}}
}

\noindent This is the second of two chapters dedicated to asymptotic Hodge theory. In the previous chapter we have seen that to each enhancement chain of singularities in the moduli space one can associate a collection of \textit{boundary data} consisting of
\begin{equation}
\{Q_\infty, N^\bullet_{i}, \delta_i\}\,,\qquad i=1,\ldots, m\,,
\end{equation}
where we recall that $Q_\infty$ denotes the boundary charge operator, which encodes the boundary Hodge structure $F_\infty$, $N^\bullet_{i}$ denotes a set of $m$ commuting real $\mathfrak{sl}(2)$-triples which are horizontal with respect to $Q_\infty$, and $\delta_i$ denotes the phase operator associated to the $i$-th limiting mixed Hodge structure. Furthermore, we have seen that, given this data, one can immediately write down the $\mathrm{Sl}(2)$-orbit approximation of the period map $h_{\mathrm{Sl}(2)}$ which determines the leading order behaviour of various physical couplings in the chosen growth sector. In a sense, on can think of these results as comprising the first part of the celebrated $\mathrm{Sl}(2)$-orbit theorem of Cattani, Kaplan, and Schmid \cite{CKS}. \\

\noindent In this chapter, we will explain how the same set of boundary data in fact completely fixes the \textit{nilpotent orbit approximation} $h_{\mathrm{nil}}$ of the period map as well, through a very intricate recursive procedure which will be referred to as a \textbf{bulk reconstruction}.\footnote{The name is inspired by a similar procedure in the context of the AdS/CFT correspondence, in which one aims to represent bulk fields purely in terms of CFT operators, see for example \cite{Kajuri:2020vxf}. In fact, this similarity lead to a proposal for a notion of ``moduli space holography'' in \cite{Grimm:2020cda}, which will be discussed in more detail in chapter \ref{chap:WZW}} One can think of this as comprising the second part of the $\mathrm{Sl}(2)$-orbit theorem. In essence, this chapter explains the core mechanisms behind the proof of this part of the theorem. Furthermore, we aim to provide a clear and concrete algorithm which can be used to actually compute $h_{\mathrm{nil}}$, and illustrate the procedure in a collection of examples. To this end, we first give a detailed explanation of the one-parameter bulk reconstruction procedure in section \ref{sec:bulk-reconstruction_single}. To exemplify the construction, we explicitly perform the bulk reconstruction procedure for all singularity types that can arise for one-parameter families of Calabi--Yau threefolds in section \ref{reconstructing_examples}, building on the same examples discussed in appendix \ref{sec:asymp_Hodge_examples}. In section \ref{sec:bulk-reconstruction_multi} we discuss how the general $m$-parameter bulk reconstruction procedure can in fact be seen as an inductive application of the one-parameter result, following the discussion in chapter \ref{chap:asymp_Hodge_I}, and illustrate the construction for the two-parameter type $\langle\mathrm{I}_1|\mathrm{IV}_2|\mathrm{IV}_1\rangle$ singularity. We additionally discuss some of the important properties of the nilpotent orbit expansion that result from the general construction, which will play a central role in chapter \ref{chap:finiteness}. Furthermore, in section \ref{sec:inner_prod} we briefly explain how this can be used to compute asymptotic expansions of Hodge inner products, and as an application we compute corrections to the central charge of certain BPS states that arise in the context of type IIB string theory. \\

\noindent Before we delve into the details of the bulk reconstruction procedure, we would like to make one important comment regarding a difference in the conventions used in this chapter versus the conventions used in the work \cite{Grimm:2021ikg} in which some of the computations were originally performed. 

\begin{redbox}{Conventions: the $\zeta$-operator}
In the original work \cite{Grimm:2021ikg} the one-parameter bulk reconstruction procedure was performed with respect to the $\mathbb{R}$-split limiting mixed Hodge structure as opposed to the $\mathrm{sl}(2)$-split limiting mixed Hodge structure which we will use in this chapter. As a result, the expressions for the boundary data differ by a factor of $\mathrm{Ad}\,e^{\zeta}$, which appears explicitly in the expression for the period map in \cite{Grimm:2021ikg}, while it does not appear in the discussion below. 
\end{redbox}
\newpage

\section{Bulk reconstruction: single-variable}\label{sec:bulk-reconstruction_single}

One of the important statements announced in section \ref{sec:SL2_orbit_one_variable} is that any one-parameter nilpotent orbit can be parametrized as
\begin{equation}\label{eq:Qref-choice}
	F^p_{\mathrm{nil}} = e^{tN}F^p_0 = h_{\mathrm{nil}}(x,y)F^p_\infty\,,
\end{equation}
where we recall that $F^p_\infty$ denotes the boundary Hodge structure defined by \eqref{eq:boundary-HS_multi} and $h_{\mathrm{nil}}(x,y)$ is the nilpotent orbit expansion of the period map. In this section, we show that indeed $h_{\mathrm{nil}}(x,y)$ enjoys an expansion
\begin{equation}\label{eq:hnil_nilp-expansion}
	h_{\mathrm{nil}}(x,y) = e^{xN}\left(1+\frac{g_1}{y}+\frac{g_2}{y^2}+\cdots\right) y^{-\frac{1}{2}N^0}\,,
\end{equation}
and provide an explicit algorithm to compute the coefficients $g_i$ in terms of the boundary data \eqref{eq:boundary_data_one_variable} associated to the singularity in question, following the seminal work of Cattani, Kaplan, and Schmid \cite{CKS}. \\

\noindent From a practical point of view, we are interested in determining solutions to the horizontality conditions \eqref{eq:horizontality_h} under the assumption that the period map satisfies
\begin{equation}
	h(x+c,y) = e^{cN}h(x,y)\,,
\end{equation}
which is the case in the nilpotent orbit approximation. The strategy will consist of three steps, which we briefly outline below.
\begin{itemize}
	\item \textbf{Step 1:}\\
	First, we recall that from $h_{\mathrm{nil}}$ one can define three real operators $\cN^0(y), \cN^\pm(y)$ that satisfy 
	Nahm's equations \eqref{eq:Nahm_N}. As the first step, we show in section \ref{CKS-recursion} that after combining $\cN^0(y), \cN^\pm(y)$ in a clever way one can translate 
	Nahm's equations into an intricate set of recursion relations, which we refer to as the \textbf{CKS recursion}. 
	\item \textbf{Step 2:}\\
	In the second step, we explain how the boundary data, namely the $\slt$-triple and the phase operator $\delta$, provide the initial conditions of the recursion. At this point also the remaining conditions coming from the $Q$-constraint \eqref{eq:Q_constraint_N} play an essential role. This recursion relation can then be solved, which yields a unique solution for $\cN^0(y), \cN^\pm(y)$.
	\item \textbf{Step 3:}\\
	Having found a solution for $\cN^0(y), \cN^\pm(y)$ we will then 
	describe in section \ref{subsec:h_reconstruction} how this leads to a solution for $h_{\mathrm{nil}}(x,y)$.
\end{itemize}

\subsection{Preliminaries}

\subsubsection*{Rotating to the complex algebra}
In a similar fashion as was done in appendix \ref{app:horizontality}, the real $\slt$-triple generated by $N^\bullet\in\mathfrak{g}_{\mathbb{R}}$ gives rise to a complex $\slt$-triple generated by $L_\bullet\in\mathfrak{g}_\mathbb{C}$ via the Cayley transform
\begin{equation}
\label{eq:rho_switch}
L_{\bullet} = \rho N^\bullet \rho^{-1}\,,
\end{equation}
where
\begin{equation} \label{def-rho}
\rho=\mathrm{exp}\Big[\frac{i\pi}{4}\big(N^+ + N^-\big)\Big]=\mathrm{exp}\Big[\frac{i\pi}{4}\left(L_{1} + L_{-1}\right)\Big]\ .
\end{equation}
The fact that the real $\slt$-triple is horizontal with respect to the boundary charge operator $Q_\infty$, recall the relations \eqref{eq:horizontal_sl2_real}, can then be phrased in terms of the following commutation relations with respect to the complex $\slt$-triple 
\begin{equation}
[Q_\infty, L_q] =q L_q \,,\quad q = 1,0,-1 \, .
\end{equation}
A crucial observation is that $L_0$ commutes with $Q_\infty$, which allows us to find a common eigenbasis for the two operators. For this reason we will mostly work with the complex algebra. There is, of course, one more operator which commutes with both $L_0$ and $Q_\infty$, namely the Casimir operator $L^2$. It is given by
\begin{equation} \label{def-Casimir}
L^2=2L_{1}L_{-1} + 2L_{-1} L_{1} +(L_0)^2 \, .
\end{equation}
Let us note that there is another way of interpreting the operators $L_q, Q_\infty$. In fact, we see that 
$\hat Q \equiv  Q_\infty - \frac{1}{2} L_0$ commutes with all $L_{q}$ and hence we have the algebra 
\beq
\mathfrak{sl}(2,\bbR) \oplus \mathfrak{u}(1): \ L_q, \hat Q\ . 
\eeq
in this work we prefer to work with the charge operator $Q_\infty$ instead of $\hat{Q}$, but note that $\hat{Q}$ does appear naturally in the bulk reconstruction in e.g.~\eqref{charges_cL}. 

\subsubsection*{Various eigenspace decompositions}
Since much of our discussion revolves around solving operator equations, it will be extremely useful to split the space of operators using the eigenvalues of $L^2, L_0$ and $Q_\infty$. Concretely, given any operator $\cO\in\mathfrak{g}_{\mathbb{C}}$ we may decompose it as
\beq \label{weight-charge-expansion-operator1}
\cO = \sum_{0 \leq d \leq D} \sum_{-d\leq s\leq d} \sum_{-D\leq q \leq D} \cO^{(d,s)}_q \ , 
\eeq
with 
\bea \label{weight-charge-expansion-operator2}
(\ad L)^2 \cO^{(d,s)}_q &=& d(d+2)\, \cO^{(d,s)}_q  \ , \nn \\
\big[L_0,\cO^{(d,s)}_q\big] &=&  s\, \cO^{(d,s)}_q\ ,  \\
\big[Q_\infty ,\cO^{(d,s)}_q \big] &=& q\, \cO^{(d,s)}_q\ ,  \nn 
\eea
where we have used the shorthand notation $(\ad L)^2$ to denote replacing each left-multiplication in \eqref{def-Casimir} with an 
adjoint action, i.e.~we have set
\beq
(\ad L)^2 \cO :=2\big[L_{1},\big[L_{-1},\cO \big]\big]+2\big[L_{-1},\big[L_{1},\cO \big]\big]+\big[L_0,\big[L_0,\cO \big]\big]\ .
\eeq
We call $d, s$ and $q$ the \textit{highest weight}, \textit{weight} and \textit{charge} of the operator, respectively. In the following, it is sometimes not necessary to perform all three decompositions \eqref{weight-charge-expansion-operator2}.
We will then employ the notation
\beq \label{weightonly}
\cO^{(l)}_q \equiv  \sum_{d \in \bbZ} \cO^{(d,l)}_q\ ,\qquad \cO^{(d,l)} \equiv  \sum_{q \in \bbZ} \cO^{(d,l)}_q \ , \qquad \cO^{[d]} \equiv  \sum_{q,l \in \bbZ} \cO^{(d,l)}_q \ ,
\eeq
when we do not perform the highest weight decomposition, the charge decomposition, or only perform the highest weight decomposition, respectively.\\

\noindent Note that in the case we do not perform a charge decomposition, one could also have chosen to perform the highest weight and weight decomposition with respect to the real $\slt$-algebra. 
These two decompositions are related precisely by $\rho$ introduced in \eqref{def-rho}. In other words, if $\cO$ is an operator with weight $s$ under $N^0$, then $\hat{\cO}=\rho\cO\rho^{-1}$ is an operator with weight $s$ under $L_0$, and similarly for the highest weight. In the following we will add a hat to an operator if it is obtained via the transformation with $\rho$. This is particularly relevant if $\cO$ is a real operator. Such operators are naturally decomposed with respect to the real $\slt$-algebra but cannot be an eigenoperator under $Q_\infty$, unless they are uncharged.  

\subsubsection*{The phase operator}
Lastly, let us recall the important properties of the phase operator $\delta\in\mathfrak{g}_{\mathbb{R}}$, as it is an essential part of the boundary data and determines crucially the complexity of the associated nilpotent orbit expansion. In accordance with the notation introduced above, we use $\delta$ to define
\begin{equation}
\hat{\delta} = \rho\delta\rho^{-1} \ ,
\end{equation} 
which is thus an element of $ \mathfrak{g}_{\bbC}$.
In terms of $\hat{\delta}$ the properties of the phase operator, recall the discussion in section \ref{sec:MHS}, can be formulated as follows.
\begin{itemize}
\item \textbf{Property 1:}\\
Firstly, it has to commute with $L_{-1}$, so that
\begin{equation}
\label{eq:delta_Lmin}
[L_{-1},\hat{\delta}]=0\ .
\end{equation}
In other words, each component of $\hat{\delta}$ in an irreducible representation of the complex $\slt$-triple should be a lowest-weight state, since its weight cannot be lowered further due to \eqref{eq:delta_Lmin}. 
\item \textbf{Property 2:}\\
Secondly, its components must all have weight less than or equal to $-2$, and charge less than or equal to $-1$. In other words, it admits an expansion
\begin{equation}\label{eq:deltaexpansion}
\hat{\delta} = \sum_{s\leq -2} \sum_{q \leq -1} \hat{\delta}^{(s)}_q\ .
\end{equation}
\end{itemize}
The first property \eqref{eq:delta_Lmin} is simply the complex analogue of \eqref{eq:delta_commutes_N}, while the second property \eqref{eq:deltaexpansion} is equivalent to the condition \eqref{eq:delta_Deligne}. Regarding the latter, one has the dictionary 
\begin{equation}
\hat{\delta}^{(s)}_q \quad \leftrightarrow \quad \delta_{s-q,q}\,,
\end{equation}
where we recall that $\delta_{p,q}$ denote the components of $\delta$ with respect to the Deligne splitting c.f.~\eqref{eq:delta_Deligne}. It will turn out that the various $(s,q)$-components of $\hat{\delta}$ will enter into the bulk reconstruction separately, so that it is of utmost importance to have control over these various components. This highlights one of the computational advantages of working in the complex algebra. Indeed, if instead one would work purely in the real algebra, one would have to keep track of the various $(p,q)$-components of $\delta$ with respect to the Deligne splitting, which cannot be described nicely in terms of eigenspaces of some auxiliary operator.

\subsection{The CKS Recursion} \label{CKS-recursion}
Having discussed some basic properties of the boundary data, let us now turn to the equations we would like to solve, which we recall here for convenience
\bea
\partial_{y} \cN^{\pm} &=& \pm \tfrac{1}{2} [\cN^{\pm},\cN^0] \ , \qquad  
\partial_{y} \cN^0 =- [\cN^{+} ,\cN^{-}]\ , \label{Nahm3}\\ 
\big[Q_{\infty},\cN^{0}\big]&=& i  (\cN^+ + \cN^-) \ , \qquad     \big[Q_{\infty},\cN^{\pm}\big] =-\tfrac{i}{2} \cN^0 \ , \label{Q_constr3}
\eea
as already given in \eqref{eq:Q_constraint_N} and \eqref{eq:Nahm_N}. 
Note that we have made a particular choice for the reference charge operator $Q_{\mathrm{ref}}$, namely
\begin{equation}
Q_{\mathrm{ref}}=Q_\infty\,,
\end{equation}
in order to match with \eqref{eq:Qref-choice}. In the following we will solve these equations for $\cN^0,\cN^\pm$ and identify solutions that match the boundary data.

\subsubsection*{Rotating to the complex algebra}
As advocated in \cite{Grimm:2020cda} it will be convenient to 
the transform the $\cN^0,\cN^\pm$ with the operator $\rho$ defined in \eqref{def-rho}, which was already employed in \eqref{eq:rho_switch} to transform from the real $\mathfrak{sl}(2,\mathbb{R})$-triple $\{N^+,N^0,N^-\}$ to the complex $\mathfrak{sl}(2,\mathbb{C})$-triple $\{L_{1},L_0,L_{-1}\}$. This transformation allows us to work in the complex algebra $\mathfrak{g}_{\mathbb{C}}$, which is necessary 
to discuss eigenoperators under $Q_\infty$ as alluded to at the beginning of this section. Concretely, we define operators 
\begin{equation} \label{def-bfLbullet}
	\mathbf{L}^\bullet = \rho \cN^\bullet\rho^{-1}\ , 
\end{equation}
where again $ \bullet $ stands for either $0$, $+$, or $-$. In terms of these operators Nahm's equations take the form
\begin{equation}
	\label{eq:Nahm_L}
	\partial_y \mathbf{L}^\pm = \pm \frac{1}{2}[\mathbf{L}^\pm, \mathbf{L}^0],\quad \partial_y\mathbf{L}^0 = -[\mathbf{L}^+,\mathbf{L}^-]\, .
\end{equation}
Moreover, the $Q$-constraint \eqref{Q_constr3} becomes 
\bea \label{charges_cL}
\big[2Q_\infty -  L_0, \mathbf{L}^{0}\big]&=& 2 i  ( \mathbf{L}^+ +  \mathbf{L}^-) +i \big[  L_1, \mathbf{L}^0\big]- i[L_{-1},  \mathbf{L}^{0}\big] \ ,  \nn \\
\big[2Q_\infty  -  L_0 , \mathbf{L}^{\pm}\big]   &=& -i  \mathbf{L}^0 + i \big[ L_1, \mathbf{L}^\pm\big] - i[L_{-1},  \mathbf{L}^{\pm}\big] \ .
\eea
Note that this $Q$-constraint does not appear in this form in \cite{CKS}. However, as was shown in \cite{Grimm:2020cda}, and will be recalled in appendix \ref{app:input}, this 
approach allows us to more easily impose it on the solution.

\subsubsection*{Setting up the CKS recursion}
We now discuss the general procedure to solve \eqref{eq:Nahm_L} under the constraint \eqref{charges_cL}. Let us recall that the $\mathbf{L}^\bullet$ are operators which act on the finite-dimensional vector space $H_{\mathbb{C}}$ and may therefore be represented by matrices. Alternatively, one may pick a basis of $\mathfrak{g}_\bbC$ and represent each $\mathbf{L}^\bullet$ as a vector with respect to this basis. The main strategy of CKS is to solve Nahm's equations by combining the $\mathbf{L}^\bullet$ into a single vector $\Phi$ as
\begin{equation} \label{def-Phi_L}
	\Phi = \begin{pmatrix}
		\mathbf{L}^+\\ \mathbf{L}^0 \\ \mathbf{L}^-\\
	\end{pmatrix}.
\end{equation}
We note that $\Phi$ can either be viewed as a 3-vector with matrices as entries, or as a $(3\times\mathrm{dim}\;\mathfrak{g})$-component vector. The former interpretation will be most useful for formal manipulations, whereas the latter will be more practical to use in concrete examples, as we will see in section \ref{reconstructing_examples}. The reason for introducing $\Phi$ is that one can construct another $\slt$-triple that acts on it, which allows one to perform further decompositions besides the ones for the separate $\mathbf{L}^\bullet$. We will introduce this triple shortly. First, in order to write down Nahm's equations in terms of $\Phi$, we introduce a bilinear $B$ acting on two vectors $\Phi$ and $\tilde{\Phi}$ as\footnote{In \cite{CKS} the notation $Q$ is used for the bilinear $B$, whose expressions differ by a choice of basis. Our basis is the same as in \cite{Pearlstein2006}.}
\begin{equation}
	\label{eq:bilinear_B}
	B(\Phi,\tilde{\Phi}) =\frac{1}{4} \begin{pmatrix}
		[\mathbf{L}^0,\tilde{\mathbf{L}}^+] - [\mathbf{L}^+,\tilde{\mathbf{L}}^0]\\
		2 [\mathbf{L}^+,\tilde{\mathbf{L}}^-] - 2 [\mathbf{L}^-,\tilde{\mathbf{L}}^+]  \\
		[\mathbf{L}^-,\tilde{\mathbf{L}}^0] - [\mathbf{L}^0,\tilde{\mathbf{L}}^-]
	\end{pmatrix}.
\end{equation}
Note that $B(\Phi,\tilde{\Phi})$ is symmetric under exchanging $\Phi$ and $\tilde{\Phi}$, hence for $\Phi=\tilde{\Phi}$ this takes the simple form
\begin{equation}
	B(\Phi,\Phi) = \half\begin{pmatrix}
		[\mathbf{L}^0,\mathbf{L}^+] \\ 2[\mathbf{L}^+, \mathbf{L}^-] \\ [\mathbf{L}^-,\mathbf{L}^0]
	\end{pmatrix}.
\end{equation}
We readily see that \eqref{eq:Nahm_L} can then be written as
\begin{equation}
	\label{eq:eom_B}
	\frac{d\Phi}{dy}=-B\left(\Phi,\Phi\right).
\end{equation}
To turn the differential equation \eqref{eq:eom_B} into an algebraic recursion relation we now perform a series expansion of $\Phi$ around $y\rightarrow\infty$. In order to match the solution to the boundary data, we impose that the leading behaviour of $\Phi$ is given by
\begin{equation} \label{leading-Phi}
	\Phi= y^{-1}\begin{pmatrix}
		L_{+1} \\ L_0 \\ L_{-1}
	\end{pmatrix}+\cO(y^{-3/2})\, .
\end{equation}
In other words, the leading behaviour of $\Phi$ is given precisely by the $\slt$-triple $(L_{0},L_{\pm 1})$ of the boundary data. Recall from the discussion in section \ref{sec:SL2_orbit_one_variable} that this simply corresponds to the statement that the $\mathrm{Sl}(2)$-orbit approximation $h_{\mathrm{Sl}(2)}$ dictates the leading order behaviour of the period map. To parametrize possible sub-leading terms in $\Phi$, we make the ansatz
\begin{equation} \label{Phi-expand}
	\Phi=\sum_{n\geq 0} \Phi_n y^{-1-n/2}= \sum_{n\geq 0} \begin{pmatrix}
		L_n^+\\ L_n^0 \\ L_n^-\\
	\end{pmatrix}y^{-1-n/2},\qquad \Phi_0:=\begin{pmatrix}
		L_{+1} \\ L_0 \\ L_{-1}
	\end{pmatrix}.
\end{equation}
In terms of the $\Phi_n$, \eqref{eq:eom_B} reduces to the following recursion relation.
\begin{subbox}{The CKS recursion}
	\begin{equation}
		\label{eq:recursion_CKS}
		(n+2)\Phi_n-4B(\Phi_0,\Phi_n)=2\sum_{0<k<n}B(\Phi_k,\Phi_{n-k})\, .
	\end{equation}
\end{subbox}

\subsubsection*{Simplifying the CKS recursion}
In principle, the relation \eqref{eq:recursion_CKS} allows one to determine the $\Phi_n$ in terms of the previous $\Phi_k$, $k<n$. However, the expression for $B(\Phi_0,\Phi_n)$ will generically be very complex. In order to simplify this, we proceed in two steps. First, we make use of the highest weight decomposition \eqref{weight-charge-expansion-operator1} to decompose the operators $L_n^\bullet$ as 
\begin{equation}
	L_n^\bullet = \sum_{d\geq 0} (L_n^\bullet)^{[d]}\ ,
\end{equation}
where each $(L_n^\bullet)^{[d]}$ is an operator of highest weight $d$ in the notation introduced in \eqref{weightonly}. Note that in the discussion of the recursion relations it will not be necessary to perform the weight or charge decomposition. Using this decomposition we have split the various components of $\Phi_n$ into different pieces, which can be collected into
\begin{equation}
	\Phi_n^d = \begin{pmatrix}
		(L_n^+)^{[d]} \\ (L_n^0)^{[d]} \\ (L_n^-)^{[d]}
	\end{pmatrix}.
\end{equation}
We can, however, perform a further decomposition of the full $\Phi_n^d$ by diagonalizing the operator $B(\Phi_0,\cdot)$ which appears in \eqref{eq:recursion_CKS}. This will lead to a great simplification of the recursion. For convenience, let us abbreviate $B_0:=B(\Phi_0,\cdot)$. To be explicit, using \eqref{eq:bilinear_B} this operator can be written as
\begin{equation}
	\label{eq:B_identity}
	B_0= \frac{1}{4}\begin{pmatrix}
		\ad L_0 & -\ad L_{1} & 0\\
		-2\ad L_{-1} & 0 & 2 \ad L_{1}\\
		0 & \ad L_{-1} & -\ad  L_0
	\end{pmatrix}.
\end{equation}
To evaluate the action of $B_0$ on $\Phi$ we use the interpretation of $\Phi$ as a 3-vector consisting of matrices $\{\mathbf{L}^+,\mathbf{L}^0,\mathbf{L}^-\}$ on which $(L_0,L_{\pm 1})$ can act via the adjoint action. In other words
\begin{equation} \label{B0Phi}
	B_0(\Phi) = \frac{1}{4}\begin{pmatrix}
		[L_0,\mathbf{L}^+] - [L_1,\mathbf{L}^0]\\
		2 [L_{1},\mathbf{L}^-]- 2 [L_{-1},\mathbf{L}^+]\\
		[L_{-1},\mathbf{L}^0] - [L_0,\mathbf{L}^-]
	\end{pmatrix}\ .
\end{equation}
Our next goal will be to find a split of $\Phi$ which diagonalizes $B_0$ and hence lets us 
evaluate the second term in \eqref{eq:recursion_CKS}. The remarkable idea of CKS is to introduce yet another $\slt$-decomposition, which now acts on the 3-vectors $\Phi^d_n$. This new decomposition 
allows us to split 
\begin{equation} \label{Phi-epsilon-split}
	\Phi_n^d= \sum_{\epsilon=-1,0,1} \Phi_n^{d,\epsilon}\ .
\end{equation}
We stress that this is \textit{not} the $(d,s)$-decomposition introduced in \eqref{weightonly} for which the indices were written with 
brackets. 
The second eigenvalue $\epsilon$ arises from an $\slt$-triple $(\Lambda^0,\Lambda^\pm)$ which acts on $\Phi$ by also 
mixing the 3-vector components $\mathbf{L}^\bullet$. This $\slt$-triple is given by 
\begin{equation}
	\label{eq:sl2_tensor}
	\begin{aligned}
		\Lambda^+ &= \begin{pmatrix}
			\ad L^+ & 0 & 0 \\
			2 & \ad L^+& 0 \\
			0 & -1 & \ad L^+ \\
		\end{pmatrix} , \\
		\Lambda^0 &= \begin{pmatrix}
			\ad L^0-2 & 0 & 0 \\
			0 & \ad L^0 & 0 \\
			0 & 0 & \ad L^0+2 \\
		\end{pmatrix} ,\\
		\Lambda^- &= \begin{pmatrix}
			\ad L^- & 1 & 0 \\
			0 & \ad L^- & -2 \\
			0 & 0 & \ad L^- \\
		\end{pmatrix}.
	\end{aligned}
\end{equation}
By slight abuse of notation the integer entries are proportional to identity matrices.\footnote{To elaborate, the $3\times 3$ matrices
	\begin{equation}
		\begin{pmatrix}
			0 & 0 & 0\\
			2 & 0 & 0\\
			0 & -1 & 0
		\end{pmatrix},\qquad \begin{pmatrix}
			-2 & 0 & 0\\
			0 & 0 & 0\\
			0 & 0 & 2
		\end{pmatrix}, \qquad \begin{pmatrix}
			0 & 1 & 0\\
			0 & 0 & -2\\
			0 & 0 & 0
		\end{pmatrix}\, .
	\end{equation}
	also form an $\slt$-triple (more precisely, they correspond to the co-adjoint representation). The $\slt$-triple $(\Lambda^+, \Lambda^0, \Lambda^-)$ is then obtained by taking the tensor product between the above generators and the $\slt$-triple $(L_{1} ,  L_0 , L_{-1})$.} The label $\epsilon$ appearing in \eqref{Phi-epsilon-split} is then related to the eigenvalue under the Casimir $\Lambda^{2}$ via
\begin{equation}
	\Lambda^{2}= 2 \Lambda^+  \Lambda^- +2\Lambda^-  \Lambda^+ +(\Lambda^0)^2: \qquad \Lambda^{2} \Phi^{d,\epsilon} = (d+2\epsilon)(d+2\epsilon+2) \Phi^{d,\epsilon}\, .
\end{equation}
In other words, for a given $d$, each $\Phi^d$ splits into three components $\Phi^{d,\epsilon}$, which have highest weight $d+2\epsilon$ with respect to the Casimir $\Lambda^2$.
Using \eqref{eq:sl2_tensor} it is straightforward to compute $\Lambda^2$ explicitly as
\begin{equation}
	\Lambda^2 = \begin{pmatrix}
		(\ad L)^2+8-4\ad L^0 & 4\ad L^+ & 0\\
		8\ad L^- & (\ad L)^2+8& -8 \ad L^+\\
		0 & -4\ad L^- & (\ad L)^2+8 +4\ad L^0 
	\end{pmatrix}\ . 
\end{equation}
We can now compare this expression with the expression \eqref{eq:B_identity} for $B_0$ and 
observe that the Casimir can also be written as
\beq
\Lambda^2   = \big( (\ad L)^2+8 \big)\mathbb{I}_{3\times 3} - 16 B_0\ .
\eeq
In other words, we see that the components $\Phi_n^{d,\epsilon}$ are also eigenvectors of $B_0$. In fact,
we evaluate 
\beq
\label{eq:B_eigenvalue}
-4B_0(\Phi_n^{d,\epsilon})=\left\{\epsilon(1+d+\epsilon)-2\right\}\Phi_n^{d,\epsilon}\ . 
\end{equation}
Returning to the recursion relation \eqref{eq:recursion_CKS}, we see that for such eigenvectors it simplifies to the following.
\begin{subbox}{The CKS recursion}
\begin{equation}
	\label{eq:CKS_recursion_epsilon}
	\big(n+\epsilon(1+d+\epsilon) \big) \Phi_n^{d,\epsilon}=2\sum_{0<k<n} B(\Phi_k,\Phi_{n-k})^{d,\epsilon} \, .
\end{equation}
\end{subbox}
\noindent The recursion \eqref{eq:CKS_recursion_epsilon} is the master equation that encodes the constraints on the 
coefficients $\Phi_n^{d,\epsilon}$ for any solution $\Phi$ of~\eqref{eq:eom_B}. 

\subsubsection*{Input data}
Let us make some further remarks regarding the structure of the CKS recursion. To begin with, we note that the 
representation theory of $\slt$ implies that the number of linearly independent operators with a given $d$ 
and $\epsilon$ is equal to $d+2\epsilon+1$. This implies, in particular, that $\Phi^{1,-1}_1 = 0$. 
Furthermore, for $n=1$ the right-hand side of \eqref{eq:CKS_recursion_epsilon} vanishes and we conclude that also 
$\Phi_1^{d,\epsilon}=0$ for $(d,\epsilon)\neq (1,-1)$. Taken together, we thus find that 
\begin{equation}
\Phi_1=0\ .
\end{equation}
Applying this result to the expansion \eqref{Phi-expand} of $\Phi$ this means that the term proportional to $y^{-3/2}$ vanishes and the first sub-leading term is of order $y^{-2}$. Inspecting the recursion \eqref{eq:CKS_recursion_epsilon} we see that for $n>1$ the $\Phi_n^{d,\epsilon}$ can be obtained recursively by computing the action of $B$ on $\Phi_k$ and $\Phi_{n-k}$, $k<n$, and projecting the result onto its $d,\epsilon$ components. The only $\Phi_n^{d,\epsilon}$ which are not determined recusively from \eqref{eq:CKS_recursion_epsilon} are those with highest weight $d=n$.
It is easy to check that $\Phi_n^{n,-1}$ actually does not appear on the left-hand side, since its pre-factor vanishes for this component. 
To obtain the constraints on $\Phi_n^{n,1}$ and $\Phi_n^{n,0}$ one has to use the properties of $B$
to realize that the right-hand side in \eqref{eq:CKS_recursion_epsilon} vanishes and therefore implies that\footnote{To be precise, one uses the fact that for two operators $S$ and $T$, one has that $B(S^{d,\epsilon}, T^{d',\epsilon'})^{d'',\epsilon''}=0$ unless the following three conditions hold: (1) $0\leq d'' \leq d+d'$, (2) $d'' \equiv d+d' \;\mathrm{mod}\;2,$ and (3) $0\leq d''+2\epsilon'' \leq d+d'+2\epsilon+2\epsilon'$. These properties can be derived from the specific expression for $B$ and its behaviour with respect to the underlying $\slt$-structure. We refer the reader to \cite{CKS} for more details. }  
\beq
\Phi_n^{n,1}=\Phi_n^{n,0}= 0 \ . 
\eeq
In conclusion, we find that we need to supply 
\beq \label{input-data}
\text{input data}:   \quad \Phi_n^{n,-1}\ 
\eeq  
for the recursion. Recalling that there 
are maximally  $d+2\epsilon+1$ independent $\Phi_n^{d,\epsilon}$, we conclude that there are generically $n-1$ linearly independent $\Phi_n^{n,-1}$ that need to be given. \\

\noindent We will now discuss a way to encode the input data  $\Phi_n^{n,-1}$ in an efficient way, which makes the $n-1$ linearly independent degrees of freedom manifest. Furthermore, note that we have so far only discussed the differential constraint \eqref{eq:Nahm_L} and it remains to impose \eqref{charges_cL} on any 
solution. In the following we will evaluate the conditions \eqref{charges_cL} imposes on the input data $\Phi_n^{n,-1}$.
From \eqref{eq:B_eigenvalue} we know that $4B_0(\Phi_n^{n,-1})=(n+2)\Phi_n^{n,-1}$. As we will recall in appendix~\ref{app:input} this equation can be solved by the following ansatz \cite{CKS}.
\begin{subbox}{Input data ansatz for the CKS recursion}
\begin{equation}
	\label{eq:input_data}
	\Phi_n^{n,-1}= \sum_{1\leq s,q \leq n-1} a^{n,s}_q \begin{pmatrix}
		-\frac{1}{n-s} \left(\mathrm{ad}\;L_1\right)^{s+1}\\
		2 \left(\mathrm{ad}\;L_1\right)^{s} \\
		(n-s+1) \left(\mathrm{ad}\;L_1\right)^{s-1}
	\end{pmatrix}\hat{\eta}^{(-n)}_{-q} \ , 
\end{equation}
where $\hat \eta \in \mathfrak{g}_{\bbC}$ has to obey
\beq \label{eta-properties}
\hat \eta= \sum_{s\leq -2} \sum_{q \leq -1} \hat{\eta}^{(s)}_q \ , \qquad [L_{-1},\hat \eta]= 0 \ .
\eeq
\end{subbox}
\noindent Note that $\Phi$ is a 3-vector made out of operators $\mathbf{L}^\bullet$, which themselves stem 
from real operators $\cN^\bullet \in \mathfrak{g}_\bbR$ by transformation with $\rho$ as given in \eqref{def-bfLbullet}. 
This implies that also $\hat \eta$ can 
be obtained from a real operator $\eta = \rho^{-1} \hat \eta \rho \in \mathfrak{g}_\bbR $. We also see in \eqref{eq:input_data} that there are indeed 
$n-1$ linearly independent contributions to $\Phi_n^{n,-1}$, which correspond to the various charge components of the operator $\hat{\eta}$. 
The operator $\hat \eta$ now 
encodes the input data of the recursion relation. 
It is important to stress, however, that the condition 
$4B_0(\Phi_n^{n,-1})=(n+2)\Phi_n^{n,-1}$ does not yet fix the complex coefficients $a^{n,s}_q$. In order to 
fix these coefficients we now impose the $Q$-constraint on the entries of $\Phi_n^{n,-1}$. A direct computation, which can be found 
in appendix~\ref{app:input}, reveals that 
\begin{equation}
a^{n,s}_q=i^{s-1} \frac{(n-s)!}{n!} b^{s-1}_{q-1,n-q-1}\, ,
\end{equation}
where the coefficients $b^{s}_{p,q}$ are defined by
\begin{equation}
(1-x)^p(1+x)^q = \sum_{s=0}^{p+q} b^s_{p,q} x^s \, ,\quad p,q\geq 0\, .
\end{equation}
To summarize, we have reduced Nahm's equations to a set of recursion relations \eqref{eq:CKS_recursion_epsilon} for the components $\Phi_n^{d,\epsilon}$ which encode the full $\Phi$ defined in \eqref{def-Phi_L}. The initial conditions are determined by the boundary data $\hat \eta$ and 
the $\slt$-triple via \eqref{eq:input_data}. In order to show this we have  imposed the $Q$-constraint \eqref{charges_cL} on $\Phi^{n,-1}_n$. Using 
the recursion relations \eqref{eq:CKS_recursion_epsilon} this ensures that the full solution obeys this constraint.  
We next discuss how one can relate a solution $\Phi$ back to period map $h_{\mathrm{nil}}$ and how $\hat \eta$ is determined in terms of the 
phase operator $\delta$ which was part of the boundary data.

\subsection{Recovering the period map and a matching condition}
\label{subsec:h_reconstruction}

In this section we will show how a solution for $\Phi$ can be used to 
obtain a solution for $h_{\mathrm{nil}}$. Furthermore, we will see how a single matching condition \eqref{eq:matching_delta} allows us to fix the input data 
$\hat \eta$ for the CKS recursion in terms of the boundary data. Additionally, this matching condition will also clarify the role played by the operator $\zeta$ that was introduced in \ref{eq:def_zeta} to perform the rotation to the $\mathrm{sl}(2)$-split Deligne splitting and, in particular, how exactly it is fixed in terms of the phase operator $\delta$. \\

\noindent Recall that the vector $\Phi$ contains $\mathbf{L}^0$ as one its components. Furthermore, we recall the relations
\begin{equation}
\label{eq:h_cN}
-2h^{-1}\partial_y h = \cN^0\, ,\qquad \cN^0 = \rho^{-1} \mathbf{L}^0\rho \, .
\end{equation}
In essence, we need to solve this equation to fix the $y$-dependence in $h_{\mathrm{nil}}(x,y)$ for a given $\mathbf{L}^0$. Note that if $h$ were simply a number, one could write this relation as $-2\partial_y \log(h)=\cN^0$ and solve it straightforwardly. However, because $h$ is matrix-valued one needs to do a bit more work. First, in anticipation of the result \eqref{eq:hnil_nilp-expansion}, let us write
\begin{equation}
\label{eq:def_g}
g(y) = e^{-xN} h_{\mathrm{nil}}(x,y) y^{\frac{1}{2} N^0}\ .
\end{equation}
Clearly, the resulting function is $x$-independent. Recalling that
\begin{equation}
\cN^0(y) = \frac{N^0}{y}+\cO(y^{-2})\, ,
\end{equation}
one sees that the factor  $y^{\frac{1}{2} N^0}$ in the definition of $g(y)$ ensures that the leading $N^0$-term in $\cN^0$ drops when
computing $g^{-1}\partial_y g$.  In other words, we find that $g^{-1}\partial_y g=\cO(y^{-2})$. Secondly, since $\Phi$ is described in terms of the complex algebra, it also convenient to introduce a rotated version of $g(y)$ using $\rho$
\begin{equation}
\label{eq:ghat}
\hat{g}(y) =\rho g(y) \rho^{-1}\, .
\end{equation}
By combining \eqref{eq:h_cN}, \eqref{eq:def_g} and \eqref{eq:ghat}, together with the fact that $\mathrm{Ad}\, {y^{\frac{1}{2} N^0}}=y^{\frac{1}{2}\ad  N^0}$ one can obtain the following relation between $\hat{g}$ and $\mathbf{L}^0$
\begin{equation}
\label{eq:invg_dg}
\hat{g}^{-1}\partial_y\hat{g} = \sum_{n\geq 2} B_n y^{-n}\, ,
\end{equation}
where the $B_n\in\mathfrak{g}_{\mathbb{C}}$ are comprised of particular $(d,s)$-components of the $L^0_n$ as follows
\begin{equation}
\label{eq:Bn}
B_n = -\frac{1}{2}\sum_{s\leq n-2} \sum_{d\leq 2n-2-s}\left(L_{2n-2-s}^0\right)^{(d,s)}.
\end{equation}
Note that we did not yet solve the differential equation \eqref{eq:invg_dg}, but have merely identified how the solution for $\Phi$ contributes to it through the $B_n$. We are now in a position to solve it, by again performing a series expansion of $\hat{g}$ around $y\rightarrow\infty$ and writing 
\begin{equation}
\hat{g}(y) = \sum_{k\geq 0} \frac{\hat{g}_k}{y^{k}}\ ,\qquad  \hat{g}_0=1\ .
\end{equation}
Inserting this ansatz into \eqref{eq:invg_dg} we find
\begin{equation}
\partial_y \hat{g} = \hat{g}(y)\sum_{n\geq 2} B_n y^{-n}=\sum_{m\geq 0} \sum_{n\geq 2} \hat{g}_m B_n y^{-m-n}=\sum_{k\geq 1} \Big[\sum_{j=1}^k \hat{g}_{k-j} B_{j+1}\Big] y^{-k-1},
\end{equation}
where in the last line we have changed summation variables. Comparing this result with the series expansion of $\partial_y \hat{g}$ we find
the following general solution.
\begin{subbox}{Recursive solution for $\hat{g}_k$}
\beq \label{eq:recursion_g_1}
\hat g_k = P_k (B_2,...,B_{k+1})\, ,
\eeq
where the $P_k$ are iteratively defined non-commutative polynomials
\begin{equation}
	\label{eq:recursion_g_2}
	P_0=1\ ,\qquad P_k=-\frac{1}{k} \sum_{j=1}^k P_{k-j} B_{j+1} \ ,
\end{equation}
and we recall that $B_n$ is defined in equation \eqref{eq:Bn}.
\end{subbox}
\noindent We see that each $\hat{g}_k$ is recursively given in terms of the $\hat{g}_{k-j}$ and $B_{j+1}$. Of course, solving this recursion may still be very complicated, but we will show that it can be done in our examples. It is also interesting to note that while the $B_n$ are elements of the algebra, the $\hat{g}_k$ are, generically, not. Indeed, the algebra is closed under the commutator, but \eqref{eq:recursion_g_1}, \eqref{eq:recursion_g_2} contains only the product of matrices. As a result the full function $\hat{g}(y)$ is also not an algebra element. 
Taking these findings together and rotating back to the real algebra, we thus arrive at the nilpotent orbit expansion of the period map
\beq \label{solution-expansion}
h_{\mathrm{nil}}(x,y) = e^{x N^-}  \Big( 1 + \frac{g_1}{y} + \frac{g_2}{y^2} + \ldots \Big)   y^{-\frac{1}{2} N^0}\,.
\eeq
Furthermore, one finds that the coefficients in this expansion satisfy 
\beq \label{g-conditions}
(\ad {N^-})^{n+1} g_n = 0 \ , \qquad ( g_n)^{(l)}_q = 0\, ,\quad l\geq n\, .
\eeq
Note that these conditions arise a as a very non-trivial consequence of this iterative process 
and the fact that the $B_i$ are determined by the CKS recursion and are part of the famous Sl(2)-orbit theorem \cite{CKS}. 

\subsubsection*{A matching condition}
It remains to show how the input data $\eta$ is fixed in terms of the boundary data. To this end, we recall that $g(y)$ can be interpreted as the interpolation between the $\mathrm{Sl}(2)$-orbit approximation and the nilpotent orbit approximation. At the level of Hodge filtrations, this is captured by the relation
\begin{align}
e^{iyN}F_0 &= g(y)e^{iyN}\hat{F}_0\,.	
\end{align}
Moving a factor of $e^{iyN}$ to the right-hand side and recalling the relation between $F_0$ and the $\mathrm{sl}(2)$-split $\hat{F}_0$, one can write this as
\begin{equation}\label{eq:matching_filtrations}
e^{i\delta}e^{-\zeta} \hat{F}_0  =\left[e^{-iy\,\mathrm{ad}\,N}g(y)\right]\hat{F}_0\,.
\end{equation}
The relation \eqref{eq:matching_filtrations} will serve as a matching condition between the coefficients $g_k$, and thus the input data $\eta$, and the phase operator $\delta$. To see this, we compute
\begin{align}
e^{-iy\,\mathrm{ad}\,N}g(y) &= \sum_{k,l=0}^\infty\frac{(-i)^k}{k!} y^{k-l} \left(\mathrm{ad}\,N\right)^k g_{l}\\
&= \sum_{k=0}^\infty\sum_{l=k}^\infty\frac{(-i)^k}{k!} y^{k-l} \left(\mathrm{ad}\,N\right)^k g_{l}\\
\label{eq:matching_subleading}
&=\sum_{k=0}^{\infty}\frac{(-i)^k}{k!} \left(\mathrm{ad}\,N\right)^k g_{k}+\mathcal{O}(y^{-1})\,,
\end{align}
where in the second step we used the relation \eqref{g-conditions}, which implies that (recall that $N=N^-$)
\begin{equation}
\left(\mathrm{ad}\,N\right)^k g_l=0\,,\qquad k\geq l\,,
\end{equation}
and in the third step we have simply extracted the leading term in $y^{-1}$. Since the matching condition \eqref{eq:matching_filtrations} relation should hold for all values of $y$, it should in particular hold in the limit $y\rightarrow\infty$. Taking this limit we may drop the sub-leading terms in \eqref{eq:matching_subleading} and thus we are left with the following constraint.
\begin{subbox}{The matching condition}
\begin{equation}\label{eq:matching_delta}
	e^{i\delta}e^{-\zeta} = \sum_{k=0}^{\infty}\frac{(-i)^k}{k!} \left(\mathrm{ad}\,N\right)^k g_{k}\,.
\end{equation}
\tcblower 
\textbf{Note:}\\
Since $N$ is nilpotent, the right-hand side of \eqref{eq:matching_delta} only contains \textit{finitely} many non-zero terms. To be precise, if $d$ is the largest integer such that $N^d\neq 0$, then $\left(\mathrm{ad}\,N\right)^{2d+1}=0$ and hence only the terms involving $g_0,\ldots, g_{2d}$ appear in the matching condition. In particular, for a variation of Hodge structure of weight $D$, at most $g_0,\ldots, g_{2D}$ can appear.
\end{subbox}
\noindent Naturally, one can also rotate the matching condition \eqref{eq:matching_delta} to the complex algebra, in which case it reads
\begin{equation}\label{eq:matching_delta-complex}
e^{i\hat{\delta}}e^{-\hat{\zeta}} = \sum_{k=0}^{\infty}\frac{(-i)^k}{k!} \left(\mathrm{ad}\,L_{-1}\right)^k \hat{g}_{k}\,.
\end{equation}
To see that this condition fixes $\hat{g}_k, \hat \zeta$ uniquely, we first note that 
\beq
C_{k+1}(\eta) \equiv \frac{(-i)^k}{k!} (\ad L_{-1})^k B_{k+1}  =  i  \sum_{l\geq k+1} \sum_{q\geq 1}\ b^{k-1}_{q-1,l-q-1} \hat\eta^{(-l)}_{-q}\ ,
\eeq
as can be shown by using \eqref{eq:Bn}, \eqref{eq:input_data}, \eqref{g-conditions} and \eqref{eta-properties}. Applying 
the properties \eqref{g-conditions} of $\hat g_k$ in the recursive solution \eqref{eq:recursion_g_1} with \eqref{eq:recursion_g_2} we  find that \eqref{eq:matching_delta-complex} can be written as
\beq \label{delta-zeta-eta}
e^{-i \hat \delta}  =e^{-\hat{\zeta}} \Big(1+\sum_{k\geq 1 } P_k(C_2,...,C_{k+1})\Big)\ .
\eeq
Note that the right-hand side only depends on $\hat \zeta$ and $\hat \eta$, while the left-hand side only depends on $\hat \delta$. Recalling 
that $\hat \delta, \hat \eta,\hat \zeta$ stem from the real counterparts $\delta,\eta,\zeta$ we realize that \eqref{delta-zeta-eta} gives a complex matrix equation determining two real unknowns $\eta$ and $\zeta$. As an illustrative example, for $D=3$ one finds the following relation between $\hat{\zeta}$ and $\hat{\delta}$
\begin{alignat}{2}\label{eq:zeta-delta_complex}
&\hat \zeta_{-1}^{(-3)} =- \frac{i}{2} \hat \delta_{-1}^{(-3)} \ ,\qquad &&\hat \zeta_{-1}^{(-4)} =  -\frac{3i}{4} \hat \delta_{-1}^{(-4)}\ ,\nonumber\\
&\hat \zeta_{-2}^{(-5)} = -\frac{3i}{8}\hat \delta_{-2}^{(-5)} -\frac{1}{8}\Big[\hat\delta_{-1}^{(-2)},\hat\delta_{-1}^{(-3)} \Big]\ , \qquad 
&&\hat \zeta_{-3}^{(-6)} = - \frac{1}{8} \Big[\hat\delta_{-1}^{(-2)},\hat\delta_{-2}^{(-4)} \Big]\ ,
\end{alignat}
with all other components either equal to zero or related to the listed component by complex conjugation.\footnote{To be precise, one has
\begin{equation*}
\overline{\hat{\zeta}^{(s)}_q} = \left(\overline{\hat{\zeta}}\right)^{s}_{q-s}\,,
\end{equation*}
and similarly for $\hat{\delta}$ and $\hat{\eta}$.} After rotating back to the real algebra using $\rho$, these conditions are equivalent to \eqref{eq:delta_zeta_threefold}. Similarly, the components of $\eta$ are fixed by 
\begin{alignat}{2}\label{eq:eta_delta}
&\hat \eta_{-1}^{(-2)}=-\hat \delta_{-1}^{(-2)}  \, ,\qquad 
&&\hat \eta_{-1}^{(-3)} =-  \hat \delta_{-1}^{(-3)} \, ,\nonumber \\
&\hat \eta_{-1}^{(-4)} =  -\frac{3}{4} \hat \delta_{-1}^{(-4)}\, , \qquad
&&\hat \eta_{-2}^{(-4)} =  -\frac{3}{2} \hat \delta_{-2}^{(-4)}\, , \\
&\hat \eta_{-2}^{(-5)} =  -\frac{3}{2} \hat \delta_{-2}^{(-5)}+\frac{i}{2} \Big[\hat \delta_{-1}^{(-2)},  \hat \delta_{-1}^{(-3)} \Big] \, ,\qquad 
&&\hat \eta_{-3}^{(-6)} =  -\frac{15}{8} \hat \delta_{-3}^{(-6)}-\frac{5i}{4} \Big[\hat \delta_{-1}^{(-3)},  \hat \delta_{-2}^{(-3)} \Big]\, . \nonumber
\end{alignat}
For further details on how to derive these relations we refer the reader to \cite{Grimm:2020cda} as well as \cite{vandeHeisteeg:2022gsp}. In particular, in the latter reference one can also find the generalization of these relations to the weight $D=4$ case. 
\newpage

\subsubsection*{Summary}
To close our general discussion on the one-parameter bulk reconstruction procedure, we provide below a short summary of the various steps one has to take in order to explicitly compute the nilpotent orbit expansion from a given set of boundary data. 

\begin{subbox}{The bulk reconstruction algorithm}
\begin{itemize}
	\item \textbf{Step 1:}\\
	Using the boundary charge operator $Q_\infty$ and the complex grading operator $L_0$, perform the charge and weight decomposition of the (rotated) phase operator
	\begin{equation}
		\hat{\delta} = \sum_{-D\leq q \leq -1} \sum_{-2D\leq s\leq -2} \hat{\delta}^{(s)}_q\,,
	\end{equation}
	and use the relation \eqref{eq:matching_delta-complex} to determine $\hat{\eta}$ and $\hat{\zeta}$. For the case $D=3$ one may directly use the result \eqref{eq:eta_delta} and \eqref{eq:zeta-delta_complex}. For explicit relations in the $D=4$ case we refer the reader to \cite{vandeHeisteeg:2022gsp}.
	\item \textbf{Step 2:}\\
	Use the relation \eqref{eq:input_data} to write down the most general collection of input data 
	\begin{equation}
		\Phi^{2,-1}_2,\ldots, \Phi^{2D,-1}_{2D}\,,
	\end{equation}
	for the CKS recursion in terms of the components of $\hat{\eta}$ computed in step 1. Note that this step will involve the raising operator $L_{1}$. 
	\item \textbf{Step 3:}\\
	Insert the input data into the CKS recursion \eqref{eq:CKS_recursion_epsilon} and compute the various $(d,\epsilon)$ components of $\Phi_n$ to any desired order in $n$. 
	\item \textbf{Step 4:}\\
	Finally, using the relations \eqref{Phi-expand}, \eqref{eq:Bn} and \eqref{eq:recursion_g_1}--\eqref{eq:recursion_g_2} one can iteratively compute the coefficients $\hat{g}_k$ to any desired order. After rotating back to the real algebra, one obtains the nilpotent orbit expansion to the period map from \eqref{solution-expansion}. 
\end{itemize}
\end{subbox}
\hspace{7cm}

\section{Examples: One-parameter families of Calabi--Yau threefolds}\label{reconstructing_examples}
In this section we explicitly work out the bulk reconstruction procedure for all singularity types arising in one-parameter weight $D=3$ models. In particular, this covers all possible singular limits in the complex structure moduli space of any one-parameter family of Calabi--Yau threefolds. Recall that there are three such limits, namely type $\mathrm{I}_1$, $\mathrm{II}_0$, and $\mathrm{IV}_1$, for which the most general form of the boundary data has been constructed in appendix \ref{sec:asymp_Hodge_examples}. In the following, we will first discuss the type $\mathrm{I}_1$ and $\mathrm{II}_0$ limits, as they are very similar in nature and relatively simple due to the restricted form of the phase operator. Subsequently, we will discuss the type $\mathrm{IV}_1$ limit, for which the bulk reconstruction will be significantly more non-trivial. 

\subsection{Type $\mathrm{I}_{1}$ and $\mathrm{II}_0$}
\label{sec:conifold}

Recall from our discussion in appendix \ref{sec:asymp_Hodge_examples} that for both the type $\mathrm{I}_1$ and the type $\mathrm{II}_0$ singularity the only possibility for the phase operator is that it is proportional to the lowering operator. In particular, after rotating to the complex basis we have
\begin{equation}
\hat{\delta} = \hat{\delta}^{(-2)}_{-1} = -c L_{-1}\,,
\end{equation}
for some constant $c$. Furthermore, we also argued that, in principle, one can effectively set $c=0$ by performing a coordinate shift $y\mapsto y+c$. In the following, we will not perform this coordinate shift, but rather go through the bulk reconstruction procedure and show that indeed one recovers the same result. Thus, let us follows the steps outline in the bulk reconstruction algorithm.

\subsubsection*{Step 1:}
\label{ssec:CKSI1}
We begin by determining the initial data of our recursion, which is fixed by the operator  $\hat{\eta}$. This operator can be expressed in terms of the phase operator $\hat{\delta}$ by using \eqref{eq:eta_delta} as
\begin{equation}
\hat{\eta} = \hat{\eta}_{-1}^{(-2)} = - \hat{\delta}_{-1}^{(-2)} = c L_{-1}\, .
\end{equation} 

\subsubsection*{Step 2:}
The initial data can then be computed from $\hat{\eta}$ by using \eqref{eq:input_data}. This gives us just $\Phi_2^{2,-1}$ as input for the recursion, since $\hat{\eta}$ only has a component with weight $n=2$. Furthermore the sums over the weights $s$ and charges $q$ only run over a single term, hence
\begin{equation}
\Phi_2^{2,-1}= \begin{pmatrix}
	(L^{+}_{2})^{[2]} \\
	(L^{0}_{2})^{[2]} \\
	(L^{-}_{2})^{[2]}
\end{pmatrix}= c\begin{pmatrix}
	L_{1}\\ 
	L_{0} \\
	L_{-1}
\end{pmatrix} = c\, \Phi_0\, .
\end{equation}
where in the last equality we noticed that the initial data of the recursion is simply proportional to the leading term $\Phi_0=(L_{1},L_{0},L_{-1})$ of the expansion \eqref{Phi-expand}. 

\subsubsection*{Step 3:}
In order to perform the CKS recursion we then need to evaluate the bilinear $B$ defined in \eqref{eq:bilinear_B} for this initial data. For our initial data it is interesting to point out that in general
\begin{equation}
B(\Phi_{0} \, , \, \Phi_{0}) = \Phi_{0}\ .
\end{equation}
This implies that each subsequent term $\Phi_{n}$ in the CKS recursion will be proportional to $\Phi_{0}$. Furthermore, from the structure of the recursion relation \eqref{eq:CKS_recursion_epsilon} we find at odd orders that $\Phi_{2n+1}=0$. Let us therefore make the ansatz
\begin{equation}
\Phi_{2n}^{2,-1}=c_n \Phi_0\, ,
\end{equation}
for some coefficients $c_{n}$ with $c_0=1$ and $c_{1}=c$. By plugging our ansatz into \eqref{eq:CKS_recursion_epsilon} we then obtain the recursion relation
\begin{equation}
(2n-2)c_{n} = 2 \sum_{0<k<n}c_k c_{n-k},\quad n>1\ .
\end{equation}
One can easily verify that under the specified initial conditions this recursion is solved by
\begin{equation}
c_{n} =c^n\ .
\end{equation}
Hence the full solution to the recursion relation is given by
\begin{equation}\label{eq:phiI1}
\Phi_{2n} =c^n\Phi_0,\quad \mathrm{or}\quad \begin{pmatrix}
	L_{2n}^+ \\
	L_{2n}^0\\ 
	L_{2n}^-
\end{pmatrix} = c^n\begin{pmatrix}
	L_{1}\\L_0\\L_{-1}
\end{pmatrix}\ .
\end{equation}
We see that the result for the $\mathrm{I}_{1}$ and $\mathrm{II}_0$ boundaries is remarkably simple. From the observation that the phase operator can be expressed in terms of the lowering operator as $\hat{\delta}=-cL_{-1}$ we have been able to solve the recursion completely algebraically. In particular we did not need to perform the highest-weight decompositions with respect to the Casimir operators $L^{2}$ and $\Lambda^{2}$ explicitly. In anticipation of next example, let us already note that for the type  $\mathrm{IV}_{1}$ boundary we do have to make use of this $\mathfrak{sl}(2)$-machinery, and hence the bulk reconstruction will be considerably more involved.

\subsubsection*{Step 4:}

Despite the apparent simplicity of the $\Phi_{2n}$ given in \eqref{eq:phiI1}, there are still some non-trivial steps to perform in order to complete the bulk reconstruction of $h_{\mathrm{nil}}(x,y)$, to which we now turn our attention. Following the procedure laid out in section \ref{subsec:h_reconstruction} we first write down the middle component $L^0_{2n}$ of $\Phi_{2n}$ as
\begin{equation}
(L^0_{2n})^{(2,0)}= c^n L_0\ .
\end{equation}
By using \eqref{eq:Bn} to compute the coefficients $B_{n}$ we then find that the sum only runs over a single term, yielding
\begin{equation}
B_n = -\frac{1}{2} L^0_{2n-2}=-\frac{1}{2}c^{n-1} L_{0}\ .
\end{equation}
In turn these coefficients can be used to recursively determine the $\hat{g}_k$ by using \eqref{eq:recursion_g_2}. Plugging our expression for the $B_{n}$ into this recursion relation we obtain
\begin{equation}\label{eq:gkrecursionI1}
-k \hat{g}_k =-\frac{1}{2} \sum_{j=1}^k \hat{g}_{k-j} c^j L_{0}\, .
\end{equation}
with the initial condition $\hat{g}_0=1$. In order to solve this recursion it is convenient to rotate back to the real basis via the transformation matrix \eqref{def-rho}. This is helpful because, in this basis, we simply need to work with a diagonal matrix $N^{0}$ instead of $L_{0}$, and consequently the terms $g_{k}$ are diagonal as well. It is at this point that we will make use of the explicit expressions for $L_0$ at the type $\mathrm{I}_1$ and $\mathrm{II}_0$ boundaries. To exemplify the computation, let us focus on the type $\mathrm{I}_1$ boundary. The analysis for the type $\mathrm{II}_0$ is very similar. Recall from the analysis in appendix \ref{sec:asymp_Hodge_examples} that the grading operator $N^0$ of the type $\mathrm{I}_1$ boundary is given by
\begin{equation}
N^0 = \begin{pmatrix}
	1 & 0 & 0 & 0\\
	0 & 0 & 0 & 0\\
	0 & 0 & 0 & 0\\
	0 & 0 & 0 & -1
\end{pmatrix}\,,
\end{equation}
up to a basis rotation with the matrix $\Lambda$ introduced in \eqref{eq:transform_I1}, which we suppress for notational clarity. To solve the recursion \eqref{eq:gkrecursionI1} we make the ansatz
\begin{equation}
g_k=c^k \, \mathrm{diag}\left(g^+_k\,0 ,\, 0 , \, g_k^- \right)\, ,
\end{equation}
where $g^\pm_k$ are arbitrary coefficients with $g^\pm_1=1$ for which we want to solve the recursion. By using this ansatz we find that \eqref{eq:gkrecursionI1} reduces to two decoupled recursion relations
\begin{equation}
2k \, g^\pm_k = \pm \sum_{j=1}^k g^\pm_{k-j}\, .
\end{equation}
Combining the equations at levels $k$ and $k+1$ one can rewrite these equations as
\begin{equation}
g^\pm_{k+1} = \frac{(2k\pm 1)}{2(k+1)} g^\pm_k\, ,
\end{equation}
which are solved by
\begin{equation}
g^+_k = \frac{1}{2}\frac{\Gamma(k+1/2)}{\Gamma(3/2)\Gamma(k+1)}\, ,  \qquad g^-_k = -\frac{1}{2}\frac{\Gamma(k-1/2)}{\Gamma(1/2)\Gamma(k+1)}\, ,
\end{equation}
where $\Gamma(n+1)=n!$ denotes the gamma function. Putting everything together, one finds
\begin{equation}\label{eq:gkI1}
g_k = -\frac{1}{2}\frac{c^k}{k!}\begin{pmatrix}
	-\frac{\Gamma(k+1/2)}{\Gamma(3/2)} & 0 & 0 & 0\\
	0 & 0  & 0 & 0\\
	0 & 0 & 0 & 0\\
	0 & 0 & 0 & \frac{\Gamma(k-1/2)}{\Gamma(1/2)}
\end{pmatrix},\quad k\geq 1\, .
\end{equation}
As a consistency check, one can verify that this solution to the CKS recursion indeed satisfies the matching condition given in \eqref{eq:matching_delta}. Resumming the series in $y^{-k}$ for the matrix-valued function $g(y)$ we find
\begin{equation}
g(y)=\sum_{k\geq 0} g_k y^{-k}=\begin{pmatrix}
	\sqrt{\frac{y}{y-c}} & 0 & 0 & 0\\
	0 & 1 & 0 & 0\\
	0 & 0 &  1& 0\\
	0 & 0 & 0 & \sqrt{\frac{y-c}{y}}
\end{pmatrix}.
\end{equation}
By multiplying with $e^{xN}$ from the left and $y^{-N^{0}/2}$ from the right we obtain final result for the nilpotent orbit approximation of the period map
\begin{equation}
\label{eq:I1_h}
\boxed{\rule[-1.1cm]{.0cm}{2.4cm} \quad
	h_{\mathrm{nil}}(x,y) = \begin{pmatrix}
		\frac{1}{\sqrt{y-c}} & 0 & 0 & 0\\
		0 & 1 & 0 & 0\\
		0 & 0 &  1 & 0\\
		-\frac{x}{\sqrt{y-c}} & 0 & 0 & \sqrt{y-c}
	\end{pmatrix} = e^{xN}(y-c)^{-\frac{1}{2}N^0}\ . \quad }
\end{equation}
As anticipated, we find that for the type $\mathrm{I}_1$ boundary the nilpotent orbit expansion of the period map is identical to the $\mathrm{Sl}(2)$-orbit approximation up to the coordinate shift $y\mapsto y+c$. This agrees with the fact that the same coordinate shift can be used to effectively set the component of the phase operator $\delta$ along $N^{-}$ to zero. 

\subsubsection*{Reconstructing the period vector for type $\mathrm{I}_1$}
From the nilpotent orbit approximation of the period map, we can immediately recover the full nilpotent orbit approximation of the underlying variation of Hodge structure via the relation
\begin{equation}
F^p_{\mathrm{nil}} = h_{\mathrm{nil}}\,F^p_\infty\,.
\end{equation}
Using this relation and the explicit form of the boundary Hodge structure as encoded in $Q_\infty$, one can show that
\begin{equation}
F_{\mathrm{nil}}(t) =e^{tN} \begin{pmatrix}
0 & 1 & 0 & 0\\
1 & 0 & 0 & 0\\
i & 0 & 0 & 1\\
0 & ic & 1 & 0
\end{pmatrix}\,,
\end{equation}	
Here the above matrix notation for the Hodge filtration means that $F^p_{\mathrm{nil}}$ is spanned by the first $4-p$ columns of the matrix. Furthermore, this Hodge filtration can be equivalently described in terms of the following period vector
\begin{equation}
\mathbf{\Pi}_{\mathrm{nil}} = e^{tN}\left[\begin{pmatrix}
0 \\ 1 \\ i \\ 0
\end{pmatrix}+A_1 e^{2\pi it}\begin{pmatrix}
1 \\ 0 \\ 0 \\ ic+\frac{1}{2\pi i}
\end{pmatrix}+A_1^2e^{4\pi i t}\begin{pmatrix}
0 \\ 1 \\ -i \\0
\end{pmatrix} \right]\,,
\end{equation}
where we have performed an additional K\"ahler--Weyl transformation in order to simplify the most general form of the period vector. Importantly, in order for $\mathbf{\Pi}_{\mathrm{nil}}$ and its derivatives to fully span $F_{\mathrm{nil}}^p(t)$ it is necessary to include the exponential corrections parametrized by the non-zero coefficient $A_1\in\mathbb{C}$, whose value is model-dependent. In \cite{Bastian:2021eom} these exponential corrections have been dubbed \textbf{essential instantons}. It is rather striking that, through the bulk reconstruction procedure, the boundary data associated to a given limit is rich enough to encode instanton corrections to the period vector. For completeness, let us also record the period vector after including the additional rotation matrix \eqref{eq:transform_I1} to include the dependence on the rigid period $\tau$ and the extension data $\gamma,\delta$
\begin{equation}
\mathbf{\Pi}_{\mathrm{nil}} = e^{tN}\left[\begin{pmatrix}
0 \\ 1 \\ \tau \\ \delta-\gamma \tau
\end{pmatrix}+B_1e^{2\pi it}\begin{pmatrix}
1 \\ \gamma \\ \delta \\ ick+\frac{k}{2\pi i}
\end{pmatrix}+B_1^2e^{4\pi i t}\begin{pmatrix}
0 \\ 1\\ \bar{\tau} \\ \delta-\gamma\bar{\tau}
\end{pmatrix} \right]\,,
\end{equation}
where $B_1 = -\frac{A_1}{\sqrt{k}}$.

\subsubsection*{Reconstructing the period vector for type $\mathrm{II}_0$}	
We stress that the final steps in the above analysis can be performed in a very similar fashion for the type $\mathrm{II}_0$ boundary, which will lead to the following result for the nilpotent orbit approximation of the period map (see \cite{Grimm:2021ikg} for more details)
\begin{equation}
\label{eq:II0_h}
\boxed{\quad \rule[-.9cm]{.0cm}{2.0cm}  h_{\mathrm{nil}}(x,y) =\sqrt{\frac{1}{y-\alpha}}   \begin{pmatrix}
		1 & 0 & 0 & 0 \\
		0 & 1& 0 & 0 \\
		x& 0 & y-\alpha & 0 \\
		0 & x & 0 & y-\alpha \\
	\end{pmatrix} = e^{xN}(y-\alpha)^{-\frac{1}{2}N^0}\ , \quad }
\end{equation}
where we have relabelled the coefficient $c$ to $\alpha$ in order to better match with the conventions used in appendix \ref{sec:asymp_Hodge_examples}, recall equation \eqref{eq:delta_II0} for the phase operator in particular. Again, we see that the nilpotent orbit expansion is equivalent to the Sl(2)-orbit expansion, up to a coordinate shift. As for the type $\mathrm{I}_1$ singularity, one can immediately infer the full nilpotent orbit approximation of the Hodge filtration to be
\begin{equation}
F_{\mathrm{nil}}(t) = e^{tN}\begin{pmatrix}
1 & 0 & 0 & 0\\
-i & 0 & 1 & 0\\
-i\alpha & 1 & 0 & 0\\
-\alpha & -i & -i\alpha & 1
\end{pmatrix}\,,
\end{equation}
Furthermore, this Hodge filtration can be recovered from the period vector
\begin{equation}
\mathbf{\Pi}_{\mathrm{nil}} = e^{tN}\left[\begin{pmatrix}
1 \\ -i \\ -i\alpha \\ -\alpha
\end{pmatrix} +A_{1}e^{2\pi i t}\begin{pmatrix}
1 \\ i \\ -i\alpha+\frac{i}{\pi} \\ \alpha-\frac{1}{\pi}
\end{pmatrix}\right]\,,
\end{equation}
where we have again performed an additional K\"ahler--Weyl transformation to simplify the expression. Importantly, we observe that also for the type $\mathrm{II}_0$ one can infer the presence of an instanton correction to the period vector, parametrized by the non-zero coefficient $A_1\in\mathbb{C}$. Using the transformation \eqref{eq:transform_II0} one may also include the dependence of $\mathbf{\Pi}_{\mathrm{nil}}$ on the extension data, though we will refrain from presenting the result here.

%%%%%%%%%%%%%%%%%%%%%%%%%%%%%%%%%%%%%%%%%%%%%%%%%%%%%%
\subsection{Type $\mathrm{IV}_{1}$}
\label{ssec:IV1}
In this section we study our final example, the type $\mathrm{IV}_1$ boundary. This case will be considerably more involved, owing to the fact that the phase operator $\delta$ is no longer proportional to $N^-$. Nevertheless we will obtain exact results for both $\Phi=(\mathbf{L}^+,\mathbf{L}^0,\mathbf{L}^-)$ and the nilpotent orbit approximation of the period map. This demonstrates the power of the CKS recursion, which provides us with a general formalism to perform this bulk reconstruction. \\

\noindent Recall that the general for of the type $\mathrm{IV}_1$ boundary data was constructed in appendix \ref{sec:asymp_Hodge_examples}, see equations \eqref{eq:sl2-triple_IV1}, \eqref{eq:delta_IV1_1}--\eqref{eq:delta_IV1_2}, and \eqref{eq:charge-operator_IV1}. However, in this section we will change conventions slightly in order to stay closer to the results of the work \cite{Grimm:2021ikg}, in which this computation was first performed. The two conventions are related by a non-symplectic basis transformation
\begin{equation}\label{eq:convention_IV1}
\Lambda_{\text{$\mathrm{IV}_1$ convention}} = \begin{pmatrix}
	1 & 0 & 0 & 0\\
	0 & 1 & 0 & 0\\
	0 & 0 & \frac{1}{2} & 0\\
	0 & 0 & 0 & -\frac{1}{6}
\end{pmatrix}\,.
\end{equation}
To be explicit, let us write down the full set of boundary data after changing to these new conventions. The real $\slt$-triple is given by
\begin{equation}\label{eq:sl2-triple_IV1-paper}
\{N^+, N^0, N^-\} = \scalebox{0.9}{$\left\{\begin{pmatrix}
		0 & 3 & 0 & 0\\
		0 & 0 & 2 & 0\\
		0 & 0 & 0 & 1\\
		0 & 0 & 0 & 0
	\end{pmatrix}\,,\begin{pmatrix}
		3 & 0 & 0 & 0\\
		0 & 1 & 0 & 0\\
		0 & 0 & -1 & 0\\
		0 & 0 & 0 & -3
	\end{pmatrix}\,,\begin{pmatrix}
		0 & 0 & 0 & 0\\
		1 & 0 & 0 & 0\\
		0 & 2 & 0 & 0\\
		0 & 0 & 3 & 0
	\end{pmatrix}\right\}$}\,,
\end{equation}
and the boundary charge operator and phase operator are given by
\begin{equation}\label{eq:Q_IV1-paper}
Q_\infty = \begin{pmatrix}
	0 & -\frac{3i}{2} & 0 & 0\\
	\frac{i}{2} & 0 & -i & 0\\
	0 & i & 0 & -\frac{i}{2}\\
	0& 0 & \frac{3i}{2} & 0
\end{pmatrix}\,,\qquad \delta = \begin{pmatrix}
	0 & 0 & 0 & 0\\
	0 & 0 & 0 & 0\\
	0 & 0 & 0 & 0\\
	\chi & 0 & 0 & 0
\end{pmatrix}\,,
\end{equation}
where we have suppressed the additional basis transformation \eqref{eq:basis-transform_IV1} for notational clarity. Furthermore, we have not included the component in $\delta$ proportional to the lowering operator $N^-$, recall equation \eqref{eq:delta_IV1_1}, since this can be set to zero by a coordinate shift, as we have explicitly demonstrated in the previous examples. Since $\delta = \delta_{-3,-3}$, we see that after rotating to the complex basis, we have
\begin{equation}
\hat{\delta} = \hat{\delta}^{(-6)}_{-3}\,.
\end{equation}

\subsubsection*{Step 1:}

Given this boundary data, we now turn to the CKS recursion in order to perform the bulk reconstruction. Again we start by determining the initial data of this recursion, for which we need to compute $\hat{\eta}$ from the phase operator $\hat{\delta}$. Since $\hat{\delta}=\hat{\delta}_{-3}^{(-6)}$ with respect to the gradings induced by $L_0$ and $Q_\infty$ we find using \eqref{eq:eta_delta} that
\begin{equation}
\hat{\eta} = \hat{\eta}^{(-6)}_{-3}   = -\frac{15}{8}\,\hat{\delta}^{(-6)}_{-3}.
\end{equation}

\subsubsection*{Step 2:}
The input data for the recursion relation can now be computed from $\hat{\eta}$. Since $\hat{\eta} = \hat{\eta}^{(-6)}_{-3}$ we see that the only terms in \eqref{eq:input_data} that contribute have $n=6$ and $q=3$, and therefore the only input is $\Phi_6^{6,-1}$. Note however, that in contrast to the type $\mathrm{I}_1$ and type $\mathrm{II}_0$ boundaries, the sum over weights now runs over $1\leq s \leq 5$, hence we expect $\Phi_6^{6,-1}= ( (L_6^+)^{[6]}, (L_6^0)^{[6]}, (L_6^-)^{[6]})$ to be significantly more complex. Explicitly, we find
\begin{align}\label{eq:initialdatamatrix}
\left(L^+_6\right)^{[6]} &=\frac{\chi}{8} 
\left(
\begin{array}{cccc}
-3 i & 6 & 0 & 0 \\
2 & 9 i & -6 & 0 \\
0 & -6 & -9 i & 2 \\
0 & 0 & 6 & 3 i \\
\end{array}\right)\,, \nn\\ 
\left(L^0_6\right)^{[6]} &= \frac{\chi}{2}\left(
\begin{array}{cccc}
0 & -3 i & 0 & 0 \\
i & 0 & 3 i & 0 \\
0 & -3 i & 0 & -i \\
0 & 0 & 3 i & 0 \\
\end{array}
\right)\ , \\
\left(L^-_6\right)^{[6]} &= \frac{\chi}{8} \left(
\begin{array}{cccc}
3 i & 6 & 0 & 0 \\
2 & -9 i & -6 & 0 \\
0 & -6 & 9 i & 2 \\
0 & 0 & 6 & -3 i \\
\end{array}
\right). \nn
\end{align}

\subsubsection*{Step 3:}
As the reader may verify, computing commutators for the action of the bilinear $B$ defined in \eqref{eq:bilinear_B} on $\Phi_6^{6,-1}$, given the above expressions, is not very enlightening. At this point, it will be more convenient to leverage the interpretation of $\mathfrak{g}$ as a vector space and choose a convenient basis in terms of which the action of $B$ is easier to handle. Recall that $\mathfrak{g}$ is 10-dimensional, with three of the generators given by the $\slt$-triple, which form an irreducible representation of highest weight $d=2$. The remaining seven generators form an irreducible representation of highest weight $d=6$ and can also be constructed explicitly using the $\slt$-triple. In other words, we have the decomposition
\begin{equation}
\mathbf{10} = \mathbf{3}\oplus\mathbf{7}\,.
\end{equation}
The basis we take for the irreducible representation of highest weight $d=6$ is constructed out of its highest weight state $(L_{1})^{3}$ via
\begin{equation}
\label{eq:basis_LCS}
\begin{aligned}
	T_{2k}  &= \frac{i n_k}{6}  \bigg( \frac{(\ad L_{-1})^k}{k!} + \frac{(\ad L_{-1})^{6-k}}{(6-k)!} \bigg) (L_{1})^3 \qquad (k=0,1,2,3) \, ,\\
	T_{2k+1}  &= \frac{n_k}{6}    \bigg( \frac{(\ad L_{-1})^k}{k!} - \frac{(\ad L_{-1})^{6-k}}{(6-k)!} \bigg) (L_{1})^3 \qquad (k=0,1,2)\, ,\\
\end{aligned}
\end{equation}
where for briefness we defined the coefficients
\begin{equation}
n_k = i^k\sqrt{\frac{1}{2} \frac{6!}{k!(6-k)!}} \times \begin{cases}
	1 \qquad &\text{ for $k=0,1,2$,}\\
	1/\sqrt{2} & \text{ for $k=3$.}
\end{cases} \, 
\end{equation}
Our basis is completed by fixing a basis for the irreducible representation of highest weight $d=2$, which we span by
\begin{equation}
T_{7}  =\frac{1}{2\sqrt{5}}\big( L_{1} + L_{-1} \big)\, , \qquad
T_{8}  = \frac{i}{2\sqrt{5}}\big( L_{1} - L_{-1} \big)\, , \qquad
T_{9}  = \frac{1}{2\sqrt{5}}L_0\, .
\end{equation}
We have included these normalization factors such that the $T_a$ form an orthonormal basis, in the sense that
\begin{equation}
\mathrm{Tr}(T_a T_b)=\delta_{ab}\, .
\end{equation}
This allows one to easily switch between expressing elements of $\mathfrak{g}$ as $4\times 4$ matrices or as 10-component vectors. Furthermore, the adjoint action of the $\mathfrak{sl}(2,\mathbb{C})$-triple $(\ad L_{1}, \ad L_0, \ad L_{-1})$ can now be realized by $10\times 10$ matrix multiplication, whose explicit expressions can be found in appendix B of \cite{Grimm:2021ikg}. In the $T_a$ basis we can write the initial data of the CKS recursion \eqref{eq:initialdatamatrix} as
\begin{equation}
\begin{aligned}
	\left(L_6^+\right)^{[6]}  &=\frac{\chi}{8}\sqrt{\frac{15}{2}}\, \big(i\sqrt{15}T_1 -\sqrt{15} T_2-3i T_5
	+ T_6\big)\ ,\\
	\left(L_6^0\right)^{[6]} &= \frac{\chi}{4} \sqrt{15}\,\big( -\sqrt{5} T_3 + \sqrt{3} T_7 \big),\\
	\left(L_6^-\right)^{[6]} &=\frac{\chi}{8}\sqrt{\frac{15}{2}}\,\big(-i\sqrt{15}T_1-\sqrt{15}T_2+ 3i T_5+ T_6\big).
\end{aligned}
\end{equation} 
The next step of the CKS recursion is to combine the data $(L^+_n, L^0_n,L^-_n)$ at each step into 30-component vectors $\Phi_n$. A natural $\slt$-triple $(\Lambda^+,\Lambda^0,\Lambda^-)$ that acts on these vectors is then constructed in \eqref{eq:sl2_tensor}. We can use its Casimir $\Lambda^2$ to decompose the 30-dimensional vector space further based on the highest weight $d+2\epsilon$. By construction the highest weights of our input data $\Phi_6^{6,-1}$ under $L^2$ and $\Lambda^2$ are $d=6$ and $\epsilon=-1$, hence let us put
\begin{equation}
\Phi_6^{6,-1} = -8 i \sqrt{\frac{2}{15}}\, \chi \, e^{6,-1}\, .
\end{equation}
The crucial observation is that there exist three more vectors $e^{6,1},\;e^{2,1}$ and $e^{2,-1}$ on which the action of the bilinear $B$ closes. Together these are given by
\begin{align}\label{eq:ebasis}
	e^{6,1} &= \begin{pmatrix}
		-4\sqrt{15} \, T_{0}+3i\sqrt{15}  \, T_{1}+12  \, T_{4}-3i  \, T_{5} \\
		6 i \sqrt{10}  \,T_{2} -6i\sqrt{6} \, T_{6} \\
		4 \sqrt{15} \, T_{0} +3 i \sqrt{15} \, T_{1} - 12  \,T_{4} - 3i  \, T_{5}
	\end{pmatrix}\, \nonumber\\
	e^{6,-1} &= \begin{pmatrix}
		\sqrt{15} \, T_{0}+i\sqrt{15} \, T_{1}-3  \, T_{4}-i  \, T_{5} \\
		2i \sqrt{10}  \, T_{2}-2i\sqrt{6}  \, T_{6} \\
		-\sqrt{15}  \, T_{0}+i \sqrt{15}  \, T_{1}+3  \,T_{4}-i  \, T_{5}
	\end{pmatrix}\,,\\
e^{2,1} &=  \sqrt{5}  \begin{pmatrix}
 	2i\, T_{7} +T_{8} \\
 	2 \,T_{9} \\
 	-2i \,T_{7}+  T_{8}
\end{pmatrix}\,,\qquad e^{2,-1} =  \sqrt{5} \begin{pmatrix}
		-i  \, T_{7}+ T_{8}\\
		2\,T_{9}\\
		i \,T_{7}+T_{8}
	\end{pmatrix}\,. \nonumber
\end{align}
We can then write our ansatz for  $\Phi_{6n}$ as\footnote{For convenience we chose to rotate the coefficients  appearing with the vectors $e^{6,1}$, $e^{6,-1}$, $e^{2,1}$ and $e^{2,-1}$ such that the 30-component $\Phi_{6n}$ does not have multiple $a_{n},b_{n},c_{n},d_{n}$ in its entries. This  makes it easier to find the solution to the coupled system of recursion relations later.} 
\begin{equation}
\label{eq:LCS_Phi_ansatz}
\Phi_{6n}= (a_{n}+b_{n})e^{6,1}+(-3a_{n}+4b_{n})e^{6,-1}+(c_{n}+d_{n})e^{2,1}+(2c_{n}-d_{n})e^{2,-1} \, ,
\end{equation}
with the initial data of the recursion given by
\begin{equation}
a_{1} = \frac{i\chi}{56} \sqrt{\frac{15}{2}} \, , \quad b_{1}=-\frac{i\chi}{56}\sqrt{\frac{15}{2}} \, , \quad c_{1}=0\, , \quad d_{1}=0\, .
\end{equation}
For arbitrary coefficients $a_n,b_n,c_n,d_n$ it can be verified that $ B(\Phi_{6k}, \Phi_{6(n-k)})$ is again spanned by the vectors $e^{6,1}$, $e^{6,-1}$, $e^{2,1}$ and $e^{2,-1}$, so the action of the bilinear $B$ is indeed closed on this subspace. This is a huge simplification, since there are now only four components of $\Phi$ that enter the recursion relation instead of thirty. It should be noted, however, that $B$ still acts very non-trivially on these four components. For example, in the basis $(e^{6,1},\, e^{6,-1},\, e^{2,1},\, e^{2,-1})$ one finds that $B(e^{6,-1},\cdot)$ acts as
\begin{equation}
B(e^{6,-1},\cdot)=-\frac{i}{14}\left(
\begin{array}{cccc}
	-\frac{i \sqrt{30}}{7} & \frac{8}{7} i \sqrt{\frac{6}{5}} & \frac{24}{5} & 0 \\
	\frac{2}{7} i \sqrt{\frac{2}{15}} & \frac{12}{7} i \sqrt{\frac{6}{5}} &
	-\frac{32}{15} & -\frac{112}{15} \\
	-\frac{3}{14} & \frac{8}{7} & 0 & 0 \\
	0 & 2 & 0 & 0 \\
\end{array}
\right).
\end{equation}
Indeed, we see that the coefficients $a_n,b_n,c_n,d_n$ all become intertwined, and already at the next step $\Phi_{12}$ each of them is non-vanishing. The recursion relation \ref{eq:CKS_recursion_epsilon} therefore reduces to a system of four coupled recursion relations, which have been given in \eqref{eq:recursion} for completeness. For illustrative purposes we also listed the coefficients that follow from iterating the recursion for the first couple of steps of the recursion in table \ref{table:recursion}. \\
\begin{table}[!ht]
\renewcommand{\arraystretch}{1.5}
\centering
\begin{tabular}{|c c c c|}
	\hline
	\rule[-.4cm]{.0cm}{1cm}
	$128i\sqrt{\frac{10}{3}}\times \left(\frac{4}{\chi}\right)^n a_n$   &  $ 28i\sqrt{\frac{10}{3}}\times \left(\frac{8}{\chi}\right)^n  b_n$ &$ 5 \left(\frac{8}{\chi}\right)^n  c_n$&$ 15 \left(\frac{4}{\chi}\right)^n  d_n$ \\ \hline
	-45 & 20 & 0 & 0 \\
	36 & -36 & -22 & 9 \\
	-126 & 224 & 48 & -9 \\
	198 & -724 & -278 & 27 \\
	-450 & 3216 & 952 & -45 \\
	846 & -12168 & -4156 & 99 \\
	-1746 & 49920 & 16000 & -189 \\
	3438 & -196788 & -65446 & 387 \\
	-6930 & 791792 & 259464 & -765 \\
	13806 & -3154456 & -1044212 & 1539 \\ \hline
\end{tabular}
\caption{Values obtained for the coefficients $a_n,b_n,c_n,d_n$ by iterating the recursion relation for the first $10$ steps, where we included overall normalization factors for convenience.}
\label{table:recursion}
\renewcommand{\arraystretch}{1.0}
\end{table}

\noindent Looking at the data listed in table \ref{table:recursion} one realizes that we are dealing with quite non-trivial series of coefficients. From the recursion relations \eqref{eq:recursion} we could have expected such complications to arise, since we need to solve a coupled system where we sum over all previous terms $0<k<n$. Nevertheless we can present an exact solution to these recursions, where the coefficients $a_n$ and $d_n$ are given by relatively simple power series, while $b_n$ and $c_n$ involve the hypergeometric function $_2F_1(\mu_1,\mu_2;\nu_1;z)$. To be precise, the coefficients are given by
\begin{align}
\label{eq:LCS_recursion_solution}
a_n &= \frac{i}{7} \sqrt{\frac{3}{10}} \left(\frac{\chi}{4}\right)^n \left(1-3 (-1)^n 2^{n-2} \right)\, ,\nonumber\\
b_n &=  -\frac{i}{28}  \sqrt{\frac{3}{10}} \left(\frac{\chi}{4}\right)^n \Big(2 \sqrt{3} -3 (-1)^{n}
2^{n}-2^{n+2} \binom{\frac{1}{2}}{n+1} \, _2F_1\big(1,n+\frac{1}{2};n+2;-2\big)  \Big)\, ,\nonumber\\
c_n &=  -\frac{1}{5} \left(\frac{\chi}{4}\right)^n \Big(\sqrt{3}+(-1)^n 2^n-2^{n+1} \binom{\frac{1}{2}}{n+1} \, _2F_1\big(1,n+\frac{1}{2};n+2;-2\big)\Big)\, , \nonumber\\
d_n &=  \frac{1}{5}\left(\frac{\chi}{4}\right)^n \left(1+(-1)^n 2^{n-1}\right)\, .
\end{align}
By resumming the series expansion in $y^{-1-3n}$ for $(\mathbf{L}^+(y),\mathbf{L}^0(y),\mathbf{L}^-(y))$ we obtain the functions
\begin{align}
a(y) &= \frac{i}{7}\sqrt{\frac{3}{10}} \frac{\chi}{4}\times  \frac{\chi+20 y^3}{y \left(4 y^3-\chi\right) \left(\chi+2 y^3\right)}, \nonumber \\ 
b(y)&=-\frac{i}{28} \sqrt{\frac{3}{10}}\times \frac{\sqrt{F(y)}}{y \left(4 y^3-\chi \right) \left(\chi +2 y^3\right)}\, , \\
c(y) &= \frac{1}{3 y}+\frac{28i}{3}  \sqrt{\frac{2}{15}} b(y) \nonumber\\
&\qquad +\frac{\sqrt{F(y)} \left(F(y)-\left(\chi +8 y^3\right) \left(\chi ^3+72 \chi ^2 y^3+96 \chi  y^6+128 y^9\right)\right)}{12 \chi  y \left(4 y^3-\chi \right) \left(\chi +2 y^3\right) \left(-\chi ^3+48 \chi ^2 y^3+96 
	\chi  y^6+640 y^9\right)} \, ,\nonumber \\
d(y) &= \frac{3 \chi^2}{10 y \left(4 y^3-\chi\right) \left(\chi+2 y^3\right)} \nonumber \, .
\end{align}
where for convenience we defined
\begin{equation}
F(y) = \chi ^4+288 \chi ^2 y^6+16 \chi ^3 y^3+128 \chi  y^9-8 \sqrt{2} y^{3/2} \left(\chi -4 y^3\right)^2 \left(\chi +2 y^3\right)^{3/2}+512 y^{12}\, .
\end{equation}

\subsubsection*{Step 4:}

Next, we determine the polynomials $g_k$ from the $L^0_{6n}$ components of $\Phi_{6n}$. For this recursion we first have to determine the coefficients $B_k$ from \eqref{eq:Bn}. From \eqref{eq:ebasis} we see that $L^0_{6n}$ is spanned by the generators $T_2$, $T_6$ and $T_9$, which means we only have to consider the weights $s=4,0,-4$. Therefore we can express the coefficients $B_k$ as
\begin{align}
B_{3k} &=-\half\left(L_{6k-6}^0\right)^{6,4}, \quad &k&\geq 1\, ,\\
B_{3k+1} &= -\half\left(L_{6k}^0\right)^{2,0}- \half \left( L_{6k}^0\right)^{6,0},\quad &k&\geq 1\, ,\\
B_{3k+2} & =-\half \left(L_{6k+6}^0\right)^{6,-4},\quad &k&\geq 0\, .
\end{align}
By inserting our expression \eqref{eq:LCS_Phi_ansatz} for $L^0_{6k}$ we find
\begin{equation}
\begin{aligned}
	B_{3k} &=  7\sqrt{\frac{5}{2}}\,  b_{k-1} \big( i \, T_2  -   T_3 \big)\, , \\
	B_{3k+1} &=  -7i\sqrt{6}\, b_{k} \, T_6  + 3\sqrt{5}\, c_{k}\, T_9 \, , \\
	B_{3k+2} &=  7\sqrt{\frac{5}{2}}\,  b_{k+1} \big( i \, T_2  +   T_3 \big)\, . \\
\end{aligned}
\end{equation}
The coefficients $g_k$ are determined by the recursion relation \eqref{eq:recursion_g_2}. As an alternative approach, we want to mention that one can also compute $\mathbf{L}^0(y)$ from the coefficients $b_n,c_n$, and then solve the differential equation \eqref{eq:invg_dg} for $\hat{g}(y)$ with boundary condition $\hat{g}(0)=1$. The first approach gets rather complicated since $\hat{g}_k$ is no longer Lie algebra-valued. This means it no longer suffices to work with the basis $T_a$, but instead we have to write out the matrices explicitly. Either way we can present an exact solution for this recursion, which is given by
\begin{equation}
\begin{aligned}
	g_{3k} &= \alpha_{3k} \begin{pmatrix}
		1+3\beta_{k} & 0 & 0 & 0 \\
		0 & 3+\gamma_{k}  & 0 & 0 \\
		0 & 0 & 3+\delta_{k}& 0 \\
		0 & 0 & 0 & \frac{ 9(1-2k)\varepsilon_{k}+6k+1 }{2k(3/2-k)} 
	\end{pmatrix}\, , \\
	g_{3k+1} &= \alpha_{3k+3}\begin{pmatrix}
		0 & 0 & 0 & 0 \\
		0 & 0 & 0 & 0 \\
		\delta_{k+1}-1 & 0 & 0 & 0 \\
		0 & \frac{9(1+2k)^2 \varepsilon_k +(6k+7)(8k^2+2k-5)}{(k+1)(4k^2-1)} & 0 & 0 
	\end{pmatrix} \, , \\
	g_{3k+2} &= \alpha_{3k}\begin{pmatrix}
		0 & 0 & 3 \beta_{k}-3 & 0 \\
		0 & 0 & 0 &  \gamma_{k}-1\\
		0 & 0 & 0 & 0 \\
		0 & 0 & 0 & 0
	\end{pmatrix} \, , \\
\end{aligned}
\end{equation}  
where for brevity we defined
\begin{equation}\label{eq:a3k}
\alpha_{3k} = \frac{(-\chi)^k}{k!} \frac{4^{-1-k}\sqrt{\pi}}{\Gamma(1/2-k)}\, ,
\end{equation}
and denoted the generalized hypergeometric functions ${}_p F_q(\mu_1,\ldots, \mu_p; \nu_1,\ldots \nu_q;z)$ by the coefficients\footnote{Note that the coefficient $\delta_{k}$ should not be confused with the phase operator $\delta$.}
\begin{equation}
\begin{aligned}
	\beta_{k} &= {}_2\mathrm{F}_1\big(\frac{1}{2},-k;\frac{1}{2}-k;-2\big)\, , \\
	\gamma_{k} &= {}_2\mathrm{F}_1\big(-\frac{1}{2},-k;\frac{1}{2}-k;-2\big)\, , \\
	\delta_{k} &= {}_3\mathrm{F}_2\big(-\frac{1}{2}, \frac{5}{6}, -k; -\frac{1}{6},\frac{1}{2}-k;-2 \big) \, , \\
	\varepsilon_{k} &= {}_2\mathrm{F}_1\big(\frac{1}{2},1-k;\frac{1}{2}-k;-2\big) \, .
\end{aligned}
\end{equation}
Note that we have chosen to rotate back to the real basis by $g_{k}=\rho^{-1} \hat{g}_{k} \rho$. \\

\noindent By resumming the coefficients $g_{k}$ as a series expansion in $y^{-k}$ one recovers the matrix-valued function $g(y)$. From there, one obtains the following expression for the nilpotent orbit approximation of the period map
\begin{equation}
\label{eq:LCS_h}
\boxed{\rule[-1cm]{.0cm}{2.2cm}\quad \Scale[0.90]{
		h_{\mathrm{nil}}(y) = \alpha(y) \left(
		\begin{array}{cccc}
			1 + 3 \beta(y)^{-1} & 0 &  3 \beta(y)^{-1} -3 & 0 \\
			0 & 3y +y \beta(y) & 0 & y  \beta(y) -y \\
			\frac{\gamma(y)}{\alpha(y)}-4 y^2  & 0 & \frac{\gamma(y)}{\alpha(y)}& 0 \\
			0 &  3\chi-3 y^3+3y^3 \beta(y) & 0 & y^3-\chi+3y^3  \beta(y) \\
		\end{array}
		\right)}\ ,\quad}
\end{equation}
where we have suppressed the overall axion-dependent factor $e^{xN}$. Furthermore, for convenience we have defined the functions\footnote{Note that the function $\alpha(y)$ corresponds to the series $\alpha(y) = y^{-3/2}\sum_k \alpha_{3k} y^{-3k}$ for the coefficients defined in \eqref{eq:a3k}.}
\begin{align}
\label{eq:LCS_normalisation}
\alpha(y)&=\frac{1}{2\sqrt{4y^3-\chi}}\, , \qquad \beta(y)=\sqrt{1+\frac{\chi }{2 y^3}}\, , \\
\gamma(y) &=\sqrt{2} \frac{y^3-\chi +3 y^3 \beta(y)}{2y^{5/2}  \sqrt{\frac{\chi  \left(2
			y^3-\chi \right)}{y^6}+8}}\, .
\end{align}
Note that, in contrast to the $\mathrm{I}_1$ and $\mathrm{II}_0$ boundaries, the parameter $\chi$ that entered through the phase operator $\delta$ cannot be removed by some coordinate redefinition here.

\subsubsection*{Reconstructing the period vector for type $\mathrm{IV}_1$}
Lastly, let us also present the result for the nilpotent orbit approximation of the Hodge filtration. Rather strikingly, one finds the following \textit{exact} result
\begin{equation}
F_{\mathrm{nil}}(t) = e^{tN}\begin{pmatrix}
1 & 0 & 0 & 0\\
0 & 1 & 0 & 0\\
0 & 0 & 1 & 0\\
i\chi & 0 & 0 & 0 
\end{pmatrix}\,,
\end{equation}
where by exact we of course mean up to exponential corrections. It is an extremely non-trivial consistency check that all the sub-leading terms in the complicated expression for $h_{\mathrm{nil}}(x,y)$ obtained in \eqref{eq:LCS_h} are exactly such that the final result is precisely the expected expression for the nilpotent orbit approximation of the Hodge filtration. Furthermore, let us also record the period vector which gives rise to the above Hodge filtration. At this point, we choose to rotate back to the conventions used in appendix \ref{sec:asymp_Hodge_examples} using the transformation \eqref{eq:basis_LCS}. Additionally, we also use the transformation \eqref{eq:basis-transform_IV1} to include the dependence on the various quantities that determine the integral structure. The result is given by
\begin{equation}
\mathbf{\Pi}_{\mathrm{nil}}(t) = e^{tN}\begin{pmatrix}
1 \\ 0 \\ \frac{f}{2a} \\ \gamma - \frac{i}{6}a^2 b \chi
\end{pmatrix} = \begin{pmatrix}
1 \\ a t \\ \frac{1}{2} a b t^2+et+\frac{f}{2a} \\ -\frac{1}{6}a^2b t^3+\gamma - \frac{i}{6}a^2 b \chi
\end{pmatrix}\,.
\end{equation}
In particular, after making the identifications
\begin{equation}\label{eq:LCS_mirror-data}
a=1\,,\quad b=\kappa\,\quad e=\sigma\,\quad f = -\frac{c_2}{12}\,,\quad \gamma=0\,,\quad \chi \mapsto -\frac{3\chi \zeta(3)}{8\pi^3\kappa}\,,
\end{equation}
we find that the period vector takes the form
\begin{equation}
\mathbf{\Pi}_{\mathrm{nil}}(t) = \begin{pmatrix}
1 \\ t \\ \frac{\kappa}{2}t^2+\sigma t - \frac{c_2}{24} \\ -\frac{\kappa}{6}t^3 - \frac{c_2}{24}t+ \frac{\chi\zeta(3)}{2(2\pi i)^3}
\end{pmatrix}\,,
\end{equation}
which is the well-known asymptotic form of the period vector near the LCS point, expressed in terms of the topological data of the mirror Calabi--Yau threefold, recall \eqref{eq:periods_LCS}. Let us in particular emphasize the appearance of the term involving the Euler characteristic, as the latter arises as an $(\alpha')^3$ correction which can alternatively be derived through a four-loop computation in the worldsheet theory \cite{Candelas:1990rm,Font:1992uk,Becker:2002nn}. It is rather remarkable that, through the bulk reconstruction procedure, the boundary data (in particular the phase operator) is rich enough to contain such information. 

\section{Bulk reconstruction: multi-variable}\label{sec:bulk-reconstruction_multi}

Having discussed the bulk reconstruction procedure in the one-parameter case, let us now return to the general $m$-parameter case, for which we have seen that
\begin{equation}\label{eq:Fnil_Finf_multi}
F_{\mathrm{nil}}= \mathrm{exp}\left[\sum_{i=1}^m x_i N_i\right]\cdot h_{\mathrm{nil}}(y_1,\ldots, y_m)\cdot F_{\infty}\,,
\end{equation}
with the map $h_{\mathrm{nil}}(y_1,\ldots, y_m)$ depending only on the saxions. Our goal will be to completely determine $h_{\mathrm{nil}}(y_1,\ldots, y_m)$ by a multi-parameter generalization of the bulk reconstruction procedure discussed in section \ref{sec:bulk-reconstruction_single}.

\subsubsection*{Iterating the one-parameter bulk reconstruction procedure}
The crucial observation of Cattani, Kaplan, and Schmid \cite{CKS}, which was already discussed in chapter \ref{chap:asymp_Hodge_I}, is that one can effectively perform the one-parameter bulk reconstruction in a clever iterative way in order to obtain the $m$-parameter result. Indeed, recall from our discussion in section \ref{sec:SL2_orbit_multi_variable} that one can recursively define one-parameter period maps $h_{\mathrm{nil},r}$, for $r=1,\ldots,m$, by the relation
\begin{equation}\label{eq:hnil_recursive_r}
\mathrm{exp}\left[i\left(\sum_{i=1}^{r}\frac{y_i}{y_{r+1}}N_i\right)\right] F^p_{(r)} =h_{\mathrm{nil},r}\,\mathrm{exp}\left[i\left(\sum_{i=1}^{r-1}\frac{y_i}{y_{r}}N_i\right)\right] F^p_{(r-1)}\,,
\end{equation}
with $y_{m+1}\equiv 1$. In particular, writing
\begin{equation}
h_{\mathrm{nil},(r)} = h_{\mathrm{nil},r}\cdots h_{\mathrm{nil},1}\,,
\end{equation}
we see that \eqref{eq:hnil_recursive_r} can be written as
\begin{equation}\label{eq:hnil_recursive_r2}
\boxed{
\mathrm{exp}\left[i\left(\sum_{i=1}^{r}\frac{y_i}{y_{r+1}}N_i\right)\right] F^p_{(r)} = h_{\mathrm{nil},r}\left[h_{\mathrm{nil},(r-1)}F^p_{\infty} \right]\,.}
\end{equation}
In other words, if $h_{\mathrm{nil},(r-1)}$ has been determined, one can compute $h_{\mathrm{nil},r}$ by performing the \textit{one-parameter bulk reconstruction} outlined in section \ref{sec:bulk-reconstruction_single}, using as ``boundary data'' 
\begin{equation}\label{eq:boundary_data_multi}
\{\mathrm{Ad}(h_{\mathrm{nil},(r-1)})Q_\infty, \mathrm{Ad}(h_{\mathrm{nil},(r-1)})N^\bullet_{(r)}, \delta_r\}\,,
\end{equation}
where we emphasize that one should ``translate'' the boundary charge operator $Q_\infty$ and the $\mathfrak{sl}(2,\mathbb{R})$-triple $N^\bullet_{(r)}$ using the map $h_{\mathrm{nil},(r-1)}$ c.f.~\eqref{eq:hnil_recursive_r2}, while for the phase operator one can simply use $\delta_r$. Furthermore, we recall the result (see Lemma 4.37 of \cite{CKS})
\begin{equation}
[N^0_{(j)},h_{\mathrm{nil},i}] = 0\,,\qquad \forall i\leq j\,,
\end{equation}
which implies that $\mathrm{Ad}(h_{\mathrm{nil},(r-1)})N^0_{(r)}=N^0_{(r)}$. In other words, the grading operator is not affected by the translation. As a result, we immediately find that each $h_{\mathrm{nil},i}$ takes the form
\begin{equation}
h_{\mathrm{nil},i} =  g_{i}\cdot\left(\frac{y_i}{y_{i+1}}\right)^{-\frac{1}{2}N^0_{(i)}}\,,\qquad g_{i}=1+\sum_{k_i=1}^\infty g_{i,k_i}\left(\frac{y_i}{y_{i+1}}\right)^{-k_i}\,.
\end{equation}
In particular, each $h_i$ admits a series expansion in $y_i/y_{i+1}$,  with the expansion coefficients $g_{i,k_i}$ being determined by the CKS recursion. In the one-parameter case, we have seen that they can be expressed as moduli-independent universal Lie polynomials involving certain projections of the (single) phase operator $\delta$ and raising operator $N^+$ onto specific eigenspaces of the grading operators $N^0$ and $Q_{\infty}$. All of this remains true in the multi-parameter case, except for the fact that the expressions will now become moduli-dependent, due to the factors of $\mathrm{Ad}(h_{\mathrm{nil},(r-1)})$ appearing in \eqref{eq:boundary_data_multi}. Specifically, one has
\begin{equation}
g_{i,k_i}=g_{i,k_i}\left(\frac{y_{i-1}}{y_i},\ldots,\frac{y_{1}}{y_2}\right)\,.
\end{equation}
In other words, for $i>1$, each $g_{i,k_i}$ is a function of all the previous $y_j$, with $j\leq i$. More precisely, each $g_{i,k_i}$ itself admits a power series expansion in $y_{i-1}/y_i,\ldots, y_1/y_2$. The final result for the map $h_{\mathrm{nil}}(y_1,\ldots, y_m)$ is simply given by
\begin{equation}
h_{\mathrm{nil}}(y_1,\ldots, y_m) = h_{\mathrm{nil},(m)} = h_{\mathrm{nil},m}\cdots h_{\mathrm{nil},1}\,,
\end{equation}
and should therefore be seen as a very non-trivial collection of nested power series. 

\subsubsection*{A useful rewriting}
Let us present a useful rewriting of the map $h_{\mathrm{nil}}(y_1,\ldots, y_m)$. In fact, for our purposes it will be more important to consider the form of the inverse operator $h_{\mathrm{nil}}^{-1}$, which admits a similar expansion
\begin{equation}
h_{\mathrm{nil}}^{-1} =h_{\mathrm{nil},1}^{-1}\cdots h_{\mathrm{nil},m}^{-1}\,,
\end{equation}
where
\begin{equation}
\label{eq:hinv_before_commutation}
h_{\mathrm{nil},i}^{-1} = \left(\frac{y_{i}}{y_{i+1}}\right)^{\frac{1}{2}N^0_{(i)}}\left[\sum_{k_i=0}^\infty \left(\frac{y_{i}}{y_{i+1}}\right)^{-k_i} f_{i,k_i}\right]\,,\qquad f_{i,0}\equiv 1\,.
\end{equation}
Of course, the expansion coefficients $f_{i,k_i}$ can be straightforwardly related to the $g_{i,k_i}$ via an order-by-order inversion. For practical purposes, it will be useful to commute the factor left of the square brackets \eqref{eq:hinv_before_commutation} to the right by employing the weight-decomposition
\begin{equation}\label{eq:expansion-coeff_decomp}
	f_{i,k_i} = \sum_{s_i} f_{i,k_i}^{s_i}\,,\qquad [N^0_{(j)},f_{i,k_i}^{s_i}] = s_i^j\,f_{i,k_i}^{s_i}\,,\qquad s_i = \left(s_i^{1},\ldots, s_i^{m}\right)\,,
\end{equation}
of the $f_{i,k_i}$ coefficients, with respect to the real garding operators $N^0_{(j)}$. This yields the following expression
\begin{equation}
\label{eq:hinv}
\boxed{ \rule[-.7cm]{0cm}{1.6cm}
\quad     h_{\mathrm{nil}}^{-1} = \sum_{k_1,\ldots, k_m=0}^\infty \sum_{s^1,\ldots, s^m}\prod_{i=1}^m \left[\left(\frac{y_i}{y_{i+1}}\right)^{-k_i+\frac{1}{2}s_i^{(m)}}f^{s^i}_{i,k_i}\right] h^{-1}_{\mathrm{Sl}(2)}\,,\quad }
\end{equation}
where we introduced the notation
\begin{equation}
s_i^{(m)} = s_i^{i}+\cdots +s_i^{m}\,,
\end{equation}
and identified
\begin{equation}
h^{-1}_{\mathrm{Sl}(2)} = e(y)=\prod_{i=1}^m \left(\frac{y_i}{y_{i+1}}\right)^{\frac{1}{2}N_{(i)}^0}\,,
\end{equation}
which corresponds to the $\mathrm{Sl}(2)$-orbit approximation of the inverse period map. It is usually denoted by $e(y)$ and we will use this notation as well. 

\subsubsection*{Properties of $f_{i,k_i}$}
We end our discussion by stating two important properties of the expansion functions $f_{i,k_i}$. Both of these properties will feature prominently in the finiteness proof of self-dual vacua, as will be explained in section \ref{subsec:self-dual_proof}. Recalling the notation introduced in \eqref{eq:expansion-coeff_decomp}, the properties are as follows.
\begin{itemize}
	\item \textbf{Property 1: weight restrictions}\\
	The first property is a restriction on the possible weight of the expansion coefficients, namely
	\begin{equation}
		\label{eq:expansion_coeff_weights}
		f_{i,k_i}^{s^i} = 0\,,\qquad \text{if $s_i^i> k_i-1$ or $s_j^i\neq 0$ for $j>i$}\,,
	\end{equation}
	which immediately follows from the property \eqref{g-conditions} that we derived in the one-parameter bulk reconstruction procedure. In words, this means that the last non-trivial weight of each $f_{i,k_i}$ is given by $s_i^i$, and that the value of this weight is restricted by the order $k_i$ at which this coefficient appears in the expansion. In particular, this implies that the second sum in \eqref{eq:hinv} only runs over $s_i^i\leq k_i-1$.
	\item \textbf{Property 2: scaling}\\
	While the first property \eqref{eq:expansion_coeff_weights} restricts the possible weights of the expansion functions, the second property gives a bound on the scaling of the expansion functions in terms of their weights
	\begin{equation}  
		\label{eq:f_bounds_new}
		f_{i,k_i}^{s^i}\prec \prod_{j=1}^{i-1} \left(\frac{y_j}{y_{j+1}}\right)^{-s_j^i}\,,\qquad 2\leq i\leq m\,.
	\end{equation}
	Note that in the case $i=1$ the coefficients are simply constant matrices, hence this constraint is only relevant for $i\geq 2$. Here the notation $\prec$ means the following: for two functions $f,g$ we write $f\prec g$ when $f$ is bounded by a constant multiple of $g$. The relation \eqref{eq:f_bounds_new} effectively describes the scaling of the various $f_{i,k_i}$ corrections in terms of their $\mathrm{sl}(2)$-weights.
\end{itemize}
We would like to mention that, to our knowledge, the bound \eqref{eq:f_bounds_new} has not been written down explicitly before, although it does follow directly from the proof of the $\mathrm{Sl}(2)$-orbit theorem of \cite{CKS}. We have therefore presented a proof of \eqref{eq:f_bounds_new} in Appendix \ref{app:additional_proofs}. In simple terms, one needs to use the details of how the expansion coefficients are derived from the boundary data and, in particular, use some of the restrictions on the $\mathrm{sl}(2)$-weights of the phase operators $\delta_{i}$.

\subsection{Example: type $\langle \mathrm{I}_1|\mathrm{IV}_2|\mathrm{IV}_1\rangle$}\label{sec:bulk-reconstruction_example_two-moduli}
To exemplify the multi-parameter bulk reconstruction procedure, let us consider a two-parameter type $\langle \mathrm{I}_1|\mathrm{IV}_2|\mathrm{IV}_1\rangle$ singularity. Geometrically, such a singularity corresponds to a large complex structure point that is obtained by intersecting a conifold singularity with a type $\mathrm{IV}_1$ singularity, and is part of a larger class of so-called coni-LCS boundaries. Let us also remark that such boundaries have been used to construct models with small flux superpotentials, see e.g.~\cite{Demirtas:2020ffz,Alvarez-Garcia:2020pxd}. In the following we will build on the computations performed in \cite{Bastian:2021eom}, see also \cite{vandeHeisteeg:2022gsp}, and we will follow the same conventions.  

\subsubsection*{Boundary data for $\mathrm{I}_1\rightarrow\mathrm{IV}_2$ enhancement}
In the following we list the boundary data that is associated to the type $\mathrm{I}_1\rightarrow\mathrm{IV}_2$ enhancement, corresponding to the growth sector $y_1>y_2$. We refer the reader to \cite{vandeHeisteeg:2022gsp,Bastian:2021eom} for more details on how this boundary data can be obtained. The two commuting real $\slt$-triples are collected in table \ref{tab:boundary-data_two-parameter}.
\begin{table}
	\centering
	\begin{tabular}{c|c|c|c}
		& $+$ & $0$ & $-$ \\ \hline
		 	
		$N^\bullet_{1}$ &  \rule[-1.2cm]{.0cm}{2.5cm} \scalebox{0.75}{$\begin{pmatrix}
				0 & 0 & 0 & 0 & 0 & 0 \\
				0 & 0 & 0 & 0 & -1 & 0 \\
				0 & 0 & 0 & 0 & 0 & 0 \\
				0 & 0 & 0 & 0 & 0 & 0 \\
				0 & 0 & 0 & 0 & 0 & 0 \\
				0 & 0 & 0 & 0 & 0 & 0 \\
			\end{pmatrix}$} & \scalebox{0.75}{$
			\begin{pmatrix}
				0 & 0 & 0 & 0 & 0 & 0 \\
				0 & 1 & 0 & 0 & 0 & 0 \\
				0 & 0 & 0 & 0 & 0 & 0 \\
				0 & 0 & 0 & 0 & 0 & 0 \\
				0 & 0 & 0 & 0 & -1 & 0 \\
				0 & 0 & 0 & 0 & 0 & 0 \\
			\end{pmatrix}$} &  \scalebox{0.75}{$
			\begin{pmatrix}
				0 & 0 & 0 & 0 & 0 & 0 \\
				0 & 0 & 0 & 0 & 0 & 0 \\
				0 & 0 & 0 & 0 & 0 & 0 \\
				0 & 0 & 0 & 0 & 0 & 0 \\
				0 & -1 & 0 & 0 & 0 & 0 \\
				0 & 0 & 0 & 0 & 0 & 0 \\
			\end{pmatrix}$} \\ \hline
			$N^\bullet_{2}$ &   \rule[-1.2cm]{.0cm}{2.5cm}\scalebox{0.75}{$\left(
				\begin{array}{cccccc}
					0 & 0 & 3 & 0 & 0 & 0 \\
					0 & 0 & 0 & 0 & 0 & 0 \\
					0 & 0 & 0 & 0 & 0 & 4 \\
					0 & 0 & 0 & 0 & 0 & 0 \\
					0 & 0 & 0 & 0 & 0 & 0 \\
					0 & 0 & 0 & -3 & 0 & 0 \\
				\end{array}
				\right)$} & \scalebox{0.75}{$\left(
				\begin{array}{cccccc}
					3 & 0 & 0 & 0 & 0 & 0 \\
					0 & 0 & 0 & 0 & 0 & 0 \\
					0 & 0 & 1 & 0 & 0 & 0 \\
					0 & 0 & 0 & -3 & 0 & 0 \\
					0 & 0 & 0 & 0 & 0 & 0 \\
					0 & 0 & 0 & 0 & 0 & -1 \\
				\end{array}
				\right)$} &  \scalebox{0.75}{$\left(
				\begin{array}{cccccc}
					0 & 0 & 0 & 0 & 0 & 0 \\
					0 & 0 & 0 & 0 & 0 & 0 \\
					1 & 0 & 0 & 0 & 0 & 0 \\
					0 & 0 & 0 & 0 & 0 & -1 \\
					0 & 0 & 0 & 0 & 0 & 0 \\
					0 & 0 & 1 & 0 & 0 & 0 \\
				\end{array}
				\right)$}
	\end{tabular}
	
	\caption{The two commuting real $\slt$-triples associated to a $\mathrm{I}_1\rightarrow\mathrm{IV}_2$ enhancement}
	\label{tab:boundary-data_two-parameter}
\end{table}
Furthermore, after possibly performing a shift of the coordinates, the most general phase operators are given by\footnote{For the particular example at hand, these expressions are valid for both the $\mathrm{I}_1\rightarrow\mathrm{IV}_2$ and the $\mathrm{IV}_1\rightarrow\mathrm{IV}_2$ enhancements.}
\begin{equation}
	\delta_{(1)}=0\,,\qquad \delta_{(2)}=\left(
	\begin{array}{cccccc}
		0 & 0 & 0 & 0 & 0 & 0 \\
		0 & 0 & 0 & 0 & 0 & 0 \\
		0 & 0 & 0 & 0 & 0 & 0 \\
		\delta_2 & \delta_1 & 0 & 0 & 0 & 0 \\
		\delta_1 & 0 & 0 & 0 & 0 & 0 \\
		0 & 0 & 0 & 0 & 0 & 0 \\
	\end{array}
	\right)\,.
\end{equation}
Finally, the boundary charge operator is given by
\begin{equation}
	Q_{(0)}=\left(
	\begin{array}{cccccc}
		0 & 0 & -\frac{3 i}{2} & 0 & 0 & 0 \\
		0 & 0 & 0 & 0 & \frac{i}{2} & 0 \\
		\frac{i}{2} & 0 & 0 & 0 & 0 & -2 i \\
		0 & 0 & 0 & 0 & 0 & -\frac{i}{2} \\
		0 & -\frac{i}{2} & 0 & 0 & 0 & 0 \\
		0 & 0 & \frac{i}{2} & \frac{3 i}{2} & 0 & 0 \\
	\end{array}
	\right)\,.
\end{equation}

\subsubsection*{Bulk Reconstruction: Step 1}
Let us perform the bulk reconstruction procedure associated to the enhancement chain $\mathrm{I}_1\rightarrow\mathrm{IV}_2$, corresponding to the sector in which $y_1>y_2$. For the first step of the bulk reconstruction, we are to use the boundary data $\{N_{(1)}^+, N^0_{(1)}, N_{(1)}^-\}$ as well as $\delta_{(1)}$. Since the phase operator $\delta_{(1)}$ vanishes, the bulk reconstruction for the coordinate $\frac{y_1}{y_2}$ is trivial, i.e.~we find
\begin{equation}
	h_{\mathrm{nil},1}\left(\frac{y_1}{y_2}\right) = \left(\frac{y_1}{y_2}\right)^{-\frac{1}{2}N^0_{(1)}}\,.
\end{equation}
\newpage 

\subsubsection*{Bulk Reconstruction: Step 2}
Next, we perform the bulk reconstruction for the $y_2$ coordinate. As discussed before, we are to use the data $h_1\{N_{(2)}^+, N^0_{(2)}, N_{(2)}^-\}h_1^{-1}$ as well as $Q_{(1)}=h_1Q_\infty h_1^{-1}$ and $\delta_{(2)}$, where we recall that
\begin{equation}
	N^\bullet_{(2)} = N^\bullet_1+N^\bullet_2\,.
\end{equation}
To be explicit, one finds
\begin{equation}
	Q_{(1)}\left(\frac{y_1}{y_2}\right)=\left(
	\begin{array}{cccccc}
		0 & 0 & -\frac{3 i}{2} & 0 & 0 & 0 \\
		0 & 0 & 0 & 0 & \frac{i}{2}\frac{y_2}{y_1} & 0 \\
		\frac{i}{2} & 0 & 0 & 0 & 0 & -2 i \\
		0 & 0 & 0 & 0 & 0 & -\frac{i}{2} \\
		0 & -\frac{i}{2}\frac{y_1}{y_2} & 0 & 0 & 0 & 0 \\
		0 & 0 & \frac{i}{2} & \frac{3 i}{2} & 0 & 0 \\
	\end{array}
	\right)\,,
\end{equation}
and
\begin{align}
	h_1 N_{(2)}^+h_{1}^{-1} &= \left(
	\begin{array}{cccccc}
		0 & 0 & 3 & 0 & 0 & 0 \\
		0 & 0 & 0 & 0 & -\frac{y_2}{y_1} & 0 \\
		0 & 0 & 0 & 0 & 0 & 4 \\
		0 & 0 & 0 & 0 & 0 & 0 \\
		0 & 0 & 0 & 0 & 0 & 0 \\
		0 & 0 & 0 & -3 & 0 & 0 \\
	\end{array}
	\right)\,,\\
	 h_1 N_{(2)}^-h_1^{-1}&=\left(
	\begin{array}{cccccc}
		0 & 0 & 0 & 0 & 0 & 0 \\
		0 & 0 & 0 & 0 & 0 & 0 \\
		1 & 0 & 0 & 0 & 0 & 0 \\
		0 & 0 & 0 & 0 & 0 & -1 \\
		0 & -\frac{y_1}{y_2} & 0 & 0 & 0 & 0 \\
		0 & 0 & 1 & 0 & 0 & 0 \\
	\end{array}
	\right)\,,
\end{align}
while $h_1N_{(2)}^0h_1^{-1}$ is simply equal to $N_{(2)}^0$. Note also that through the relation
\begin{equation}
	h_1 N_{(2)}^-h_1^{-1} = \frac{y_1}{y_2}N_1+N_2\,,
\end{equation}
one can immediately read off the log-monodromy matrices associated to this singularity. From here, computing the expansion of the map $h_{\mathrm{nil},2}\left(y_2;\frac{y_1}{y_2}\right)$ is straightforward and follows from simply applying the CKS recursion to the above data. However, in contrast to the one-parameter examples discussed before, we have not attempted to algebraically solve the resulting recursion relations. Nevertheless, one can straightforwardly compute the expansion to any (realistic) desired order. To give an idea of the result, we give the expansion of the full period map to first order in  $\delta_1,\delta_2$
\begin{equation}\label{eq:hnil_two-parameter}
	h_{\mathrm{nil}}=h_{\mathrm{Sl}(2)}+\begin{pmatrix}
		\frac{3\delta_2}{8y_2^{9/2}} & \frac{3\delta_1}{4y_1^{1/2}y_2^3} & 0 & 0 & 0 & \frac{9\delta_2}{4y_2^{9/2}}\\
		\frac{3\delta_1}{8y_1 y_2^{3/2}} & 0 & 0 & 0 & 0 & \frac{3\delta_1}{4y_1 y_2^{3/2}}\\
		0 & 0 & -\frac{9\delta_2}{8y_2^{7/2}} & \frac{9\delta_2}{4 y_2^{7/2}} & \frac{3\delta_1}{4y_1^{1/2}y_2^2} & 0\\
		0 & 0 & \frac{15\delta_2}{16y_2^{3/2}} & -\frac{3\delta_2}{8y_2^{3/2}} & -\frac{3\delta_1}{8y_1^{1/2}} & 0\\
		0 & 0& \frac{9\delta_1}{8y_2^{3/2}} & -\frac{3\delta_1}{4y_2^{3/2}} & 0 & 0\\
		\frac{15\delta_2}{16 y_2^{5/2}} & \frac{9\delta_1}{8 y_1^{1/2}y_2} & 0 & 0 & 0 &\frac{9\delta_2}{8y_2^{5/2}}
	\end{pmatrix}\,,
\end{equation}
where $h_{\mathrm{Sl}(2)}$ denotes the $\mathrm{Sl}(2)$-orbit approximation of the period map. 

\subsubsection*{Reconstructing the period vector for type $\langle\mathrm{I}_1|\mathrm{IV}_2|\mathrm{IV}_1\rangle$}

As a very non-trivial check, we have verified that the expression \eqref{eq:hnil_two-parameter} correctly yields the expected result for the nilpotent orbit approximation of the Hodge filtration, namely
\begin{equation}
	F_{\mathrm{nil}}(t_1,t_2) = \begin{pmatrix}
		1 & 0 & 0 & 0 & 0 & 0\\
		0 & 1 & 0 & 0 & 0 & 0\\
		t_2& 0 & 1 & 0 & 0 & 0\\
		-\frac{1}{6}t_2^3+i\delta_2 & i\delta_1 &-\frac{1}{2}t_2^2 & -t_2 & 0 & 1\\
		i\delta_1 & -t_1 & 0 & 0 & 1 & 0\\
		\frac{1}{2}t_2^2 & 0 & t_2 & 1 & 0 & 0
	\end{pmatrix}\,,
\end{equation}
where we have reinstated the axion-dependence using the monodromy matrices. We have checked that the above relation remains satisfied up to $\mathcal{O}\left(\delta_1^4\right)$ and $\mathcal{O}\left(\delta_2^3\right)$. This gives us good confidence that we have performed the two-parameter bulk reconstruction procedure correctly. Finally, the most general asymptotic form of the period vector that gives rise to this Hodge filtration is given by
\begin{equation*}\label{eq:Pi-nil_I1-IV2}
	\mathbf{\Pi}_{\mathrm{nil}}(t_1,t_2) =e^{t_1 N_1+t_2N_2}\left[\begin{pmatrix}
		1 \\ 0 \\ 0 \\ i\delta_2 \\ i\delta_1\\ 0
	\end{pmatrix}+A_1e^{2\pi it_1}\begin{pmatrix}
		0 \\ 1 \\ 0 \\ i\delta_1 \\ \frac{1}{2\pi i} \\ 0
	\end{pmatrix}+A_1^2 e^{4\pi i t_1}\begin{pmatrix}
	0 \\ 0 \\ 0 \\ \frac{1}{4\pi i} \\ 0 \\ 0
	\end{pmatrix} \right]\,,
\end{equation*}
for some non-zero $A_1\in\mathbb{C}$. Here we again note the appearance of essential instantons, whose presence is expected due to the underlying type $\mathrm{I}_1$ singularity. As the reader may verify, the result we find agrees with the results of \cite{Bastian:2021eom}, though they are obtained through different methods. \\

\noindent To close the discussion on this example, let us note that one can similarly perform the bulk reconstruction procedure for the $\mathrm{IV}_1\rightarrow\mathrm{IV}_2$ enhancement, corresponding to the sector in which $y_2>y_1$. As another non-trivial consistency check, one finds the exact same expression \eqref{eq:hnil_two-parameter} for the nilpotent orbit expansion of the period map. This is as expected, since the nilpotent orbit approximation does not rely on any hierarchy between the saxions. 

\section{Application: Asymptotic Hodge inner products}
\label{sec:inner_prod}

With the result \eqref{eq:hinv} for the nilpotent orbit approximation of the period map at hand, it is possible to evaluate Hodge inner products in great generality. Indeed, given two elements $v,w\in H$, the nilpotent orbit approximation of their Hodge inner product can be expressed as
\begin{equation}
\langle v, w\rangle_{\mathrm{nil}} = \langle h_{\mathrm{nil}}^{-1}v, h_{\mathrm{nil}}^{-1}w\rangle_\infty\,.
\end{equation}
In particular, all the moduli-dependence of the Hodge inner product is captured by the factors of $h^{-1}$. In general, the expressions that result from inserting \eqref{eq:hinv} can become rather involved, in particular due to the fact that many cross-terms can emerge. 

\subsubsection*{Central charge of D3-particles}
There is at least one case, however, in which the resulting expressions simplify nicely. This is the case where one is considering generic Hodge inner products between the period vector $\mathbf{\Pi}$, or equivalently the $(D,0)$-form $\Omega$, and an element of a definite $\mathrm{sl}(2)$-weight, denoted by $q$ in the following. This is of particular relevance when studying BPS states that arise from D3-branes wrapping a particular class of three-cycles in a Calabi--Yau threefold, in the context of type IIB string theory compactifications. Indeed, for a given BPS state, parametrized by a charge vector $q\in H^3(Y_3,\mathbb{Z})$, its mass is given by
\begin{equation}
\label{eq:central_charge}
|\mathcal{Z}| = \frac{|\langle q,\Omega\rangle|}{||\Omega||}\,,
\end{equation}
where $\mathcal{Z}$ denotes the central charge of the BPS state, and $\Omega$ the holomorphic $(3,0)$-form on $Y_3$. Let us evaluate \eqref{eq:central_charge} in the nilpotent orbit approximation. First, since $\Omega\in H^{3,0}$, we may apply the relation \eqref{eq:Fnil_Finf_multi} and write
\begin{equation}
h_{\mathrm{nil}}^{-1}\Omega = f\cdot \Omega_\infty\,,
\end{equation}
for some moduli-independent element $\Omega_\infty\in H^{3,0}_\infty$ (suppressing the axions for simplicity). Since \eqref{eq:Fnil_Finf_multi} is a vector-space identity, there is of course the freedom of an overall (possibly moduli-dependent) scaling, which we parametrize by the function $f$. Importantly, in the expression \eqref{eq:central_charge} for the central charge, this overall factor will drop out. Indeed, one finds
\begin{equation}
|\mathcal{Z}| = \frac{|\langle h_{\mathrm{nil}}^{-1}q, h_{\mathrm{nil}}^{-1}\Omega\rangle_\infty|}{||h_{\mathrm{nil}}^{-1}\Omega||_\infty} = \frac{|\langle h_{\mathrm{nil}}^{-1}q, \Omega_\infty\rangle_\infty|}{||\Omega_\infty||_\infty}\,.
\end{equation}
Assuming then, for simplicity, that $q$ has a definite $\mathrm{sl}(2)$ weight given by $\ell$, and applying the result \eqref{eq:hinv} one finds (in the nilpotent orbit approximation)
\begin{align}
\label{eq:central_charge_nilpotent}
| \mathcal{Z} |&= \prod_{i=1}^m \left(\frac{y_i}{y_{i+1}}\right)^{\frac{1}{2}\ell_i}\Big|\sum_{k_1,\ldots, k_m=0}^\infty\sum_{s^1,\ldots, s^m} \left[\prod_{i=1}^m \left(\frac{y_i}{y_{i+1}} \right)^{-k_i+\frac{1}{2}s_i^{(m)}}\right]\nn \\
&\hspace{4cm} \frac{\langle f_{1,k_1}^{s^1}\cdots f_{m,k_m}^{s^m} q, \Omega_\infty\rangle_\infty}{||\Omega_\infty||_\infty}\Big|
\end{align}
Note that the overall prefactor in \eqref{eq:central_charge_nilpotent} comes from evaluating the action of $h^{-1}_{\mathrm{Sl}(2)}$ on the weight-eigenvector $q$. Focusing for the moment on the leading term in \eqref{eq:central_charge_nilpotent}, one may write
\begin{equation}
|\mathcal{Z}| = \prod_{i=1}^m \left(\frac{y_i}{y_{i+1}}\right)^{\frac{1}{2}\ell_i}\cdot \left|\frac{\langle q, \Omega_\infty\rangle_\infty}{||\Omega_\infty||_\infty}+\mathrm{corrections}\right|\,.
\end{equation}
In particular, the asymptotic behaviour of the mass of the BPS state, given by $|\mathcal{Z}|$, is determined by weight of the charge vector $q$. This first-order expression has been used in \cite{Bastian:2020egp}, see also \cite{Gendler:2020dfp}, to compute the charge-to-mass ratio for this class of BPS states, under the assumption that the pairing $\langle q,\Omega_\infty\rangle_\infty$ is non-vanishing. Physically, this assumption can be interpreted as the statement that the asymptotic coupling of the BPS state to the graviphoton is non-zero. However, if the pairing $\langle q,\Omega_\infty\rangle_\infty$ does vanish, then it becomes necessary to consider the correction terms coming from the nilpotent orbit expansion in order to properly characterize the scaling of the mass. Notably, depending on the choice of $q$, as well as the type of boundary that one is expanding around, it may happen that a state which naively appears to become massive in fact becomes massless as one approaches the boundary. It would be interesting to generalize the analysis of \cite{Bastian:2020egp} to include the sub-leading corrections using the multi-variable bulk reconstruction procedure.

\subsection{Example: type $\langle\mathrm{I}_1|\mathrm{IV}_2|\mathrm{IV}_1\rangle$}
Let us briefly illustrate the above results by revisiting the two-parameter type $\langle\mathrm{I}_1|\mathrm{IV}_2|\mathrm{IV}_1\rangle$ singularity discussed in section \ref{sec:bulk-reconstruction_example_two-moduli}. For definiteness, we work within the region $y_1>y_2$, but the same results will hold for the region $y_2>y_1$. 

\subsubsection*{Choosing a charge vector}
Let us first choose an appropriate charge vector $q$ for which we will compute the charge-to-mass ratio. Due to the singularity structure, the weights of $\Omega_\infty$ are given schematically by
\begin{equation}
\Omega_\infty = |0,3\rangle+|0,1\rangle+|0,-1\rangle+|0,-3\rangle\,.
\end{equation}
In other words, $\Omega_\infty$ always has $N^0_{(1)}$ weight equal to 0, while its $N^0_{(2)}$ weights range over $-3,-1,1,3$, in accordance with the fact that this is a $\mathrm{I}\rightarrow\mathrm{IV}$ enhancement. To obtain a non-trivial result, we would like to choose a charge vector such that
\begin{equation}
\langle q,\Omega_\infty\rangle_\infty = 0\,.
\end{equation}
By checking which possible states are present in the current setup, one finds that there are two possibilities for $q$, namely
\begin{equation}
q_+ = |1,1\rangle\,,\qquad q_- = |-1,-1\rangle\,.
\end{equation}
Explicitly, these states are respectively given by
\begin{equation}
\mathbf{q}_+ = \begin{pmatrix}
0\\1\\0\\0\\0\\0
\end{pmatrix}\,,\qquad \mathbf{q}_-=\begin{pmatrix}
0\\0\\0\\0\\1\\0
\end{pmatrix}\,,\qquad \mathbf{\Pi}_\infty = \begin{pmatrix}
-2 \\ 0 \\ -2i \\ -\frac{i}{3} \\ 0 \\ 1
\end{pmatrix}\,,
\end{equation}
where we have included the matrix representation $\mathbf{\Pi}_\infty$ for the boundary $(3,0)$-form. 

\subsubsection*{The charge-to-mass ratio}
Using the nilpotent orbit expansion of the period map \eqref{eq:hnil_two-parameter}, one can straightforwardly compute the nilpotent orbit approximation of the Weil operator via the relation
\begin{equation}
C_{\mathrm{nil}} = h_{\mathrm{nil}} C_\infty h_{\mathrm{nil}}^{-1}\,.
\end{equation}
For transparency, we present the expression for $C_{\mathrm{nil}}$ to first order in $\delta_1$ and $\delta_2$
\begin{equation}
C_{\mathrm{nil}} = \begin{pmatrix}
0 & 0 & \frac{27\delta_2}{4y_2^4} & -\frac{6}{y_2^3}-\frac{9\delta_2}{2y_2^6} & -\frac{3\delta_1}{y_1 y_2^3}& 0\\
0 & 0 & \frac{3\delta_1}{2y_1 y_2} & -\frac{3\delta_1}{y_1 y_2^3}& -\frac{1}{y_1} & 0\\
\frac{9\delta_2}{4y_2^2} & \frac{3\delta_1}{y_2^2} & 0 & 0 & 0 & -\frac{2}{y_2}+\frac{9\delta_2}{2y_2^4}\\
\frac{y_2^3}{6}-\frac{\delta_2}{8} & -\frac{\delta_1}{2} & 0 & 0 & 0 & -\frac{9\delta_2}{4y_2^2}\\
-\frac{\delta_1}{2} & y_1 & 0 & 0 & 0& -\frac{3\delta_1}{y_2^2}\\
0 & 0 & \frac{y_2}{2}+\frac{9\delta_2}{8y_2^2}& -\frac{27\delta_2}{4y_2^4} & -\frac{3\delta_1}{2y_1 y_2}&0
\end{pmatrix}\,.
\end{equation}
Using this expression, as well as the expression for the holomorphic $(3,0)$-form, one can immediately compute to charge-to-mass ratio for the two BPS states described by the charge vectors $q_{\pm}$. We find\footnote{Recall that the mass $\mathcal{M}$ is given in terms of the central charge \eqref{eq:central_charge} via $\mathcal{M}=|\mathcal{Z}|$.}
\begin{equation}
\left(\frac{\mathcal{Q}}{\mathcal{M}}\right)^{-1} = \frac{\langle q|\Omega\rangle}{||q||\,||\Omega||} = \begin{cases}
\frac{\sqrt{3}\delta_2}{2}(y_1y_2^{3})^{-1}+\cdots \,, & q=q_+\,,\\
0\,,& q=q_-\,.
\end{cases}
\end{equation}
In other words, we find that for $q_+$, the charge-to-mass ratio diverges polynomially as
\begin{equation}
q_+:\qquad \frac{\mathcal{Q}}{\mathcal{M}}\sim \sqrt{y_1 y_2^3}\,,
\end{equation}
while for $q_-$ it turns out that the charge-to-mass ratio vanishes to polynomial order. In particular, in order to properly describe its scaling, it is necessary to include the essential instanton corrections to the period vector, as described in \eqref{eq:Pi-nil_I1-IV2}. After including these corrections, one finds
\begin{equation}
q_-:\qquad \frac{\mathcal{Q}}{\mathcal{M}}\sim \sqrt{\frac{y_2^3}{y_1}}e^{2\pi y_1}\,,
\end{equation}
which diverges polynomially in $y_2$ and exponentially in $y_1$.

\begin{subappendices}

\section{Computations on the $Q$-constraint and the CKS input data}
\label{app:input}

In this section some properties of the input data $\Phi_n^{n,-1}$ are discussed in terms of the boundary data. We first explain how the $Q$-constraint can be written in terms of the complex algebra and the boundary charge operator $Q_\infty$. The resulting constraint is then used to fix the coefficients of the input data. We also show that the expression \eqref{eq:input_data} for the input data indeed has the desired properties, i.e.~it has $(d,\epsilon)=(n,-1)$.

\subsection{Rewriting the $Q$-constraint}

We recall the $Q$-constraint, as given in \eqref{eq:Q_constraint_N}
\begin{equation}
[Q_\infty,\cN^0]=i\left(\cN^+ + \cN^-\right)\, ,\qquad [Q,\cN^\pm] = -\frac{i}{2}\cN^0\, .
\end{equation}
Let us first write the above equations in the complex algebra by recalling \eqref{def-bfLbullet}
\begin{equation}
\label{eq:rho_Q_rhoinv}
[\rho Q_\infty \rho^{-1},\mathbf{L}^0]=i\left(\mathbf{L}^+ + \mathbf{L}^-\right)\, ,\qquad [\rho Q_\infty\rho^{-1},\mathbf{L}^\pm] = -\frac{i}{2}\mathbf{L}^0\, .
\end{equation}
By using \eqref{def-rho} and the commutation relations between $Q_\infty$ and the $L_\alpha$, one computes
\begin{align}
\rho Q_\infty \rho^{-1}&= e^{\frac{i\pi}{4}\ad\left(L_{1}+L_{-1}\right)} Q_\infty\nn \\
&= Q_\infty+\sum_{n=1}^\infty 4^{n-1}\left[\frac{(i\pi /4)^{2n}}{(2n)!} 2L_0-\frac{(i\pi /4)^{2n-1}}{(2n-1)!}\left(L_{1}-L_{-1}\right) \right]\nn \\
&= Q_\infty - \frac{1}{2}L_0 - \frac{i}{2}L_{1} + \frac{i}{2} L_{-1}\, .
\end{align}
Inserting this result in \eqref{eq:rho_Q_rhoinv} then immediately yields the $Q$-constraint as presented in \eqref{charges_cL}
\begin{align}
\label{eq:Q_constr_coeff}
[2Q_\infty-L_0, \mathbf{L}^0] &= 2i(\mathbf{L}^++\mathbf{L}^-)+i[L_{1},\mathbf{L}^0]-i[L_{-1},\mathbf{L}^0]\, ,\\
[2Q_\infty-L_0, \mathbf{L}^\pm] &= -i\mathbf{L}^0 +i [L_{1}, \mathbf{L}^\pm]-i[L_{-1}, \mathbf{L}^\pm]\, .
\end{align} 
\newpage 

\subsection{Imposing the $Q$-constraint on the CKS input data}

We recall that the input data for the CKS recursion is given by vectors $\Phi^{n,-1}_{n}$ which have $d=n$ and $\epsilon=-1$. In the main text the following expression for the input data $\Phi_n^{n,-1}$ was given in terms of the boundary data
\begin{equation}
\Phi_n^{n,-1}= \sum_{1\leq s,q \leq n-1} a^{n,s}_q \begin{pmatrix}
	-\frac{1}{n-s} \left(\mathrm{ad}\;L_{1}\right)^{s+1}\\
	2 \left(\mathrm{ad}\;L_{1}\right)^{s} \\
	(n-s+1) \left(\mathrm{ad}\;L_{1}\right)^{s-1}
\end{pmatrix}\hat{\eta}^{(-n)}_{-q}\, .
\end{equation}
In this section we will show that indeed this $\Phi_n^{n,-1}$ has eigenvalue $\epsilon=-1$. By the characterization \eqref{eq:B_eigenvalue} of the eigenvalue $\epsilon$, it suffices to show that
\begin{equation}
4B_0(\Phi_n^{n,-1})=(n+2)\Phi_n^{n,-1}\, ,
\end{equation}
where 
\begin{equation}
4B_0= \begin{pmatrix}
	\ad L_0 & -\ad L_{1} & 0\\
	-2\ad L_{-1} & 0 & 2 \ad L_{1}\\
	0 & \ad L_{1} & -\ad  L_0
\end{pmatrix}.
\end{equation}
Inserting the expression for $\Phi_n^{n,-1}$ into the above equation, we obtain
\begin{equation*}
4B_0(\Phi_n^{n,-1}) = \sum_{s,q} a^{n,s}_q \begin{pmatrix}
	-\frac{1}{n-s} \ad L_0 (\ad L_{1})^{s+1} - 2 (\ad L_{1})^{s+1}\\
	\frac{2}{n-s} \ad L_{-1} (\ad L_{1})^{s+1} +2(n-s+1)(\ad L_{1})^s\\
	2 \ad L_{-1}(\ad L_{1})^s - (n-s+1) \ad L_0 (\ad L_{1})^{s-1}
\end{pmatrix}\hat{\eta}^{(-n)}_{-q}\, .
\end{equation*}
Next, we use the Jacobi identity and the $\slt$-algebra to write
\begin{align}
\ad L_0 (\ad L_{1})^{s}&= (\ad L_{1})^s(\ad L_0 +2s)\, ,\\
\ad L_{-1}(\ad L_{1})^s &= s (\ad L_{1})^{s-1}(-\ad L_0-s+1)+(\ad L_{1})^s \ad L_{-1}\, .
\end{align}
Recalling that $\hat{\eta}^{(-n)}_{-q}$ satisfies
\begin{equation}
[L_{-1},\hat{\eta}^{(-n)}_{-q}]=0,\quad [L_0, \hat{\eta}^{(-n)}_{-q}] = -n\; \hat{\eta}^{(-n)}_{-q}\, ,
\end{equation}
it follows that
\begin{align}
\ad L_0 (\ad L_{1})^{s} \hat{\eta}^{(-n)}_{-q}&= (2s-n)(\ad L_{1})^s\hat{\eta}^{(-n)}_{-q}\, ,\\
\ad L_{-1} (\ad L_{1})^s \hat{\eta}^{(-n)}_{-q} &= s (n-s+1)(\ad L_{1})^{s-1}\hat{\eta}^{(-n)}_{-q}\, .
\end{align} 
Using these two results, one may simplify	
\begin{align*}
4B_0(\Phi_n^{n,-1}) &=(n+2)\sum_{s,q} a^{n,s}_q \begin{pmatrix}
	-\frac{1}{n-s}  (\ad L_{1})^{s+1}\\
	2   (\ad L_{1})^{s}\\
	(n-s+1)  (\ad L_{1})^{s-1}
\end{pmatrix}\hat{\eta}^{(-n)}_{-q}= (n+2) \Phi_n^{n,-1}\, .
\end{align*}
We conclude that the input data indeed has $\epsilon=-1$. Note that this derivation is valid for any choice of $a^{n,s}_q$. To fix the $a^{n,s}_q$, we evaluate the $Q$-constraint at level $n$ and insert the expression for $\Phi_n^{n,-1}$. We follow the computation in \cite{Grimm:2020cda}. Using the fact that
\begin{equation}
[2Q_\infty - L_0, (\ad L_{1})^s\hat{\eta}^{(-n)}_{-q}] = (n-2q)(\ad L_{1})^s\hat{\eta}^{(-n)}_{-q}\, ,
\end{equation}
and as well as the previous relations we find that \eqref{eq:Q_constr_coeff} reduces to
\begin{align*}
0&=\sum_{1\leq s,q\leq n-1} \left(i(n-2q)a^{n,s}_q+ \frac{n-s}{n-s+1} a^{n,s-1}_q-s(n-s) a^{n,s+1}_q \right) (\ad L_{1})^{s}\hat{\eta}^{(-n)}_{-q}\, ,
\end{align*} 
where in the second line we collected terms according to their power in $\ad L_{1}$. Since this relation must hold for every $s$ separately, we find
\begin{equation}
\label{eq:Q_constraint_coef}
i(n-2q)a^{n,s}_q+ \frac{n-s}{n-s+1} a^{n,s-1}_q-s(n-s) a^{n,s+1}_q=0\, , \quad a^{n,1}_q=\frac{1}{n}\, .
\end{equation}
The particular normalization is due to the fact that one should recover $\hat{\eta}$ from the $L_n^-$ as follows:\footnote{The component of $L_n^-$ which has weight $-n$ corresponds to the $s=1$ part of the sum.}
\begin{equation}
\hat{\eta} = \sum_{n\geq 2} \sum_{q\geq 1} (L_n^-)^{(n,-n)}_{-q}=\sum_{n\geq 2} \sum_{q\geq 1} na^{n,1}_q \hat{\eta}^{(-n)}_{-q}\, .
\end{equation}
In \cite{Grimm:2020cda} it was stated that \eqref{eq:Q_constraint_coef} together with the reality condition $\overline{a^{n,s}_q}=a^{n,s}_{n-q}$ has a unique solution given by
\begin{equation}
a^{n,s}_q = i^{s-1} \frac{(n-s)!}{n!} b^{s-1}_{q-1,n-q-1}\, ,
\end{equation}
where the coefficients $b^k_{p,q}$ are determined by
\begin{equation}
(1-x)^p(1+x)^q = \sum_{k=0}^{p+q} b^k_{p,q} x^k \, .
\end{equation}
\newpage 

\section{Expressions for the bulk reconstruction of $\mathrm{IV}_1$  boundaries}
\label{app:LCS_basis}
In this appendix we collect some relevant expressions for the bulk reconstruction of the $\mathrm{IV}_1$ boundary that complements the derivation in section \ref{ssec:IV1}. In particular, we consider the recursion relation \eqref{eq:recursion_CKS} for $\Phi_n=(\mathbf{L}^+, \mathbf{L}^0, \mathbf{L}^-)$ for $\mathrm{IV}_1$ boundaries. By using ansatz \eqref{eq:LCS_Phi_ansatz} we find that the CKS recursion reduces to a system of four coupled recursion relations for the coefficients $a_n$, $b_n$, $c_n$ and $d_n$, given by
\begin{align}\label{eq:recursion}
	a_{n}+b_{n} &= \frac{2}{6n+8}\sum_{0<k<n} \Big[a_{n-k} \Big(9 c_{k}+14 i \sqrt{\frac{6}{5}} b_{k}\Big) +3 b_{n-k} \Big( \frac{28 i}{3}
	\sqrt{\frac{2}{15}} b_k +4 c_k+d_k \Big) \Big] ,  \nonumber \\
	3a_{n}-4b_{n} &= \frac{2}{6n-6}\sum_{0<k<n} \Big[12 b_{n-k} \Big(7 i \sqrt{\frac{2}{15}}
	b_{k}+3 c_{k}-d_{k} \Big)-4a_{n-k} \Big(9 c_{k}+14 i \sqrt{\frac{6}{5}} b_{k}\Big)\Big] , \nonumber \\
	c_{n}+d_{n} &= \frac{2}{6n+4}\sum_{0<k<n} \frac{1}{30} \Big[90 ( c_{k}+d_{k}) c_{n-k}-1568 b_{k} (3a_{n-k}+b_{n-k}) \Big] ,\nn \\
	2c_{n}-d_{n} &= \frac{2}{6n-2}\sum_{0<k<n} \Big[\frac{784}{15} b_{n-k}(b_k-6a_k) +3c_{n-k}(2d_k- c_{n-k} )\Big]\, . 
\end{align}
Let us remind the reader that these coupled recursion relations are solved by \eqref{eq:LCS_recursion_solution}.

\section{Properties of $f_{i,k_i}$}
\label{app:additional_proofs}
In this section we elaborate on the properties of the expansion functions $f_{i,k_i}$ appearing in the nilpotent orbit expansion, see for example equation \eqref{eq:hinv}. We focus on the property
\begin{equation}   
	\label{eq:f_bounds_app}
	f_{i,k_i}^{s^i}\prec \prod_{j=1}^{i-1} \left(\frac{y_j}{y_{j+1}}\right)^{-s_j^i}\,,\qquad 2\leq i \leq m\,,
\end{equation}
see also equation \eqref{eq:f_bounds_new} and the surrounding discussion. We present the proof of \eqref{eq:f_bounds_app} for the case $m=2$ to give the general idea. For arbitrary $m$, the argument will be very similar but becomes more cumbersome to write down. Setting $m=2$, we need to show that
\begin{equation}
	f^{(s_1^2,s_2^2)}_{2,k_2}\prec \left(\frac{y_1}{y_2}\right)^{-s_1^2}\,.
\end{equation}
It suffices to consider the case $s_1^2>0$, since it is certainly the case that $f_{2,k_2}\prec 1$. It follows from the general mechanism of the multi-variable bulk reconstruction that $f_2$ is a Lie polynomial of the form
\begin{equation}
	f_2\left(\frac{y_1}{y_2}\right) = P\left(\mathrm{Ad}\left(h_1\left(\frac{y_1}{y_2}\right)\right)N_{(2)}^+, \delta_{(2)}\right)\,.
\end{equation}
Importantly, the $y_1/y_2$-dependence of $f_2$ is due entirely to $h_1N^+_{(2)}h_1^{-1}$. It is therefore crucial to understand the scaling of this object. To this end, we recall that, according to the $\mathrm{Sl}(2)$-orbit theorem, we have the relation (see Lemma 4.37 of \cite{CKS})
\begin{equation}
	h_1 N^-_{(2)}h_1^{-1} = \frac{y_1}{y_2}N_1+N_2\,.
\end{equation}
In particular, this implies that $N_{(2)}^-$ can only have weights $0$ and $-2$ with respect to $N^0_{(1)}$. Therefore, since $N^0_{(2)}$ has weight $0$ with respect to $N^0_{(1)}$ (the two commute) it must be that $N_{(2)}^+$ has weights $0$ and $+2$ with respect to $N_{(1)}^0$. In other words, we have the decomposition
\begin{equation}
	h_1 N_{(2)}^+h_1^{-1} = \left(\frac{y_1}{y_2}\right)^{-1}\left(N_{(2)}^+\right)^{(2,2)}+\left(N_{(2)}^+\right)^{(0,2)}+\text{subleading terms}\,.
\end{equation}
Note that the subleading terms can also only have weights $(2,2)$ and $(0,2)$. 
To continue the proof, we recall that the phase operator $\delta_{(2)}$ satisfies 
\begin{equation}
	\delta_{(2)} = \sum_{\tilde{s}_1^ 2}\sum_{\tilde{s}_2^2\leq -2} \delta_{(2)}^{(\tilde{s}_1^2,\tilde{s}_2^2)}\,\qquad [N_{(1)}^-,\delta_{(2)}]=[N_{(2)}^-,\delta_{(2)}]=0\,.
\end{equation}
In words, $\delta_{(2)}$ is a lowest-weight operator, with weight $\tilde{s}_2^2$ at most $-2$. By the enhancement rules for limiting mixed Hodge structures, see e.g.~\cite{Grimm:2018cpv}, this implies that also $\tilde{s}_1^2\leq 0$. Importantly, since $\delta_{(2)}$ is lowest weight, we have that
\begin{equation}
	\mathrm{ad}\left(h_1 N_{(2)}^+h_1^{-1}\right)^s\delta_{(2)}^{(\tilde{s}_1^2,\tilde{s}_2^2)}=0\,,
\end{equation}
whenever $s>\mathrm{min}(\tilde{s}_1^2,\tilde{s}_2^2)$. The argument now proceeds as follows. In order for $f_2$ to have a component with positive weight $s_1^2$, this can only happen due to $\mathrm{ad}\left(N^+_{(2)}\right)^{(2,2)}$ acting on $\delta_{(2)}$, thereby raising the $N^0_{(1)}$ weight by two. However, this goes at the expense of an additional factor $\left(y_1/y_2\right)^{-1}$. Moreover, since $\delta_{(2)}$ is of lowest weight, the most conservative way to get an $s_1^2$ component is by acting exactly $s_1^2$ times with $\mathrm{ad}\left(N^+_{(2)}\right)^{(2,2)}$ on the $-s_1^2$ component of $\delta_{(2)}$ (if it is present). This therefore comes with an additional factor of $\left(y_1/y_2\right)^{-s_1^2}$. This concludes the proof of property \eqref{eq:f_bounds_app} for the case $m=2$. 
\end{subappendices}
%\subfile{bulkreconstruction.tex}

%%%%%%%%%%%%%%%%%%%%%%%%%%%%%%%%%%%%%%%%%%%%%%%%%%%%%%
\setpartpreamble[u][\textwidth]{
\vspace*{1cm}
\hrulefill 
\vspace*{0.5cm}

In this part of the thesis we apply the machinery of asymptotic Hodge theory to study the landscape of four-dimensional effective theories arising from type F-theory compactifications on Calabi--Yau fourfolds. In particular, in chapter \ref{chap:finiteness} we revisit known results on the finiteness of self-dual flux vacua and show how the multi-variable Sl(2)-orbit theorem, together with the multi-variable bulk reconstruction procedure discussed in chapter \ref{chap:asymp_Hodge_II}, provides a complimentary perspective on this matter. Additionally, we motivate and formulate a set of three conjectures which are meant to address finer features of the flux landscape, based on expectations coming from Hodge theory and recent results on tame/o-minimal geometry. These conjectures greatly constrain the enumeration of certain classes of self-dual flux vacua, as well as the dimensionality and geometric complexity of the vacuum locus.

\vspace*{0.5cm}
\hrulefill }

\part{Applications in the String Landscape}\label{part3}
\chapter{Finiteness Theorems and Counting Conjectures for the Flux Landscape}\label{chap:finiteness}
\epigraph{This chapter is based on: Thomas W. Grimm, Jeroen Monnee: \emph{Finiteness Theorems and Counting Conjectures for the Flux Landscape}, \href{https://arxiv.org/abs/2311.09295}{\textbf{[arXiv: 2311.09295]}}
}

\noindent String theory is known to have a plethora of solutions around which effectively four-dimensional quantum field theories coupled to classical gravity can be determined. The space of such lower-dimensional effective theories is often referred to as the \textbf{string theory landscape}. With this understanding, one might then inquire which of these theories can possibly describe our Universe. On a more basic level one might wonder if the number of such theories, after appropriately identifying equivalent theories, is at all finite. If this is not the case, one should seriously question the predictive capabilities of string theory. These issues were addressed at length in the seminal works of Douglas et al.~\cite{Douglas:2003um,Ashok:2003gk,Denef:2004ze,Acharya:2006zw}, which led to the general expectation that the string landscape is, in an appropriate sense, finite. This expectation is further corroborated by efforts in the swampland program \cite{Hamada:2021yxy,Grimm:2021vpn}, which aims to identify the fundamental properties an effective theory coupled to gravity should satisfy in order to admit a UV-completion, see \cite{Palti:2019pca,vanBeest:2021lhn} for reviews. Concurrently, there have been some major developments in the fields of algebraic geometry and logic that have lifted this expectation to the level of a mathematical theorem, at least in specific settings. The aim of this chapter is to provide a collection of finiteness results, coming from the fields of Hodge theory and tame geometry, in a way that is hopefully accessible to physicists. In particular, we hope to clarify what has/has not been shown and to give some insights and new perspectives on the various proofs. We then draw from this knowledge to put forward a number of structural conjectures about the landscape.\\

\noindent To prove something about the whole string landscape is a daunting task. Therefore, we will focus our attention on a particular corner of the string landscape, namely those four-dimensional low-energy effective theories that arise from flux compactifications of type IIB/F-theory \cite{Dasgupta:1999ss,Giddings:2001yu,Grimm:2004uq,Grimm:2010ks}, see \cite{Grana:2005jc,Douglas:2006es,Denef:2008wq} for reviews. The essential features of this setting have also been introduced and summarized in chapter \ref{chap:string}. Indeed, recall that these compactifications, viewed from the dual M-theory perspective \cite{Becker:1996gj}, are specified by a family of Calabi--Yau fourfolds varying in moduli, together with a background flux $G_4$. The moduli are generically stabilized at the critical points of the scalar potential induced by the flux, leading to a typically large landscape of flux vacua. Such vacua are of great phenomenological interest, as they feature spontaneous supersymmetry breaking down to $\mathcal{N}=1$ or even $\mathcal{N}=0$, and may eventually lead to de Sitter solutions \cite{Kachru:2003aw,Balasubramanian:2005zx} with a small cosmological constant \cite{Demirtas:2019sip,Alvarez-Garcia:2020pxd,Demirtas:2020ffz,Honma:2021klo,Demirtas:2021ote,Broeckel:2021uty,Bastian:2021hpc}. A crucial point is that the flux has to satisfy a number of consistency conditions, as the effective theory originates from a UV complete theory of quantum gravity. These conditions include a quantization condition and the tadpole cancellation condition. Consequently, the central question is whether there exists only a finite number of fluxes and associated critical points that simultaneously satisfy these consistency conditions. To be clear, we will consider the issue of finiteness within a given family of Calabi--Yau fourfolds, varying in moduli. In particular, we will not address whether there exist only finitely many distinct topological classes of Calabi--Yau fourfolds, which is an interesting question on its own. \\

\noindent In the context of IIB/F-theory flux compactification, initial evidence for this suggested finiteness was presented in the works \cite{Ashok:2003gk,Denef:2004ze,Acharya:2006zw}, which where later formalized in the mathematical works \cite{Douglas:2004zu,Douglas:2004kc,Douglas:2005df}. The underlying approach in these studies involved approximating the total number or index of flux vacua by integrating a suitable distribution of flux vacua over the moduli space, and showing that the latter is finite \cite{Eguchi:2005eh,Douglas:2006zj,Lu:2009aw}. From this distribution one could also obtain rough estimates for the total number of flux vacua, leading to the infamous number $10^{500}$. However, one critical limitation in their analysis was the relaxation of the quantization condition on the flux. Indeed, in order to give a complete proof of the finiteness of flux vacua, one expects that the quantization condition is crucial. \\

\noindent In the mathematics literature, a number of precise finiteness results have been established, which we will review in section \ref{sec:finiteness_intro}. Recalling the terminology introduced in section \ref{ssec:N=1-sugra}, we start by considering Hodge vacua, or vacua with $W_{\mathrm{flux}}=0$. From a Hodge-theoretic point of view, such vacua correspond to so-called ``Hodge classes''. One of the major milestones of Hodge theory is a theorem of Cattani, Deligne, and Kaplan which states that the locus of Hodge classes is a countable union of algebraic varieties \cite{CDK}. Interestingly, the same result can also be derived by assuming the Hodge conjecture to be true. For this reason, the result of Cattani, Deligne, and Kaplan is often viewed as the strongest evidence for the Hodge conjecture. Furthermore, if the flux satisfies the tadpole cancellation condition, meaning it has a bounded self-intersection, then the locus is in fact a \textit{finite} union of algebraic varieties. In particular, its number of connected components, which counts the number of Hodge vacua with possibly flat directions, is finite. Subsequently, we will discuss the more general class of self-dual flux vacua, for which initial finiteness results were presented in \cite{Schnellletter}, see also \cite{Grimm:2020cda}, for the case of a single complex structure modulus. In these works a detailed description of the Hodge norm of the $G_4$ flux was obtained by employing the one-variable $\mathrm{Sl}(2)$-orbit theorem of Cattani, Kaplan and Schmid \cite{CKS}, which has been discussed at length in chapters \ref{chap:asymp_Hodge_I} and \ref{chap:asymp_Hodge_II}. The finiteness of self-dual vacua in the general multi-variable case was proven recently in \cite{Bakker:2021uqw}, see also \cite{Grimm:2021vpn}. In contrast to the one-variable case, the proof of the multi-variable case is much more involved and relied heavily on recent advances in the field of tame geometry, such as the definability of the period map \cite{BKT}. \\

\noindent In section \ref{sec:self_dual_locus} we will discuss the main technical result of this chapter. Namely, we provide another perspective on the finiteness of self-dual vacua in the multi-variable case, without relying on methods from tame geometry. Instead, we generalize the analysis performed in \cite{Grimm:2020cda,Schnellletter} by considering the $\mathrm{Sl}(2)$-orbit theorem in its full multi-variable glory, utilizing the machinery outlined in chapters \ref{chap:asymp_Hodge_I} and \ref{chap:asymp_Hodge_II}. In particular, we prove the finiteness of self-dual flux vacua within the nilpotent orbit approximation. This provides a good intuition for why finiteness is likely to persist, even when there are multiple moduli at play.\\

\noindent Finally, in section \ref{sec:future_questions} we turn to our attention to some exciting speculations and future prospects regarding more detailed features of the locus of flux vacua, i.e.~beyond just its finiteness. In particular, we outline a set of three concrete mathematical conjectures which may be addressed in the near future by combining techniques from asymptotic Hodge theory, (sharply) o-minimal geometry and the theory of unlikely intersections. The first two conjectures concern the enumeration of flux vacua, in particular Hodge vacua, as well as a candidate notion of geometric complexity, as developed by Binyamini and Novikov in \cite{binyamini2022}, for the locus of self-dual flux vacua. The third conjecture is a modified version of the tadpole conjecture of \cite{Bena:2020xrh}, adapted to the special class of Hodge vacua and is instead concerned with the dimensionality of the vacuum locus. In other words, it is related to the existence of a flat direction in the scalar potential. For related work on the tadpole conjecture, we refer the reader to \cite{Braun:2020jrx,Bena:2021wyr,Marchesano:2021gyv,Lust:2021xds,Plauschinn:2021hkp,Grana:2022dfw,Lust:2022mhk,Coudarchet:2023mmm,Braun:2023pzd}.

\section{Finiteness of flux vacua}
\label{sec:finiteness_intro}

In section \ref{ssec:N=1-sugra} we have reviewed the conditions on the four-form flux $G_4$ and the complex structure moduli $z^i$ that determine the locus of self-dual flux vacua. In the remainder of this work, we will be interested in gaining a more detailed understanding of what this locus looks like. In particular, our aim is to ascertain whether it consists of a finite number of points (or, more precisely, a finite number of connected components). The purpose of this section is two-fold. First, we provide a general discussion to emphasize the main non-trivial aspects of the problem, illustrated with a simple example of a rigid $Y_3\times T^2$ compactification. Second, we formulate the problem within the broader framework of Hodge theory and present a number of exact finiteness theorems that have been established in the literature. The subsequent sections will delve into a more detailed examination of these theorems and their proofs.  

\subsection{Why finiteness is non-trivial}
\label{subsec:finiteness_intro}

\subsubsection*{Infinite tails of vacua?}
First, let us emphasize again that we are investigating the finiteness of vacua within a fixed topological class of Calabi--Yau fourfolds, but varying in complex structure moduli. In this setting, we recall from section \ref{ssec:N=1-sugra} that a self-dual flux vacuum consists of a pair $(z^i,G_4)$, where $z^i$ are the complex structure moduli and $G_4$ is the four-form flux, satisfying the following three conditions.
\begin{subbox}{Self-dual flux vacuum}
	\begin{equation}
		\label{eq:vacuum_conditions}
	G_4\in H^4\left(Y_4,\mathbb{Z}\right)\,,\qquad G_4=\star\,G_4\,,\qquad \int_{Y_4}G_4\wedge G_4 \leq L\,.
	\end{equation}
\tcblower
Here we recall that $\star$ denotes the Hodge star operator on the Calabi--Yau fourfold $Y_4$, which is to be evaluated at $z^i$, and $L$ is some positive integer that reflects the tadpole bound. 
\end{subbox}
\noindent Note that, for some choices of the flux, it may happen that not all $z^i$ are stabilized, meaning that the scalar potential has flat directions, in which case we count each connected component of the higher-dimensional vacuum locus as a single vacuum. The question, then, is how many solutions to \eqref{eq:vacuum_conditions} exist as one varies over all possible choices of $G_4$. Naively, it appears that $G_4$ varies over an infinite lattice. However, upon combining the self-duality condition and the tadpole condition, one finds the relation
\begin{equation}\label{eq:tadpole_self-dual}
    \int_{Y_4}G_4\wedge\star\,G_4 \leq L\,,
\end{equation}
which involves the Hodge norm of $G_4$. Hence, at a non-singular point in the moduli space, the left-hand side of \eqref{eq:tadpole_self-dual} is a manifestly positive-definite quadratic form in the fluxes. Therefore, at a fixed point $z^i$ the constraint \eqref{eq:tadpole_self-dual} restricts the fluxes to lie in the interior of some ellipsoid inside the flux lattice, whose exact shape depends on the chosen value of the moduli. Clearly such a region contains only finitely many discrete lattice points and hence finitely many self-dual flux vacua. Furthermore, this remains true as long as one varies the moduli $z^i$ over a compact subset of the moduli space. 
\\

\noindent However, it is not at all obvious what happens as the moduli vary over an unbounded set, as is typically the case in the context of Calabi--Yau compactifications. In other words, one might find an accumulation of vacua as one approaches a boundary of the moduli space. Along such limits the Hodge star operator may degenerate, causing some directions of the ellipsoid to become arbitrarily large and thus include arbitrarily many lattice points. In order to address the fate of these potentially infinite tails of vacua, one has to deal with the following two major roadblocks:
\begin{itemize}
    \item \textbf{Roadblock (1): Hodge star behaviour} \\
    It is necessary to understand all possible ways in which the Hodge star can degenerate as one approaches an arbitrary boundary in the moduli space of any Calabi--Yau fourfold, in particular with an arbitrary number of moduli. 
    \item \textbf{Roadblock (2): Path-dependence} \\
    When there are multiple moduli at play, the degeneration of the Hodge star is highly dependent on how one approaches a given boundary in the moduli space. 
\end{itemize}
Fortunately, the possible degenerations of the Hodge star are well-studied in the field of asymptotic Hodge theory, as we have reviewed in detail in chapters \ref{chap:asymp_Hodge_I} and \ref{chap:asymp_Hodge_II}. Consequently, the results of these chapters will play a major role in the following. The issue of path-dependence is, however, a bit more subtle. For Hodge vacua this issue can in fact be dealt with using just Hodge theoretic techniques. Essentially, one applies a clever inductive reasoning to range over all possible hierarchies between the moduli. In section \ref{sec:self_dual_locus} we employ a new strategy to tackle this issue, which is valid within the nilpotent orbit approximation. However, in order to address the fate of self-dual vacua in full generality, i.e.~including exponential corrections to the nilpotent orbit approximation, these techniques are likely to be insufficient. Recently, these issues were overcome by incorporating deep results in o-minimal geometry on the tameness of Hodge theory \cite{BKT, Bakker:2021uqw}. 

\subsubsection*{Example: Rigid $Y_3\times T^2$}

So far our discussion has been rather abstract. In order to illustrate some of the points we have made above, let us consider a simple example. The point of the example will be to highlight the possible presence of infinite tails of vacua and to give an idea why such tails nevertheless cannot appear. However, we stress that, due to its simplicity, the example will not give an adequate indication for the complexity of the general problem. In particular, the issue of path-dependence will not play a role here.\\

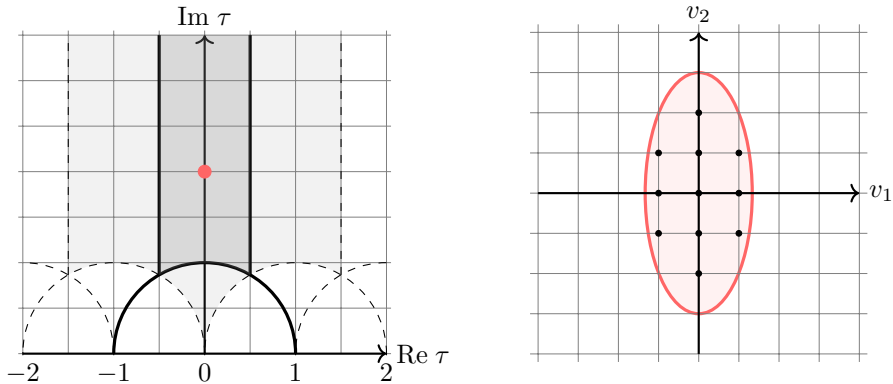
\begin{figure}[t!]
	\centering
		\begin{tikzpicture}[scale=0.6]
			
			% MODULI SPACE
			
			\draw[step=1cm, gray, very thin] (-4.1,0) grid (4.1,7.1);
			\draw[thick,->] (-4,0) -- (4,0) node[anchor =  west] {$\mathrm{Re}\;\tau$};
			\draw[thick,->] (0,0) -- (0,7) node[anchor = south ] {$\mathrm{Im}\;\tau$};   
			
			\draw[very thick] (2,0) arc (180:360: -2);
			\draw[dashed] (4,0) arc (180:360: -2);
			\draw[dashed] (0,0) arc (180:360: -2);
			\draw[dashed,thin] (-2,0) arc (180:270: -2);
			\draw[dashed,thin] (4,2) arc (270:360: -2);
			
			\draw[very thick] (1,1.73) -- (1,7);
			\draw[very thick] (-1,1.73) -- (-1,7);
			
			\draw[dashed] (3,1.73) -- (3,7);
			\draw[dashed] (-3,1.73) -- (-3,7);
			
			\fill[gray,opacity=0.3] (1,1.73) arc(60:120:2) -- (-1,7) -- (1,7) -- (1,1.73);
			
			\fill[gray,opacity=0.1] (3,1.73) arc(60:120:2) -- (1,7) -- (3,7) -- (3,1.73);
			\fill[gray,opacity=0.1] (-1,1.73) arc(60:120:2) -- (-3,7) -- (-1,7) -- (-1,1.73);
			\fill[gray,opacity=0.1] (1,1.73) arc(60:120:2) arc(60:0:2) arc(0:-60:-2);
			
			\foreach \x in {-2,-1,0,1,2}
			\draw (2*\x,1pt) -- (2*\x,-1pt) node[anchor=north]{$\x$};
			
			\filldraw[red!60] (0,4) circle (4pt);
			
		\end{tikzpicture}
		\qquad        
		\raisebox{0.4cm}{
			\begin{tikzpicture}[scale=0.53]
				
				% Tadpole constraint #2
				
				\draw[color=red!60, fill=red!5,very thick] (0,0) ellipse (4.0/3.0 and 3.0);
				
				\draw[step=1cm, gray, very thin] (-4.2,-4.2) grid (4.2,4.2);
				\draw[thick,->] (-4,0) -- (4,0) node[anchor=west]{$v_1$};
				\draw[thick,->] (0,-4) -- (0,4) node[anchor=south]{$v_2$};
				
				\filldraw[black] (0,0) circle (2pt);
				\filldraw[black] (1,0) circle (2pt);
				\filldraw[black] (0,1) circle (2pt);
				\filldraw[black] (1,1) circle (2pt);
				\filldraw[black] (-1,-1) circle (2pt);
				\filldraw[black] (-1,0) circle (2pt);
				\filldraw[black] (0,-1) circle (2pt);
				\filldraw[black] (1,-1) circle (2pt);
				\filldraw[black] (-1,1) circle (2pt);
				\filldraw[black] (0,-2) circle (2pt);
				\filldraw[black] (0,2) circle (2pt);
				
			\end{tikzpicture}
		}
		\caption{A geometric depiction of the tadpole bound \eqref{eq:tadpole_torus} for the two-torus. Left: a fundamental domain for the Teichm\"uller parameter $\tau$. Right: the corresponding region inside the flux lattice where the tadpole bound is satisfied. For simplicity, we have considered a point with $\mathrm{Re}\,\tau=0$.}
		\label{fig:tadpole_torus}		
\end{figure}

\noindent We take $Y_4$ to be a direct product
\begin{equation}
	Y_4 = Y_3\times T^2\,,
\end{equation}
with $Y_3$ a rigid Calabi--Yau threefold (i.e.~having no moduli) and $T^2$ a two-torus, whose complex structure modulus will be denoted by $\tau$, with $\mathrm{Im}\,\tau>0$. We consider a one-form flux on the torus 
\begin{equation}
	v\in H^1(T^2,\mathbb{Z}[i])\,,\qquad v=\begin{pmatrix} v_1\\v_2\end{pmatrix}\,,
\end{equation}
where $v_1,v_2\in\mathbb{Z}[i]$ are Gaussian integers. The vector representation of $v$ is taken with respect to the standard basis of 1-cycles on the torus, in terms of which the period vector is given simply by $(1,\tau)$. Then one readily computes
\begin{equation}\label{eq:tadpole_torus}
	||v||^2 = \int_{T^2} v\wedge\star\,\bar{v} = |v_1|^2\,\mathrm{Im}\,\tau+\frac{|v_2-v_1\mathrm{Re}\,\tau|^2}{\mathrm{Im}\,\tau}\,.
\end{equation}
As expected, for a fixed value of $\tau$ a region inside the flux lattice of bounded $||v||$ corresponds to the interior of an ellipsoid. Furthermore, the semi-major and semi-minor axes of the ellipsoid scale as $\mathrm{Im}\,\tau$ and $1/\mathrm{Im}\,\tau$, respectively. The situation is illustrated in figure \ref{fig:tadpole_torus}.\footnote{It should be noted that not necessarily all fluxes choices depicted in figure \ref{fig:tadpole_torus} satisfy the vacuum conditions.} Indeed, as one approaches the weak-coupling point $\mathrm{Im}\,\tau\rightarrow\infty$, corresponding to the boundary of the moduli space, one of the axes of the ellipsoid blows up, while the other shrinks. Therefore, by letting $\mathrm{Im}\,\tau$ become arbitrarily large, it appears that one can reach an infinite amount of different fluxes and thus an infinite number of vacua. \\

\noindent The crucial point, however, is that when $\mathrm{Im}\,\tau$ becomes too large, it becomes impossible to satisfy both the self-duality condition and the tadpole condition. This can be seen as follows. Since the fluxes are quantized, the quantity $|v_1|$ cannot become arbitrarily small. Therefore, as $\mathrm{Im}\,\tau$ increases, at some point one must set $v_1=0$ in order to satisfy the tadpole bound $||v||^2<L$. At this point, one is left with
\begin{equation}
    ||v||^2 = \frac{|v_2|^2}{\mathrm{Im}\,\tau}\,.
\end{equation}
It appears that $|v_2|$ can become arbitrarily large, without $||v||$ exceeding the tadpole bound. However, at this point we should recall the self-duality condition\footnote{To be precise, the analogous condition is that $v$ is imaginary anti self-dual, i.e.
\begin{equation}
    \star\,v = -i v\,.
\end{equation}}, which can be solved explicitly to give
\begin{equation}
    v_1 = \frac{v_2}{\bar{\tau}}\,. 
\end{equation}
Indeed, one immediately sees that if $v_1=0$, then the only solution to the self-duality condition is that also $v_2=0$. In other words, beyond some critical value of $\mathrm{Im}\,\tau$, the only possible vacuum is the trivial one, hence no infinite tails of vacua can occur. Furthermore, the critical value is around $\mathrm{Im}\,\tau = L$. The situation is depicted in figure \ref{fig:tadpole_torus_2}.

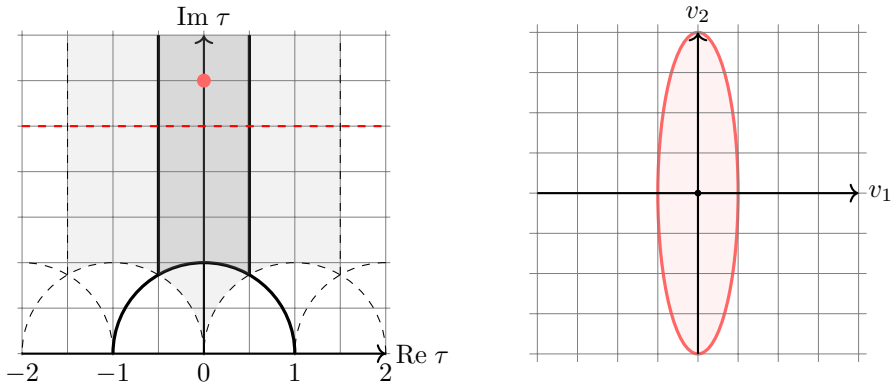
\begin{figure}[t!]
		\begin{tikzpicture}[scale=0.6]
			
			% MODULI SPACE
			
			\draw[step=1cm, gray, very thin] (-4.1,0) grid (4.1,7.1);
			\draw[thick,->] (-4,0) -- (4,0) node[anchor =  west] {$\mathrm{Re}\;\tau$};
			\draw[thick,->] (0,0) -- (0,7) node[anchor = south ] {$\mathrm{Im}\;\tau$};   
			
			\draw[very thick] (2,0) arc (180:360: -2);
			\draw[dashed] (4,0) arc (180:360: -2);
			\draw[dashed] (0,0) arc (180:360: -2);
			\draw[dashed,thin] (-2,0) arc (180:270: -2);
			\draw[dashed,thin] (4,2) arc (270:360: -2);
			
			\draw[very thick] (1,1.73) -- (1,7);
			\draw[very thick] (-1,1.73) -- (-1,7);
			
			\draw[dashed] (3,1.73) -- (3,7);
			\draw[dashed] (-3,1.73) -- (-3,7);
			
			\fill[gray,opacity=0.3] (1,1.73) arc(60:120:2) -- (-1,7) -- (1,7) -- (1,1.73);
			
			\fill[gray,opacity=0.1] (3,1.73) arc(60:120:2) -- (1,7) -- (3,7) -- (3,1.73);
			\fill[gray,opacity=0.1] (-1,1.73) arc(60:120:2) -- (-3,7) -- (-1,7) -- (-1,1.73);
			\fill[gray,opacity=0.1] (1,1.73) arc(60:120:2) arc(60:0:2) arc(0:-60:-2);
			
			\foreach \x in {-2,-1,0,1,2}
			\draw (2*\x,1pt) -- (2*\x,-1pt) node[anchor=north]{$\x$};
			
			\filldraw[red!60] (0,6) circle (4pt);

            \draw[dashed, red, thick] (-4,5)--(4,5);
			
		\end{tikzpicture}
		\qquad        
		\raisebox{0.4cm}{
			\begin{tikzpicture}[scale=0.53]
				
				% Tadpole constraint #2
				
				\draw[color=red!60, fill=red!5,very thick] (0,0) ellipse (1.0 and 4.0);
				
				\draw[step=1cm, gray, very thin] (-4.2,-4.2) grid (4.2,4.2);
				\draw[thick,->] (-4,0) -- (4,0) node[anchor=west]{$v_1$};
				\draw[thick,->] (0,-4) -- (0,4) node[anchor=south]{$v_2$};
				
				\filldraw[black] (0,0) circle (2pt);
				
			\end{tikzpicture}
		}
		\caption{The same setup is depicted as in figure \ref{fig:tadpole_torus}, but now $\mathrm{Im}\,\tau$ has exceeded the value of $L$, shown with the red dashed line. Correspondingly, there are no non-trivial self-dual fluxes beyond this point.}
		\label{fig:tadpole_torus_2}		
\end{figure}

\subsection{Finiteness theorems: global}
\label{subsec:finiteness_theorems_global}
Having discussed some general features of the problem of finiteness, let us now turn to a concrete description of the known results. This will first be done from a global point of view, meaning we focus on properties such as algebraicity and definability. We introduce the locus of Hodge classes and the locus of self-dual classes using the language of variations of Hodge structures. We briefly recall the important definitions, but refer the reader to \cite{Grimm:2018cpv,Grimm:2021ckh} for a more detailed introduction.

\subsubsection*{Locus of Hodge classes}
\label{subsubsec:locus_Hodge}
Recall from the discussion in section \ref{ssec:N=1-sugra} that a self-dual flux vacuum is called a Hodge vacuum if the flux $G_4$ only has a (2,2)-component. In other words,
\begin{equation}
    G_4\in H^4(Y_4,\mathbb{Z})\cap H^{2,2}\,.
\end{equation}
Classes of this type are so special that they have a name: they are referred to as \textit{Hodge classes}. More generally, given a variation of Hodge structure of even weight $2k$, a Hodge class is an integral class of type $(k,k)$. In view of the tadpole condition, it is natural to consider the subset of Hodge classes whose self-intersection is bounded, for which we introduce the notation
\begin{equation}\label{eq:H_bounded}
	H_{\mathbb{Z}}(L):=\{v\in H_{\mathbb{Z}}: (v,v)\leq L\}\,.
\end{equation}
The set of all Hodge classes with self-intersection bounded by $L$ then defines the following subspace of the Hodge bundle $E$ (recall equation \eqref{eq:Hodge_bundle} and the surrounding discussion).
\begin{subbox}{The locus of Hodge classes}
	\begin{equation}\label{eq:Hodge-locus}
		E_{\mathrm{Hodge}}(L) = \{(z^i, v)\in E\,|\,v\in H_{\mathbb{Z}}(L)\cap H^{k,k}  \}\,,\qquad D=2k\,.
	\end{equation}
\tcblower 
We will refer to $E_{\mathrm{Hodge}}(L)$ as the \textbf{locus of (bounded) Hodge classes}. The full locus of Hodge classes is then the countable union of $E_{\mathrm{Hodge}}(L)$ over all integers $L$ and is denoted simply by $E_{\mathrm{Hodge}}$.
\end{subbox}
\noindent It is relatively easy to see that $E_{\mathrm{Hodge}}$ defines a complex-analytic subspace of $E$. There are two ways to see this:
\begin{itemize}
    \item \textbf{Superpotential:}\\
     In the F-theory setting, a Hodge vacuum is alternatively defined by the equations $\partial_i W = W=0$, which are holomorphic in the complex structure moduli.
    \item \textbf{Hodge filtration:}\\
     More generally, it follows from the relation \eqref{eq:decomp_filtration} that
    \begin{equation}
        H_{\mathbb{Z}}\cap H^{k,k} = H_{\mathbb{Z}}\cap F^k\,.
    \end{equation}
    Note that the reality condition is crucial here. By definition of a variation of Hodge structure, the filtration $F^p$ depends holomorphically on the moduli, recall the condition \eqref{eq:holomorphicity}.
\end{itemize}
The fact that the locus of Hodge classes is complex-analytic is already quite special, as this property is not retained for generic self-dual vacua, as will be explained later. At the same time, due to the additional condition $W=0$, the locus is defined by $h^{3,1}+1$ generically independent equations, hence one expects solutions to be relatively rare. Said differently, in order for a vacuum to exist, something special must occur in order for some of the equations to become dependent. The special thing that needs to happen is captured by the following striking theorem of Cattani, Deligne and Kaplan.
\begin{subbox}{\begin{thm}[Cattani, Deligne, Kaplan \cite{CDK}]
		\label{thm:CDK}\end{thm}}
	$E_{\mathrm{Hodge}}(L)$ is an algebraic variety, finite over $\mathcal{M}$.  
\end{subbox}
\noindent By the phrase `finite over $\mathcal{M}$' it is meant that restriction of the projection $p:E\rightarrow\mathcal{M}$ to $E_{\mathrm{Hodge}}(L)$ has finite fibers. In other words, for each $z\in\mathcal{M}$ the fiber over $z$ consists of finitely many points. Furthermore, the algebraicity of $E_{\mathrm{Hodge}}(L)$ means that it can each be represented by a finite set of algebraic equations in $E$, i.e.~\textit{polynomials} in the moduli and the fluxes. In other words, it is of the form
\begin{equation}
    P_i(x_1,\ldots, x_k)=0\,,
\end{equation}
for some polynomials $P_i$. It should be stressed that this is truly remarkable, as the superpotential $W$ itself is typically a complicated transcendental function in the moduli. Nevertheless, the locus where $\partial_i W=W=0$ enjoys a comparatively simple description. This can be made very explicit in concrete examples, and we refer the reader to the work \cite{Grimm:2024fip} where this is investigated in detail. \\

\noindent For the purpose of the present work, the crucial observation is that the algebraicity of $E_{\mathrm{Hodge}}(L)$ automatically implies the finiteness of Hodge vacua. Indeed, it is clear that the zero-set of a finite collection of polynomials has only finitely many connected components. This should be contrasted with the full locus of Hodge classes $E_{\mathrm{Hodge}}$, which is only a countable union of algebraic varieties and hence does not have such a finiteness property.\footnote{See however \cite{baldi2022distribution} for recent refinements of this statement.} In this regard, it is interesting to point out that when the variation of Hodge structure under consideration comes from a family of smooth projective varieties, the same conclusion follows from the famous Hodge conjecture. However, the Hodge conjecture does not predict the stronger statement that $E_{\mathrm{Hodge}}(L)$ is algebraic. In other words, it does not predict the finiteness of Hodge vacua. It is therefore rather curious that the string-theoretic setting imposes the additional crucial constraint, namely the tadpole condition, to exactly ensure finiteness.\\

\noindent For the interested reader, let us give a very rough idea of how one would approach a proof Theorem \ref{thm:CDK}, following the original work of Cattani, Deligne, and Kaplan. In particular, we focus on how one would reduce this to a local statement, which will then be discussed in more detail in section \ref{subsec:finiteness_theorems_local} and appendix \ref{app:Hodge_locus}. The reduction is performed by employing a comparison theorem which connects algebraic geometry and analytic geometry known as Chow's theorem, which states that any closed analytic subspace of a complex projective space is algebraic.\footnote{This now falls within the broader domain of so-called GAGA results, which encompasses various types of comparison results between algebraic and analytic geometry in terms of comparisons of categories of sheaves. Here GAGA stands for \textit{G\'eometrie Alg\'ebrique et G\'eom\'etrie Analytique}.} Very roughly, this means that if some closed analytic subspace is well-behaved enough in the asymptotics, then it is in fact algebraic. Indeed, we have seen that the Hodge locus is complex-analytic on $\mathcal{M}$. Furthermore, it is well-known that $\mathcal{M}$ is quasi-projective, so that its closure can be embedded in a complex projective space \cite{Viehweg}. The strategy, then, is to show that the closure of the Hodge locus in $\overline{\mathcal{M}}$ is analytic as well and to then apply Chow's theorem to establish the desired algebraicity. Hence, one reduces the question to a study of the Hodge locus locally at the divisor $\overline{\mathcal{M}}\setminus \mathcal{M}$, which brings one into the realm of degenerations of Hodge structures and asymptotic Hodge theory. Physically, this means one is studying the structure of Hodge vacua as one approaches the boundary of the moduli space, which, following our initial discussion in section \ref{subsec:finiteness_intro}, is exactly the question we are interested in.\\

\noindent Finally, let us mention a generalization of Theorem \ref{thm:CDK} by Schnell, who introduced the ``extended locus of Hodge classes'' \cite{schnell2014extended}. The rough goal was construct a natural compactification of the Hodge locus to also incorporate so-called ``limit Hodge classes''. These are, as the name suggests, integral classes that become Hodge in an appropriate limit and should therefore lie on the boundary of the Hodge locus. 

\subsubsection*{Locus of self-dual classes}
As soon as one moves towards generic self-dual flux vacua, the situation becomes more complicated. Indeed, since the $G_4$ flux is now allowed to have also $(4,0)$ and $(0,4)$ components, it no longer corresponds to a Hodge class. In a similar fashion as before, let us introduce the following notation for the \textbf{locus of self-dual classes}.
\begin{subbox}{The locus of self-dual classes}
	\begin{equation}
		E_{\text{self-dual}}(L) = \{(z^i,v)\in E\,:\,v\in H_{\mathbb{Z}}(L),\,C(z)v = v \}\,.
	\end{equation}
Here we recall that $C(z)$ denotes the Weil operator defined by the relation \eqref{eq:def-Weil}.
\end{subbox}
\noindent In contrast to the locus of Hodge classes, the locus of self-dual classes is a priori only a \textit{real}-analytic subspace of $E$. Again, one can see this by noting that a generic self-dual vacuum is defined by the equation $D_iW_{\mathrm{flux}}=0$, which now involves the real K\"ahler potential $K$. Nevertheless, in analogy with the algebraicity of the locus of bounded Hodge classes, it was shown in \cite{Bakker:2021uqw} that the locus of bounded self-dual classes has a lot more structure than one might at first expect, as captured in the following result. 
\begin{subbox}{\begin{thm}[Bakker, Grimm, Schnell, Tsimerman \cite{Bakker:2021uqw}]\label{thm:finiteness_self_dual}\end{thm}}
	The set $E_{\rm {self\text{-}dual}}(L)$ is a definable in the o-minimal structure $\mathbb{R}_{\mathrm{an,exp}}$. Furthermore, it is a closed, real-analytic subspace of $E$, finite over $\mathcal{M}$.
\end{subbox}
\noindent Let us briefly elaborate on the phrase `definable in the o-minimal structure $\mathbb{R}_{\mathrm{an,exp}}$'. For a more detailed explanation we refer the reader to \cite{Grimm:2021vpn}. Roughly, this means that the locus of bounded self-dual classes can be described by a finite set of polynomial equations and inequalities that involve not only the moduli and fluxes, but also any restricted analytic function and real exponential function of the moduli. More precisely, the o-minimal structure $\mathbb{R}_{\mathrm{an,exp}}$ is generated (through finite products, unions, intersections and projections) by sets of the form
\begin{align}
    P(x_1,\ldots, x_k, f_1,\ldots, f_m, e^{x_1},\ldots, e^{x_k})=0\,,
\end{align}
where the $f_i$ are restricted analytic functions and $P$ is a polynomial.  \\

\noindent Importantly for our purposes, the fact that the locus of bounded self-dual classes is definable implies that it also has an inherent finiteness property, which we briefly explain in three steps. 
\begin{enumerate}
    \item The fact that the restriction of $p:E\rightarrow\mathcal{M}$ to $E_{\text{self-dual}}(L)$ has finite fibers means that for each point $z\in\mathcal{M}$, its preimage under this map consists of a finite number of points. In other words, for fixed $z$ the size of the fiber $p^{-1}(z)$ is bounded. This is, of course, not enough to prove finiteness completely, since $z$ itself ranges over an infinite set.
    \item Due to the special properties of definable functions, one can show that in fact the size of the fiber is uniformly bounded. Hence, there exists an integer $N_{\mathrm{max}}$ such that
    \begin{equation}
        |p^{-1}(z)|\leq N_{\mathrm{max}}\,,
    \end{equation}
    for all $z\in\mathcal{M}$.
    \item Finally, one can show that since the map $p$ is itself definable, the set
    \begin{equation}
        \{z\in\mathcal{M}: |p^{-1}(z)|\leq N_{\mathrm{max}}\}\,,
    \end{equation}
    is definable as well. In particular, it cannot contain infinitely many discrete points. 
\end{enumerate}
As a final remark, let us mention that Theorem \ref{thm:finiteness_self_dual} actually implies Theorem \ref{thm:CDK}, namely that the locus of bounded Hodge classes is algebraic. This was shown in \cite{BKT} by Bakker, Klingler and Tsimerman using the so-called definable Chow theorem of Peterzil and Starchenko \cite{Peterzil:2009}. The latter is an alternative version of Chow's theorem adapted to the setting of o-minimal geometry and roughly states that a complex-analytic set which is also definable is in fact algebraic. Recalling that the locus of Hodge classes is clearly complex-analytic, one recovers Theorem \ref{thm:CDK}.

\subsection{Finiteness theorems: local}
\label{subsec:finiteness_theorems_local}
In this section we discuss some local manifestations of the finiteness theorems presented in section \ref{subsec:finiteness_theorems_global}. Arguably, when it comes to developing further intuition for the finiteness of vacua, the local analysis is more illuminating. Indeed, in section \ref{subsec:finiteness_intro} it was argued that, as far as finiteness is concerned, the main question is whether it is possible for vacua to accumulate near the boundaries of the moduli space. Furthermore, in section \ref{subsubsec:locus_Hodge} we gave a rough idea of how the proof of the theorem of Cattani, Deligne, and Kaplan heavily relies on a local analysis near the boundaries of the moduli space. This brings us into the realm of asymptotic Hodge theory.

\subsubsection*{Finiteness of Hodge classes}

We can now formulate local versions of the finiteness theorems discussed in section \ref{subsec:finiteness_theorems_global}. In this section, we focus on the case of Hodge vacua. Our goal is to consider a sequence of such vacua that approaches the boundary of $\mathcal{M}$ and ask whether this sequence can take on infinitely many values. To this end, we state the following result.
\begin{subbox}{\begin{thm}[{{\cite[Theorem 3.3]{CDK}}}]
			\label{thm:finiteness_Hodge_loci_local}\end{thm}}
	Let $t^i(n)\in\mathbb{H}^m$
	be a sequence of points such that $x^i(n)$ is bounded and $y^i(n)\rightarrow\infty$ as $n\rightarrow\infty$. Suppose furthermore that
	\begin{equation*}
		v(n)\in  H_{\mathbb{Z}}(L)\cap H^{k,k}\,,\qquad D=2k\,,
	\end{equation*}
	is a sequence of integral bounded Hodge classes. Then $v(n)$ can only take on finitely many values. 
\end{subbox}  
\noindent Here we stress that the Hodge decomposition $H^{k,k}$ is itself a function of the moduli. However, in order not to clutter the notation we will often omit this dependence. The upshot of Theorem \ref{thm:finiteness_Hodge_loci_local} is that it is indeed impossible to have an accumulation of Hodge vacua near the boundary of $\mathcal{M}$. In appendix \ref{app:Hodge_locus} we will describe the proof of Theorem \ref{thm:finiteness_Hodge_loci_local} in some detail. 

\subsubsection*{Finiteness of self-dual classes}
\label{subsubsec:finiteness_self-dual_classes}
Finally, let us come to the finiteness of self-dual vacua. In contrast to Theorem \ref{thm:finiteness_Hodge_loci_local}, there has not yet appeared a fully general directly local proof for the finiteness of self-dual flux vacua. Nevertheless, the following statement clearly follows as a corollary of the global statement given in Theorem \ref{thm:finiteness_self_dual}.
\begin{subbox}{\begin{cor}
		\label{cor:finiteness_self-dual_local}\end{cor}}
	Let $t^i(n)\in\mathbb{H}^m$
	be a sequence of points such that $x^i(n)$ is bounded and $y^i(n)\rightarrow\infty$ as $n\rightarrow\infty$. Suppose furthermore that $v(n)\in H_{\mathbb{Z}}(L)$ is a sequence of integral fluxes with bounded self-intersection, such that
	\begin{equation}
		C(t(n))v(n) = v(n)\,,
	\end{equation}
	for all $n$. Then $v(n)$ can only take on finitely many values. 
\end{subbox}
\noindent An independent proof of Corollary \ref{cor:finiteness_self-dual_local} was given in \cite{Grimm:2020cda,Schnellletter} for the case of a single variable using methods from asymptotic Hodge theory. In section \ref{sec:self_dual_locus} we will extend these methods to the multi-variable setting in order to give some intuition for the finiteness of self-dual vacua in the general case, without using results from o-minimality. To be precise, we will provide a proof within the nilpotent orbit approximation. To be absolutely clear, we will prove the following
\begin{subbox}{
		\begin{thm}\label{thm:finiteness_selfdual_nilpotent}\end{thm}}
	Let $t^i(n)\in\mathbb{H}^m$
	be a sequence of points such that $x^i(n)$ is bounded and $y^i(n)\rightarrow\infty$ as $n\rightarrow\infty$. Suppose furthermore that $v(n)\in H_{\mathbb{Z}}(L)$ is a sequence of integral fluxes with bounded self-intersection, such that
	\begin{equation}
		C_{\mathrm{nil}}(t(n))v(n) = v(n)\,,
	\end{equation}
	for all $n$. Then $v(n)$ can only take on finitely many values. 
\end{subbox}
\noindent In particular, note the replacement of the general Weil operator $C$ by its nilpotent orbit approximation $C_{\mathrm{nil}}$. Of course, this will, therefore, not quite constitute a full independent proof of Corollary \ref{cor:finiteness_self-dual_local}. Nevertheless, the discussion will provide some valuable intuition for the asymptotic behaviour of vacua.

\subsection{Summary}
We close this section by providing the reader with an overview of the various theorems we have discussed, see figure \ref{fig:theorems_overview}. Let us also highlight the variety of strategies that are employed in the proofs of these various theorems. For Hodge vacua, both in the single-variable and multi-variable case, the proof relies heavily on the machinery of mixed Hodge structures, as is explained in appendix \ref{app:Hodge_locus}. Instead, our analysis of the self-dual vacua in the nilpotent orbit approximation makes use of the asymptotic expansion of the Weil operator, as is described in chapters \ref{chap:asymp_Hodge_I} and \ref{chap:asymp_Hodge_II} and \ref{sec:self_dual_locus}. Finally, for the general proof of the finiteness of self-dual flux vacua the recent advances in o-minimal geometry have played an essential role. 

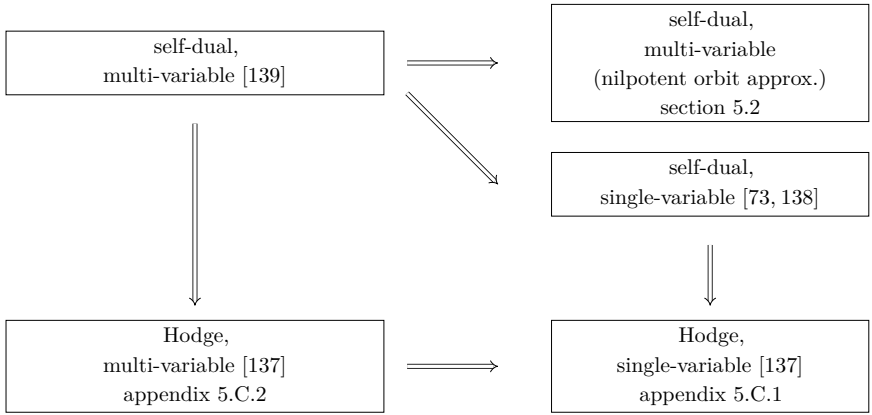
\begin{figure}[h!]
    \centering
    
\scalebox{0.8}{\begin{tikzpicture}

\node[draw, text width=6cm, align=center] at (0,0) {self-dual,\\ multi-variable \cite{Bakker:2021uqw}};

\draw[-implies,double equal sign distance] (3.5,0) -- (5,0);

\draw[-implies,double equal sign distance] (3.5,-5) -- (5,-5);

\draw[-implies,double equal sign distance] (0,-1) -- (0,-4);

\draw[-implies,double equal sign distance] (8.5,-3) -- (8.5,-4);

\draw[-implies,double equal sign distance] (3.5,-0.5) -- (5,-2);

\node[draw, text width=6cm, align=center] at (0,-5) {Hodge,\\ multi-variable \cite{CDK}\\ appendix \ref{subsubsec:proof_Hodge_locus_multi}};

\node[draw, text width=5cm,align=center] at (8.5,0) {self-dual,\\ multi-variable\\ (nilpotent orbit approx.)\\ section \ref{sec:self_dual_locus}};

\node[draw, text width=5cm,align=center] at (8.5,-2) {self-dual,\\ single-variable \cite{Grimm:2020cda,Schnellletter}};

\node[draw, text width=5cm,align=center] at (8.5,-5) {Hodge,\\ single-variable \cite{CDK}\\ appendix \ref{subsubsec:proof_Hodge_locus_single}};

\node[] at (4,2) {\Large Web of Finiteness Theorems};
 
\end{tikzpicture}}
    \caption{An overview of the various finiteness theorems discussed in this work, including the implications between them. }
    \label{fig:theorems_overview}
\end{figure}

\section{The asymptotic self-dual locus}
\label{sec:self_dual_locus}
In section \ref{sec:finiteness_intro} we have presented a number of finiteness theorems for both Hodge vacua and self-dual vacua, from both a global and a local perspective. The aim of this section is to apply the machinery of asymptotic Hodge theory to prove Theorem \ref{thm:finiteness_selfdual_nilpotent}, and address the finiteness of self-dual vacua in the nilpotent orbit approximation. Before discussing the general proof, we first restrict to a simple one-variable setting in section \ref{subsec:self-dual_locus_example} in order to exemplify some important features of the vacuum locus. In section \ref{subsec:self-dual_proof}, we present the full proof of Theorem \ref{thm:finiteness_selfdual_nilpotent}.

\subsection{Example: one-variable $\mathrm{Sl}(2)$-orbit}
\label{subsec:self-dual_locus_example}
Before delving into the detailed proof of the finiteness result, let us first consider a very simple case in which there is just a single modulus, so $m=1$, and all the expansion coefficients (except the leading ones) in the nilpotent orbit expansion \eqref{eq:hinv} vanish. Effectively, this case will correspond to a generalization of the example discussed in section \ref{subsec:finiteness_intro}, though written in more abstract language. Mathematically, the stated assumptions imply that the variation of Hodge structure under consideration is given by a one-variable $\mathrm{Sl}(2)$-orbit
\begin{equation}
	\label{eq:sl2_orbit}
	F^p = e^{xN}y^{-\frac{1}{2}N^0}F_{\infty}^p\,.
\end{equation}
We would like to investigate the set of points in the moduli space where a given $v\in H_{\mathbb{Z}}$ is self-dual and $v$ has a bounded Hodge norm, as imposed by the tadpole condition. In this simple setting, it is straightforward to evaluate these two conditions explicitly.
\subsubsection*{Tadpole constraint}
The Hodge norm of $v$ is given by
\begin{equation}
	||v||^2 = \sum_{\ell} y^\ell \,||\hat{v}_\ell||^2_\infty\,,\qquad \hat{v}=e^{-xN}v\,,
\end{equation}
where we have performed the usual weight-decomposition of $\hat{v}$ with respect to $N^0$. We are interested in the properties of vacua close to the boundary, i.e.~for large values of the saxion $y$. Clearly, for sufficiently large $y$, it is necessary to impose that $\hat{v}_\ell=0$ for all $\ell>0$, in order for $||v||^2$ to not exceed the tadpole bound. In other words, beyond some critical value of $y$, the flux is only allowed to have non-positive weights with respect to the $\mathfrak{sl}(2)$ grading operator $N^0$. 

\subsubsection*{Self-duality condition}
Using the fact that $C_\infty$ interchanges the $+\ell$ and $-\ell$ eigenspaces of $N^0$, it is straightforward to see that the self-duality condition, projected onto a weight $\ell$ component, can be written as
\begin{equation}
	\label{eq:self-duality_sl2_one-variable}
	y^{\frac{1}{2}\ell} \hat{v}_\ell = y^{-\frac{1}{2}\ell} C_\infty \hat{v}_{-\ell}\,.
\end{equation}
In particular, whenever $\hat{v}_\ell=0$ for $\ell>0$ the self-duality condition imposes that additionally $\hat{v}_\ell=0$ for $\ell<0$ (note that $C_\infty$ is invertible). Therefore, for sufficiently large $y$ it must be that $\hat{v}$ only has an $\ell=0$ component, so $\hat{v}=\hat{v}_0$. In particular, the tadpole condition reduces to
\begin{equation}
	||\hat{v}_0||_\infty^2 \leq L\,.
\end{equation}
Recalling the fact that $\hat{v}=e^{-x N}v$, that the axion $x$ is bounded and that the flux $v$ is integral, it is clear that there are only finitely many choices of $v$ which satisfy the tadpole condition. Hence, there are finitely many vacua. 

\subsubsection*{The vacuum locus}
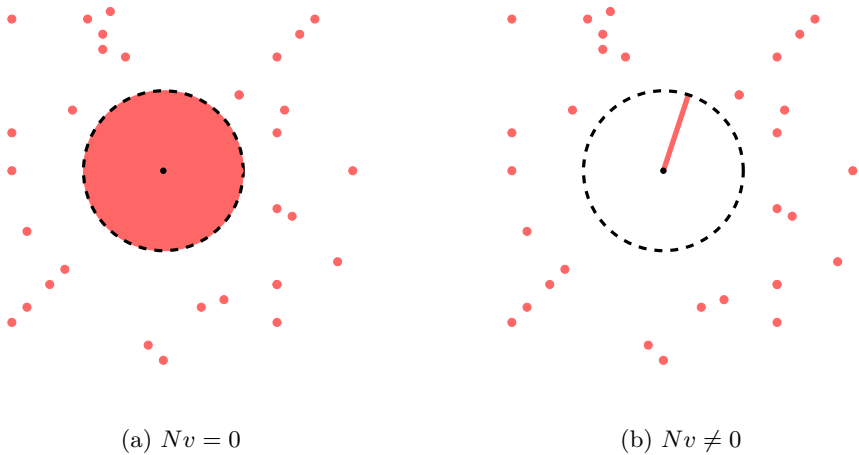
\begin{figure}[t]
	\centering
	\begin{subfigure}[b]{0.5\textwidth}
		\centering
		\begin{tikzpicture}
			\filldraw[color=red!60] (0,0) circle (30pt);    
			\filldraw[black] (0,0) circle (1pt);
			\draw[color=black, dashed, very thick] (0,0) circle (30pt);
			
			\foreach \i\j in {1.5 / 1.5, -0.7 / 2.1, 2.3 / -1.2, -1.8 / -1.8, 2.0 / 2.0,
				1.0 / 1.0, -1.2 / 0.8, 1.7 / -0.6, -0.6 / -0.6, 1.8 / 1.8,
				-0.2 / -2.3, 0.8 / -1.7, -2.0 / 0.0, 1.0 / 0.0, 0.0 / -2.5,
				-1.0 / 2.0, 2.5 / 0.0, -1.3 / -1.3, 1.5 / -0.5, -0.5 / 1.5,
				1.5 / -1.5, -2.0 / -2.0, 1.5 / 0.5,
				-0.8 / 1.8, 1.5 / -2.0, 1.5 / -1.5, -0.5 / -0.5, -1.8 / -0.8,
				1.0 / 1.0, 0.5 / -1.8, -0.8 / 0.0, -2.0 / 2.0, 0.5 / -0.5,
				-1.5 / -1.5, -2.0 / 0.5, 1.6 / 0.8, -0.8 / 1.6}
			{
				\filldraw[color=red!60] (\i,\j) circle (1.5pt);
			}
		\end{tikzpicture}
		\caption{$Nv=0$}
		\label{fig:Hodge_locus}
	\end{subfigure}
	\hfill
	\begin{subfigure}[b]{0.45\textwidth}
		\centering
		\begin{tikzpicture}
			\draw[color=red!60, line width = 0.7mm] (0,0)--(0.33,1);
			\filldraw[black] (0,0) circle (1pt);
			\draw[color=black, dashed, very thick] (0,0) circle (30pt);
			
			\foreach \i\j in {1.5 / 1.5, -0.7 / 2.1, 2.3 / -1.2, -1.8 / -1.8, 2.0 / 2.0,
				1.0 / 1.0, -1.2 / 0.8, 1.7 / -0.6, 1.8 / 1.8,
				-0.2 / -2.3, 0.8 / -1.7, -2.0 / 0.0, 0.0 / -2.5,
				-1.0 / 2.0, 2.5 / 0.0, -1.3 / -1.3, 1.5 / -0.5, -0.5 / 1.5,
				1.5 / -1.5, -2.0 / -2.0, 1.5 / 0.5,
				-0.8 / 1.8, 1.5 / -2.0, 1.5 / -1.5,  -1.8 / -0.8,
				1.0 / 1.0, 0.5 / -1.8, -2.0 / 2.0, 
				-1.5 / -1.5, -2.0 / 0.5, 1.6 / 0.8, -0.8 / 1.6}
			{
				\filldraw[color=red!60] (\i,\j) circle (1.5pt);
			}
		\end{tikzpicture}
		\caption{$Nv\neq 0$}
		\label{fig:self-dual_locus}
	\end{subfigure}
	\caption{Schematic illustration of the distribution of self-dual vacua near a punctured disk (shown in red). Close enough to the singularity, dictated by the tadpole bound (indicated by the black dashed line), either (a) both axion and saxion are unstabilized, or (b) only the axion is stabilized, the latter case corresponding to a radial ray.}
\end{figure}
It is important to stress that for this simple example the only possible vacua close to the boundary have an unstabilized saxion. Indeed, since $\hat{v}$ only has an $\ell=0$ component, the self-duality condition \eqref{eq:self-duality_sl2_one-variable} simply becomes
\begin{equation}
	v = \left(e^{x N}C_\infty e^{-xN}\right)v\,,
\end{equation}
in which the saxion does not appear. Furthermore, one can make the following case distinction
\begin{enumerate}
	\item $Nv = 0$: In this case the self-duality condition reduces further to $v=C
	_\infty v$ and also the axion is unstabilized. The resulting vacuum locus is illustrated in figure \ref{fig:Hodge_locus}. In fact, it turns out that in this case $v$ actually corresponds to a Hodge class, as is explained in Appendix \ref{app:Hodge_locus}.
	\item $Nv\neq 0$: In this case a choice of $v$ uniquely fixes a value for the axion $x$. The vacuum locus therefore corresponds to a single angular ray in the disk. This is illustrated in figure \ref{fig:self-dual_locus}.  
\end{enumerate}
The general lesson of this simple one-variable example is the following: as one approaches the boundary of the moduli space, one is more and more restricted in the allowed fluxes, i.e.~the allowed $\mathrm{sl}(2)$ components of the fluxes, that can possibly satisfy the self-duality condition and the tadpole constraint. Eventually, the restrictions become so severe that one can directly show that there are only finitely many possibilities. As will be explained in the next section, a similar phenomenon happens in the multi-variable case. However, the restrictions on the fluxes become dependent on the sector of the moduli space in which the vacua are located. 
\newpage 

\subsection{Proof: finiteness of self-dual vacua}
\label{subsec:self-dual_proof}
For ease of reference, we repeat the exact theorem we aim to prove.
\setcounter{thm}{3}
\begin{subbox}{\begin{thm}\end{thm}}
	Let $t^i(n)\in\mathbb{H}^m$
	be a sequence of points such that $x^i(n)$ is bounded and $y^i(n)\rightarrow\infty$ as $n\rightarrow\infty$. Suppose furthermore that $v(n)\in H_{\mathbb{Z}}(L)$ is a sequence of integral fluxes with bounded self-intersection, such that
	\begin{equation}
		\label{eq:self-duality_nil}
		C_{\mathrm{nil}}(t(n))v(n) = v(n)\,,
	\end{equation}
	for all $n$. Then $v(n)$ can only take on finitely many values.
\end{subbox}
\noindent We emphasize again that our proof is restricted to the case where the variation of Hodge structure under consideration is described by a nilpotent orbit. In the one-variable case it is relatively straightforward to reduce to this case from the general setting of an arbitrary variation of Hodge structure by including an appropriate exponentially small correction term to the flux sequence $v(n)$ \cite{Schnellletter}. However, in the multi-variable case it is not clear whether a similar strategy can be applied.\\

\noindent In the following, we will often omit the argument in $y(n)$ and simply write $y$, to avoid cluttering the notation. As has been described a few times already, combining the self-duality condition with the tadpole bound on the self-intersection of $v(n)$ gives the following bound on the Hodge norm
\begin{equation}
	\label{eq:bound_v(n)}
	||v(n)||
	^2\leq L\,.
\end{equation}
The strategy of the proof will be to show that in fact $v(n)$ is bounded with respect to the boundary Hodge norm $||\cdot||_\infty$. Then the desired finiteness follows from the fact that $v(n)$ is integral. We will divide the proof in several steps. 

\subsubsection*{Step 1: Boundedness of $\mathrm{Sl}(2)$-norm}
For the first step of the proof, we would like to translate the bound \eqref{eq:bound_v(n)} into a more detailed statement on the various $\mathrm{sl}(2)$-components of $v(n)$. Indeed, the fact that the Hodge norm of $v(n)$ is bounded implies that also its $\mathrm{Sl}(2)$-norm is bounded. This is reasonable, since one can view the latter as providing the leading approximation to the full Hodge norm. Recalling the result \eqref{eq:growth_theorem_multi}, the crucial point is that the latter is also straightforward to evaluate explicitly and allows one to obtain the following bound (after possibly enlarging $L$)
\begin{equation}
	\label{eq:boundedness_sl2}    \sum_{\ell}\left[\prod_{i=1}^m\left(\frac{y_i}{y_{i+1}}\right)^{\ell_i}\right]||\hat{v}_{\ell}(n)||_\infty^2\leq L\,,
\end{equation}
where we have introduced the notation
\begin{equation}
	\hat{v}(n) = \mathrm{exp}\left[-\sum_{i=1}^m x_i N_i\right]v(n)\,,
\end{equation}
and we recall that the $\hat{v}_\ell(n)$ denote the weight-components of $\hat{v}(n)$, as defined in \eqref{eq:sl2_decomp_vector}. Since the left-hand side of \eqref{eq:boundedness_sl2} consists of a sum of positive terms, this bound in fact applies for each $\ell$ separately. Because of this it will be natural to prove boundedness of each individual $\hat{v}_\ell$ component. In other words, we have the following
\begin{equation}
	\text{Goal:}\qquad \text{Show that $||\hat{v}_\ell(n)||_\infty$ is bounded, for each $\ell$.}
\end{equation}
Since the axions are assumed to take values on a bounded interval, this immediately implies that also each $||v_\ell(n)||_\infty$ is bounded.\\ 

\noindent In the one-variable case ($m=1$) the relation \eqref{eq:boundedness_sl2} yields a natural separation of weight-components into the classes $\ell<0,\,\ell=0,\,\ell>0$, corresponding to fluxes whose Hodge norm tends to zero, stays constant, or grows as one approaches the boundary of the moduli space. In fact, this is the strategy that was used in \cite{Grimm:2020cda,Schnellletter} to prove finiteness for the one-variable case. However, in the multi-variable case such a separation is not available, as the scaling of the various terms in \eqref{eq:boundedness_sl2} highly depends on the exact hierarchy between $y_1,\ldots, y_m$, which in turn is highly path-dependent. For example, even though we do assume that $y_1>y_2>\cdots >y_m$, recall the growth sector \eqref{eq:growth_sector}, it is not necessarily the case that also $y_1>y_2^2$. Indeed, this is one of the main difficulties that were mentioned in section \ref{subsec:finiteness_intro}. In order to tackle the multi-variable case, we proceed in a different way by introducing a \textit{finite} partition of the moduli space $\mathbb{H}^m$ into subsectors on which the scaling behaviour of the various $\ell$-components is under control. Subsequently, the proof will proceed by considering the different types of subsectors individually. 

\subsubsection*{Step 2: Reduction to subsectors}
The key insight is to use the quantization of the fluxes, together with tadpole bound as formulated in \eqref{eq:boundedness_sl2}, to construct the desired partition of the moduli space. First, we note that since the flux $v(n)$ is integral, there exists a constant $\lambda>0$ such that
\begin{equation}
	\label{eq:lower_bound_Hodge_infty}
	||\hat{v}_{\ell}(n)||_\infty^2>\lambda\,,
\end{equation}
for all $n$, whenever $\hat{v}_\ell(n)$ is non-zero. In practice, one can construct $\lambda$ from the smallest eigenvalue of the boundary Hodge norm, and furthermore take $0<\lambda\leq 1$.\footnote{The reason is that, in general, the eigenvalues of the boundary Hodge norm come in pairs $(\lambda_i, \lambda_i^{-1})$, with each $\lambda_i> 0$.} In order to illustrate this rather abstract statement, we have included some basic examples of boundary Hodge norms in Appendix \ref{app:Hodge_norms}, for which one can write down the constant $\lambda$ explicitly. \\

\noindent We now combine the relation \eqref{eq:lower_bound_Hodge_infty} together with the tadpole bound to define the desired partition of the moduli space into sectors. For each $\ell$, we define a sector in $\mathbb{H}^m$ by
\begin{equation}    R^{\mathrm{heavy}}_{\ell} = \left\{(y_1,\ldots,y_m):\,\prod_{i=1}^m \left(\frac{y_i}{y_{i+1}}\right)^{\ell_i}> \frac{L}{\lambda}\right\}\,,
\end{equation}
and we define another sector $R^{\mathrm{light}}_{\ell}$ as the complement of $R^{\mathrm{heavy}}_{\ell}$. In other words, for each choice of $\ell$, we split up the moduli space into two disjoint pieces. See figure \ref{fig:subsectors} for an illustration of these sectors in the case of two moduli. The motivation for this definition is as follows. Clearly, if the sequence $y(n)$ lies entirely inside the region $R^{\mathrm{heavy}}_{\ell}$ and also $v_\ell(n)$ is non-zero, for some fixed $\ell$, then 
\begin{equation}
	\sum_{\ell'} \left[\prod_{i=1}^m \left(\frac{y_i}{y_{i+1}}\right)^{\ell'_i}\right]||\hat{v}_{\ell'}(n)||^2_\infty\geq \prod_{i=1}^m \left(\frac{y_i}{y_{i+1}}\right)^{\ell_i}||\hat{v}_{\ell}(n)||^2_\infty> \frac{L}{\lambda}\lambda=L\,,
\end{equation}
which is in contradiction with the bound \eqref{eq:boundedness_sl2}. In other words, inside the region $R^{\mathrm{heavy}}_{\ell}$, the weight-$\ell$ component of $v(n)$ would have a Hodge norm that exceeds the tadpole bound. By passing to a subsequence, we may therefore assume that the sequence $y(n)$ lies entirely inside the region $R^{\mathrm{light}}_{\ell}$.\\

\begin{figure}
	\centering
	\includegraphics[scale=0.45]{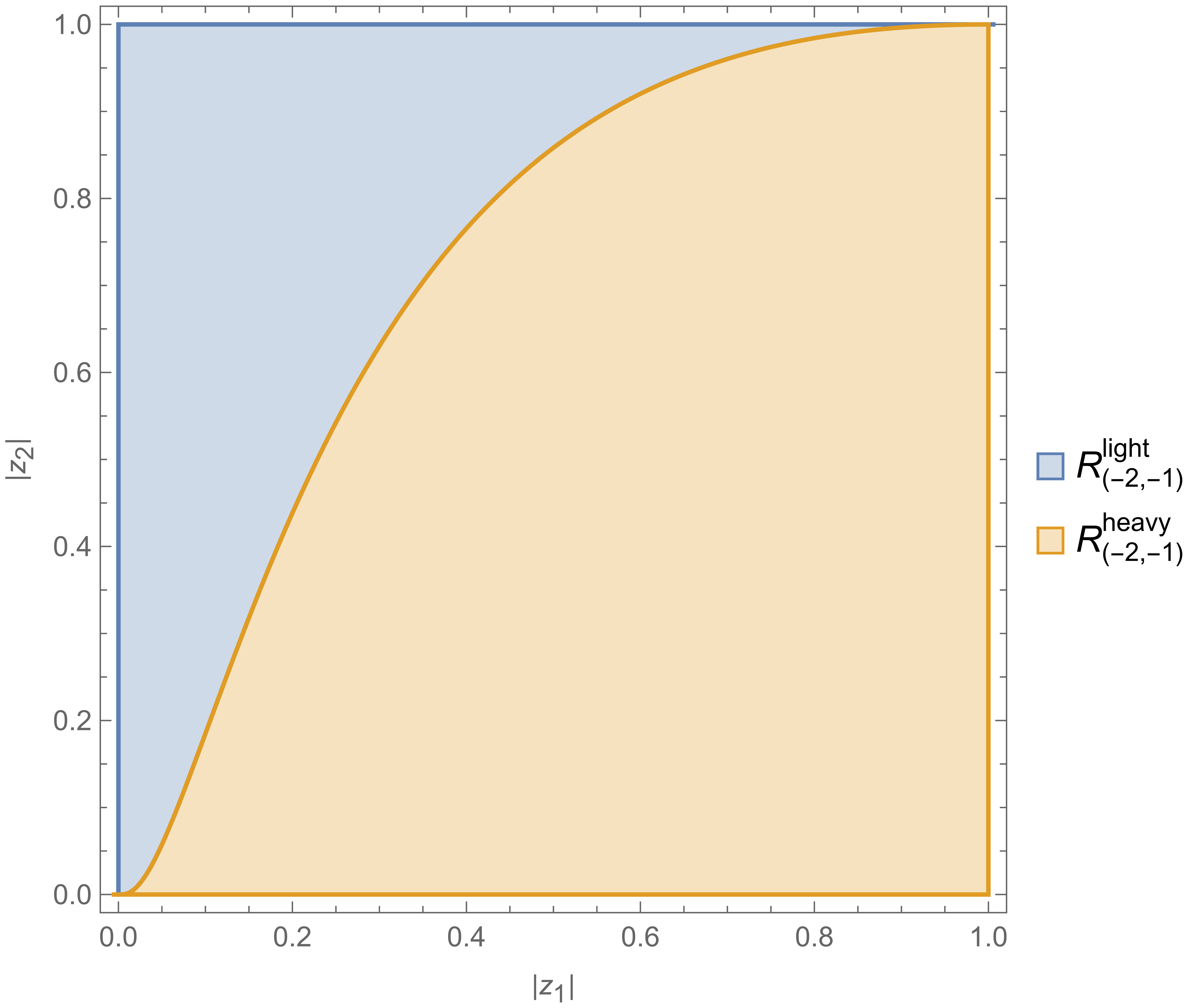}
	\caption{Depiction of the subsectors $R^{\mathrm{light}}_\ell$ (in blue) and $R^{\mathrm{heavy}}_\ell$ (in orange), in terms of the disk coordinates. Here we chose the weight vector to be $\ell=(-2,-1)$ and took $\frac{L}{\lambda}=2$. Note that the growth sector \eqref{eq:growth_sector} translates to the region $|z_1|< |z_2|$.}
	\label{fig:subsectors}
\end{figure}

\noindent We now come to the central point of the proof. Since we may assume that $y(n)$ lies entirely inside $R^{\mathrm{light}}_\ell$, we have the upper bound
\begin{equation}
	\prod_{i=1}^m \left(\frac{y_i}{y_{i+1}}\right)^{\ell_i}\leq \frac{L}{\lambda}\,.
\end{equation}
However, since there appears to be no obvious lower bound, the scaling factor that accompanies each $||\hat{v}_\ell(n)||^2_\infty$ in \eqref{eq:boundedness_sl2} could become arbitrarily small. As a result, it appears that some $\hat{v}_\ell(n)$ can be made arbitrarily large, without exceeding the tadpole bound. Note that this is very similar in spirit to the toy example discussed in section \ref{subsec:finiteness_intro}, as well as the example discussed in section \ref{subsec:self-dual_locus_example}. Of course, the missing ingredient that we have not yet exploited is the self-duality condition. To this end, it will be useful to make a further case distinction:
\begin{enumerate}[a.)]
	\item $y(n)\in R^{\mathrm{light}}_{-\ell}$.
	\item $y(n)\in R^{\mathrm{heavy}}_{-\ell}$.
\end{enumerate}
Clearly this covers all possible cases. The motivation for this additional case distinction is that, in order to address the fate of a $\hat{v}_{\ell}$ component, it is actually necessary to consider the $\hat{v}_{-\ell}$ component as well. This is because these two components are related via the self-duality condition. The first case is easy hence we will discuss it first. 

\subsubsection*{Step 3a: Case $y(n)\in R^{\mathrm{light}}_{-\ell}$}
Since $y(n)\in R_{-\ell}^{\mathrm{light}}$ we have that
\begin{equation}
	\prod_{i=1}^m \left(\frac{y_i}{y_{i+1}}\right)^{-\ell_i}\leq \frac{L}{\lambda}\quad\text{or, equivalently} \quad \prod_{i=1}^m \left(\frac{y_i}{y_{i+1}}\right)^{\ell_i}\geq  \frac{\lambda}{L}\,.
\end{equation}
In other words, inside the region $R^{\mathrm{light}}_{-\ell}$ we do get a lower bound on the scaling factors. In turn, this provides an upper bound on $||\hat{v}_\ell(n)||_\infty^2$, as is seen explicitly as follows
\begin{equation}
	L\geq \sum_{\ell'} \prod_{i=1}^m \left(\frac{y_i}{y_{i+1}}\right)^{\ell'_i}||\hat{v}_{\ell'}(n)||^2_\infty\geq \prod_{i=1}^m \left(\frac{y_i}{y_{i+1}}\right)^{\ell_i}||\hat{v}_{\ell}(n)||^2_\infty\geq \frac{\lambda}{L}||v_{\ell}(n)||_\infty^2\,.
\end{equation}
Hence, one finds the upper bound
\begin{equation}
	\label{eq:bound_Hodge_case_a}
	||\hat{v}_{\ell}(n)||_\infty^2 \leq \frac{L^2}{\lambda}\,.
\end{equation}
Therefore, within the region $R^{\mathrm{light}}_{-\ell}$, we immediately obtain the desired bound on the boundary Hodge norm of $\hat{v}_\ell(n)$, in terms of an `effective tadpole bound' given by the combination $L^2/\lambda$. We stress that this bound depends on both the original tadpole bound $L$, as well as the constant $\lambda$ introduced in \eqref{eq:lower_bound_Hodge_infty}, where the latter reflects the quantization condition of the fluxes and depends on the properties of the boundary Hodge structure. It is also important to note that, in the derivation of the bound \eqref{eq:bound_Hodge_case_a} it has not been necessary to use the self-duality condition. Indeed, in this particular case the resulting finiteness of the fluxes should not be viewed as a property of only the vacuum locus. In particular, it is not obvious whether one could use the refined tadpole bound $L^2/\lambda$ to obtain an accurate estimate for the number of flux vacua in this region of the moduli space. 

\subsubsection*{Step 3b: Case $y(n)\in R^{\mathrm{heavy}}_{-\ell}$ }
This case is more involved and comprises the most difficult part of the whole proof. This is because in this case there is no obvious lower bound for the scaling factor. Instead, the only information at our disposal is that $y(n)$ lies inside $R^{\mathrm{heavy}}_{-\ell}$, which implies that $v_{-\ell}(n)=0$ by the reasoning in step 2. The strategy will be to combine this fact together with an explicit evaluation of the self-duality condition using the nilpotent orbit expansion \eqref{eq:hinv}. Indeed, recalling the definition of the map $h(y_1,\ldots, y_m)$, and using the fact that the boundary Weil operator $C_\infty$ exchanges the $+\ell$ and $-\ell$ eigenspaces, the self-duality condition \eqref{eq:self-duality_nil} can be written as 
\begin{equation}
	\label{eq:duality_weights}
	\left(h^{-1}\hat{v}(n)\right)_{\ell}=C_\infty\left(h^{-1}\hat{v}(n)\right)_{-\ell}\,,
\end{equation}
The strategy, then, will be to first evaluate $\left(h^{-1}\hat{v}(n)\right)_{-\ell}$ explicitly to derive its scaling with the moduli and then use the relation \eqref{eq:duality_weights} to infer information about $\left(h^{-1}\hat{v}(n)\right)_\ell$ and subsequently $\hat{v}_{\ell}(n)$ itself. To proceed, we therefore apply the result for $h^{-1}$ stated in equation \eqref{eq:hinv} to find
\begin{equation}
	\label{eq:hinv_b}
	\left(h^{-1}\hat{v}(n)\right)_{-\ell}=\sum_{k_1,\ldots, k_m=0}^\infty \sum_{s^1,\ldots, s^m}\prod_{i=1}^m \left[\left(\frac{y_i}{y_{i+1}}\right)^{-k_i+\frac{1}{2}s_i^{(m)}}f^{s^i}_{i,k_i}\right]\left(e(y)\hat{v}(n)\right)_{-\ell-s^{(m)}}\,.
\end{equation}
where we recall the notation
\begin{equation}    s^{(m)}=\left(s_1^{(m)},\ldots, s_m^{(m)}\right)\,,\qquad s_i^{(m)}=s_i^i+\cdots +s_i^m\,,
\end{equation}
and note that the second sum in \eqref{eq:hinv_b} runs over all possible values of the $m$ weight vectors $s^j$. Due to the particular weight properties of the expansion coefficients $f_{i,k_i}$, the sum only runs over $s_i^i\leq k_i-1$ and $s_j^i=0$ for $j>i$. The leading contribution to \eqref{eq:hinv_b} is given by the $k_1,\ldots, k_m=0$ term, and is proportional to $\hat{v}(n)_{-\ell}$. The crucial point is that, because we are considering the case where $y(n)\in R^{\mathrm{heavy}}_{-\ell}$, this leading contribution vanishes. In other words, in order to properly assess the scaling of $\left(h^{-1}\hat{v}(n)\right)_{-\ell}$ it is necessary to understand the scaling of the correction coefficients $f_{i,k_i}$. In particular, we will make use of the result \eqref{eq:f_bounds_new}.\\

\noindent We now come to the main technical computation, namely the estimation of the scaling of the term in brackets in \eqref{eq:hinv_b}. To this end we make two additional observations. 
\begin{itemize}
	\item \textbf{Observation (1):}\\
	First, note that
	\begin{equation}
		\left(\frac{y_i}{y_{i+1}}\right)^{-k_i+\frac{1}{2}s_i^i}\prec \left(\frac{y_i}{y_{i+1}}\right)^{-\frac{1}{2}s_i^i}\cdot\begin{cases}
			\left(\frac{y_i}{y_{i+1}}\right)^{-1}\,,&k_i\neq 0\,,\\
			1\,, & k_i=0\,.
		\end{cases}
	\end{equation}
	This follows from the fact that for $k_i\neq 0$, there is the restriction $s_i^i\leq k_i-1$, while for $k_i=0$ one automatically has $s_i^i=0$. 
	\item \textbf{Observation (2):}\\
	Second, we note that $\hat{v}_{-\ell-s^{(m)}}(n)$ is only non-zero if the sequence $y(n)$ lies in $R^{\mathrm{light}}_{-\ell-s^{(m)}}$, or, in other words, when the factor
	\begin{equation*}    \prod_{i=1}^m\left[\left(\frac{y_i}{y_{i+1}}\right)^{-\left(\ell_i+s_i^{(m)}\right)}\right]
	\end{equation*}
	is bounded by a constant. In particular, we may apply this to all the terms appearing in \eqref{eq:hinv_b}.
\end{itemize}
Now suppose, for the moment, that all $k_i$ are non-zero, then combining these observations with the bounds stated in \eqref{eq:f_bounds_new}, we find the following estimate
\begin{align*}    &\quad\,\, \prod_{i=1}^m\left[\left(\frac{y_i}{y_{i+1}}\right)^{-k_i+\frac{1}{2}s_i^{(m)}}f^{s^i}_{i,k_i}\right]\\
	&\stackrel{\text{(a)}}{\prec}  \prod_{i=1}^m\left[\left(\frac{y_i}{y_{i+1}}\right)^{-1-\frac{1}{2}s_i^{i}+\frac{1}{2}(s_i^{i+1}+\cdots +s_i^m)}\prod_{j=1}^{i-1} \left(\frac{y_j}{y_{j+1}}\right)^{-s_j^i}\right]\\
	&\stackrel{\text{(b)}}{=} \prod_{i=1}^m\left[\left(\frac{y_i}{y_{i+1}}\right)^{-1-\frac{1}{2}s_i^{(m)}}\right]\\    &\stackrel{\text{(c)}}{=}\prod_{i=1}^m\left[\left(\frac{y_i}{y_{i+1}}\right)^{\frac{1}{2}\ell_i}\right]\cdot \underbrace{\prod_{i=1}^m\left[\left(\frac{y_i}{y_{i+1}}\right)^{-\frac{1}{2}\left(\ell_i+s_i^{(m)}\right)}\right]}_{\prec 1}\cdot \prod_{i=1}^m\left[\left(\frac{y_i}{y_{i+1}}\right)^{-1}\right]\\
	&\stackrel{\text{(d)}}{\prec} \prod_{i=1}^m\left[\left(\frac{y_i}{y_{i+1}}\right)^{\frac{1}{2}\ell_i}\right]\cdot y_1^{-1}\,.
\end{align*}
To be clear, in step (a) we used \eqref{eq:f_bounds_new} and applied the first observation, in step (b) we simply collected all the terms, in step (c) we expanded the product to uncover the middle term and in step (d) we applied the second observation stating that the middle term is bounded. \\

\noindent If, in contrast, $k_{i}=0$ for some $i$, the only difference is that the corresponding factor of $\left(y_{i}/y_{i+1}\right)^{-1}$ will not be present, see again the first observation. For example, if $k_1=0$ but all other $k_i$ are non-zero, one will instead get
\begin{equation}
	k_1=0:\qquad \prod_{i=1}^m\left[\left(\frac{y_i}{y_{i+1}}\right)^{-k_i+\frac{1}{2}s_i^{(m)}}f^{s^i}_{i,k_i}\right]\prec \prod_{i=1}^m\left[\left(\frac{y_i}{y_{i+1}}\right)^{\frac{1}{2}\ell_i}\right]\cdot y_2^{-1}\,.
\end{equation}
In particular, the factor of $y_1^{-1}$ is now replaced by a factor of $y_2^{-1}$. A similar thing happens when multiple $k_i$'s are equal to zero. The important point is that one always ends up with some rational factor which goes to zero as all $y_i\rightarrow\infty$.\footnote{Here it is important to recall that sequence $y(n)$ is restricted to lie inside the growth sector \eqref{eq:growth_sector}.} It remains to consider the term in which all $k_i$ are zero. However, as said before, this leading term vanishes since we have assumed $y(n)\in R^{\mathrm{heavy}}_{-\ell}$. To summarize, we have argued that
\begin{equation}
	||(h^{-1}\hat{v}(n))_{-\ell}||_\infty \prec \prod_{i=1}^m\left[\left(\frac{y_i}{y_{i+1}}\right)^{\frac{1}{2}\ell_i}\right]\cdot\alpha(y_1,\ldots,y_m)\,,
\end{equation}
where $\alpha(y_1.\ldots,y_m)$ is a rational function of $y_1,\ldots, y_m$ that goes to zero as $n\rightarrow\infty$. To complete the argument, we now apply the duality condition \eqref{eq:duality_weights} to find
\begin{equation}
	||(h^{-1}\hat{v}(n))_{\ell}||_\infty \prec \prod_{i=1}^m\left[\left(\frac{y_i}{y_{i+1}}\right)^{\frac{1}{2}\ell_i}\right]\cdot\alpha(y_1,\ldots,y_m)\,,
\end{equation}
Moving the term in square brackets to the left-hand side and noting that $e(y)h^{-1}\sim 1$, we find the result
\begin{equation}
	||v_{\ell}(n)||_\infty\prec \alpha(y_1,\ldots,y_m)\,,
\end{equation}
where we have again used the fact that the axions $x_i(n)$ are bounded to remove the hat. In particular, we have shown that the sequence $v_{\ell}(n)$ is bounded with respect to the boundary Hodge norm. In fact, since $v(n)$ is integral, it cannot become arbitrarily small, hence after some finite $n$ we must in fact have that $v_{\ell}(n)=0$.

\subsubsection*{Step 4: Finishing the proof}
Let us collect the results so far. For a fixed $\ell$, we have effectively shown that
\begin{itemize}
	\item $y(n)\in R^{\mathrm{heavy}}_{\ell}$: $v_{\ell}(n) =0$. 
	\item  $y(n)\in R^{\mathrm{light}}_{\ell}\cap R^{\mathrm{light}}_{-\ell}$: $||v_{\pm \ell}(n)||_\infty$ is bounded. More precisely,
	\begin{equation}
		||\hat{v}_{\pm\ell}(n)||_\infty^2\leq\frac{L^2}{\lambda}\,.
	\end{equation}
	\item  $y(n)\in R^{\mathrm{light}}_{\ell}\cap R^{\mathrm{heavy}}_{-\ell}$: $v_{-\ell}(n)=0$. Furthermore, there exists an $n'$ such that for all $n>n'$ we have that  $v_{\ell}(n)$ vanishes.
\end{itemize}
\begin{figure}
	\centering
	\includegraphics[scale=0.45]{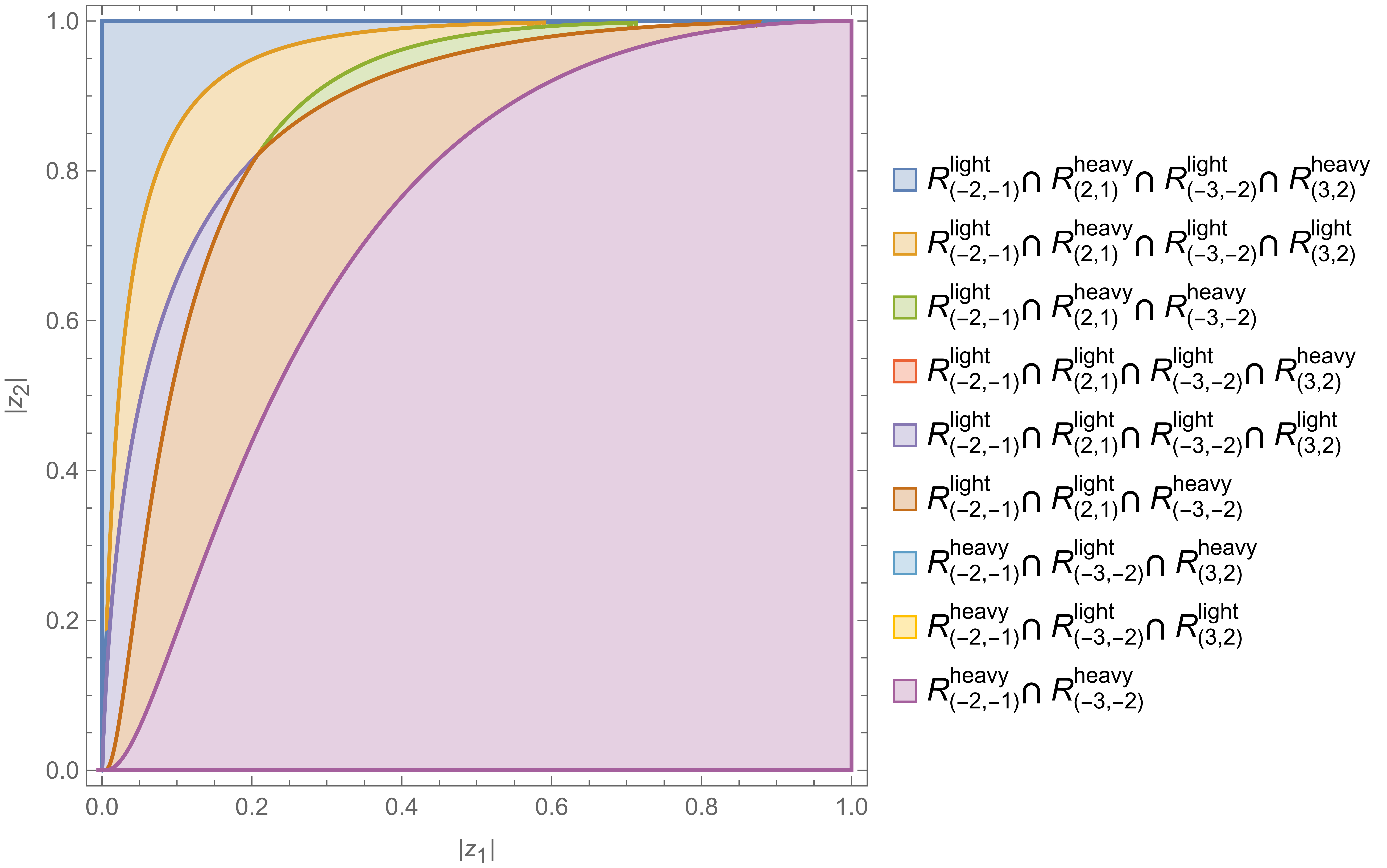}
	\caption{Depiction of the nine disjoint subsectors that arise from taking intersections between $R^{\mathrm{light}}_\ell$ and $R^{\mathrm{heavy}}_\ell$ for various values of $\ell$. Again we have taken $\frac{L}{\lambda}=2$. Note that some of the possible intersections, such as the one in red, cover only a region near small values of $|z_1|$ and $|z_2|$, and are therefore not visible due to the limited resolution. }
	\label{fig:subsectors2}
\end{figure}
This covers all possibilities. Therefore, the sequence $v_{\ell}(n)$ is bounded with respect to the boundary Hodge norm in all sectors. Furthermore, for $n$ sufficiently large, the only way in which it can attain non-zero values is if $y(n)\in R^{\mathrm{light}}_{\ell}\cap R^{\mathrm{light}}_{-\ell}$. One may now simply apply this argument for all possible values of $\ell$, by considering subsequences $y(n)$ lying in all possible intersections of subsectors. For example, one might start with the weight $\ell=(-2,-1)$  and consider the three spaces
\begin{equation}
	R^{\mathrm{light}}_{(-2,-1)}\cap R^{\mathrm{heavy}}_{(2,1)}\,,\qquad R^{\mathrm{light}}_{(-2,-1)}\cap R^{\mathrm{light}}_{(2,1)}\,,\qquad R^{\mathrm{heavy}}_{(-2,-1)}\,,
\end{equation}
which exactly cover the three cases listed above. Then one considers the same three spaces but for $\ell=(-3,-2)$, and constructs all nine pairwise intersections between these spaces. This is illustrated in figure \ref{fig:subsectors2}. One then continues this process ranging over the total number $\#$ of possible values of $\ell$, yielding at most $3^{\#}$ disjoint sectors.\footnote{Note that some intersections may be empty.} Importantly, this procedure always results in a finite partition. Therefore, it suffices to consider a finite number of subsequences of $y(n)$, each lying in a fixed intersection. In this way we conclude that $||v_{\ell}(n)||_\infty$ is bounded for all $\ell$ throughout all sectors. Combining this with the fact that $v(n)$ is integral completes the proof.

\section{Conjectures about the flux landscape}
\label{sec:future_questions}

In the preceding sections we have focused our attention on relatively rudimentary properties of the flux landscape, in particular with regards to its finiteness. In this section we would like to point out some additional questions that could feasibly be addressed in the near-future, whose answers would further elucidate more precise features of the flux landscape, and formulate them into precise mathematical conjectures. These conjectures would pose interesting challenges which can likely be tackled by the application and development of techniques in asymptotic Hodge theory and o-minimality. 

\subsection{Recounting flux vacua}

Having established that the number of self-dual flux vacua is finite, a natural follow-up question would be: how many are there? The early works of Douglas et al.~\cite{Ashok:2003gk,Denef:2004ze} suggest that such numbers could be very large, giving rough estimates of the order $10^{500}$ to $10^{272,000}$, see also \cite{Taylor:2015xtz}. At the same time, it has also been pointed out that these analyses have their shortcomings. In particular, it is possible that the smearing approximation used to effectively ignore the quantization condition significantly affects the precise counting of vacua. It is a challenging task to establish robust mathematical counting results. \\

\noindent One might ask if this problem becomes attainable for the case of Hodge vacua. Here one faces the fact that the approximations of \cite{Ashok:2003gk,Denef:2004ze} are likely even less reliable. As discussed also in section \ref{subsubsec:locus_Hodge}, a Hodge vacuum is expected to be relatively rare. The main reason for this is the fact that a Hodge vacuum has to satisfy $h^{3,1}+1$ equations for only $h^{3,1}$ variables, hence the system is overdetermined. Importantly, after solving the $D_i W_{\mathrm{flux}}=0$ equations for the complex structure moduli in terms of the fluxes and inserting the result into the remaining $W_{\mathrm{flux}}=0$ equation, one is left with a highly transcendental equation for the fluxes. This transcendentality originates from the fact that the flux-induced superpotential is expressed in terms of the periods of the Calabi--Yau fourfold. The crucial point is that, due to the quantization condition, this highly transcendental equation needs to be solved over the integers, hence its solutions are expected to be rare. Indeed, in the context of o-minimal geometry, some intuition for this is provided by the celebrated counting theorem of Pila and Wilkie \cite{Pila:2006}. Very roughly speaking, the Pila--Wilkie theorem states that there are very few rational points on the transcendental part of a definable set. More precisely, the number of such points grows slower than any positive power of their multiplicative height.\footnote{For an integral flux $v=(v_1,\ldots, v_k)\in H_{\mathbb{Z}}$, its multiplicative height is simply $\mathrm{max} |v_i|$.}\textsuperscript{,}\footnote{In \cite{barroero2013counting} this theorem was applied to provide bounds on the number of lattice points in the fibers of definable families.}\\

\noindent Based on the above considerations, one would expect that for those variations of polarized Hodge structure which are ``sufficiently transcendental'' (dictated by a property called the ``level'' \cite{baldi2022distribution}), the number of connected components in the locus of Hodge classes with a fixed self-intersection $L$ should grow sub-polynomially in $L$. In particular, this would imply that the number of $W_{\mathrm{flux}}=0$ vacua in F-theory grows much slower than expected. However, it was recently shown in an explicit example investigated in \cite{Grimm:2024fip} that this conclusion is not quite correct, but for a very interesting reason. Namely, it can happen that a collection of Hodge classes actually lies on a higher-dimensional locus where additional Hodge \textit{tensors} appear, see appendix \ref{app:Hodge-tensors} for a basic introduction to Hodge tensors. The important point is that, because of the presence of these additional Hodge tensors, the restriction of the variation of Hodge structure to this higher-dimensional locus typically has a reduced level and thus becomes ``less transcendental'', such that the original logic based on the Pila--Wilkie counting theorem may not apply. As discussed in \cite{Grimm:2024fip} this reduction in transcendentality on these loci indicates the presence of an underlying symmetry in the compactification manifold. In order to take into account these subtle matters, we therefore propose the following refined version of the counting conjecture.
\begin{subbox}{\begin{conjecture}    \label{conjecture:scaling_Hodge_vacua} Counting of Hodge vacua
	\end{conjecture}}
	Consider a variation of polarized Hodge structure $E\rightarrow\mathcal{M}$ of even weight $D=2k$. Fix a positive integer $L$ and consider the locus of Hodge classes with a fixed self-intersection $L$,
	\begin{equation}
		\hat{E}_{\mathrm{Hodge}}(L) = \{(z^i,v)\in E: v\in H^{k,k}\cap H_{\mathbb{Z}}\,,(v,v)= L\}\,.
	\end{equation}
	Furthermore, denote by $\hat{E}_{\mathrm{Hodge}}^{\mathrm{iso}}(L)$ the subset of points $(z^i,v)\in\hat{E}_{\mathrm{Hodge}}(L)$ for which $z^i$ are isolated points in the locus of Hodge tensors, see appendix \ref{app:Hodge-tensors}.\\
	
	\noindent We claim that if the level of the variation of Hodge structure is at least 3, then the number points in $\hat{E}^{\mathrm{iso}}_{\mathrm{Hodge}}(L)$ grows sub-polynomially in $L$. More precisely, for every $\epsilon>0$ there exists a $C>0$, such that
	\beq
	\# \hat{E}^{\mathrm{iso}}_{\mathrm{Hodge}}(L) < C L^\epsilon\ ,  
	\eeq
	where $\# \hat{E}^{\mathrm{iso}}_{\mathrm{Hodge}}(L)$ is the number points in $\hat{E}^{\mathrm{iso}}_{\mathrm{Hodge}}(L)$ and $C$ is independent of $L$.
\end{subbox} 
\noindent Some remarks are in order. First, we note that in \cite{baldi2022distribution} a related conjecture has been proposed. The latter states that, under similar conditions, the number of points in $\hat{E}^{\mathrm{iso}}_{\mathrm{Hodge}}$ is in fact finite, \textit{without fixing the self-intersection}. It is important to stress that while a similar statement for higher-dimensional loci has, rather strikingly, been proven in \cite{baldi2022distribution}, the statement for isolated points, which is the case of interest for us, is still a wide open problem. \\
  
\noindent Second, let us briefly elaborate on the notion of the `level' of a variation of Hodge structure. The precise definition is somewhat technical and is explained in \cite{baldi2022distribution}. Roughly speaking, it is related to the length of the Hodge filtration and serves as a measure of its `complexity'. However, it should not be confused with the weight $D$ of the Hodge structure. For example, while the Hodge structure on the middle cohomology of a K3 surface is of weight $D=2$, its level is in fact equal to one. As another example, while one generically expects that the middle cohomology of a Calabi--Yau fourfold has level equal to four, one can show that for special cases such as $Y_3\times T^2$ or $\mathrm{K3}\times \mathrm{K3}$ the level is again equal to one. In particular, Conjecture \ref{conjecture:scaling_Hodge_vacua} does not apply to these cases. \\

\noindent To elaborate on this point, consider the weak-coupling limit corresponding to type IIB orientifold compactifications, in which case one effectively reduces to a direct product $Y_4=Y_3\times T^2$ and hence the level reduces to one. In this setting, known scans of vacua in one-parameter and two-parameter Calabi--Yau manifolds, defined as hypersurfaces in weighted projective space, indicate that the number of vacua with $W_{\mathrm{flux}}=0$ in fact scales polynomially in $L$ \cite{DeWolfe:2004ns,Giryavets:2004zr,Conlon:2004ds}. This is confirmed by the recent work \cite{Plauschinn:2023hjw} in which a complete counting of vacua, including $W_{\mathrm{flux}}=0$ vacua, was performed for the mirror octic. To be clear, this is not in contradiction with Conjecture \ref{conjecture:scaling_Hodge_vacua}, due to the reduction in the level in the weak coupling limit. We believe, however, that this counting is actually not representative for the number of exact Hodge vacua in the non-perturbative setting of F-theory. Indeed, the observed polynomial scaling in the type IIB setting should be viewed as an artifact of truncating the axio-dilaton dependence to the polynomial, i.e.~algebraic, level. To emphasize this point, recall that the axio-dilaton $\tau$ can trivially be solved for in terms of the $F_3$ and $H_3$ fluxes as
\begin{equation}
\label{eq:solution_tau_IIB}
    \bar{\tau} = \frac{\int \Omega\wedge F_3}{\int \Omega\wedge H_3}\,.
\end{equation}
In contrast, as soon as one includes exponential corrections in $\tau$ it is clear that this is no longer so straightforward and we expect that the transcendental nature of the equations greatly restricts the number exact Hodge vacua.\footnote{Of course, there can also be perturbative corrections which break the simple relation \eqref{eq:solution_tau_IIB}, but these do not affect the transcendentality of the equations.} Put shortly, one should perform the counting of $W_{\mathrm{flux}}=0$ vacua in the full F-theory setting, which, in particular, requires a non-trivial elliptic fibration. Mathematically, this is captured by the condition that the level of the variation of Hodge structure should be at least three. A further motivation for this comes from the recent work \cite{baldi2022distribution}, in which it was shown that, when the level is at least three, the locus of Hodge classes corresponds to an atypical intersection, reflecting the fact that it is expected to occur only rarely.\\

\noindent Finally, let us mention some recent developments in mathematics concerning the issues of algebraicity and transcendentality in a Hodge-theoretic context. From a more number-theoretic point of view, a Hodge vacuum effectively requires that some of the $h^{3,1}+1$ equations are no longer algebraically independent. It is a long-standing question when there exist algebraic relations among transcendental numbers, which lies at the heart of the Schanuel conjecture. More concretely, given a collection of complex numbers $\alpha_1,\ldots,\alpha_n$ which are algebraically independent over $\mathbb{Q}$, the Schanuel conjecture gives a bound on the number of algebraic relations among the numbers $\alpha_1,\ldots, \alpha_n, e^{\alpha_1},\ldots, e^{\alpha_n}$. A functional analogue of this question, where one is considering algebraic relations between $f_1(x),\ldots, f_n(x), e^{f_1(x)},\ldots, e^{f_n(x)}$, is addressed by the Ax--Schanuel theorem \cite{Ax:1971}, which has also been generalized for certain transcendental functions besides the exponential function. Recently, techniques from o-minimal geometry and the theory of atypical/unlikely intersections have lead to great developments in this field as well as a proof of the Ax--Schanuel conjecture in the Hodge-theoretic setting \cite{klingler2017hodge,bakker2017axschanuel}. Very roughly speaking, the latter relates the appearance of an atypical intersection, meaning the existence of additional algebraic relations among e.g.~the periods, to a reduction of the so-called Mumford--Tate group. In a similar spirit, the recent work \cite{baldi2022distribution} has elucidated further properties of the Hodge locus using the theory of unlikely intersections. It would be very interesting to further investigate these techniques in the context of F-theory flux compactifications and ascertain whether they could lead to improved quantitative results on the counting of Hodge vacua and possibly prove or disprove Conjecture \ref{conjecture:scaling_Hodge_vacua}. Whether these techniques could also be applied to study self-dual vacua is not so clear.     

\subsection{Complexity of the flux landscape}

Another exciting avenue to explore with regards to the counting of flux vacua is using a certain notion of complexity that has recently been developed in the context of sharp o-minimality, which moreover may be applicable to study both Hodge vacua and self-dual vacua. The basic idea of sharp o-minimality, introduced by Binyamini and Novikov \cite{binyamini2022,binyamini2022sharply}, is to endow definable sets, and thereby definable functions, with some additional positive integers $(F,D)$, called the ``format'' $F$ and ``degree'' $D$ , that reflect the inherent geometric complexity of that set/function. This is in analogy with the degree of a polynomial, which clearly gives the number of zeroes of said polynomial over the complex numbers, but can also be used to give bounds on the number of its zeroes over the real numbers.\footnote{More generally, this falls under Khovanskii's theory of fewnomials \cite{Khovanskii:1980}.} Roughly speaking sharply o-minimal structure are defined in such a way that the functions arising in these structures have similar bounds on their number of zeros \cite{binyamini2022sharply}. Recently, the concept of sharp o-minimality has been explored in a variety of quantum mechanical systems in order to assign a well-defined notion of complexity to various physical observables \cite{Grimm:2023xqy}, see also \cite{Grimm:2021vpn,Douglas:2022ynw, Douglas:2023fcg}. It is natural to ask if a similar strategy can be applied to assign a complexity to e.g.~the F-theory flux scalar potential, which may then provide a new method of estimating the number of flux vacua. In this regard, we propose the following
\begin{subbox}{\begin{conjecture}
			\label{conjecture:complexity} Complexity of self-dual flux vacua
		\end{conjecture}}
	We conjecture that the locus of self-dual flux vacua is definable in a sharply o-minimal structure. Furthermore, we expect that its associated sharp complexity $(F,D)$ depends on the tadpole bound $L$ and the number of moduli $h^{3,1}$ in the following way: 
	\begin{equation}
		D=\mathrm{poly}(L)\,,\qquad F=\mathcal{O}(h^{3,1})\,.
	\end{equation}
\end{subbox}
\noindent Our expectation for the scaling of $D$ and $F$ is rather conservative, and is motivated by the form of the Ashok--Douglas index density \cite{Ashok:2003gk,Denef:2004ze}. Indeed, the latter grows as $L^{h^{3,1}}$, while generically the number of zeroes of functions that are definable in a sharply o-minimal structure depends polynomially on $D$ and exponentially on $F$. Since the sharp complexity $(F,D)$ only gives upper bounds on the number of such zeroes, it could also be the case that already for self-dual vacua, the scaling is in fact more restricted. Certainly, this is expected for the special class of Hodge vacua, as captured by Conjecture \ref{conjecture:scaling_Hodge_vacua}. \\

\noindent Nevertheless, we stress that the statement of Conjecture \ref{conjecture:complexity} is highly non-trivial. Indeed, while Theorem \ref{thm:finiteness_self_dual} establishes that the locus of self-dual flux vacua is definable in the o-minimal structure $\mathbb{R}_{\mathrm{an},\mathrm{exp}}$, it has been shown that this structure is not sharply o-minimal. Roughly speaking, a generic restricted analytic function does not have a well-defined notion of complexity, because one has too much freedom in specifying the coefficients in its series expansion. Nevertheless, it is currently conjectured \cite{binyamini2022}, that period integrals are in fact definable in a sharply o-minimal structure, meaning that they actually live in a much smaller o-minimal structure than $\mathbb{R}_{\mathrm{an},\mathrm{exp}}$. This would, in particular, imply a positive answer to the first part of Conjecture \ref{conjecture:complexity}. Lastly, let us mention the recent work \cite{binyamini2022wilkies} in which a proof was given for Wilkie's conjecture \cite{Pila:2006} when restricting to certain sharply o-minimal structures. Together with Conjecture \ref{conjecture:complexity}, the latter suggests that the scaling in Conjecture \ref{conjecture:scaling_Hodge_vacua} may be even more restricted by replacing the sub-polynomial scaling with a logarithmic scaling in $L$. It would be very interesting to investigate this further. 

\subsection{A generalized tadpole conjecture for the Hodge locus}

In the previous points we have focused on counting the number of flux vacua or, more precisely, the number of connected components of the vacuum locus. A related question concerns the dimension of the various connected components, in particular whether it can be zero. In other words, one might ask whether all complex structure moduli can always be stabilized for a suitable choice of flux. When one is only solving the vacuum conditions, it is reasonable to expect that this can indeed be achieved, since one imposes at least $h^{3,1}$ complex conditions for the same number of complex variables. However, it is not obvious whether this can be done whilst also imposing the tadpole condition. Indeed, the tadpole conjecture postulates that one cannot stabilize a large number of complex structure moduli within the tadpole bound, i.e.~when $h^{3,1}$ is much larger than all other Hodge numbers \cite{Bena:2020xrh}. More concretely, it states that for large $h^{3,1}$ and all moduli stabilized, one has
\begin{equation}
    \frac{1}{2}\int_{Y_4}G_4\wedge G_4 > \alpha h^{3,1}\,,
\end{equation}
with $\alpha>1/3$. Recalling that $\frac{\chi(Y_4)}{24}\sim \frac{1}{4}h^{3,1}$, this implies that for large $h^{3,1}$ the tadpole grows too quickly to be contained within the tadpole bound. We refer the reader to \cite{Braun:2020jrx,Bena:2021wyr,Marchesano:2021gyv,Lust:2021xds,Plauschinn:2021hkp,Grana:2022dfw,Lust:2022mhk,Coudarchet:2023mmm,Braun:2023pzd} for related works on the tadpole conjecture. \\

\noindent Let us attempt to formulate a version of the tadpole conjecture in a more mathematical fashion. Let $v\in H_{\mathbb{Z}}$ be an integral class, playing the role of the flux, and denote by 
\begin{equation}
    (v,v) = L\,,
\end{equation}
its self-intersection. The spirit of the tadpole conjecture is that when the flux $v$ defines a vacuum in which all moduli are stabilized, one necessarily has $L>\mathcal{O}(1)\cdot\mathrm{dim}\,\mathcal{M}$, where we recall that $\mathcal{M}$ denotes the complex structure moduli space. Conversely, if $\mathrm{dim}\,\mathcal{M}>\mathcal{O}(1)\cdot L$, then it must be that not all moduli are stabilized. The latter statement can be formalized as follows. Generically, the vacuum locus consists of several connected components, each having a well-defined notion of dimension.\footnote{See also \cite{Becker:2022hse} for a related discussion.} For the locus of Hodge classes this is immediately clear, since it is algebraic. For the locus of self-dual classes this follows from its definability in an o-minimal structure, since a natural notion of dimension is provided by the cell decomposition \cite{dries_1998}. Within this locus, some components may correspond to points, having dimension zero, while other components may correspond to higher-dimensional loci, having strictly positive dimension. The statement that not all moduli are stabilized then means that all components of the vacuum locus of with a fixed self-intersection $L$ have strictly positive dimension. For the class of Hodge vacua, such a special feature of the vacuum locus appears to be more plausible. Thus we are lead to the following
\begin{subbox}{\begin{conjecture}
			\label{conjecture:tadpole} Generalized tadpole conjecture \end{conjecture}}
	Consider a variation of polarized Hodge structure $E\rightarrow\mathcal{M}$ of even weight $D=2k$. Fix a positive integer $L$ and recall the notation
	\begin{equation}
		E_{\mathrm{Hodge}}(L) = \{(z^i,v)\in E: v\in H^{k,k}\cap H_{\mathbb{Z}}\,,(v,v)\leq L\}\,,
	\end{equation}
	for the locus of Hodge classes with self-intersection bounded by $L$. We conjecture that for certain positive constants $C_1, C_2$, which are independent of $L$ and $\mathrm{dim}\,\mathcal{M}$ (but may depend on other details of the variation of Hodge structure, such as the weight $D$), the following holds: if
	\begin{equation}   
		\label{eq:conj3_conditions}\mathrm{dim}\,\mathcal{M}>C_1\,\qquad \text{and}\qquad \mathrm{dim}\,\mathcal{M}>C_2\cdot L\,,
	\end{equation}
	then every connected component of $E_{\mathrm{Hodge}}(L)$ has strictly positive dimension.\footnote{Note that since $L\geq 1$ for non-trivial fluxes, the two conditions in \eqref{eq:conj3_conditions} reduce to a single condition whenever $C_2\geq C_1$.} Furthermore, when the variation of Hodge structure comes from the middle cohomology of a family of Calabi--Yau fourfolds, we expect that the constant $C_2$ is of order one. 
\end{subbox}
\noindent On the one hand, the statement of Conjecture \ref{conjecture:tadpole} is more general than the original tadpole conjecture of \cite{Bena:2020xrh}, as it is formulated for a general variation of Hodge structure. On the other hand, it should be emphasized that, in the specific setting of Calabi--Yau fourfold compactifications, the statement of Conjecture \ref{conjecture:tadpole} is weaker than the original tadpole conjecture, for a number of reasons. Firstly, Conjecture \ref{conjecture:tadpole} is formulated for Hodge vacua only, corresponding to vacua with $W_{\mathrm{flux}}=0$, while the original tadpole conjecture  applies to all self-dual vacua. Additionally, in the formulation of Conjecture \ref{conjecture:tadpole} there is no restriction on how many moduli are left unstabilized, as long as there is at least one. Finally, the exact values of the constants $C_1$ and $C_2$ are left undetermined. Especially for the physical application of studying the landscape of fully stabilized Hodge vacua, it is of utmost importance to quantify the exact values of $C_1,C_2$.\\

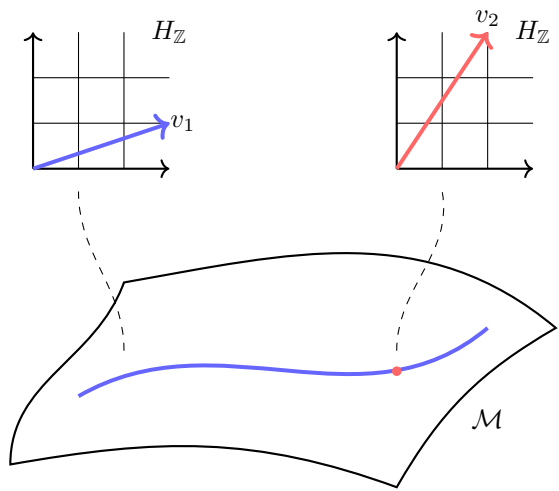
\begin{figure}[t]
    \centering
    \begin{tikzpicture}[scale=0.6]
\draw[->, thick] (1.5, 0) to[out=10, in=160] (10, -0.5) to [out=60, in=210] (13.5, 3)
    to[out=140,in=10] (4,4) to[out=250, in=90] cycle;

\draw[line width=1.5pt, color=blue!60] (3,1.5) to[in=220, out=30] (12,3);

\fill[, color=red!60] (10,2.05) circle[radius=3pt];

\draw[dashed] (4,2.5) to[in=270, out=90] (3,6);

\draw[dashed] (10,2.5) to[in=280,out=90] (11,6);

\draw[->,thick] (10,6.5) -- (10,9.5);
\draw[->,thick] (10,6.5) -- (13,6.5);
\draw[thin] (11,6.5) -- (11,9.5);
\draw[thin] (12,6.5) -- (12,9.5);
\draw[thin] (10,7.5) -- (13,7.5);
\draw[thin] (10,8.5) -- (13,8.5);
\draw[->, line width=1.5pt, color=red!60] (10,6.5) -- (12,9.5);
\node at (12,9.8) {$v_2$};

\draw[->,thick] (2,6.5) -- (5,6.5);
\draw[->,thick] (2,6.5) -- (2,9.5);
\draw[thin] (3,6.5) -- (3,9.5);
\draw[thin] (4,6.5) -- (4,9.5);
\draw[thin] (2,7.5) -- (5,7.5);
\draw[thin] (2,8.5) -- (5,8.5);
\draw[->, line width=1.5pt, color=blue!60] (2,6.5) -- (5,7.5);
\node at (5.3,7.5) {$v_1$};

\node at (5,9.5) {$H_{\mathbb{Z}}$};
\node at (13,9.5) {$H_{\mathbb{Z}}$};

\node at (12,1) {$\mathcal{M}$};
\end{tikzpicture}
    \caption{Schematic illustration of the Hodge bundle, in which the flux lattice $H_{\mathbb{Z}}$ is fibered over the moduli space $\mathcal{M}$. In blue and in red we have depicted two components of the vacuum locus with different dimensionality, corresponding to two choices of flux $v_1$ and $v_2$, respectively.}
    \label{fig:Hodge_bundle}
\end{figure}
 
\noindent In figure \ref{fig:Hodge_bundle} we have illustrated two possible components of the locus of Hodge classes to exemplify the statement of Conjecture \ref{conjecture:tadpole}, for the case of a two-dimensional flux lattice $H_{\mathbb{Z}}$ and a real two-dimensional moduli space $\mathcal{M}$, so that $\mathrm{dim}\,E=4$. Suppose that $v_1$ is a choice of flux with sufficiently small tadpole $L_1$ so that Conjecture \ref{conjecture:tadpole} applies. Then the  vacuum locus corresponding to $v_1$ inside the full Hodge bundle may, for example, be a one-dimensional curve. Thus, this component of $E_{\mathrm{Hodge}}(L_1)$ has positive dimension. If instead $v_2$ is another choice of flux, with tadpole $L_2$, for which the corresponding vacuum is simply a point, then $L_2$ must be sufficiently large, in particular $L_2>L_1$. Note that it could additionally happen that this point lies on the component of the vacuum locus corresponding to $v_1$, as indicated in figure \ref{fig:Hodge_bundle}. In order to disentangle the two components, one should always consider the vacuum locus within the full Hodge bundle. To conclude, we believe a positive or negative answer to Conjecture \ref{conjecture:tadpole} would be an important step towards proving or disproving the tadpole conjecture. Although the conjecture remains rather speculative, it is conceivable that at least for Hodge classes a definite answer can be given in the near future.

\begin{subappendices}
	
\section{Basics of Hodge tensors}\label{app:Hodge-tensors}
In this section we introduce some basic concepts regarding so-called Hodge tensors, as well as the corresponding locus of Hodge tensors, see e.g.~\cite{Moonen:2004,Filippini_2015} for further details and references. We recall that these notions play an important role in the formulation of conjecture \ref{conjecture:scaling_Hodge_vacua}.\\

\noindent Let $H_{\mathbb{Z}}$ be a free abelian group of finite rank, and suppose that 
\begin{equation}\label{eq:Hodge_structure}
	H_{\mathbb{C}} = \bigoplus_{p+q=D} H^{p,q}\,,
\end{equation}
is a Hodge structure of weight $D$ on $H_{\mathbb{Z}}$. There are three basic operations one can perform to construct new Hodge structures from \eqref{eq:Hodge_structure}.
\begin{itemize}
	\item \textbf{Direct sum:}\\
	Given two Hodge structures $\left(H_{\mathbb{Z}},H^{p,q}\right)$ and $\left(H'_{\mathbb{Z}}, H'^{p,q}\right)$ of the same weight $D$, the direct sum of the two lattices
	\begin{equation}
		H_{\mathbb{Z}}\oplus H'_{\mathbb{Z}}\,,
	\end{equation}
	carries a Hodge structure of weight $D$ given by
	\begin{equation}
		\left(H\oplus H'\right)^{p,q}:=H^{p,q}\oplus H'^{p,q}\,,\qquad p+q=D\,.
	\end{equation}
	\item \textbf{Dual:}\\
	The dual lattice 
	\begin{equation}        H^\vee_{\mathbb{Z}}:=\mathrm{Hom}\left(H_{\mathbb{Z}},\mathbb{Z}\right)\,,
	\end{equation}
	consisting of homomorphisms from the lattice $H_{\mathbb{Z}}$ to $\mathbb{Z}$, carries a Hodge structure of weight $-D$ given by
	\begin{equation}
		\left(H^\vee\right)^{p,q}:= \left(H^{-p,-q}\right)^\vee = \mathrm{Hom}\left(H^{-p,-q},\mathbb{C}\right)\,,\qquad p+q=-D\,.
	\end{equation}
	\item \textbf{Tensor product:}\\
	Given two Hodge structures $\left(H_{\mathbb{Z}},H^{p,q}\right)$ and $\left(H'_{\mathbb{Z}}, H'^{p,q}\right)$ of weights $D$ and $D'$, respectively, the tensor product
	\begin{equation}
		H_{\mathbb{Z}}\otimes H '_{\mathbb{Z}}\,,
	\end{equation}
	naturally carries a Hodge structure of weight $D+D'$ given by
	\begin{equation}
		\left(H\otimes H'\right)^{p'',q''}= \bigoplus_{\substack{p+p'=p''\\ q+q'=q''}} H^{p,q}\otimes H'^{p',q'}\,,\qquad p''+q''=D+D'\,.
	\end{equation}
\end{itemize}
Combining the last two operations one finds that the space 
\begin{equation*}
	H^{\otimes a}\otimes \left(H^\vee\right)^{\otimes b}:=\underbrace{H_{\mathbb{Z}}\otimes\cdots\otimes H_{\mathbb{Z}}}_{\text{$a$ copies}}\otimes \underbrace{H^\vee_{\mathbb{Z}}\otimes\cdots\otimes H^\vee_{\mathbb{Z}}}_{\text{$b$ copies}}
\end{equation*}
of $(a,b)$-tensors on $H_{\mathbb{Z}}$ carries a Hodge structure of weight $(a-b)D$. Note that different values of $a,b$ can give rise to a Hodge structure of the same weight. In particular, for each $k\in\mathbb{Z}$, we can apply the first operation and collect all these Hodge structures into one direct sum
\begin{equation}
	\bigoplus_{\substack{a,b\in\mathbb{N}\\a-b=k}}H^{\otimes a}\otimes \left(H^\vee\right)^{\otimes b}\,,
\end{equation}
which carries a Hodge structure of weight $kD$. Finally, one can consider the formal direct product of all these Hodge structures by summing over $k$, which results in the space
\begin{equation}\label{eq:complete_tensor_product}
	H^{\otimes} := \bigoplus_{k=-\infty}^\infty \bigoplus_{\substack{a,b\in\mathbb{N}\\a-b=k}}H^{\otimes a}\otimes \left(H^\vee\right)^{\otimes b}\,.
\end{equation}
Finally, we come to central object we wish to study.

\begin{subbox}{Definition: Hodge tensor}
	A \textbf{Hodge tensor} is a Hodge class in $H^{\otimes}$. More precisely, a type $(p,p)$ Hodge tensor is a Hodge class in the component of $H^{\otimes}$ that carries a Hodge structure of weight $2p$, i.e.~in the component
	\begin{equation}\label{eq:Hodge_tensor_component}
		\bigoplus_{\substack{a,b\in\mathbb{N}\\(a-b)D=2p}} H^{\otimes a}\otimes \left(H^\vee\right)^{\otimes b}\,.
	\end{equation}
\end{subbox}

\subsubsection*{Examples}
\begin{itemize}
	\item \textbf{Type $\left(\frac{D}{2},\frac{D}{2}\right)$ Hodge tensors}\\
	In the following we assume that $D$ is even. Setting $p=D/2$ in \eqref{eq:Hodge_tensor_component} and working through the definitions, we are searching for $(p,p)$ classes in
	\begin{align}
		\bigoplus_{\substack{a,b\in\mathbb{N}\\a-b=1}} H^{\otimes a}\otimes \left(H^\vee\right)^{\otimes b}&= H\oplus\left[ H\otimes H\otimes H^\vee\right]\oplus\cdots\,.
	\end{align}
	Focusing on the first term on the right-hand side, we see that an example of a $\left(\frac{D}{2},\frac{D}{2}\right)$ tensor is simply a Hodge class in the original Hodge structure $H^{p,q}$. However, the notion of a $\left(\frac{D}{2},\frac{D}{2}\right)$ tensor is more general, as it can also arise from the other summands. As an example, one can also construct a $(p,p)$ tensor via
	\begin{equation}
		\left[ H\otimes H\otimes H^\vee\right]^{p,p} = \left[H^{p+1,p-1}\otimes H^{p-2,p+2}\otimes \left(H^\vee\right)^{-p+1,-p-1}\right]\oplus\cdots\,,
	\end{equation}
	where we have just chosen one of the terms that could appear to illustrate the resulting structure. 
	\item \textbf{Type (0,0) Hodge tensors}\\
	Another illuminating example is given by considering type $(0,0)$ Hodge tensors. Setting $p=0$ in \eqref{eq:Hodge_tensor_component} and working through the definitions, we are searching for $(0,0)$ classes in
	\begin{align}
		\bigoplus_{\substack{a,b\in\mathbb{N}\\a-b=0}} H^{\otimes a}\otimes \left(H^\vee\right)^{\otimes b}&= \left[H\otimes H^\vee\right]\oplus\left[ H\otimes H\otimes H^\vee\otimes H^\vee\right]\oplus\cdots\,.
	\end{align}
	Let us focus on the first term, for which we find
	\begin{align}
		\left(H\otimes H^\vee\right)^{0,0}& = \bigoplus_{\substack{p+p^\vee=0\\q+q^\vee=0}} H^{p,q}\otimes \left(H^\vee\right)^{p^\vee, q^\vee}\,,\\
		&=\bigoplus_{p,q} H^{p,q}\otimes \left(H^{p,q}\right)^{\vee}\,,\\
		&\cong \bigoplus_{p,q} \mathrm{End}\left(H^{p,q}, H^{p,q}\right)\,.
	\end{align}
	In other words, such a type $(0,0)$ Hodge tensor can be interpreted as an endomorphism of the various $H^{p,q}$ spaces, i.e.~a map which preserves the original Hodge structure we started with.
\end{itemize}

\subsubsection*{The locus of Hodge tensors}
The above considerations naturally generalize to the setting of variations of Hodge structure, where the Hodge decomposition varies over a moduli space $\mathcal{M}$. Recall that, given an integral class $v\in H_{\mathbb{Z}}$, it is a non-trivial condition on the moduli whether $v$ is a Hodge class (which may or may not have a solution). Similarly, one might ask which points in the moduli space admit Hodge tensors. To be precise, one should consider the locus of points $z\in\mathcal{M}$ for which the Hodge structure admits more Hodge tensors than the general fibre, see for example \cite{Moonen:2004} for further details. This locus will be referred to as the locus of Hodge tensors. Note that, by the first example discussed above, the locus of Hodge tensors contains the locus of Hodge classes, recall also \eqref{eq:Hodge-locus}.

\section{Examples of boundary Hodge norms}
\label{app:Hodge_norms}
In this section we present two examples of boundary Hodge norms. For simplicity, we restrict to the case of Calabi--Yau threefolds having a single complex structure modulus, and consider the large complex structure point and the conifold point, which have been discussed in appendix \ref{sec:asymp_Hodge_examples} based on the results of \cite{Bastian:2023shf}, to which we refer the reader for further details. 

\subsubsection*{Type $\mathrm{IV}_1$: LCS point}
Using the result for the boundary charge operator $Q_\infty$ for the type $\mathrm{IV}_1$ singularity, see equation \eqref{eq:charge-operator_IV1}, one immediately finds that the boundary Hodge inner product can be represented by the matrix
\begin{equation}
    S\cdot C_\infty = \left(
\begin{array}{cccc}
 \frac{2 \hat{c}_2^2}{\kappa }+\frac{\kappa }{6} & -\frac{2 \hat{c}_2 \sigma }{\kappa } & -\frac{2 \hat{c}_2}{\kappa } & 0 \\
 -\frac{2 \hat{c}_2 \sigma }{\kappa } &\frac{12 \hat{c}_2^2+\kappa ^2+4 \sigma ^2}{2 \kappa } & \frac{2 \sigma }{\kappa } & -\frac{6
   \hat{c}_2}{\kappa } \\
 -\frac{2 \hat{c}_2}{\kappa } & \frac{2 \sigma }{\kappa } & \frac{2}{\kappa } & 0 \\
 0 & -\frac{6 \hat{c}_2}{\kappa } & 0 & \frac{6}{\kappa } \\
\end{array}
\right)\,,\qquad \hat{c}_2=\frac{c_2}{24}\,,
\end{equation}
where we have made the identification \eqref{eq:LCS_mirror-data}. Here we recall that $S$ denotes the symplectic pairing and $C_\infty$ is the Weil operator associated to the boundary Hodge structure. It is instructive to compute the boundary Hodge norm of the various weight-components of an integral flux. Let $G_3=(g_1,g_2,g_3,g_4)\in H_{\mathbb{Z}}$. Since we are working in an integral basis, this means that $g_1,\ldots, g_4\in\mathbb{Z}$. For the $\mathrm{IV}_1$ singularity, the only possible weights are $\ell=3,1,-1,-3$. By projecting $G_3$ on the various weight eigenspaces of $N^0$, recall equation \eqref{eq:sl2-triple_IV1}, one straightforwardly computes
\begin{align}
    ||(G_3)_{3}||^2_\infty &= \frac{\kappa}{6}g_1^2\,,\\
    ||(G_3)_{1}||^2_\infty &= \frac{\kappa}{2}g_2^2\,,\\
    ||(G_3)_{-1}||^2_\infty &= \frac{2
    }{\kappa}\left(\hat{c}_2 g_1-g_3+\sigma g_2\right)^2\,,\\
    ||(G_3)_{-3}||^2_\infty &= \frac{6}{\kappa}\left(\hat{c}_2 g_2-g_4\right)^2\,.
\end{align}
One sees that, for example, the boundary Hodge norm of a non-zero $(G_3)_3$ is bounded from below by $\kappa/6$. On the other hand, for the $\ell=-1,-3$ components the exact bound will depend on the values of the coefficients $c_2$ and $\sigma$.

\subsubsection*{Type $\mathrm{I}_1$: conifold point}
As for the case of the LCS point, one can write down the boundary Hodge norm associated to the type $\mathrm{I}_1$ limiting mixed Hodge structure using the result \eqref{eq:I1-Q}. This yields
\begin{equation}
    S\cdot C_\infty = \left(
\begin{array}{cccc}
 \frac{|\tau|^2}{\tau_2}+\frac{\delta ^2}{k} & \frac{\delta  \tau_1-\gamma  |\tau|^2}{\tau_2} & -\frac{\delta }{k} &
   -\frac{\tau_1}{\tau_2}-\frac{\gamma  \delta }{k} \\
 \frac{\delta \tau_1-\gamma  |\tau|^2}{\tau_2} & \frac{\gamma ^2 |\tau|^2+\delta ^2+k\tau_2-2 \gamma  \delta 
   \tau_1}{\tau_2} & 0 & \frac{\gamma  \tau_1-\delta }{\tau_2} \\
 -\frac{\delta }{k} & 0 & \frac{1}{k} & \frac{\gamma }{k} \\
 -\frac{\tau_1}{\tau_2}-\frac{\gamma  \delta }{k} & \frac{\gamma \tau_1-\delta }{\tau_2} & \frac{\gamma }{k} &
   \frac{1}{\tau_2}+\frac{\gamma ^2}{k} \\
\end{array}
\right)
\end{equation}
Let $G_3=(g_1,g_2,g_3,g_4)\in H_{\mathbb{Z}}$ again be an integral flux, with $g_1,\ldots, g_4\in\mathbb{Z}$. For the $\mathrm{I}_1$ singularity, the only possible weights are $\ell=1,0,-1$. By projecting $G_3$ on the various weight eigenspaces of $N^0$, recall \eqref{eq:I1-sl2}, one straightforwardly computes
\begin{align}
    ||(G_3)_{1}||^2_\infty &= k g_2^2\,,\\
    ||(G_3)_{0}||^2_\infty &=\frac{1}{\tau_2}\left|g_4-\delta g_2 -(g_1-\gamma g_2)\tau\right|^2\,,\\
    ||(G_3)_{-1}||^2_\infty &=\frac{1}{k}\left(g_3+\gamma g_4-\delta g_1\right)^2
\end{align}
For example, we see that if the $\ell=1$ component of $G_3$ is non-zero, then its boundary Hodge norm is bounded by $k$ (since $g_2^2\geq 1$). For the other components the exact bound will depend on the details of the geometry, namely the values of $\gamma,\delta,\tau$. To avoid possible confusion, we stress that for example
\begin{equation}
    (G_3)_{-1} = (0,0,g_3+\gamma g_4-\delta g_1,0)\,,
\end{equation}
so that indeed $||(G_3)_{-1}||_\infty=0$ if and only if $(G_3)_{-1}=0$. 

\section{Finiteness of Hodge vacua}
\label{app:Hodge_locus}

In this section we will describe the proof of Theorem \ref{thm:finiteness_Hodge_loci_local} in some detail. In contrast to the proof of Theorem \ref{thm:finiteness_selfdual_nilpotent}, which relied heavily on the nilpotent orbit expansion, the proof discussed here relies more on understanding in which space a sequence of Hodge classes ends up when approaching the boundary. Hence, we let
\begin{equation}
        \left(t^i(n), v(n)\right)\in \mathbb{H}^m\times H_{\mathbb{Z}}(L)\,,
\end{equation}
be a sequence of points such that $\mathrm{Re}\,t^i(n)$ is bounded and $\mathrm{Im}\,t^i(n)\rightarrow\infty$ as $n\rightarrow\infty$.
Furthermore, we assume that
\begin{equation}
        v(n)\in  H^{k,k}\,,\qquad D=2k\,,
    \end{equation}
is a sequence of Hodge classes. Our goal is to show that $v(n)$ can only take on finitely many values. To this end, it suffices to show that there exists a constant subsequence of $v(n)$. For this reason, we may and will freely pass to a subsequence of $v(n)$ whenever possible, without changing the notation to avoid unnecessary cluttering. To ease the reader into the proof, we start by considering the simpler one-variable case ($m=1$). Afterward, we explain how one may inductively apply the one-variable result to obtain the general result. 
\newpage

\subsection{One variable}
\label{subsubsec:proof_Hodge_locus_single}
The proof will be divided into five main steps. 

\subsubsection*{Step 1: Incorporating exponential corrections}
It will be helpful to effectively reduce the problem to the case where the variation of Hodge structure in question is a nilpotent orbit. This can be done as follows. As explained in \cite{CDK,schnell2014extended}, one may assume, without loss of generality, that the flux is parametrized as
\begin{equation}
    v(n) = v_{\mathrm{nil}}(n)+v_{\mathrm{inst}}(n)\,,
\end{equation}
where $v_{\mathrm{nil}}(n)\in F^k_{\mathrm{nil}}$ is a sequence of Hodge classes with respect to a nilpotent orbit $F_\mathrm{nil}$, while $v_{\mathrm{inst}}(n)$ is a series of exponentially small corrections, satisfying
\begin{equation}
    \frac{||v_{\mathrm{inst}}(n)||_\infty}{||v(n)||_\infty}\sim e^{-\alpha y(n)}\,,
\end{equation}
for some constant $\alpha$. 

\subsubsection*{Step 2: Boundedness and $\mathfrak{sl}(2)$-weights}
The start of the proof is identical to the discussion in section \ref{subsec:self-dual_proof}. Indeed, we recall the important result that the self-duality condition and the tadpole cancellation condition together imply that the Hodge norm $||v(n)||$ is bounded. Consequently, the relation \begin{equation}\label{eq:Fsharp_F_relation_limit}
	\lim_{n\rightarrow\infty} e(n)F^p(t(n)) = F_\infty^p\,,
\end{equation}
implies that also the boundary Hodge norm $||e(n)v(n)||_\infty$ is bounded. In the one-variable case, this means that
\begin{equation}
    \sum_{\ell}y^\ell \,||\hat{v}_\ell(n)||^2_\infty < L\,,\qquad \hat{v}(n) = e^{-x(n)N}v(n)\,.
\end{equation}
As discussed in section \ref{subsec:self-dual_locus_example}, this implies that $\hat{v}(n)$, for $n$ sufficiently large, can only have non-zero weight components for $\ell\leq 0$. This means that, after passing to a subsequence, the sequence $\hat{v}(n)$ lies inside the $W_{2k}$ component of the weight filtration induced by the single monodromy operator $N$ via \eqref{eq:def-weight_filtration}. Additionally, note that the $\ell=0$ component $\hat{v}_0(n)$ can only take finitely many values, due to the quantization condition. Therefore, after passing to another subsequence, we may write
\begin{equation}
\label{eq:hatv_limit}
    \hat{v}(n) \equiv \hat{v}_0\quad \mathrm{mod}\,W_{2k-1}\,,
\end{equation}
where $\hat{v}_0$ is a constant. Intuitively, one can think of the $\hat{v}_0$ component of $\hat{v}(n)$ as the part of $\hat{v}(n)$ that remains in the limit $n\rightarrow\infty$. The next step of the proof amounts to showing that this component is rather restricted, owing to the fact that $\hat{v}(n)$ is a sequence of Hodge classes.

\subsubsection*{Step 3: Show that $N\hat{v}_0=0$}
For the moment, let us denote
\begin{equation}
    w = \lim_{n\rightarrow\infty} e(n)v(n)=\lim_{n\rightarrow\infty} e(n)v_{\mathrm{nil}}(n)\,,
\end{equation}
where the second relation follows from the fact that the contribution from $v_{\mathrm{inst}}(n)$ is sub-leading. Recalling the relation \eqref{eq:Fsharp_F_relation_limit} and using the fact that $v_{\mathrm{nil}}(n)\in F_{\mathrm{nil}}^k$, one finds that $w\in F^k_\infty$. At the same time, since $v(n)\in W_{2k}$ and $v(n)$ is real, the same is true for $w$. This is because $e(n)$ is a real operator that does not change the weights. In other words, we have
\begin{equation}
    w\in F^k_\infty\cap W_{2k}\cap H_{\mathbb{R}}\,.
\end{equation}
Such elements are very restricted, as captured by the following
\begin{lem}[{{\cite[Lemma 4.4]{CDK}}}]
\label{lem:Fsharp_W0}
    If $w\in F^k_{\infty}\cap W_{2k}(N)\cap H_{\mathbb{R}}$, then $N^0 w = Nw=0$.
\end{lem}
\begin{proof}
We follow the proof of Schnell, see Lemma 12.4 of \cite{schnell2014extended}. We proceed in two steps:
\begin{enumerate}
\item First, we show that $Nw=0$. Since $N$ acts on the weight filtration as $N W_{2k}\subseteq W_{2k-2}$ and $F_\infty = e^{iN}\tilde{F}$, with $\tilde{F}$ the limiting filtration, one finds that
\begin{equation}
    e^{-iN}w\in \tilde{F}^k\cap W_{2k}(N)\,.
\end{equation}
In particular, it follows that both $w$ and $Nw$ lie inside $\tilde{F}^{k}\cap W_{2k}(N)$. At the same time, however, noting that $N \tilde{F}^{k}\subseteq \tilde{F}^{k-1}$, it follows that $Nw\in \tilde{F}^{k-1}\cap W_{2k-2}(N)$. Combining these results, together with the fact that $w$ and $N$ are real, gives the condition
\begin{equation}
    Nw \in \tilde{F}^{k}\cap W_{2k-2}(N)\cap H_{\mathbb{R}}\,.
\end{equation}
Using the fact that $(W,\tilde{F})$ defines a mixed Hodge structure, the space on the right-hand side is empty, hence $Nw=0$. To give some feeling for this property, we have illustrated the relevant spaces in the case of a weight $D=4$ limiting mixed Hodge structure (so $k=2$) in figure \ref{fig:Deligne1}.
\item Next, we show that $N^0w=0$. This follows straightforwardly from the fact that
\begin{equation}
    W_{2k}\cap \tilde{F}^k\cap \overline{\tilde{F}^k}= \tilde{I}^{k,k}\,,
\end{equation}
where $\tilde{I}^{k,k}$ denotes the $(k,k)$-component of the Deligne splitting associated to the $\mathbb{R}$-split mixed Hodge structure $(W,\tilde{F})$, recall the discussion in section \eqref{sec:MHS}. In particular, using the earlier result that $Nw=0$, one finds that $w\in \tilde{I}^{k,k}$. Using the fact that $N^0$ acts on $\tilde{I}^{p,q}$ as multiplication by $p+q-D$ the result follows. 
\end{enumerate}
\end{proof}
\begin{figure}[h!]
\centering
\scalebox{0.8}{\begin{tikzpicture}[baseline={([yshift=-.5ex]current bounding box.center)},scale=1,cm={cos(45),sin(45),-sin(45),cos(45),(15,0)}]

\filldraw[color=orange!50] (2,0) node[below right]{$W_{2}$} -- (0,2) -- (0,0);
\filldraw[color=blue!30] (4,4) -- (0,4) node[left]{$\tilde{F}^2$}-- (0,2) -- (4,2);

\draw[step = 1, gray, ultra thin] (0, 0) grid (4, 4);

  \foreach \i\j in {0/0, 0/1, 0/2, 0/3, 0/4,
1/0, 1/1, 1/2, 1/3, 1/4,
2/0, 2/1, 2/2, 2/3, 2/4,
3/0, 3/1, 3/2, 3/3, 3/4,
4/0, 4/1, 4/2, 4/3, 4/4}
{
  \draw[fill] (\i,\j) circle[radius=0.04] ;
}

\draw[fill] (2,2) circle[radius=0.04] node[above]{$\tilde{I}^{2,2}$};

\draw[fill, color=red!60] (0,2) circle[radius=0.06] node[left]{$Nw$};
\draw[->,color=red!60, very thick] (0.9,2.9) -- (0.15,2.15); 

\end{tikzpicture}}
\caption{Arrangement of the Deligne splitting $\tilde{I}^{p,q}$ for a weight four limiting mixed Hodge structure $(W,\tilde{F})$. In blue: $\tilde{F}^2$ component of the limiting Hodge filtration. In orange: $W_{2}$ component of the weight filtration. In red: the potential location of $Nw$, with the arrow denoting the action of the log-monodromy matrix $N$. Since complex conjugation acts on the Deligne splitting as reflection in the vertical axis, the vector $Nw$ cannot be real unless it is zero. Following the proof of Lemma \ref{lem:Fsharp_W0}, the only possible location for $w$ is in the space $\tilde{I}^{2,2}$.
\label{fig:Deligne1}}
\end{figure}
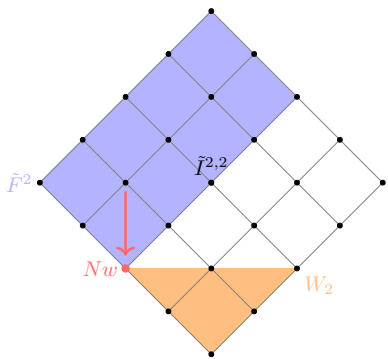

\noindent As a result of Lemma \ref{lem:Fsharp_W0}, we find that 
\begin{equation}
    N^0w = 0\,,\qquad Nw = 0\,.
\end{equation}
One can summarize this result in the statement that $w$ should be a singlet under the $\mathfrak{sl}(2)$ action. Returning to our original sequence $\hat{v}(n)$, recall equation \eqref{eq:hatv_limit}, and projecting the congruence
\begin{equation}
    e(n)\hat{v}(n)\equiv \hat{v}_0\quad \mathrm{mod}\,W_{2k-1} 
\end{equation}
to the eigenspace $E_0$, one finds that indeed $w=\hat{v}_0$ and hence $N\hat{v}_0=0$. Additionally, as mentioned in the proof of Lemma \ref{lem:Fsharp_W0}, the limiting element $\hat{v}_0$ lies inside the space $\tilde{I}^{k,k}$. In other words, it lies exactly in the center of the Deligne diamond. 

\subsubsection*{Step 4: Show that $N\hat{v}(n)=0$}
Having shown that $N\hat{v}_0=0$, the next step is to show that in fact this relation applies to the whole sequence $\hat{v}(n)$, after passing to a subsequence. Indeed, since $N\hat{v}_0=0$ and $N W_{2k-1}\subseteq W_{2k-3}$, it follows that 
\begin{equation}
    N\hat{v}(n)\in W_{2k-3}\,.
\end{equation}
Now, let us write
\begin{equation}
    e(n)N\hat{v}(n)=e(n)N\hat{v}_{\mathrm{nil}}(n)+e(n)N\hat{v}_{\mathrm{inst}}(n)\,.
\end{equation}
Our goal will be to first show that both terms on the right-hand side are in fact exponentially small. Indeed, suppose not, then the ratio
\begin{equation}
    \frac{||e(n)N\hat{v}_{\mathrm{inst}}(n)||_\infty}{||e(n)\hat{v}(n)||_\infty}
\end{equation}
would go to zero as $n\rightarrow\infty$. Consequently, one would find that
\begin{equation}
    \lim_{n\rightarrow\infty}\frac{e(n)N\hat{v}(n)}{||e(n)N\hat{v}(n)||_\infty}=\lim_{n\rightarrow\infty}\frac{e(n)N\hat{v}_{\mathrm{nil}}(n)}{||e(n)N\hat{v}_{\mathrm{nil}}(n)||_\infty}\,,
\end{equation}
and furthermore the resulting limit would give a unit vector in the space $NF^k_\infty\cap W_{2k-3} \cap H_{\mathbb{R}}$. To see the latter, one uses the fact that $e(n)N = y(n)^{-1}Ne(n)$ to swap the order of $e(n)$ and $N$ (note that these are vector space identities, so an overall rescaling is irrelevant) and again applies the relation \eqref{eq:Fsharp_F_relation_limit}. The important observation is that $NF^k_\infty\cap W_{2k-3} \cap H_{\mathbb{R}}=\{0\}$, due to the fact that $(W,F_\infty)$ defines a mixed Hodge structure (recall also that $N F^k_\infty\subseteq F^{k-1}_\infty$). This is again illustrated in figure \ref{fig:Deligne2} for the case $D=4$, or $k=2$. We have therefore arrived at a contradiction and must conclude that in fact $||e(n)\hat{v}(n)||_\infty$ is bounded by a constant multiple of $||e(n)N\hat{v}_{\mathrm{inst}}(n)||_\infty$, in particular it is exponentially small. Since $e(n)$ grows at most polynomially, we conclude that $N\hat{v}(n)$ itself becomes exponentially small as $n\rightarrow\infty$. Since $\hat{v}(n)$ is quantized, this means that for sufficiently large $n$, we must have $N\hat{v}(n)=0$, as desired.

\begin{figure}[t!]
	\centering
	\scalebox{0.8}{\begin{tikzpicture}[baseline={([yshift=-.5ex]current bounding box.center)},scale=1,cm={cos(45),sin(45),-sin(45),cos(45),(15,0)}]
			
			\filldraw[color=orange!50] (1,0) node[below right]{$W_{1}$} -- (0,1) -- (0,0);
			\filldraw[color=blue!30] (4,4) -- (0,4) node[left]{$F_\infty^1$}-- (0,1) -- (4,1);
			
			\draw[step = 1, gray, ultra thin] (0, 0) grid (4, 4);
			
			\foreach \i\j in {0/0, 0/1, 0/2, 0/3, 0/4,
				1/0, 1/1, 1/2, 1/3, 1/4,
				2/0, 2/1, 2/2, 2/3, 2/4,
				3/0, 3/1, 3/2, 3/3, 3/4,
				4/0, 4/1, 4/2, 4/3, 4/4}
			{
				\draw[fill] (\i,\j) circle[radius=0.04] ;
			}
			
			\draw[fill, color=red!60] (0,1) circle[radius=0.06] node[left]{$e(n)N\hat{v}_{\mathrm{nil}}(n)$};
			
	\end{tikzpicture}}
	\caption{Arrangement of the Deligne splitting $I^{p,q}$ for a weight four limiting mixed Hodge structure $(W,F_\infty)$. In blue: $F_\infty^1$ component of the boundary Hodge filtration. In orange: $W_{1}$ component of the weight filtration. In red: the potential location of the limit of the sequence $e(n)N\hat{v}_{\mathrm{nil}}(n)$ as $n\rightarrow\infty$. Since complex conjugation acts on the Deligne splitting as reflection in the vertical axis, the vector $e(n)N\hat{v}_{\mathrm{nil}}(n)$ cannot be real unless it is zero. 
		\label{fig:Deligne2}}
\end{figure}
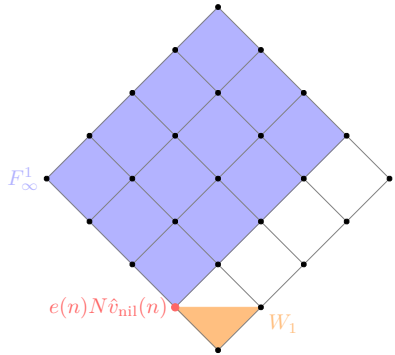 

\subsubsection*{Step 5: $v(n)$ is bounded}
Since $N\hat{v}(n)=0$, it immediately follows that also $Nv(n)=0$. This is a huge simplification, because it effectively allows us to remove the moduli-dependence from the problem. Indeed, recall that the original nilpotent orbit is of the form
\begin{equation}
    F^p_{\mathrm{nil}}(t) = e^{t(n)N}F^p_0\,.
\end{equation}
Now choose some fixed $t_*$ with imaginary part large enough, and note that
\begin{equation}
    v(n) = e^{t_*N-t(n)N}v(n)\in F^k_\mathrm{nil}(t_*)\,.
\end{equation}
In particular, due to the tadpole condition, $v(n)$ is bounded in Hodge norm with respect to the fixed filtration $F^k_{\mathrm{nil}}(t_*)$. Together with the quantization condition this implies that $v(n)$ can take on only finitely many values. This finishes the proof in the one-variable case.

\subsection{Multiple variables}
\label{subsubsec:proof_Hodge_locus_multi}
In this section we present the general multi-variable proof of finiteness of Hodge classes, based on the original work of Cattani, Deligne, and Kaplan \cite{CDK}. We also draw heavily from the formulation of the proof in \cite{schnell2014extended}. The proof is based on an inductive application of the one-variable result discussed in the previous section. To start, we set up the induction and briefly describe in which stage of the proof the induction step is used.  

\subsubsection*{Step 0: Setting up the induction}
The major complicating factor of the multi-variable proof is the fact that there are many possible hierarchies between the saxions $y_i$ that become large as one approaches a boundary of the moduli space. The strategy of \cite{CDK} is to inductively iterate over all such possible hierarchies. Indeed, it is argued that, after passing to a subsequence, one can always parametrize the sequence of saxions as
\begin{equation}
\label{eq:parametrization_saxions}
    y_k(n) = i \sum_{l=1}^d A_{kl}\,\lambda_l(n)+b_k(n)\,,
\end{equation}
where the $A_{kl}$ are real and non-negative constants, comprising the entries of an $m\times d$ matrix, and $b(n)\in\mathbb{R}^m$ is a bounded sequence. Furthermore, the sequence $\lambda(n)\in\mathbb{R}^d$ has the property that
\begin{equation}
    \frac{\lambda_k(n)}{\lambda_{k+1}(n)}\rightarrow\infty\,,\qquad \text{as $n\rightarrow\infty$}\,,
\end{equation}
for all $1\leq k\leq d$, where we put $\lambda_{d+1}(m)=1$. Next, we note that \eqref{eq:parametrization_saxions} leads to the following relation
\begin{equation}
    i\sum_{j=1}^m y_j(n)N_i = i\sum_{k=1}^d \lambda_k(n) M_k+\sum_{j=1}^m b_j(n)N_j\,,
\end{equation}
where the new monodromy matrices $M_k$ are related to the original ones by $M_k = A_{kj}N_j$. Effectively, the integer $1\leq d\leq m$ parametrizes the number of different hierarchies between the saxions. Consequently, the proof will proceed by induction on $d$. To be precise, we will show that the sequence $v(n)$ has a constant subsequence, which we denote by $v$, such that 
\begin{equation}
    M_k v=0\,,\qquad 1\leq k\leq d\,.
\end{equation}
Finally, to avoid cluttering the notation, we will set the axions $x_i$ to zero for the rest of the proof. As in the one-variable case, the axion-dependence can be straightforwardly incorporated using the monodromy matrices.

\subsubsection*{Step 1: Incorporating exponential corrections}
As in the one-variable case, it will be useful to write
\begin{equation}
    v(n) = v_{\mathrm{nil}}(n)+v_{\mathrm{inst}}(n)\,,
\end{equation}
where $v_{\mathrm{nil}}\in F^k_{\mathrm{nil}}$ is a sequence of Hodge classes with respect to a nilpotent orbit $F_{\mathrm{nil}}$, and the sequence $v_{\mathrm{inst}}(n)$ satisfies
\begin{equation}
    \frac{||v_{\mathrm{inst}}(n)||_\infty}{||v(n)||_\infty}\sim e^{-\alpha\, \mathrm{sup}(y_i)}\,,
\end{equation}
for some constant $\alpha$.

\subsubsection*{Step 2: Boundedness and $\mathfrak{sl}(2)$-weights}

Again, as in the one-variable case, the starting point of the proof is the statement that the Hodge norm $||v(n)||$ is bounded, and therefore the boundary Hodge norm $||e(n)v(n)||_\infty$ is bounded. In the one-variable case, one could immediately conclude from the latter that $v(n)$ lies inside $W_{2k}$, meaning that it only has weight components $\ell\leq 0$. In the multi-variable case, it similarly turns out to be true that $v(n)$ lies inside $W^{(1)}_{2k}$, meaning that its weight-components with respect to the first $\mathfrak{sl}(2)$ grading operator $N^0_{(1)}$ must have $\ell_1\leq 0$. However, the proof of this statement is significantly more involved. Below we present the general idea in a number of steps, but refer the reader to \cite{CDK} for the details of some of the statements.
\begin{enumerate}
    \item Suppose $v(n)\in W^{(1)}_{D+\ell_1}$ and define $\tilde{H}_{\mathbb{C}}=\mathrm{Gr}_{D+\ell_1}^{(1)}$, which supports a polarized variation of Hodge structure of weight $D+\ell_1$. Denote by $\tilde{v}(n)$ the projection of $v(n)$ onto $\tilde{H}_{\mathbb{C}}$. Suppose that $\ell_1\geq 1$, then one can show that
    \begin{equation}
    \label{eq:relation_norms_projection}
        ||\tilde{e}(n)\tilde{v}(n)||_\infty^2\leq \lambda_1(n)^{-\ell_1}||e(n)v(n)||_\infty^2\leq  \lambda_2(n)^{-\ell_1}||e(n)v(n)||_\infty^2\,,
    \end{equation}
    where similarly $\tilde{e}(n)$ denotes the projection of the operator $e(n)$ onto $\tilde{H}_{\mathbb{C}}$. Intuitively, the first relation in \eqref{eq:relation_norms_projection} holds because the projection removes the $\lambda_1(n)$ dependence, while the second relation follows from the fact that $\lambda_1(n)> \lambda_2(n)$. 
    \item As a result of \eqref{eq:relation_norms_projection} and the fact that $||e(n)v(n)||_\infty$ is bounded, it follows that the expression 
    \begin{equation}
        \lambda_2(n)^{\ell_1}||\tilde{e}(n)\tilde{v}(n)||^2_\infty
    \end{equation}
    must be bounded as well. We will now argue that the highest weights of $\tilde{v}(n)$ satisfy
    \begin{equation}
    \label{eq:weights_projection}
    \ell_i=0\,,\qquad i>1\,.
    \end{equation} 
    This can be seen as follows. Since the unit vector
    \begin{equation}
        \frac{\tilde{e}(n)\tilde{v}(n)}{||\tilde{e}(n)\tilde{v}(n)||_\infty}
    \end{equation}
    converges to an element in $W^{(2)}_{D+\ell_2}\cap F^k_\infty\cap H_{\mathbb{R}}$ and $(W^{(2)}, F_\infty)$ defines a mixed Hodge structure, one has that $\ell_2\geq 0$. Now one can once more project $\tilde{v}(n)$ onto the graded space $\mathrm{Gr}_{D+\ell_2}^{(2)}$, which carries a polarized variation of Hodge structure of weight $D+\ell_2$. Denoting this projection by $\tilde{v}(n)'$ and noting that
    \begin{align}
        \lambda_3(n)^{\ell_2} ||\tilde{e}(n)' \tilde{v}(n)'||_\infty^2 &\leq \lambda_2(n)^{\ell_2} ||\tilde{e}(n)' \tilde{v}(n)'||_\infty^2\\
        &\leq \lambda_2(n)^{\ell_1}||\tilde{e}(n)\tilde{v}(n)||_\infty^2\\
        &\leq ||e(n)v(n)||^2_\infty\,,
    \end{align}
    one concludes that $\lambda_3(n)^{\ell_2} ||\tilde{e}(n)' \tilde{v}(n)'||_\infty^2$ is bounded. Proceeding by induction, it follows that $\ell_3=\ldots=\ell_d=0$. But then
    \begin{equation}
        \lambda_2(n)^{\ell_1}||\tilde{e}(n)\tilde{v}(n)||_\infty^2\geq \lambda_2(n)^{\ell_2}||\tilde{v}(n)^{(\ell_2,0,\ldots, 0)}||_\infty^2\,,
    \end{equation}
    and boundedness of the left-hand side implies that $\ell_2\leq 0$, hence $\ell_2=0$.     
    \item Returning to the sequence $v(n)$, and applying \eqref{eq:weights_projection}, one finds that
    \begin{equation}
        ||e(n)v(n)||_\infty^2\geq ||e(n)v(n)^{(\ell_1,\ldots, \ell_d)}||_\infty^2 = \left(\frac{\lambda_1(n)}{\lambda_2(n)
        }\right)^{\ell_1}||v(n)^{(\ell_1,0,\ldots, 0)}||_\infty^2\,.
    \end{equation}
    This is in contradiction with the fact that $||e(n)v(n)||_\infty^2$ must remain bounded, hence the assumption that $\ell_1\geq 1$ is false. Therefore, we conclude that $\ell_1\leq 0$.
\end{enumerate}

\subsubsection*{Step 3: Restricting the limit of $v(n)$}
The idea is now to apply the induction hypothesis to projection $\tilde{v}(n)$ of $v(n)$ onto $\tilde{H}_{\mathbb{C}}=\mathrm{Gr}_{D}^{(1)}$, which again supports a variation of Hodge structure of weight $D$. This effectively projects the relation \eqref{eq:parametrization_saxions} to
\begin{equation}
    \tilde{y}_k(n) = \sum_{l=2}^d A_{kl}\tilde{\lambda}_l(n)+\tilde{b}(n)\,.
\end{equation}
This expansion is very similar to \eqref{eq:parametrization_saxions}, but with $d-1$ terms instead of $d$. Therefore, after passing to a subsequence, the induction step implies that $\tilde{v}(n)$ has a constant value $\tilde{h}\in \tilde{H}_{\mathbb{Z}}$ and that $\tilde{M}_k \tilde{h}=0$ for $k=2,\ldots, d$. Lifting this result back to the sequence $v(n)$, we may write
\begin{equation}
    v(n) \equiv v_0\quad\mathrm{mod}\,W_{D-1}^{(1)}\,,
\end{equation}
where 
\begin{equation}
    v_0 = \sum_{\ell_2,\ldots, \ell_d\leq 0} v(n)^{(0,\ell_2,\ldots, \ell_d)}
\end{equation}
is a constant sequence that projects to $\tilde{h}$. The strategy is now similar to the one-variable case, where we have shown that $v_0$ is annihilated by the monodromy operator. In the multi-variable case, the result is not quite as strong, because the element $v_0$ contains multiple $\mathfrak{sl}(2)$-components. Indeed, we will instead show that $M_1 v_0^{(0,\ldots, 0)}=0$. To this end, let us again introduce the limiting vector
\begin{equation}
    w:=\lim_{n\rightarrow\infty}e(n)v(n)\in F^k_\infty\cap W_{2k-1}^{(1)}\cap H_{\mathbb{R}}\,.
\end{equation}
Applying the result of Lemma \ref{lem:Fsharp_W0}, one finds 
\begin{equation}
    N_1^0w = M_1w=0\,.
\end{equation}
Projecting the congruence
\begin{equation}
    e(n)v(n)\equiv e(m)v_0\quad \mathrm{mod}\,W_{2k-1}^{(1)}\,,
\end{equation}
to the weight $\ell_1=0$ eigenspace of $N_{(1)}^0$, we get that indeed $w=v_0^{(0,\ldots, 0)}$ and hence $M_1v_0^{(0,\ldots, 0)}=0$ as desired.

\begin{figure}[t!]
\centering
\scalebox{0.8}{\begin{tikzpicture}[baseline={([yshift=-.5ex]current bounding box.center)},scale=1,cm={cos(45),sin(45),-sin(45),cos(45),(15,0)}]

\filldraw[color=blue!30] (4,4) -- (0,4) node[left]{$F_\infty^1$}-- (0,1) -- (4,1);
\filldraw[color=orange!50] (2,0) node[below right]{$W^{(1)}_{2}$} -- (0,2) -- (0,0);

\draw[step = 1, gray, ultra thin] (0, 0) grid (4, 4);

  \foreach \i\j in {0/0, 0/1, 0/2, 0/3, 0/4,
1/0, 1/1, 1/2, 1/3, 1/4,
2/0, 2/1, 2/2, 2/3, 2/4,
3/0, 3/1, 3/2, 3/3, 3/4,
4/0, 4/1, 4/2, 4/3, 4/4}
{
  \draw[fill] (\i,\j) circle[radius=0.04] ;
}

\draw[fill] (1,1) circle[radius=0.06] node[above]{$u$};
\end{tikzpicture}}
\caption{Arrangement of the Deligne splitting $\tilde{I}^{p,q}_{(1)}$ associated to the mixed Hodge structure $(W^{(1)},\tilde{F}_\infty)$. In blue: $\tilde{F}^1_{(1)}$ component of the $\tilde{F}^p_{(1)}$ filtration. In orange: $W_1^{(1)}$ component of the weight filtration $W_\ell^{(1)}$. The only possible location for the limiting vector $u=\lim_{n\rightarrow\infty} u(n)$ is also indicated. Note that indeed $u$ necessarily has weight $\ell_1=-2$.}
\label{fig:Deligne_diamond_multi_var}
\end{figure}
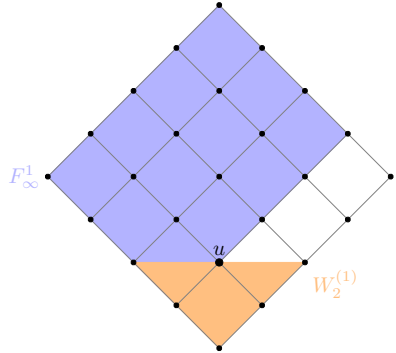

\subsubsection*{Step 4: Showing that $M_1 v(n)=0$}
Having shown that $M_1v_0^{(0,\ldots, 0)}=0$, we next show that in fact $M_1 v(n)=0$, after passing to a subsequence. The argumentation is again analogous to the one-variable case. However, it is important to note that now it is not immediately obvious whether $M_1 v_0=0$. So far we have only shown this for the $v_0^{(0,\ldots, 0)}$ component. As a result, the analysis needs to be slightly modified, but is very similar in spirit. Let us again write
\begin{equation}
    e(n)M_1 v(n)=e(n)M_1 v_{\mathrm{nil}}(n)+e(n)M_1 v_{\mathrm{inst}}(n)\,,
\end{equation}
and suppose that the ratio
\begin{equation}
    \frac{||e(n)M_1 v_{\mathrm{inst}}(n)||_\infty}{||e(n)M_1 v(n)||_\infty}
\end{equation}
goes to zero as $n\rightarrow\infty$. Consequently, one would find that
\begin{equation}
    u(n):=\frac{e(n)M_1v(n)}{||e(n)M_1v(n)||_\infty}\in W^{(1)}_{2k-2}\cap H_{\mathbb{R}}
\end{equation}
defines a sequence of unit vectors (note the appearance of $W_{2k-2}^{(1)}$ as opposed to $W_{2k-3}$ in the one-variable case!), which would converge to a unit vector $u\in F^{k-1}_\infty\cap W^{(1)}_{2k-2}\cap H_{\mathbb{R}}$. Because the index on the weight filtration is increased by one compared to the one-variable case, it is no longer the case that this space is immediately trivial. It is, however, very restricted. Indeed, another application of Lemma \ref{lem:Fsharp_W0} gives that $u$ necessarily has weight $\ell_1=-2$. This is illustrated in figure \ref{fig:Deligne_diamond_multi_var} for the case $D=4$. \\

\noindent At the same time, we may recall that $v(n)\equiv v_0\,\mathrm{mod}\, W^{(1)}_{2k-1}$, hence by projecting onto the $\ell_1=-2$ component, we have
\begin{equation}
    u = \frac{e(n)M_1v_0}{||e(n)M_1v(n)||_\infty}\in W^{(1)}_{2k-2}\cap \cdots \cap W^{(d)}_{2k-2}\,.
\end{equation}
Hence, we find that
\begin{equation}
    u\in F^k_\infty\cap W^{(i)}_{2k-2}\cap H_{\mathbb{R}}\,,
\end{equation}
for every $i=1,\ldots, d$. Yet another application of Lemma \ref{lem:Fsharp_W0} implies that $u$ necessarily has weights $\ell_i=-2$ for all $i=1,\ldots, d$. However, recalling that $M_1$ lowers all the weights by exactly $-2$, it must be that
\begin{equation}
    u(m)^{(-2,\ldots, -2)}=\frac{e(n)M_1 v_0^{(0,\ldots, 0)}}{||e(n)M_1v(n)||_\infty}=0\,.
\end{equation}
Hence, this is in contradiction to the fact that $u$ is a unit vector. Consequently, it must be the case that $||e(n)M_1v(n)||_\infty$ is bounded by a constant multiple of $||e(n)M_1v_{\mathrm{inst}}(n)||_\infty$. In particular, the former is also exponentially small. Since $e(n)$ grows at most polynomially, this implies that $M_1v(n)$ is exponentially small. Due to the quantization condition, we may therefore assume that $M_1v(n)=0$ after passing to a subsequence. 

\subsubsection*{Step 5: Finishing the proof}
Having argued that $M_1 v(n)=0$ the final step of the proof proceeds in the same spirit as the one-variable case. Indeed, in the one-variable case this relation allowed us to completely remove the moduli dependence of the problem. In the multi-variable case, it instead allows us to remove one of the moduli from the problem. Indeed, since the sequence of filtrations
\begin{equation}
    e^{-i\lambda_1(n)M_1} F^p_{\mathrm{nil}}
\end{equation}
no longer depends on $\lambda_1(n)$, we have effectively reduced the value of $d$ to $d-1$. By induction, we may pass to a subsequence of $v(n)$ which is constant and lies in the kernel of $M_2\,\ldots, M_d$. Since we already have $M_1 v(n)=0$ we have indeed shown that $M_k v=0$ for all $1\leq k \leq d$. Finally, when $d=1$ one may simply apply the one-variable proof. This concludes the proof of the multi-variable case. 
\end{subappendices}

\setpartpreamble[u][\textwidth]{
	\vspace*{1cm}
	\hrulefill 
	\vspace*{0.5cm}
	
	In this part of the thesis we describe how the tools of (asymptotic) Hodge theory also find applications in the context of integrable systems. More precisely, we will be concerned with the classical dynamics of two classes of two-dimensional integrable non-linear $\sigma$-models, and argue for an intricate underlying connection to the defining equations of variations of Hodge structure and their asymptotic behaviour. \\
	
	\noindent In chapter \ref{chap:WZW} we provide some general motivation for studying this connection, which will also apply to the subsequent chapter, and show how the horizontality conditions of the period map can be formulated in terms of the equations of motion of the \textbf{$\lambda$-deformed gauged Wess--Zumino--Witten model}. Physically, this can be understood as a reformulation of the mathematics of variations of Hodge structures in terms of the classical dynamics of a string which propagates on the corresponding classifying space. \\
	
	\noindent In chapter \ref{chap:biYB} we instead use results from asymptotic Hodge theory, in particular the Sl(2)-orbit theorem, to generalize known results on classical solutions of another class of integrable models. More precisely, we consider the so-called \textbf{critical bi-Yang--Baxter model} and show that the asymptotic form of the Weil operator, as described by an Sl(2)-orbit, provides a solution for any target space group that admits a horizontal $\mathfrak{sl}(2)$-triple.

	\vspace*{0.5cm}
	\hrulefill }

\part{Applications in Integrable Systems}\label{part4}
\chapter{Hodge theory and $\lambda$-deformed WZW models}
\label{chap:WZW}
\epigraph{This chapter is based on: Thomas W. Grimm, Jeroen Monnee: \emph{Deformed WZW models and Hodge theory. Part I}, \textbf{JHEP 05 (2022) 103},  \href{https://arxiv.org/abs/2112.00031}{\textbf{[arXiv: 2112.00031]}}}

\noindent In this chapter we discuss a connection between the abstract mathematical machinery of (asymptotic) Hodge theory , discussed in part \ref{part2}, and the intriguing landscape of integrable non-linear $\sigma$-models. As we have already seen in chapter \ref{chap:string}, integrable non-linear $\sigma$-models have played a major role across numerous branches of physics, ranging from string theory to condensed matter physics. Such models possess a large amount of symmetries and are constrained to the extent that they are often solvable, for example, via the Bethe ansatz. Nevertheless, despite these constraints there has emerged a vast and intricate network of different integrable models, many of which are obtained by suitably deforming known models such as the Wess-Zumino-Witten (WZW) model and its gauged extension \cite{Wess:1971,Witten:1983_current_algebra,Witten:1983_bosonization,Gawedzki:1988_I,Gawedzki:1988_II,Witten:1991}, as well as the principal chiral model. At the same time, it has recently been suggested to study non-linear $\sigma$-models in the context of string moduli spaces \cite{Grimm:2020cda,Grimm:2021ikg}, see also \cite{Cecotti:2020uek} for related ideas. These works propose to use an auxiliary field theory on the moduli space to provide a physical reformulation of the intricate mathematical structures that arise, as described by (asymptotic) Hodge theory. The purpose of this chapter, as well as the next chapter, is to further develop a possible relationship between integrable models and the physics of string compactifications via variations of Hodge structures. \\

\noindent In section \ref{sec:lambda-motivation} we discuss some of the initial motivation for studying this connection. Then, in section \ref{sec:lambda_WZW} we present a basic review of the class of integrable models we will consider in this chapter: the gauged WZW model and its $\lambda$-deformation. Finally, in section \ref{sec:strings_Hodge_theory} the connection between the $\lambda$-deformed $G/G$ model and the notion of a variation of Hodge structure is made precise, by showing that the Weil operator of a variation of Hodge structure realizes a particular solution to the equations of motion of the $\lambda$-deformed $G/G$ model. We also make some additional comments regarding integrability and provide an in-depth analysis of more general solutions at the end of the section. Finally, there are three appendices which contain some computational details.

\section{Motivation}\label{sec:lambda-motivation}

\noindent It was suggested in  \cite{Grimm:2020cda,Cecotti:2020uek,Grimm:2021ikg} that one might be able to describe the properties of a variation of Hodge structure, in particular the horizontality of the period map, in terms of an auxiliary field theory that is defined on the moduli space. In other words, in this field theory the spacetime of the model is identified with the field space of the effective theory obtained from string compactification. Let us elaborate on this. 

\subsubsection*{An action principle for horizontality}
In the following we restrict to one-parameter variations of Hodge structure. Let us recall some results from appendix \ref{subsec:Nahm}, where it was argued, based on earlier results of \cite{Donaldson:1984} and \cite{Grimm:2020cda,Cecotti:2020uek}, that one can encode Nahm's equations \eqref{eq:Nahm_N} in terms of the following action principle for the period map $h$.
	\begin{equation}\label{eq:action_Nahm2}
		S_{\mathrm{Nahm}}[h] = \frac{1}{4}\int_{\mathcal{M}}\mathrm{Tr}\left|h^{-1}\mathrm{d}h+\left(h^{-1}\mathrm{d}h\right)^\dagger \right|^2\,,
	\end{equation}
\noindent Indeed, by computing the variation of $S_{\mathrm{Nahm}}[h]$ with respect to $h$, one readily shows that the resulting equations of motion are equivalent to Nahm's equations \eqref{eq:Nahm_N}, as done explicitly in appendix \ref{subsec:Nahm}. However, there are two shortcomings with this particular approach. 
\begin{itemize}
	\item \textbf{Shortcoming (1): Nilpotent orbit approximation}\\
	First, one only recovers Nahm's equations from the action \eqref{eq:action_Nahm2} if one additionally imposes the condition that $\partial_x(h^{-1}\partial_xh) = \partial_x(h^{-1}\partial_y h)=0$, which, in particular, is satisfied in the nilpotent orbit approximation but may not be satisfied beyond this approximation. 
	\item \textbf{Shortcoming (2): Nahm's equations vs.~horizontality}\\
	Secondly, while Nahm's equations play an essential role in describing the asymptotic form of the period map, as discussed at length in chapter \ref{chap:asymp_Hodge_II}, they do not comprise the full set of constraints coming from the horizontality of the period map. Thus, the action \eqref{eq:action_Nahm2} does not properly describe the full dynamics of the period map. 
\end{itemize}
The upshot is that it would be desirable to formulate an action principle which encodes the full horizontality conditions of the period map, which moreover does not rely on the nilpotent orbit approximation. There still remains much to understand about the precise formulation of the associated non-linear $\sigma$-model. For example, restricting to two-dimensional field spaces, one might expect the model to be among the plethora of two-dimensional \textit{integrable} field theories that have been studied extensively over the last decennia. This expectation stems from the well-known results of \cite{Cecotti:1991me}, in which it is shown that the defining equations of variations of Hodge structure can be written in the form of $tt^*$ equations, highlighting an underlying integrable structure. It is the main purpose of this chapter to identify the appropriate field theory as an integrable deformation of the gauged WZW model known as the $\lambda$-deformation. This results in the exciting possibility to use techniques of integrability to study the field spaces of string compactifications and, conversely, to use existing methods of %e.g. asymptotic 
Hodge theory and the study of period mappings to obtain explicit solutions to integrable field theories. 

\subsubsection*{Moduli space holography}
Another intriguing motivation for studying the aforementioned auxiliary field theory on the moduli space is to better understand a proposed principle of \textbf{moduli space holography} \cite{Grimm:2020cda}. Originally, this proposed principle is built on the expectation -- a variant of the Swampland distance conjecture \cite{Ooguri2007} -- that along any infinite distance limit in the moduli space a global symmetry arises \cite{Banks:1988yz,Banks:2010zn}, whose presence dictates much of the asymptotic structure of the physical couplings of the effective theory, see for example \cite{Grimm:2018ohb,Grimm:2018cpv,Corvilain:2018lgw,Grimm:2019wtx,Grimm:2020cda,Grimm:2022sbl}. The latter mechanism has been illustrated explicitly in the bulk reconstruction procedure discussed in chapter \ref{chap:asymp_Hodge_II}. It has furthermore been suggested that the abstract boundary data coming from asymptotic Hodge theory should be described by an actual theory living \textit{on} the boundary of the moduli space. It is then an interesting question whether this notion of holography can be understood in terms of more conventional approaches to holography, and what exactly this boundary theory should be. That this may be the case is further corroborated by the observation that the physical metric on the moduli space, i.e.~the Weil--Petersson metric, always asymptotes to the Poincar\'e metric near any infinite distance boundary \cite{Wang97onthe}. The latter can be understood as a patch of Euclidean $\mathrm{AdS}_2$, whose isometry algebra is intricately related to the emergent $\mathfrak{sl}(2,\mathbb{R})$-symmetry present in asymptotic Hodge theory.\footnote{It should be noted that this $\mathfrak{sl}(2,\mathbb{R})$ also arises at finite distance boundaries, as long as the monodromy has a non-trivial unipotent part.} The conventional wisdom of the AdS/CFT correspondence then suggests that the boundary $\mathfrak{sl}(2,\mathbb{R})$ should be understood as a global symmetry of some putative boundary theory. Let us also remark that the existence of such a boundary theory matches well with the result of \cite{Lee:2019wij,Lanza:2020qmt,Lanza:2021udy}. The results of this chapter should be viewed as a first step towards making these matters more precise, by focussing first on an appropriate formulation of the bulk theory. Indeed, while we will succeed in writing down a non-linear $\sigma$-model which correctly encodes the full dynamics of the period map, it remains to find an appropriate coupling of this theory to gravity, such that also the physical metric on the moduli space is realized dynamically, as well as its associated boundary theory.

\section{$\lambda$-deformed WZW models}
\label{sec:lambda_WZW}
Let us now turn to a description of the model we will consider in the remainder of this section, which arises as a certain integrable deformation of the gauged WZW model. From a more string-theoretic perspective, one of the motivations for studying integrable deformations of two-dimensional $\sigma$-models has been to find a worldsheet action principle that describes the $q$-deformation of the S-matrix of the $AdS_5\times S^5$ superstring \cite{Hoare:2011wr,Delduc:2013qra,Delduc:2014kha}. Such deformations are of interest, since they reduce the amount of supersymmetry while retaining integrability, allowing for a more general study of the AdS/CFT correspondence beyond $\mathcal{N}=4$ SYM \cite{Borsato:2013qpa,Hoare:2015kla,Pachol:2015mfa}, see also \cite{Beisert:2010jr} for an elaborate list of references. Nevertheless, even in the bosonic setting integrable deformations of two-dimensional $\sigma$-models have received much attention \cite{Klimcik:2002zj,Klimcik:2008eq,Delduc:2013fga,Kawaguchi:2013gma,Sfetsos:2014_integrability,Hollowood:2014,Hollowood:2014qma,Hoare:2014pna,Klimcik:2014bta,Delduc:2014uaa,Kawaguchi:2014qwa,vanTongeren:2015uha,Osten:2016dvf} and a vast web of connections and dualities between them has been uncovered over the years \cite{Klimcik_2015,Klimcik:2016rov,Hoare:2016wsk,Borsato:2016pas,Borsato:2017qsx}, see also \cite{Georgiou:2021pbd,Klimcik:2021bjy,Hoare:2021dix} for recent reviews. \\

\noindent The precise deformations we will consider in this work are the so-called $\lambda$-deformations, first discovered by Sfetsos in \cite{Sfetsos:2014_integrability} for the principal chiral model and WZW model, and later generalized to symmetric and semi-symmetric spaces in \cite{Hollowood:2014,Hollowood:2014qma}. In their simplest form, one can view these as a one-parameter family of integrable $\sigma$-models that interpolate between the WZW model and the non-Abelian T-dual of the principal chiral model. Also the $\lambda$-deformations have seen considerable development and generalizations to e.g.~multi-parameter/asymmetric deformations \cite{Sfetsos:2015nya,Georgiou:2016urf,Georgiou:2018gpe,Driezen:2019ykp}. Additionally, recently great progress has been made on formulating integrable $\lambda$-deformations on worldsheets with boundaries and studying D-brane configurations, see e.g.~\cite{Alekseev:1998mc,Felder:1999ka,Figueroa-OFarrill:1999cmq,Driezen:2018glg,Sfetsos:2021pcs}. However, for the purpose of this work we will restrict our attention to the so-called $\lambda$-deformed $G/G$ model, which corresponds to the $\lambda$-deformation of the fully gauged WZW model. Its field content consists of a group-valued field and a gauge field. By appropriately identifying these fields with the moduli-dependent charge operator $Q(t,\bar{t})$ c.f.~equation \eqref{eq:charge-operator_bulk}, we show that the resulting equations of motion precisely describe a variation of Hodge structure. Conversely, this implies that any one-parameter variation of Hodge structure yields a solution to the $\lambda$-deformed $G/G$ model. We begin in section \ref{subsec:gWZW} with a standard review on the WZW model and the $G/G$ model. The $\lambda$-deformation of the latter is then discussed in section \ref{subsec:lambda_G/G}, where a detailed analysis of the resulting equations of motion is given. The important equations that will be used in subsequent sections are \eqref{eq:A_ginvdg}, \eqref{eq:Abar_ginvdbarg} and \eqref{eq:eomg_full}, together with \eqref{eq:starA_closed}. 

\subsection{Gauged WZW models}
\label{subsec:gWZW}

We start by reviewing classical aspects of the WZW model and its gauged extension, mostly to set the notation and fix our conventions. For an in-depth discussion on WZW models, also beyond the classical description, we refer the reader to e.g.~\cite{Gawedzki_1999}.

\subsubsection*{The WZW action}
A WZW model is a non-linear $\sigma$-model defined on a two-dimensional worldsheet $\Sigma$ with a Lie group $G$ as target space, which we assume to be semi-simple. It describes the dynamics of a group-valued field 
\begin{equation}
	g: \Sigma\rightarrow G\,,
\end{equation}
whose action is given by the sum of the principal chiral model and the celebrated Wess-Zumino action  \cite{Wess:1971,Witten:1983_current_algebra,Witten:1983_bosonization}
\begin{equation}
\label{eq:WZW_action}
	S_{\mathrm{WZW}}[g]= \frac{k}{8\pi} \int_\Sigma \mathrm{Tr}\left(g^{-1}\dd g\wedge\star\, g^{-1}\dd g \right) + \frac{k}{12\pi i}\int \mathrm{Tr}\left(g^{-1}\dd g\wedge g^{-1}\dd g\wedge g^{-1}\dd g  \right)\,.
\end{equation}
Here $\star$ denotes the Hodge star on $\Sigma$ and the second term features an integration of a suitable extension of $g$ over a three-manifold whose boundary is the worldsheet $\Sigma$. Furthermore, $\mathrm{Tr}$ denotes any non-degenerate $\mathrm{Ad}$-invariant bilinear form on the Lie algebra $\mathfrak{g}$ of $G$. Finally, the constant $k$ is referred to as the level of the model. It is restricted to take integer values when $G$ is compact so that the path integral is well-defined. In the current work however, $k$ will not play an important role as we are merely concerned with classical features. 

\subsubsection*{The gauged WZW model}
By virtue of the trace, the action \eqref{eq:WZW_action} enjoys a global $G\times G$ symmetry 
\begin{equation}
\label{eq:global_symmetry}
	g\mapsto g_L\cdot g\cdot g_R^{-1}\,,\qquad (g_L, g_R) \in G\times G\,.
\end{equation}
One may now proceed to define a gauged WZW model by gauging a particular subgroup of this global symmetry, as reviewed in e.g.~\cite{Chung:1993}. In the current work, we will consider the vector gauged WZW model in which the diagonal subgroup of $G\times G$ is gauged, corresponding to the transformation $g\mapsto hgh^{-1}$, for $h\in G$. The resulting action is most conveniently written down using local coordinates $x,y$ on $\Sigma$, which we take have Euclidean signature, and passing to complex coordinates $t,\bar{t}$ via $t=x+iy$. Furthermore, by a suitable conformal transformation the metric on $\Sigma$ can be taken to be the flat metric, i.e.
\begin{equation}
	ds^2 = \dd x^2+\dd y^2 = \dd t\,\dd \bar{t}\,.
\end{equation}
Finally, the gauge field will be denoted by
\begin{equation}
	\mathbf{A} = A\,\dd t+\bar{A}\,\dd\bar{t}\,,
\end{equation}
and its components $A,\bar{A}$ are fields on $\Sigma$ taking values in $\mathfrak{g}$. Then the action of the vector gauged WZW model reads\footnote{Here $d^2t = \frac{i}{2}\dd t\wedge \dd \bar{t}=\dd x\wedge \dd y$ and furthermore $\partial = \partial_t$, $\bar{\partial} = \partial_{\bar{t}}$.} \cite{Gawedzki:1988_I,Gawedzki:1988_II,Witten:1991}
\begin{equation}
\label{eq:gWZW_action}
S_{G/G}[g,\mathbf{A}] = S_{\mathrm{WZW}}[g]+\frac{k}{\pi}\int_\Sigma d^2t\;\mathrm{Tr}\left(A\Bar{\partial}g g^{-1}-\bar{A}g^{-1}\partial g -Ag\Bar{A}g^{-1}+A\Bar{A} \right)\,.
\end{equation}
Indeed, one may verify that \eqref{eq:gWZW_action} is invariant under the gauge transformation
\begin{equation}
	g\mapsto hgh^{-1}\,,\qquad \mathbf{A}\mapsto h\left(\dd+\mathbf{A}\right)h^{-1}\,,\qquad h\in G\,.
\end{equation}
Since the resulting action is gauge-invariant under conjugation by the full group $G$, this model is also referred to as the $G/G$ model.

\subsubsection*{Equations of motion}
For completeness and later reference, we record the equations of motion 
\begin{align}
\label{eq:eom_gWZW_barA}	&\delta\bar{A}:\qquad g^{-1}Dg = 0\,, \\
	&\delta A:\qquad \bar{D}g g^{-1}=0\,, \\
\label{eq:eom_gWZW_g}	&\delta g:\qquad F-\bar{D}(g^{-1}Dg)=0\,, 
\end{align}
which are obtained from the variation of \eqref{eq:gWZW_action} w.r.t. the various fields. Here $D,\bar{D}$ denote the covariant derivatives, defined by
\begin{equation}
	D = \partial + [A,-]\,,\qquad \bar{D} = \bar{\partial} + [\bar{A},-]\,,
\end{equation}
and $F$ denotes the field-strength, which is given by 
\begin{equation}
	F = \partial \bar{A} - \bar{\partial}A + [A,\bar{A}]\,.
\end{equation}
An important property of the $G/G$ model is the on-shell vanishing of the field strength, which is easily seen by combining \eqref{eq:eom_gWZW_barA} and \eqref{eq:eom_gWZW_g}. As a result, the gauge field is (locally) pure-gauge and hence it does not contain any physical degrees of freedom. As will become apparent, this is no longer true for the $\lambda$-deformed $G/G$ model.

\subsection{$\lambda$-deformations}
\label{subsec:lambda_G/G}

Let us now introduce the $\lambda$-deformed $G/G$ model, following the discussion in \cite{Hollowood:2014}. This model was first described by Sfetsos in \cite{Sfetsos:2014_integrability} as an interpolation between the exact CFT WZW model and the non-Abelian T-dual of the principal chiral model. It can be constructed by employing a particular gauging of the combined action for the gauged principal chiral model and the gauged WZW model. For the purpose of this work we are mostly interested in the final result of this procedure, and refer the interested reader to \cite{Sfetsos:2014_integrability,Hollowood:2014} for more details on the construction of the model itself. The action that describes the $\lambda$-deformed $G/G$ model is that of the $G/G$ model plus an additional deformation term and reads as follows.
\begin{subbox}{$\lambda$-deformed $G/G$ model}
	\begin{equation}
		\label{eq:lambda_gWZW_action}
			\quad S_{\lambda}[g,\mathbf{A}] = S_{G/G}[g,\mathbf{A}]+\frac{k}{\pi} \int d^2t\;\mathrm{Tr}\left(\gamma A\bar{A} \right)\,,\qquad \,,
	\end{equation}
	where the $\gamma$ is an arbitrary deformation parameter. A standard convention is to introduce another parameter $\lambda$ via the relation
	\begin{equation}
		\gamma = \lambda^{-1}-1\,.
	\end{equation} 
	\tcblower 
	Clearly, for $\gamma=0$ (or, equivalently, $\lambda=1$) one recovers the ordinary $G/G$-model, while for $\gamma\rightarrow\infty$ (or, equivalently, $\lambda = 0)$ one recovers the WZW model when integrating out the gauge field.\footnote{To recover the non-Abelian T-dual of the principal chiral model, one should combine the limit $\lambda\rightarrow 1$ with the limit $k\rightarrow \infty$ in a particular way, see \cite{Sfetsos:2014_integrability}.}
\end{subbox}
\noindent  Some remarks are in order. To begin with, we stress that due to the deformation term in \eqref{eq:lambda_gWZW_action} the action $S_\lambda$ is no longer gauge invariant. As a result, the field $\mathbf{A}$ should no longer be interpreted as a gauge field, but rather as simply a constraint. However, we will still informally refer to $\mathbf{A}$ as the gauge field. Secondly, it should be noted that while the $\lambda$-deformed $G/G$ model is still conformal at the classical level, this is no longer true at the quantum level, in contrast to the $G/G$ model. This can be seen, for example, by analyzing the running of the coupling $\gamma$, whose $\beta$-function turns out to be non-trivial \cite{Tseytlin:1994,Sfetsos:2014_betafunction,Appadu:2015}. Nevertheless, the $\lambda$-deformed $G/G$ model retains some remarkable properties such as (strong) integrability \cite{Hollowood:2014,Sfetsos:2014_integrability,Georgiou:2019} and renormalizability \cite{Hoare:2019}. As a last remark, let us note that the action \eqref{eq:lambda_gWZW_action} also enjoys a discrete $\mathbb{Z}_2$ symmetry \cite{Kutasov:1989,Itsios:2014,Hoare:2015}
\begin{equation}\label{eq:Z2_symmetry}
	\lambda\mapsto \lambda^{-1}\,,\quad k\mapsto -k\,,\quad g\mapsto g^{-1}\,,\quad A\mapsto g(\partial + A)g^{-1}\,,\quad \bar{A}\mapsto \lambda^{-1}\bar{A}\,,
\end{equation}
which will be relevant in section \ref{sec:strings_Hodge_theory}. Physically, this symmetry can be viewed as a duality between a strong coupling regime $(\lambda\rightarrow \infty$) and a perturbative regime ($\lambda\rightarrow 0)$.

\subsubsection*{Equations of motion}
Let us now turn to the equations of motion of \eqref{eq:lambda_gWZW_action}, which are modified slightly compared to those of the $G/G$ model due to the deformation term. The equations of motion of $A$ and $\bar{A}$ are given by
\begin{align}
	\label{eq:eom_Abar_lambda}
	&\delta \bar{A}:\qquad g^{-1}Dg = \gamma A\,,\\
	\label{eq:eom_A_lambda}
	&\delta A:\qquad \bar{D}g g^{-1} = -\gamma \bar{A}\,,
\end{align}
while equation of motion of $g$ remains unchanged, but we record it here for convenience
\begin{equation}
	\label{eq:eom_g_lambda}
	\delta g:\qquad F-\bar{D}(g^{-1}Dg)=0\qquad \iff \qquad  F-D(\bar{D}g g^{-1}) = 0\,,
\end{equation}
where the equivalence of the two expressions can easily be verified.\footnote{This can be seen, for example, by writing the left-hand side as $[\partial+A+g^{-1}Dg, \bar{\partial}+\bar{A} ]=0$ and then conjugating this expression with $g$.} Our goal will be to simplify the above equations as much as possible. First, we note that \eqref{eq:eom_Abar_lambda} and \eqref{eq:eom_A_lambda} may be written as
\begin{align}
\label{eq:A_ginvdg}	&\delta\bar{A}:\qquad A = \frac{1}{\lambda^{-1}-\,\mathrm{Ad}_{g^{-1}}} g^{-1}\partial g\,,\\
\label{eq:Abar_ginvdbarg}	&\delta A:\qquad \bar{A} = \frac{1}{1-\lambda^{-1}\,\mathrm{Ad}_{g^{-1}}} g^{-1}\bar{\partial}g\,.
\end{align}
To be clear, the fraction in \eqref{eq:A_ginvdg} denotes the inverse of the linear operator $\lambda^{-1}-\mathrm{Ad}_{g^{-1}}$, and similarly in \eqref{eq:Abar_ginvdbarg}. While not immediately apparent, this form of the equations of motion will be extremely useful later on. \\

\noindent Next, we consider the equation of motion of $g$. By inserting \eqref{eq:eom_Abar_lambda} into the left-hand side of \eqref{eq:eom_g_lambda} and \eqref{eq:eom_A_lambda} into the right-hand side of \eqref{eq:eom_g_lambda} one finds two equations
\begin{align}
	\label{eq:F_lambda}	&F-\gamma \bar{D}A = 0\qquad \iff \qquad  \lambda\partial\bar{A} - \bar{\partial}A+[A,\bar{A}] = 0\,,\\
	\label{eq:Fbar_lambda}	&F+\gamma D\bar{A}= 0\qquad \iff \qquad \partial\bar{A} - \lambda\bar{\partial}A+[A,\bar{A}] = 0\,.
\end{align}
When $\lambda=1$ (or $\gamma=0$, i.e.~no deformation) the equations \eqref{eq:F_lambda} and \eqref{eq:Fbar_lambda} coincide and one recovers the vanishing of the field-strength of the $G/G$ model. However, for $\lambda\neq 1$ (which we will henceforth assume) the field-strength need not vanish and the two equations are independent. By adding or subtracting the two equations appropriately one may rewrite them as follows
\begin{equation}\label{eq:eomg_full}
	\delta g:\qquad \partial\bar{A}+\bar{\partial}A = 0\,,\qquad \partial\bar{A} = \mu[A,\bar{A}]\,,\qquad \mu=-\frac{1}{1+\lambda}\,,
\end{equation}
where we also assume $\lambda\neq -1$. Passing to differential form notation, the equation on the left-hand side states that the one-form $\star\mathbf{A}$ is closed. In the remainder of this work, we will assume the worldsheet $\Sigma$ to be simply-connected by passing to the universal covering space. As a result, the one-form $\star\mathbf{A}$ is also exact, so that $\mathbf{A}$ must be of the form
\begin{equation}\label{eq:starA_closed}
	\mathbf{A} = \star\;\dd U\,,
\end{equation}
for some Lie-algebra valued function $U$. While not crucial, this observation will prove useful in section \ref{sec:strings_Hodge_theory}, where it will facilitate a natural ansatz for the gauge field.

\subsubsection*{Generalizations}
We close this section by pointing out some possible generalizations of the (comparatively) simple $\lambda$-deformation we have considered that have been studied in the literature. One possibility is to consider $\lambda$-deformations of the $G/G$ model which do retain some of the gauge symmetry. These were first constructed for the $\mathrm{SU}(2)/\mathrm{U}(1)$ coset CFT in \cite{Sfetsos:2014_integrability}, and were later generalized to symmetric spaces $G/H$ in \cite{Hollowood:2014} and then applied to study the $AdS_5\times S^5$ superstring in \cite{Hollowood:2014qma}. Here the restriction to (semi)-symmetric spaces is crucial in order to retain properties such as (strong) integrability and renormalizability. Along a different vein, one may also consider deformations of multiple WZW models at different levels \cite{Georgiou:2018gpe}, as well as multi-parameter deformations \cite{Sfetsos:2015nya} and asymmetric deformations \cite{Georgiou:2016urf,Driezen:2019ykp}. For a recent overview of deformed $\sigma$-models that have arisen over the years and a discussion on their integrable structure, one may also consult e.g.~\cite{Georgiou:2021pbd}. Finally, in recent years great progress has been made on formulating integrable $\lambda$-deformations on worldsheets with boundaries and studying D-brane configurations, see e.g.~\cite{Alekseev:1998mc,Driezen:2018glg,Sfetsos:2021pcs}.

\section{Strings on classifying spaces}
\label{sec:strings_Hodge_theory}

In section \ref{sec:lambda_WZW} we have introduced the $\lambda$-deformed $G/G$ model, a particular $\sigma$-model for a group-valued field $g$ and gauge field $\mathbf{A}$. This model describes the classical propagation of a string on a group manifold, in the presence of particular background fields. At the same time, we recall from the discussion in section \ref{sec:period_map} that a variation of Hodge structure can be described in terms of a period map which takes values in the classifying space $D$, which can be realized as a quotient space $G/V$. A natural step, then, is to consider the motion of the string \textit{on the classifying space}, as dictated by the equations of motion of the $\lambda$-deformed $G/G$ model, and ask whether this describes a variation of Hodge structures. In other words, whether its dynamics are described by the horizontality of the period map. The purpose of this section is to study precisely this question. Indeed, our main result is that for a suitable choice of gauge field and $\lambda$, the equations of motion of the string indeed describe a variation of Hodge structure. Conversely, this implies that any one-parameter variation of Hodge structure provides a solution to the $\lambda$-deformed $G/G$ model. Along the way we will also identify a more general set of solutions, which satisfy a generalized set of Nahm's equations.

\subsubsection*{The VHS ansatz}
We recall the properties of the bulk charge operator $Q(t,\bar{t})$, introduced at the end of section \ref{sec:period_map}. It is a grading element of a semisimple Lie algebra $\mathfrak{g}_\mathbb{C}$, meaning that it is a semisimple operator that acts on $\mathfrak{g}_{\mathbb{C}}$ via the adjoint action by integer eigenvalues, see e.g.~\cite{Kerr2017}. Furthermore, it must satisfy $Q\in i \mathfrak{g}$. Whenever such a $Q$ is available, one can consider the following ansatz for the fields of the $\lambda$-deformed $G/G$ model.
\begin{subbox}{The VHS ansatz}
	\begin{equation}\label{eq:ansatz}
		g = e^{i\beta Q}\,,\qquad U = \alpha\, iQ\,,
	\end{equation}
	for some and non-zero parameters $\alpha,\beta\in\mathbb{C}$.
\end{subbox}
\noindent The precise result of this section, then, is that for a suitable choice of $\alpha,\beta$, this provides a solution to the equations of motion of the $\lambda$-deformed $G/G$ model if and only if $Q$ satisfies the horizontality condition \eqref{eq:horizontality_Q}.  Note moreover that one can consider this ansatz without reference to Hodge theory as the above conditions merely pertain to the algebra $\mathfrak{g}_{\mathbb{C}}$, although there is a natural correspondence between a $Q_{\mathrm{ref}}$ with the above properties and a Hodge structure \cite{robles2012schubert}.\\

\noindent Of course, the ansatz \eqref{eq:ansatz} is motivated by the idea of considering strings on classifying spaces of Hodge structures. Indeed, we have argued that $Q$ itself contains all the information necessary to describe a VHS. Therefore the ansatz \eqref{eq:ansatz} is quite natural, as it is the most general ansatz one can make using only $Q$ as input. It should be mentioned that if one desires the fields $g$ and $U$ to be real (which is required for the action \eqref{eq:lambda_gWZW_action} to be real) one must restrict these parameters to $\alpha,\beta\in\mathbb{R}$. However, we will not make this restriction and leave the parameters arbitrary for now. In fact, it will turn out to be crucial to allow $\alpha$ to take complex values.\\

\noindent To interpret this ansatz from a physical perspective, we recall the passage of an action of the type we consider \eqref{eq:lambda_gWZW_action} to a more traditional $\sigma$-model action. One writes
\begin{equation}\label{eq:g_to_X}
	g^{-1}\dd g = T_A \tensor{e}{^A_\mu}\dd X^\mu\,,
\end{equation}
where the $T_A$ denote a set of generators of the Lie algebra $\mathfrak{g}$ and $\tensor{e}{^A_\mu}$ the left-invariant	 vielbein on $G$, which map the tangent bundle $TG$ to $\mathfrak{g}$. Finally, $X^\mu$ denote a set of local coordinates on $G$. One can then show that the action of the $\lambda$-deformed $G/G$ model, evaluated on the constraints \eqref{eq:A_ginvdg},\eqref{eq:Abar_ginvdbarg}, can be written as \cite{Driezen:2018glg}
\begin{equation}\label{eq:sigma_model}
	S_\lambda[g] = \frac{1}{4\pi\alpha'}\int_\Sigma \mathcal{G}_{\mu\nu}\,\dd X^\mu\wedge\star \,\dd X^\nu-\mathcal{B}_{\mu\nu} \,\dd X^\mu\wedge \dd X^\nu\,,
\end{equation}
with $\alpha'=2$ and
\begin{align}
	ds^2 &= k\,\tensor{e}{^A_\mu}\tensor{e}{^B_\nu} \mathrm{Tr}\left(T_A\cdot\left(\mathcal{O}_{g^{-1}}+\mathcal{O}_g-1\right)T_B\right)\,\dd X^\mu\otimes\dd X^\nu\,,\\
	\mathcal{B} &= \mathcal{B}_{\mathrm{WZW}}+k \,\tensor{e}{^A_\mu}\tensor{e}{^B_\nu}\mathrm{Tr}\left(T_A\cdot\left(\mathcal{O}_{g^{-1}}-\mathcal{O}_g\right)T_B\right)\dd X^\mu\wedge\dd X^\nu\,.
\end{align}
Here $\mathcal{B}_{\mathrm{WZW}}$ is such that locally $\dd \mathcal{B}_{\mathrm{WZW}} = -\frac{2k}{3}g^{-1}\dd g\wedge g^{-1}\dd g\wedge g^{-1}\dd g$ and we have used the short-hand $\mathcal{O}_g = (1-\lambda\mathrm{Ad}_g)^{-1}$. Note that $\mathcal{O}_g\rightarrow 1$ for $\lambda\rightarrow 0$, from which one readily sees that the above $\sigma$-model reduces to that of the ordinary WZW model in this limit. \\

\noindent Clearly, equation \eqref{eq:sigma_model} describes the propagation of a classical string in a background metric $\mathcal{G}_{\mu\nu}$ and $\mathcal{B}_{\mu\nu}$ field, with the $X^\mu$ describing the embedding of the worldsheet, recall also our discussion in section \ref{sec:string_theory_basics}. Moreover, inserting the ansatz \eqref{eq:ansatz} into \eqref{eq:g_to_X} one sees that the bulk charge operator $Q$ plays the role of the coordinates $X^\mu$. In other words, the ansatz \eqref{eq:ansatz} simply proposes that the moduli dependence of the Hodge structure is described by the embedding of the string worldsheet in the classifying space. However, in order for the motion of the string to correctly yield a variation of Hodge structures one must consider a particular value of $\lambda$, which amounts to a specific choice of the background fields. It is the purpose of the rest of this section to make this statement precise, by evaluating the equations of motion on the ansatz \eqref{eq:ansatz} and analysing the possible solutions. 

\subsection{The VHS solution}
\label{sec:VHS_solution}

While we have formulated the ansatz \eqref{eq:ansatz} in terms of the bulk charge operator $Q$, it will be convenient to rephrase it in terms of the period map $h$ and the reference charge operator $Q_{\mathrm{ref}}$, by recalling that $Q=h Q_{\mathrm{ref}}h^{-1}$. First, for notational convenience, let us introduce the parameter
\begin{equation}
	z = e^{i\beta}\,,
\end{equation}
such that $g=z^{Q}$. Then one may compute
\begin{align}
	g^{-1}\dd g &= h z^{-Q_{\mathrm{ref}}}h^{-1} \dd h z^{Q_{\mathrm{ref}}}h^{-1}-\dd h h^{-1}\\
	&= h\left( z^{-\mathrm{ad}\,Q_{\mathrm{ref}}} \mathbf{B} - \mathbf{B}\right)h^{-1}\\
	&= \sum_{q\neq 0}(z^{-q}-1) h\mathbf{B}_q h^{-1}
\end{align}
where we recall $\mathbf{B}=h^{-1}\dd h$ and in the last line we have decomposed $\mathbf{B}$ into its charge-modes with respect to $Q_{\mathrm{ref}}$ as in \eqref{eq:q_mode}. Similarly, we compute
\begin{align}
	\mathbf{A} &= \star\,\dd (h Q_{\mathrm{ref}} h^{-1})\\
	&= -i\alpha\star h[Q_{\mathrm{ref}},h^{-1}\dd h]h^{-1}\\
	&=-i\alpha\sum_{q\neq 0} q \star h\mathbf{B}_q h^{-1}\,.
\end{align}
Together, these two identities provide us with the following dictionary
\begin{align}
\label{eq:g_to_B}	g^{-1}\dd g &= -\sum_{q\neq 0}(1-z^{-q})\, h\mathbf{B}_q h^{-1}\,,\\
\label{eq:A_to_B}	\mathbf{A} &= -i\alpha \sum_{q\neq 0} q\, \star h\mathbf{B}_{q}h^{-1}\,,
\end{align}
which will be very useful in evaluating the equations of motion. The formulation in terms of charge modes w.r.t. $Q_{\mathrm{ref}}$ is especially helpful combined with the observation that
\begin{equation}
\mathrm{Ad}_{g} = h z^{\mathrm{ad}\,Q_{\mathrm{ref}}}h^{-1},
\end{equation}
which acts as multiplication by $z^{q}$ on $h \mathbf{B}_q h^{-1}$.\\

\noindent The strategy in the remainder of this work is to use the dictionary \eqref{eq:g_to_B} and \eqref{eq:A_to_B} to translate the equations of motion of the $\lambda$-deformed $G/G$ model into statements about the charge-modes of $\mathbf{B}$. The goal, then, is to show that these statements imply the horizontality condition \eqref{eq:horizontality_B}, so that any solution to the equations of motion describes a variation of Hodge structures. However, it will turn out that this only happens for a particular value of $z$, namely $z=-1$ (or equivalently $\beta=\pi$). This is not unexpected, since in that case $g=(-1)^Q$, i.e.~it is precisely given by the Weil operator, previously denoted by $C$. Therefore, in the following we will first analyse this solution, dubbed the `VHS solution'. For completeness, we also present a more general analysis (i.e.~for all values of $z$) in the next section. 

\subsubsection*{Equations of motion of $A$ and $\bar{A}$}

For $z=-1$, equation \eqref{eq:g_to_B} simplifies to
\begin{equation}
	g^{-1}\dd g = -2\sum_{q\, \mathrm{ odd}} h \mathbf{B}_q h^{-1}\,,
\end{equation}
and furthermore $\mathrm{Ad}_g$ acts as multiplication by minus one on $\mathbf{B}_q$, for $q$ odd. With these observations at hand one can easily evaluate \eqref{eq:eom_Abar_lambda} and \eqref{eq:eom_A_lambda} to find
\begin{align}
\label{eq:eom_Abar_B}\delta\bar{A}:\qquad	\alpha \sum_{q\neq 0} q B_{q}= -\frac{2\lambda}{1+\lambda}\sum_{q\, \mathrm{ odd}}B_q\,,\\
\label{eq:eom_A_Bbar}\delta A:\qquad	\alpha \sum_{q\neq 0} q \bar{B}_{q}=+\frac{2\lambda}{1+\lambda}\sum_{q\, \mathrm{ odd}}\bar{B}_q\,.
\end{align}
These equations are best understood by projecting them onto a specific charge-mode, which we denote by $q$. Then one finds 
\begin{align}
	\text{$q$ non-zero and even}:\qquad &B_{q} = \bar{B}_{q}=0\,,\\
	\text{$q$ odd}:\qquad &b(q)\bar{B}_{q}=b(q)B_{-q}=0\,,
\end{align}
where we have introduced the function
\begin{equation}
	b(q)=\alpha q - \frac{2\lambda}{1+\lambda}\,.
\end{equation}
The upshot of this is the following. First, $\mathbf{B}$ cannot contain any non-zero even charge-modes. Second, it can only contain an odd charge-mode $q$ if $b(q)=0$. Furthermore, since $b(q)$ is linear in $q$, if $b(q_i)=0$ for two distinct odd $q_i$, $i=1,2$, then it must be that $\alpha=\lambda=0$. In that case the ansatz for the gauge field becomes trivial, hence we exclude this solution. Therefore $b(q)=0$ for at most one odd $q$. In short, we have shown that $\mathbf{B}$ must be of the form
\begin{equation}\label{eq:B_simplified}
	B = B_0 + B_{-q}\,,\qquad \bar{B}=\bar{B}_0+\bar{B}_q\,,
\end{equation}
for some odd $q$, and furthermore the parameters $\alpha,\lambda$ are restricted by
\begin{equation}\label{eq:sol_alpha}
	\alpha q = \frac{2\lambda}{1+\lambda}\,,
\end{equation}
so that indeed $b(q)=0$. This is then the most general solution to the equations of motion of $A$ and $\bar{A}$, for the ansatz \eqref{eq:ansatz} when imposing $\beta=\pi$ or $z=-1$. As promised, \eqref{eq:B_simplified} is precisely the desired horizontality condition \eqref{eq:horizontality_B}. To make the match precise, one should also fix $\alpha$ to be $2\lambda/(1+\lambda)$, which effectively enforces $q=1$. 

\subsubsection*{Equation of motion of $g$}

While the equations of motion of $A$ and $\bar{A}$ restrict the possible charge-modes in $B$ and $\bar{B}$, as just discussed, the equation of motion of $B$ will determine the remaining dynamics and fix the value of $\lambda$ uniquely. Of course, as already described in section \ref{sec:period_map}, once the horizontality condition is satisfied one can use the Bianchi identity to determine the dynamics completely, which resulted in Nahm's equations. Here we will show that the equations of motion of $g$ essentially reduce to the same equations.\\

\noindent First, using \eqref{eq:B_simplified}, one sees that the gauge field simplifies to
\begin{equation}
	A = -\alpha q\, hB_{-q}h^{-1}\,,\qquad \bar{A} = -\alpha q\, h \bar{B}_{q}h^{-1}\,.
\end{equation}
Inserting these expressions into the equation of motion of $g$ \eqref{eq:eomg_full} yields the following equation
\begin{equation}
	\delta g:\qquad \partial\bar{B}_{q}+[B_0,\bar{B}_q]+(1+q\alpha\mu)[B_{-q},\bar{B}_q]=0\,.
\end{equation}
Again, this equation is best understood by considering its various charge components. In particular, the first two terms have charge $q$, while the last term has charge zero. In other words, it reduces to the following two equations
\begin{align}
	&\text{charge $q$}:\qquad \partial\bar{B}_{q}+[B_0,\bar{B}_q]=0\,,\\
	&\text{charge 0}:\qquad (1+q\alpha\mu)[B_{-q},\bar{B}_q]=0\,.
\end{align}
Naturally, by taking the complex conjugate of the first equation, one may equivalently derive
\begin{equation}
	\text{charge $-q$}:\qquad \bar{\partial}B_{-q}+[\bar{B}_0,B_{-q}]=0\,.
\end{equation}
As expected, the equation of motion of $g$ has reduced to the same equations that follow from the Bianchi identity, i.e.~Nahm's equations \eqref{eq:charge_plus1} and \eqref{eq:charge_min1}, together with an additional constraint on the parameters, namely $1+q\alpha \mu=0$. Combined with the constraint \eqref{eq:sol_alpha} this results in a unique solution (up to a sign) for $\alpha$ and $\lambda$, given by
\begin{equation}\label{eq:alpha_sol}
	q\alpha = 1+\lambda\,,\qquad \lambda = \pm i.
\end{equation} 
Furthermore, one can show that these two solutions are precisely related via the $\mathbb{Z}_2$-symmetry \eqref{eq:Z2_symmetry}. \\

\noindent To close this discussion, we stress that we have not actually solved the equations of motion in full, but have merely rewritten them into the horizontality condition and Nahm's equations. However, as described at length in chapters \ref{chap:asymp_Hodge_I} and \ref{chap:asymp_Hodge_II}, the powerful formalism of asymptotic Hodge theory allows one to find the proper solutions to these equations for a particular set of boundary conditions.

\subsubsection*{Integrability and the on-shell action}

Now that we have identified for which ansatz the equations of motion of the $\lambda$-deformed $G/G$ model reduce to the horizontality condition, let us return to some properties of the $\sigma$-model. Firstly, as mentioned before, this model is integrable, i.e.~it has a Lax connection 
\begin{equation}
	\nabla = \dd +\mathscr{L}(\zeta)\,,
\end{equation}
with $\mathscr{L}(\zeta)$ given by (for notational clarity we use $\pm$ to denote the (anti)-holomorphic components of the one-form $\mathscr{L}$ and $\mathbf{A}$)
\begin{equation}\label{eq:Lax}
	\mathscr{L}_\pm(\zeta)=\frac{2}{1\mp \zeta} \frac{A_{\pm}}{1+\lambda}\,.
\end{equation}
Concretely, this means that the curvature of the connection $\nabla$ vanishes if and only if $A_\pm$ satisfy \eqref{eq:F_lambda} and \eqref{eq:Fbar_lambda}, for every value of the spectral parameter $\zeta\in\mathbb{C}\mathbb{P}^1$. It is natural, then, to evaluate \eqref{eq:Lax} on the VHS solution to find an expression for $\mathscr{L}(\zeta)$ in terms of $B$ and $\bar{B}$. The result of this computation is the following
\begin{equation}
	\mathscr{L}_\pm(\zeta) = -\frac{2}{1\mp\zeta} h \left(B_{\pm}\right)_{\pm q}h^{-1}\,,
\end{equation}
where we have used \eqref{eq:alpha_sol}, for arbitrary $q$. To elucidate this expression, we perform a gauge transformation by the period mapping itself, i.e.
\begin{equation}
	\mathscr{L}_\pm(\zeta)\rightarrow \mathscr{L}_{\pm}^{h^{-1}}(\zeta) = h^{-1}\mathscr{L}_{\pm}(\zeta)h + h^{-1}\partial_\pm h\,,
\end{equation} 
which yields
\begin{equation}\label{eq:Lax_gauge_transform}
	\mathscr{L}^{h^{-1}}_\pm(\zeta) = (B_\pm)_0 + \frac{\zeta\pm 1}{\zeta\mp 1} (B_{\pm})_{\pm q}\,.
\end{equation}
Interestingly, \eqref{eq:Lax_gauge_transform} is precisely the Lax connection of the $G/K$ principal chiral model \cite{Hollowood:2014}, where we remind the reader that in the current setting $K$ is generated by all operators of even charge. In other words, when the horizontality condition is satisfied, the remaining dynamics of the period mapping are described by the integrable $G/K$ principal chiral model. This is not surprising and has long been known from the perspective of $tt^*$-geometry \cite{Cecotti:1991me,Cecotti:2020rjq} and was more recently noted by \cite{Cecotti:2020uek} and \cite{Grimm:2020cda}, see also the last paragraph of appendix \ref{subsec:Nahm}.\\

\noindent To further elucidate the relation to these previous works, it is instructive to compute the on-shell action (i.e.~by imposing \eqref{eq:A_ginvdg} and \eqref{eq:Abar_ginvdbarg}) and rephrasing the result in terms of $B$ and $\bar{B}$. The result of this computation is (see appendix \ref{sec:app_action} for details)
\begin{equation}\label{eq:on-shell_action}
	S_\lambda[g] = \pm \frac{2k}{\pi i}\int_\Sigma d^2t \,\mathrm{Tr}\left(B_{-1}\bar{B}_{+1}\right)\,,
\end{equation}
where we have put $q=1$ for simplicity and have suppressed the Wess--Zumino term. Apart from the imaginary prefactor, this agrees precisely with the actions put forward in \cite{Cecotti:2020uek} and \cite{Grimm:2020cda,Grimm:2021ikg}. Therefore, the formulation in terms of the $\lambda$-deformed $G/G$ model can be seen as a generalization of these earlier works, which reduces to them when $A$ and $\bar{A}$ are on-shell, i.e.~when the horizontality condition is satisfied. \\

\noindent As a final remark, we comment on the imaginary prefactor in \eqref{eq:on-shell_action}, which is a result of the fact that both $\alpha$ and $\lambda$ are complex. Interestingly, a similar situation arises in the study of a duality between $\lambda$-deformations and so-called $\eta$-deformations \cite{Klimcik_2015}, where we remark that also in our setting $|\lambda|=1$, as is required for this duality to apply. In fact, a more general analysis of the equations of motion shows that whenever $|z|=1$ also $|\lambda|=1$, as is discussed in the next section. It would be interesting to understand this duality from the perspective of variations of Hodge structures, which may provide new insights into the structure of classifying spaces. Some initial results in this direction are discussed in chapter \ref{chap:biYB}, see also the thesis \cite{Lagendijk:2024}.

\subsection{General analysis of the equations of motion}

In the following, we relax the assumption that $z=-1$ and go through the same analysis as before, which now becomes considerably more involved. In particular, it becomes possible for $B$ and $\bar{B}$ to have more than one charge-mode, hence this solution will not satisfy the horizontality condition. Nevertheless, it turns out that this solution is still rather constrained, with its various charge-modes independently satisfying a variant of Nahm's equations. 

\subsubsection*{Equations of motion of $A$ and $\bar{A}$}

We recall the relations \eqref{eq:g_to_B} and \eqref{eq:A_to_B} and insert them into the equations of motion of $A$ and $\bar{A}$, see \eqref{eq:eom_Abar_lambda} and \eqref{eq:eom_A_lambda}. Then one finds
\begin{align}
	&\delta\bar{A}:\qquad \alpha \sum_{q\neq 0} q B_{q}= -\sum_{q\neq 0}\frac{1-z^{-q}}{\lambda^{-1}-z^{-q}}B_q\,,\\
	&\delta A:\qquad\alpha \sum_{q\neq 0} q \bar{B}_{q}=+\sum_{q\neq 0}\frac{1-z^{-q}}{1-\lambda^{-1}z^{-q}}\bar{B}_q\,.
\end{align}
Note that for $z=-1$ this indeed reduces to \eqref{eq:eom_Abar_B} and \eqref{eq:eom_A_Bbar}. Since both equations should hold for each $q$ separately, this yields the following constraints
\begin{align}\label{eq:beta}
	b(q)B_{-q} & =0\,, \qquad b(q)\bar{B}_q = 0\,,\qquad q\neq 0\,,
\end{align}
where we have defined the function
\begin{equation}
	b(q)=\alpha q - \frac{1-z^{q}}{\lambda^{-1}-z^{q}}\,.
\end{equation}
By a slight abuse of notation we use the same symbol $b(q)$ as in section \ref{sec:VHS_solution}, but of course the two agree for $z=-1$. However, the important observation is that for $z\neq -1$, one can actually solve $b(q)=0$ simultaneously for multiple values of $q$. \\

\noindent Let us therefore denote by $\mathbf{q}=(q_1,\ldots, q_n)$ a collection of distinct non-zero charges $q_i$ for which $b(q_i)=0$. Clearly, \eqref{eq:beta} implies that $\bar{B}_q=B_{-q}=0$ for all $q\not\in\mathbf{q}$. Therefore, to understand which combinations of charge-modes in $B$ and $\bar{B}$ are allowed, one should study the zeroes of $b(q)$. At this point, it is convenient to distinguish the trivial solutions to $b(q)=0$, corresponding to $\alpha=0$ and $\lambda=1$, or $\alpha=0$ and $z^{q}=1$. Furthermore, as alluded to in section \ref{sec:lambda_WZW}, we also assume $\lambda\neq -1$ and similarly refer to this as a trivial solution. In the following, whenever we speak of a solution, it is implicitly implied to be non-trivial.\\

\noindent Below one finds a number of properties of $b(q)$ that will be useful in the remainder of this section. The first property is obvious, whereas the latter properties are elaborated upon in appendix \ref{sec:app_beta}. 
\begin{itemize}
	\item \textbf{Property (1):}\\
	If $q$ is such that both $b(q)=0$ and $b(-q)=0$, then this immediately implies that $\lambda=-1$.
	\item \textbf{Property (2):}\\
	For $q_1,q_2$ distinct, with $b(q_1)=b(q_2)=0$, the following relations hold.
	\begin{align}
		&\text{(2a)}:\qquad b(2q_2)=0\quad\hspace{0.6cm}\iff\quad b(2q_2-q_1)=0\,,\\
		\label{eq:2b}	&\text{(2b)}:\qquad b(q_1-q_2)=0\quad\iff\quad 1+q_1\alpha \mu=0\,\\
		\label{eq:2c}	&\text{(2c)}:\qquad b(q_1+q_2)=0\quad\hspace{0.05cm}\iff\quad 1+(q_1+q_2)\alpha\mu=0\,.
	\end{align}	
\end{itemize}
Note that property (2a) can be derived from (2b) and (2c), but we still write it down separately for easy reference later on. Property (1) implies that it is impossible for both $b(q)$ and $b(-q)$ to be zero, when restricting to non-trivial solutions. Therefore, it is impossible to have $q\in\mathbf{q}$ and $-q\in\mathbf{q}$. One must therefore take $B$ and $\bar{B}$ to be of the form
\begin{equation}\label{eq:B_form}
	B = B_0 + B_{-q_1}+\cdots B_{-q_n}\,,\qquad \bar{B}=\bar{B}_0 + \bar{B}_{q_1}+\cdots+\bar{B}_{q_n}\,,
\end{equation}
where $q_i\neq 0$ and $q_i\neq \pm q_j$ for any distinct $i,j$. Furthermore, we remind the reader that all these charges should satisfy $b(q_i)=0$. We stress that this solves the equations of motion of $A$ and $\bar{A}$, and that all solutions respecting the ansatz \eqref{eq:ansatz} must be of this form. \\

\noindent A natural question is whether there is an upper bound on $n$, i.e.~the number of non-zero charge components of $B$ and $\bar{B}$, above which no solutions exist. Intuitively, this is expected since $b(q_1)=\ldots=b(q_n)=0$ constitutes $n$ complex equations, whereas there are only three free complex parameters $\alpha,\lambda, z$. Naively, one therefore expects $n\leq 3$. However, property (2a) implies that the system of equations can degenerate when $q_3=2q_2$ and $q_4=2q_2-q_1$. Hence solutions with $n=4$ do exist. In fact, a numerical scan\footnote{This constitutes 11176 potential charge vectors $\mathbf{q}$, of which 57 yield non-trivial solutions.} indicates that for $D\leq 15$ all solutions with $n=4$ are of this form. Furthermore, another numerical scan\footnote{This constitutes 48913 potential charge vectors $\mathbf{q}$, of which 0 yield non-trivial solutions.} shows that for $D\leq 15$, no solutions with $n=5$ exist. In summary, we have made the following two claims
\begin{align}
	&\text{claim (1)}:\qquad \text{For $n=4$, all solutions have $\mathbf{q}=(q_1, q_2, 2q_2, 2q_2-q_1)$}\,,\\
	&\text{claim (2)}:\qquad \text{No solutions exist with $n>4$.}
\end{align}
In the remainder of this work, we will assume that these two claims are true for all $D$, though we stress that the claims have been verified for all $D\leq 15$. 

\subsubsection*{Equation of motion of $g$}

Let us now discuss the remaining equation of motion, which we recall for convenience
\begin{equation}
	\delta g:\qquad \partial\bar{A} = \mu[A,\bar{A}]\,,\qquad \mu = -\frac{1}{1+\lambda}\,.
\end{equation}
Again, our goal is to evaluate this equation on the ansatz \eqref{eq:ansatz} to translate it into a condition on the various charge modes of $\mathbf{B}$. In appendix \ref{sec:app_eomg} this computation is performed in detail, here we simply record the result (see equations \eqref{eq:Bbar} and \eqref{eq:B})
\begin{align}
\label{eq:eomg_qstar_1}	0&=q^*\left(\partial\bar{B}_{q^*}+[B_0,\bar{B}_{q^*}]\right) + \sum_{q\neq 0,q^*}q (1+(q-q^*)\alpha\mu)[B_{q^*-q}, \bar{B}_{q}]\,,\\
\label{eq:eomg_qstar_2}	0&=q^*\left(\bar{\partial}B_{q^*}+[\bar{B}_0,B_{q^*}]\right)+\sum_{q\neq 0,q^*}(q-q^*)(1+q\alpha\mu)[B_{q^*-q}, \bar{B}_{q}]\,.
\end{align}
To be clear, here we have projected the resulting equations onto a general charge-mode, denoted by $q^*$. In principle, these two equations are not independent as one can derive one from the other by using the flatness of $\mathbf{B}$, i.e.~the fact that $\partial\bar{B}-\bar{\partial}B+[B,\bar{B}]=0$. However, in practice it will be useful to use both equations as opposed to the flatness condition. One sees that the equation of motion of $g$ has reduced to a set of differential equations for the various charge components of $B$ and $\bar{B}$ that is similar to Nahm's equations \eqref{eq:charge_plus1}-\eqref{eq:charge_min1}, but is spoiled by the additional cross-terms appearing in the sums. We will now argue that these additional terms take a very simple form. To do so, it is convenient to distinguish the four types of charge-modes that can appear in \eqref{eq:eomg_qstar_1} and \eqref{eq:eomg_qstar_2}.

\subsubsection*{Type (1): $q^*=0$}

When $q^*=0$, the equations \eqref{eq:eomg_qstar_1} and \eqref{eq:eomg_qstar_2} simply reduce to a single equation, namely
\begin{equation}
	0 =\sum_{q\neq 0} q(1+q\alpha\mu)[B_{-q},\bar{B}_{q}]\,.
\end{equation}
Note that it is impossible for the prefactor $1+q\alpha\mu$ to vanish for more than one value of $q$, so this yields a proper constraint on the commutators of the form $[B_{-q},\bar{B}_q]$. 

\subsubsection*{Type (2): $q^*\in\mathbf{q}$}

Suppose there exists a $q\in\mathbf{q}$ such that $\bar{B}_q\neq 0$ and also $B_{q^*-q}\neq 0$. In particular, this requires $q,q-q^*\in\mathbf{q}$. By property (2b), it follows that $1+q\alpha\mu=0$. Therefore, if such a $q$ exists it must be unique. In fact, there can be at most one $q^*\in\mathbf{q}$ for which this is the case. Finally, since $q^*\in\mathbf{q}$ it is impossible to have $-q^*\in\mathbf{q}$, hence it must be that $B_{q^*}=0$. As a result, equation \eqref{eq:eomg_qstar_2} is automatically satisfied, while \eqref{eq:eomg_qstar_1} reduces to
\begin{equation}\label{eq:eomg_case2}
	\partial\bar{B}_{q^*}+[B_0,\bar{B}_{q^*}]+[B_{q^*-q},\bar{B}_q]=0\,.
\end{equation}
Of course, if a $q$ with the above properties does not exist, the last term in this expression is simply absent. 

\subsubsection*{Type (3): $-q^*\in\mathbf{q}$}

As in case (2), suppose there exists a $q\in\mathbf{q}$ such that $\bar{B}_q\neq 0$ and also $B_{q^*-q}\neq 0$, which implies $q,q-q^*\in\mathbf{q}$. By property (2c), it follows that $1+(q-q^*)\alpha\mu=0$, hence the sum in \eqref{eq:eomg_qstar_1} is zero. Furthermore, since already $-q^*\in\mathbf{q}$ it is impossible to have $q^*\in\mathbf{q}$, hence $\bar{B}_{q^*}=0$. In other words, in this case equation \eqref{eq:eomg_qstar_1} is automatically satisfied. Moreover, equation \eqref{eq:eomg_qstar_2} reduces to
\begin{equation}
	\bar{\partial}B_{q^*}+[\bar{B}_0,B_{q^*}]-[B_{q^*-q},\bar{B}_q]=0\,.
\end{equation}
Note that this is simply the complex conjugate of \eqref{eq:eomg_case2} with $q^*\mapsto -q^*$ and $q\mapsto q-q^*$, as expected. 

\subsubsection*{Type (4): $\pm q^*\not\in\mathbf{q}$ and $q^*\neq 0$}

Finally, we discuss the remaining case, which will turn out to be most complex. First, since $\pm q^*\not\in\mathbf{q}$ and $q^*\neq 0$, it follows that $\bar{B}_{q^*}=0$ and $B_{q^*}=0$, so \eqref{eq:eomg_qstar_1} and \eqref{eq:eomg_qstar_2} reduce to 
\begin{align}
	\label{eq:eomg_qstar_1_case4}	0&= \sum_{q\neq 0,q^*}q (1+(q-q^*)\alpha\mu)[B_{q^*-q}, \bar{B}_{q}]\,,\\
	\label{eq:eomg_qstar_2_case4}	0&=\sum_{q\neq 0,q^*}(q-q^*)(1+q\alpha\mu)[B_{q^*-q}, \bar{B}_{q}]\,.
\end{align}
To continue, it will be useful to introduce the `multiplicity' $m(q^*)$ of $q^*$, by which we mean the number of distinct pairs $q_i,q_j\in\mathbf{q}$ whose difference is equal to $q^*$. In other words, $m(q^*)$ counts the number of terms appearing in the sums in \eqref{eq:eomg_qstar_1_case4} and \eqref{eq:eomg_qstar_2_case4}. It is clear that for $n\leq 3$, one can have at most multiplicity two. Furthermore, if $n=4$ one may apply claim (1) to see that the only way to have $m(q^*)=3$ is for $\mathbf{q}$ to be of the form $\mathbf{q}=(2q,q,4q,3q)$. However, one can check that such a $\mathbf{q}$ cannot yield non-trivial solutions to $b(\mathbf{q})=0$. Therefore, since we assume $n\leq 4$ following claim (2), we may restrict to the cases $m(q^*)=1,2$. These will now be discussed in turn. 

\subsubsection*{Type (4a): $m(q^*)=1$}

Let $q_i,q_j\in\mathbf{q}$ and suppose that the difference $q^*=q_i-q_j$ has multiplicity one. In the following, we again assume $\bar{B}_q\neq 0$ for $q=q_i,q_j$, otherwise \eqref{eq:eomg_qstar_1_case4} and \eqref{eq:eomg_qstar_2_case4} are trivially satisfied. Since $m(q^*)=1$, they simply reduce to
\begin{align}
	(1+q_j\alpha \mu)[B_{-q_j},\bar{B}_{q_i}]&=0\,,\\
	(1+q_i\alpha\mu)[B_{-q_j},\bar{B}_{q_i}]&=0\,.
\end{align}
Clearly, since $q_i\neq q_j$ this can hold only when $[B_{-q_j},\bar{B}_{q_i}]=0$. 

\subsubsection*{Type (4b): $m(q^*)=2$}

Next, let $q_i,q_j,q_k,q_l\in\mathbf{q}$ such that the difference $q^*=q_i-q_j=q_k-q_l$ has multiplicity two, which requires $q_i\neq q_k$ and $q_j\neq q_l$. In the following, we again assume $\bar{B}_{q}\neq 0$ for $q=q_i,q_j,q_k,q_l$, otherwise \eqref{eq:eomg_qstar_1_case4} and \eqref{eq:eomg_qstar_2_case4} are trivially satisfied. We turn to \eqref{eq:eomg_qstar_1_case4} and \eqref{eq:eomg_qstar_2_case4}, which reduce to
\begin{align}
	\text{$\bar{B}_{q^*} = 0$}:\qquad q_i(1+q_j\alpha \mu)[B_{-q_j},\bar{B}_{q_i}]+ q_k(1+q_l\alpha \mu)[B_{-q_l},\bar{B}_{q_k}]&=0\,,\\
	\text{$B_{q^*}=0$}:\qquad q_j(1+q_i\alpha\mu)[B_{-q_j},\bar{B}_{q_i}]+q_l(1+q_k\alpha\mu)[B_{-q_l},\bar{B}_{q_k}]&=0\,.
\end{align}
Since both equations must hold, one may combine them succinctly as a matrix equation
\begin{equation}\label{eq:eomg_M}
	M\cdot \begin{pmatrix}
		[B_{-q_j},\bar{B}_{q_i}]\\ [B_{-q_l},\bar{B}_{q_k}] 
	\end{pmatrix}=0\,,\qquad M = \begin{pmatrix}
		q_i(1+q_j\alpha\mu) & q_k(1+q_l\alpha\mu)\\
		q_j(1+q_i\alpha\mu) & q_l(1+q_k\alpha\mu)
	\end{pmatrix}\,,
\end{equation}
with
\begin{align}
	\det M &= (q_i q_l - q_j q_k) - \left[q_i q_j q_k+q_j q_k q_l - q_i q_j q_l - q_i q_k q_l\right]\alpha\mu\,.
\end{align}
Using the fact that $q_i-q_j=q_k-q_l$, one finds that
\begin{equation}
	\det M = 0\quad\iff\quad 1+(q_j+q_k)\alpha\mu = 0\,.
\end{equation}
From property (2c) it then follows that $\det M=0$ if and only if $q_j+q_k\in\mathbf{q}$. In particular, the equation \eqref{eq:eomg_M} only has a non-trivial solution when $q_j+q_k\in\mathbf{q}$. In summary, we have shown that
\begin{equation}
		[B_{-q_j},\bar{B}_{q_i}] = [B_{-q_l},\bar{B}_{q_k}]=0\quad\text{unless}\quad q_j+q_k\in\mathbf{q}\,.
\end{equation}
It is interesting to note that this precisely occurs in the case $n=4$, for which the charge vectors take the form (recall claim (1))
\begin{equation}
	\mathbf{q} = (q_1, q_2, 2q_2, 2q_2-q_1)\,.
\end{equation}
Indeed, one has $(2q_2-q_1)-(q_2) = q_2-q_1$ and additionally $2q_2\in\mathbf{q}$, so that a non-trivial solution exists. 

\subsubsection*{Summary of the analysis}

For the convenience of the reader, let us collect the results of this section. To simplify the overview, we only consider the $\bar{B}$ component. Statements about the $B$ component are naturally obtained by complex conjugation. First, by studying the equations of motion of $A$ and $\bar{A}$, it was argued that the charge decomposition of $\bar{B}$ must take the following form
\begin{equation}\label{eq:sol_barB}
	\bar{B} = \bar{B}_0+\bar{B}_{q_1}+\cdots+\bar{B}_{q_n}\,,
\end{equation}
where $q_i\neq 0$ and furthermore $q_i\neq\pm q_j$ for any $i\neq j$. In order to satisfy the equations of motion, the integers $q_i$ and parameters $\alpha,\lambda,z$ must be such that $b(q_i)=0$ for all $i$. Also, due to the particular structure of $b(q)$, it appears impossible to have $n>4$ based on numerical analysis, while for $n=4$ only special combinations of charges are allowed (recall claim (1)), although we have not rigorously proven this. \\

\noindent Second, the dynamics of the $\bar{B}_{q_i}$ modes are restricted via the equation of motion of $g$. Concretely, we have found that they must satisfy
\begin{equation}\label{eq:dynamics_B}
	\partial\bar{B}_{q_i}+[B_0,\bar{B}_{q_i}]+[B_{q_i-q_j},\bar{B}_{q_j}]=0\,.
\end{equation}
Additionally, there exists at most one value of $q_i$ for which there exists a $q_j$ such that the last term is non-zero. In that case, also $q_j$ is unique. Strikingly, this means that all the $\bar{B}_{q_i}$ modes satisfy a variant of Nahm's equations \eqref{eq:charge_plus1}-\eqref{eq:charge_min1} apart from possibly one of the modes, for which this one additional term appears. \\

\noindent Lastly, from the discussion of type (4), all the commutators of the form $[B_{-q_j},\bar{B}_{q_i}]$ with $q_i\neq q_j$ and $q_i-q_j\not\in\mathbf{q}$ must in fact vanish, except when $q_i-q_j=q_k-q_l$, for $q_i\neq q_k$, and additionally $q_j+q_k\in\mathbf{q}$. We stress that this conclusion only holds for $n\leq 4$. Moreover, when $q_i=q_j$ corresponding to type (1), then the commutators are restricted to satisfy
\begin{equation}\label{eq:constraint_commutator}
	\sum_{i=1}^n q_i(1+q_i\alpha\mu)[B_{-q_i},\bar{B}_{q_i}]=0\,.
\end{equation}
This concludes the analysis of the equations of motion. 

\subsubsection*{Properties of the solutions}

We close this section with some final remarks on properties of solutions to the equations of motion, which shed further light on the properties we encountered for the VHS solution. To this end, it is convenient to distinguish the cases $n=1$ and $n\geq 1$, with $n$ counting the number of non-zero charge-modes of $B$ and $\bar{B}$, as in \eqref{eq:sol_barB}.

\subsubsection*{Case (I): $n=1$}

In this case one simply has $\bar{B} = \bar{B}_0 + \bar{B}_q$, as for the VHS solution, except for generic $z$ the condition $b(q)=0$ also allows for even values of $q$. In fact, one can solve the conditions $b(q)=0$ and $1+q\alpha\mu=0$ algebraically to find
\begin{equation}
	\alpha q = 1+\lambda\,\qquad \lambda = \pm z^{-q/2}\,.
\end{equation}
Note that for $z=-1$ and $q$ odd this precisely reduces to the VHS solution, as expected. Again the two solutions we find are related via the $\mathbb{Z}_2$ symmetry \eqref{eq:Z2_symmetry}. The main observation we would like to make is that whenever $|z|=1$, which was required for $g$ to be real, it must again be that also $|\lambda| =1$.

\subsubsection*{Case (II): $n\geq 2$}

When $n\geq 2$, one needs to solve $b(q_1)=\cdots=b(q_n)=0$. It is actually possible to solve $b(q_1)=b(q_2)=0$ algebraically for $\alpha$ and $\lambda$, with the result given by
\begin{align}
	b(q_1)=b(q_2)=0\quad\iff\quad \alpha = -\frac{F(z^{q_1},z^{q_2})}{F(q_2,q_1)}\,,\qquad \lambda  = \frac{F(z^{q_1},z^{q_2})}{F(1,1)}\,,
\end{align}
where, for convenience, we have introduced the function
\begin{equation}
	F(A,B) = A q_1(1-z^{q_2})-B q_2 (1-z^{q_1})\,.
\end{equation}
Unfortunately, it is generically not possible to additionally solve $b(q_3)=0$ analytically to obtain a closed expression\footnote{For the interested reader, we record the actual equation that needs to be solved:
\begin{equation*}
b(-q_3)=0\quad\iff\quad	\sum_{\sigma\in S_3} (-1)^{\mathrm{sign}\,\sigma} q_{\sigma(1)}q_{\sigma(2)} z^{q_{\sigma(1)}}\left(1-z^{q_{\sigma(3)}}\right) = 0\,,
\end{equation*}
which is a degree $\mathrm{max}_{i,j=1,2,3}(q_i+q_j)$ polynomial in $z$, which is furthermore fully anti-symmetric in $q_1,q_2,q_3$.} for $z$. Nevertheless, the above solution has a striking feature. Namely, one can show that whenever $|z|=1$ it must again be true that $|\lambda|=1$, as in case (I). \\

\noindent From the above discussion one concludes that in general
\begin{equation}
	|z|=1\quad\implies\quad |\lambda|=1\,.
\end{equation}
As remarked earlier, this is an interesting observation when compared with the study of a duality between $\lambda$-deformations and $\eta$-deformations \cite{Klimcik_2015}. For completeness, we also remark that for this general set of solutions, the on-shell action again takes the form (see appendix \ref{sec:app_action} for more details)
\begin{equation}
	S_\lambda[g] = i \sum_{q\neq 0}\left[ f(q,\arg z,\arg \lambda) \int_\Sigma d^2t \,\mathrm{Tr}\left(B_{-q}\bar{B}_{+q}\right)\right]\,,
\end{equation}
where $f(q,\arg z,\arg \lambda)$ is a real function of $q$ and the phases of $z$ and $\lambda$ and we have omitted the WZ-term.

\begin{subappendices}

\section{Properties of $b(q)$}
\label{sec:app_beta}

In this section we present and derive some algebraic properties of the function
\begin{equation}
	\tilde{b}(q) = z^q - \frac{1-\alpha q\lambda^{-1}}{1-\alpha q}\,.
\end{equation}
Note that this differs from the function $b(q)$ in \eqref{eq:beta} by a factor of $(1-\alpha q)/(\lambda^{-1}-z^q)$. However, since we assume $\alpha q\neq 1$ and $\lambda\neq z^{-q}$, and are only interested in the zeroes of $b(q)$, it suffices to consider $\tilde{b}(q)$. This will turn out to be slightly more convenient. As discussed in the main text, we additionally exclude the following values of the parameters: $z=0,z^q=1$, $\alpha=0$, $\lambda=\pm 1$. \\

\noindent We are interested in the case where $\tilde{b}(q_1)=\tilde{b}(q_2)=0$ for two distinct integers $q_1,q_2$. One can solve this algebraically to find
\begin{align}\label{eq:sol_alpha_lambda}
	\tilde{b}(q_1)=\tilde{b}(q_2)=0\quad\iff\quad \alpha = -\frac{F(z^{q_1},z^{q_2})}{F(q_2,q_1)}\,,\qquad \lambda  = \frac{F(z^{q_1},z^{q_2})}{F(1,1)}\,,
\end{align}
where, for convenience, we have introduced the function
\begin{equation}
	F(A,B) = A q_1(1-z^{q_2})-B q_2 (1-z^{q_1})\,.
\end{equation}
Using this result, we will now provide a proof of the properties (2b) and (2c), see also \eqref{eq:2b} and \eqref{eq:2c}.

\subsubsection*{Property (2b)}
First, we prove property (2b), which states that
\begin{equation}\label{eq:prop2b}
	1+q_1\alpha\mu = 0\quad\iff\quad b(q_1-q_2)=0\,,
\end{equation}
To this end, one computes both the left-hand side and right-hand side, with $\alpha$ and $\lambda$ given by \eqref{eq:sol_alpha_lambda}. For the right-hand side, this yields
\begin{equation}
	\tilde{b}(q_1-q_2) = z^{-q_2}\frac{\left[z^{q_1}q_1^2(1-z^{q_2})^2-2q_1 q_2(1-z^{q_2})^2+q_2^2(1-z^{q_1})(z^{q_1}-z^{2q_2})\right]}{F(2q_2-q_1, q_2)}
\end{equation}
while the left-hand side becomes
\begin{equation}
	1+q_1\alpha\mu = \frac{\left[z^{q_1}q_1^2(1-z^{q_2})^2-2q_1 q_2(1-z^{q_2})^2+q_2^2(1-z^{q_1})(z^{q_1}-z^{2q_2}) \right]}{q_2(z^{q_1}-z^{q_2})F(1+z^{q_1}, 1+z^{q_2})}\,.
\end{equation}
Indeed, one sees that both the left-hand side and right-hand side are proportional to the same expression (in square brackets), hence we have proven \eqref{eq:prop2b}.

\subsubsection*{Property (2c)}
Finally, we prove property (2c), which states that
\begin{equation}\label{eq:prop2c}
	1+(q_1+q_2)\alpha\mu = 0\quad\iff\quad b(q_1+q_2)=0\,.
\end{equation}
To this end, one first computes the right-hand side, which yields
\begin{equation}
	\tilde{b}(q_1+q_2) = -\frac{q_1^2z^{q_1}\left(1-z^{q_2}\right)^2-q_2^2z^{q_2}\left(1-z^{q_1}\right)^2}{F(q_1,q_2)}\,,
\end{equation}
while the left-hand side is equal to
\begin{equation}
	1+(q_1+q_2)\alpha\mu = \frac{q_1-q_2}{q_1q_2(z^{q_1}-z^{q_2})}\times \frac{q_1^2z^{q_1}\left(1-z^{q_2}\right)^2-q_2^2z^{q_2}\left(1-z^{q_1}\right)^2}{F(1+z^{q_1},1+z^{q_2})}\,.
\end{equation}
Again, one sees that the left-hand side and right-hand side are proportional to the same factor, hence 
\eqref{eq:prop2c} follows. 

\section{Equation of motion of $g$}\label{sec:app_eomg}

In this section we present some explicit computations regarding the equation of motion of $g$, i.e.
\begin{equation}
	\partial\bar{A}=\mu[A,\bar{A}]\,.
\end{equation}
Using the relations \eqref{eq:g_to_B} and \eqref{eq:A_to_B} one readily computes
\begin{align}
	\partial\bar{A} &= \alpha \sum_{q\neq 0} q\, h\left( \partial\bar{B}_q + [h^{-1}\partial h, \bar{B}_q]\right)h^{-1} = \alpha \sum_{q\neq 0} q\,h \left( \partial\bar{B}_q + \sum_{q'} [B_{q'},\bar{B}_q]\right) h^{-1}\,.
\end{align}
It follows that
\begin{align}
	\partial\bar{A} = \mu[A,\bar{A}]\quad &\iff \quad \alpha \sum_{q\neq 0} q\, \left( \partial\bar{B}_q + \sum_{q'} [B_{q'},\bar{B}_q]\right) = \alpha^2 \mu \sum_{q\neq 0} \sum_{q'\neq 0} q q' [B_{q'},\bar{B}_q]\,,\\
	&\iff\quad  0= \sum_{q\neq 0}q\left[ \partial\bar{B}_q+ \sum_{q'}(1-q'\alpha\mu)[B_{q'},\bar{B}_q]\right]\,.
\end{align}
Projecting the final equation onto $q^*$ modes and relabeling $q'\rightarrow q$ gives
\begin{align}\label{eq:Bbar}
	 0=q^*\left(\partial\bar{B}_{q^*}+[B_0,\bar{B}_{q^*}]\right) + \sum_{q\neq 0,q^*}q (1+(q-q^*)\alpha\mu)[B_{q^*-q}, \bar{B}_{q}]\,.
\end{align}
Using the Bianchi identity
\begin{align}
	0 &= (\partial\bar{B}-\bar{\partial}B+[B,\bar{B}])_{q^*}\\
	&=\left(\partial\bar{B}_{q^*}+[B_0,\bar{B}_{q^*}]\right)-\left(\bar{\partial}B_{q^*}+[\bar{B}_0,B_{q^*}]\right)+\sum_{q\neq 0,q^*} [B_{q^*-q},\bar{B}_q]\,,
\end{align}
and subtracting it from \eqref{eq:Bbar} yields the similar equation
\begin{equation}\label{eq:B}
	0=q^*\left(\bar{\partial}B_{q^*}+[\bar{B}_0,B_{q^*}]\right)+\sum_{q\neq 0,q^*}(q-q^*)(1+q\alpha\mu)[B_{q^*-q}, \bar{B}_{q}]\,.
\end{equation}
In the main text, equations \eqref{eq:Bbar} and \eqref{eq:B} are studied further.

\section{Evaluating the on-shell action}\label{sec:app_action}

By evaluating the action on the constraints \eqref{eq:Abar_ginvdbarg} and \eqref{eq:A_ginvdg} one obtains the on-shell action
\begin{equation}
	S_\lambda[g] =\frac{k}{\pi}\int_\Sigma d^2t\,\mathrm{Tr}\left(g^{-1}\partial g\left[\frac{1}{2}+\frac{\lambda\,\mathrm{Ad}_g}{1-\lambda \,\mathrm{Ad}_g} \right]g^{-1}\bar{\partial}g \right)+ S_{\mathrm{WZ}}[g]\,.
\end{equation}
By recalling that
\begin{equation}
	g^{-1}\bar{\partial}g = -\sum_{q\neq 0} (1-z^{-q})h \bar{B}_q h^{-1}\,,
\end{equation}
one readily sees that
\begin{equation}
	\frac{\lambda\,\mathrm{Ad}_g}{1-\lambda\, \mathrm{Ad}_g} g^{-1}\bar{\partial}g = -\sum_{q\neq 0}(1-z^{-q}) \frac{\lambda z^{q}}{\lambda-z^{q}}h \bar{B}_q h^{-1}\,.
\end{equation}
Furthermore, since for non-zero $q,q'$
\begin{equation}
	\mathrm{Tr}\left(\mathcal{O}_q\cdot\mathcal{O}_{q'}\right)=\frac{1}{q'}\mathrm{Tr}\left(\mathcal{O}_q\cdot[Q_{\mathrm{ref}},\mathcal{O}_{q'}]\right) = -\frac{1}{q'}\mathrm{Tr}\left([Q_{\mathrm{ref}},\mathcal{O}_q]\cdot\mathcal{O}_{q'}\right)=-\frac{q}{q'}\mathrm{Tr}\left(\mathcal{O}_q\cdot\mathcal{O}_{q'}\right)\,,
\end{equation}
it follows that $\mathrm{Tr}\left(\mathcal{O}_q\cdot\mathcal{O}_{q'}\right)$ is only non-zero for $q'=-q$. Therefore the on-shell action simplifies to
\begin{equation}
	S_\lambda[g] =\frac{k}{\pi}\sum_{q\neq 0}\left[(1-z^{q})(1-z^{-q})\left(\frac{1}{2}-\frac{1}{1-z^{-q}/\lambda}\right)\int_\Sigma d^2t\,\mathrm{Tr}\left(B_{-q}\bar{B}_{+q} \right)\right] +S_{\mathrm{WZ}}[g]
\end{equation}
The term inside the integral is positive definite, while the coefficients satisfy
\begin{equation}
	\mathrm{Re}\left[\frac{1}{2}-\frac{1}{1-z^{-q}/\lambda}\right] = 0 \quad \iff \quad |\lambda| = |z|^{-q}. 
\end{equation}
In particular, when $|z|=|\lambda|=1$ this is satisfied for all $q$, hence the action is purely imaginary-valued. Concretely, introducing the angles
\begin{equation}
	z = e^{2i\theta}\,,\qquad \lambda = e^{2i\phi}\,,
\end{equation} 
one finds
\begin{equation}
	S_\lambda[g] =\frac{2i k}{\pi}\sum_{q\neq 0}\left[\sin^2(q\theta)\cot(q\theta+\phi)\int_\Sigma d^2t\,\mathrm{Tr}\left(B_{-q}\bar{B}_{+q} \right)\right] +S_{\mathrm{WZ}}[g]
\end{equation}
Moreover, for the VHS solution one has $\theta=\pi/2$ and $\phi = \pm \pi/4$ and the summand only runs over $q=1$. In that case, the on-shell action reduces to
\begin{equation}
	S_\lambda[g] = \mp\frac{2i k}{\pi}\int_\Sigma d^2t\,\mathrm{Tr}\left(B_{-1}\bar{B}_{+1} \right) +S_{\mathrm{WZ}}[g]\,.
\end{equation}
\end{subappendices}

\chapter{Bi-Yang--Baxter models and Sl(2)-orbits}
\label{chap:biYB}
\epigraph{This chapter is based on: Thomas W. Grimm, Jeroen Monnee: \emph{Bi-Yang--Baxter models and Sl(2)-orbits}, \textbf{JHEP 11 (2023) 123},  \href{https://arxiv.org/abs/2212.03893}{\textbf{[arXiv: 2212.03893]}}}

\noindent In this chapter we will focus on another 
class of two-dimensional integrable non-linear $\sigma$-models, different from the $\lambda$-deformations discussed in the previous chapter, and show that once again one can employ the powerful tools of Hodge theory
to determine some of their classical solutions. Our findings will further strengthen the connection between integrable models and this vast field of mathematics observed in \cite{Grimm:2021idu}.\\

\noindent The class of integrable models considered in this chapter are known as bi-Yang--Baxter models, which were introduced in \cite{Klimcik:2008eq}. To construct these models one starts with the principal chiral model, which is a
non-linear $\sigma$-model encoding the dynamics of a field valued in a group $G$. The bi-Yang--Baxter model is then defined as a two-parameter deformation that depends on a Yang--Baxter operator $R$ satisfying a modified classical Yang--Baxter equation. These models can be abstractly defined for any group $G$ when making sure that an appropriate $R$-matrix is constructed, e.g.~following the classical work of Drinfel'd--Jimbo \cite{Drinfeld:466366,Jimbo:466362}, see also \cite{Belavin:1982}. Despite its general definition, the study of solutions of such models has so far been restricted to only the simplest choices of $G$. In particular, the $G=\mathrm{SU}(2)$ bi-Yang--Baxter model has been investigated in \cite{Schepers:2020ehn}. Our first aim is to provide a new perspective on the classical 
solutions found in \cite{Schepers:2020ehn} that allows for a natural generalization to higher-rank groups $G$. \\

\noindent Our study of solutions to the bi-Yang--Baxter model will be restricted to a special one-parameter subspace of the two-parameter moduli space, where the symmetries of the model enhance and additional dualities to other theories emerge \cite{Klimcik:2008eq,Klimcik:2014bta,Delduc:2013qra,Hoare:2014oua}. 
These models will be referred to as critical bi-Yang--Baxter models following \cite{Schepers:2020ehn}. In the simplest 
situation, namely the critical SL$(2,\bbR)$ bi-Yang--Baxter model, we will show that the equations of motion are solved by the Weil operator $C$ associated to a two-torus. In analogy to the approach taken in chapter \ref{chap:WZW}, it is thus key to identify the complex structure deformation space of the two-torus, namely the upper half-plane, with part of the two-dimensional space-time. Furthermore, it is a general fact that the Weil operator of a two-torus obeys $C^2=-1$, and hence that the solutions must be a special class of all possible solutions. Exactly these types of solutions were called uniton solutions in the literature on integrable models and were first considered for the SU$(N)$ principal chiral model in \cite{Uhlenbeck:1989}. We explicitly show 
that the Weil operator is equivalent to a certain complex uniton solution of the $\mathrm{SU}(2)$ bi-Yang--Baxter
model that has been constructed in \cite{Schepers:2020ehn}, see also \cite{Demulder:2016mja}. \\

\noindent A particular feature of the Weil operator of a two-torus is that it is exactly equal to its $\mathrm{Sl}(2)$-orbit approximation. Let us recall from chapter \ref{chap:asymp_Hodge_I} that, roughly speaking, this means that the Weil operator changing over space-time can be written as an orbit of some special fixed Weil operator. The orbit is derived by picking 
a distinguished element of $\mathrm{SL}(2,\mathbb{R})$ and parameterizing the transformations by this element with a complex parameter labelling the space-time position. Due to the result of the $\mathrm{Sl}(2)$-orbit theorem, discussed in chapter \ref{chap:asymp_Hodge_I}, it is now well-known that $\mathrm{Sl}(2)$-orbits are in fact ubiquitous in the study of general variations of Hodge structures. In geometric settings, the two-dimensional space-time of the bi-Yang--Baxter model would then be identified with the complex structure deformation space of the underlying geometry.\footnote{Due to the underlying complex structure it will be most natural to work in Euclidean signature.} From an abstract point of view, the central ingredient underlying an $\mathrm{Sl}(2)$-orbit is a horizontal $\mathfrak{sl}(2)$-triple, recall equation \eqref{eq:horizontal_sl2_real}. Given such a horizontal $\mathfrak{sl}(2)$-triple, we argue that it selects a particular class of $R$-matrices and we explicitly show that the associated $\mathrm{Sl}(2)$-orbit approximation of the Weil operator solves the equations of motion of the corresponding critical bi-Yang--Baxter model. In general, the Weil operator associated to a weight $D$ variation of Hodge structure satisfies $C^2=(-1)^D$. Furthermore, we argue that the resulting solution has finite action. It can therefore be thought of as a generalization of the complex uniton solution of the $\mathrm{SL}(2,\mathbb{R})$ model to groups of higher rank. Typical examples are $G=\mathrm{Sp}(2n,\mathbb{R})$ and $G=\mathrm{SO}(r,s)$, as discussed in section \ref{sec:VHS}, or more generally reductive subgroups thereof \cite{Robles:2015}. For illustrative purposes, we have included an explicit example of a solution to the critical  $\mathrm{Sp}(4,\mathbb{R})$ bi-Yang--Baxter model based on the type $\mathrm{IV}_1$ boundary data constructed in section \ref{sec:asymp_Hodge_examples}.\\

\noindent Let us also remark that horizontal $\mathfrak{sl}(2)$-triples have been classified in the mathematics literature \cite{Robles:2015,Kerr2017}. Therefore, these general mathematical results provide a concrete classification of uniton solutions of the bi-Yang--Baxter model and further indicates the significance of Hodge theory in the study of integrable models. In this chapter, we do not aim to develop this latter point of view in full, but rather aim to lay the foundations for further studying this connection. \\

\noindent The chapter is organized as follows. In section \ref{sec:bi_YB_unitons} we introduce the bi-Yang--Baxter model. The action of this theory depends on the choice of an $R$-matrix and we recall how the Drinfel'd--Jimbo solution for $R$ indeed satisfies the classical modified Yang-Baxter equation. We then discuss the SU$(2)$ example in detail and introduce its uniton solutions. We give a detailed account on how this solution can be related to the Weil operator of a two-torus and point out that the solution can be described purely in terms of a horizontal $\mathfrak{sl}(2)$-triple. In section \ref{sec:Hodge_Theory} the solution is then extended to higher-rank groups by identifying the group-valued field with the general $\mathrm{Sl}(2)$-orbit of special fixed Weil operators. We argue that the horizontal $\mathfrak{sl}(2)$-triple selects a particular class of $R$-matrices and explicitly show that the equations of motion of the corresponding bi-Yang--Baxter model are solved at the critical point. We end with an illustrative example and comment on possible generalizations of the proposed solutions. In the appendix we have included some computational details and elaborate on some of the expressions used in the example of section~\ref{subsec:example}.

\section{The bi-Yang--Baxter model and unitons}
\label{sec:bi_YB_unitons}

In this section we analyze classical aspects of the bi-Yang--Baxter model. In section \ref{subsec:bi_YB} we introduce the model and establish our notation and conventions. Then, in section \ref{subsec:unitons}, we study the $\mathrm{SU}(2)$ model in more detail and consider a class of finite action solutions known as \textit{unitons}. Finally, in section \ref{subsec:Weil_SL2} we observe a relation between the complex uniton solution and the Weil operator of a two-torus. This observation will then lead us to consider more general solutions for Weil operators of arbitrary variations of Hodge structure in section \ref{sec:Hodge_Theory}. 

\subsection{Bi-Yang--Baxter model}
\label{subsec:bi_YB}

Let us start by introducing the basics of the bi-Yang--Baxter model. The reader who is already familiar with the topic can safely skip this section. The model was originally introduced by Klim\v{c}\'ik in \cite{Klimcik:2008eq} as a two-parameter integrable deformation of the principal chiral model. In particular, it is a non-linear $\sigma$-model for a group-valued field
\begin{equation}
    g:\Sigma\rightarrow G\,,
\end{equation}
where $\Sigma$ is the two-dimensional worldsheet and $G$ is a real Lie group, whose Lie algebra will be denoted by $\mathfrak{g}$. We will take the worldsheet to have Euclidean signature, and introduce complex coordinates $z,\bar{z}$ with $z=x+iy$. Additionally, we will assume $\mathfrak{g}$ to be simple and denote by 
\begin{equation}
    (\cdot,\cdot):\mathfrak{g}\times\mathfrak{g}\rightarrow\mathbb{R}
\end{equation}
the (up to an overall scaling) unique invariant symmetric bilinear form on $\mathfrak{g}$.

\subsubsection*{$R$-matrix}
The bi-Yang--Baxter model lies in the class of Yang--Baxter deformations of non-linear $\sigma$-models, all of which involve an object called the (classical) $R$-matrix. It is an endomorphism of the Lie algebra $\mathfrak{g}$, i.e.~a linear map
\begin{equation}
    R:\mathfrak{g}\rightarrow\mathfrak{g}\,,
\end{equation}
satisfying the following equation.
\begin{subbox}{The modified classical Yang--Baxter equation}
	\begin{equation}
		\label{eq:YBE}
		[RX, RY]-R\left([RX, Y]+[X,RY]\right)=-c^2[X,Y]\,,
	\end{equation}
	for all $X,Y\in\mathfrak{g}$. Here $c^2$ is a real constant. 
	\tcblower
	\textbf{Note:}\\
	The word `modified' refers to the fact that $c$ is allowed to be non-zero. Note that by a real rescaling of $R$ we may restrict to the cases $c\in\{0,1,i\}$. 
\end{subbox}
\noindent In the following, we will additionally impose the condition that $R$ is skew-symmetric with respect to the chosen bilinear form on $\mathfrak{g}$. In other words
\begin{equation}
\label{eq:skew}
    (RX,Y)+(X,RY)=0\,,\qquad \forall X,Y\in\mathfrak{g}\,.
\end{equation}
Such $R$-matrices are also referred to as Yang--Baxter operators and have been classified in the mathematics literature \cite{Belavin:1982}, see also \cite{Chari:1995}. 

\subsubsection*{Drinfel'd--Jimbo $R$-matrix}

There is a standard solution to the modified classical Yang--Baxter equation, given by the Drinfel'd--Jimbo solution \cite{Drinfeld:466366,Jimbo:466362}. In order to write it down, let us first consider the complexification $\mathfrak{g}_\mathbb{C}$ of $\mathfrak{g}$ and let $\mathfrak{h}$ be a Cartan subalgebra of $\mathfrak{g}_\mathbb{C}$. Then $\mathfrak{g}_\mathbb{C}$ enjoys a root space decomposition
\begin{equation}
    \mathfrak{g}_\mathbb{C}=\mathfrak{h}\oplus\bigoplus_\alpha\mathfrak{g}_\alpha\,,
\end{equation}
where $\alpha\in\mathfrak{h}^*$ runs over all the roots and each $\mathfrak{g}_\alpha$ denotes the root space associated to a root $\alpha$. A choice of simple roots fixes a notion of positive roots, denoted by $\alpha>0$. Let $\{H^\mu, E^{\pm\alpha}\}$ denote a Cartan--Weyl basis of $\mathfrak{g}_\mathbb{C}$, where $\alpha$ runs over the positive roots. These generators satisfy the usual commutation relations
\begin{equation}
    [H^\mu, H^\nu]=0\,,\qquad [H^\mu, E^{\pm\alpha}]=\pm\alpha(H^\mu)E^{\pm\alpha}\,.
\end{equation}
In terms of this basis, the Drinfel'd--Jimbo $R$-matrix is defined as\footnote{The overall sign of $R$ is, of course, a matter of convention. Here we have chosen the sign to match with \cite{Hoare:2021dix}.}
\begin{equation}
\label{eq:def_R_DJ}
    RH^\mu=0\,,\qquad RE^{\pm\alpha}=\mp c\,E^{\pm\alpha}\,.
\end{equation}
One can verify by direct computation that \eqref{eq:def_R_DJ} solves the modified classical Yang--Baxter equation \eqref{eq:YBE} and that it satisfies the skew-symmetry condition \eqref{eq:skew}. However, it is important to keep in mind that the above $R$-matrix is defined on the complexified algebra $\mathfrak{g}_\mathbb{C}$. Depending on the choice of real form $\mathfrak{g}$ and the constant $c$, it may happen that $R$ is not a real endomorphism of $\mathfrak{g}$. For further details we refer the reader to the lecture notes \cite{Hoare:2021dix}.

\subsubsection*{Action}
For a given choice of Yang--Baxter operator $R$, the action of the associated bi-Yang--Baxter model is given as follows.
\begin{subbox}{Bi-Yang--Baxter model}
	\begin{equation}
		\label{eq:biYB}
		S = \int_\Sigma d^2\sigma\,\left(g^{-1}\partial_+g\,,\frac{1}{1-\eta R-\zeta R^g}g^{-1}\partial_- g\right)\,,
	\end{equation}
	where $\eta$ and $\zeta$ are two constants parametrizing the deformation, and we have introduced the notation
	\begin{equation}
		R^g:= \mathrm{Ad}_{g^{-1}}\circ R\circ\mathrm{Ad}_g\,.
	\end{equation}
\end{subbox}
\noindent Clearly, for $\eta=\zeta=0$ one recovers the action of the principal chiral model, which enjoys a global $G_L\times G_R$ symmetry. Keeping $\zeta=0$ but letting $\eta$ be non-zero, the introduction of the operator $R$ breaks the global $G_R$ symmetry down to the $\mathrm{U}(1)_R^{\mathrm{rk}\,G}$ subgroup. In this case, one retrieves the action of the (single-parameter) Yang--Baxter model, which is also referred to as the $\eta$-model \cite{Klimcik:2002zj}. Upon letting $\zeta$ be non-zero also the global $G_L$ symmetry is broken down to the $\mathrm{U}(1)_L^{\mathrm{rk}\,G}$ subgroup. \\

\noindent There is a special point in the parameter space where the symmetry of the bi-Yang--Baxter model is enhanced \cite{Klimcik:2014bta}. Indeed, whenever $\zeta=\eta$, which we will refer to as the \textit{critical line} following \cite{Schepers:2020ehn}, one has an additional symmetry given by $g\mapsto g^{-1}$. The critical bi-Yang--Baxter model will play a crucial role in this work, as we will find a set of solutions to the model which solve the model precisely when $\zeta=\eta$. There have been a number of observations that indicate that the critical bi-Yang--Baxter model can be related to other integrable models. For example, the critical $\mathrm{SU}(2)$ bi-Yang--Baxter model is equivalent to the coset $\mathrm{SO}(4)/\mathrm{SO}(3)$ $\eta$-model \cite{Delduc:2013qra,Hoare:2014oua}. Furthermore, at the conformal point $\zeta=\eta=\frac{i}{2}$ the target space geometry coincides with that of the $\mathrm{SU}(1,1)/\mathrm{U(1)}$ gauged WZW model with an additional $\mathrm{U}(1)$ boson \cite{Hoare:2014oua}.

\subsubsection*{Equations of motion}
In order to write down the equations of motion of the bi-Yang--Baxter model, it is convenient to introduce the currents (we follow the conventions of \cite{Klimcik:2014bta})
\begin{equation}
\label{eq:def_Jpm}
    J_\pm= \mp\frac{1}{1\pm \eta R\pm \zeta R^g}j_\pm\,,\qquad j_\mu = g^{-1}\partial_\mu g\,.
\end{equation}
In terms of $J_\pm$, the equations of motion read
\begin{equation}
\label{eq:eom_biYB}
    \partial_+ J_--\partial_-J_+ - \eta[J_+,J_-]_R=0\,,
\end{equation}
where
\begin{equation}
\label{eq:R_bracket}
    [X,Y]_R:=[RX,Y]+[X,RY]\,,\qquad \forall X,Y\in\mathfrak{g}\,,
\end{equation}
defines a second Lie-bracket on $\mathfrak{g}$, by virtue of the classical Yang--Baxter equation \eqref{eq:YBE}. It is referred to as the $R$-bracket and is central for the underlying Poisson-Lie symmetry of the bi-Yang--Baxter model, see e.g.~\cite{Sfetsos:2015nya}.  

\subsubsection*{Integrability and Lax connection}
It was shown by Klim\v{c}\'ik in \cite{Klimcik:2014bta} that the bi-Yang--Baxter model is classically integrable. Vital in this regard is the condition that $R$ satisfies the modified classical Yang--Baxter \eqref{eq:YBE} equation and is anti-symmetric. The integrability condition means that the equations of motion of the bi-Yang--Baxter model can be reformulated as the zero-curvature condition of a Lax connection
\begin{equation}
    \mathscr{L}_\pm(\lambda) = \left(\eta(R-i)+\frac{2i\eta\pm(1-\eta^2+\zeta^2)}{1\pm \lambda}\right)J_\pm\,,
\end{equation}
where $\lambda\in\mathbb{C}$ is the spectral parameter (not to be confused with the deformation parameter $\lambda$ of chapter \ref{chap:WZW}). More precisely, introducing the connection
\begin{equation}
    \nabla_\pm = \partial_\pm+\mathscr{L}_\pm(\lambda)\,,
\end{equation}
the zero-curvature condition simply states that
\begin{equation}
    [\nabla_+,\nabla_-]=0\,,
\end{equation}
for all $\lambda$. This is equivalent to the equations of motion \eqref{eq:eom_biYB} together with the Bianchi identities for $J_\pm$. The flatness of the Lax connection ensures the existence of an infinite tower of conserved charges. The Hamiltonian integrability of the model, i.e. the condition that all these charges in fact Poisson-commute with each other, was established in \cite{Delduc:2015xdm}.

\subsection{$\mathrm{SU}(2)$ unitons}
\label{subsec:unitons}

In this section, we restrict to the case where $G=\mathrm{SU}(2)$ and study a class of finite action solutions to the classical theory. These solutions are referred to as \textit{unitons}, owing to the fact that they are analogous to instantons and additionally satisfy the requirement that $g^2=-1$. They were originally constructed by K.~Uhlenbeck as solutions to the $\mathrm{SU}(N)$ principal chiral model \cite{Uhlenbeck:1989}, and were later extended to the Yang--Baxter and bi-Yang--Baxter models in \cite{Demulder:2016mja,Schepers:2020ehn} for $N=2$. Our discussion closely follows the works \cite{Demulder:2016mja,Schepers:2020ehn}.

\subsubsection*{$\mathfrak{su}(2)$ $R$-matrix}

For $\mathfrak{su}(2)$, the solution to the modified Yang--Baxter equation is essentially unique and is given by the Drinfel'd--Jimbo solution discussed in the previous section. Let us go through the construction of the $R$-matrix in some detail. The complexification of $\mathfrak{su}(2)$ is $\mathfrak{sl}(2,\mathbb{C})$, which has a Cartan--Weyl basis given by
\begin{equation}
    H = \begin{pmatrix}
    1&0\\
    0&-1
    \end{pmatrix}\,,\qquad E_+ = \begin{pmatrix}
    0&1\\
    0&0
    \end{pmatrix}\,,\qquad E_- = \begin{pmatrix}
    0&0\\
    1&0
    \end{pmatrix}\,,
\end{equation}
satisfying
\begin{equation}
    [H,E_\pm]=\pm 2 E_\pm\,.
\end{equation}
Using \eqref{eq:def_R_DJ} we obtain a solution to the modified classical Yang--Baxter equation, at the level of $\mathfrak{sl}(2,\mathbb{C})$. In order to see if this descends to a real solution when restricting to the real form $\mathfrak{su}(2)$, let us fix a basis of $\mathfrak{su}(2)$ as $T_j=i\sigma_j$, where $\sigma_i$ denote the Pauli matrices. Then one readily finds that $R$ acts on this basis as
\begin{equation}
    R T_1=-ic\,T_2\,,\qquad R T_2 = ic\, T_1\,,\qquad RT_3=0\,. 
\end{equation}
In particular, for $R$ to be a real endomorphism we require $c=i$, in which case $R$ can be represented as a matrix in the $T_i$ basis as
\begin{equation}
\label{eq:R_su2}
    R = \begin{pmatrix}
    0&-1&0\\
    1&0&0\\
    0&0&0
    \end{pmatrix}\,.
\end{equation}
In the remainder of this section, we will implicitly use this $R$-matrix. 

\subsubsection*{$\mathrm{SU}(2)$ bi-Yang--Baxter model}
We will adopt the following parametrization of the $\mathrm{SU}(2)$ group element
\begin{equation}
\label{eq:g_SU2}
    g = \begin{pmatrix}
    \cos\theta\,e^{i\phi_1} & i\sin\theta\,e^{i\phi_2}\\
    i\sin\theta\,e^{-i\phi_2} & \cos\theta\,e^{-i\phi_1}
    \end{pmatrix}\,,\qquad \theta,\phi_1\in[0,\pi)\,,\quad \phi_2\in[0,2\pi)\,.
\end{equation}
As a non-linear $\sigma$-model, the bi-Yang--Baxter model can be characterized by the metric and $B$-field it induces on the target space. These follow from inserting the ansatz \eqref{eq:g_SU2} for $g$, together with the $R$-matrix \eqref{eq:R_su2}, into the action \eqref{eq:biYB}. The resulting metric reads
\begin{align*}
    ds^2&=\frac{1}{\Delta}\left[\mathrm{d}\theta^2+\cos^2\theta\left(1+(\eta+\zeta)^2\cos^2\theta\right)\mathrm{d}\phi_1^2+\sin^2\theta\left(1+(\eta-\zeta)^2\sin^2\theta\right)\mathrm{d}\phi_2^2\right]\\
    &\qquad +\frac{\sin^2(2\theta)}{2\Delta}(\eta-\zeta)(\eta+\zeta)\mathrm{d}\phi_1\mathrm{d}\phi_2\,,
\end{align*}
where we have defined
\begin{equation}
    \Delta = 1+\eta^2+\zeta^2+2\eta\zeta\cos(2\theta)\,.
\end{equation}
Moving on, the $B$-field is found to be pure-gauge, and is given by
\begin{equation}
    B=\mathrm{d}A\,,\qquad A=\frac{\log\Delta}{4\eta\zeta}\left[(\eta-\zeta)\mathrm{d}\phi_1-(\eta+\zeta)\mathrm{d}\phi_2\right]\,.
\end{equation}
The form of the metric and $B$-field indicate that at the points $\zeta=\eta$ and $\zeta=-\eta$ the model simplifies significantly (for example, the metric becomes diagonal). We recall that the point $\zeta=\eta$ is referred to as the critical line. In contrast, the point $\zeta=-\eta$ is referred to as the co-critical line. As alluded to before, on the critical line this simplification is due to the emergence of the $\mathbb{Z}_2$ symmetry $g\mapsto g^{-1}$. In the parametrization \eqref{eq:g_SU2} this corresponds to
\begin{equation}
    \phi_1\mapsto -\phi_1\,,\qquad \phi_2\mapsto \phi_2+\pi\,.
\end{equation}
Note that for the $\mathrm{SU}(2)$ model in particular there is an additional $\mathbb{Z}_2$-symmetry given by the transformations
\begin{equation}
\label{eq_Z2_cocritical}
    \theta\mapsto \theta+\frac{\pi}{2}\,,\qquad \phi_1\leftrightarrow\phi_2\,,\qquad \zeta\mapsto -\zeta\,.
\end{equation}
This exactly maps the critical line $\zeta=\eta$ to the co-critical line $\zeta=-\eta$.

\subsubsection*{Real and complex unitons}
The unitons are finite action solutions to the classical equations of motion of the $\mathrm{SU}(2)$ bi-Yang--Baxter model, which additionally satisfy $g^2=-1$. In the parametrization \eqref{eq:g_SU2}, this condition imposes that either $\theta=\frac{\pi}{2}$ or $\phi_1=\frac{\pi}{2}$. We will consider the latter case. There are two types of unitons, dubbed the real and complex unitons. Both are determined by the choice of a holomorphic function $f(z)$ of the worldsheet coordinate. The expressions for $\phi_1$ and $\phi_2$ are the same for both unitons, and are given by
\begin{equation}
    \phi_1=\frac{\pi}{2}\,,\quad \phi_2=\pi+\frac{i}{2}\log\left(\frac{f}{\bar{f}}\right)\,,
\end{equation}
while the expressions for $\theta$ differ and are respectively given by
\begin{align}
\label{eq:real_uniton}
    \text{real uniton}:\qquad \sin^2\theta &= \frac{4|f|^2}{(1+|f|^2)^2+(\eta-\zeta)^2(1-|f|^2)^2}\,,\\
\label{eq:complex_uniton}
    \text{complex uniton}:\qquad \theta&=\frac{\pi}{2}+i\arctanh\left(\frac{1}{2}\left(|f|+\frac{1}{|f|}\right)\sqrt{(\eta-\zeta)^2+1} \right)\,.
\end{align}
The nomenclature `real' vs. `complex' is due to the fact that for the real uniton, $\theta$ is manifestly real and hence the group-valued field $g$ indeed lies in $\mathrm{SU}(2)$. In contrast, for the complex uniton $\theta$ is complex-valued and hence $g$ takes values in the complexified group $\mathrm{SL}(2,\mathbb{C})$.\footnote{Note that for real $x$
\begin{equation*}
    \mathrm{Im}\,\arctanh\,x=\begin{cases}
    0\,,& |x|<1\,,\\
    -\frac{\pi}{2}\mathrm{sign}(x)\,,& |x|>1\,.
    \end{cases}
\end{equation*}
Therefore, one finds that for the complex uniton $\mathrm{Re}\,\theta=\pi$, hence $g$ in fact takes values in $\mathrm{SU}(1,1)$. See also the discussion in section \ref{subsec:Weil_SL2}.}\\

\noindent For completeness, we also record the metric of the $\mathrm{SU}(2)$ bi-Yang--Baxter model when evaluated on the real and complex unitons. Introducing polar coordinates $f=r e^{i\alpha}$ and writing $R=r^2$ one finds\footnote{Here we have used the fact that
\begin{equation*}
    4R^2 \left(\frac{\mathrm{d}\theta}{\mathrm{d}R}\right)^2 = \sin^2\theta+(\zeta-\eta)^2\sin^4\theta\,,
\end{equation*}
which holds for both the real and complex uniton and can be verified by explicit computation.}
\begin{equation}
\label{eq:ds2_onshell}
    ds^2 = \frac{1}{\Delta}\left(\frac{\mathrm{d}\theta}{\mathrm{d}R}\right)^2\left(\mathrm{d}R^2+4R^2\mathrm{d}\alpha^2\right)\,,
\end{equation}
for both unitons. We see that in the target space the uniton solutions correspond to a squashed two-sphere inside $\mathrm{SU}(2)$. From here it follows from explicit integration that the unitons have finite action and we refer the reader to \cite{Schepers:2020ehn} for further details. One finds that the on-shell actions evaluate to (we neglect the overall factor coming from the angular integration)
\begin{equation}
    S_{\text{real uniton}} = \frac{1}{2\eta\zeta}\left[(\eta+\zeta)\arctan(\eta+\zeta)-(\eta-\zeta)\arctan(\eta-\zeta) \right]\,,
\end{equation}
and\footnote{It should be noted that $S_{\text{complex uniton}}$ is not finite in the limit $\eta,\zeta\rightarrow 0$, hence this point should be excluded.}
\begin{equation}
    S_{\text{complex uniton}} = \frac{1}{2\eta\zeta}\left[(\eta+\zeta)\arccot(\eta+\zeta)-(\eta-\zeta)\arccot(\eta-\zeta) \right]\,.
\end{equation}
Here we have assumed the domain of integration to be the entire complex plane. In other words, $R$ ranges from 0 to infinity. In the next section we will encounter a situation where $f(z)$ instead takes values in the unit disk, in which case $R\in[0,1]$. One can verify that this only changes the above results by a factor of 1/2.

\subsection{Weil operator as a complex uniton}
\label{subsec:Weil_SL2}

The unitons discussed in the previous section are solutions to the $\mathrm{SU}(2)$ bi-Yang--Baxter model. In this section, we will instead be concerned with the $\mathrm{SL}(2,\mathbb{R})$ bi-Yang--Baxter model. This model has a solution which naturally arises from the study of variations of Hodge structure, applied to the simplest example of a torus. More precisely, the solution is given by the so-called Weil operator which, roughly speaking, corresponds to the Hodge star when viewed as an operator on the middle de Rham cohomology of the torus. The Weil operator is a function of the Teichm\"uller parameter $\tau$ of the torus. This parameter is then reinterpreted as the worldsheet coordinate in the bi-Yang--Baxter model. Interestingly, this solution also satisfies $g^2=-1$, which suggests that it might be related to the uniton solutions. Indeed, we show that the Weil operator can be mapped to the \textit{complex} uniton, for a particular choice of holomorphic function $f(z)$, via a Cayley transformation.

\subsubsection*{Weil operator}
Let us start by introducing the Weil operator of the torus. A convenient description of the torus $\mathbb{T}$ is as a lattice 
\begin{equation}
    \mathbb{T} = \mathbb{C}/(\mathbb{Z}+\tau\mathbb{Z})\,,\qquad \mathrm{Im}\,\tau>0\,,
\end{equation}
where $\tau$ is the Teichm\"uller parameter taking values in the complex (strict) upper half-plane. We parametrize the torus by two periodic coordinates $\xi_1,\xi_2$ with $\xi_i\sim\xi_i+1$. Then the metric on the torus can be written as
\begin{equation}
\label{eq:metric_torus}
    ds^2 = \frac{|\mathrm{d}\xi_1+\tau\,\mathrm{d}\xi_2|^2}{\mathrm{Im}\,\tau}\,.
\end{equation}
Here we have normalized the metric so that the torus has unit volume. The Weil operator is closely related to the Hodge star operator on the torus. The action of the Hodge star on the one-forms $\mathrm{d}\xi_1$ and $\mathrm{d}\xi_2$ follows directly from the metric \eqref{eq:metric_torus} and is given by
\begin{equation}
\label{eq:Hodge_star_torus}
    \star\,\mathrm{d}\xi_1 =\frac{\mathrm{Re}\,\tau}{\mathrm{Im}\,\tau}\mathrm{d}\xi_1+\frac{|\tau|^2}{\mathrm{Im}\,\tau}\mathrm{d}\xi_2\,,\qquad \star\,\mathrm{d}\xi_2 = -\frac{1}{\mathrm{Im}\,\tau}\mathrm{d}\xi_1-\frac{\mathrm{Re}\,\tau}{\mathrm{Im}\,\tau}\mathrm{d}\xi_2\,.
\end{equation}
To obtain the Weil operator, one should view this as an action on the middle de Rham cohomology $H^1(\mathbb{T},\mathbb{C})$ of the torus.\footnote{Recall that, as a real manifold, the torus is two-dimensional, hence its cohomology groups $H^n(\mathbb{T},\mathbb{C})$ run over $n=0,1,2$. This explains the name `middle cohomology' for $H^1(\mathbb{T},\mathbb{C})$.} Indeed, in the basis $\{[\mathrm{d}\xi_1],[\mathrm{d}\xi_2]\}$, where $[\omega]$ denotes the equivalence class of a one-form $\omega$, the action of the Hodge star can be represented as a matrix
\begin{equation}
\label{eq:Weil_torus}
    C(x,y) = \frac{1}{y}\begin{pmatrix}
    x & -1\\
    x^2+y^2 & -x
    \end{pmatrix}\,,
\end{equation}
where we have set $\tau=x+iy$. We will refer to \eqref{eq:Weil_torus} as the Weil operator of the torus. Note that it is an element of $\mathrm{SL}(2,\mathbb{R})$.

\subsubsection*{Relation to the complex uniton}
An interesting property of \eqref{eq:Weil_torus} is that it satisfies\footnote{More generally, the origin of this relation is the fact that $\star\star$ evaluates to $\pm 1$, with the sign determined by the degree of the differential form it acts on and the dimension of the spacetime in question.}
\begin{equation}
    C^2=-1\,,
\end{equation}
which is shared by the unitons solutions discussed in section \ref{sec:bi_YB_unitons}. In fact, we will now argue that the Weil operator can be viewed as a complex uniton for a specific choice of the holomorphic function $f(z)$.\\ 

\noindent As a preliminary remark, we stress that one cannot simply compare the expressions \eqref{eq:g_SU2} and \eqref{eq:Weil_torus}, as the two lie in different groups, namely $\mathrm{SU}(2)$ and $\mathrm{SL}(2,\mathbb{R})$, respectively. There is, however, a natural two-step procedure to pass between the two groups by combining a so-called Cayley transformation with an analytic continuation. Let us first elaborate on the former. We introduce the matrix
\begin{equation}
\label{eq:Cayley_sl2}
    \rho = \frac{1}{\sqrt{2}}\begin{pmatrix}
    1 & i\\
    i & 1
    \end{pmatrix}\,,
\end{equation}
which is an element of $\mathrm{SL}(2,\mathbb{C})$. Then it is straightforward to show that the adjoint action
\begin{equation}
    \mathrm{Ad}_\rho:\mathrm{SL}(2,\mathbb{R})\rightarrow\mathrm{SU}(1,1)
\end{equation}
is an isomorphism of real Lie groups, which is commonly referred to as a Cayley transformation. Indeed, it provides an interpolation between the two real forms $\mathrm{SL}(2,\mathbb{R})$ and $\mathrm{SU}(1,1)$ of $\mathrm{SL}(2,\mathbb{C})$. For the second step of the proposed procedure, one interpolates between $\mathrm{SU}(1,1)$ and $\mathrm{SU}(2)$ via an analytic continuation. In the parametrization \eqref{eq:g_SU2} this is straightforwardly given by sending $\theta\mapsto -i\theta$.\\

\noindent We now apply the above procedure to compare the Weil operator of the torus to a generic element in $\mathrm{SU}(2)$. Practically, it is easiest to take the expression in \eqref{eq:g_SU2}, analytically continue it to $\mathrm{SU}(1,1)$ by setting $\theta= i\tilde{\theta}$ and then apply $\mathrm{Ad}_{\rho^{-1}}$ to end up in $\mathrm{SL}(2,\mathbb{R})$. The result of this computation is
\begin{equation}
\label{eq:g_Sl2}
    g_{\mathrm{SL}(2,\mathbb{R})} = \begin{pmatrix}
    \cosh\tilde{\theta} \cos\phi_1 - \sinh\tilde{\theta} \sin\phi_2 & -\cosh\tilde{\theta}\sin\phi_1-\sinh\tilde{\theta}\cos\phi_2\\
    \cosh\tilde{\theta}\sin\phi_1-\sinh\tilde{\theta}\cos\phi_2 & \cosh\tilde{\theta}\cos\phi_1+\sinh\tilde{\theta}\sin\phi_2
    \end{pmatrix}\,,
\end{equation}
which is indeed an element of $\mathrm{SL}(2,\mathbb{R})$. Comparing \eqref{eq:Weil_torus} and \eqref{eq:g_Sl2} and solving for $\tilde{\theta},\phi_1,\phi_2$ in terms of $x,y$ gives
\begin{equation}
\label{eq:Weil_sol}
    \phi_1 = \frac{\pi}{2}\,,\qquad \phi_2 = \pi+\frac{i}{2}\log \left(\frac{f}{\Bar{f}}\right)\,,\qquad \tilde{\theta} = \frac{i\pi}{2}+\arctanh\left[\frac{1}{2}\left(|f|+\frac{1}{|f|}\right)\right]\,,
\end{equation}
where $f(z)$ is the following holomorphic function of the complexified worldsheet coordinates
\begin{equation}
\label{eq:f_mobius}
    f(z) = \frac{z-i}{z+i}\,,\qquad z=x+iy\,.
\end{equation}
Indeed, identifying the Teichm\"uller parameter $\tau$ with the worldsheet coordinate $z$ and recalling that $\theta=i\tilde{\theta}$, one sees that this solution is precisely of the form of a complex uniton \eqref{eq:complex_uniton} with additionally $\zeta=\eta$.\footnote{Strictly speaking, an exact match is obtained after sending $\theta\mapsto\theta+\pi$, corresponding to $C\mapsto -C$. Of course, the overall sign of $C$ is simply a convention.} The function $f(z)$ is a special type of M\"obius transformation that conformally maps the upper half-plane to the unit disc. Note that it is holomorphic on the upper half-plane, but has a first order pole at $z=-i$. Let us thus emphasize the following important observation. One can invert \eqref{eq:f_mobius} to find
\begin{equation}\label{eq:f_mobius_inverse}
    z = i\,\frac{f(z)+1}{f(z)-1}\,.
\end{equation}
In particular, if one were to start the analysis of this section with a two-torus whose complex structure parameter $\tau$ is parametrized as \eqref{eq:f_mobius_inverse}, then one would obtain exactly the solution \eqref{eq:Weil_sol} for an arbitrary function $f(\tau)$ that is holomorphic on the upper half-plane. From this point of view, the origin of the family of solutions to the critical bi-Yang--Baxter model therefore follows is naturally explained in terms of the freedom in the choice of a holomorphic coordinate on the moduli space of the two-torus. Furthermore, this clarifies the importance of restricting to the critical line $\zeta=\eta$. Indeed, one may verify that if one were to extend the original uniton solution \eqref{eq:real_uniton} beyond the critical line, it would lead to a non-holomorphic change of coordinates. \\

\begin{figure}
	\centering
	\begin{subfigure}[b]{0.3\textwidth}
		\centering
		\includegraphics[width=\textwidth]{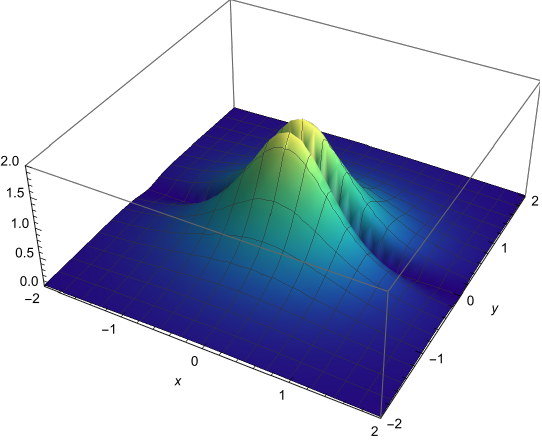}
		\caption{$\eta=0.7$, $\zeta=0.6$}
		\label{fig:plot1}
	\end{subfigure}
	\hfill
	\begin{subfigure}[b]{0.3\textwidth}
		\centering
		\includegraphics[width=\textwidth]{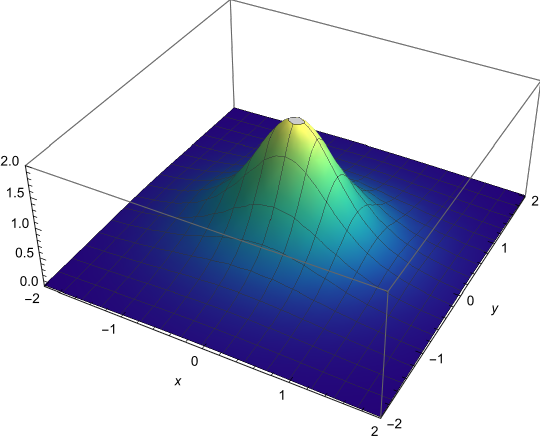}
		\caption{$\eta=0.7$, $\zeta=0.7$}
		\label{fig:plot2}
	\end{subfigure}
	\hfill
	\begin{subfigure}[b]{0.3\textwidth}
		\centering
		\includegraphics[width=\textwidth]{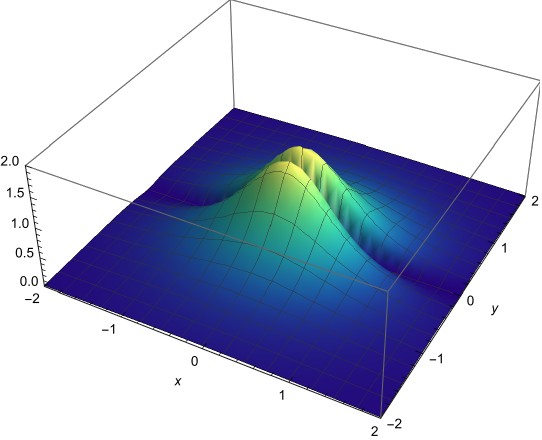}
		\caption{$\eta=0.7$, $\zeta=0.8$}
		\label{fig:plot3}
	\end{subfigure}
	\caption{Plot of the Lagrangian density of the $\mathrm{SU}(2)$ bi-Yang--Baxter model evaluated on the complex uniton solution with $f(z) = \frac{z-i}{z+i}$, for three values of the deformations parameters $(\eta,\zeta)$. For $\eta\neq\zeta$ there are two bumps, which coalesce on the critical line $\eta=\zeta$.}
	\label{fig:unitons}
\end{figure}

\noindent In figure \ref{fig:unitons} we have illustrated the Lagrangian density for the complex uniton defined by the particular holomorphic function \eqref{eq:f_mobius}. It is interesting to contrast this with the plots in \cite{Schepers:2020ehn}, where instead a linear function $f(z)= \frac{z}{2}$ was used. In both cases there is a clear transition at the critical point $\zeta=\eta$. On the other hand, the concentric valley structure in \cite{Schepers:2020ehn} is not present here. Rather, in our case the density is not rotationally invariant, but only invariant under reflections $x\mapsto -x$ and $y\mapsto -y$. Of course, this can be explained by noting that the Lagrangian density is a function of $|f(z)|$. \\

\noindent It is also interesting to recall the $\mathbb{Z}_2$-symmetry \eqref{eq_Z2_cocritical} which maps the critical line to the co-critical line. Indeed, one finds that it acts on the Weil operator as
\begin{equation}
    C(x,y)\mapsto \frac{i}{y}\begin{pmatrix}
    -1 & -x\\
    x & x^2+y^2
    \end{pmatrix} = C(x,y)\cdot\begin{pmatrix}
    0 & i\\
    i & 0
    \end{pmatrix}\,.
\end{equation}
Interestingly, the $\mathbb{Z}_2$ transformation can be described as a right-multiplication of $C(x,y)$ by an element in $\mathrm{SL}(2,\mathbb{C})$ (even in $\mathrm{SU}(2)$). However, as a result the transformed Weil operator is no longer real-valued. Therefore, it will define a solution to the co-critical $\mathrm{SL}(2,\mathbb{C})$ bi-Yang--Baxter model, as can be verified by explicit computation.\\

\noindent As a final comment, one can use the result \eqref{eq:ds2_onshell}, taking $f(z)$ as above, to find that the on-shell metric is given by
\begin{equation}
    ds^2 = \frac{4|\mathrm{d}z|^2}{(z-\bar{z})^2-4\eta^2(1+|z|^2)^2}\,.
\end{equation}
Note that in the limit $\eta\rightarrow 0$ one recovers the standard metric on the Poincar\'e upper half-plane.

\subsubsection*{$R$-matrix revisited}
As a result of the identification between the Weil operator and a complex uniton, one expects that the Weil operator provides a solution to the $\mathrm{SL}(2,\mathbb{R})$ bi-Yang--Baxter model. To make this precise, one should also identify the appropriate $R$-matrix by taking the $\mathrm{SU}(2)$ $R$-matrix in \eqref{eq:R_su2} and translating it to an endomorphism of $\mathrm{SL}(2,\mathbb{R})$ via the Cayley transform. Explicitly, this gives\footnote{One can check that the $R$-matrix remains unchanged under the analytic continuation from $\mathrm{SU}(2)$ to $\mathrm{SU}(1,1)$.}
\begin{equation}
    R_{\mathrm{SL}(2,\mathbb{R})} = \mathrm{Ad}_{\rho^{-1}}\circ R_{\mathrm{SU}(2)}\circ\mathrm{Ad}_\rho\,.
\end{equation}
One can verify that the resulting $R$-matrix acts as
\begin{equation}
\label{eq:R_Weil}
    R_{\mathrm{SL}(2,\mathbb{R})}\,H = -(E_++E_-)\,,\qquad R_{\mathrm{SL}(2,\mathbb{R})}\,E_\pm = \frac{1}{2}H\,,
\end{equation}
and is manifestly real. The reader is invited to check that indeed the Weil operator \eqref{eq:Weil_sol} is a solution to the critical $\mathrm{SL}(2,\mathbb{R})$ bi-Yang--Baxter model defined by the $R$-matrix \eqref{eq:R_Weil}. One may wonder how this $R$-matrix differs from the Drinfeld'd--Jimbo solution we started with. The answer is that the $R$-matrix \eqref{eq:R_Weil} is also a Drinfel'd--Jimbo solution, but for a different choice of Cartan generators. This will be explained further in section \ref{subsec:horizontal_sl2}.

\subsubsection*{An underlying $\mathrm{Sl}(2)$-orbit}
Let us return to the Weil operator \eqref{eq:Weil_torus} to elucidate a very particular underlying structure, which may not be immediately apparent from its matrix representation. Indeed, note that the Weil operator can be factorized in the following way
\begin{equation}
    C(x,y) = h(x,y)C_\infty h(x,y)^{-1}\,,
\end{equation}
where we have introduced
\begin{equation}
\label{eq:h_Cinfty}
    h(x,y) = \frac{1}{\sqrt{y}}\begin{pmatrix}
    1 & 0\\
    x & y
    \end{pmatrix}\,,\qquad C_\infty = \begin{pmatrix}
    0 & 1\\
    -1 & 0
    \end{pmatrix}\,.
\end{equation}
More abstractly, both $h(x,y)$ and $C_\infty$ can be written in terms of $\mathfrak{sl}(2,\mathbb{R})$-valued objects as
\begin{equation}
\label{eq:h_Cinfty_sl2}
    h(x,y) = e^{xN^-}y^{-\frac{1}{2}N^0}\,,\qquad C_\infty = (-1)^{Q_\infty}\,,
\end{equation}
where we have chosen to change our notation for the $\mathfrak{sl}(2,\mathbb{R})$ generators to
\begin{equation}
    N^\pm = E_\pm\,,\qquad N^0=H\,,\qquad 
\end{equation}
and introduced another operator
\begin{equation}
\label{eq:Q_sl2}
    Q_\infty = \frac{i}{2}\begin{pmatrix}
    0 & -1\\
    1 & 0
    \end{pmatrix}\,,
\end{equation}
for which $iQ_\infty$ is valued in $\mathfrak{sl}(2,\mathbb{R})$. In light of the results of chapter \eqref{chap:asymp_Hodge_I}, we immediately recognize the operator $h(x,y)$ in \eqref{eq:h_Cinfty} as the $\mathrm{Sl}(2)$-orbit of the period map. In other words, we indeed see that for the two-torus the $\mathrm{Sl}(2)$-orbit approximation is exact. Based on the simple form of the period vector, recall equation \eqref{eq:periods_flat-torus}, this is to be expected. Furthermore, we recognize the operator $Q_\infty$ as the corresponding boundary charge operator. Indeed, one can verify that it satisfies the following commutation relations with the real $\mathfrak{sl}(2)$ triple $N^+,N^0,N^-$
\begin{equation}
\label{eq:Q_commutation_sl2}
    [Q_\infty, N^0] = i\left(N^+ + N^-\right)\,,\qquad [Q_\infty, N^\pm] = -\frac{i}{2}N^0\,,
\end{equation}
such that, as expected, the triple is in fact a horizontal $\mathfrak{sl}(2)$-triple. \\

\noindent As a final comment, one may compare \eqref{eq:Q_commutation_sl2} with the $R$-matrix \eqref{eq:R_Weil} to find that
\begin{equation}
    R_{\mathrm{SL}(2,\mathbb{R})}=\mathrm{ad}_{iQ_\infty}\,.
\end{equation}
In other words, we have found that the solution we have obtained for the critical $\mathrm{SL}(2,\mathbb{R})$ bi-Yang--Baxter model can be completely characterized by an operator $Q_\infty$ and an $\mathfrak{sl}(2,\mathbb{R})$-triple $N^+, N^0,N^-$ that is horizontal with respect to $Q_\infty$. It turns out that by studying the abstract properties of horizontal $\mathfrak{sl}(2)$-triples one can greatly generalize the simple solution we have considered here to groups of larger rank. This is the topic of the next section.
\newpage

\section{Horizontal $\mathfrak{sl}(2)$-triples and generalized unitons}
\label{sec:Hodge_Theory}

In the previous section we have argued that the Weil operator of the torus provides a solution to the critical $\mathrm{SL}(2,\mathbb{R})$ bi-Yang--Baxter model. In this section we show that this solution can be generalized to arbitrary groups as long as they admit at least one horizontal $\mathfrak{sl}(2)$-triple. Given a horizontal $\mathfrak{sl}(2)$-triple we discuss how it defines a particular class of $R$-matrices and then write down the generalized solution in section \ref{subsec:sl2_orbits}. Subsequently, we explicitly show that it solves the equations of motion of the bi-Yang--Baxter model associated to the mentioned class of $R$-matrices. Finally, in section \ref{subsec:example} we have included a very explicit example of a horizontal $\mathfrak{sl}(2)$-triple in $\mathfrak{sp}(4,\mathbb{R})$ and the associated $\mathrm{Sl}(2)$-orbit approximation of the Weil operator, based on the type $\mathrm{IV}_1$ boundary data.   

\subsection{Horizontal $\mathfrak{sl}(2)$-triples}
\label{subsec:horizontal_sl2}

\subsubsection*{$R$-matrix from horizontal $\mathfrak{sl}(2)$-triples}

For completeness, let us briefly recall the abstract features of horizontal $\mathfrak{sl}(2)$-triples, without making reference to a possible underlying variation of Hodge structure. To this end, let $\mathfrak{g}_{\mathbb{R}}$ be a real Lie algebra. The first important ingredient is a standard $\mathfrak{sl}(2)$-triple, which is a set of three elements in $\mathfrak{g}_{\mathbb{R}}$, denoted by $\{N^+,N^0,N^-\}$, satisfying the commutation relations
\begin{equation}
\label{eq:sl2_algebra}
    [N^0,N^\pm] =\pm 2 N^{\pm}\,,\qquad [N^+,N^-]=N^0\,.
\end{equation}
For classical Lie algebras $\mathfrak{g}_\mathbb{R}$, there is a classification of $\mathfrak{sl}(2)$-triples in $\mathfrak{g}_\mathbb{R}$ in terms of signed Young diagrams, see e.g.~\cite{McGovern:1993}. The second important ingredient is the charge operator $Q_\infty$, which we recall should satisfy two important properties. First, it should be purely imaginary-valued, i.e.
\begin{equation}
    \bar{Q}_{\infty} = -Q_{\infty}\,,
\end{equation}
where the bar denotes complex conjugation. Second, $Q_{\infty}$ should be a so-called grading element. This means that the adjoint action $\mathrm{ad}\,Q_{\infty}$ has only integer eigenvalues. \\

\noindent Given a standard $\mathfrak{sl}(2)$-triple and a charge operator $Q_{\infty}$, we say that the triple is \textit{horizontal} (with respect to $Q_{\infty}$) if it satisfies the following commutation relations
\begin{equation}
\label{eq:Q_commutation}
    [Q_{\infty},N^0] = i\left(N^+ +N^-\right)\,,\qquad [Q_{\infty},N^{\pm}] = -\frac{i}{2}N^0\,.
\end{equation}
In the following, we will argue that a horizontal $\mathfrak{sl}(2)$-triple provides a solution to the critical bi-Yang--Baxter model, for a particular class of $R$-matrices. These $R$-matrices must be compatible with the $\mathfrak{sl}(2)$-triple in the sense that they must act in the same way as we found in the simple example of section \ref{subsec:Weil_SL2}. Let us make this more precise.\\

\noindent Given a charge operator $Q_{\infty}$ and a horizontal $\mathfrak{sl}(2)$-triple in a Lie algebra $\mathfrak{g}$, we impose that the $R$-matrix acts as
\begin{equation}
\label{eq:def_R_general}
    R\,Q_{\infty}=0\,,\qquad R\,N^0 = -(N^++N^-)\,,\qquad R\,N^\pm = \frac{1}{2}N^0\,.
\end{equation}
One readily checks that the modified classical Yang--Baxter equation is then satisfied for this $\mathfrak{sl}(2)$-triple. In terms of the complex generators $L_\alpha$ defined via the Cayley transform, recall equation \eqref{eq:rho_switch}, it acts as
\begin{equation}
    R\,Q_{\infty}=0\,,\qquad R\,L_0 = 0\,,\qquad R\,L_{\pm 1} = \pm i L_{\pm 1}\,.
\end{equation}
Comparing this to \eqref{eq:def_R_DJ}, we see that this choice of $R$-matrix is nothing but the Drinfel'd--Jimbo solution, where we have chosen $L_0$ and $Q_{\infty}$ as the generators of a Cartan subalgebra of $\mathfrak{sl}(2,\mathbb{C})\times\mathfrak{u}(1)$ and $L_{\pm 1}$ correspond to the positive and negative roots.
However, in general this does not yet specify the full $R$-matrix, as its action on the remaining generators of $\mathfrak{g}$ is not yet fixed. Indeed, if the rank of $\mathfrak{g}$ is greater than two, one must identify additional Cartan generators beyond $L_0$ and $Q_{\infty}$ to complete the full Drinfel'd--Jimbo solution.

\subsection{$\mathrm{Sl}(2)$-orbits and generalized unitons}
\label{subsec:sl2_orbits}

In this section we will introduce a generalization of the Weil operator of the torus discussed in section \ref{subsec:Weil_SL2} using purely the data of a horizontal $\mathfrak{sl}(2)$-triple. Hodge-theoretically, the operators we discuss correspond to so-called $\mathrm{Sl}(2)$-orbits. In the following, we will purely use the abstract properties of the horizontal $\mathfrak{sl}(2)$-triple, without making reference to any underlying Hodge structure, to show that such operators also solve the critical bi-Yang--Baxter model and therefore provide a generalization of the complex unitons to groups beyond $\mathrm{SU}(2)$. 

\subsubsection*{$\mathrm{Sl}(2)$-orbits}

In section \ref{subsec:Weil_SL2} we highlighted a particular underlying structure in the Weil operator of the torus in terms of two elements $h(x,y)$ and $C_\infty$, see \eqref{eq:h_Cinfty} and \eqref{eq:h_Cinfty_sl2}. Let us recall that this can be generalized straightforwardly by considering the $\mathrm{Sl}(2)$-orbit approximation of the period map
\begin{equation}
\label{eq:def_hnil_biYB}
    h(x,y) = e^{xN^-}y^{-\frac{1}{2}N^0}\,,\qquad C_\infty = (-1)^{Q_\infty}
\end{equation}
and defining
\begin{equation}
\label{eq:def_Weil_general}
    g(x,y) = h(-1)^{Q_\infty}h^{-1}\,.
\end{equation}
We will refer to \eqref{eq:def_Weil_general} as simply the Weil operator. Note that $g(x,y)$ satisfies a `twisted boundary condition' 
\begin{equation}
    g(x+1,y) = e^{N^-} g(x,y) e^{-N^-}\,,
\end{equation}
which is similar to the adiabatic reduction used in \cite{Schepers:2020ehn} to reduce the $\mathrm{SU}(2)$ bi-Yang--Baxter model to a model of quantum mechanics. 

\subsubsection*{Generalized unitons}

We now explicitly show that the Weil operator $g$ in \eqref{eq:def_Weil_general}, together with the $R$-matrix defined in \eqref{eq:def_R_general} solves the equations of motion of the critical bi-Yang--Baxter model. Here we will focus on the main steps in the calculation, and refer the reader to appendix \ref{app:computations} where further details are given. More precisely, throughout the calculation it is necessary to know how $\mathrm{Ad}_h$, $\mathrm{Ad}_{h^{-1}}$ and $\mathrm{Ad}_g$ act on the $\mathfrak{sl}(2,\mathbb{R})$-triple. This can be straightforwardly computed using the commutation relations, and the results are listed in equations \eqref{eq:Adh_N+}-\eqref{eq:Adhinv_N-} and \eqref{eq:Adg_N+}-\eqref{eq:Adg_N-}.\\

\noindent To start, it will be most convenient to present the equations of motion in real coordinates $x,y$ in terms of which they read
\begin{equation}
    \partial_x J_y - \partial_y J_x - \eta[J_x,J_y]_R=0\,,
\end{equation}
where we recall the $R$-bracket in \eqref{eq:R_bracket} and denote
\begin{equation}
    J_x = J_+ + J_-\,,\qquad J_y = -i\left(J_+ - J_-\right)\,,
\end{equation}
with $J_\pm$ defined in \eqref{eq:def_Jpm}. For clarity of presentation, we will divide the main computation into three steps.

\subsubsection*{Step 1: $j_\mu$}
As a first step, let us compute the simpler objects
\begin{equation}
    j_\mu = g^{-1}\partial_\mu g\,.
\end{equation} 
Using \eqref{eq:def_Weil_general} one has
\begin{equation}
    j_\mu = h\left[(-1)^{\mathrm{ad}\,Q_\infty}h^{-1}\partial_\mu h - h^{-1}\partial_\mu h\right]h^{-1}\,.
\end{equation}
Then, using the expression \eqref{eq:def_hnil_biYB} for $h$, one readily computes
\begin{equation}
    h^{-1}\partial_x h = \frac{1}{y}N^-\,,\qquad h^{-1}\partial_y h = -\frac{1}{2y}N^0\,.
\end{equation}
Therefore, to compute $j_\mu$ it remains to apply the commutation relations \eqref{eq:Q_commutation} and evaluate the action of $\mathrm{Ad}_h$. For the first step we use the relations \eqref{eq:dagger_N+}-\eqref{eq:dagger_N-} and for the second step we use \eqref{eq:Adh_N+}-\eqref{eq:Adh_N-}. The result of this computation is
\begin{align}
\label{eq:jx}
    j_x &= -\frac{1}{y^2}\left[N^+-x N^0-(x^2-y^2)N^- \right]\,,\\
\label{eq:jy}
    j_y &= \frac{1}{y}\left[N^0+2x N^-\right]\,.
\end{align}

\subsubsection*{Step 2: $J_\mu$}
To proceed, we would like to compute the currents $J_\mu$, for which we must know the action of $R$ and $R_g$ on the currents $j_\mu$. The general action of these two operators on the $\mathfrak{sl}(2)$-triple follows from the definition \eqref{eq:def_R_general} and the action of $\mathrm{Ad}_g$ on the $\mathfrak{sl}(2)$-triple, which we record in \eqref{eq:Adg_N+}-\eqref{eq:Adg_N-}. Applying this to $j_x$ and $j_y$ gives
\begin{align}
    \left(\eta R+\zeta R^g\right)j_x &= \eta \frac{\alpha}{y}\left[N^0+2x N^-\right]+\frac{\zeta-\eta}{y^2}\left[xN^++\left(\frac{\alpha}{2}-x^2\right)N^0+x(1+2\alpha)N^-\right]\,,\\
     \left(\eta R+\zeta R^g\right)j_y&=\eta\frac{\alpha}{y^2}\left[N^+-x N^0-(x^2-y^2)N^-\right]\\
     &\qquad +\frac{\zeta-\eta}{y^3}\left[-(1+x^2)N^++x(1+x^2)N^0+(x^2+x^4-y^4)N^-\right]\,.\nonumber
\end{align}
where we have introduced
\begin{equation}
    \alpha=\alpha(x,y) = \frac{1+x^2+y^2}{y}\,.
\end{equation}
Crucially, one sees that the result simplifies considerably when evaluated on the critical line $\zeta=\eta$ and becomes
\begin{align}
    \left(R+R^g\right)j_x &= \frac{\alpha(x,y)}{y}\left[N^0+2x N^-\right]\,,\\
    \left(R+R^g\right)j_y &= \frac{\alpha(x,y)}{y^2}\left[N^+-x N^0-(x^2-y^2)N^-\right]\,,
\end{align}
The main observation is now the following. Comparing the above result with the expressions \eqref{eq:jx}-\eqref{eq:jy} for $j_x$ and $j_y$, we see that
\begin{equation}
    (R+R^g)j_x = \alpha(x,y)\,j_y\,,\qquad (R+R^g)j_y = -\alpha(x,y)\,j_x\,.
\end{equation}
In other words, for this particular choice of $g$ the currents $j_+$ and $j_-$ are eigenvectors of $R+R_g$. As a result, computing $J_x$ and $J_y$ is straightforward and gives
\begin{equation}
\label{eq:Jmu_result}
    J_x = \beta(x,y)j_y\,,\qquad J_y = -\beta(x,y)j_x\,,
\end{equation}
where we have introduced the function
\begin{equation}
\label{eq:def_beta}
    \beta(x,y) = \frac{1}{i+\alpha(x,y)\eta}\,.
\end{equation}

\subsubsection*{Step 3: equations of motion}
For the final step, we show that \eqref{eq:Jmu_result} solves the equations of motion
\begin{equation}
    \partial_x J_y - \partial_y J_x - \eta[J_x,J_y]_R=0\,.
\end{equation}
Indeed, inserting \eqref{eq:Jmu_result} one effectively needs to show that
\begin{equation}
    \beta\left(\partial_x j_x+\partial_y j_y\right) + \left(\partial_x\beta j_x+\partial_y\beta j_y + \eta\beta^2 [j_x,j_y]_R\right) = 0\,.
\end{equation}
This follows from straightforwardly using the definition of $\beta(x,y)$, inserting our results for $j_x$ and $j_y$, see \eqref{eq:jx}-\eqref{eq:jy}, and evaluating the $R$-bracket using \eqref{eq:R_bracket} and \eqref{eq:def_R_general}. In fact, the two terms in brackets vanish separately. Note that for the first term this is simply the statement that the one-form $\star j$ is closed, i.e.
\begin{equation}
\label{eq:star_j}
    \mathrm{d}\,\star\,j=0\,.
\end{equation}
Stated differently, $g$ also solves the principal chiral model. This is of course not surprising if one expects the solution to hold for all values of $\eta$, since in the limit $\eta\rightarrow 0$ the critical bi-Yang--Baxter model reduces precisely to the principal chiral model. Furthermore, it agrees with the results stated at the end of appendix \ref{subsec:Nahm}. 

\subsubsection*{Discussion}

To summarize, we have shown that for any real Lie group $G$ whose Lie algebra $\mathfrak{g}$ contains a horizontal $\mathfrak{sl}(2)$-triple, the Weil operator defined in \eqref{eq:def_Weil_general} provides a solution to the critical bi-Yang--Baxter model associated to the class of $R$-matrices satisfying \eqref{eq:def_R_general}. As this constitutes the main result of this chapter, let us make some additional comments.  

\subsubsection*{Finite action}
As a first remark, we note that it is straightforward to show that our solutions have finite action. Indeed, one finds that
\begin{equation}
    S_{\mathrm{on-shell}} \sim \int d^2\sigma\,\beta \left(j_+,j_-\right)\,,
\end{equation}
where we recall that $\beta$ is defined by \eqref{eq:def_beta}. We stress that our solution is completely defined in terms of the horizontal $\mathfrak{sl}(2)$-triple and that the pairing $(j_+,j_-)$ is independent of the particular representation of the triple, except for possibly an overall coordinate-independent factor. Therefore, it suffices to compute the on-shell action for the fundamental representation, where it was shown that the result is finite, as discussed in section \ref{subsec:unitons}. 

\subsubsection*{Multiple $\mathfrak{sl}(2)$-triples}

As a second remark, we note that it is entirely possible that $\mathfrak{g}$ contains multiple inequivalent (commuting) $\mathfrak{sl}(2)$-triples that are all horizontal with respect to the same charge operator. Of course, one can construct a corresponding Weil operator as in \eqref{eq:def_Weil_general} for each of these triples and obtain multiple inequivalent solutions to the corresponding bi-Yang--Baxter models. Interestingly, as discussed in chapter \ref{chap:asymp_Hodge_I}, in Hodge theory the presence of multiple $\mathfrak{sl}(2)$-triples gives rise to so-called multi-variable $\mathrm{Sl}(2)$-orbits, which take the similar form
\begin{equation}
    g = h(-1)^{Q_\infty}h^{-1}\,,
\end{equation}
but with $h$ given by
\begin{equation}
    h(x_1,\ldots, x_n, y_1,\ldots, y_n) = \prod_{i=1}^n h_i(x_i, y_i)\,,\qquad h_i(x_i,y_i) = e^{x_i N^-_i}y_i^{-\frac{1}{2}N^0_i}\,,
\end{equation}
where $i$ enumerates the various horizontal $\mathfrak{sl}(2)$-triples. Of course, to view this as a solution of the bi-Yang--Baxter models one has to view one of the coordinates $t_i=x_i+iy_i$, for a fixed $i$, as the worldsheet coordinate and the others as some additional parameters. In this case one readily sees that this also defines a solution by exactly the same arguments above, as the transformation
\begin{equation}
    g\mapsto a\cdot g\cdot a^{-1}\,,\qquad R\mapsto \mathrm{Ad}_a\circ R\circ \mathrm{Ad}_a^{-1}\,,\qquad a\in G\,,
\end{equation}
is a global symmetry of the bi-Yang--Baxter model. It would, however, also be interesting to see if there exists a natural extension of the model to higher dimensions for which the multi-variable $\mathrm{Sl}(2)$-orbits provide a solution. 

\subsubsection*{Generalizations and relations to other integrable models}

As a third remark, we would like to stress that we have shown that the Weil operator solves the \textit{critical} bi-Yang--Baxter model, i.e. when $\zeta=\eta$. Additionally, it is straightforward to check that it does not solve the non-critical model, as we showed explicitly in section \ref{subsec:Weil_SL2} for the Weil operator of the torus. However, it is possible that by a suitable generalization of the ansatz one can also find solutions of the non-critical model. For example, for the $\mathrm{SU}(2)$ model one can apply the $\mathbb{Z}_2$-symmetry \eqref{eq_Z2_cocritical} to obtain instead a solution of the co-critical model and the same can be done for the $\mathrm{SL}(2,\mathbb{R})$ model. It would therefore be interesting to see if this symmetry, or an appropriate generalization thereof, also applies for the bi-Yang--Baxter model based on other groups.\\

\noindent Another point that is worth emphasizing is the fact that the critical bi-Yang--Baxter model is related to other integrable models. A trivial example is the limit $\eta\rightarrow 0$, for which it reduces to the principal chiral model. As mentioned earlier, since our solution holds for any value of $\eta$, this implies that the Weil operator also solves the principal chiral model, for which the equations of motion are simply the harmonicity condition \eqref{eq:star_j}. Relations to other integrable models have been established as well in the literature. For example, as mentioned before, the critical $\mathrm{SU}(2)$ bi-Yang--Baxter model is equivalent to the coset $\mathrm{SO}(4)/\mathrm{SO}(3)$ $\eta$-model \cite{Delduc:2013qra,Hoare:2014oua}. Furthermore, at the conformal point $\eta=\frac{i}{2}$ it coincides with the $\mathrm{SU}(1,1)/\mathrm{U}(1)$ gauged WZW model with an additional $\mathrm{U}(1)$ boson \cite{Hoare:2014oua}. Therefore the Weil operator (of the torus) also provides a solution to these models. There has also been work on relating the bi-Yang--Baxter model to generalized $\lambda$-deformations via Poisson-Lie T-duality. This was first worked out for the $\mathrm{SU}(2)$ model in \cite{Sfetsos:2015nya} and later proved in general in \cite{Klimcik:2016rov}, see also \cite{Klimcik:2015gba,Vicedo:2015pna,Hoare:2015gda}. Therefore, via this duality it is reasonable to expect that our solution can be mapped to solutions of generalized $\lambda$-deformations. This points to the striking idea that Hodge theory allows one to construct solutions to many integrable models. \\

\noindent Indeed, as a last comment let us compare the results obtained in this chapter with those of the previous chapter, which was based on the work \cite{Grimm:2021idu}. There it was shown that the Weil operator of an arbitrary variation of Hodge structure solves the $\lambda$-deformed $G/G$ model when $|\lambda|=1$. Indeed, in \cite{Grimm:2021idu} we considered the ansatz
\begin{equation}
\label{eq:g_previous}
    g = h\,z^{Q_\infty}h^{-1}\,,
\end{equation}
where $h$ reduces to \eqref{eq:def_hnil_biYB} in the $\mathrm{Sl}(2)$-orbit approximation. In that case, when also $z=-1$, the expression \eqref{eq:g_previous} coincides with \eqref{eq:def_Weil_general}. Furthermore, one can easily check that what was referred to as the `horizontality condition' in \cite{Grimm:2021idu} reduces to the condition that the $\mathfrak{sl}(2)$-triple is horizontal with respect to $Q_\infty$. It would be interesting to investigate how the relation among the solutions translates into a relation of the underlying integrable models.

\subsection{Example: Type $\mathrm{IV}_1$}
\label{subsec:example}

The preceding discussion has been rather abstract, so let us end this section by providing an explicit example of a horizontal $\mathfrak{sl}(2)$-triple in $\mathfrak{sp}(4,\mathbb{R})$ and write down the corresponding solution to the bi-Yang--Baxter model. We also end with some speculative comments regarding possible generalizations to the nilpotent orbit approximation discussed in chapter \ref{chap:asymp_Hodge_I}. We refer the reader to \cite{Grimm:2021ikg} for further details on this example.

\subsubsection*{Weil operator}

Using the results of appendix \ref{sec:asymp_Hodge_examples}, where the boundary data for the type $\mathrm{IV}_1$ singularity was constructed, one can immediately write down the corresponding $\mathrm{Sl}(2)$-orbit approximation of the Weil operator. Explicitly, it is given by\footnote{Here we have suppressed the additional transformation \eqref{eq:basis-transform_IV1} and we are using the conventions of chapter \ref{chap:asymp_Hodge_II}, see section \ref{ssec:IV1}, in which we performed the additional basis transformation \eqref{eq:convention_IV1}. In particular, we have used the $\mathfrak{sl}(2)$-triple \eqref{eq:sl2-triple_IV1-paper} and charge operator \eqref{eq:Q_IV1-paper}.}
\newcommand\scalemath[2]{\scalebox{#1}{\mbox{\ensuremath{\displaystyle #2}}}}
\begin{equation}
\label{eq:Weil_IV1}
    g(x,y) = \frac{1}{y^3}\scalemath{0.90}{\left(
\begin{array}{cccc}
 -x^3 & 3 x^2 & -3 x & 1 \\
 -x^2 \left(x^2+y^2\right) & 3 x^3+2 x y^2 & -3 x^2-y^2 & x \\
 -x \left(x^2+y^2\right)^2 & \left(x^2+y^2\right) \left(3 x^2+y^2\right) & -3 x^3-2 x y^2 & x^2 \\
 -\left(x^2+y^2\right)^3 & 3 x \left(x^2+y^2\right)^2 & -3 x^2 \left(x^2+y^2\right) & x^3 \\
\end{array}
\right)}\,.
\end{equation}
By our general arguments, the operator \eqref{eq:Weil_IV1} together with the $R$-matrix \eqref{eq:def_R_general} provide a solution to the critical $\mathfrak{sp}(4,\mathbb{R})$ bi-Yang--Baxter model, as can be verified by explicit computation. For some further details on how we constructed a full $R$-matrix for $\mathfrak{sp}(4,\mathbb{R})$ we refer the reader to appendix \ref{app:Rmat}, see in particular \eqref{eq:R_IV1}. 

\subsubsection*{Nilpotent orbit approximation}

We end this example with an observation regarding the nilpotent orbit approximation. As a rough summary, let us recall that, for general variations of Hodge structure, the $\mathrm{Sl}(2)$-orbit approximation \eqref{eq:Weil_IV1} will only provide the first order approximation of the full variation of Hodge structure. To obtain a better approximation one must consider the nilpotent orbit approximation \eqref{eq:hnil_nilp-expansion} by incorporating an infinite tower of $y^{-1}$ corrections. In section \ref{ssec:IV1} this computation was done explicitly for this particular example. The resulting Weil operator is given by (we have set $x=0$ for simplicity)

\begin{equation}
\label{eq:Weil_IV1_nilpotent}
    \frac{g(y)}{N(y)} = \scalemath{0.85}{\begin{pmatrix}
    0 & -9y^2\chi & 0 & 8y^3+\chi\\
    -\frac{3}{2}y(2y^3+\chi)\chi & 0 & -\frac{(2y^3+\chi)(8y^3+\chi)}{2y} & 0\\
    0 & y(8y^6-y^3\chi+2\chi^2) & 0 & -3y^2\chi\\
    -\frac{1}{2}(2y^3+\chi)(8y^6-y^3\chi+2\chi^2) & 0 & -\frac{9}{2}y(2y^3+\chi)\chi & 0
    \end{pmatrix}}\,,
\end{equation}
with 
\begin{equation}
    N(y)=\frac{1}{(4y^3-\chi)(2y^3+\chi)}\,,
\end{equation}
and we recall that the parameter $\chi$ originates from the expression of the phase operator \eqref{eq:delta_IV1_2} and, in the geometric setting, when this boundary Hodge structure arises in the large complex structure regime of a Calabi--Yau threefold $Y_3$, the parameter $\chi$ is proportional to the Euler characteristic of the mirror of $Y_3$.\\

\noindent The curious reader may wonder whether also the expression \eqref{eq:Weil_IV1_nilpotent} solves the bi-Yang--Baxter model, since it at least does so in the $\chi\rightarrow 0$ limit. If one naively takes the same $R$ matrix \eqref{eq:R_IV1} as was used for the $\mathrm{Sl}(2)$-orbit approximation, one will find that it does not provide a solution. However, it is an interesting possibility that an appropriate dependence of $R$ on the parameter $\chi$ alleviates this issue, thus promoting the full nilpotent orbit approximation to a solution of the associated bi-Yang--Baxter model. If this is indeed the case, this would imply a remarkable connection between the $R$ matrix and the phase operator $\delta$. \\

\noindent At present, there is no concrete evidence that this will indeed be the case. There are, however, two indications. The first is that sometimes the nilpotent orbit and $\mathrm{Sl}(2)$-orbit are related in a rather simple manner. Indeed, it was found in \cite{Grimm:2021ikg} for the type $\mathrm{I}_1$ and $\mathrm{II}_0$ boundaries, that after appropriately resumming the corrections in the nilpotent orbit approximation, the two are related by a simple coordinate shift $y\mapsto y+y_0$, recall also the results of section \ref{reconstructing_examples}. Of course, such a shift will not spoil the solution. The second indication is the result of chapter \ref{chap:WZW}, in which it was shown that in fact the full Weil operator (hence also the nilpotent orbit approximation) solves the equations of motion of the $\lambda$-deformed $G/G$ model. Therefore, there is already an established relationship between objects appearing in Hodge theory and deformations of integrable non-linear $\sigma$-models. It appears plausible, then, that this relation runs deeper and also applies to the bi-Yang--Baxter model beyond just the $\mathrm{Sl}(2)$-orbit approximation. This is, however, still rather speculative and we hope to return to this question in future work.

\begin{subappendices}
\section{Overview of Formulae}
\label{app:computations}

In this section we have collected some formulae that are used in the computations of section \ref{subsec:sl2_orbits}. 

\subsubsection*{Action of $\mathrm{Ad}_h$ and $\mathrm{Ad}_{h^{-1}}$}
In the $\mathrm{Sl}(2)$-orbit approximation, the period map is given by
\begin{equation}
    h = e^{xN^-}y^{-\frac{1}{2}N^0}\,.
\end{equation}
Using the commutation relations \eqref{eq:sl2_algebra}, it follows by direct computation that
\begin{align}
\label{eq:Adh_N+}
    \mathrm{Ad}_h N^+ &= \frac{1}{y}N^+-\frac{x}{y}N^0-\frac{x^2}{y}N^-\,,\\
\label{eq:Adh_N0}
    \mathrm{Ad}_h N^0 &=N^0+2xN^-\,,\\
\label{eq:Adh_N-}
    \mathrm{Ad}_h N^- &= yN^-\,.
\end{align}
In a similar fashion, one finds
\begin{align}
\label{eq:Adhinv_N+}
    \mathrm{Ad}_{h^{-1}}N^+ &= yN^++xN^0-\frac{x^2}{y}N^-\,,\\
\label{eq:Adhinv_N0}
    \mathrm{Ad}_{h^{-1}}N^0 &= N^0-\frac{2x}{y}N^-\,,\\
\label{eq:Adhinv_N-}
    \mathrm{Ad}_{h^{-1}}N^- &= \frac{1}{y}N^-\,.
\end{align}

\subsubsection*{Action of $(-1)^{\mathrm{ad}\,Q_{\infty}}$}
It will be convenient to denote
\begin{equation}
    \mathcal{O}^\dagger = -(-1)^{\mathrm{ad}\,Q_{\infty}}\mathcal{O}\,.
\end{equation}
Then using the commutation relations \eqref{eq:sl2_algebra} and \eqref{eq:Q_commutation}, one finds
\begin{align}
\label{eq:dagger_N+}
    \left(N^+\right)^\dagger &= N^-\,,\\
\label{eq:dagger_N0}
    \left(N^0\right)^\dagger &= N^0\,,\\
\label{eq:dagger_N-}
    \left(N^-\right)^\dagger &= N^+\,.
\end{align}

\subsubsection*{Action of $\mathrm{Ad}_g$}
We recall that 
\begin{equation}
    g = h(-1)^{Q_{\infty}}h^{-1}\,,
\end{equation}
so that
\begin{equation}
    \mathrm{Ad}_g = -\mathrm{Ad}_h\circ \dagger\circ\mathrm{Ad}_{h^{-1}}\,.
\end{equation}
In order to compute the action of $\mathrm{Ad}_g$ on the $\mathfrak{sl}(2)$-triple, we can use our earlier results \eqref{eq:Adh_N+}-\eqref{eq:Adhinv_N-} as well as \eqref{eq:dagger_N+}-\eqref{eq:dagger_N-}.  Then one finds
\begin{align}
\label{eq:Adg_N+}
    \mathrm{Ad}_g N^+ &= \frac{1}{y^2}\left[x^2 N^+-x(x^2+y^2)N^0-(x^2+y^2)^2N^- \right]\,,\\
\label{eq:Adg_N0}
    \mathrm{Ad}_g N^0 &= \frac{2}{y^2}\left[x N^+-\left(x^2+\frac{y^2}{2}\right)N^0-x(x^2+y^2)N^- \right]\,,\\
\label{eq:Adg_N-}
    \mathrm{Ad}_g N^-&=-\frac{1}{y^2}\left[N^+-x N^0-x^2 N^- \right]\,.
\end{align}
Note that since $g^2=\pm 1$, we also have $\mathrm{Ad}_g=\mathrm{Ad}_{g^{-1}}$.

\section{$R$-matrix for type $\mathrm{IV}_1$}
\label{app:Rmat}

In this section we explicitly write down the $R$-matrix used in section \ref{subsec:example}, where the type $\mathrm{IV}_1$ Weil operator is discussed. Recall that $\mathfrak{sp}(4,\mathbb{R})$ consists of real $4\times 4$ matrices $X$ satisfying
\begin{equation}
    X^T\cdot S+S\cdot X=0\,,
\end{equation}
where $S$ is a non-singular skew-symmetric matrix and $T$ denotes the transpose. In this particular example we have made the following (non-standard) choice
\begin{equation}
    S = \begin{pmatrix}
    0 & 0 & 0 & -1\\
    0 & 0 & 3 & 0\\
    0 & -3 & 0 & 0\\
    1 & 0 & 0 & 0
    \end{pmatrix}\,.
\end{equation}
A basis of $\mathfrak{sp}(4,\mathbb{R})$ is given by
\begin{equation}
    T_k = \frac{1}{k!}\left(\mathrm{ad}\,N^+\right)^k \left(N^-\right)^3\,,\qquad k=0,\ldots, 6\,,
\end{equation}
together with
\begin{equation}
    T_7 = N^+\,,\qquad T_8 = N^0\,,\qquad T_{9}=N^-\,,
\end{equation}
with $N^+,N^0,N^-$ given in \eqref{eq:sl2-triple_IV1-paper}. Note that this particular embedding of $\mathfrak{sl}(2,\mathbb{R})$ into $\mathfrak{sp}(4,\mathbb{R})$ corresponds to the $\mathbf{10}=\mathbf{3}\oplus\mathbf{7}$ representation. \\

\noindent With the charge operator $Q_\infty$ in \eqref{eq:Q_IV1-paper} at hand, it is straightforward to define an $R$-matrix that furthermore satisfies the condition \eqref{eq:def_R_general}. Indeed, one simply demands that $R$ commutes with $Q_\infty$ and $L_0$ and acts as $\pm c$ on the eigenvectors of $\mathrm{ad}\,Q_\infty$ with positive/negative eigenvalues, respectively. In other words, we put
\begin{equation*}
    R\,Q_\infty=0\,,\qquad R\,L_0 = 0\,,\qquad R\,\mathcal{O}_q = c\, \mathrm{sign}(q)\mathcal{O}_q\,,\qquad [Q_\infty, \mathcal{O}_q] = q\,\mathcal{O}_q\,.
\end{equation*}
For this particular example, these equations have a unique solution if one furthermore demands that $R$ is anti-symmetric with respect to the Killing form on $\mathfrak{sp}(4,\mathbb{R})$, as is required in the bi-Yang--Baxter model. Of course, the resulting solution is essentially the Drinfel'd--Jimbo solution.\footnote{In principle, there is another Cartan generator besides $Q_\infty$, but one can check that taking this into account does not change the solution in this case.} In the basis of $\{T_k\}$ it reads
\begin{equation}
\label{eq:R_IV1}
    R=ic \left(
\begin{array}{cccccccccc}
 0 & \frac{5}{16} & 0 & \frac{1}{16} & 0 & \frac{1}{16} & 0 & 0 & 0 & 0 \\
 -\frac{15}{8} & 0 & \frac{3}{8} & 0 & \frac{1}{8} & 0 & \frac{3}{8} & 0 & 0 & 0 \\
 0 & -\frac{15}{16} & 0 & \frac{9}{16} & 0 & \frac{5}{16} & 0 & 0 & 0 & 0 \\
 -\frac{5}{4} & 0 & -\frac{3}{4} & 0 & \frac{3}{4} & 0 & \frac{5}{4} & 0 & 0 & 0 \\
 0 & -\frac{5}{16} & 0 & -\frac{9}{16} & 0 & \frac{15}{16} & 0 & 0 & 0 & 0 \\
 -\frac{3}{8} & 0 & -\frac{1}{8} & 0 & -\frac{3}{8} & 0 & \frac{15}{8} & 0 & 0 & 0 \\
 0 & -\frac{1}{16} & 0 & -\frac{1}{16} & 0 & -\frac{5}{16} & 0 & 0 & 0 & 0 \\
 0 & 0 & 0 & 0 & 0 & 0 & 0 & 0 & \frac{1}{2} & 0 \\
 0 & 0 & 0 & 0 & 0 & 0 & 0 & -1 & 0 & -1 \\
 0 & 0 & 0 & 0 & 0 & 0 & 0 & 0 & \frac{1}{2} & 0 \\
\end{array}
\right)\,.
\end{equation}
One may verify that \eqref{eq:R_IV1} satisfies the modified classical Yang--Baxter equation \eqref{eq:YBE}. Furthermore, one sees that when $c=i$ the $R$-matrix is indeed a real endomorphism of $\mathfrak{sp}(4,\mathbb{R})$. By examining the $3\times 3$ block on the bottom right-hand side, one verifies that it acts on the real $\mathfrak{sl}(2)$-triple as required by \eqref{eq:def_R_general}. In contrast, it acts on the remaining $7\times 7$ block in a more complicated fashion.
\end{subappendices}

\addchap{Summary and Outlook}
\subsection*{Summary}
In this thesis we have studied the beautiful framework of asymptotic Hodge theory, together with various applications in the context of the string landscape and the landscape of two-dimensional integrable field theories. We have shown how this mathematical framework allows for a general characterization of the asymptotic behaviour of physical couplings in low-energy effective theories that originate from (flux) compactification of type IIB/F-theory, and applied this knowledge to investigate the finiteness and geometric structure of the flux landscape. At the same time, we have found that the defining equations of variations of Hodge structure also arise in the context of certain integrable field theories, which opened the way to finding new classes of very general solutions to said models.\\

\noindent In part \ref{part1} of the thesis we have laid the groundwork by reviewing the basics of type IIB/F-theory flux compactifications. Here we started with a general worldsheet perspective and subsequently focussed on the broad class of supersymmetric non-linear $\sigma$-model backgrounds, with particular emphasis on type IIB string theory. We described its ten-dimensional low-energy effective description and emphasized the importance of including localized sources such as branes and orientifold planes in order to evade the Maldacena--Nu\~nez no-go theorem. Finally, we discussed the compactification of type IIB on Calabi--Yau threefolds, as well as the compactification of F-theory on elliptically fibered Calabi--Yau fourfolds, and characterized the resulting four-dimensional low-energy effective theories in terms of the geometry of the internal manifold. In doing so, we have found that important physical features of the vector multiplet sector such as the field space metric, gauge-kinetic coupling matrices and scalar potential, depend on the periods of the Calabi--Yau manifold. These, in turn, are well-described by the machinery of (asymptotic) Hodge theory. \\

\noindent In part \ref{part2} of the thesis we have discussed the mathematical machinery of asymptotic Hodge theory at great length. After a basic introduction to Hodge theory in chapter \ref{chap:Hodge}, employing perspectives of both period integrals, variations of Hodge structure, and finally the period map, our main goal was to describe the asymptotic behaviour of variations of Hodge structure near singular loci in the complex structure moduli space. To this end, we introduced a successive approximation scheme in chapter \ref{chap:asymp_Hodge_I} by reviewing the results of the nilpotent orbit theorem of Schmid, and the Sl(2)-orbit theorem of Cattani, Kaplan, and Schmid. The former describes the asymptotic form of a variation of Hodge structure up to exponential corrections, and naturally gives rise to a set of limiting structures that can be associated to the boundary of the moduli space: a web of limiting mixed Hodge structures. In turn, the latter theorem highlights how these limiting structures give rise to so-called horizontal $\mathfrak{sl}(2,\mathbb{R})$-triples that dictate the leading-order behaviour of physical couplings, when neglecting sub-polynomial corrections within a certain hierarchy. In chapter \ref{chap:asymp_Hodge_II} we delved even deeper into the mechanism of the Sl(2)-orbit theorem by describing a bulk reconstruction procedure with which the full multi-variable nilpotent orbit approximation can be recovered recursively from a simple set of boundary data associated to a given singularity. We carried out this procedure explicitly for all singularity types that arise in one-dimensional complex structure moduli spaces of Calabi--Yau threefold, and additionally worked out a two-modulus example to illustrate how the procedure generalizes to arbitrary multi-moduli limits. As a first application, we have shown how this allows one to compute general expansions for Hodge inner products, going beyond the Sl(2)-orbit approximation. This is of particular relevance when computing charge-to-mass ratios of BPS states,  arising from certain wrapped D3-branes, that have an asymptotically vanishing coupling to the graviphoton. \\

\noindent In part \ref{part3} of the thesis we applied the aforementioned expressions for the nilpotent orbit expansion of generic Hodge inner products to study the finiteness of the F-theory flux landscape. Using the quantization of the four-form flux, the self-duality condition, and the tadpole cancellation condition, we showed on general grounds that, within the nilpotent orbit approximation, it is impossible to have infinite tails of vacua that approach the boundary of the moduli space. In this way we provide an alternative point of view on the recent finiteness results of Bakker, Grimm, Schnell, and Tsimerman. Furthermore, motivated by recent advances in the field of o-minimal geometry and the theory of unlikely intersections, in the context of Hodge theory, we have proposed three conjectures which aim to address finer features of the flux landscape. These conjectures are concerned with the counting of vacua as well as the dimensionality of the vacuum loci, with the latter forming a mathematically formulated version of the well-known tadpole conjecture. Taken together, these conjectures severely constrain the F-theory flux landscape, suggesting that in fact the landscape of exact and isolated vacua could be significantly smaller than was first expected based on statistical arguments. \\

\noindent Finally, in part \ref{part4} of the thesis we have turned our attention towards a different landscape within the realm of physics: the landscape of two-dimensional integrable non-linear $\sigma$-models. This was motivated by the search for an auxiliary field theory, formulated on the complex structure moduli space, that encodes the information of a variation of Hodge structure in a more physical fashion. Somewhat surprisingly, we have in fact found multiple integrable field theories whose classical equations of motion admit solutions that are built from the Weil operator associated to a variation of Hodge structure. Notably, for the $\lambda$-deformed $G/G$ model discussed in chapter \ref{chap:WZW} we have found that, for a suitable ansatz and value of the deformation parameter $\lambda$, the equations of motion effectively reduce to the horizontality condition of the period map. In contrast, for the bi-Yang--Baxter model discussed in chapter \ref{chap:biYB} we have only managed to show that the Sl(2)-orbit approximation of the Weil operator solves the equations of motion of said model, by making use of the abstract properties of horizontal $\mathfrak{sl}(2)$-triples and identifying an appropriate $R$-matrix. Nevertheless, our findings indicate a further relation between Hodge theory and integrability beyond those observed before, which deserves further attention. This is especially interesting in light of the proposal of moduli space holography and the possible existence of a physical theory on the boundary of the moduli space. 

\subsection*{Outlook}
Let us conclude by highlighting some open questions which have been raised throughout this thesis, as well as a number of future research avenues that naturally follow from the results presented here. 

\subsubsection*{Systematics of sub-leading corrections and the Swampland}
Although there has been a tremendous effort in applying the results from asymptotic Hodge theory to test various Swampland conjectures, in the context of four-dimensional $\mathcal{N}=2$ and $\mathcal{N}=1$ low-energy effective theories arising from compactifications of the type II superstring theories and M-/F-theory, it is fair to say that a large portion of the results rely on a leading-order analysis using the Sl(2)-orbit approximation. One of the central results of this thesis is a concrete and algorithmic procedure to compute corrections to general Hodge norms/inner products, going far beyond the Sl(2)-orbit approximation. It would thus be interesting to use our results to perform a systematic study of the effects of these sub-leading corrections in the context of the Swampland program. For example, one could try to classify the allowed corrections to F-theory scalar potentials, and study possible implications for the de Sitter conjecture as well as the possibility to realize certain cosmological scenarios such as accelerated expansion. In a similar spirit, it may be insightful to revisit earlier works on the Swampland Distance Conjecture and the Weak Gravity Conjecture, and extend known results beyond the Sl(2)-orbit approximation. In this regard, we believe our novel strategy of partitioning the boundary region of the moduli space into finitely many special subsectors in order to control the path-dependence of Hodge norms, as introduced in part \ref{part3}, may play an important role.

\subsubsection*{Moduli space holography and coupling to gravity}
From a string-theoretic perspective, the main motivation for the analysis performed in part \ref{part4} of the thesis is the speculative proposal of ``moduli space holography'', in which the relevant features of a variation of Hodge structure are encoded in terms of an auxiliary field theory on the moduli space, which is furthermore supposed to be dual to a boundary theory that dynamically fixes the boundary data employed in the reconstruction of the period map.  While we have managed to write down concrete physical models on the moduli space that indeed encode the properties of a variation of Hodge structure in terms of the dynamics of a non-linear $\sigma$-model, there are still many puzzles regarding the proposed bulk-boundary
correspondence. First, we have not yet determined what this boundary theory exactly is. In this regard, it is probably crucial to
couple the theory to gravity. For the two-dimensional setting we considered, we recall
that Einstein gravity is of topological nature. This fact, together with the desire to fix the
background metric to be the Weil--Petersson metric, lead to the proposal of \cite{Grimm:2020cda} to
couple the action to an alternative gravitational theory in two dimensions known as JT-gravity \cite{Jackiw:1984je,Teitelboim:1983ux}. It would therefore be interesting to study integrable two-dimensional non-linear $\sigma$-models, such as the $\lambda$-model and the bi-Yang--Baxter model, and their coupling to gravity, as well as their corresponding boundary theories. Furthermore, in order to extend these models to more general settings, another immediate question is to find a generalization to higher-dimensional moduli spaces. 

\subsubsection*{Hodge theory and integrability}
Another question that arises is how much the concepts in Hodge theory underlie the study
of solutions to integrable non-linear $\sigma$-models in general. Indeed, it is now clear that the Weil
operator plays a special role in this regard, as it can be used to construct solutions to the $\lambda$-deformed G/G model in general, as well as for the critical bi-Yang–Baxter model when restricting to the Sl(2)-orbit approximation. Clearly, it would also be desirable to ascertain whether the latter can be generalized to the nilpotent orbit solution and beyond. It is expected that this can indeed be done, based on the fact that these two models are dual to each other via Poisson--Lie T-duality, although a precise mapping between the two solutions is expected to be very involved. It is likely that this relies on a deeper understanding of a potential relation between the R-matrix of the bi-Yang--Baxter model and the phase operator of the boundary Hodge structure, which is an intriguing question on its own that may shed light on the physical interpretation of the boundary data. More generally, it would be interesting to investigate how far our
proposed solutions coming from Hodge theory extend across the duality web of integrable field theories. Of perhaps even
greater importance is to establish the exact relevance of integrability in this regard. For example, it would be interesting to investigate the precise role of the infinite towers of conserved charges from a Hodge-theoretic perspective, and whether (from a string-theoretic point of view) their presence indicates a hidden symmetry of the associated low-energy effective theory. 

\subsubsection*{The structure and future of the string landscape}
Lastly, let us zoom out a bit and return to the main premise of the introduction of this thesis. The overarching purpose of our work has been to investigate certain corners of the string landscape, focusing in particular on the possible low-energy effective theories that can arise from string compactifications. It is rather striking that, even within the well-controlled and well-studied setting of type IIB/F-theory flux compactifications, it appears that the resulting mathematical structures underlying such theories may be more restrictive than first appreciated. Indeed, already the (previously established) finiteness of this part of the landscape is an incredibly non-trivial result, and now, through statements like the tadpole conjecture and the other conjectures we have proposed in part \ref{part3} of this thesis, it may very well be that the actual landscape is much smaller than expected. Clearly, one of the most exciting avenues for further study is to ascertain whether these conjectures are actually true. Ultimately, a possible underlying reason for all these observations may be the intricate interplay between the world of complicated transcendental equations that arise from geometry, and the world of integer quantities forced upon us by quantum mechanics. \\

\noindent As a last point, we would like to acknowledge and emphasize the fact that there are many more corners of string theory to explore, ranging from non-supersymmetric formulations of string theory to non-geometric backgrounds and beyond. At the same time, there are many fundamental aspects of string theory itself, such as a proper non-perturbative formulation as well as the issue of background independence, which still remain elusive. Thus, we believe that there is still plenty of reason to traverse the mountains and valleys of the string landscape, with an occasional dive into the swampland, in search of treasure and uncharted territory.

\addchap{Samenvatting}
In dit proefschrift hebben we het prachtige raamwerk van asymptotische Hodge-theorie bestudeerd, tezamen met verscheidene applicaties in de context van het snaarlandschap en het landschap van twee-dimensionale integreerbare veldentheorie\"en. We hebben laten zien hoe dit wiskundige raamwerk een algemene karakterisering mogelijk maakt van het asymptotische gedrag van fysische koppelingen in effectieve theorie\"en die voortkomen uit (flux) compactificatie van type IIB/F-theorie, en hebben deze kennis toegepast om de eindigheid en geometrische structuur van het fluxlandschap te onderzoeken. Tegelijkertijd hebben we ontdekt dat de bepalende vergelijkingen van de variaties van een Hodge-structuur ook voorkomen in de context van bepaalde integreerbare veldentheorie\"en, wat de weg opende voor het vinden van nieuwe klassen van zeer algemene oplossingen voor de genoemde modellen.\\

\noindent In deel \ref{part1} van dit proefschrift hebben we de fundamenten gelegd door de basisprincipes van type IIB/F-theorie fluxcompactificaties samen te vatten. Hier zijn we begonnen vanuit een algemeen perspectief van het snaarwereldvlak en hebben we ons vervolgens geconcentreerd op de brede klasse van supersymmetrische niet-lineaire $\sigma$-modelachtergronden, met bijzondere nadruk op type IIB-snaartheorie. We beschreven de tiendimensionale, effectieve beschrijving van de lage energie theorie en benadrukten het belang van het opnemen van gelokaliseerde bronnen zoals branen en orientivouwvlakken (Engels: orientifold planes) om de Maldacena--Nu\~nez no-go-stelling te omzeilen. Ten slotte bespraken we de compactificatie van type IIB op Calabi--Yau-drievari\"eteiten, evenals de compactificatie van F-theorie op elliptisch gevezelde Calabi--Yau-viervari\"eteiten, en karakteriseerden we de resulterende vierdimensionale effectieve theorie\"en in termen van de geometrie van de interne vari\"eteit. Daarbij hebben we ontdekt dat belangrijke fysische kenmerken van de vectormultipletsector, zoals de veldruimtemetriek, ijkkinetische koppelingsmatrices en scalaire potentiaal, afhankelijk zijn van de perioden van de Calabi--Yau-vari\"eteit. Deze worden op hun beurt goed beschreven door de machinerie van de (asymptotische) Hodge-theorie. \\

\noindent In deel \ref{part2} van dit proefschrift hebben we uitgebreid de wiskundige machinerie van de asymptotische Hodge-theorie besproken. Na een basisintroductie tot de Hodge-theorie in hoofdstuk \ref{chap:Hodge}, waarbij perspectieven van zowel periode-integralen, variaties van Hodge-structuur, en uiteindelijk de periode-afbeelding worden gebruikt, was ons hoofddoel om het asymptotische gedrag van variaties van Hodge-structuur nabij singuliere locaties in de complexe structuur moduli-ruimte te beschrijven. Hiervoor introduceerden we in hoofdstuk \ref{chap:asymp_Hodge_I} een opeenvolging van benaderingen door de resultaten van de nilpotente baan stelling van Schmid en de Sl(2)-baan stelling van Cattani, Kaplan, en Schmid te bespreken. De eerste beschrijft de asymptotische vorm van een variatie van Hodge-structuur tot op exponentiële correcties, en geeft op natuurlijke wijze aanleiding tot een reeks limiterende structuren die kunnen worden geassocieerd met de rand van de moduli-ruimte: een web van limiterende gemengde Hodge-structuren. De laatste stelling laat zien hoe deze limiterende structuren op hun beurt zogenaamde horizontale $\mathfrak{sl}(2,\mathbb{R})$-algebras voortbrengen die het leidende gedrag van fysische koppelingen beschrijven, wanneer sub-polynomische correcties binnen een bepaalde hi\"erarchie worden verwaarloosd. In hoofdstuk \ref{chap:asymp_Hodge_II} doken we nog dieper in het mechanisme van de Sl(2)-baan stelling door een procedure voor bulkreconstructie te beschrijven waarmee de volledige multi-variabele nilpotente baanbenadering recursief kan worden hersteld uit een eenvoudige set randgegevens die geassocieerd zijn met een gegeven singulariteit. We voerden deze procedure expliciet uit voor alle soorten singulariteiten die voorkomen in één-dimensionale complexe structuur moduli-ruimtes van Calabi-Yau-drievariëteiten, en werkten bovendien een voorbeeld uit met twee moduli om te illustreren hoe de procedure generaliseert naar willekeurige multi-moduli limieten. Als eerste toepassing hebben we laten zien hoe dit het mogelijk maakt om algemene reeksontwikkelingen voor Hodge-inproducten te berekenen, die verder gaan dan de Sl(2)-baan benadering. Dit is van bijzonder belang bij het berekenen van de lading-tot-massa verhoudingen van BPS-toestanden, die voortkomen uit bepaalde gewikkelde D3-branen, die een asymptotisch verdwijnende koppeling hebben met het gravifoton.\\

\noindent In deel \ref{part3} van het proefschrift pasten we de eerder genoemde uitdrukkingen voor de nilpotente baanbenadering van generieke Hodge-inproducten toe om de eindigheid van het F-theorie fluxlandschap te bestuderen. Door gebruik te maken van de kwantisatie van de vier-vormige flux, de zelf-dualiteitsvoorwaarde en de voorwaarde van de tadpole-annulering, toonden we op algemene gronden aan dat het binnen de nilpotente baanbenadering onmogelijk is om oneindige reeksen van vacua te hebben die de rand van de moduli-ruimte benaderen. Op deze manier bieden we een alternatief perspectief op de recente eindigheidsresultaten van Bakker, Grimm, Schnell en Tsimerman. Verder, gemotiveerd door recente ontwikkelingen op het gebied van o-minimale meetkunde en de theorie van onwaarschijnlijke doorsnedes, in de context van de Hodge-theorie, hebben we drie vermoedens voorgesteld die als doel hebben om fijnere kenmerken van het fluxlandschap te karakteriseren. Deze vermoedens hebben betrekking op het tellen van vacua en de dimensionaliteit van de vacuümloci, waarbij de laatste een wiskundig geformuleerde versie vormt van het bekende tadpole-vermoeden. Samen genomen beperken deze vermoedens het F-theorie fluxlandschap aanzienlijk, aangezien zij suggereren dat het landschap van exacte en geïsoleerde vacua in feite aanzienlijk kleiner zou kunnen zijn dan aanvankelijk werd verwacht op basis van statistische argumenten.\\

\noindent Tenslotte, in deel \ref{part4} van het proefschrift hebben we onze aandacht gericht op een ander landschap binnen het domein van de natuurkunde: het landschap van tweedimensionale integreerbare niet-lineaire $\sigma$-modellen. Dit werd gemotiveerd door de zoektocht naar een extra veldentheorie, geformuleerd op de complexe-structuur moduliruimte, die de informatie van een variatie van Hodge-structuur op een meer fysische manier codeert. Verassend genoeg hebben we in feite meerdere integreerbare veldtheorieën gevonden waarvan de klassieke bewegingsvergelijkingen oplossingen toelaten die zijn opgebouwd uit de Weil-operator geassocieerd met een variatie van Hodge-structuur. Met name voor het $\lambda$-gedeformeerde $G/G$-model, besproken in hoofdstuk \ref{chap:WZW}, hebben we gevonden dat, voor een geschikte ansatz en waarde van de vervormingsparameter $\lambda$, de bewegingsvergelijkingen effectief gereduceerd kunnen worden tot de horizontaaliteitsvoorwaarde van de periode-afbeelding. Daarentegen hebben we voor het bi-Yang--Baxter-model, besproken in hoofdstuk \ref{chap:biYB}, alleen kunnen laten zien dat de Sl(2)-baanbenadering van de Weil-operator de bewegingsvergelijkingen van het model oplost, door gebruik te maken van de abstracte eigenschappen van horizontale $\mathfrak{sl}(2)$-algebras en het identificeren van een geschikte $R$-matrix. Desalniettemin duiden onze bevindingen op een diepere relatie tussen de Hodge-theorie en integreerbaarheid dan voorheen is waargenomen, die verdere aandacht verdient. Dit is vooral interessant in het licht van het voorstel van moduliruimte holografie en het mogelijke bestaan van een fysische theorie op de rand van de moduliruimte.

%\appendix
%\subfile{appendices.tex}

\addchap{Acknowledgements}
With closing this thesis comes the great pleasure and privilege of thanking all the people who accompanied and supported me on this exciting, challenging, inspiring, and overall wonderful journey as a PhD student.  \\

\noindent Thomas, it has been over six years since I first contacted you as a young bachelor student looking for a supervisor. Since then, you have guided me through my bachelor thesis, master thesis, and now also my PhD. I deeply admire your ability to connect seemingly distant concepts, and your ambition in trying to unravel the deepest mathematical structures in string theory and beyond, while at the same time being a pleasant person to discuss and share ideas with. Thanks for everything; I have had an absolutely wonderful time as your student. I would also like thank Stefan for being my second promotor and for always livening up the conversations during lunch, as well as the discussions during talks and seminars. Let me also thank Erik here, for always being quick to give advice about careers matters, as well as for the occasional reminder to enjoy life as a student.  \\

\noindent Next, I would like to thank a few people who have been especially important to me during my time in Utrecht. Damian, you were my daily supervisor during my master thesis project, and subsequently my first collaborator. You have been a role model to me, and it was great to work with you. Mick, it has been an absolute pleasure to have you around as both a colleague, as well as a friend and fellow bouldering enjoyer. Lorenz, I am still amazed at the depth of your knowledge, and have learned so many things from you over the many chats we have had. I also greatly appreciate the movie nights you organized! Arno, you were the first student I supervised and I am happy you decided to join us as a PhD student thereafter. You were also my gym buddy for some time, which was a lot of fun (and served as a good motivation to fix my sleep schedule...). Stefano, although our project on black hole attractors is yet to be completed, I very much enjoyed learning about this new area of physics with you, and admire your enthusiasm in finding new things to learn and think about. \\
\newpage 

\noindent After spending the first one-and-a-half years of the PhD mostly at home due to the COVID-19 lockdowns, I have come to cherish the many small interactions at the institute even more. From the many discussions over lunch on the pinnacle of Dutch cuisine (a peanut butter sandwich), the emergence of AI, German politics, and, of course, tameness, to the canonical invasion of the coffee room afterwards, this could not have been possible without my many wonderful colleagues. In particular, I would like to thank: Brice, Casey, Cesar, Claire, Chongchuo, David, Edwan, Filippo, George, Guim,  Guoen, Javi, Nico, Ra\'ul, Shradha, Thorsten, Umut, and Wilke. Here I would also like to thank my other students Christoph, Hanneke, Lara, Lars, and Luuk whom I had the pleasure of supervising during their thesis projects.\\

\noindent Moving outside of Utrecht, I have had the privilege of meeting many great people at conferences and workshops all over the world. Though there are too many to name them all,  I would like to thank a few of them in particular: Adam, Bernardo, Bjoern, Bruno, Chad, Chris, Fien, Gonzalo, Joan, Marco, Mario, Muthu, Nicole, Rafa, Roberta, Simon, Thomas, and Vincent. \\

\noindent Going beyond academia, I would like to extend my sincerest appreciation to a few of the Eindhoven boys: Martijn, Sam, and Vlad. I am deeply thankful to Martijn and Vlad in particular, for being my closest friends for so many years, and for their genuine attempts at trying to understand this weird string theory stuff I keep telling them about. Additionally, I wish to thank Vlad again for designing the awesome cover art for this thesis. Also, despite the fact that I have not seen them as much as I would have liked these past years, I want to thank the other members of the Eindhoven boys: Brent, Clemens, and Max. Let me also thank Damian, Rauand, Robin, and Toon for joining and livening up many game nights over the weekends. \\

\noindent Some others I would like to thank for their company, as well as keeping the meme spam server alive, are my friends from my time as a bachelor/master student: Arjan, Cas, Elise, Erin, Fedor/Io, Koen, and Pepijn.\\

\noindent Another group of people I wish to thank are my fellow volunteers at Vierkant voor Wiskunde. You are some of the kindest and most accepting people I have ever met, and I am deeply grateful to you all for letting me be a part of this wonderful community. I want to especially thank Robin and Willeke for being such great company both during and outside of the yearly summer camps, and for making sure I get my yearly dose of laughter. \\

\noindent Finally, and most importantly, I wish to extend my sincerest gratitude to my parents and my sister, for their unconditional and unwavering love and support, their advice and wisdom, and for keeping an open mind when I chose to embark and continue on this journey. Thank you for everything. 

\addchap{About the author}
%\begin{wrapfigure}{r}{0.35\linewidth}
%\begin{center}
%\vspace*{-20pt}
%\includegraphics[width=0.35\textwidth]{photo.jpeg}
%\vspace*{-30pt}
%\end{center}
%\end{wrapfigure}
Jeroen Monnee was born on the 18th of October 1997 in Eindhoven. He obtained his high school diploma from the Stedelijk College Henegouwelaan in 2015. In that same year he started with a double bachelor's program in mathematics and physics at Utrecht University, which he completed cum laude in 2018. His bachelor's thesis on differential geometry in physics was supervised by Thomas Grimm. He then continued his studies with the master's program in theoretical physics, graduating cum laude in 2020. The research of his master's thesis focused on string theory, specifically the asymptotic distribution of flux vacua in type IIB/F-theory flux compactifications, and was completed under the supervision of Thomas Grimm. \\

\noindent After obtaining his master's degree he began as a PhD candidate advised by Thomas Grimm at the Institute for Theoretical Physics in Utrecht. His research focused on the study of asymptotic Hodge theory and its applications in the context of string compactifications and integrable systems, culminating in this doctoral thesis. He will continue as a postdoctoral researcher at both Ben-Gurion University of the Negev and the University of Hamburg.

\bibliographystyle{JHEP}
\bibliography{references.bib}

\backmatter

\end{document}